\documentclass[a4paper,twoside,10pt]{extbook}

\usepackage[latin1,utf8]{inputenc}
\usepackage[LGR,T1]{fontenc}
\usepackage[polutonikogreek,english,frenchb]{babel}%
\usepackage[sectionbib,square,numbers,sort&compress]{natbib}
\usepackage[centerlast,small]{caption}
\usepackage[curve,frame,cmtip,line,rotate,color,matrix,arrow,graph]{xy}
\usepackage[french]{minitoc}
\usepackage[a4paper]{geometry}
\usepackage[dvipsnames]{xcolor}
\usepackage{numprint,graphicx,eurosym,amssymb,amsmath,amstext,amsfonts,amsthm,%
longtable,setspace,chapterbib,multicol,array,multirow,color,expl3}%

\usepackage[linktocpage=true,linkbordercolor=cyan,citebordercolor=green,runbordercolor=Lavender]{hyperref}
\hypersetup{pdfauthor={Pierre de Thier (Université de Lorraine)},%
pdftitle={Adhésion des IgG sur une surface hydrophobe: Théorie, modélisations et application à l'ELISA.},%
pdfsubject={Physico-chimie des colloïdes, interfaces et protéines.},%
pdfkeywords={ELISA} {anticorps} {hydrophobie} {RSA} {relaxation} {AFM}}

\dominitoc
\makeatletter
\renewcommand{\fnum@figure}{\normalfont\bfseries\small{Figure~\thefigure}}
\addto\captionsenglish{}
\addto\captionsfrenchb{}
\renewcommand{\@evenhead}{\sffamily\small{\thepage \hfill \bfseries{\leftmark}}}
\renewcommand{\@oddhead}{\sffamily\small{\rightmark \hfill \thepage}}

\theoremstyle{definition}
\newtheorem{axiome}{Axiome}
\theoremstyle{definition}
\newtheorem{definition}{D\'{e}finition}
\theoremstyle{plain}
\newtheorem{proposition}{Proposition}
\newcommand{\dbar}{\ensuremath{\,\mathchar'26\mkern-12mu d}}
\makeatother

\title{\textbf{Adhésion des IgG}\\%
\textbf{sur une surface hydrophobe}\\%
Théorie, modélisations et application à l'ELISA}
\author{\textbf{Pierre de Thier}}
\date{13 mars 2015}

\begin{document}
\frontmatter
\newgeometry{left=2cm,right=2cm,bottom=3cm,top=3cm}

\thispagestyle{empty}

\begin{titlepage}
\begin{center}

\centerline{
\begin{tabular}{p{8.5cm}p{0.05cm}|p{0.05cm}p{8.5cm}}
                                                                                     &&& \multirow{6}*{\raisebox{-0.95cm}[0pt][0pt]{\includegraphics*[width=6cm]{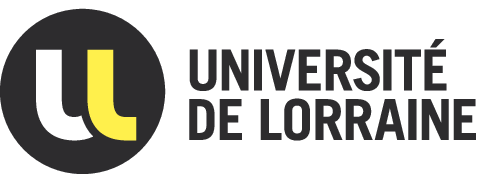}}} \\
\raggedleft{\begin{large}{\sffamily\textbf{\'{E}COLE DOCTORALE SESAMES}}\end{large}} &&& \\
                                                                                     &&& \\
\raggedleft{\begin{Large}{\sffamily LCPME --- UMR 7564 du CNRS}\end{Large}}          &&& \\
\raggedleft{\begin{Large}{\sffamily LEMTA --- UMR 7563 du CNRS}\end{Large}}          &&& \\
                                                                                     &&& \\
\end{tabular}}

\vspace*{4cm}
\begin{LARGE}
\textbf{Adhésion des IgG}\\
\vspace*{0.5mm}
\textbf{sur une surface hydrophobe}\\
\vspace*{0.5mm}
Théorie, modélisations et application à l'ELISA\\
\end{LARGE}

\vspace*{2.5cm}
\begin{large}Mémoire soutenu par\end{large}\\
\vspace*{5mm}
\begin{Large}\textbf{Pierre de Thier}\end{Large}\\
\vspace*{5mm}
\begin{large}en vue de l'obtention du grade de\end{large}\\
\vspace*{5mm}
\begin{large}\textbf{Docteur en Chimie}\end{large}\\
\vspace*{1.5cm}
\begin{large}devant le jury composé de\end{large}\\
\vspace*{1.4cm}

\centerline{
\begin{tabular}{>{\raggedleft}p{8.5cm}p{8.5cm}}
Dr G. Francius (HDR, dir. de thèse):       &LCPME (UMR 7564), Nancy          \\
Dr S. Skali-Lami (HDR, co-dir. de thèse):  &LEMTA (UMR 7563), Nancy          \\
Dr J.F.L. Duval (HDR):             &LIEC (UMR 7360), Nancy           \\
Dr J. Landoulsi:              &UPMC (Paris VI), Paris            \\
Pr C. Dupont-Gillain:        &UCLouvain, Louvain-la-Neuve  \\
Dr B. Senger (HDR):        &INSERM (UMRS 1121), Strasbourg  \\
\end{tabular}}

\vspace*{1.4cm}
\begin{large}Nancy\end{large}\\
\vspace*{0.5mm}
\begin{large}{\textbf{13 mars 2015}}\end{large}
\end{center}

\end{titlepage}

\cleardoublepage\thispagestyle{empty}
\

\restoregeometry

\maketitle
\chapter*{Remerciements}
\markboth{}{}

Ce travail a été réalisé grâce à l'accueil désintéressé qui m'a été réservé au sein du Laboratoire de Chimie Physique et Microbiologie pour l'Environnement (UMR 7564 du CNRS) à Nancy. C'est donc avec beaucoup de reconnaissance que j'adresse mes remerciements à M. Alain \textsc{Walcarius}, directeur du LCPME. \`{A} ce titre, mes remerciements s'adressent aussi à M. Xavier \textsc{Assfeld}, directeur de l'\'{E}cole doctorale SESAMES (ED 412), pour la confiance qui m'a été accordée.

\vspace*{0.75mm}

S'il m'a été possible d'arriver jusqu'au bout de ce parcours doctoral, c'est bien grâce à la générosité et à la bienveillance de MM. Grégory \textsc{Francius} (LCPME) et Salaheddine \textsc{Skali-Lami} (LEMTA) qui ont accepté la direction et la co-direction de ma thèse. Je tiens tout particulièrement à remercier M. Grégory \textsc{Francius} pour son encadrement, son ouverture d'esprit et la mise à disposition du matériel nécessaire à la réalisation du travail de recherche lors de mon séjour au LCPME.
\chapter*{Résumé}
\markboth{R\'{E}SUM\'{E}}{}
\addstarredchapter{Résumé}

Les ELISA (\textit{Enzyme-Linked ImmunoSorbent Assay}) sont parmi les protocoles d'analyses biochimiques les plus couramment utilisés dans la recherche et les technologies (bio)médicales. Ces tests permettent en effet la détection et le dosage d'une espèce antigènique contenue dans un fluide d'origine biologique en tirant parti de la spécificité antigène-anticorps. Typiquement, les ELISA sont réalisés dans les puits de plaques multipuits en polystyrène, polymère particulièrement hydrophobe, ce qui le dote d'une forte capacité à fixer les protéines. L'ELISA consiste à fonctionnaliser la surface des puits en y faisant adhérer un film d'anticorps spécifiques de l'antigène à détecter/doser. Ces antigènes pourront alors être captés et immobilisés sur la surface et l'on pourra ensuite quantifier cette fixation grâce à des anticorps conjugués à une enzyme catalysant une réaction produisant une espèce chromogénique.

L'étape la plus importante de l'ELISA consiste donc à construire un film d'anticorps sur la surface des puits en polystyrène. Afin que le film d'anticorps puisse immobiliser les antigènes, il devra contenir des anticorps intacts, c'est-à-dire non ou peu dénaturés; les anticorps le constituant devront être, du moins partiellement, convenablement orientés afin d'exposer les parties variables (F(ab')$_2$) vers la suspension contenant les antigènes, c'est-à-dire que les anticorps devront adhérer sur la surface par leurs parties constantes (Fc) et, finalement, les anticorps devront adhérer de manière suffisamment forte à la surface afin de ne pas en être décollés lors des étapes suivantes du protocole ELISA.

L'amélioration des protocoles ELISA peut donc passer par l'étude et l'optimisation de la formation des films d'anticorps sur les surfaces hydrophobes. Le présent travail se donne dès lors pour objectif de répondre aux questions: \emph{Quel est le taux de dénaturation des anticorps fixés sur la surface?} \emph{Quelle est l'orientation préférentielle des anticorps sur la surface?} \emph{Les anticorps adhèrent-ils suffisamment sur la surface?} Dans un premier temps, les anticorps doivent nécessairement atteindre la surface sur laquelle ils viendront adhérer. Les aspects hydrodynamique et diffusionnel présents dans la littérature tendent à montrer que les anticorps devraient arriver à la surface selon une orientation de type \textit{end-on}, orientation recherchée dans le cadre de l'ELISA où l'anticorps présente sa partie constante vers la surface et ses parties variables vers la suspension antigènique.

Ce n'est toutefois pas parce que les anticorps sont supposés se présenter à la surface dans une orientation de type \textit{end-on} qu'ils y conserveront cette orientation une fois fixés. Le thermodynamique des processus irréversibles vient alors expliciter les raisons physico-chimiques de l'adhésion des anticorps sur la surface hydrophobe et, ce faisant, l'orientation préférentielle qu'ils devront \textit{a priori} y adopter. Les anticorps, et les protéines en général, sont entourés d'une enveloppe d'espèces chimiques adsorbées depuis le milieu liquide qui les suspend (molécules du solvant et du cosolvant) en fonction de leur caractère hydrophile/hydrophobe. Il en est de même pour la surface solide sur laquelle ils devront venir adhérer. Le développement du terme de production d'entropie dans le cadre de la thermodynamique des processus irréversibles montre que les anticorps devraient venir spontanément adhérer sur le polystyrène car la création d'une aire de contact entre eux nécessite l'expulsion des espèces adsorbées sur leurs surfaces, facteur produisant de l'entropie. De plus, l'adhésion pouvant être subdivisée en deux phénomènes, addition (l'anticorps vient adhérer sur le polystyrène tel qu'il s'y présente) et relaxation (l'anticorps optimise son aire de contact avec le polystyrène en changeant d'orientation et/ou de conformation), l'anticorps devrait s'additionner et se relaxer spontanément dès qu'il arrive à la surface. Arrivant préférentiellement selon une orientation \textit{end-on}, l'aire de contact qu'il développera avec la surface ne sera pas très élevée et il aura alors tendance à relaxer spontanément afin d'augmenter cette aire de contact afin d'adopter des orientations de types \textit{side-on}, \textit{flat} ou se dénaturer (\textit{flat$+$}). L'orientation \textit{flat} et l'état de dénaturation \textit{flat$+$} maximise cette aire de contact, laissant à penser que l'anticorps devrait spontanément tendre à l'adopter sur la surface. Cette orientation \textit{flat} n'est toutefois pas de nature à rencontrer les conditions posées dans le cadre de l'ELISA puisque les parties variables de l'anticorps deviennent indisponibles.

La thermodynamique explique la tendance spontanée des anticorps à s'additionner et se relaxer sur la surface mais n'aborde pas les effets d'ordres cinétiques dus au remplissage progressif de la monocouche. Afin de comprendre les conséquences du remplissage sur la structure de la monocouche, le modèle des additions séquentielles aléatoires (RSA) est mis à profit grâce à l'environnement Matlab. Le modèle RSA montre qu'il existe un effet d'exclusion par la taille pour les anticorps venant s'additionner à la monocouche. Cet effet d'exclusion a pour conséquence de sélectionner, au fur et à mesure du remplissage, les anticorps ayant des orientations prenant moins de place dans la monocouche. De ce fait, l'orientation \textit{end-on} sera de plus en plus favorisée au cours du remplissage et ce sont les anticorps ayant cette orientation qui viendront saturer la surface. L'effet d'exclusion par la taille mis en lumière par les simulations RSA permet dès lors d'expliquer la présence naturelle des anticorps \textit{end-on} dans la monocouche saturée.

Le modèle RSA ne tenant compte que du seul phénomène d'addition, il est étendu à un modèle des additions et relaxations séquentielles aléatoires (RSA+R) tenant compte, comme son nom l'indique, du phénomène de relaxation, et ce, de telle sorte qu'il est possible d'en préciser l'intensité. Les simulations RSA+R réalisées sur un système de boîtes de conserve ayant la capacité de relaxer leur conformation en s'aplatissant sur la surface montrent qu'il est possible d'obtenir des monocouches saturées contenant des nombres variables de boîtes de conserve. Non seulement le nombre total de boîtes de conserve varie mais la fraction de boîtes de conserve demeurées dans un état natif est variable. En extrapolant ce raisonnement aux anticorps et en constatant que le paramètre qui est utilisé est fonction de la concentration en anticorps dans la suspension de dépôt, il apparaît que, plus la concentration est élevée, plus les monocouches seront densément peuplées et que les anticorps qu'elles contiennent présenteront majoritairement une orientation \textit{end-on}. Inversement, lorsque la concentration est faible, les monocouches seront pauvrement peuplées et les anticorps présenteront plus rarement l'orientation \textit{end-on}. La raison de cette observation tient au fait que, plus la concentration est élevée, plus la vitesse à laquelle les anticorps arriveront à la surface sera élevée, ce qui provoquera une inhibition du phénomène de relaxation par l'addition.

Des mesures expérimentales réalisées par AFM et ELISA permettent de corroborer les résultats théoriques et ceux des modélisations RSA+R. Deux anticorps monoclonaux de souris anti-interleukine-2 humaine (mauvaises capacités de fixation sur les surfaces) et anti-interleukine-6 humaine (très bonnes capacités de fixation sur les surfaces) sont utilisés afin de construire des séries de monocouches selon plusieurs concentrations dans la suspension de dépôt. Les mesures des épaisseurs et des activités immunologiques de ces monocouches montrent pour l'anti-IL-6 que, plus la concentration est élevée, plus la monocouche est épaisse et active du point de vue immunologique, alors même qu'elle demeure latéralement homogène et assez lisse indépendamment de la concentration. Pour cet anticorps, la concentration dans la suspension semble donc favoriser les monocouches peuplées en anticorps \textit{end-on}, induisant une épaisseur plus élevée et une activité immunologique plus importante. Ceci tend à démontrer le réalisme des modélisations RSA+R réalisées mais ne peut expliquer le comportement d'adhésion moins favorable à l'ELISA de l'anti-IL-2.

En conclusion, ce travail montre comment l'effet hydrophobe pousse les anticorps à adhérer, c'est-à-dire s'additionner et se relaxer, spontanément sur une surface de polystyrène. Pour les effets cinétiques, le modèle RSA montre premièrement pourquoi les anticorps d'orientation \textit{end-on} finiront par être favorisés, et ce, malgré les implications de la thermodynamique. Deuxièmement, le modèle RSA+R met en lumière l'interdépendance entre l'augmentation de la concentration d'anticorps dans la suspension de dépôt et l'augmentation de la fraction d'anticorps \textit{end-on} que contiendra la monocouche saturée. Ces liens semblent finalement être corroborés, pour certains anticorps, par les résultats expérimentaux, permettant dès lors d'envisager une meilleure maîtrise des techniques de dosage par ELISA.\\

\vspace*{2cm}
\noindent\textbf{Mots-clés:} ELISA, anticorps, hydrophobie, RSA, relaxation \& AFM.
\chapter*{Avant-propos}
\markboth{AVANT-PROPOS}{}
\addstarredchapter{Avant-propos}

Les ELISA (\textit{Enzyme-Linked ImmunoSorbent Assay}) sont des protocoles analytiques faisant partie de la grande famille des techniques immunologiques et demeurent, par leur efficacité, d'un usage particulièrement courant dans la recherche et l'industrie (bio)médicale. De manière générale, ils permettent la détection et/ou le dosage d'une espèce antigènique présente dans un fluide \textit{via} une cascade d'interactions permettant \textit{in fine} une mesure spectroscopique du produit d'une réaction catalysée par une enzyme, mesure dont l'intensité devra être corrélée à la quantité d'antigènes d'intérêt. Certaines techniques ELISA nécessitent la construction d'un film d'anticorps (anticorps spécifiques d'un autre anticorps ou de l'antigène à détecter) sur les parois d'un puits d'une plaque multipuits en polystyrène. Grâce à son caractère fortement hydrophobe, le polystyrène est un substrat sur lequel les anticorps viendront spontanément se fixer et y créer un film, film dont les propriétés devront rendre possibles les étapes ultérieures du protocole d'analyse ELISA. Cette étape de fixation et de construction du film d'anticorps sur le substrat étant critique pour l'efficacité du test, elle a engendré un volume particulièrement important de travaux, lesquels visent essentiellement à caractériser les films formés par certains types d'anticorps ou encore à élaborer de nouveaux protocoles permettant de rendre plus efficiente cette étape de fixation. Or, fort souvent, on doit remarquer que ces protocoles inspirés par les techniques nanobiotechnologiques engendrent de très nombreuses étapes intermédiaires, parfois très contraignantes, rendant alors les possibilités d'un \textit{scaling-up} assez hasardeuses.

De surcroît, les approches utilisées afin d'améliorer cette étape de formation du film d'anticorps sur le substrat hydrophobe demeurent très empiriques voire, pour certaines, peu efficaces. De nombreuses questions restent sans réponse, en particulier celles visant à une meilleure compréhension des mécanismes de formation d'un tel film. Or, une meilleure compréhension de ces mécanismes pourrait très certainement faciliter les recherches expérimentales directement orientées vers l'amélioration des techniques ELISA en ce qu'elle permettrait de cibler plus précisément les paramètres à ajuster. Ce travail ne se donne dès lors pas pour objectif d'énoncer de façon équivoque \textit{la} bonne manière de réaliser un ELISA mais d'offrir, par une meilleure compréhension des mécanismes et concepts fondamentaux à l'{\oe}uvre lors de la construction du film, une série de pistes de réflexions au chercheur désireux d'améliorer concrètement les techniques ELISA. De même, les concepts qui seront développés pourraient aussi être appliqués lors d'études plus fondamentales des films de protéines globulaires. Au cours de ce travail, une proposition de mécanisme menant à la formation d'un film d'anticorps sur un substrat hydrophobe sera dès lors élaborée en faisant référence aux propriétés attendues d'un tel film dans le cadre de l'ELISA, propriétés qui seront énoncées dans l'introduction. Bien que de nombreuses pistes se trouvent déjà exposées dans la littérature, le mécanisme général qui sera élaboré ne semble pas encore avoir été proposé.

\`{A} cette fin, le présent travail est structuré en quatre chapitres. Les trois premiers visent à développer, à partir des connaissances actuelles, une théorie explicative sur la formation des monocouches d'IgG. Des simulations numériques y sont associées afin d'en illustrer les potentialités. Partant de cette tentative de théorisation et des simulations associées, le quatrième chapitre illustre le lien entre le modèle développé et l'expérience.

Dans le \textbf{Chapitre \ref{Chap1}}, les notions de la thermodynamique chimique sont développées afin de tenir compte des transformations irréversibles. Ces notions permettent de conceptualiser d'un point de vue physico-chimique le phénomène d'adhésion hydrophobe entre les IgG et la surface de polystyrène. Ensuite, la relation obtenue est étendue pour rendre compte de différents phénomènes tels que l'addition simple d'une IgG sur la surface et sa relaxation. Ainsi, la thermodynamique permet d'expliciter les conditions expérimentales qui mènent les IgG, protéines hydrophiles, à venir spontanément adhérer à une surface hydrophobe. Ce phénomène est à la base de la formation d'une monocouche (ou film) complète de ces IgG sur une telle surface.

Les aspects cinétiques sont souvent considérés comme des facteurs limitant ceux d'origine thermodynamique. Afin de tenir compte de ceux pouvant influencer la formation des monocouches d'IgG, le \textbf{Chapitre \ref{SectionRSA}} introduit le modèle des additions séquentielles aléatoires (RSA). Le modèle RSA, grâce aux hypothèses sur lesquelles il se fonde, permet de construire \textit{in silico} des monocouches d'IgG et de <<~boîtes de conserve~>> en tentant de les additionner séquentiellement et aléatoirement sur la surface. Ces monocouches sont construites en tenant compte du volume exclu, notion conceptualisant le remplissage progressif et donc la difficulté croissante d'additionner une nouvelle IgG. Ce faisant, la croissance du volume exclu donne une première explication à la présence d'IgG \textit{end-on} (orientation recherchée dans le cadre de l'ELISA) dans les monocouches. Un lien entre le transport par diffusion des IgG vers la surface et le modèle RSA est discuté par la suite.

Le \textbf{Chapitre \ref{ChapIsoth}} est une synthèse des deux premiers car il développe un modèle des additions \emph{et} relaxations séquentielles aléatoires (RSA+R). Ce modèle, vu comme une extension du modèle RSA en y incluant le phénomène de relaxation des protéines, permet de construire des monocouches en faisant varier de manière extrêmement précise l'intensité du phénomène de relaxation (la quantité d'IgG relaxant) par rapport à l'addition. La proportion d'IgG ayant pu relaxer sera liée à leur concentration dans la solution de dépôt et mène à l'obtention de courbes comparables aux <<~isothermes d'adsorption~>> des protéines. De manière générale, la gestion qu'offre le modèle RSA+R des deux phénomènes que sont l'addition et la relaxation permet de construire des monocouches plus ou moins riches en protéines relaxées/natives.

Dans le \textbf{Chapitre \ref{ChapExp}}, les notions théoriques développées dans les trois premiers chapitres sont confrontées à une série d'estimations expérimentales fournies par la technique ELISA et la microscopie à force atomique (AFM). Des monocouches d'IgG sont construites sur des substrats hydrophobes (h-SAMs) pour diverses concentrations dans la suspension de dépôt et des propriétés telles que leur activité immunologique (obtenue par ELISA) ou leur épaisseur (obtenue par AFM) peuvent être évaluées. Ces résultats expérimentaux sont discutés compte tenu du modèle théorique développé dans les trois premiers chapitres. Les aspects expérimentaux de ce dernier chapitre ont fait l'objet d'une publication\footnote{P. de \textsc{Thier}, J. \textsc{Bacharouche}, J.F.L. \textsc{Duval}, S. \textsc{Skali-Lami} et G. \textsc{Francius}, \textit{Atomic force microscopy analysis of IgG films at hydrophobic surfaces: a promising method to probe IgG orientations and optimise ELISA tests performance.} BBA -- Proteins and Proteomics, 2015, 1854(2), 138-145. \href{http://dx.doi.org/10.1016/j.bbapap.2014.12.001}{doi:10.1016/j.bbapap.2014.12.001}.} qui est présentée \textit{in extenso} à l'\textbf{Annexe \ref{Ann4}}.

Après la \textbf{Conclusion}, les \textbf{Annexes \ref{AnnRSA}} et \textbf{\ref{Ann2}} détaillent la structure des algorithmes écrits dans l'environnement Matlab afin de modéliser, selon les hypothèses des modèles RSA et RSA+R, la construction des diverses monocouches d'IgG et de boîtes de conserve, modélisations ayant servi à illustrer le propos. Quelques codes Matlab importants sont présentés à l'\textbf{Annexe \ref{Ann3}} de ce travail.

\begin{flushright}
\textit{Pierre de Thier}\\
\textit{Nancy, mars 2015.}\\
\end{flushright}

\setcounter{tocdepth}{1}
\tableofcontents
\markboth{TABLE DES MATI\`{E}RES}{}
\addstarredchapter{Table des matières}
\chapter*{Notations}
\markboth{NOTATIONS}{}
\addstarredchapter{Notations}

\begin{longtable}{r@{\quad :\quad}p{10cm}}
$A_r$ & affinité du $r$-ième processus physico-chimique (J$\cdot$mol$^{-1}$)\\
$\mathrm{A}_\mathrm{pls}$ & constante de Hamaker dans un système à trois corps <<~p~>>, <<~l~>> et <<~s~>> (J)\\
$C$ & concentration molaire d'un colloïde dans la suspension de dépôt (mol$\cdot$m$^{-3}$)\\
$C_j$ & concentration molaire d'une espèce chimique $j$ (mol$\cdot$m$^{-3}$)\\
$\mathrm{d}$ & opérateur de différentiation totale (ou <<~différentielle exacte~>>)\\
$\mathrm{\dbar}$ & quantité élémentaire (ou <<~différentielle inexacte~>>)\\
$\mathrm{d}_x$ & opérateur de différentiation partielle (sur la variable $x$)\\
$D$ ou $\mathbf{D}$ & coefficient de diffusion d'une espèce chimique ou colloïdale (m$^2\cdot$s$^{-1}$)\\
$\frac{\partial}{\partial x}$ & opérateur de dérivation partielle\\
$\nabla$ & opérateur gradient\\
$\nabla\cdot$ & opérateur divergence\\
$\nabla^2$ & opérateur laplacien (divergence d'un gradient)\\
$E$ & énergie interne d'un système thermodynamique (J)\\
$E^\ddagger_A$ & énergie d'activation de l'addition (J)\\
$E^\ddagger_R$ & énergie d'activation de la relaxation (J)\\
$F$ & force exercée sur un corps (N)\\
$G$ & enthalpie libre d'un système thermodynamique (J)\\
$\gamma_{\alpha\mathrm{l}}$ & tension superficielle d'une interface <<~$\alpha\mathrm{l}$~>> (J$\cdot$m$^{-2}$)\\
$\Gamma_{j(\alpha\mathrm{l})}$ & adsorption de l'espèce chimique $j$ à l'interface <<~$\alpha\mathrm{l}$~>> (mol$\cdot$m$^{-2}$)\\
$H$ & enthalpie d'un système thermodynamique (J)\\
$\mathcal{H}_{\alpha\mathrm{l}\cdot\mathrm{l}}$ & hydrophobie d'un corps <<~$\alpha$~>> suspendu dans un liquide <<~l~>> (J$\cdot$m$^{-2}$)\\
$\mathcal{H}_{\mathrm{l}\cdot\alpha\mathrm{l}}$ & hydrophilie d'un corps <<~$\alpha$~>> suspendu dans un liquide <<~l~>> (J$\cdot$m$^{-2}$)\\
$\eta$ & viscosité d'un fluide monophasique (Pa$\cdot$s)\\
$I$ & force ionique d'une solution (mol$\cdot$L$^{-1}$)\\
$j$ ou $\mathbf{j}$ & vecteur densité de courant des espèces chimiques ou colloïdales (mol$\cdot$m$^{-2}\cdot$s$^{-1}$)\\
$k_A^\circ$ & vitesse initiale d'addition des colloïdes à la monocouche (mol$\cdot$m$^{-2}\cdot$s$^{-1}$)\\
$k_A$ & vitesse instantanée d'addition des colloïdes à la monocouche (mol$\cdot$m$^{-2}\cdot$s$^{-1}$)\\
$k_B$ & constante de Boltzmann (J$\cdot$K$^{-1}$)\\
$k_{ex}$ & vitesse instantanée d'exclusion (interdiction d'addition) des colloïdes de la monocouche (mol$\cdot$m$^{-2}\cdot$s$^{-1}$)\\
$k_R^\circ$ & vitesse initiale de relaxation des colloïdes dans la monocouche (mol$\cdot$m$^{-2}\cdot$s$^{-1}$)\\
$k_R$ & vitesse instantanée de relaxation des colloïdes dans la monocouche (mol$\cdot$m$^{-2}\cdot$s$^{-1}$)\\
$k_\phi$ & vitesse instantanée de recouvrement de la surface (mol$\cdot$m$^{-2}\cdot$s$^{-1}$)\\
$K^\circ$ & ratio des vitesses initiales d'addition et de relaxation\\
$K$ & ratio des vitesses instantanées d'addition et de relaxation\\
$\kappa^{-1}$ & longueur de Debye (m)\\
$\mathcal{L}_{\alpha\mathrm{l}\cdot\mathrm{l}}$ & lyophobie d'un corps <<~$\alpha$~>> suspendu dans un liquide <<~l~>> (J$\cdot$m$^{-2}$)\\
$\mathcal{L}_{\mathrm{l}\cdot\alpha\mathrm{l}}$ & lyophilie d'un corps <<~$\alpha$~>> suspendu dans un liquide <<~l~>> (J$\cdot$m$^{-2}$)\\
$\ln$ & opérateur du logarithme naturel\\
$M$ ou $\mathbf{M}$ & mobilité d'une espèce chimique ou colloïdale (m$^2\cdot$s$^{-1}\cdot$J$^{-1}$)\\
$\mu_j$ & potentiel chimique de l'espèce chimique $j$ (J$\cdot$mol$^{-1}$)\\
$n_j$ & quantité molaire d'une espèces chimique $j$ (mol)\\
$\mathcal{N}$ & quantité molaire d'un colloïde (mol)\\
$p$ & pression (Pa ou J$\cdot$m$^{-3}$)\\
$P$ & fonction de production d'entropie superficielle (J$\cdot$K$^{-1}\cdot$m$^{-2}\cdot$s$^{-1}$)\\
$P(A)$ & probabilité que l'événement <<~$A$~>> se réalise\\
$P(A^\ast)$ & probabilité que l'événement <<~$A$~>> ne se réalise pas\\
$\phi$ & taux de recouvrement d'une surface solide\\
$\Phi$ & fraction volumique d'un système occupée par une phase suspendue dispersée ou d'un seul tenant\\
$\psi_\alpha$ & potentiel électrostatique de la surface d'un corps <<~$\alpha$~>> (V ou mV)\\
$Pe$ & nombre adimensionnel de Péclet (transfert par diffusion/transfert massique)\\
$\boldsymbol{q}$ & pseudovecteur rassemblant $\boldsymbol{x}$ et $\boldsymbol{\omega}$\\
$Q$ & chaleur échangée par un système thermodynamique (J)\\
$Q^\prime$ & chaleur non compensée de Clausius (J)\\
$\rho$ & masse volumique d'un corps (kg$\cdot$m$^{-3}$)\\
$R$ & rayon d'un corps sphérique (m)\\
$s$ & entropie superficielle (J$\cdot$K$^{-1}\cdot$m$^{-2}$)\\
$S$ & entropie (J$\cdot$K$^{-1}$)\\
$\sigma$ & empreinte d'une phase colloïdale sur une surface solide (m$^2\cdot$mol$^{-1}$)\\
$\sigma(\boldsymbol{\omega})$ & empreinte caractéristique pour une orientation-conformation particulière ($\boldsymbol{\omega}$) d'un colloïde (m$^2\cdot$mol$^{-1}$)\\
$\Sigma_{\alpha\mathrm{l}}$ & aire d'une interface $\alpha\mathrm{l}$ entre deux corps (m$^2$)\\
$\Sigma_\mathrm{s}$ & aire d'une surface solide (m$^2$)\\
$Sc$ & nombre adimensionnel de Schmidt (viscosité/transfert par diffusion)\\
$Sh$ & nombre adimensionnel de Sherwood (transfert massique/transfert par diffusion)\\
$t$ & temps (s)\\
$T$ & température (K)\\
$\Theta$ & quantité molaire de colloïdes accumulés dans une monocouche (mol$\cdot$m$^{-2}$ ou pmol$\cdot$cm$^{-2}$)\\
$\Theta_\infty$ & quantité molaire de colloïdes accumulés dans une monocouche saturée (mol$\cdot$m$^{-2}$ ou pmol$\cdot$cm$^{-2}$)\\
$\mathbf{u}$ & pour un fluide ou un solide, vitesse d'un point matériel (m$\cdot$s$^{-1}$)\\
$U$ & énergie potentielle (J)\\
$V$ & volume d'un corps (m$^3$)\\
$W$ & travail exercé sur un système thermodynamique (J)\\
$\boldsymbol{\omega}$ & pseudovecteur des coordonnées angulaires eulériennes $\vartheta$ ,$\varphi$ et $\psi$. Par extension, il s'agit de toutes les coordonnées permettant de décrire l'orientation et la conformation d'une protéine.\\
$\boldsymbol{x}$ & pseudovecteur des coordonnées cartésiennes $x$, $y$ et $z$\\
$\mathcal{X}$ & fraction molaire d'un colloïde\\
$\xi_r$ & degré d'avancement du $r$-ième processus physico-chimique (mol)\\
\end{longtable}

\mainmatter
\begin{cbunit}
\chapter*{Introduction}
\addstarredchapter{Introduction}
\minitoc

\section*{Les colloïdes}
\addcontentsline{toc}{section}{Les colloïdes}
\markboth{INTRODUCTION}{LES COLLOÏDES}

Un nuage de brouillard formé par un froid matin au dessus d'une étendue d'eau ne semble pas avoir grand-chose en commun avec les gommes arabique et d'adragante, la gélatine ou bien même une fine couche d'hydrocarbure s'étirant à la surface d'un liquide. De prime abord, rien non plus ne pourrait rapprocher ce nuage matinal des protéines circulant dans notre sang telles l'albumine ou les diverses immunoglobulines (IgG, IgM, etc.) acteurs de notre immunité.

Leurs propriétés communes n'ont pas échappé au chimiste écossais Thomas Graham qui, dans un article paru en 1861 \citep{graham1961} visant à classer une série de substances selon leur capacité à traverser une paroi par diffusion, regroupa pour la première fois quelques-unes de ces manifestations de la matière sous l'appellation de \emph{colloïde} (du grec \textgreek{\`{h} k\`{o}lla}, la colle et \textgreek{t\'{o} e\'{i}dos}, l'aspect). Outre de piètres propriétés diffusionnelles et une tendance extrêmement faible à la cristallisation, Th. Graham relève le caractère gluant de leurs formes hydratées et le fait qu'ils sont, pour une large part, solubles dans l'eau bien qu'ils y seraient maintenus par une force beaucoup plus faible que celles responsables de la stricte solubilisation. Leur relative inertie chimique associée à une tendance propre à l'agrégation font aussi affirmer à Th. Graham qu'ils représenteraient l'état de la matière requis dans les processus vivants. Pour ces quelques raisons, il procède à une répartition de la matière entre les \emph{crystalloïdes} et les \emph{colloïdes}.

Avec l'avancement des recherches, il apparaît au début du XX\ieme{} siècle que certains cristalloïdes peuvent présenter des comportements similaires à ceux des colloïdes, et ce, dans certains milieux solvants \citep{lajusteargile3}. Les colloïdes semblent donc être un état de la matière correspondant à un certain degré de division plutôt qu'une classe de substances à part entière. Dès lors, on attribue aux colloïdes leur caractère fondamental: la taille. La taille des colloïdes se situe classiquement (normes IUPAC) entre $10^{-9}$ et $10^{-6}$ mètre (1 nanomètre ou 10 \r{a}ngströms à 1 micromètre) \citep{hiemenz1997tout}. Ces dimensions en font des entités plus grosses que les molécules de bas poids moléculaires comme l'eau ($\sim 0,3$~nm) ou le phosphate ($\sim 0,4$~nm) mais suffisamment petites afin d'être peu sensibles à la gravité, leur permettant de rester suspendues au sein d'une phase liquide (la diffusion compense la gravité \citep{lajusteargile3}). Ces dimensions les rendent aussi inobservables au microscope optique traditionnel. La fine division de la phase colloïdale en fait un ensemble d'entités dont les propriétés ne seront ni attachées à leur chimie ni à leur masse mais plutôt à l'interface qu'elles vont développer vis-à-vis du milieu qui les baigne. En considérant une masse totale $m_\mathrm{tot}$ de colloïdes sphériques de rayon $R$, leur surface étant fournie par $4\pi R^2$ et leur volume par $\frac{4}{3}\pi R^3$, la surface spécifique (m$^2\cdot$kg$^{-1}$) développée par la phase colloïdale est donnée par la relation
\begin{equation}\label{AqAireSpec}
A_\mathrm{sp}=\frac{A_\mathrm{tot}}{m_\mathrm{tot}}=\frac{\mathcal{N}4\pi R^2}{\mathcal{N}\frac{4}{3}\pi R^3 \rho}=\frac{3}{\rho R}
\end{equation}
dans laquelle $\mathcal{N}$ est le nombre de colloïdes constituant la totalité de la phase dispersée et $\rho$ la masse volumique de cette dernière. La formule \ref{AqAireSpec} de l'aire spécifique donnée par P.~C. Hiemenz et R. Rajagopalan \citep{hiemenz1997tout} montre que, plus le rayon $R$ des sphères colloïdales diminue, plus l'aire de l'interface avec le milieu suspendant sera élevée. Un rayon faible donne une aire spécifique élevée indiquant donc une prédominance des propriétés superficielles sur les aspects inertiels, une aire spécifique trop faible indique que le colloïde aura tendance à être emporté par sa masse et donc à ne pas rester suspendu, on parle de sédimentation.

L'état colloïdal autorise donc à réunir des substances ayant des comportements aussi divers que l'opacité et l'aspect blanc voire laiteux d'une nappe de brouillard, l'iridescence manifestée par une lame mince d'hydrocarbure, l'aspect mou et visqueux de la gélatine <<~solidifiée~>>, le comportement rhéologique particulier des solutions de gommes arabique et d'adragante ou encore l'augmentation de la viscosité effective $\eta$ d'une suspension de protéines (albumine, IgG, etc.) selon la loi d'A. Einstein \citep{einstein1906a}: \begin{equation}\label{EqViscoEinstein}
\eta/\eta^\ast=1+\tfrac{5}{2}\,\Phi
\end{equation}
dans laquelle $\eta^\ast$ est la viscosité du fluide suspendant et $\Phi$ la fraction volumique occupée par les protéines\footnote{Cette loi est valable pour des valeurs de $\Phi$ inférieures à $0,1$. Elle a, par ailleurs, été étendue jusqu'au deuxième degré par G.~K. Batchelor \citep{batchelor1977}: $\eta/\eta^\ast=1+\tfrac{5}{2}\,\Phi+\tfrac{19}{4}\,\Phi^2$ couvrant un domaine plus large tel que $\Phi<0,3$.}.

Puisque l'on parle de deux phases, le terme de solubilité est en fait impropre pour décrire le comportement des colloïdes vis-à-vis de l'eau ou d'autres liquides. Comme illustré à la figure \ref{FigBerlins}.A, deux phases sont en présence: la phase suspendante (l'eau) et la phase suspendue. De plus, l'enchâssement d'une phase dans une autre implique nécessairement la présence d'une interface entre les deux. \`{A} la figure \ref{FigBerlins}.B, lorsque la phase suspendue est dispersée ou divisée en très fines entités ayant chacune les dimensions requises par la définition d'un colloïde (diamètre entre 1 nm et 1 \textgreek{m}m), on obtiendra une phase suspendue dispersée ou divisée qui pourra être qualifiée de colloïdale. Cette phase colloïdale, divisée en petites entités, est donc discontinue par opposition à la phase suspendante qualifiée de continue.
On remarquera aussi que la dispersion de la phase suspendue en fines entités augmente considérablement l'aire de l'interface entre les deux phases car, comme le montre l'équation \ref{AqAireSpec}, la diminution du rayon des entités de la phase dispersée est inversement proportionnelle à l'aire de l'interface développée.

\begin{figure}[t]\centering
\includegraphics[width=11cm]{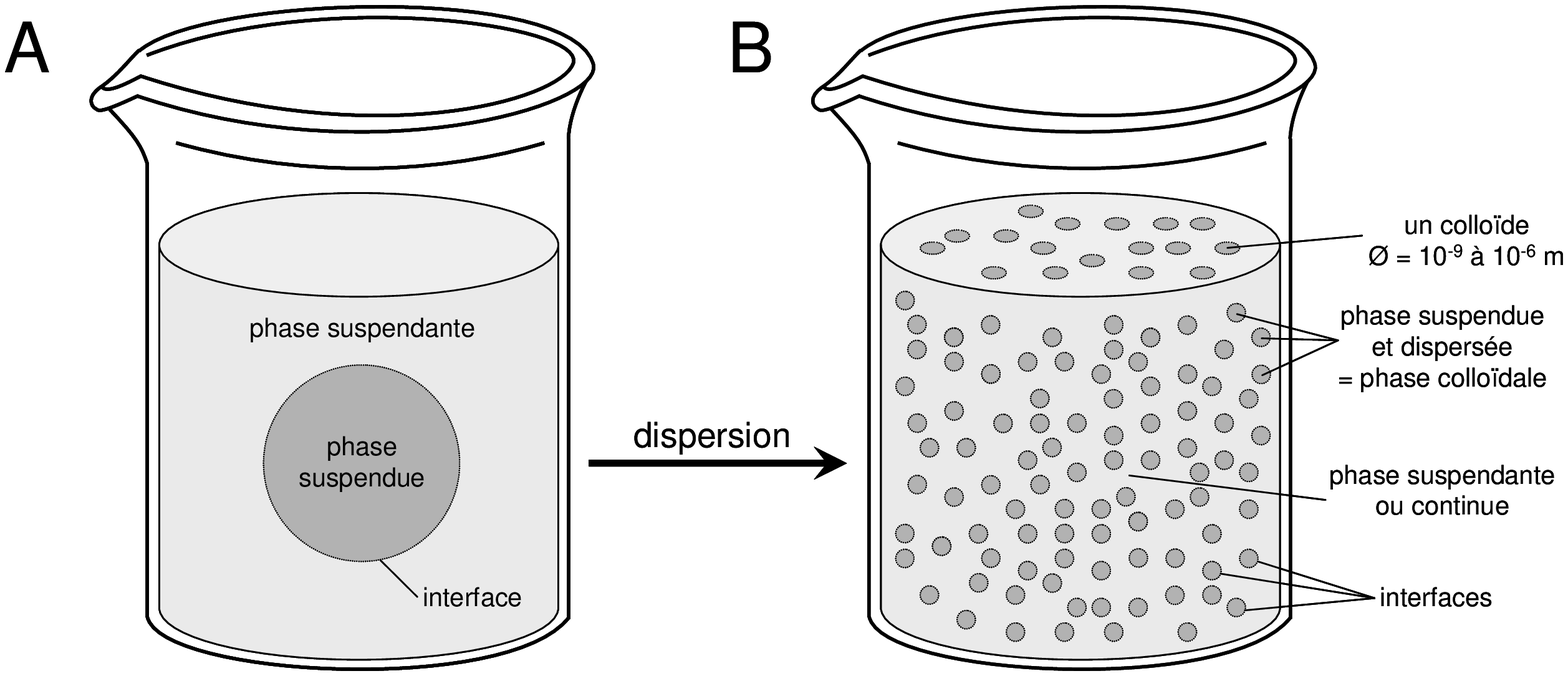}
\caption[Illustration de la dispersion d'une phase suspendue]{Passage d'une phase suspendue d'un seul tenant et visible à l'{\oe}il nu à une phase dispersée dont les entités, invisibles à l'{\oe}il nu, sont de dimensions colloïdales (1 nm à 1 \textgreek{m}m). Cette dispersion s'accompagne d'une augmentation remarquable de l'aire de l'interface entre les deux phases, une caractéristique des liquides colloïdaux.}\label{FigBerlins}
\end{figure}

De plus, cette phase colloïdale tient en suspension dans la phase aqueuse, sans s'y solubiliser, grâce aux propriétés de l'interface colloïde/eau. Lorsqu'une phase dispersée présente une bonne affinité pour la phase continue, on la qualifie de lyophile (hydrophile lorsque la phase continue est aqueuse), le cas inverse étant appelé lyophobie (hydrophobie). L'albumine, les IgG ou la gomme arabique sont hydrophiles au contraire d'un hydrocarbure qui est, lui, hydrophobe \citep{everett1972,hunter1987}. La cause de cette propriété doit être recherchée au niveau de la nature de l'interface entre la phase suspendante et la phase colloïdale. La stabilité des colloïdes, lyophobes en particulier, a fait l'objet d'études intensives dans un ouvrage de E. Verwey et J. Overbeek \citep{verwey1948} paru en 1948, dont le contenu reste largement à jour. Ils y exposent la façon dont celle-ci est déterminée par les propriétés de l'interface entre les phases continue et dispersée. S'ils ne sont pas stables en suspension, les colloïdes s'agrégeront en formant des ensembles plus importants mais au sein desquels ils conserveront leur individualité. S'ils ne la conservaient pas et fusionnaient comme le ferait une phase dispersée de fines gouttelettes d'hydrocarbure dans de l'eau, on parlerait de coalescence. L'agrégation, synonyme de coagulation, conduit à des structures plus grosses qui seront, de ce fait, beaucoup plus sensibles à des forces comme la gravité. Les agrégats dépassant une certaine taille critique finiront par couler selon le même processus que celui se produisant dans les embouchures fluviales: la sédimentation\footnote{Inversement, il est aussi possible de <<~resuspendre~>> les éléments du \textit{coagulum} en changeant les propriétés de la phase continue; on parle de peptisation.}. Cette sédimentation peut être provoquée artificiellement par ajout d'électrolytes dans la phase continue, elle est alors qualifiée de floculation. Dans les estuaires, l'eau douce d'un cours d'eau se mélange à l'eau de mer, bien plus chargée en sels, de telle sorte que les éléments initialement suspendus dans l'eau douce du fleuve se trouvent déstabilisés et s'agrègent en particules plus grosses finissant, elles-mêmes, par sédimenter, participant à l'enlisement de l'embouchure du fleuve \citep{Mayer2011}.

Les données expérimentales concernant la floculation des suspensions colloïdales ont fini par montrer (\textit{cf.} E. Verwey et J. Overbeek \citep{verwey1948}) que la raison pour laquelle les colloïdes tiennent en suspension est due à l'interface qu'ils possèdent en commun avec la phase continue et, en particulier, les propriétés électrostatiques de cette dernière. Il s'agit donc plutôt de la qualité lyophobe ou lyophile de la seule surface qui détermine la stabilité de la suspension colloïdale. Le c{\oe}ur du colloïde ne doit pas nécessairement être lyophile si sa surface l'est déjà. Le cas des protéines (albumine et IgG) et des micelles lipidiques formant les émulsions en est une illustration frappante: surface hydrophile et c{\oe}ur hydrophobe \citep{maibaum2004}. La chaîne peptidique, polymère d'acides aminés constituant les protéines, augmentera son hydrophilie, et donc sa capacité à tenir en suspension dans l'eau, en se repliant d'une manière qui masque vis-à-vis de l'eau les parties hydrophobes qu'elle contient (chaînes latérales des acides aminés). De même, les acides gras se regrouperont en micelles en exposant leur tête hydrophile vers la phase aqueuse, les queues hydrophobes étant, quant à elles, accumulées dans le c{\oe}ur du colloïde. Les macromolécules hydrophobes montrent donc une tendance à l'agrégation entrainant la formation d'ensembles plus gros et plus complexes, agrégation dont la force motrice semble être l'expulsion de l'eau vers la phase continue, phénomène communément appelé <<~effet hydrophobe~>>.

Cette <<~phobie~>> de l'eau peut sembler contradictoire au regard du lien intime et bien connu entre l'eau et la vie sur Terre, lien d'autant plus accepté par les scientifiques que la recherche de la vie dans le système solaire (et ailleurs) se fait en fonction des endroits où l'on peut en trouver à l'état liquide \citep{ball2004,seager2013}. En effet, les biomolécules (protéines, ADN, etc.) sont, certes, suspendues dans un milieu aqueux mais leur structure se fonde sur l'absence d'eau, son expulsion vers la phase continue. La vie semblerait donc avoir besoin d'eau pour mieux la détester, détestation qui lui permet de créer des structures complexes telles que les protéines ou des phases micellaires ou lamellaires, toutes de dimensions colloïdales. Ch. Tanford souligne \citep{tanford1978} que l'eau est plus qu'un simple environnement permettant la réalisation d'une certaine chimie, elle force certains composés qui s'y forment à s'organiser en structures de plus en plus complexes comme les protéines. La chimie permet, par ses réactions, de créer des macromolécules hydrophobes dans un environnement aqueux, lesquelles, afin de minimiser leur énergie superficielle, sont obligées de s'agréger en micelles, en protéines, enzymes, etc.

De par son importance, le rôle de l'eau dans la conformation des protéines est un sujet très étudié \citep{tanford1978,levy2006,ball2013}. De plus en plus, elle apparaît comme une matrice et non comme un environnement purement chimique. Une matrice agissant de manière active dans le repliement de la protéine sur elle-même. Non contente de jouer un rôle moteur dans la conformation de la protéine, l'eau joue aussi un rôle stabilisateur grâce à son interface avec la partie hydrophile du colloïde. Afin d'assurer ce rôle stabilisateur, il est nécessaire que la surface du colloïde soit suffisamment hydrophile; autrement, le colloïde demeurera dans une conformation lâche sans structure bien déterminée et ne pouvant dès lors remplir une tâche précise dans les processus biochimiques. Par ailleurs, une faible hydrophilie (hydrophobie) de la surface du colloïde rendra l'atmosphère de ce dernier faite de molécules du solvant peu enclines à empêcher le rapprochement de deux colloïdes, rapprochement pouvant se solder par l'agrégation de ceux-ci.

La relative stabilité des colloïdes en suspension est donc due à la nature de l'interface qu'ils développent avec la phase continue. Celle-ci peut en effet présenter une barrière d'énergie potentielle plus ou moins répulsive dont la forme déterminera la stabilité du colloïde vis-à-vis de l'agrégation. Les c{\oe}urs de deux colloïdes distincts mais de mêmes natures auront toutefois tendance à s'attirer en raison des forces de London-van der Waals, très intenses à faible distance. Toutefois, ceux-ci peuvent rester suspendus à cause du fort potentiel électrostatique, éventuellement répulsif, porté par leurs surfaces. Par exemple, deux micelles ou deux protéines distinctes mais identiques porteront les mêmes charges et donc se repousseront l'une et l'autre restant ainsi stables dans l'état dispersé. Si ces charges avaient été absentes ou trop faibles comme dans le cas de deux pelotes d'un même polymère aliphatique (du polyéthylène par exemple), les forces de London-van der Waals ne se seraient pas vues contrecarrées et les pelotes auraient fusionné entrainant la coalescence de la phase dispersée. La stabilité des colloïdes, due aux phénomènes d'attraction et de barrière d'énergie potentielle, est dite cinétique car la hauteur de la barrière d'énergie potentielle déterminera la lenteur (\textit{i.e.} barrière élevée) ou la rapidité (\textit{i.e.} barrière faible à nulle) du processus d'agrégation/coalescence.

\section*{Des films de colloïdes à l'ELISA}
\addcontentsline{toc}{section}{Des films de colloïdes à l'ELISA}
\markboth{INTRODUCTION}{DES FILMS DE COLLOÏDES \`{A} L'ELISA}

En vertu de leurs intéressantes propriétés et de leur omniprésence, les colloïdes sont utilisés dans de très nombreuses applications technologiques. Et, lorsqu'ils sont des protéines comme les IgG (immuno-\textgreek{g}-globulines), ces applications sont souvent à usage médical ou plus généralement analytique. La fabrication de dispositifs spécifiques tire souvent parti de la capacité intrinsèque à l'agrégation voire l'adhésion des colloïdes. Par exemple, la synthèse de certains biocapteurs peut nécessiter la fixation en surface de protéines (ELISA, \textit{cf.} ci-après), d'ADN (\textit{DNA microarray} ou puce à ADN), etc. On considère classiquement que les biocapteurs sont composés d'au moins trois éléments: l'échantillon à étudier, un capteur d'origine biologique et un élément électronique associé s'il s'agit d'un biocapteur ampérométrique \citep{kissinger2005}. Le dispositif permet, par exemple, la réalisation d'un enregistrement dans le temps de ce que le capteur mesure de l'échantillon (par exemple, une électrode fonctionnalisée qui aura été implantée dans l'organisme). Dans le cas des puces à ADN, il s'agira plutôt de détecter par une différence de potentiel électrique l'hybridation d'un brin d'ADN complémentaire provenant du milieu analysé par celui fixé au capteur \citep{templin2002,dorazio2011}.

Un dispositif d'analyse biochimique extrêmement courant est l'ELISA (\textit{Enzyme-Linked ImmunoSorbent Assay}). Cet ELISA est constitué de deux éléments des biocapteurs: un échantillon d'antigène et un capteur immunologique. Basé sur la très haute spécificité anticorps-antigène, l'ELISA vise à détecter et souvent à doser la quantité d'antigènes qui pourrait être présente dans une suspension d'origine biologique. Ce dosage s'effectue en mettant la suspension potentiellement antigénique en contact avec une surface sur laquelle sont fixés des anticorps spécifiquement dirigés vers l'antigène à détecter \citep{butler2000,porstmann1992}. Deux protocoles ELISA classiquement utilisés sont illustrés sur la figure \ref{FigELISAsandwich}. On y voit le montage moléculaire nécessaire au dosage d'un antigène.

\begin{figure}[t]\centering
\includegraphics[width=13cm]{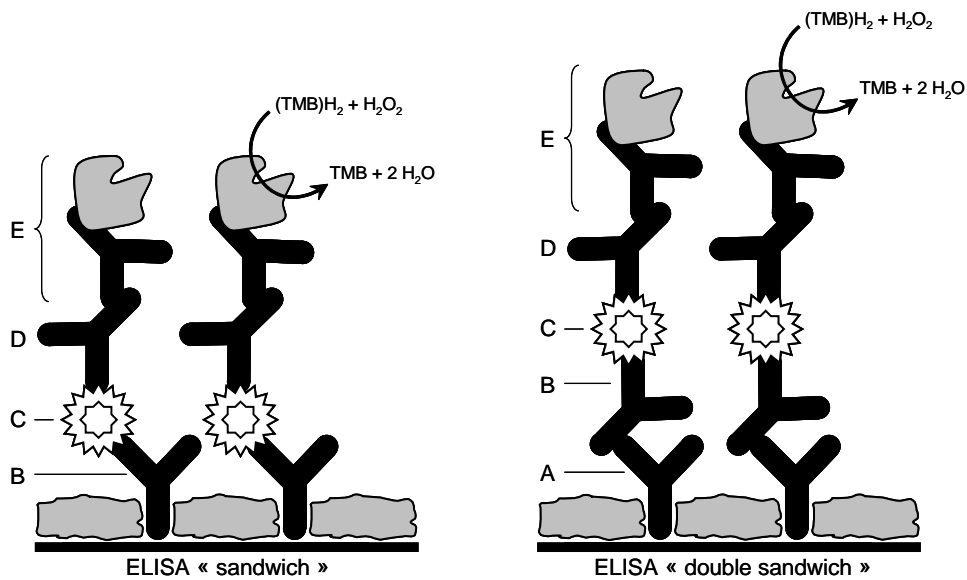}
\caption[Deux types de protocoles ELISA]{Illustrations de deux types de protocoles ELISA classiques: l'ELISA <<~sandwich~>> et l'ELISA <<~double sandwich~>>. \textbf{A}: anticorps de capture (se fixe sur la surface solide supportant le test en formant un film); \textbf{B}: anticorps spécifique (détecte l'antigène ou toute autre biomolécule à doser); \textbf{C}: antigène à détecter; \textbf{D}: anticorps spécifique intermédiaire (facultatif) entre l'antigène et l'anticorps conjugué (permet l'utilisation d'anticorps conjugués standards); \textbf{E}: anticorps conjugué à une peroxydase (catalyse l'oxydation de la 3,3',5,5'-tétraméthylbenzidine (TMB) ou de toute autre susbstance chromogénique, substance produisant une coloration du mélange permettant de quantifier l'antigène par lecture optique). Dans le protocole <<~sandwich~>>, l'anticorps de capture et l'anticorps spécifique ne font qu'un, c'est-à-dire que l'on fixe directement l'anticorps spécifique sur la surface solide supportant le test. On remarquera aussi que le support solide est saturé par des protéines quelconques afin d'éviter toute fixation non-spécifique de l'antigène pouvant fausser les résultats du test. D'après S.~E. Hornbeck \textit{et al.} \citep{hornbeck2001} et les protocoles de DIAsource ImmunoAssays (Louvain-la-Neuve, Belgique).}\label{FigELISAsandwich}
\end{figure}

Dans le cadre de l'ELISA, la figure \ref{FigELISAsandwich} montre qu'il est nécessaire de réaliser un film d'anticorps (de capture et/ou spécifique) sur la surface solide supportant le test. La formation de ce film d'anticorps à la surface du support du test est toutefois un phénomène complexe faisant, lui aussi, intervenir des notions touchant aux sciences colloïdales. Il faudra, en effet, coller une grande quantité d'anticorps sur une surface solide telle que du polystyrène. Dans la pratique, ceci se fait très simplement par la mise en contact d'une suspension d'anticorps et d'une surface de polystyrène (PS). Le film, spontanément constitué, devra répondre à plusieurs exigences afin d'être efficace lors de l'étape de reconnaissance, c'est-à-dire de la fixation de l'antigène sur l'anticorps spécifique \citep{butler2000}:
\begin{enumerate}
\item les propriétés immunologiques des anticorps collés sur la surface doivent demeurer intactes, il faudra donc éviter qu'ils se dénaturent,
\item les anticorps doivent être convenablement orientés de telle sorte qu'ils puissent fixer des antigènes provenant de la phase liquide lors de l'étape de reconnaissance et
\item le film doit être fixé sur le support d'une manière suffisamment forte afin qu'il ne soit pas détérioré lors des étapes suivantes de l'ELISA, étapes nécessitant divers rinçages (sollicitation mécanique).
\end{enumerate}
\'{E}tant donné son intérêt pour l'industrie, ces tests sont largement utilisés et commercialisés par des firmes spécialisées. Ils ont, dès lors, fait l'objet d'abondantes recherches sous des angles tant biochimiques que physico-chimiques. Ces quelques exigences alliées à son intérêt industriel illustrent à quel point la compréhension et la maîtrise de la construction du film d'IgG est importante dans la synthèse de dispositifs biosensibles tels que les ELISA.

\`{A} ce titre, la fabrication de biomatériaux (compatibles avec les tissus du corps humain comme dans le cas de l'électrode mentionnée ci-dessus) font un grand usage de ces propriétés en créant des interfaces appropriées entre une pièce amenée à jouer un rôle de soutien mécanique dans l'organisme et les tissus dans lesquels ils sont insérés. La nature des surfaces de ces éléments devant être insérés dans l'organisme est cruciale; en effet, ces surfaces y seront directement exposées à des milieux très concentrés en protéines tels que le plasma sanguin (albumine, fibrinogène, IgG, etc.) \citep{hlady1996}. La réponse de l'organisme vis-à-vis de ces interfaces dépendra de leur capacité à adsorber les protéines\footnote{Dans ce cas précis, le verbe \emph{adsorber} semble être un abus de langage. En effet, comme nous le verrons dans le chapitre consacré à la thermodynamique de l'adhésion, il faudrait plutôt parler d'\emph{adhésion} des protéines. Cette nuance est fondamentale mais, vu son utilisation très fréquente dans la littérature, il serait malaisé de s'en départir à ce stade. Par fidélité aux sources ici consultées, l'usage du mot \emph{adsorption} sera confondu avec celui d'\textit{adhésion} avant que ne soit définitivement levée l'ambigu\"{i}té.} et il est dès lors préférable de rendre les surfaces non adsorbantes, l'adsorption pouvant provoquer de la coagulation sanguine, une réaction du système immunitaire à travers l'activation du complément ou encore causer l'adhésion des plaquettes et des globules blancs \citep{vroman1977b}. Rendre une surface résistante à l'adsorption peut toutefois ne pas être judicieux si l'on souhaite y favoriser l'adhésion cellulaire, phénomène nécessitant la disponibilité de sites d'ancrage de nature protéique \citep{hlady1996}.

Au-delà de son intérêt pour les technologies analytiques, la formation des films d'IgG revêt aussi divers attraits beaucoup plus fondamentaux. En effet, la formation de ce genre d'ensemble colloïdal implique une cascade d'événements qui, pour certains, sont bien décrits dans la littérature et qui, pour d'autre, le sont beaucoup moins voire pas du tout. Premièrement, les IgG devront atteindre la surface et cela impliquera la prise en considération de l'hydrodynamique des suspensions colloïdales et de la diffusion des colloïdes. Deuxièmement, les IgG étant arrivées à la surface, des considérations à plus petite échelle devront être faites. Si une surface de polystyrène est envisagée, il faudra tenter de comprendre comment l'effet hydrophobe agira pour y faire adhérer l'IgG. S'intéresser à l'effet hydrophobe, c'est s'intéresser aux mêmes causes que celles impliquées dans la formation d'ensembles de première importance dans les structures biologiques telles que les bicouches lipidiques, les complexes de protéines (anticorps-antigène), etc. \citep{nagarajan1991}. Il est aussi bien connu, qu'une fois collées sur la surface, les protéines peuvent s'y dénaturer \citep{buijs1996a}. Discuter de la dénaturation des protéines sur quelques bases thermodynamiques, c'est aussi manipuler, certes à l'envers et de façon plus vague, les phénomènes à l'{\oe}uvre lors du repliement de ces mêmes protéines vers leur état natif. Connaître les causes et la manière dont pourrait se déplier ou changer de conformation une protéine semble très instructif quant à leur repliement natif.

Comprendre comment se colle une IgG sur une surface hydrophobe comme le polystyrène est une chose, comprendre comment se forme un film complet d'IgG sur cette même surface en est une autre. Comme le rappelle E.~A. Vogler \citep{vogler2012}, les films de protéines se forment dans un espace tridimensionnel et il faut dès lors discuter de la façon dont cet espace pourra être rempli jusqu'à saturation, moment où le film sera complet. L'organisation qu'adopteront les IgG et l'ordre dans lequel elles le feront sont intimement liés à leurs dimensions et à leur forme. Les notions sur lesquelles fonder une discussion pourront sûrement être extrapolées à la formation d'autres ensembles de corps colloïdaux et les IgG sont, à ce titre, un très bon modèle. En effet, elles présentent une forme relativement complexe en <<~Y~>>, certes liée à leur fonction immunitaire, mais devant aussi leur attribuer des propriétés particulières dans leur manière de remplir l'espace. \`{A} partir de ces caractéristiques, on pourra voir l'adsorption des IgG comme une généralisation et aisément déduire des renseignements pour l'adsorption d'autres protéines ou colloïdes de formes plus simples: ellipsoïdes, sphères, etc. Cet optimisme pourrait paraître excessif car, selon W. Norde \citep{norde2008}, on est encore bien loin d'une théorie unifiée de l'adsorption des protéines, chacune de celles-ci ayant leur propre <<~personnalité~>>, multipliant alors à l'envi les comportements possibles.

\section*{Les immuno-\textgreek{g}-globulines (IgG)}
\addcontentsline{toc}{section}{Les immuno-\textgreek{g}-globulines (IgG)}
\markboth{INTRODUCTION}{LES IMMUNO-\textgreek{G}-GLOBULINES}

De par son intérêt dans le cadre des dosages immunologiques (ELISA, RIA, \textit{latex agglutination assays}, etc.), l'étude de l'adsorption des immuno-\textgreek{g}-globulines (IgG) sur différents types de surfaces a donné lieu à un très grand nombre de publications dont d'intéressantes synthèses ont été réalisées par C.~E. Giacomelli 
\citep{giacomelli2006}, J.~N. Herron \textit{et al.} 
\citep{herron1998} ou encore F. Caruso 
\citep{caruso2000}. Malgré le fait que l'adsorption des IgG ait été étudiée sur divers types de surfaces, s'intéresser à une surface hydrophobe telle que le polystyrène a plusieurs avantages: en plus d'être très facile à obtenir, elle manifeste une forte affinité pour les IgG et les protéines en général. Le polystyrène, réputé fortement hydrophobe, n'est pas intrinsèquement porteur de groupements chimiques chargés ou ionisables. Toutefois, lorsqu'il est plongé dans de l'eau, il présente un potentiel-\textgreek{z} (charge acquise par le polystyrène dans une solution, ici l'eau pure) légèrement négatif \citep{dupont2000}. D'autre part, il peut être traité chimiquement afin de devenir porteur de charges négatives ou positives \citep{elgersma1990}.

Les immunoglobulines sont, avec d'autres récepteurs des lymphocytes T, responsables de la reconnaissance des antigènes par le système immunitaire. Ce sont des glycoprotéines présentes dans les liquides biologiques de tous les mammifères. Les antigènes sont, quant à eux, des protéines étrangères à un organisme de plus d'un kDa ($=1$ kg$\cdot$mol$^{-1}$), des polysaccharides, des glycoprotéines, des lipides, des lipoprotéines ou de simples sucres \citep{dehoux2010}. On distingue habituellement chez les mammifères cinq classes d'immunoglobulines: IgG, IgA, IgM, IgD et IgE auxquelles il faut additionner un certain nombre de sous-classes. Les différences entre classes se marquent par leur masse moléculaire, leur charge, leur composition en acides aminés et en polysaccharides. Totalisant entre 70 et 75 $\%$ des immunoglobulines contenues dans le sérum normal, les immuno-\textgreek{g}-globulines ou IgG sont les plus abondantes. Leur structure schématisée à la figure \ref{FigIgG} est faite de deux chaînes polypeptidiques dites lourdes de 50 kDa (chaîne \textgreek{g}) associées à deux autres de 25 kDa dites légères (chaînes \textgreek{k} et \textgreek{l}). Ces chaînes sont associées entre elles par des ponts disulfures afin de former un ensemble d'environ 150 kDa ayant l'aspect d'un trèfle plutôt que d'un <<~Y~>> \citep{roitt1993}.

\begin{figure}[ht]\centering
\includegraphics[width=6cm]{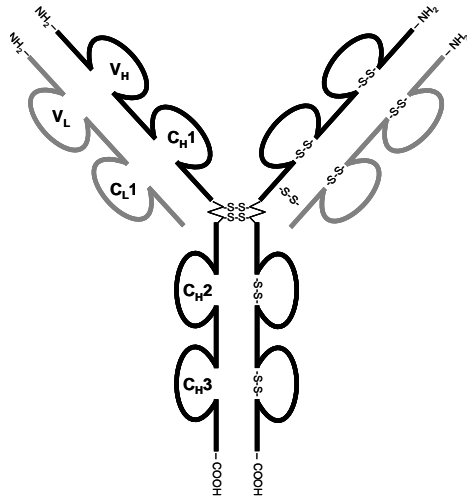}
\caption[Schéma d'une immuno-\textgreek{g}-globuline (IgG)]{Schéma d'une immuno-\textgreek{g}-globuline (IgG) d'après B. Mach et V. Ossipow \citep{mach2009}. Les chaînes lourdes sont représentées en noir alors que les chaînes légères le sont en gris. En plus des ponts disulfures $-\textrm{S}-\textrm{S}-$, on y représente les extrémités N-terminales $-\textrm{NH}_2$ et C-terminales $-\textrm{COOH}$. Chaque domaine de la structure secondaire est représenté par une boucle.}\label{FigIgG}
\end{figure}

Les chaînes polypeptidiques composant les immunoglobulines révèlent un schéma de repliement qui leur est typique. Ce motif, ou domaine, est présent quatre fois sur les chaînes lourdes et deux fois sur les chaînes légères de telle sorte qu'une IgG totalise douze de ces domaines. D'après P. Bork \textit{et al}. \citep{bork1994}, ces domaines appelés repliement d'immunoglobuline (\textit{immunoglobulin fold}, Ig-\textit{fold}) sont composés de neuf brins \textgreek{b} repliés selon un schéma analogue aux tonneaux \textgreek{b}. Ils s'en démarquent toutefois car on peut y distinguer deux feuillets \textgreek{b} de telle sorte que le domaine soit en fait un sandwich \textgreek{b}. Les structures secondaires dominantes dans les IgG sont donc les feuillets \textgreek{b} antiparallèles et quelques portions d'hélices \textgreek{a} dans certains tournants \citep{padlan1994} de l'Ig-\textit{fold}.

Les IgG se distinguent entre elles de manière très subtile au niveau de leur parties variables (V), le reste de l'IgG étant qualifié de constant (C) car demeurant relativement bien conservé d'une IgG à l'autre \citep{mach2009}. Seuls les domaines situés aux extrémités N-terminales des chaînes lourdes (H) et légères (L) présentent cette variabilité et seront dès lors notés V$_\textrm{H}$ et V$_\textrm{L}$ (\textit{cf}. figure \ref{FigIgG}). C'est par cette région variable que l'IgG pourra lier un antigène. Le domaine constant de la chaîne légère est noté C$_\textrm{L}$ tandis que les trois domaines constants appartenant aux chaînes lourdes le sont par C$_\textrm{H}1$, C$_\textrm{H}2$ et C$_\textrm{H}3$.

Les IgG montrent une charnière (\textit{hinge}) à l'endroit où les deux chaînes lourdes se séparent pour former les deux bras du <<~Y~>>. Cette charnière qui varie en longueur et flexibilité selon la classe et l'isotype concerné peut être clivée par une enzyme, la papaïne, pour donner deux fragments: le Fc contenant les deux domaines constants C-terminaux C$_\textrm{H}2$ et C$_\textrm{H}3$ et le F(ab')$_2$ composé de deux plus petits fragments Fab. Les deux fragments Fab, identiques, sont constitués de la chaîne légère et des deux premiers domaines de la chaîne lourde V$_\textrm{H}$ et C$_\textrm{H}1$. Le site de liaison de l'antigène (paratope) est situé à l'extrémité du fragment Fab \citep{padlan1994}.

\section*{Adhésion des IgG et des protéines}
\addcontentsline{toc}{section}{Adhésion des IgG et des protéines}
\markboth{INTRODUCTION}{ADH\'{E}SION DES IGG ET DES PROT\'{E}INES}

Brièvement décrite, la configuration des IgG permettra de mieux cerner l'adsorption des IgG. Celle-ci a, entre autre, fait l'objet de nombreuses études par le groupe de W. Norde. Deux premiers travaux d'A.~V. Elgersma \textit{et al}. \citep{elgersma1991,elgersma1992} montrent, d'une part, les variations observées dans le comportement d'adsorption de différents types d'IgG\footnote{Plus précisément, il s'agit d'un isotype 1 d'une IgG de souris anti-hCG (hormone chorionique gonadotrope humaine).}\label{hohbdhfbvkjhbfvkjhdfbvkjbd} et, d'autre part, les effets de la compétition entre les immuno-\textgreek{g}-globulines et l'albumine, cette dernière étant aussi présente dans le plasma et largement utilisée comme agent bloquant dans les ELISA \citep{hornbeck2001}. La première étude vise à montrer la corrélation entre les potentiels-\textgreek{z} pour différents pH et la quantité d'IgG adsorbée au plateau de l'isotherme. L'isotherme illustrée à la figure \ref{FigIsothermeIntro} est un phénomène caractéristique de l'adsorption des protéines: il se manifeste lorsque des surfaces de même nature sont soumises à une série de solutions d'une même protéine mais en concentrations différentes (mais invariables au cours de la formation du film). Plus la concentration en protéines est élevée, plus les quantités adsorbées seront fortes. Habituellement, cette <<~isotherme~>> présentant la quantité adsorbée en fonction de la concentration engagée prend la forme d'une courbe plafonnant à une valeur dite de <<~plateau~>>. Ce plateau est donc la quantité maximale dont on peut saturer une surface de protéines. Pour ces valeurs de plateau d'isotherme des diverses IgG, A.~V. Elgersma \textit{et al}. \citep{elgersma1991} montrent qu'elles sont fonctions du pH sous la forme d'une cloche dont le maximum se situe autour d'un pH correspondant au point isoélectrique du complexe entre l'IgG et un latex de polystyrène. Ceci montre que le pH doit permettre une compensation adéquate des charges, à la fois de l'IgG et de la surface de polystyrène, afin que l'adsorption soit optimale. La corrélation entre le pH (agissant sur la charge du complexe IgG-PS) et la quantité adsorbée au plateau montrée par A.~V. Elgersma \textit{et al}. \citep{elgersma1991} est remarquable. Dans un travail assez fouillé, M. Bremer \textit{et al}. \citep{bremer2004} fournissent une série d'éléments explicatifs de l'adsorption des IgG basée sur une étude théorique des interactions électrostatiques entre l'IgG et la surface, d'une part, et les IgG d'autre part.

\begin{figure}[h]\centering
\includegraphics[width=8cm]{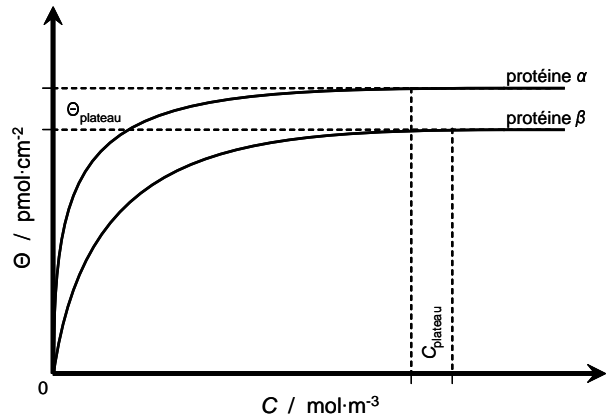}
\caption[Allures d'isothermes d'adsorption]{D'après W. Norde et C.~A. Haynes \citep{nordehaynes}, allures d'isothermes d'adsorption de deux protéines fictives <<~$\alpha$~>> et <<~$\beta$~>>. L'isotherme d'adsorption est une courbe reliant des points expérimentaux représentant les densités superficielles $\Theta$ (en pmol$\cdot$cm$^{-2}$) d'une série de films aux concentrations en protéines $C$ (en mol$\cdot$m$^{-3}$) des suspensions à partir desquels ils se sont formés. La valeur au <<~plateau~>> de l'isotherme, \textit{i.e.} $\Theta_\text{plateau}$, est, en fonction de la concentration $C$, la valeur maximale des densités superficielles des films que l'on pourra former à partir d'une protéine donnée.}\label{FigIsothermeIntro}
\end{figure}

Plusieurs isothermes sont présentées par J. Buijs \textit{et al}. \citep{buijs1995} pour des films d'IgG monoclonales déposés sur des surfaces de polystyrène chargé positivement ou négativement. Ils observent des \textit{maxima} pour la quantité adsorbée au plateau de l'isotherme analogues à ceux cités ci-avant. Le travail de J. Buijs \textit{et al}. \citep{buijs1995} portant aussi sur l'orientation des IgG adsorbées, une discussion utilisant ces résultats d'adsorption auxquels ils adjoignent d'autres résultats d'adsorption pour les seuls fragments Fc et F(ab')$_2$ des mêmes IgG donne quelques indications à ce sujet. Illustrées à la figure \ref{FigIntroOri}, quatre orientations sont identifiées et discutées: l'orientation \textit{end-on} (IgG <<~debout~>> adsorbée sur son fragment Fc) avec les fragments Fab collés l'un à l'autre, l'orientation \textit{end-on} avec des fragments Fab se repoussant faiblement ou fortement et l'orientation \textit{side-on} (l'IgG est <<~allongée~>> sur la surface)\footnote{On trouve un certain nombre d'orientations possibles dans la littérature: \textit{end-on} (adsorbée debout par le fragment Fc), \textit{side-on} (adsorbée sur le côté), \textit{head-on} (adsorbée par le fragment F(ab')$_2$), \textit{flat} (couchée sur le flanc), etc.}. En terme d'orientation, les surfaces hydrophobes (polystyrène) sembleraient favoriser l'orientation \textit{end-on} puisque les quantités adsorbées au plateau de l'isotherme sont les plus élevées \citep{buijs1995} et que cette orientation dispose d'une faible emprise à la surface \citep{bremer2004}. \`{A} partir de l'épaisseur du film d'IgG formé, M. Malmsten \citep{malmsten1995} rapporte, quant à lui, des adsorptions majoritairement \textit{end-on} tant sur des surfaces hydrophobes qu'hydrophiles, ce qui dénoterait une tendance intrinsèque aux IgG à s'adsorber de la sorte.

\begin{figure}[t]\centering
\includegraphics[width=7cm]{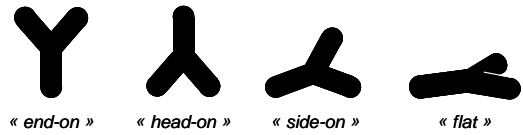}
\caption[Orientations des IgG sur une surface]{Les orientations possibles des IgG sur une surface telles qu'elles sont décrites par J. Buijs \textit{et al.} \citep{buijs1995}. Dans la suite de l'exposé, les écarts entre les <<~bras~>> des IgG seront considérés comme identiques, \textit{i.e.} 120$^\circ$, de telle sorte que les orientations \textit{head-on} et \textit{side-on} puissent être confondues.}\label{FigIntroOri}
\end{figure}

Toutefois, déduire des informations quant aux mécanismes d'adsorption à partir des seules données issues des états finaux atteints (\textit{i.e.} le plateau de l'isotherme) par les films d'IgG formés sur diverses surfaces peut, fort souvent, se révéler assez spéculatif. C'est pourquoi, de nombreuses études visent à caractériser la formation de ces films \textit{in situ} et surtout en temps réel. Dans ce but, J. Buijs \textit{et al}. \citep{buijs1996b} utilisent la réflectométrie (méthode de J.~C. Dijt \textit{et al}. \citep{dijt1994}) afin de mesurer les changements en temps réel se produisant sur une surface et liés à l'adsorption des IgG. Ce genre de dispositif se base sur une cellule dans laquelle la solution contenant les IgG vient impacter perpendiculairement la surface adsorbante et s'en échappe par les côtés (\textit{cf}. \citep{dijt1994}). Les courbes d'adsorption en fonction du temps montrées par J. Buijs \textit{et al}. \citep{buijs1996b} illustrent à quel point le phénomène est complexe. En effet, ces courbes révèlent des formes assez diverses et difficiles à interpréter étant donné la présence d'un écoulement constant affectant très certainement l'adsorption et de phénomènes dus au remplissage du dispositif expérimental. Il est en effet possible que ce remplissage de la cellule puisse mener à des artefacts tels que des \textit{overshoot} sur les courbes de cinétiques que J. Buijs \textit{et al}. \citep{buijs1996b} interprètent comme un réarrangement du film voire une transition entre un régime où l'adsorption serait dominante à un autre où ce serait la relaxation des IgG adsorbées qui le serait. Ce genre de phénomène d'\textit{overshoot} a aussi été observés par V. Hlady \citep{hlady1991} et largement discutés dans les travaux récents de M. Rabe \textit{et al}. \citep{rabe2007,rabe2009,rabe2011}.

La conception de dispositifs de mesure de l'adsorption est rendue très délicate de par l'intrication existant entre les phénomènes d'adsorption, de diffusion et de convection. \`{A} ce titre, M. J\"{o}nsson \textit{et al}. \citep{jonsson2007} ont modélisé ce à quoi pourrait ressembler les écoulements dans des dispositifs de formes diverses, en ce notamment celui utilisé par la microbalance à cristal de quartz (QCM-D) commercialisée par la firme Q-Sense (G\"{o}teborg, Suède). En plus de l'écoulement, ils illustrent très clairement les gradients de concentrations lorsque le solvant, initialement présent dans le dispositif, est remplacé par la solution de protéines devant s'adsorber. Il est fort probable que cette transition puisse mener à des effets dont l'interprétation sur la seule base de l'adsorption des protéines ne soit pas représentative de la réalité. Tout comme l'outil réflectométrique utilisé par J. Buijs \textit{et al}. \citep{buijs1996b}, la QCM-D est sensible en temps réel et \textit{in situ} à l'adsorption des protéines \citep{hook1998un} et est de ce fait devenue un outil largement utilisé afin de mesurer l'adsorption des protéines. Toutefois, dans le cadre des études utilisant la QCM-D, il arrive souvent que les considérations hydrodynamiques soient négligées (\textit{cf}. H\"{o}\"{o}k \textit{et al}. \citep{hook1998un} dans une étude sur l'adsorption des IgG). Au-delà de cette approximation qui pourrait biaiser les résultats (même dans le cas d'un écoulement laminaire), la QCM-D ne semble pas encore tout-à-fait maîtrisée du point de vue physique car les modèles, assez complexes, n'arrivent toujours pas à rendre compte des données expérimentales en faisant un lien clair entre les signaux enregistrés et les quantités adsorbées \citep{johannsmann2007a,McEvoy2012}. L'amélioration de la QCM-D passera sûrement par une amélioration de ces deux aspects: l'hydrodynamique et une meilleure connaissance théorique de sa sensibilité aux quantités adsorbées. Il n'en demeure pas moins que la QCM-D permet de comparer efficacement des films construits dans un certain cadre et qu'elle reste dès lors une technique de choix dans l'étude des monocouches de protéines.

Un autre phénomène se produisant lors de l'adsorption et dont il est très important de tenir compte est celui de la relaxation. La relaxation est la possibilité pour une protéine de se dénaturer afin d'optimiser ses interactions avec la surface ou bien, simplement, de changer son orientation. Processus complexe, il dépend à la fois de la nature de la surface et des propriétés de la protéine adsorbante. En pointant le fait que cette tendance à la relaxation était liée à la cohésion interne de la protéine, W. Norde \citep{norde2008} a qualifié de \textit{soft} (molles) les protéines pouvant se dénaturer relativement facilement, alors que celles présentant une certaine résistance à cette dénaturation l'ont été de \textit{hard} (dures). La relaxation est un phénomène important pour au moins deux raisons: premièrement, elle peut participer au remplissage de la surface et, d'une certaine manière, s'opposer à l'adsorption d'autre protéines; et, deuxièmement, par le changement de conformation de la protéine qu'elle occasionne, elle peut lui faire perdre un certain nombre de propriétés. Parmi celles-ci, les propriétés immunologiques des IgG (leur capacité à reconnaître un antigène) pourraient être détériorées par la relaxation. En effet, si la relaxation imposait une orientation défavorable, les sites de reconnaissance des antigènes pourraient ne pas être disponibles. Si de surcroît la relaxation provoquait une dénaturation de ces sites, les antigènes ne pourraient pas non plus être reconnus.

Cette nécessité de maîtriser \textit{via} les conditions physico-chimiques du milieu et de comprendre comment la relaxation pourrait se produire a poussé J. Buijs \textit{et al.} \citep{buijs1997un} à étudier les effets de l'adsorption des IgG sur la reconnaissance d'un antigène, l'hormone hCG. Ils montrent que le taux de reconnaissance de l'hCG par les IgG est d'autant plus élevé que les conditions dans lesquelles se fait l'adsorption des IgG permettent d'entraver l'adsorption du fragment F(ab')$_2$ qu'ils testent par ailleurs. Inversement, les conditions favorisant l'attraction du F(ab')$_2$ sont corrélées à un faible taux de reconnaissance de l'hCG. Outre le changement de conformation, les changements de conformation, eux aussi dépendants des conditions physico-chimiques (pH et force ionique), sont susceptibles de provoquer une inactivation des sites de reconnaissance de l'anticorps. Il a été mis en évidence par J. Buijs \textit{et al}. \citep{buijs1996a} que les surfaces hydrophobes étaient responsables d'importants changements de conformation des IgG, se dont témoigne une transition d'une structure où les feuillets \textgreek{b} sont majoritaires vers une structure où les hélices \textgreek{a} le deviennent.

Pour M. Bremer \textit{et al}. \citep{bremer2004}, les IgG s'adsorbant dans une orientation \textit{end-on} peuvent encore relaxer leur orientation afin d'optimiser leurs interactions avec la surface, menant à une orientation de type plutôt \textit{side-on}. Des IgG adsorbées dans cette dernière orientation plus gourmande en espace devraient, en toute logique, induire des quantités adsorbées plus faibles au plateau de l'isotherme. D'autre part, toujours selon les mêmes auteurs, ce genre de réorientation ne pourrait avoir lieu que lorsqu'il demeure encore suffisamment de place sur la surface, ce qui découlerait sur le fait que, si la vitesse initiale d'adsorption est élevée par rapport à la relaxation, les possibilités de réorientation sont minimisées et la quantité maximale adsorbée pourra être élevée. Cette idée a récemment été précisée par W. Norde et J. Lyklema \citep{Norde2012} sous la forme d'un ratio entre un temps caractéristique de la relaxation $\tau_r$ et un autre, caractéristique du remplissage de la surface, $\tau_f$. Lorsque le ratio $\tau_r/\tau_f$ serait très supérieur à l'unité, la relaxation serait totalement inhibée.

Comme nous le verrons dans la suite de ce travail, l'introduction de ce ratio par W. Norde et J. Lyklema \citep{Norde2012} est une idée intéressante et prometteuse dans l'optique d'une formalisation du mécanisme d'adsorption des protéines. Elle permet en effet de simplifier le problème de la compétition entre relaxation et adsorption en un seul paramètre beaucoup plus manipulable que les seules vitesses. Il n'en demeure pas moins qu'I. Lundstr\"{o}m et H. Elwing \citep{lundstrom1990} soulignent à quel point les mécanismes par lesquels un film de protéines peut se former restent complexes. Ils notent, en particulier, qu'en plus des contraintes cinétiques habituelles dues à la seule diffusion des protéines vers la surface, les interactions protéine-surface contiennent de nombreux phénomènes dynamiques et dépendants du temps dont il est nécessaire de tenir compte. En particulier, il existerait une dépendance au temps dans le développement des interactions entre la protéine et la surface, de même que dans une certaine mobilité de la protéine dans le plan de la surface ou encore dans le changement de conformation de la protéine. I. Lundstr\"{o}m et H. Elwing \citep{lundstrom1990} ont développé un modèle cinétique en faisant l'hypothèse que certaines suites d'événements pouvaient se produire: adsorption-désorption, dénaturation-désorption, désorption-renaturation, etc. Quoique critiquable car ne tenant pas bien compte de l'irréversibilité de l'adsorption des protéines de même que de considérations d'ordre microscopique, cet élégant modèle cinétique peut, d'une certaine manière, faire penser à l'effet Vroman. L'effet Vroman, se manifeste lorsqu'une dilution de plasma sanguin est mise en contact avec une surface hydrophile ou hydrophobe. Les concentrations adsorbées des diverses protéines de la dilution que l'on observe alors montrent une variabilité dans le temps, en particulier pour le fibrinogène \citep{wojciechowski1986,slack1989,leduc1995}. De façon générale, cet effet est résumé par C.~A. LeDuc \textit{et al.} \citep{leduc1995} de la façon suivante:
\begin{enumerate}
\item le type de protéine qui sera dominant (\textit{i.e.} albumine, fibrinogène, facteur VII ou autre) sur la surface dépendra de la durée d'exposition de cette surface à la dilution de plasma;
\item la plupart des protéines apparaîtront sur la surface et puis en disparaîtront et la séquence qui sera observée sera plus importante dans le cas d'une surface hydrophile par rapport à une hydrophobe.
\end{enumerate}
Dans leur explication des mécanismes d'adhésion des protéines appliquée à l'effet Vroman, ces mêmes auteurs complètent I. Lundstr\"{o}m et H. Elwing \citep{lundstrom1990} en ajoutant aux stricts phénomènes de la surface, l'évolution des concentrations en protéines dans la solution. Cette évolution est due aux diverses géométries et poids moléculaires, à la présence ou non de convection, de la dilution du plasma, etc. \citep{leduc1995}. L'effet Vroman est un beau problème en ce sens qu'il associe une grande partie des caractéristiques de l'adsorption des protéines et qu'il s'applique à un mélange de protéines qui se comporteront dans le processus d'adsorption de l'ensemble du film selon leur propre <<~personnalité~>> \citep{norde2008}.

La thermodynamique de l'adsorption est un autre aspect important. De par leur nature, M. Malmsten \citep{malmsten1998} insiste sur le fait que les protéines font aussi partie des polymères et des colloïdes et devraient donc, à ce titre, être soumises à la même thermodynamique. Peu d'études traitent exhaustivement de la thermodynamique de l'adsorption des protéines. De même, on pourra constater, malgré les importants volumes, un relatif manque de notions efficaces dans le traitement de la thermodynamique des colloïdes. La description qu'en fait R. Hunter \citep{hunter1987} est certainement une des plus abouties. Pour les protéines, un article restant largement à jour de W. Norde et J. Lyklema \citep{norde1979} abordent certains aspects. Bien que critiquables, il est aussi nécessaire de citer les importants travaux de C.~J. van Oss \textit{et al}. \citep{vanoss1986} dans lequel il demeure une confusion entre les énergies potentielles dues aux forces de van der Waals (considérations microscopiques) et les variations d'enthalpie libre (considérations macroscopiques). De façon générale, H. Reiss \citep{reiss1965} souligne à quel point la thermodynamique semble être une matière assez mal comprise; entre autre, on citera comme exemple cette application à l'adsorption des protéines que fait E.~A. Vogler \citep{vogler2012} d'une formule (5.56.4) de E.~A. Guggenheim \citep{guggenheim1949} dévolue aux milieux continus (par opposition à colloïdal). On notera aussi le mauvais usage qui est fait du mot \emph{irréversible}. L'adsorption des protéines est un phénomène irréversible comme tous les phénomènes naturels, nous le verrons plus loin \citep{planck1913}, mais le fait que, lors du rinçage d'un film ou d'un changement de concentration dans la suspension qui lui est adjacente, rien (ou presque rien) ne se désorbe n'en est pas une manifestation puisque le système ne demeure pas dans les mêmes conditions (l'irréversibilité se manifeste pour des systèmes laissés à eux-même \citep{planck1913}). Les raisonnements invoqués dans une théorisation de ce qui semblerait être une isotherme de Langmuir pour les protéines tels que ceux de C.~A. Haynes \citep{nordehaynes} semblent ainsi plus que discutables. Pour l'aspect thermodynamique, on pourra retenir de l'étude de W. Norde et J. Lyklema \citep{norde1979} que la production d'entropie lors du processus d'adsorption se fait essentiellement à partir de la déshydratation de la surface adsorbante, cet événement comprenant un réarrangement structural à l'intérieur de la protéine induisant, lui aussi, une déshydratation de la protéine. Il est utile de retenir ces conclusions car elles sont assez proches de ce qui sera développé dans la suite de ce travail.

Les données contenues dans la littérature sont fort nombreuses et très disparates car récoltées dans des systèmes expérimentaux très divers. Il est dès lors assez difficile de tirer des conclusions claires et précises sur l'adsorption des IgG, ce qui transparait bien de la synthèse de C.~E. Giacomelli \citep{giacomelli2006}. Fixer un cadre efficace et réaliste, c'est-à-dire pouvant être mis en {\oe}uvre expérimentalement, est nécessaire si l'on veut tenter une étude théorique de l'adsorption des protéines. Afin de répondre à la question: \textit{Comment les IgG s'organisent-elles sur une surface hydrophobe?}; on commencera, dans la section suivante, par passer en revue quelques aspects relevant du transfert des IgG vers la surface.

\section*{Transport des IgG dans le dispositif expérimental}
\addcontentsline{toc}{section}{Transport des IgG dans le dispositif expérimental}
\markboth{INTRODUCTION}{TRANSPORT DES IGG}

Le mode de production le plus standard des ELISA se fait dans des plaques multipuits illustrées à la figure \ref{ELISAplate}, le support sur lequel les anticorps sont fixés étant simplement les faces internes des parois de polystyrène de chaque puits constituant la plaque. Le modèle expérimental qui permettra de développer la formation du film d'anticorps est illustré à la figure \ref{FigModExp}. Il consiste en un volume cylindrique découpé d'une plaque multipuits dont le côté plat supérieur est ouvert afin de permettre l'injection et le retrait des divers fluides nécessités par le protocole. Les parois du cylindre sur lesquelles viennent se fixer les anticorps sont supposées parfaitement lisses à l'échelle macroscopique de telle sorte que les effets éventuels dus à la rugosité puissent être négligés.

\begin{figure}[h]\centering
\includegraphics[width=9cm]{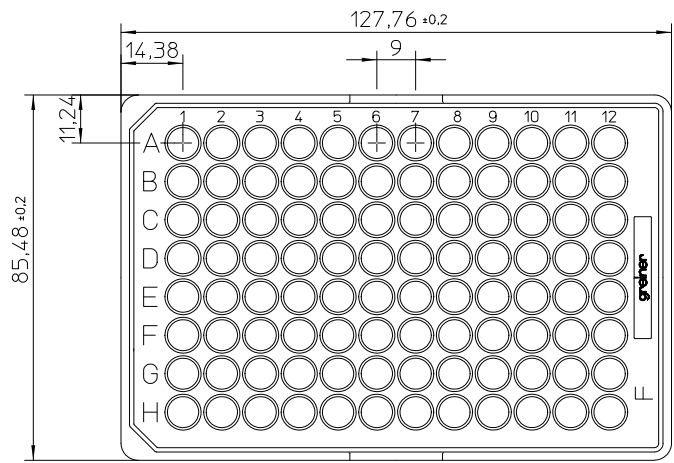}
\caption[Dessin d'une microplaque ELISA]{Dessin (vue du dessus) d'une microplaque 96 puits en polystyrène utilisée pour l'ELISA. Les dimensions sont en millimètres. Référence: 655061, Greiner Bio-One GmbH.}\label{ELISAplate}
\end{figure}

\begin{figure}[h]\centering
\includegraphics[width=7cm]{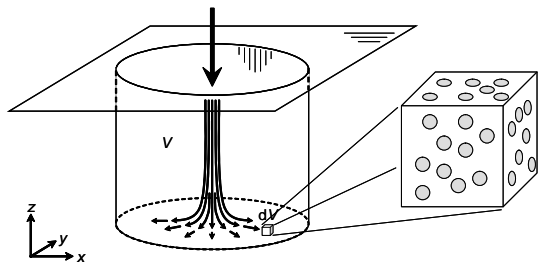}
\caption[Le dispositif expérimental modèle composé d'un puits de volume $V$]{Le dispositif expérimental composé d'un puits de volume $V$ dont les parois sont en polystyrène. L'injection d'une suspension colloïdale (protéines, IgG, etc.) se réalise par l'ouverture supérieure du puits (flèche épaisse). L'espace des coordonnées cartésiennes $\boldsymbol{x}=\lbrace (x,y,z)\in\mathbb{R}^3\rbrace$ est divisé en volumes infinitésimaux $\mathrm{d}V$ dont une représentation est montrée à droite de la figure et au sein duquel les colloïdes sont représentés.}\label{FigModExp}
\end{figure}

On étudiera en particulier le film d'anticorps se formant sur la paroi inférieure du puits afin de bénéficier d'une surface macroscopiquement plane, c'est-à-dire ne présentant pas de rayon de courbure. La planéité de cette surface permet, en outre, de la substituer par un senseur piézoélectrique tels que ceux utilisés dans une microbalance à cristal de quartz (QCM-D) ou de la soumettre à l'analyse d'un microscope à force atomique (AFM).

La première étape de la fabrication du film est l'injection de la suspension colloïdale dans le réservoir cylindrique. Typiquement, cette suspension contient 5 \textgreek{m}g d'IgG par mL dispersés dans le PBS (\textit{phosphate buffer saline} ou tampon phosphate), un tampon phosphate préparé par dissolution de $0,2$ g de KCl, $8$ g de NaCl, $0,88$ g de KH$_2$PO$_4$ et $1,28$ g de Na$_2$HPO$_4\cdot$12H$_2$O dans de l'eau distillée. La force ionique est évaluée à environ $0,1$ mol$\cdot$L$^{-1}$ et le pH est porté à $7,4$ avec de la soude caustique. L'expérimentateur prélève un millilitre de la suspension en utilisant une micropipette et en éjecte ensuite le liquide dans le réservoir en faisant que le jet vienne frapper perpendiculairement et en son centre le fond du puits. Le liquide, considéré comme incompressible (masse volumique invariable), manifeste alors un mouvement qu'il est possible de conceptualiser grâce à l'équation de Navier-Stokes donnée par l'équation \ref{EqNavierStokes} d'après L.~D. Landau et E.~M. Lifshitz \citep{landau6}.
\begin{equation}\label{EqNavierStokes}
\frac{\partial\mathbf{u}}{\partial t}+\mathbf{u}\cdot\nabla\mathbf{u}=\frac{1}{\rho}\nabla p+\frac{\eta}{\rho}\nabla^2\mathbf{u}
\end{equation}
où le membre de gauche représente l'accélération d'un élément de fluide sous la forme de la dérivée lagrangienne de la vitesse $\mathbf{u}$ de celui-ci. L'opérateur $\partial/\partial t$ est la dérivée partielle en fonction du temps, $\nabla^2$ l'opérateur laplacien et $\nabla$ l'opérateur gradient. La pression est symbolisée par $p$, la masse volumique par $\rho$ et la viscosité par $\eta$. Pour les fluides inviscides (ou lorsque les effets de la viscosité sont faibles comme dans le cas des grands systèmes), le second terme du membre de droite se simplifie, fournissant une nouvelle équation du mouvement dite d'Euler. Dans le cas des puits d'une microplaque utilisée pour l'ELISA, les puits ont un diamètre allant de quelques millimètres à un centimètre, une faible taille qui aura pour conséquence d'accroître les effets dus à la viscosité. L'approximation d'Euler n'est dès lors pas viable puisque le membre de droite (terme de Stokes) de l'équation \ref{EqNavierStokes} sera prépondérant.

Les aspects rhéologiques ont certes une influence importante mais ne sont pas les seuls à pouvoir décrire l'accumulation des IgG sur la surface du support. En effet, l'équation de Navier-Stokes n'invite qu'à considérer le mouvement des éléments du fluide $\mathrm{d}V$ (voir figure \ref{FigModExp}) qui ne font que transporter la matière (dont la phase dispersée) vers des lieux de l'espace, pour certains, proches de la surface. Le film que l'on souhaite considérer est une accumulation sur la surface d'un grand nombre d'éléments, les anticorps, de la phase dispersée. Considérer le transport comme étant celui d'un fluide monophasique n'est donc pas vraiment tenable et il est donc nécessaire d'étudier la manière dont la phase dispersée se désolidarisera en partie de la phase continue pour venir se coller sur la surface en polystyrène du support. Le point de vue adopté dans l'équation de Navier-Stokes est celui d'un système monophasique qu'il serait bien hardi de vouloir abandonner compte tenu des difficultés potentiellement engendrées. Par contre, à une échelle différente, celle de l'élément infinitésimal $\mathrm{d}V$, il est possible de considérer la suspension colloïdale comme deux phases séparées, ce qu'elle est réellement.

Afin de décrire le déplacement des colloïdes dans la phase liquide suspendante, il sera fait l'hypothèse que le fluide ne subit aucun mouvement de convection susceptible de modifier la distribution des IgG dans l'espace. Cette hypothèse peut revenir à focaliser notre attention sur un petit volume de fluide proche de la surface solide sur laquelle se produira l'adhésion. Dans ce petit espace, une série de coordonnées seront utilisées afin de décrire le déplacement et les changements d'orientation d'une IgG. Ces coordonnées généralisées, au nombre de six, seront décrites par un pseudovecteur $\boldsymbol{q}$ et l'on parlera alors du déplacement des IgG dans le $\boldsymbol{q}$-espace. Ce pseudovecteur $\boldsymbol{q}$ est lui-même un ensemble de coordonnées $\lbrace\boldsymbol{x},\boldsymbol{\omega}\rbrace$ permettant de décrire la position des particules dans l'espace des coordonnées cartésiennes $\boldsymbol{x}=\lbrace(x,y,z)\in\mathbb{R}^3\rbrace$ mais aussi leur orientation \textit{via} les coordonnées angulaires eulériennes $\boldsymbol{\omega}=\lbrace(\vartheta,\varphi,\psi):0\leqslant\vartheta\leqslant\pi,0\leqslant\varphi\leqslant2\pi,0\leqslant\psi\leqslant2\pi\rbrace$ \citep{goldstein1980}.

Lorsque le transport convectif ($\mathbf{u}\cdot\nabla\mathbf{u}$) est négligé, la description des flux de matière passe naturellement par la nécessaire conservation de cette dernière, amenant alors l'équation de continuité \ref{EqMécaConti} où $\textbf{j}(\boldsymbol{q},t)$ est le flux à un instant $t$ dans l'espace des coordonnées généralisées \citep{batchelor1967}.
\begin{equation}\label{EqMécaConti}
\frac{\partial}{\partial t}C(\boldsymbol{q},t)=-\nabla\cdot\textbf{j}(\boldsymbol{q},t)
\end{equation}
L'équation de continuité formalise le fait que la variation de $C(\boldsymbol{q},t)$, la concentration en particules en un point matériel $\boldsymbol{q}$, est l'opposé de la divergence (opérateur <<~$-\nabla\cdot$~>>) de $\textbf{j}(\boldsymbol{q},t)$, le flux de particules, en ce point. En l'absence de mouvement convectif, l'équation constitutive du flux de particules $\textbf{j}$ est donnée, d'après H. Brenner et L. G. Leal \citep{brenner1977b}, par
\begin{equation}\label{EqMécaFlux}
\textbf{j}(\boldsymbol{q},t)=-\mathbf{D}\cdot\nabla C(\boldsymbol{q},t)+\mathbf{M}\cdot C(\boldsymbol{q},t)\cdot\textbf{F}(\boldsymbol{q},t)
\end{equation}
dans laquelle $\mathbf{D}$ est la matrice des coefficients de diffusion (m$^2$.s$^{-1}$) dans chaque direction du $\boldsymbol{q}$-espace, $\mathbf{M}$ la matrice des mobilités liées aux coefficients de diffusion \textit{via} l'équation d'Einstein-Sutherland \citep{bretonnet2010}:
\begin{equation}\label{EqMécaEinsteinSutherland}
\mathbf{D}=k_BT\,\mathbf{M}
\end{equation}
et $\textbf{F}(\boldsymbol{q},t)$ la force résultante à laquelle sont soumises les particules suspendues en chaque point du $\boldsymbol{q}$-espace. Ensuite, en injectant l'équation \ref{EqMécaFlux} constitutive du flux de matière dans l'équation \ref{EqMécaConti} de continuité et en remarquant que la force (conservative) est liée au potentiel par la relation $\textbf{F}(\boldsymbol{q},t)=-\nabla U(\boldsymbol{q},t)$, on obtient l'équation de la diffusion généralisée de M. von Smoluchowski \citep{bretonnet2010}:
\begin{equation}\label{EqMécaSmol}
\frac{\partial}{\partial t}C(\boldsymbol{q},t)=\mathbf{D}\cdot\nabla^2 C(\boldsymbol{q},t)+\mathbf{M}\cdot\nabla\cdot\big[C(\boldsymbol{q},t)\cdot\nabla U(\boldsymbol{q},t)\big].
\end{equation}

Il est possible de voir l'équation de Smoluchowski comme une généralisation de la loi de diffusion de Fick par l'ajout d'un terme tenant compte des forces dérivant des champs de potentiels auxquels peuvent être soumis les particules diffusantes. Ce terme est, bien entendu, le second du membre de droite de l'équation \ref{EqMécaSmol} sans lequel on retrouverait l'équation de Fick.

Le second terme de l'équation \ref{EqMécaSmol} contenant le gradient de potentiel, c'est-à-dire les forces <<~externes~>> auxquelles sont soumises les particules, n'est rien d'autre que l'ensemble des forces dues à la présence de la surface solide (et des surfaces des autres protéines) dont les conséquences seront décrites ci-après. Elles s'appliquent aux particules diffusantes selon leur position dans l'espace.

Si les forces externes sont essentiellement dues à la surface solide, il est commode de séparer la diffusion en deux zones: la première, assez éloignée de la surface, où l'intensité des forces dérivant du champ de potentiel dû à la surface est négligeable et une deuxième zone, proche de la surface, où l'échelle de longueur caractéristique des variations par cause des forces $\textbf{F}(\boldsymbol{q},t)$ est large par rapport au déplacement quadratique moyen des particules soumises à un mouvement brownien typique \citep{brenner1977b} ($\sim0,03$ nm pour les IgG \citep{vandeven1989}), ces forces domineront donc la diffusion. La première zone, éloignée de la surface, revient à négliger le second terme du membre de droite de l'équation \ref{EqMécaSmol} afin de ne retenir que la diffusion <<~fickienne~>> (diffusion répondant à la loi de Fick). La seconde zone, proche de la surface, est caractérisée par le fait que la diffusion fickienne est négligée car totalement dominée par les forces externes, on y tiendra compte des forces caractéristiques des interfaces: forces électrostatiques, forces de Hamaker issues des forces de van der Waals et forces de volume exclu.

La diffusion des particules selon la loi de Fick s'obtient lorsque le mouvement brownien auquel elles sont soumises surpasse en intensité les forces externes. De telles conditions permettant de négliger le second terme du membre de droite de l'équation \ref{EqMécaSmol} s'obtiennent loin de la surface.
Ce processus diffusionnel, guidé par l'aplanissement des gradients de concentration, se fait en déplacement et rotation de telle sorte qu'il se décrive efficacement dans le $\boldsymbol{q}$-espace. L'équation de la diffusion loin de la surface se résume au premier terme du membre de droite de l'équation \ref{EqMécaSmol} qui n'est autre que l'équation de la diffusion de Fick:
\begin{equation}\label{EqFickIntro}
\frac{\partial}{\partial t}C(\boldsymbol{q},t)=-\nabla^2 C(\boldsymbol{q},t)
\end{equation}
qui, comme il le sera décrit ci-après, pourra se développer selon les coordonnées $\boldsymbol{x}$ ou $\boldsymbol{\omega}$. Avant d'aller plus loin il est important de remarquer que, le fluide baignant les colloïdes étant susceptible de mouvements convectifs, la diffusion n'est \textit{a fortiori} pas seule à agir dans leur dispersion. En présence de ces deux phénomènes de convection et de diffusion, la convection favorise l'action de la diffusion et finit par accélérer le mélange dans son ensemble\footnote{L'accélération du transport des colloïdes par convection est un phénomène justifiant l'utilisation d'une cuillère afin de mélanger un café dans lequel on viendrait d'additionner quelques gouttes d'un liquide colloïdal comme le lait.}. La diffusion est bien la seule cause du mélange entre les molécules proprement dites, cette dernière ayant pour action d'aplanir les gradients de concentration (\textit{cf.} équation \ref{EqFickIntro}) mais, la convection, en faisant se mouvoir les uns par rapport aux autres les points constitutifs du système et déplaçant donc la concentration, créera des gradients de concentration à de multiples endroits donnant ainsi globalement plus de grains à moudre à la diffusion, augmentant alors l'efficacité de cette dernière \citep{stocker2012,taylor2012}. Ce problème qui lie les événements appartenant à deux échelles différentes (l'échelle de la convection et l'échelle de la diffusion) peut se formaliser dans le cadre de la théorie généralisée de Taylor \citep{frankel1989}.

Une protéine n'est pas l'autre et si l'on s'intéresse à l'adsorption des IgG, il est nécessaire de se pencher sur ses spécificités. Or, les IgG disposent d'une structure quaternaire bien particulière ayant l'aspect d'un <<~Y~>>. Cette structure, quoique dictée par ses fonctions immunologiques, lui conférera des propriétés fort singulières lors de l'adsorption. Afin d'en tenir compte, il est nécessaire de les considérer comme des particules anisotropiques, à l'inverse d'une sphère dont l'orientation dans l'espace n'a pas vraiment de sens (particule isotropique).

On voit dès lors l'intérêt d'introduire les angles d'Euler \citep{goldstein1980} $\boldsymbol{\omega}=\lbrace\vartheta,\varphi,\psi\rbrace$, ceux-ci permettant de paramétriser la position et l'orientation des IgG dans l'espace des coordonnées généralisées $\boldsymbol{q}=\lbrace\boldsymbol{x},\boldsymbol{\omega}\rbrace$. Les IgG ne diffuseront donc pas seulement selon les trois dimensions $\boldsymbol{x}=\lbrace x,y,z\rbrace$ de l'espace cartésien classique mais bien dans un espace en contenant six, certaines de ces directions pouvant être corrélées entre elles sous l'effet de la traînée hydrodynamique, le solvant étant considéré comme visqueux.

Dans deux articles importants traitant du mouvement brownien d'une particule ellipsoïdale, F. Perrin \citep{perrin1934,perrin1936} montre, en résolvant l'équation de la diffusion de Fick, comment les particules finissent par atteindre un large spectre d'orientations mais aussi à quel point il est difficile d'entièrement séparer le mouvement de translation de celui de rotation. Il s'avère en effet impossible, d'après F. Perrin \citep{perrin1936}, de déterminer les translations d'une molécule non sphérique sans tenir compte de ses rotations. 

La résolution de ce problème purement mécanique est donnée, en plus de F. Perrin, par T. van de Ven \citep{vandeven1989} et une discussion pour des particules de formes plus arbitraires est faite par H. Brenner \citep{brenner1965,brenner1967}. Ce phénomène de couplage des composantes de la diffusion se manifestera dans la matrice des coefficients de diffusion $D$ par des éléments non diagonaux \citep{vandeven1989}. En effet, comme tout corps se déplaçant dans un fluide visqueux, les IgG sont sensibles aux frottements qui se manifesteront et, lors d'un de ses mouvements de translation et/ou de rotation vers la surface sous l'effet d'un gradient de concentration suffisamment fort \citep{perrin1936}, il aura tendance à les minimiser respectivement par un autre mouvement de rotation et/ou de translation. Ce transport dans l'espace des coordonnées généralisées $\boldsymbol{q}$ commence par l'énonciation de l'équation de continuité, développement de l'équation \ref{EqMécaConti}, donnée par J. M. Nitsche et H. Brenner \citep{nitsche1990}:

\begin{equation}\label{EqMécaConti2}
\frac{\partial}{\partial t}C(\boldsymbol{q},t)=-\nabla_{\boldsymbol{x}}\cdot\textbf{j}(\boldsymbol{x},t)-\nabla_{\boldsymbol{\omega}}\cdot\textbf{j}(\boldsymbol{\omega},t).
\end{equation}

Ensuite, les équations constitutives des flux de matière dans l'espace des coordonnées $\boldsymbol{q}$ s'écrivent selon l'équation \ref{EqMécaFluxTrans} pour le flux de translation et l'équation \ref{EqMécaFluxOri} pour le flux de rotation \citep{nitsche1990}.

\begin{flalign}
\textbf{j}(\boldsymbol{x},t)&=-\textbf{D}_{\boldsymbol{xx}}\cdot\nabla_{\boldsymbol{x}}C(\boldsymbol{q},t)-^\dagger\textbf{D}_{\boldsymbol{\omega}\boldsymbol{x}}\cdot\nabla_{\boldsymbol{\omega}}C(\boldsymbol{q},t)\label{EqMécaFluxTrans}\\
\textbf{j}(\boldsymbol{\omega},t)&=-\textbf{D}_{\boldsymbol{\omega x}}\cdot\nabla_{\boldsymbol{x}}C(\boldsymbol{q},t)-\textbf{D}_{\boldsymbol{\omega\omega}}\cdot\nabla_{\boldsymbol{\omega}}C(\boldsymbol{q},t)\label{EqMécaFluxOri}
\end{flalign}

Les éléments $\textbf{D}_{\boldsymbol{xx}}$, $\textbf{D}_{\boldsymbol{\omega\omega}}$ et $\textbf{D}_{\boldsymbol{\omega x}}$ sont, respectivement, les tenseurs de translation, de rotation et de couplage entre la translation et la rotation. L'opérateur <<~$\dagger$~>> est l'opérateur de transposition. Pour une particule anisotropique, le terme de couplage sera généralement non nul de telle sorte que toute force appliquée au centre de masse de la particule aura sa contrepartie sous forme d'un couple lui faisant ajuster son orientation. En d'autres mots, un déplacement dans l'espace des coordonnées $\boldsymbol{x}$ sera compensé par un déplacement dans l'espace des coordonnées $\boldsymbol{\omega}$. Comme montré à la figure \ref{FigMécaFick}, un ellipsoïde se déplaçant sous l'influence de son gradient de concentration finira par aligner son grand axe parallèlement au sens de son déplacement afin de minimiser la traînée hydrodynamique que lui impose ce déplacement.

\begin{figure}[h]\centering
\includegraphics[height=3cm]{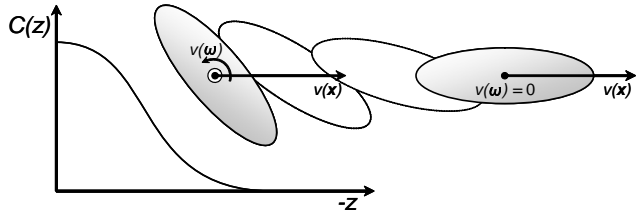}
\caption[Couplage des mouvements browniens de translation et de rotation]{Effet du couplage des mouvements browniens de translation et de rotation sur l'orientation d'un ellipsoïde se déplaçant à une vitesse $v(\boldsymbol{x})$ par diffusion dans le sens opposé à son gradient de concentration $C$. Il subit un couple $\boldsymbol{\tau}$ l'animant d'une vitesse de rotation $v(\boldsymbol{\omega})$ lui faisant aligner son grand axe parallèlement au sens de son déplacement afin de minimiser les forces de frottement.}\label{FigMécaFick}
\end{figure}

\`{A} l'instar de l'ellipsoïde de la figure \ref{FigMécaFick}, la diffusion translationnelle du centre d'inertie de l'IgG étant dominante puisque l'on doit transporter l'IgG d'un point de l'espace des coordonnées $\boldsymbol{x}$ à un autre proche de la surface, l'IgG subira un couple $\boldsymbol{\tau}$ sous l'effet de la traînée hydrodynamique de frottement. En effet, le fragment F(ab')$_2$ de l'IgG étant deux fois plus gros que son fragment Fc, le premier sera soumis de manière bien plus intense aux frottements que le second. La résultante des forces de frottements s'exerçant dans le sens opposé au mouvement du centre d'inertie de l'IgG, le Fc aura tendance à précéder le F(ab')$_2$ dans le mouvement de translation (le Fc fera face au gradient de concentration alors que le F(ab')$_2$ sera, en quelque sorte, retenu vers l'arrière).

Ce phénomène ne se produira que lorsque la diffusion dans l'espace des coordonnées $\boldsymbol{x}$ est dominante par rapport à la diffusion les coordonnées $\boldsymbol{\omega}$ sous l'effet du gradient de concentration, en opposition à l'effet d'étalement orientationnel dû au mouvement brownien \citep{perrin1936,batchelor1977} rencontré sur des temps plus longs. On peut donc penser que, grâce au transport et aux interactions hydrodynamiques, les IgG arrivant à l'interface s'y présenteront majoritairement selon une orientation \textit{end-on} (le fragment constant Fc en premier), et ce, d'autant plus que le voyage qu'elles auront effectué aura été suffisamment long pour leur donner le temps de s'aligner de la sorte.

Ces aspects relatifs au transport des IgG permettent de se faire une première idée de l'état (orientation) selon lequel elles arriveront à la surface, juste avant d'y adhérer. Au-delà de cet instant, il faudra tenir compte d'aspects relevant du second terme du membre de droite de l'équation de Smoluchowski (équation \ref{EqMécaSmol}), c'est-à-dire des forces exercées par la surface sur l'IgG: les forces électrostatiques, les forces de van der Waals et la force de résistance de la surface et des autres IgG (impossibilité pour l'IgG de traverser la surface ou une autre IgG). Les aspects liés à la thermodynamique seront aussi à considérer. Quels seront les effets sur la structure de la monocouche et le devenir de cette IgG arrivant selon une orientation \textit{end-on} à la surface? \`{A} cette question, le présent travail tentera d'apporter quelques pistes explicatives ainsi qu'une preuve expérimentale.

\begin{footnotesize}

\end{footnotesize}\end{cbunit}
\begin{cbunit}
\chapter[Thermodynamique de l'adhésion des protéines]{Thermodynamique de l'adhésion des protéines}\label{Chap1}
\markboth{Chapitre \ref{Chap1}: THERMODYNAMIQUE}{}
\minitoc

\section{Généralités}\label{hb35d8f}

Les notions exposées dans le chapitre introductif permettent de se donner une première idée de la façon dont les IgG, et les protéines en général, sont transportés vers les parois hydrophobes des puits d'une microplaque ELISA (voir figures \ref{ELISAplate} et \ref{FigModExp}). D'après les données fournies par la littérature, il semblerait que les IgG arrivent à la surface de polystyrène dans une orientation \textit{end-on}. Une fois arrivées en contact très intime avec le polystyrène, les IgG subiront l'influence des potentiels qui y sont associés en premier lieu desquels le potentiel électrostatique dû aux charges portées par la surface des IgG et du polystyrène ainsi que le potentiel de Hamaker émanant des forces de London-van der Waals. Relevant d'aspects d'ordres cinétiques, ces potentiels (théorie DLVO), typiques des systèmes colloïdaux et dont dérivent des forces susceptibles d'influencer le comportement des IgG vis-à-vis de la surface de polystyrène, seront discutés dans ce chapitre après avoir développé les aspects thermodynamiques de l'adhésion des IgG sur le polystyrène.

La thermodynamique est une branche de la physique classique (macroscopique) permettant de conceptualiser et de prévoir l'évolution des systèmes physico-chimiques. Elle s'y emploiera en faisant usage de fonctions, au sens mathématique du terme, dont on pourra décrire l'évolution en un ou plusieurs points de leur domaine de définition, et ce, à l'aide d'outils tels que la différentielle, notée <<~$\mathrm{d}$~>>. Dans le cadre de la thermodynamique, ces fonctions sont nommées \emph{fonctions d'état} car elles permettent de décrire l'état du système physico-chimique d'intérêt à un moment donné de son évolution, et ce, en fonction de diverses variables nommées \emph{variables d'état}. Une de ces fonctions, bien connue des chimistes et introduite par J.~W. Gibbs \citep{Gibbs1878}, est la fonction d'état $G$ donnant l'\emph{enthalpie libre} (en J) du système et dont le domaine de définition peut être décrit selon deux variables d'état lorsque le système est fermé: la température $T$ (en K) et la pression $p$ (en Pa ou J$\cdot$m$^{-3}$). Cette fonction
\begin{equation}
G(T,p)
\end{equation}
est différentiable de telle sorte que sa \emph{différentielle totale} s'écrive
\begin{equation}\label{DiffTot1E}
\mathrm{d}G=\bigg(\frac{\partial G}{\partial T}\bigg)_p\mathrm{d}T+\bigg(\frac{\partial G}{\partial p}\bigg)_T\mathrm{d}p
\end{equation}
dans laquelle on retrouve les \emph{dérivées partielles} $(\partial G/\partial T)_p$ et $(\partial G/\partial p)_T$ représentant respectivement la variation de $G$ lors d'une variation infinitésimale de la variable $T$ ou $p$, et ce, de façon exclusive. Les ouvrages généraux de thermodynamique \citep{guggenheim1949,kondepudi1998} montrent que ces dérivées partielles $(\partial G/\partial T)_p$ et $(\partial G/\partial p)_T$ correspondent respectivement à l'opposé de la fonction d'état \emph{entropie} (en J$\cdot$K$^{-1}$) et à la fonction d'état \emph{volume} (en m$^3$) du système (voir section \ref{35st14dw3b1}). La différentielle totale \ref{DiffTot1E} peut être réécrite selon
\begin{equation}\label{DiffTot2E}
\mathrm{d}G=\mathrm{d}_TG+\mathrm{d}_pG
\end{equation}
dans laquelle $\mathrm{d}_TG$ et $\mathrm{d}_pG$ sont des \emph{différentielles partielles}\footnote{Les différentielles partielles permettent de séparer les contributions à un changement dont est affecté le système thermodynamique. Le système thermodynamique demeure une entité totale, une et indivisible, caractérisé par une fonction d'état $G$ (ou autre) qui l'est tout autant. La différentielle totale de la fonction d'état $G$ ($\mathrm{d}G$) conceptualise un changement infinitésimal du système, lequel changement peut être attribué à différentes contributions liées aux variables d'état le caractérisant. La somme des changements attribués à chaque changement des variables d'état, changements conceptualisés par les différentielles partielles, donne le changement total affectant le système.}. Ces différentielles partielles sont la part de la différentielle totale $\mathrm{d}G$ due respectivement à une variation infinitésimale de la variable $T$ et de la variable $p$. Par comparaison des relations \ref{DiffTot1E} et \ref{DiffTot2E}, on aura alors
\begin{equation}
\mathrm{d}_TG=\bigg(\frac{\partial G}{\partial T}\bigg)_p\mathrm{d}T\qquad\text{et}\qquad\mathrm{d}_pG=\bigg(\frac{\partial G}{\partial p}\bigg)_T\mathrm{d}p
\end{equation}
donnant la signification des deux différentielles partielles. Au contraire de la différentielle totale $\mathrm{d}G$, les différentielles partielles $\mathrm{d}_TG$ et $\mathrm{d}_pG$ ne sont pas intégrables. Ensuite, c'est en divisant la différentielle $\mathrm{d}G$ par un intervalle de temps infinitésimal $\mathrm{d}t$ pendant lequel a lieu la transformation que l'on obtiendra tout naturellement $\mathrm{d}G/\mathrm{d}t$, la \emph{dérivée} de l'enthalpie libre $G$ en fonction du temps (en J$\cdot$s$^{-1}$). Ces diverses relations ne seront pas d'une grande utilité dans le raisonnement physico-chimique qui sera suivi dans ce chapitre mais elles permettent de rappeler, avant leur usage intensif, ce que sont l'une par rapport à l'autre les notions de \emph{fonction d'état} et \emph{variable d'état} ainsi que leur \emph{différentielle totale}, \emph{dérivée partielle}, \emph{différentielle partielle} et \emph{dérivée}.

La thermodynamique se sert des outils susmentionnés afin de décrire les systèmes physico-chimiques mais elle dispose aussi d'une série d'axiomes à partir desquels tout le raisonnement qui suivra pourra être déduit. Ces axiomes dont la valeur est essentiellement empirique sont au nombre de quatre dont on ne rapportera que les trois premiers, le quatrième (principe de Nernst) n'étant pas utile dans le cadre de ce travail. Le premier axiome de la thermodynamique permet de définir l'\emph{équilibre} d'un système, état dans lequel il n'évoluera plus; il s'agit du <<~principe zéro~>> de la thermodynamique.
\begin{axiome}[Principe zéro de la thermodynamique]
Deux systèmes en équilibre avec un même troisième sont en équilibre entre eux \citep{guggenheim1949}.
\end{axiome}
La conséquence de ce principe est l'absence de flux d'énergie (sous forme de chaleur, travail, transferts de matière) entre le système et son environnement lorsqu'un système est à l'équilibre. Par extension, on admettra que cet équilibre se matérialise aussi à l'intérieur du système par l'absence de tout flux d'énergie entre les corps le constituant (flux causés par des transferts de chaleur, un travail, des transferts de matière ou même des transformations de celle-ci). De manière générale, on retiendra qu'à l'équilibre statique toutes les variables d'état caractérisant le système seront fixées, fait qui implique nécessairement l'impossibilité de tout transfert, tandis qu'à l'équilibre thermodynamique seules les variables d'état caractérisant les degrés d'avancement des processus pouvant se produire indépendamment de tout processus d'échange entre le système et son environnement seront fixées.

L'énoncé du principe zéro de la thermodynamique amène aussi à approfondir la notion de \emph{système thermodynamique}, de ce dont il est constitué et de ce qui fonde sont état d'équilibre ou de déséquilibre thermodynamique. La thermodynamique utilise trois types de systèmes: les systèmes \emph{isolés}, \emph{fermés} et \emph{ouverts}. Un système \emph{isolé} est un système qui ne peut rien échanger avec son environnement; ni matière, ni énergie. Un système \emph{fermé} peut échanger de l'énergie avec son environnement mais pas de matière. Pour terminer, le système \emph{ouvert} est le système qui a la possibilité d'échanger matière et énergie avec son environnement \citep{prigogine1968}.

De plus, ces systèmes, quelles que soient leurs capacités d'échanges avec leurs environnements, peuvent être constitués de plusieurs \emph{corps}. Entre ces corps, dont il est fait référence dans l'interprétation du principe zéro, peuvent exister des flux de chaleur pouvant être liés à des transferts d'énergie. L'existence de tels flux indiquera que le système n'est pas à l'équilibre thermodynamique tandis que leur absence sera un critère d'équilibre du système. Ce transfert de chaleur entre les corps d'un même système sera noté $Q^\prime$ (en J) par opposition à la notation $Q$ (en J) réservée aux échanges de chaleur entre le système et son environnement. Ces notions seront réexaminées lors de l'introduction du second principe de la thermodynamique.

Toutes ces définitions donnent maintenant la possibilité de délimiter clairement le système thermodynamique  sur base duquel le raisonnement de ce chapitre sera focalisé. Le système thermodynamique, illustré à la figure \ref{Fig_Chap1_1}, dispose d'un volume $V$ (en m$^3$) délimité par une frontière fermée, c'est-à-dire que le système pourra échanger de l'énergie avec son environnement mais pas de matière. L'illustration de la figure \ref{Fig_Chap1_1} peut être vue comme une stylisation de la suspension illustrée à la figure \ref{FigBerlins}.A, cette suspension étant constituée d'une phase aqueuse suspendante <<~l~>> et d'une phase suspendue <<~p~>> d'un seul tenant mais figurant l'ensemble des protéines. En plus de ces deux corps <<~l~>> et <<~p~>>, est additionné un troisième corps <<~s~>> représentant la surface en polystyrène sur laquelle la phase suspendue viendra adhérer. Chacun de ces corps aura son propre volume dont la somme est égale au volume total du système:
\begin{equation}
V=V_\mathrm{s}+V_\mathrm{p}+V_\mathrm{l}.
\end{equation}
D'autre part, ces corps constitutifs du système sont délimités par des interfaces (frontières fermées) <<~sl~>> et <<~pl~>> dont les aires (en m$^2$) sont respectivement notées $\Sigma_\mathrm{sl}$ et $\Sigma_\mathrm{pl}$. Ce système correspond à un \emph{système capillaire} car constitué de plusieurs phases, système exhaustivement décrit par R. Defay et I. Prigogine \citep{defay1951}.

\begin{figure}[t] \centering
\includegraphics[width=3cm]{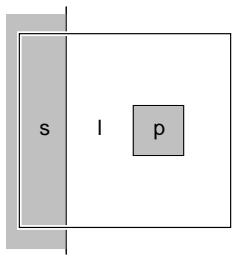}
\caption[Le système thermodynamique]{Le système thermodynamique, de volume total $V$, composé de troix corps: la phase suspendante <<~l~>> de volume $V_\mathrm{l}$, la phase suspendue <<~p~>> de volume $V_\mathrm{p}$ et la surface solide <<~s~>> de volume $V_\mathrm{s}$. Cette représentation est une stylisation (surface solide incluse) du système biphasique montré à la figure \ref{FigBerlins}.A.}\label{Fig_Chap1_1}
\end{figure}

Il paraît contradictoire de s'intéresser à la simple adhésion de la phase suspendue <<~p~>> sur la surface <<~s~>> alors que le sujet de ce travail porte sur l'adhésion des IgG sur cette surface. Toutefois, ces IgG sont des colloïdes et font donc partie de la phase suspendue <<~p~>> justifiant de l'intérêt que l'on peut porter à ce genre d'adhésion. Cette phase suspendue sera plus tard dispersée (voir figures \ref{FigBerlins}.A et B) afin d'obtenir une phase colloïdale de façon à permettre la conceptualisation de l'adhésion d'IgG sur la surface.

Dans un premier temps, les équations des potentiels de Gibbs seront développées lorsque le système évolue dans le cadre de l'équilibre thermodynamique. Cette évolution à l'équilibre est qualifiée de \emph{réversible} et l'on obtiendra dès lors des équations décrivant l'adhésion réversible de la phase suspendue sur la surface solide.

Dans un second temps, les équations obtenues seront étendues afin de tenir compte des processus \emph{irréversibles}, c'est-à-dire ceux se produisant en dehors de l'équilibre thermodynamique. En effet, un système évolue nécessairement hors de l'équilibre car un système à l'équilibre n'aurait, par définition, aucune raison de s'en éloigner. \'{E}tendre les équations caractéristiques de l'équilibre aux cas hors de l'équilibre est le seul moyen permettant de conceptualiser l'évolution spontanée du système d'intérêt. Pour ce faire, la notion d'\emph{affinité} sera introduite avant que la phase suspendue ne soit dispersée et d'autres notions telles que l'\emph{empreinte}, l'\emph{accumulation} et l'\emph{hydrophobie} ne soient définies. Toutes ces notions, alliées à la potentialité de la thermodynamique et de l'outil mathématique sur lequel elle se fonde, permettront d'envisager l'évolution spontanée d'un système colloïdal au sein duquel se formera une monocouche d'IgG sur du polystyrène.

\section{Potentiels de Gibbs à l'équilibre}

Les potentiels thermodynamiques sont des expressions introduites par J.~W. Gibbs \citep{Gibbs1878} afin de décrire les changements des diverses fonctions d'état caractérisant l'état dans lequel se trouve le système à un moment donné de son existence. Ces potentiels prennent la forme de différentielles de fonctions telles que l'énergie interne $E$, l'enthalpie $H$ et l'entropie $S$, fonctions intimement associées à l'état interne du système, système pouvant être affecté d'un changement. Afin de pouvoir différentier ces fonctions d'état et ainsi obtenir les potentiels thermodynamiques recherchés, il est nécessaire de les exprimer, à l'instar de toute fonction, sur base d'une série de variables d'état (température, pression, volumes, etc.). Ce raisonnement pourra se faire à partir de l'introduction du premier principe de la thermodynamique et d'une série de définitions.

Après l'introduction du principe zéro donnant les conditions de l'équilibre, le premier principe de la thermodynamique, dû à S.~Carnot \citep{poincare1908}, permet d'exprimer pour un système fermé l'équivalence entre le travail $W$ (en J) et la chaleur $Q$ (en J) sous forme d'énergie interne $E$ (en J). 
\begin{axiome}[Premier principe de la thermodynamique]
L'accroissement de l'énergie interne d'un système est égal à l'énergie que le système reçoit de son environnement pendant la durée de la transformation considérée \citep{prigogine1950}. Cette variation d'énergie $\Delta E$ est la somme de la chaleur échangée à travers les frontières du système et du travail mécanique effectué sur le système (on comptera positivement la chaleur et le travail reçus par le système) \citep{poincare1908,kondepudi1998,lebon2008a}.
\end{axiome}
Ce principe se traduit en équation selon $\Delta E=Q+W$ et, pour un changement infinitésimal, la variation d'énergie interne se déroulant de l'instant $t$ à $t+\mathrm{d}t$ devient \citep{prigogine1950,prigogine1968,lebon2008a}
\begin{equation}\label{EqPremierPrincipe1}
\mathrm{d}E=\mathrm{\dbar}Q+\mathrm{\dbar}W
\end{equation}
où $\mathrm{\dbar}Q$ et $\mathrm{\dbar}W$ représentent respectivement la quantité élémentaire de chaleur reçue et le travail élémentaire effectué sur le système au cours de cette transformation produisant un accroissement infinitésimal de la fonction énergie interne. Il est important de noter que la chaleur $Q$ et le travail $W$ ne sont pas des fonctions d'état du système mais seulement des quantités échangées avec son environnement et que, dès lors, $\mathrm{\dbar}Q$ et $\mathrm{\dbar}W$ en représentent des quantités élémentaires et non des différentielles.

Lorsque la surface limitrophe du système considéré subit une pression $p$, le travail mécanique élémentaire effectué sur le système et dû à cette même pression s'exprime par
\begin{equation}
\mathrm{\dbar}W=-p\mathrm{d}V
\end{equation}
où $\mathrm{d}V$ est l'accroissement infinitésimal du volume du système. Cette expression du travail mécanique est largement utilisée en physico-chimie et s'applique aux gaz et aux fluides compressibles. En injectant cette expression du travail dans l'équation \ref{EqPremierPrincipe1} du premier principe, on obtiendrait une expression bien connue de l'accroissement de l'énergie interne du système \citep{guggenheim1949}.

L'évolution du système au cours de laquelle se produit l'adhésion de la phase suspendue <<~p~>> sur la surface solide est montrée à la figure \ref{Fig_Chap1_2}. On y constate que le volume du système demeure constant, rendant cette expression du travail peu adéquate au traitement de l'adhésion de la phase suspendue sur la surface, évolution pour laquelle il s'agirait plutôt de tenir compte de changements aux interfaces: augmentation de l'aire de l'interface <<~sp~>>, diminution des aires des interfaces <<~sl~>> et <<~pl~>>. D'autre part, ces interfaces étant associées à des tensions superficielles $\gamma_\mathrm{ps}$, $\gamma_\mathrm{sl}$ et $\gamma_\mathrm{pl},$ il semble nécessaire que le potentiel thermodynamique d'une telle transformation tienne compte de ces termes. Les tensions superficielles sont des quantités ayant les dimensions d'une énergie par unité de surface, \textit{i.e.} J$\cdot$m$^{-2}$, et caractérisent la traction mécanique exercée sur une mince couche dite \emph{superficielle} séparant deux milieux tels que le liquide et la protéine \citep{defay1951}.

\begin{figure}[t]\centering
\includegraphics[width=7cm]{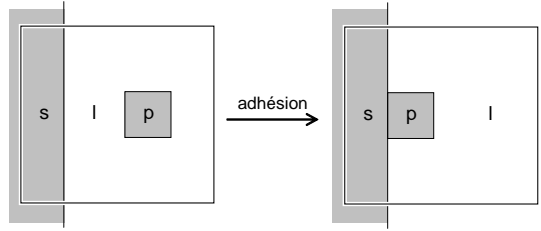}
\caption[\'{E}volution du système au cours de laquelle se produit l'adhésion]{\'{E}volution du système au cours de laquelle se produit l'adhésion de la phase <<~p~>>, initialement suspendue dans le milieu liquide <<~l~>>, sur la surface solide <<~s~>>.}\label{Fig_Chap1_2}
\end{figure}

Afin d'obtenir l'expression du travail mécanique se rapportant au phénomène d'adhésion illustré à la figure \ref{Fig_Chap1_2}, le terme de travail élémentaire est posé \citep{hunter1987} tel que
\begin{equation}\label{EqTravailAdhésion1}
\mathrm{\dbar}W=\gamma_\mathrm{pl}\mathrm{d}\Sigma_\mathrm{pl}
+\gamma_\mathrm{sl}\mathrm{d}\Sigma_\mathrm{sl}
+\gamma_\mathrm{sp}\mathrm{d}\Sigma_\mathrm{ps}.
\end{equation}
Sous cette forme, il ne se rapporte qu'aux changements des aires des diverses interfaces. Ensuite, comme montré à la figure \ref{Fig_Chap1_2}, l'adhésion de la phase suspendue sur la surface détruit autant d'aire $\Sigma_\mathrm{pl}$ et $\Sigma_\mathrm{sl}$ prises séparément qu'elle ne crée d'aire $\Sigma_\mathrm{ps}$, ce qui autorise l'écriture de la double égalité:
\begin{equation}\label{EqEgaliteAires}
\mathrm{d}\Sigma_\mathrm{pl}=\mathrm{d}\Sigma_\mathrm{sl}=-\mathrm{d}\Sigma_\mathrm{ps}.
\end{equation}
Ces considérations permettent, après tous calculs et réarrangements, d'exprimer \citep{defay1951,hunter1987} le travail mécanique élémentaire sous la forme
\begin{equation}\label{EqTravailAdhésion2}
\mathrm{\dbar}W=\big(\gamma_\mathrm{ps}-\gamma_\mathrm{sl}-\gamma_\mathrm{pl}\big)\mathrm{d}\Sigma_\mathrm{ps}.
\end{equation}
dans laquelle l'équation de Dupré est obtenue entre les parenthèses, ce terme donnant le travail d'adhésion dû à l'accroissement d'une unité de l'aire de contact $\Sigma_\mathrm{ps}$ \citep{hunter1987,hiemenz1997tout}.

Ensuite, afin d'éliminer le terme $\mathrm{\dbar}Q$ de la quantité élémentaire de chaleur reçue par le système et de le remplacer par la différentielle d'une fonction d'état, il est nécessaire de définir une nouvelle fonction d'état du système: l'entropie $S$ (en J$\cdot$K$^{-1}$). L'entropie est définie telle que
\begin{equation}\label{EqDefinitionEntropie}
\mathrm{d}S=\frac{\mathrm{\dbar}Q}{T},
\end{equation}
c'est-à-dire que sa différentielle est équivalente au ratio de la quantité élémentaire de chaleur reçue par le système et de sa température au même instant. Finalement, la substitution des termes $\mathrm{\dbar}Q$ et $\mathrm{\dbar}W$ de l'équation \ref{EqPremierPrincipe1} par les expressions \ref{EqTravailAdhésion2} et \ref{EqDefinitionEntropie} mène facilement à l'obtention d'un premier potentiel thermodynamique, celui de l'énergie interne, tel que
\begin{equation}\label{EqPotentielGibbs:EnergieInterne1}
\mathrm{d}E=T\mathrm{d}S+\big(\gamma_\mathrm{ps}-\gamma_\mathrm{sl}-\gamma_\mathrm{pl}\big)\mathrm{d}\Sigma_\mathrm{ps}.
\end{equation}
Ce premier potentiel permet de décrire un changement d'état du système (différentielle de l'énergie interne) en fonction de changements de l'entropie ($\mathrm{d}S$) et de l'aire de l'interface protéine-solide ($\mathrm{d}\Sigma_\mathrm{ps}$).

Un second potentiel vient immédiatement mais, cette fois-ci, défini grâce à l'entropie:
\begin{equation}\label{EqPotentielGibbs:Entropie1}
\mathrm{d}S=\frac{\mathrm{d}E}{T}-\frac{1}{T}\big(\gamma_\mathrm{ps}-\gamma_\mathrm{sl}-\gamma_\mathrm{pl}\big)\mathrm{d}\Sigma_\mathrm{ps}.
\end{equation}
Ces deux potentiels thermodynamiques, ou potentiels de Gibbs \citep{prigogine1968}, conceptualisent l'évolution instantanée (signification de la différentielle) du système à un moment précis de son existence, ce système étant un système capillaire fermé. \`{A} ce stade, cette évolution peut se faire dans le sens de l'accroissement (différentielles positives) et de la décroissance (différentielles négatives) de $E$ et de $S$ car il s'agit de changements se produisant exclusivement dans le cadre de l'équilibre, c'est-à-dire que ces évolutions sont réversibles.

L'entropie est la fonction d'état la plus importante car elle ouvre la voie à une description exhaustive de l'évolution du système et ce, notamment, en dehors de l'équilibre.

\section{Potentiels de Gibbs hors de l'équilibre}

L'introduction du premier principe et la substitution des termes de chaleur et de travail a permis la déduction de potentiels thermodynamiques, différentielles décrivant une évolution du système, dans le cadre de l'équilibre thermodynamique. De cette manière, les équations \ref{EqPotentielGibbs:EnergieInterne1} et \ref{EqPotentielGibbs:Entropie1}, potentiels thermodynamiques décrivant l'adhésion de la phase suspendue sur la surface solide, rendent possible une description de l'évolution du système le long d'un chemin \emph{réversible}, c'est-à-dire ne s'éloignant jamais de l'équilibre thermodynamique. Dès lors, ces équations \ref{EqPotentielGibbs:EnergieInterne1} et \ref{EqPotentielGibbs:Entropie1} peuvent être vues comme une continuation de l'établissement de l'équilibre mécanique venant donc s'additionner aux énergies potentielle et cinétique du système \citep{perez1993}.

Le principe zéro fixe la nature de l'équilibre thermodynamique par l'absence de flux entre les différents corps du système. En effet, on observera que le potentiel \ref{EqPotentielGibbs:Entropie1}, décrivant une évolution réversible, ne tient compte que des échanges entre le système et son environnement mais nullement des échanges entre les corps qui le constituent, c'est-à-dire ses processus internes. \`{A} ce titre, on remarquera que la relation \ref{EqTravailAdhésion2} n'est \textit{a priori} pas à comptabiliser parmi les processus internes car il s'agit d'un travail mécanique qui est exercé sur le système par son environnement suite à l'adhésion de la phase suspendue sur la surface.

De ces considérations, on comprend que les processus internes seront liés à l'établissement de l'équilibre thermodynamique et que les potentiels s'y rapportant permettront de caractériser des processus hors de cet équilibre thermodynamique. Il sera alors nécessaire de tenir compte des flux de chaleur et de matière entre les parties du système. Ces processus internes parmi lesquels il faudra considérer les réactions chimiques et d'adsorption ainsi que les échanges entre phases sont caractéristiques des transformations \emph{irréversibles} ou \emph{naturelles} \citep{prigogine1950}. Parmi ces processus produisant des flux à l'intérieur du système, il pourrait y avoir des transferts de chaleur entre phases, des réactions chimiques, des passages de certaines espèces chimiques d'une phase à l'autre ou encore des processus d'adsorption. Si l'on considère le système comme thermiquement homogène, les flux de chaleur seront inexistants. Le système d'intérêt n'étant pas \textit{a priori} le siège d'une quelconque réaction chimique, ni même du passage d'une espèce chimique d'une phase à une autre (les phases <<~p~>> et <<~s~>> sont fermées), on considérera que seules les adsorptions entreront en jeu afin de décrire les flux caractéristiques de l'état de non-équilibre.

\begin{figure}[t]\centering
\includegraphics[width=7cm]{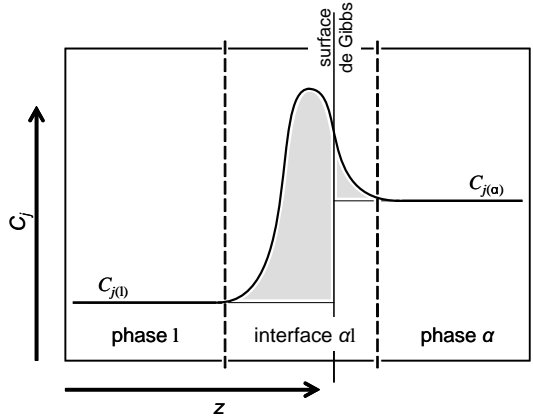}
\caption[Système capillaire constitué de deux phases <<~$\alpha$~>> et <<~$\mathrm{l}$~>>]{D'après \citep{everett1972, hunter1987}, système capillaire constitué de deux phases <<~$\alpha$~>> et <<~$\mathrm{l}$~>> séparées par une région interfaciale <<~$\alpha\mathrm{l}$~>>. La concentration $C_j$ de l'espèce chimique $j$ varie en fonction de la coordonnée $z$ et montre une singularité dans la région interfaciale qui sera associée à l'adsortion $\Gamma_{j(\alpha\mathrm{l})}$. Les deux phases sont conventionnelement séparées par une limite arbitraire (différente de la surface de tension): la surface de Gibbs (ou de division).}\label{Fig_Chap1_inter}
\end{figure}

Malgré l'inconnue sur la manière dont les réactions d'adsorption peuvent contribuer à l'adhésion de la protéine sur la surface, leur implication dans l'établissement de l'équilibre thermodynamique semble \textit{a priori} le phénomène causant l'irréversibilité de la transformation d'adhésion. L'adsorption, quantité définie par J.~W. Gibbs \citep{Gibbs1878}, est une quantité très importante que l'on associe à la présence d'au moins une interface dans un système thermodynamique, système alors qualifié de capillaire \citep{defay1951} et montré à la figure \ref{Fig_Chap1_inter}. La quantité totale d'une espèce $j$ contenue dans le système est donnée par le bilan fait sur les différentes phases du système (<<~$\alpha$~>> et <<~$\mathrm{l}$~>>) et l'interface entre celles-ci (<<~$\alpha\mathrm{l}$~>>). Il vient que $n_j$ (en mol), la quantité molaire de cette espèce $j$ contenue dans l'entièreté du système, se développe selon
\begin{equation}
n_j=n_{j(\alpha\mathrm{l})}+V_{\alpha}C_{j(\alpha)}+V_{\mathrm{l}}C_{j(\mathrm{l})}
\end{equation}
où $n_{j(\alpha\mathrm{l})}$ est la quantité molaire de $j$ présente à l'interface <<~$\alpha\mathrm{l}$~>>, $V_{\alpha}$ et $V_{\mathrm{l}}$ les volumes (en m$^3$) des phases <<~$\alpha$~>> et <<~$\mathrm{l}$~>> et $C_{j(\alpha)}$ et $C_{j(\mathrm{l})}$ les concentrations molaires (mol$\cdot$m$^{-3}$) de cette espèce dans chacune des phases.
Il vient donc de ce bilan que la quantité totale en $j$ ne peut être déduite de la seule somme des quantités présentes dans les phases dont est constitué le système. La quantité restante $n_{j(\alpha\mathrm{l})}$, associée à l'interface est dite quantité \emph{adsorbée} à l'interface <<~$\alpha\mathrm{l}$~>>. On définit alors l'adsorption comme une <<~concentration superficielle~>> en rapportant la quantité adsorbée $n_{j(\alpha\mathrm{l})}$ à l'aire $\Sigma_{\alpha\mathrm{l}}$ de la surface qui la porte \citep{everett1972}. Il vient que l'adsorption de l'espèce chimique $j$ à l'interface <<~$\alpha\mathrm{l}$~>> s'obtient selon
\begin{equation}\label{DefAdsorb}
\Gamma_{j(\alpha\mathrm{l})}=\frac{n_{j(\alpha\mathrm{l})}}{\Sigma_{\alpha\mathrm{l}}}
\end{equation}
où $\Gamma_{j(\alpha\mathrm{l})}$ est définie comme l'adsorption de $j$ (en mol$\cdot$m$^{-2}$). Comme suggéré ci-dessus, l'implication de l'adsorption dans le processus d'adhésion de la protéine est assez évident. Toutefois, la manière dont intervient cette adsorption dépend fortement du point de vue depuis lequel il est possible de se placer, point de vue dont la détermination fera l'objet de la suite de cet exposé.

Le premier de celui-ci est celui qui est classiquement considéré dans la littérature: la phase suspendue (les protéines) s'adsorbe sur la surface. Bien que le traitement thermodynamique d'une telle façon de voir le phénomène d'adhésion de la phase suspendue puisse être plus simple et intuitif, il implique un paradoxe qu'il est difficile d'ignorer: si la phase suspendue s'adsorbe, elle doit pouvoir être assimilée à une simple espèce chimique à laquelle on pourrait alors attribuer un potentiel chimique $\mu_j$ (incrément d'énergie interne qu'apporte au système une quantité infinitésimale $\mathrm{d}n_j$ de l'espèce chimique $j$ exprimé en J$\cdot$mol$^{-1}$) afin de l'impliquer dans un processus quelconque (réaction) d'adsorption.

Réduire la phase suspendue (les protéines) à une espèce chimique (pouvant s'adsorber), ce que font de très nombreux auteurs \citep{nagarajan1991,israelachvili2011}, déboucherait aussi sur le fait que son addition à un système thermodynamique n'aurait d'autre apport à l'énergie interne que son potentiel chimique. La phase suspendue aurait été \emph{dissoute} et non \emph{suspendue} dans la phase liquide \citep{everett1972}. Or, intrinsèquement, une espèce chimique --- pensons à une molécule d'eau ou un cation Na$^+$ --- ne peut pas être considérée comme une phase à part entière et sa présence n'implique dès lors pas l'existence d'une région interfaciale. Sachant que, par définition et comme montré sur la figure \ref{Fig_Chap1_inter}, l'interface est une région de l'espace séparant deux phases, région dans laquelle les propriétés changent très rapidement mais de manière continue \citep{hunter1987}, on en déduit que, sans interface, la tension superficielle n'a pas lieu d'être et la formule \ref{EqTravailAdhésion1} de Dupré ne pourrait pas s'appliquer aux protéines.

En outre, cette assimilation d'une protéine à une espèce chimique ne pourrait pas tenir longtemps car il s'agirait de nier la nature des protéines en tant que colloïdes dont les remarquables propriétés liées à la présence d'interfaces ont été soulignées dans l'introduction. Dans la continuité de ce qui a été décrit dans la section précédente, et il s'agit du second point de vue adopté, les protéines seront traitées comme des colloïdes, c'est-à-dire de petits corps auxquels volumes et interfaces peuvent être associés et faisant partie d'une phase suspendue traitée comme un tout. De ce fait, en découle un autre d'une importance cruciale: si les protéines présentent une interface avec le milieu suspendant, alors les espèces chimiques provenant de ce dernier peuvent s'adsorber sur la surface de ces protéines. Cette façon de conceptualiser le système est illustrée à la figure \ref{Fig_Chap1_3}, figure montrant l'adhésion de la phase suspendue (faite de protéines) sur la surface tout en tenant compte des adsorptions.

\begin{figure}[t]\centering
\includegraphics[width=7cm]{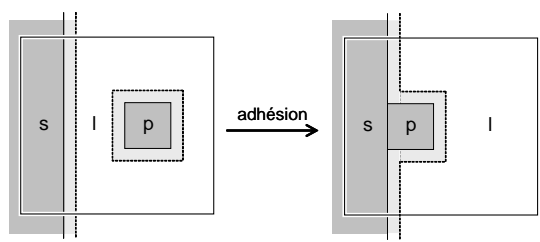}
\caption[\'{E}volution d'un système où se produit une adhésion irréversible]{\'{E}volution du système au cours de laquelle se produit l'adhésion de la phase <<~p~>> initialement suspendue dans le milieu liquide <<~l~>> sur la surface solide <<~s~>>. Les phases <<~s~>> et <<~p~>> sont entourées de couches adsorbées <<~sp~>> et <<~sl~>>, figurées par une phase ouverte. Lors de l'adhésion, une partie de ces couches adsorbées sont détruites et leur contenu libéré dans le milieu liquide <<~l~>>.}\label{Fig_Chap1_3}
\end{figure}

La cause de l'adhésion de la phase suspendue sur la surface ne devrait-elle pas être recherchée au niveau du contenu de son interface avec le milieu liquide? L'adsorption-désorption des espèces chimiques du milieu suspendant sur la surface de la phase suspendue ne permettrait-elle pas de clarifier l'irréversibilité de l'adhésion des protéines sur une surface hydrophobe? Ces questions auxquelles nous allons tenter de répondre par la suite sont soutenues par divers éléments présents dans la littérature. W. Norde \citep{norde1979} souligne, qu'entre autres choses, l'augmentation de l'entropie, faisant partie des forces motrices de l'adsorption (\textit{i.e.} l'adhésion d'une protéine sur une surface), semble due à la déshydratation du polystyrène et de la protéine. Ph.~Ball \citep{ball2013} relève encore que l'interaction entre deux particules se collant l'une à l'autre serait causée par le fait que les enveloppes d'eau structurée (entourant les deux particules, c'est-à-dire les molécules adsorbées) se recouvrent, ce qui a pour effet de les libérer, produisant ainsi de l'entropie. D'autres éléments sont rapportés par les références \citep{timasheff2002a,scharnagl2005,levy2006}. Cette attribution de l'adhésion de la phase suspendue sur la surface à l'effet hydrophobe, effet dont l'origine serait essentiellement entropique, permettrait de relier efficacement cette adhésion à l'irréversibilité qui la caractérise. Sur la figure \ref{Fig_Chap1_3}, on voit clairement que l'adhésion de la phase suspendue sur la surface solide nécessite la disparition d'une partie des enveloppes constituées des espèces initialement adsorbées.

La mise en équation de cette évolution du système en tenant compte des enveloppes d'espèces adsorbées sur les corps passera par l'expression de l'entropie. La fonction d'état entropie est en effet synonyme de changement au sein du système et l'obtention de son expression devrait nous amener à une conceptualisation précise du phénomène d'adhésion des corps <<~s~>> et <<~p~>>. \`{A} la fin de la section précédente, la variation de l'entropie au cours d'un processus a été reliée à la chaleur par la relation \ref{EqDefinitionEntropie} dans laquelle la différentielle de l'entropie était entièrement déterminée par la quantité de chaleur échangée $Q$ entre le système et son environnement. Cette relation a été développée pour les systèmes subissant une transformation réversible et doit dès lors être étendue afin de tenir compte de l'irréversibilité. \`{A} cette fin, la relation \ref{EqDefinitionEntropie} est précisée sous la forme d'une inégalité:
\begin{equation}\label{EqSecondPrincipe1}
\mathrm{d}S\geqslant\frac{\mathrm{\dbar}Q}{T}
\end{equation}
dont l'intégrale le long d'une transformation cyclique, strictement inférieure à zéro, est connue sous le nom d'inégalité de Clausius \citep{kondepudi1998}. Afin d'obtenir une égalité, un second terme est ajouté au membre de droite. On obtient
\begin{equation}\label{EqSecondPrincipe2}
\mathrm{d}S=\frac{\mathrm{\dbar}Q}{T}+\frac{\mathrm{\dbar}Q^\prime}{T}
\end{equation}
dans laquelle $Q^\prime$ est la chaleur non compensée de Clausius et $\mathrm{\dbar}Q^\prime$ sa quantité élémentaire. Cette chaleur non compensée (par les échanges avec l'environnement) est due aux flux d'énergie entre les différents corps constitutifs du système, système dont on différentie l'entropie. Sachant maintenant que la chaleur $Q$ est une chaleur échangée entre le système et son environnement et que $Q^\prime$ est une chaleur due aux flux à l'intérieur du système, il est possible d'écrire la différentielle totale de l'entropie sous la forme \citep{prigogine1968}
\begin{equation}\label{EqSecondPrincipe3}
\mathrm{d}S=\mathrm{d}_eS+\mathrm{d}_iS
\end{equation}
dans laquelle $\mathrm{d}_eS$ et $\mathrm{d}_iS$ sont des différentielles partielles de l'entropie. La première de ces différentielles partielles est réalisée sur les variables d'état caractérisant les échanges <<~$e$~>> entre le système et l'environnement tandis que la seconde est réalisée sur les variables caractérisant les processus physico-chimiques internes <<~$i$~>> au système. Par comparaison avec \ref{EqSecondPrincipe2}, il vient que
\begin{equation}\label{EqSecondPrincipe6}
\mathrm{d}_eS=\frac{\mathrm{\dbar}Q}{T}\qquad\text{et}\qquad\mathrm{d}_iS=\frac{\mathrm{\dbar}Q^\prime}{T}.
\end{equation}

Compte tenu de ces différentes notions, le second principe de la thermodynamique pourra être énoncé et adéquatement exploité. Il est par ailleurs illustré à la figure \ref{FigSecondPrincipe}.
\begin{axiome}[Second principe de la thermodynamique]
La production d'entropie dans tout système est positive ou nulle
\begin{equation}\label{EqThermoCroissanceEntrop}
\mathrm{d}_iS\geqslant0
\end{equation}
et ce quels que soient les échanges avec le monde extérieur \citep{prigogine1959}.
\end{axiome}

\begin{figure}[t]\centering
\includegraphics[width=4cm]{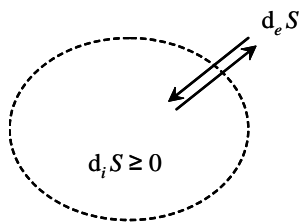}
\caption[Illustration du second principe de la thermodynamique]{Illustration du second principe de la thermodynamique dans un système pouvant échanger au moins de l'énergie avec son environnement. D'après I. Prigogine \citep{prigogine1959}.}\label{FigSecondPrincipe}
\end{figure}

En se rappelant que, selon le principe zéro de la thermodynamique, l'équilibre est atteint lorsque les flux de chaleur $Q^\prime$ sont nuls, la seconde relation \ref{EqSecondPrincipe6} permet de déduire que la différentielle $\mathrm{d}_iS$ sera nulle à l'équilibre. De ce fait, les transformations d'un système se déroulant lorsque $\mathrm{d}_iS=0$ sont des transformations se déroulant à l'équilibre et sont qualifiées de \emph{réversibles}. Inversement, lorsque $\mathrm{d}_iS>0$, les flux de chaleur $Q^\prime$ entre les corps constitutifs du système seront non nuls de telle sorte que le système se trouve dans un état hors de l'équilibre, état qui, selon le second principe, est associé à une transformation. Les transformations se déroulant hors du cadre de l'équilibre sont qualifiées d'\emph{irréversibles}.

Pour qu'une transformation soit possible, il y a donc nécessairement une production de chaleur $Q^\prime$, quantité toujours positive. Cette production se fait par frottement, conduction de chaleur \citep{planck1913}, réactions chimiques \citep{prigogine1968} ou processus d'adsorptions et il s'agit donc d'une perte pour le système (énergie rendue inutilisable puisque transformée en chaleur). \`{A} ce titre, I. Prigogine et I. Stengers \citep{prigogine1979} soulignent que l'entropie devient ainsi un <<~indicateur d'évolution~>>, et traduit l'existence physique d'une <<~flèche du temps~>>: pour tout système isolé, le futur est la direction dans laquelle l'entropie augmente.

Afin qu'une transformation soit irréversible, il faut que, après l'achèvement de celle-ci, il ne se trouve aucun moyen de ramener la nature dans l'état dans lequel elle se trouvait initialement \citep{planck1913}. Si l'on inversait une telle transformation, il se trouvera toujours un agent, dans la nature, qui aura dû être modifié. Toutes les autres transformations sont réversibles. M. Planck \citep{planck1913} va plus loin en affirmant que tous les processus naturels sont en réalité irréversibles alors que les processus réversibles forment un cas limite, une stylisation théorique.

De manière générale, E.~A. Guggenheim \citep{guggenheim1949} souligne la distinction faite par M. Planck entre les processus naturels, non-naturels et réversibles. Les processus naturels sont ceux qui se produisent dans la réalité en procédant dans une direction vers l'état d'équilibre. Ils sont irréversibles et caractérisés par une variation strictement positive. Les processus non-naturels se caractérisent par une évolution non dirigée vers l'équilibre et sont dès lors impossibles. Quant aux transformations réversibles, elles sont caractérisées par l'absence de dissipation et la variation d'entropie due aux processus internes est donc nulle. Ces transformations ne se produisent pas dans la nature comme l'énonce le second principe. Il s'agit d'une transformation dont toutes les étapes sont chacune un état d'équilibre de telle sorte qu'elle puisse être inversée à chaque instant.

La conséquence du second principe tel qu'il vient d'être décrit et interprété est la possibilité de prévoir le sens d'évolution d'un système grâce au développement de la différentielle partielle $\mathrm{d}_iS$ (inégalité \ref{EqThermoCroissanceEntrop}). La méthode, due à Th. De Donder \citep{prigogine1950}, est basée sur la calcul
\begin{equation}\label{EqSecondPrincipe5}
\mathrm{d}_iS=\sum_r\frac{A_r}{T}\mathrm{d}\xi_r\geqslant0
\end{equation}
permettant de développer $\mathrm{d}_iS$ et d'ainsi mettre un signe sur les $\mathrm{d}\xi_r,$ différentielles des variables caractérisant les degrés d'avancement des $r$ processus physico-chimiques. Dans cette méthode, le système est supposé être le siège de $r$ processus physico-chimiques (réactions chimiques, adsorptions ou passage d'une phase à une autre), chacun de ces processus étant caractérisé par une affinité $A_r$ (en J$\cdot$mol$^{-1}$) et un degré d'avancement $\xi_r$ (en mol). Lorsqu'elle est divisée par la température, l'affinité est une force chimique, c'est-à-dire une contrainte exercée sur le système, contrainte à même d'y produire un changement \citep{prigogine1968}.

Le développement \ref{EqSecondPrincipe5} de la production d'entropie par le système permet l'écriture complète du potentiel de Gibbs relatif à l'entropie. En effet, en se rappelant que l'équation \ref{EqPotentielGibbs:Entropie1} a été obtenue à partir des termes d'échanges (chaleur reçue et travail mécanique exercé sur le système) avec le milieu environnant, la différentielle \ref{EqPotentielGibbs:Entropie1} doit être une différentielle partielle de l'entropie faite sur les seules variables d'état rendant compte des échanges avec l'environnement du système, de telle sorte qu'elle sera écrite plus précisément selon
\begin{equation}\label{EqPotentielGibbs:Entropie2}
\mathrm{d}_eS=\frac{\mathrm{d}E}{T}-\frac{1}{T}\big(\gamma_\mathrm{ps}-\gamma_\mathrm{sl}-\gamma_\mathrm{pl}\big)\mathrm{d}\Sigma_\mathrm{ps}.
\end{equation}
Cette expression peut dès lors être substituée dans \ref{EqSecondPrincipe3} avec la relation \ref{EqSecondPrincipe5} afin d'obtenir la différentielle totale de l'entropie
\begin{equation}\label{EqPotentielGibbs:Entropie3}
\mathrm{d}S=\frac{\mathrm{d}E}{T}-\frac{1}{T}\big(\gamma_\mathrm{ps}-\gamma_\mathrm{sl}-\gamma_\mathrm{pl}\big)\mathrm{d}\Sigma_\mathrm{ps}+\sum_r\frac{A_r}{T}\mathrm{d}\xi_r.
\end{equation}
La différentielle totale obtenue est dès lors à même de tenir compte à la fois des processus réversibles et irréversibles, c'est-à-dire causant une augmentation de l'entropie. Cette expression complète est le potentiel de Gibbs pour l'entropie et décrit l'évolution instantanée de la fonction d'état entropie à un moment de l'existence du système, celui-ci étant affecté par des changements se déroulant dans ou en dehors du cadre de l'équilibre.

\section{Les affinités d'adsorption et de désorption}

L'affinité est une notion fondamentale en thermodynamique du non-équilibre et son calcul repose sur les potentiels chimiques des différentes espèces présentes dans le système susceptibles de prendre part à une transformation physico-chimique. Parmi ces transformations, on citera les réactions chimiques, les adsorptions et les passages d'une phase vers une autre. En toute logique, dans le cadre d'un système capillaire, l'affinité d'adsorption sera celle à laquelle on s'intéressera. L'adsorption d'une espèce chimique $j$ à la surface d'une phase suspendue <<~$\alpha$~>> peut être vue comme le passage de cette espèce de la phase suspendante <<~l~>> à l'interface <<~$\alpha\mathrm{l}$~>> entre ces deux phases \citep{defay1951} tel que
\begin{equation}
j_{(\mathrm{l})}\quad\longrightarrow\quad j_{(\alpha\mathrm{l})}.
\end{equation}
Ensuite, les potentiels chimiques, respectivement notés $\mu_{j(\mathrm{l})}$ (potentiel chimique de l'espèce $j$ dans la phase suspendante <<~l~>> en J$\cdot$mol$^{-1}$) et $\mu_{j(\alpha\mathrm{l})}$ (potentiel chimique de la même espèce $j$ mais adsorbée à l'interface <<~$\alpha\mathrm{l}$~>>), permettent l'écriture de l'affinité d'adsorption $A_{j(\mathrm{l}\cdot\alpha\mathrm{l})}$ selon la formule
\begin{equation}\label{AffiniteAdsorption}
A_{j(\mathrm{l}\cdot\alpha\mathrm{l})}=\mu_{j(\mathrm{l})}-\mu_{j(\alpha\mathrm{l})}.
\end{equation}
Cette affinité représente l'excès d'énergie de l'espèce $j$ (à rapprocher d'une différence de potentiel) dans la phase suspendante par rapport à la même espèce adsorbée à l'interface. Une affinité d'adsorption positive sera la cause d'une transformation du système par adsorption, effet qui se matérialisera par le degré d'avancement $\xi_{j(\mathrm{l}\cdot\alpha\mathrm{l})}$ dont la différentielle est reliée à la quantité molaire de l'espèce $j$ adsorbée et dissoute telle que
\begin{equation}\label{DiffDegreAvancement}
\mathrm{d}\xi_{j(\mathrm{l}\cdot\alpha\mathrm{l})}=-\mathrm{d}n_{j(\mathrm{l})}=\mathrm{d}n_{j(\alpha\mathrm{l})}.
\end{equation}
La substitution de \ref{AffiniteAdsorption} et \ref{DiffDegreAvancement} dans \ref{EqSecondPrincipe5} mène directement à la formulation de la production d'entropie lorsque le système est affecté par des processus d'adsorption:
\begin{equation}\label{DefEntropieAdsorption1}
\mathrm{d}_iS=\frac{1}{T}\sum_{\alpha}\sum_j \big(\mu_{j(\mathrm{l})}-\mu_{j(\alpha\mathrm{l})}\big)\mathrm{d}n_{j(\alpha\mathrm{l})}
\end{equation}
Il est maintenant commode de faire apparaître les concepts directement liés à ces transformations aux interfaces, c'est-à-dire les adsorptions $\Gamma_{j(\alpha\mathrm{l})}$ et les aires des interfaces $\Sigma_{\alpha\mathrm{l}}$. La définition \ref{DefAdsorb} de l'adsorption fournit
\begin{equation}
n_{j(\alpha\mathrm{l})}=\Gamma_{j(\alpha\mathrm{l})} \Sigma_{\alpha\mathrm{l}}\quad\therefore\quad
\mathrm{d}n_{j(\alpha\mathrm{l})}=\Gamma_{j(\alpha\mathrm{l})}\mathrm{d}\Sigma_{\alpha\mathrm{l}}+\Sigma_{\alpha\mathrm{l}}\mathrm{d}\Gamma_{j(\alpha\mathrm{l})}
\end{equation}
dont la différentielle est substituée dans \ref{DefEntropieAdsorption1} afin d'obtenir
\begin{equation}\label{DefEntropieAdsorption2}
\mathrm{d}_iS=\frac{1}{T}\sum_{\alpha}\sum_j \big(\mu_{j(\mathrm{l})}-\mu_{j(\mathrm{l}\alpha)}\big)\Gamma_{j(\alpha\mathrm{l})}\mathrm{d}\Sigma_{\alpha\mathrm{l}}+
\frac{1}{T}\sum_{\alpha}\sum_j \big(\mu_{j(\mathrm{l})}-\mu_{j(\alpha\mathrm{l})}\big)\Sigma_{\alpha\mathrm{l}}\mathrm{d}\Gamma_{j(\alpha\mathrm{l})}
\end{equation}
donnant la production d'entropie due aux variations des aires des interfaces $\Sigma_{\alpha\mathrm{l}}$ et des adsorptions sur celles-ci $\Gamma_{j(\alpha\mathrm{l})}$. Cette relation \ref{DefEntropieAdsorption2} représente à elle-seule la façon dont doit évoluer le système afin de respecter le second principe. En effet, en vertu de l'inégalité $\mathrm{d}_iS>0$ et en fonction des signes des parenthèses $\big(\mu_{j(\mathrm{l})}-\mu_{j(\alpha\mathrm{l})}\big)$, c'est-à-dire des affinités d'adsorption, il sera possible de prévoir si le système doit augmenter/diminuer les aires de ses interfaces et les quantités adsorbées sur celles-ci.

La température $T$, les adsorptions $\Gamma_{j(\alpha\mathrm{l})}$ et les aires des interfaces $\Sigma_{\alpha\mathrm{l}}$ sont, par définition, des quantités positives ou nulles de telle sorte que seuls les signes des affinités dicteront le sens des différentielles $\mathrm{d}\Sigma_{\alpha\mathrm{l}}$ et $\mathrm{d}\Gamma_{j(\alpha\mathrm{l})},$ c'est-à-dire la façon dont évoluera le système. Lorsque les parenthèses $\big(\mu_{j(\mathrm{l})}-\mu_{j(\alpha\mathrm{l})}\big)$ seront positives, les différentielles $\mathrm{d}\Sigma_{\alpha\mathrm{l}}$ et $\mathrm{d}\Gamma_{j(\alpha\mathrm{l})}$ devront, toutes deux être positives afin de respecter l'inégalité $\mathrm{d}_iS>0$ posée par le second principe. En d'autres termes, des affinités d'adsorption positives seront les causes d'augmentations des aires des interfaces (l'aire augmente afin de faire de la place aux espèces ayant tendance à s'adsorber) et des adsorptions sur celles-ci. Inversement, lorsque les parenthèses $\big(\mu_{j(\mathrm{l})}-\mu_{j(\alpha\mathrm{l})}\big)$ seront négatives (affinités d'adsorption négatives ou affinités de désorption positives), les différentielles $\mathrm{d}\Sigma_{\alpha\mathrm{l}}$ et $\mathrm{d}\Gamma_{j(\alpha\mathrm{l})}$ devront, cette fois-ci, être négatives afin de respecter le second principe, c'est-à-dire que les aires des interfaces auront tendance à diminuer et les espèce adsorbées à se désorber.

Ces changements affecteront le système lors d'une période transitoire hors de l'équilibre. D'après le principe zéro, l'équilibre sera rencontré lorsque les flux de chaleur (liés aux flux de matière) entre les différents corps du système seront nuls. Le flux de l'espèce $j$ de la phase <<~$\mathrm{l}$~>> vers la phase <<~$\alpha\mathrm{l}$~>> (impliquant un transfert de chaleur $Q^\prime$) s'annulera lorsque cette espèce aura un potentiel chimique égal aux deux endroits:
\begin{equation}
\mu_{j(\mathrm{l})}=\mu_{j(\alpha\mathrm{l})}
\end{equation}
et il s'ensuivra que l'affinité du processus s'annulera:
\begin{equation}
A_{j(\mathrm{l}\cdot\alpha\mathrm{l})}=0\qquad\mathrm{quels\ que\ soient}\ j\ \mathrm{et}\ \alpha.
\end{equation}
La valeur de l'affinité détermine donc l'état de déséquilibre thermodynamique du système et sa tendance à évoluer dans un sens ou dans un autre. Des affinités non nulles seront le signe que le système se situe en dehors de l'équilibre thermodynamique, celui-ci ayant dès lors tendance à y retourner en annulant toutes les affinités des $r$ processus. \`{A} l'équilibre, \emph{toutes} les affinités seront nulles de telle sorte que la production d'entropie $\mathrm{d}_iS$ sera nulle \citep{prigogine1968}.

L'équation \ref{DefEntropieAdsorption2} permettant de prévoir l'évolution d'un système muni d'interfaces, il est maintenant logique de l'appliquer au système ici étudié, système comprenant des interfaces <<~sl~>> (surface-liquide) et <<~pl~>> (protéines-liquide). Par application directe, il vient que
\begin{equation}\label{DefEntropieAdsorption3}
\begin{split}
\mathrm{d}_iS=&
\frac{1}{T}
\sum_j \big(\mu_{j(\mathrm{l})}-\mu_{j(\mathrm{sl})}\big)\Gamma_{j(\mathrm{sl})}\mathrm{d}\Sigma_{\mathrm{sl}}
+\frac{1}{T}
\sum_j \big(\mu_{j(\mathrm{l})}-\mu_{j(\mathrm{pl})}\big)\Gamma_{j(\mathrm{pl})}\mathrm{d}\Sigma_{\mathrm{pl}}\\
&+\frac{1}{T}
\sum_j \big(\mu_{j(\mathrm{l})}-\mu_{j(\mathrm{sl})}\big)\Sigma_{\mathrm{sl}}\mathrm{d}\Gamma_{j(\mathrm{sl})}
+\frac{1}{T}
\sum_j \big(\mu_{j(\mathrm{l})}-\mu_{j(\mathrm{pl})}\big)\Sigma_{\mathrm{pl}}\mathrm{d}\Gamma_{j(\mathrm{pl})}
\end{split}
\end{equation}
dans laquelle on retrouve les aires des interfaces <<~sl~>> et <<~pl~>> ($\Sigma_{\mathrm{sl}}$ et $\Sigma_{\mathrm{pl}}$) de même que les adsorptions sur celles-ci ($\Gamma_{j(\mathrm{sl})}$ et $\Gamma_{j(\mathrm{pl})}$). Lors de l'adhésion de la phase suspendue sur la surface, l'égalité \ref{EqEgaliteAires} montrait que la quantité de surface créée entre la phase suspendue et la surface étant équivalente à celle détruite entre la phase suspendue et le liquide et aussi entre la surface et le liquide ($\mathrm{d}\Sigma_{\mathrm{sl}}=\mathrm{d}\Sigma_{\mathrm{pl}}=-\mathrm{d}\Sigma_{\mathrm{sp}}$). De plus, il faut aussi constater que les différentielles des adsorptions restent nulles au cours de cette adhésion. En effet, comme le montre l'équation \ref{DefAdsorb}, l'adsorption est un rapport entre une quantité molaire présente à l'interface et l'aire de celle-ci et, lors de l'adhésion, la quantité de surface détruite provoque \textit{ipso facto} l'expulsion des espèces qui étaient adsorbées: il y a toujours autant d'espèces adsorbées par unité de surface ($\mathrm{d}\Gamma_{j(\alpha\mathrm{l})}=0$ quels que soient $j$ et $\alpha$). L'utilisation de \ref{EqEgaliteAires} et de la nullité des différentielles des adsorptions donne que
\begin{equation}\label{DefEntropieAdsorption4}
\mathrm{d}_iS=
-\frac{1}{T}
\sum_j \big(\mu_{j(\mathrm{l})}-\mu_{j(\mathrm{sl})}\big)\Gamma_{j(\mathrm{sl})}\mathrm{d}\Sigma_{\mathrm{sp}}
-\frac{1}{T}
\sum_j \big(\mu_{j(\mathrm{l})}-\mu_{j(\mathrm{pl})}\big)\Gamma_{j(\mathrm{pl})}\mathrm{d}\Sigma_{\mathrm{sp}}.
\end{equation}

En définissant les affinités de désorption
\begin{equation}
A_{j(\mathrm{sl}\cdot\mathrm{l})}=\mu_{j(\mathrm{sl})}-\mu_{j(\mathrm{l})}
\quad\mathrm{et}\quad
A_{j(\mathrm{pl}\cdot\mathrm{l})}=\mu_{j(\mathrm{pl})}-\mu_{j(\mathrm{l})}
\end{equation}
et les affinités d'adsorption
\begin{equation}
A_{j(\mathrm{l}\cdot\mathrm{sl})}=\mu_{j(\mathrm{l})}-\mu_{j(\mathrm{sl})}
\quad\mathrm{et}\quad
A_{j(\mathrm{l}\cdot\mathrm{pl})}=\mu_{j(\mathrm{l})}-\mu_{j(\mathrm{pl})}
\end{equation}
telles que les relations d'oppositions
\begin{equation}\label{v154658gb4fx7b47}
A_{j(\mathrm{sl}\cdot\mathrm{l})}=-A_{j(\mathrm{l}\cdot\mathrm{sl})}
\quad\mathrm{et}\quad
A_{j(\mathrm{pl}\cdot\mathrm{l})}=-A_{j(\mathrm{l}\cdot\mathrm{pl})}
\end{equation}
puissent être vérifiées, on obtiendra l'équation \ref{DefEntropieAdsorption5} en y substituant les différences de potentiels chimiques par les affinités de désorption.

\begin{proposition}[Production d'entropie lors de l'adhésion]
La production d'entropie dans un système au sein duquel une phase <<~p~>> initialement suspendue dans un milieu aqueux <<~l~>> vient à adhérer à une surface solide <<~s~>> est donnée par
\begin{equation}\label{DefEntropieAdsorption5}
\mathrm{d}_iS
=\frac{1}{T}\Big(
\sum_j\Gamma_{j(\mathrm{sl})}A_{j(\mathrm{sl}\cdot\mathrm{l})}+\sum_j
\Gamma_{j(\mathrm{pl})}A_{j(\mathrm{pl}\cdot\mathrm{l})}\Big)\mathrm{d}\Sigma_{\mathrm{sp}}
\end{equation}
dans laquelle les interfaces entre les différents corps sont notées <<~sl~>>, <<~pl~>> et <<~sp~>>.
\end{proposition}

Cette équation venant d'être démontrée sera d'une grande utilité dans l'interprétation des phénomènes d'adhésion et leur prévision. Le développement de la différentielle \ref{DefEntropieAdsorption5} offre maintenant la possibilité de prévoir le signe de la différentielle $\mathrm{d}\Sigma_{\mathrm{sp}}$ à partir des conditions expérimentales et du second principe de la thermodynamique.

\section{Adhésion et hydrophobie}

Le processus décrit par l'équation \ref{DefEntropieAdsorption5} sera spontané si et seulement s'il respecte la condition émanant du second principe: $\mathrm{d}_iS>0$. Deux processus sont en fait possibles: l'adhésion de la phase suspendue sur la surface et le décollement de celle-ci de la surface. Ceux-ci sont reliés au signe de la différentielle de l'aire de contact entre la phase suspendue et la surface par les deux relations:
\begin{equation}
\begin{split}
\mathrm{d}\Sigma_{\mathrm{sp}}>0&\quad\Longrightarrow\quad\text{adhésion de <<~s~>> et <<~p~>>}\\
\mathrm{d}\Sigma_{\mathrm{sp}}<0&\quad\Longrightarrow\quad\text{décollement de <<~s~>> et <<~p~>>}
\end{split}
\end{equation}
Le caractère non nul de la différentielle $\mathrm{d}\Sigma_{\mathrm{sp}}$ matérialise la transformation du système, transformation qui est l'effet des deux sommes de l'équation \ref{DefEntropieAdsorption5}, ces dernières en étant les causes.

Les causes de l'adhésion, attribuées aux deux sommes de l'équation \ref{DefEntropieAdsorption5}, méritent que l'on s'y attarde car elles peuvent revêtir un sens particulier.
L'affinité est intimement liée à une notion de force chimique \citep{nicolis2005}, c'est-à-dire qu'elle est une certaine quantité d'énergie qui devra être dissipée, car en excès, lors du processus irréversible menant à la transformation du système, ici l'adhésion.
Pour mémoire, elle est exprimée en J$\cdot$mol$^{-1}$. Lorsqu'elle est multipliée par l'adsorption $\Gamma$, en mol$\cdot$m$^{-2}$, on obtiendra une quantité en J$\cdot$m$^{-2}$.
La somme faite sur toutes les $j$ espèces chimiques adsorbées à l'interface considérée donne alors un excès d'énergie par unité d'aire, excès lié à la présence d'une couche adsorbée. Cet excès a, par nature, vocation à être dissipé. Lorsque la somme contient des affinités de désorption, comme à l'équation \ref{DefEntropieAdsorption5}, cet excès d'énergie aura tendance à être dissipé par la désorption indifférente des espèces chimiques se trouvant à l'interface. Inversement, lorsque la somme contient des affinités d'adsorption, l'excès d'énergie aura tendance à être dissipé par l'adsorption d'espèces chimiques en provenance du solvant et du cosolvant. Ceci semble indiquer qu'il s'agit d'une tendance qu'a le corps en question à repousser les espèces chimiques de l'environnement dans lequel il est suspendu loin de la surface qui le délimite, c'est-à-dire à les désorber de sa surface. On pourra alors dire que le corps présentant cette propension dispose d'une certaine <<~phobie~>> vis-à-vis des espèces contenues dans son environnement. Posons alors que ces sommes représentent la \emph{lyophobie} (la somme contient des affinités de \emph{désorption}) et, par extension, la \emph{lyophilie} (la somme contient des affinités d'\emph{adsorption}) du corps suspendu dans la phase aqueuse.

\begin{definition}[Lyophobie/lyophilie d'un corps]
Le caractère lyophobe/lyophile d'un corps est une propriété des interfaces qui le délimitent. Premièrement, la \emph{lyophobie} d'un corps <<~$\alpha$~>> suspendu dans une phase liquide <<~l~>> est définie par la somme
\begin{equation}
\mathcal{L}_{\alpha\mathrm{l}\cdot\mathrm{l}}=\sum_j\Gamma_{j(\alpha\mathrm{l})}A_{j(\alpha\mathrm{l}\cdot\mathrm{l})}
\end{equation}
et, deuxièmement, sa \emph{lyophilie} l'est par
\begin{equation}
\mathcal{L}_{\mathrm{l}\cdot\alpha\mathrm{l}}=\sum_j\Gamma_{j(\alpha\mathrm{l})}A_{j(\mathrm{l}\cdot\alpha\mathrm{l})}
\end{equation}
avec la propriété (voir relations \ref{v154658gb4fx7b47})
\begin{equation}\label{EqLyophylieLyophobie}
\mathcal{L}_{\alpha\mathrm{l}\cdot\mathrm{l}}=-\mathcal{L}_{\mathrm{l}\cdot\alpha\mathrm{l}}
\end{equation}
que la lyophobie de ce corps <<~$\alpha$~>> est l'opposée de sa lyophilie. L'unité de la lyophobie/lyophilie est le J$\cdot$m$^{-2}$. La lyophobie est un potentiel de désorption par unité d'aire, toutes espèces confondues. La lyophilie est, quant à elle, un potentiel d'adsorption par unité d'aire.
\end{definition}

\'{E}tant donné le fait que le milieu suspendant est de caractère aqueux, le terme \emph{hydrophobe} sera employé sous la notation $\mathcal{H}_{\alpha\mathrm{l}\cdot\mathrm{l}}$ en lieu et place du terme \emph{lyophobe} noté $\mathcal{L}_{\alpha\mathrm{l}\cdot\mathrm{l}}$. De même, l'\emph{hydrophilie} $\mathcal{H}_{\mathrm{l}\cdot\alpha\mathrm{l}}$ sera préférée à la \emph{lyophilie} $\mathcal{L}_{\mathrm{l}\cdot\alpha\mathrm{l}}$. Par application de cette définition, il vient, à partir de la proposition démontrée à la section précédente, une autre proposition liant la production d'entropie d'un système au caractère hydrophobe des interfaces des corps qu'il contient, et ce, lors de l'adhésion/décollement de ces deux corps.

\begin{proposition}[Production d'entropie lors de l'adhésion hydrophobe]
La production d'entropie dans un système au sein duquel une phase <<~p~>> initialement suspendue dans un milieu liquide <<~l~>> vient à adhérer à une surface solide <<~s~>> est donnée par
\begin{equation}\label{DefEntropieAdsorption6}
\mathrm{d}_iS
=\frac{1}{T}\big(
\mathcal{H}_{\mathrm{sl}\cdot\mathrm{l}}+
\mathcal{H}_{\mathrm{pl}\cdot\mathrm{l}}\big)\mathrm{d}\Sigma_{\mathrm{sp}}
\end{equation}
dans laquelle les interfaces entre les différents corps sont notées <<~sl~>>, <<~pl~>> et <<~sp~>>. $\mathcal{H}_{\mathrm{sl}\cdot\mathrm{l}}$ et $\mathcal{H}_{\mathrm{pl}\cdot\mathrm{l}}$ sont respectivement les hydrophobies des corps <<~s~>> et <<~p~>>.
\end{proposition}

De l'équation \ref{DefEntropieAdsorption6} et application de l'inégalité \ref{EqThermoCroissanceEntrop} du second principe, il vient facilement que
\begin{equation}
\begin{split}
\text{si}\quad\mathcal{H}_{\mathrm{sl}\cdot\mathrm{l}}+
\mathcal{H}_{\mathrm{pl}\cdot\mathrm{l}}>0&\quad\text{alors}\quad\mathrm{d}\Sigma_{\mathrm{sp}}>0\\
\text{si}\quad\mathcal{H}_{\mathrm{sl}\cdot\mathrm{l}}+
\mathcal{H}_{\mathrm{pl}\cdot\mathrm{l}}<0&\quad\text{alors}\quad\mathrm{d}\Sigma_{\mathrm{sp}}<0
\end{split}
\end{equation}
montrant que la combinaison des hydrophobies des corps en présence détermine le signe de la différentielle $\mathrm{d}\Sigma_{\mathrm{sp}}$, c'est-à-dire la tendance du système à produire une adhésion ou un décollement de la surface solide <<~s~>> et de la phase suspendue <<~p~>> constituée de protéines.

Directement, lorsque les deux corps <<~s~>> et <<~p~>> sont hydrophobes, c'est-à-dire que les termes $\mathcal{H}_{\mathrm{sl}\cdot\mathrm{l}}$ et $\mathcal{H}_{\mathrm{pl}\cdot\mathrm{l}}$ sont tous deux positifs (donnant que $\mathcal{H}_{\mathrm{sl}\cdot\mathrm{l}}+
\mathcal{H}_{\mathrm{pl}\cdot\mathrm{l}}>0$), le système évoluera spontanément vers une adhésion de <<~s~>> et <<~p~>>. Inversement, lorsque les deux corps sont hydrophiles, c'est-à-dire que les termes $\mathcal{H}_{\mathrm{sl}\cdot\mathrm{l}}$ et $\mathcal{H}_{\mathrm{pl}\cdot\mathrm{l}}$ sont tous deux négatifs (donnant que $\mathcal{H}_{\mathrm{sl}\cdot\mathrm{l}}+
\mathcal{H}_{\mathrm{pl}\cdot\mathrm{l}}<0$), il y aura décollement des deux corps. Plus simplement exprimé, on dira que, lorsque deux corps hydrophobes sont suspendus dans un milieux aqueux, ils auront tendance à adhérer l'un à l'autre et, inversement, lorsque les deux corps suspendus en milieu aqueux sont hydrophiles, ils auront tendance à se décoller l'un de l'autre.

De manière plus précise, l'inégalité $\mathcal{H}_{\mathrm{sl}\cdot\mathrm{l}}+\mathcal{H}_{\mathrm{pl}\cdot\mathrm{l}}>0$ permet d'obtenir que
\begin{equation}
\mathcal{H}_{\mathrm{sl}\cdot\mathrm{l}}
>-\mathcal{H}_{\mathrm{pl}\cdot\mathrm{l}}
\end{equation}
donnant, grâce à la relation \ref{EqLyophylieLyophobie} entre l'hydrophobie et l'hydrophilie, l'inégalité suivante
\begin{equation}
\mathcal{H}_{\mathrm{sl}\cdot\mathrm{l}}
>\mathcal{H}_{\mathrm{l}\cdot\mathrm{pl}}.
\end{equation}
Cette inégalité exprime que l'adhésion des corps <<~s~>> et <<~p~>> se produira spontanément lorsque l'hydrophobie $\mathcal{H}_{\mathrm{sl}\cdot\mathrm{l}}$ de la surface solide <<~s~>> sera plus grande que l'hydrophilie $\mathcal{H}_{\mathrm{l}\cdot\mathrm{pl}}$ de la phase suspendue <<~p~>> (les protéines). On aura donc intérêt, afin de faire adhérer une phase suspendue hydrophile sur une surface solide, à faire usage d'une surface solide particulièrement hydrophobe. Cela permettra de contrer l'hydrophilie de la phase suspendue, hydrophilie qui la retient dans son environnement aqueux. L'usage du polystyrène, particulièrement hydrophobe, comme surface solide paraît dès lors tout-à-fait indiqué afin de provoquer l'adhésion des IgG, protéines globulaires et <<~hydrosolubles~>>.

\section{L'empreinte et l'accumulation des protéines}

La propension à l'adhésion ou au décollement spontané de la phase suspendue à la surface vient d'être développée et se conceptualise par le signe de la différentielle de l'aire de contact entre cette phase et la surface $\Sigma_{\mathrm{sp}}$. Il serait maintenant utile et souhaitable de relier cette aire de contact aux phénomènes propres aux interactions entre les protéines et les surfaces solides.

Pour ce faire, il faut considérer la phase suspendue comme un tout pouvant être divisé en un très grand nombre d'entités (corps) séparées les unes des autres et de dimensions colloïdales. Il s'agit donc de passer d'une phase suspendue d'un seul tenant à une phase suspendue finement divisée, démarche schématisée dans l'introduction à la figure \ref{Fig_Chap1_1} et transposée sur la figure \ref{Fig_Chap1_4}. En supposant que les petites entités colloïdales de la phase suspendue qui aura ainsi été conceptuellement divisée sont stables du point de vue de leur géométrie, on aura affaire à une phase colloïdale (phase suspendue et dispersée) pour laquelle la relation \ref{DefEntropieAdsorption5} demeure valable et dont un certain nombre d'éléments qui peuvent être des protéines pourront se désolidariser de leurs congénères et venir adhérer sur la surface solide selon différentes orientations

\begin{figure}[t]\centering
\includegraphics[width=7cm]{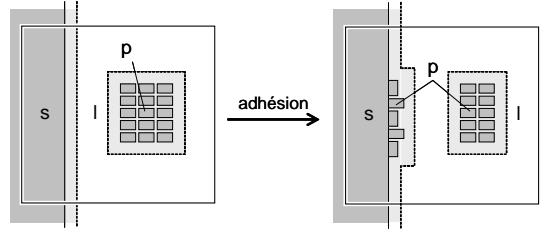}
\caption[Le système contenant une phase suspendue dispersée]{Le système thermodynamique composé de plusieurs phases: une surface solide <<~s~>> et une phase <<~p~>> divisée en une série de petites entitées de dimensions colloïdales (figurant des protéines) suspendues dans un milieu liquide <<~l~>>. Les colloïdes (protéines) peuvent alors se désolidariser de l'ensemble afin de venir adhérer séparement et selon différentes orientations sur la surface solide.}\label{Fig_Chap1_4}
\end{figure}

Lorsqu'une protéine, considérée comme une des entités de la phase colloïdale, viendra adhérer sur la surface, elle contribuera à l'accroissement de $\Sigma_\mathrm{sp}$, l'aire de contact entre la phase suspendue et la surface. Il vient alors une première quantité: l'empreinte.
\begin{definition}[L'empreinte]
Si l'on considère qu'un certain nombre $\mathcal{N}$ (en mol) de protéines adhèrent à la surface, il est possible de définir une quantité $\sigma$ telle que
\begin{equation}\label{DefinitionEmpreinte}
\sigma=\frac{\Sigma_{\mathrm{sp}}}{\mathcal{N}}.
\end{equation}
Cette quantité $\sigma$, que l'on qualifiera d'\emph{empreinte}, est l'aire de contact moyenne ou molaire entre les protéines et la surface. L'unité de l'empreinte est donc le m$^2\cdot$mol$^{-1}$.
\end{definition}
L'empreinte $\sigma$ est une quantité moyenne caractérisant l'ensemble des protéines appartenant à la monocouche. \`{A} côté de celle-ci, il est aussi judicieux de définir des empreintes caractéristiques pour chacune des orientations et/ou conformations (leur état de manière générale\footnote{Le pseudovecteur $\boldsymbol{\omega}$ est un ensemble de variables permettant de caractériser l'état dans lequel se trouve une protéine lorsqu'elle est collée sur la surface solide. Il pourra s'agir de variables décrivant la position du centre de masse de la protéine ($x, y, z$) mais aussi de variables décrivant l'orientation de la protéine par rapport à son centre de masse telles que les angles d'Euler ($\vartheta,\varphi,\psi$) relevés dans le chapitre introductif. Par ailleurs, les protéines ayant la possibilité de se dénaturer, c'est-à-dire de changer de conformation, le pseudovecteur $\boldsymbol{\omega}$ pourra aussi contenir un ensemble de variables décrivant sa conformation. De façon générale, $\boldsymbol{\omega}$ contiendra donc un ensemble de variables permettant de caractériser la position, l'orientation et la conformation d'une protéine.

Plusieurs ensembles de ces variables seront utilisés: $\boldsymbol{\omega}_1$, $\boldsymbol{\omega}_2$, etc. Ils représentent chacun un état (position, orientation et conformation) particulier dans lequel la protéine peut se trouver sur la surface.}) que peuvent adopter les protéines sur la surface. En effet, comme illustré sur la figure \ref{Fig_Chap1_4}, les protéines pourront, selon les orientations $\boldsymbol{\omega}_1$, $\boldsymbol{\omega}_2$, etc. qu'elles adopteront en adhérant sur la surface, avoir des aires de contact différentes. Lorsque $\mathcal{N}(\boldsymbol{\omega}_1)$ protéines adhèrent sur la surface selon une orientation $\boldsymbol{\omega}_1$, $\mathcal{N}(\boldsymbol{\omega}_2)$ selon une orientation $\boldsymbol{\omega}_2$, etc., l'empreinte est une combinaison linéaire (moyenne pondérée) se développant comme suit:
\begin{equation}\label{EmpreinteVSempreintesLimites}
\sigma=
\sigma(\boldsymbol{\omega}_1)\mathcal{X}(\boldsymbol{\omega}_1)+
\sigma(\boldsymbol{\omega}_2)\mathcal{X}(\boldsymbol{\omega}_2)+\dots
\end{equation}
dans laquelle $\sigma(\boldsymbol{\omega}_1)$ et $\sigma(\boldsymbol{\omega}_2)$ sont les empreintes caractéristiques pour chacune des orientations $\boldsymbol{\omega}_1$ et $\boldsymbol{\omega}_2$. Sur la figure \ref{Fig_Chap1_4}, on voit bien que les protéines, en adoptant deux orientations différentes, présentent des empreintes caractéristiques différentes. Les $\mathcal{X}(\boldsymbol{\omega}_1)$, $\mathcal{X}(\boldsymbol{\omega}_2)$, etc. sont les fractions molaires des protéines présentes à la surface selon chaque orientation, fractions définies par
\begin{equation}\label{FracionsOri}
\mathcal{X}(\boldsymbol{\omega}_1)=\frac{\mathcal{N}(\boldsymbol{\omega}_1)}{\mathcal{N}},\quad
\mathcal{X}(\boldsymbol{\omega}_2)=\frac{\mathcal{N}(\boldsymbol{\omega}_2)}{\mathcal{N}},\dots
\end{equation}
Clairement, l'empreinte caractéristique d'une orientation particulière apparaît comme l'aire de contact développée entre une mole de protéines adhérant dans cette orientation et la surface. Ces empreintes caractéristiques seront abondamment utilisées par la suite, notamment dans le cas des IgG qui présentent un certain nombre d'orientations caractéristiques.

Une deuxième grandeur peut être déduite: la \textit{quantité accumulée} que l'on nommera aussi plus simplement \textit{accumulation}.
\begin{definition}[La quantité accumulée]
La quantité accumulée, notée $\Theta$, est définie comme le nombre de moles de protéines $\mathcal{N}$ (en mol) adhérant sur la surface rapporté à l'aire totale de cette surface $\Sigma_{\mathrm{s}}$ (en m$^2$). Il vient
\begin{equation}\label{DefinitionTheta}
\Theta=\frac{\mathcal{N}}{\Sigma_{\mathrm{s}}}
\end{equation}
dans laquelle il est évident que l'unité de $\Theta$ est la mol$\cdot$m$^{-2}$. Par commodité, la pmol$\cdot$cm$^{-2}$ sera aussi utilisée par la suite.
\end{definition}
Il est par ailleurs utile de noter que cette quantité peut être développée sous la forme d'une somme telle que
\begin{equation}\label{2d4fg5db3654xgf6b65fx48}
\Theta=\Theta(\boldsymbol{\omega}_1)+\Theta(\boldsymbol{\omega}_2)+\dots
\end{equation}
où les $\Theta(\boldsymbol{\omega}_1)$ et $\Theta(\boldsymbol{\omega}_2)$ sont les quantités de protéines accumulées sur la surface en fonction des différentes orientations $\boldsymbol{\omega}_1$ et $\boldsymbol{\omega}_2$ qu'elles pourront y adopter. Ces quantités sont reliées aux fractions molaires et à la quantité totale selon les relations
\begin{equation}\label{f564desf874gs7}
\Theta(\boldsymbol{\omega}_1)=\mathcal{X}(\boldsymbol{\omega}_1)\Theta,\quad
\Theta(\boldsymbol{\omega}_2)=\mathcal{X}(\boldsymbol{\omega}_2)\Theta,\dots
\end{equation}

Ces deux quantités $\sigma$ (m$^2\cdot$mol$^{-1}$) et $\Theta$ (mol$\cdot$m$^{-2}$) seront abondamment utilisées dans la suite de ce travail. Elles permettront en effet de caractériser de manière précise et relativement simple l'état d'une monocouche. $\Theta$ et $\sigma$ apparaissent donc comme de nouvelles variables d'état caractérisant la monocouche. La grandeur $\Theta$ donnera la quantité de protéines contenues dans la monocouche tandis que $\sigma$ indiquera une valeur moyenne relative à l'orientation de ces protéines.

\section{Adhésion, addition et relaxation des protéines}

Afin d'injecter ces notions d'empreinte $\sigma$ et de quantité accumulée $\Theta$ dans la formulation de la différentielle de l'entropie et ainsi de se faire une idée de la façon dont elles pourraient évoluer au cours du processus de construction de la monocouche, il est nécessaire de définir une troisième notion: le taux de recouvrement de la surface par la monocouche.
\begin{definition}[Le taux de recouvrement]
Le taux de recouvrement, noté $\phi$, est défini par le rapport
\begin{equation}\label{DéfinitionTauxDOccupation}
\phi=\frac{\Sigma_\mathrm{sp}}{\Sigma_{\mathrm{s}}}
\end{equation}
dans lequel $\Sigma_\mathrm{sp}$ est l'aire de contact de l'ensemble des protéines avec la surface dont l'aire totale est $\Sigma_{\mathrm{s}}$. Le taux de recouvrement n'a pas d'unité puisque les aires $\Sigma_\mathrm{sp}$ et $\Sigma_{\mathrm{s}}$ sont toutes deux en m$^2$. 
\end{definition}
Ce taux de recouvrement est donc la proportion de la surface en contact avec les protéines de la monocouche. L'aire de la surface accueillant la monocouche restant constante au cours de la croissance de la monocouche, la différentielle du taux de recouvrement s'exprimera selon
\begin{equation}
\mathrm{d}\phi=\frac{\mathrm{d}\Sigma_\mathrm{sp}}{\Sigma_{\mathrm{s}}}
\end{equation}
dont on déduit l'égalité $\mathrm{d}\Sigma_\mathrm{sp}=\Sigma_{\mathrm{s}}\,\mathrm{d}\phi$ qui, une fois substituée dans \ref{DefEntropieAdsorption5} et division des deux membres par $\Sigma_{\mathrm{s}}$, fournit une nouvelle proposition corollaire à la précédente.

\begin{proposition}[Production d'entropie superficielle lors de l'adhésion]
La production d'entropie superficielle en chaque point d'une surface solide <<~s~>> d'hydrophobie $\mathcal{H}_{\mathrm{sl}\cdot\mathrm{l}}$ sur laquelle vient adhérer une phase suspendue <<~p~>> d'hydrophobie $\mathcal{H}_{\mathrm{pl}\cdot\mathrm{l}}$ est donnée par
\begin{equation}\label{DiffEntropieVSphi}
\mathrm{d}_is
=\frac{1}{T}\big(
\mathcal{H}_{\mathrm{sl}\cdot\mathrm{l}}+
\mathcal{H}_{\mathrm{pl}\cdot\mathrm{l}}\big)\mathrm{d}\phi
\end{equation}
en fonction de la différentielle du taux de recouvrement. L'unité de l'entropie superficielle $s$ est le J$\cdot$K$^{-1}\cdot$m$^{-2}$.
\end{proposition}

Cette différentielle de la production d'entropie superficielle (entropie produite par unité d'aire de la surface solide) due à l'accroissement du taux de recouvrement de la surface par les protéines pourra encore être développée en remarquant que la multiplication membre à membre des rapports \ref{DefinitionEmpreinte} et \ref{DefinitionTheta} donne
\begin{equation}\label{ProduitTauxDOccupation}
\phi=\sigma\,\Theta
\end{equation}
dont la différentielle s'écrit
\begin{equation}\label{DifferentiellePhi}
\mathrm{d}\phi=\sigma\,\mathrm{d}\Theta+\Theta\,\mathrm{d}\sigma.
\end{equation}

Quoique d'apparence assez banale, cette différentielle \ref{DifferentiellePhi} mérite quelques commentaires car elle permettra d'interpréter correctement la différentielle \ref{DiffEntropieVSsigmaTheta}. Celle-ci représente en effet, grâce à des valeurs intensives, l'accroissement de la monocouche selon deux contributions: les premier et second termes du membre de droite. Le premier terme, $\sigma\,\mathrm{d}\Theta$, apparaît comme la contribution de l'accroissement de $\Theta$ à l'accroissement de $\phi$ lorsque $\sigma$ demeure constant. Le second terme, $\Theta\,\mathrm{d}\sigma$, apparaît, quant à lui, comme la contribution de l'accroissement de $\sigma$ à l'accroissement de $\phi$ lorsque $\Theta$ demeure constant. Au-delà de ces considérations mathématiques, la signification de cette différentielle peut s'avérer particulièrement parlante:
\begin{itemize}
\item La différentielle $\mathrm{d}\phi$, lorsqu'elle est positive, symbolise l'\emph{adhésion} (accroissement de l'aire de contact entre les protéines et la surface par unité d'aire de cette surface). Lorsqu'elle est négative, elle symbolise le décollement de la monocouche.
\item La différentielle $\mathrm{d}\Theta$, lorsqu'elle est positive, symbolise l'\emph{addition} (augmentation du nombre de moles de protéines par unité de surface). Lorsqu'elle est négative, elle symbolise la déplétion (décrochement des protéines) de la monocouche.
\item La différentielle $\mathrm{d}\sigma$, lorsqu'elle est positive, symbolise la \emph{relaxation} (augmentation de l'empreinte des protéines accumulées dans la monocouche alors que leur nombre y reste constant).
\end{itemize}

Fort de telles significations, il apparaît clairement que l'obtention de la différentielle de la production d'entropie en fonction de ces différentielles serait à même de fournir un moyen efficace de prédire la spontanéité des phénomènes d'\emph{addition} et de \emph{relaxation}. Dès lors, après substitution de cette différentielle \ref{DifferentiellePhi} dans la relation \ref{DiffEntropieVSphi} et distribution des termes qu'elle contient, il vient une nouvelle proposition:

\begin{proposition}[Production d'entropie superficielle lors de l'adhésion]
La production d'entropie superficielle en chaque point d'une surface solide <<~s~>> d'hydrophobie $\mathcal{H}_{\mathrm{sl}\cdot\mathrm{l}}$ sur laquelle vient adhérer une phase suspendue <<~p~>> d'hydrophobie $\mathcal{H}_{\mathrm{pl}\cdot\mathrm{l}}$ est donnée par
\begin{equation}\label{DiffEntropieVSsigmaTheta}
\mathrm{d}_is
=\frac{\sigma}{T}\big(\mathcal{H}_{\mathrm{sl}\cdot\mathrm{l}}+
\mathcal{H}_{\mathrm{pl}\cdot\mathrm{l}}\big)\mathrm{d}\Theta
+\frac{\Theta}{T}\big(\mathcal{H}_{\mathrm{sl}\cdot\mathrm{l}}+
\mathcal{H}_{\mathrm{pl}\cdot\mathrm{l}}\big)\mathrm{d}\sigma
\end{equation}
en fonction des différentielles de l'empreinte $\sigma$ (en m$^2\cdot$mol$^{-1}$) et de la quantité accumulée $\Theta$ (mol$\cdot$m$^{-2}$).
\end{proposition}

La différentielle \ref{DiffEntropieVSsigmaTheta} se développe en deux termes: le premier terme rend compte de la production d'entropie superficielle lors d'une augmentation du nombre de protéines venant adhérer sur la surface solide mais à empreinte constante tandis que le second membre le fait lors d'une augmentation de l'empreinte à nombre de protéines dans le film constante. En d'autres mots, le premier terme comptabilise la production d'entropie superficielle lors de l'addition ($\mathrm{d}\Theta>0$), c'est-à-dire l'augmentation de la quantité accumulée, alors que le second comptabilise la production d'entropie superficielle lors de la relaxation ($\mathrm{d}\sigma>0$), c'est-à-dire l'augmentation de l'empreinte.

En se référant à l'équation \ref{DiffEntropieVSphi}, l'adhésion ($\mathrm{d}\phi>0$) des protéines ne se produit, en vertu du second principe ($\mathrm{d}_is>0$), que lorsque la somme des hydrophobies est positive, c'est-à-dire que l'hydrophobie de la surface est supérieure à l'hydrophilie des protéines devant y adhérer. Étant donné que ces hydrophobies sont elles-mêmes présentes dans la différentielle \ref{DiffEntropieVSsigmaTheta} et que les quantités $\sigma$ et $\Theta$ sont nécessairement positives (il n'y a en effet pas d'aire de contact négative ni de nombre de protéines accumulées négatif), on en déduira que, si l'addition est spontanée, la relaxation le sera tout autant. L'addition et la relaxation apparaissent donc comme les deux parties d'un seul et même phénomène: l'adhésion.

Lorsque les interfaces présentent des hydrophobies adéquates afin que les protéines suspendues dans le milieu liquide qui lui est adjacent viennent y adhérer (créer une aire de contact entre les protéines et la surface), la formation de la monocouche par addition des protéines se fera spontanément. De même, dans ces conditions, la relaxation des protéines (augmentation de l'aire de contact des protéines de la monocouche avec la surface) se fera spontanément, c'est-à-dire qu'elle suivra automatiquement l'addition. Cette relaxation pourra se faire de plusieurs façons dont le changement d'orientation de la protéine et l'étalement (écrasement) de celle-ci sur la surface.

\section{Discussion}

\subsection{Implications du potentiel DLVO}

\paragraph*{Le potentiel de Hamaker (forces de London-van der Waals).}
Tout le développement qui vient d'être réalisé se base uniquement sur la thermodynamique et sur le fait acquis que les protéines s'approcheraient suffisamment de la surface afin d'y interagir suivant les règles déduites. Toutefois, une condition nécessaire afin que l'IgG puisse adhérer sur la surface est de passer outre l'éventuelle barrière de potentiel pouvant résulter des forces électrostatiques englobées dans le modèle DLVO (Derjaguin, Landau, Verwey et Overbeek), modèle constitué par la somme du potentiel électrostatique et du potentiel de Hamaker. Les forces électrostatiques dues à la présence d'ions dans le milieu et de charges en surface des constituants (PS et IgG) résultent en la formation de la double couche diffuse qui peut amener une répulsion ou une attraction entre les deux surfaces chargées. Les forces de van der Waals sont, quant à elles, toujours attractives et sont une propriété des éléments du milieu. Le potentiel de Hamaker est la résultante macroscopique de l'ensemble des interactions dues à la présence des forces de London-van der Waals et, tout comme le potentiel de double couche, il peut être attractif ou répulsif. Dans le cas du système PS-IgG, il sera nettement attractif \citep{roth1996}.

Les forces de van der Waals sont des forces qui s'exercent entre les molécules (elles s'attirent l'une et l'autre) et peuvent être de trois types: Keesom (dipôle-dipôle), Debye (dipôle-dipôle induit) et London (dipôle induit-dipôle induit). Les forces de London-van der Waals (forces de dispersion) sont les plus intenses et elles sont toujours présentes; s'exerçant entre des paires de dipôles s'induisant l'un et l'autre, elles seront d'autant plus fortes que les molécules seront dotées d'un nuage électronique <<~polarisable~>>, c'est-à-dire contenant des électrons délocalisables. Ceci est particulièrement le cas du polystyrène, celui-ci étant extrêmement riche en noyaux aromatiques lesquels sont dotés d'électrons \textgreek{p}. Les forces de London-van der Waals s'exercent lorsque les particules se trouvent à une très faible distance mais, en contrepartie, elles sont violemment attractives \citep{lajusteargile3}.

Le potentiel dû aux forces de van der Waals auquel serait soumise une sphère figurant une IgG <<~p~>> de rayon $r$ approchant à une distance $z$ perpendiculaire à la surface <<~s~>> dans un milieu <<~l~>> (\textit{cf.} figure \ref{FigMécaVdW}) est donné par l'équation \ref{EqHamaker} d'après H. C. Hamaker \citep{hamaker1937} et H. Lyklema \citep{lyklema1} sur base d'une théorie purement macroscopique. La distance entre le centre de la sphère et la surface étant donnée par $z$, l'équation \ref{EqHamaker} est valable pour $z>r$, le domaine $z<r$ correspondant à la pénétration de la sphère dans la surface, ce qui est impossible dans le cas de corps solides et indéformables, cas posé ici. L'approximation dite de Hamaker-De Boer qui y est faite se base sur la simple sommation des forces de London-van der Waals agissant entre tous les points matériels des corps macroscopiques composant le système \citep{lyklema1}.

\begin{figure}[t]\centering
\includegraphics[height=2cm]{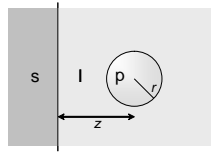}
\caption[Le système colloïdal surface <<~s~>> / fluide <<~l~>> / sphère <<~p~>>]{Un système colloïdal composé d'une sphère <<~p~>> de rayon $r$ dont le centre se trouve à une distance $z$ de la surface <<~s~>>. La sphère et la surface sont séparées par une masse du fluide <<~l~>>.}\label{FigMécaVdW}
\end{figure}

\begin{equation}\label{EqHamaker}
U_{\mathrm{vdw}}(z)=\tfrac{1}{6}\,\mathrm{A}_\mathrm{pls}\bigg[\frac{r}{z+r}+\frac{r}{z-r}+\ln\frac{z-r}{z+r}\bigg]
\end{equation}

La constante de Hamaker $\mathrm{A}_\mathrm{pls}$ dont le signe caractérise la nature attractive ($+$) ou répulsive ($-$) du potentiel $U_{\mathrm{vdw}}(z)$ peut être approximée indépendamment de la distance $z$ séparant les deux corps par l'équation \ref{EqLifshitz} issue de la théorie DLP (Dzyaloshinskii, Lifshitz et Pitaevskii \citep{dzyaloshinskii1961}) des forces de van der Waals.

\begin{equation}\label{EqLifshitz}
\begin{split}
\mathrm{A}_\mathrm{pls}=&\,
\tfrac{1}{2}\,k_BT\bigg[\frac{\varepsilon_\mathrm{l}(i\nu_0)-\varepsilon_\mathrm{p}(i\nu_0)}{\varepsilon_\mathrm{l}(i\nu_0)+\varepsilon_\mathrm{p}(i\nu_0)}\bigg]\bigg[\frac{\varepsilon_\mathrm{s}(i\nu_0)-\varepsilon_\mathrm{p}(i\nu_0)}{\varepsilon_\mathrm{s}(i\nu_0)+\varepsilon_\mathrm{p}(i\nu_0)}\bigg]\\
&\,+\tfrac{2}{3}\,k_BT{\sum_{n=1}^\infty}\bigg[\frac{\varepsilon_\mathrm{l}(i\nu_n)-\varepsilon_\mathrm{p}(i\nu_n)}{\varepsilon_\mathrm{l}(i\nu_n)+\varepsilon_\mathrm{p}(i\nu_n)}\bigg]\bigg[\frac{\varepsilon_\mathrm{s}(i\nu_n)-\varepsilon_\mathrm{p}(i\nu_n)}{\varepsilon_\mathrm{s}(i\nu_n)+\varepsilon_\mathrm{p}(i\nu_n)}\bigg]
\end{split}
\end{equation}

On retrouve dans l'équation \ref{EqLifshitz} \citep{israelachvili2011}, la constante diélectrique du milieu <<~m~>> le long de l'axe imaginaire $i$ en fonction des fréquences $\nu_n$ sont symbolisées par $\varepsilon_\mathrm{m}$. Une estimation grossière de la valeur de la constante $\mathrm{A}_\mathrm{pls}$ est donnée par C.~M. Roth \textit{et al.} \citep{roth1996} pour un système polystyrène-eau-albumine: $1,54\times k_BT$. Cette constante, positive, montre que la force résultante entre la surface de polystyrène et l'IgG est attractive.

\paragraph*{Le potentiel de double couche (forces électrostatiques).}
La solution dans laquelle sont suspendues les IgG est une solution saline et contient donc des ions en grandes quantités: H$_3$O$^+$, OH$^-$, Na$^+$, K$^+$, Cl$^-$, H$_2$PO$_4^-$, HPO$_4^{2-}$, PO$_4^{3-}$, etc. Les IgG, en tant que protéines, présentent à leur surfaces des groupements chimiques ionisables (l'amide de la liaison peptidique) par réaction acide-base et sont donc potentiellement chargées en fonction des conditions physico-chimiques du milieu. Quant au polystyrène, il ne présente pas de charge formelle, il n'est pas ionisé, mais il est capable d'attirer ou de lier ces ions \textit{via} d'autres interactions. D'autres espèces chimiques pourront, en fonction de leurs propriétés, être repoussées des interfaces. Ces attractions et répulsions vis-à-vis des différents ions auront pour conséquence la création d'un nuage ionique par adsorption sur la surface et mènera \textit{in fine} à la formation de doubles couches diffuses amplement étudiées par E. Verwey et J. Overbeek \citep{verwey1948}.

La présence de doubles couches est importante car elle aura une grande influence sur la nature répulsive ou attractive des forces à longue distance s'exerçant entre le polystyrène et les IgG mais aussi entre les IgG elles-mêmes. Elles détermineront la stabilité cinétique du système. La formation de doubles couches est responsable de la solubilité des IgG dans le PBS et le fait qu'elles ne s'agrègent pas mais pourraient venir s'opposer à l'adhésion de celles-ci sur le polystyrène. En effet, le rapprochement des surfaces entraînera une superposition des doubles couches qui y sont associées, ce qui résultera en une force dont le signe et l'intensité dépendront essentiellement des épaisseurs respectives des doubles couches en interaction et des charges de celles-ci. L'obtention d'une expression analytique de la force d'attraction entre une surface et une sphère chargée dont les doubles couches se superposent est une tâche très délicate non encore résolue (\textit{cf.} H. Ohshima \citep{ohshima2006}). Il existe toutefois des solutions donnant un potentiel approché de l'interaction entre une sphère et une paroi dont les surfaces ne sont pas identiquement chargées. Le phénomène est assez compliqué du fait des perturbations réciproques entre les deux doubles couches qui empêchent de faire une simple addition des potentiels de chacune. Lorsque, par approximation, les potentiels de surface de l'IgG $\psi_\mathrm{p}(z)$ et du polystyrène $\psi_\mathrm{s}(z)$ ne se perturbent pas mutuellement en fonction de $z$, le potentiel dû à l'approche des doubles couches est donné par l'équation \ref{EqDoubleC} \citep{hiemenz1997tout}.

\begin{equation}\label{EqDoubleC}
U_\mathrm{\acute{e}lec}(z)=2\frac{\psi_\mathrm{s}(0)\psi_\mathrm{p}(0)}{\psi_\mathrm{s}^2(0)+\psi_\mathrm{p}^2(0)}\frac{1+\exp[-\kappa z]}{1-\exp[-\kappa z]}-\ln\Big(1-\exp[2\kappa z]\Big)
\end{equation}

La longueur de Debye, caractéristique de l'épaisseur de la double couche, est donnée par $\kappa^{-1}=\sqrt{\varepsilon k_BT/2I}$ où $I$ est la force ionique.
Dans du tampon phosphate typique, elle se situe entre 1 et 2 nm. Les points isoélectriques approximatifs du polystyrène (potentiel d'écoulement dans le KNO$_3$ \citep{dupont2000}) et des IgG (électrophorèse sur gel dans le PBS \citep{buijs1995,bremer2004,buijs1996b,buijs1997un,elgersma1992,elgersma1991}) sont respectivement de 4 et 6,3, laissant supposer que $\psi_\mathrm{s}(z)<\psi_\mathrm{p}(z)\leqslant0$ mV. Le potentiel électrocinétique $\psi(z)$ est rarement égal au potentiel $\psi(0)$ et son interprétation demeure assez délicate surtout en présence d'ions multivalents comme le phosphate \citep{lyklema2}, rendant ces chiffres peu concluants. Il semble dès lors nécessaire de poursuivre les investigations dans ce domaine surtout que les forces électrostatiques semblent largement dominantes; le potentiel DLVO estimé en sommant les équations \ref{EqLifshitz} et \ref{EqDoubleC} devient répulsif à partir d'une distance de 1 nm ($\kappa^{-1}\approx2$ nm). \'{E}tant donné leur forme approximative, ces potentiels doivent, eux aussi, être observés avec une certaine prudence en particulier pour la partie attractive. M. Bremer \textit{et al. } \citep{bremer2004} suggèrent d'ailleurs que l'influence des doubles couches devraient simplement être déterminée à partir des caractéristiques chimiques et électrostatiques des deux surfaces.

D'autre part, on peut aussi s'attendre à ce que les interactions d'origine électrostatique jouent elles-mêmes un rôle dans l'orientation qu'adoptera l'IgG à la surface. Malgré la complexité susmentionnée, il est toutefois possible de raisonner efficacement en considérant les deux situations suivantes: potentiels PS-IgG répulsif ou attractif. Pour le cas attractif, la conclusion est la même que pour les interactions de London-van der Waals. Concernant le cas répulsif, les forces électrostatiques créent une barrière de potentiel dans la cadre de la théorie DLVO compromettant l'approche de la surface par l'IgG et rendant superflue toute spéculation quant à son orientation. Malgré la rareté de l'événement, on ne peut toutefois pas exclure que les IgG puissent passer outre cette barrière de potentiel, tombant alors dans le champ d'action des forces de van der Waals qui réaliseront alors leur {\oe}uvre en maximisant son empreinte $\sigma(\boldsymbol{\omega})$.

\paragraph*{Aspects cinétiques de l'hydrophobie.}
Le commentaire réalisé permet de synthétiser l'action de la diffusion, du potentiel DLVO et de la thermodynamique. La diffusion, conséquence du mouvement brownien dont sont animées les protéines, permet à la protéine et à la surface solide de s'approcher à une distance à laquelle elles pourront interagir. \`{A} partir d'une certaine distance, les potentiels dus aux doubles couches entourant chaque surface se superposeront et la forme du potentiel résultant déterminera le devenir du système. Si le potentiel résultant de cette superposition est très positif, alors il constitue une barrière difficilement franchissable rendant une approche plus intime des deux corps très improbable. Inversement, lorsque le potentiel est négatif, dû au fait que les doubles couches sont globalement chargées selon des signes opposés, la barrière de potentiel sera absente rendant improbable le maintient à distance des deux surfaces. Les deux corps pourront s'approcher à une distance telle que le potentiel de Hamaker, très attractif, pourra à son tour agir. La situation où les doubles couches sont toutes deux porteuses du même signe mais dont les charges globales sont peu élevées aboutira à une probabilité d'approche intermédiaire, ni nulle, ni très élevée. Quoi qu'il en soit, après que la protéine et la surface ont pu s'approcher selon une probabilité déterminée par la barrière de potentiel due à l'approche de leurs doubles couches respectives, le potentiel de Hamaker, agissant à plus courte distance mais de manière très intense, les fera adhérer l'une à l'autre.

Le potentiel électrostatique détermine la stabilité du système et le potentiel de Hamaker agira, lui, toujours dans le sens de l'agrégation lorsque les corps sont réputés lyophobes. Ce potentiel de Hamaker, dû aux forces de London-van der Waals, est très intense et provoquera, au fur et à mesure de l'approche, une compression des doubles couches au niveau des surfaces approchantes (surface de la protéine et surface de dépôt). La compression des doubles couches induira nécessairement une augmentation des potentiels chimiques $\mu_{j(\mathrm{pl})}$ et $\mu_{j(\mathrm{sl})}$ des espèces qui y sont adsorbées éloignant le système de l'équilibre thermodynamique. Cette rupture de l'équilibre mécanique due, premièrement, à la faible barrière de potentiel électrostatique et, deuxièmement, au potentiel attractif de Hamaker induira une augmentation des affinités de désorption $A_{j(\mathrm{pl\cdot l})}$ et $A_{j(\mathrm{sl\cdot l})}$ des espèces $j$. Le retour à l'équilibre thermodynamique, défini par l'annulation des affinités, se fera dès lors par désorption des espèces qui étaient initialement adsorbées aux interfaces IgG/PBS et PS/PBS. Accompagné d'une production nette d'entropie, ce processus de désorption est irréversible.

C'est l'effet hydrophobe. Celui-ci sera d'autant plus intense que les éléments impliqués seront faiblement chargés afin d'éviter la répulsion entre les doubles couches, fortement ressemblants afin de favoriser les forces de London-van der Waals et que l'adsorption initiale soit élevée augmentant la production entropique lors de l'adhésion. Le polystyrène ne porte pas de charge électrique formelle, la barrière d'énergie potentielle qu'il peut dès lors développer vis-à-vis d'un autre élément comme les IgG demeurera donc faible. L'IgG peut alors suffisamment approcher le polystyrène de façon à subir la très forte influence des forces de London-van der Waals qui la forceront à adhérer au polystyrène tout en éjectant une partie des espèces qu'ils avaient tous deux adsorbée, rendant leur adhésion irréversible. 

\subsection{Adhésion des IgG sur la surface}

Lorsque le potentiel électrostatique dû à la superposition des doubles couches des IgG et de la surface permettent une approche suffisante de l'IgG et de la surface, le potentiel de Hamaker, très attractif, {\oe}uvrera très nettement dans le sens de l'adhésion de l'IgG sur la surface. Cette attraction conduira, \textit{via} l'accroissement des affinités de désorption, à une production nette d'entropie dont il est maintenant possible d'examiner les conséquences dans le cas des IgG et en particulier sur les deux événements que sont l'addition et la relaxation.

De manière générale, lorsqu'une protéine viendra adhérer sur la surface, elle y présentera une certaine portion de la surface qui la délimite afin d'y créer un contact (\textit{cf}. figure \ref{Fig_Chap1_4}). Cette portion de sa surface dépendra de son orientation voire de sa conformation, état déterminé par le pseudovecteur $\boldsymbol{\omega}$, et a été qualifiée d'empreinte caractéristique $\sigma(\boldsymbol{\omega})$.

\begin{figure}[t]\centering
\includegraphics[width=8cm]{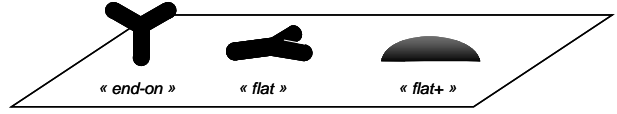}
\caption[Orientations et conformations possibles pour des IgG]{Quelques orientations et conformations possibles pour des IgG adhérant sur une surface solide. Lorsque l'IgG est attachée à la surface par un de ses fragments, son orientation est dite \textit{end-on}, lorsqu'elle est allongée de tout son long, son orientation est dite \textit{flat} et lorsqu'elle est dénaturée, on la qualifiera de \textit{flat$+$}.}\label{Fig_Chap1_5}
\end{figure}

Comme on peut le voir sur le schéma de la figure \ref{Fig_Chap1_5}, les IgG ne vont pas avoir la même surface de contact selon leur orientation: \textit{end-on} (l'IgG est debout, collée par sa partie constante) et \textit{flat} (l'IgG est entièrement couchée sur la surface). La dénaturation étant vue comme un étalement sur la surface, lorsque l'IgG sera dénaturée (\textit{flat$+$}), son aire de contact sera encore plus grande. Cela se traduit par la série d'empreintes caractéristiques
\begin{equation}\label{SérieEmpreintesLimitesIgG}
\sigma(\boldsymbol{\omega}_\text{end-on})<\dots<
\sigma(\boldsymbol{\omega}_\text{flat})<\dots<
\sigma(\boldsymbol{\omega}_\text{flat+})
\end{equation}
conjecturant que l'empreinte caractéristique est la plus grande pour les IgG dénaturées.

Grâce à la différentielle \ref{DifferentiellePhi} du taux d'occupation de la surface et du développement \ref{EmpreinteVSempreintesLimites} de l'empreinte à partir des empreintes caractéristiques, il est possible de démontrer\footnote{\`{A} partir de l'expression \ref{EmpreinteVSempreintesLimites} de l'empreinte, il vient facilement la différentielle
\begin{equation}\label{vjglsvbwdg}
\mathrm{d}\sigma=
\sigma(\boldsymbol{\omega}_1)\mathrm{d}\mathcal{X}(\boldsymbol{\omega}_1)+
\sigma(\boldsymbol{\omega}_2)\mathrm{d}\mathcal{X}(\boldsymbol{\omega}_2)+\dots
\end{equation}
puisque les empreintes caractéristiques sont des constantes au cours du remplissage. En substituant dans la différentielle \ref{DifferentiellePhi} du taux de recouvrement la différentielle $\mathrm{d}\sigma$ et l'empreinte $\sigma$ par leurs expressions \ref{vjglsvbwdg} et \ref{EmpreinteVSempreintesLimites}, on obtient la relation
\begin{equation}
\mathrm{d}\phi
=\big[\sigma(\boldsymbol{\omega}_1)\mathcal{X}(\boldsymbol{\omega}_1)+
\sigma(\boldsymbol{\omega}_2)\mathcal{X}(\boldsymbol{\omega}_2)+\dots\big]\mathrm{d}\Theta
+\Theta\big[
\sigma(\boldsymbol{\omega}_1)\mathrm{d}\mathcal{X}(\boldsymbol{\omega}_1)+
\sigma(\boldsymbol{\omega}_2)\mathrm{d}\mathcal{X}(\boldsymbol{\omega}_2)+\dots
\big]
\end{equation}
Après distribution, il vient
\begin{equation}
\mathrm{d}\phi
=\sigma(\boldsymbol{\omega}_1)\mathcal{X}(\boldsymbol{\omega}_1)\mathrm{d}\Theta+
\sigma(\boldsymbol{\omega}_2)\mathcal{X}(\boldsymbol{\omega}_2)\mathrm{d}\Theta+
\sigma(\boldsymbol{\omega}_1)\Theta\mathrm{d}\mathcal{X}(\boldsymbol{\omega}_1)+
\sigma(\boldsymbol{\omega}_2)\Theta\mathrm{d}\mathcal{X}(\boldsymbol{\omega}_2)+\dots
\end{equation}
qui, après factorisation des empreintes caractéristiques, fournit
\begin{equation}\label{d4v24wdb54b5464}
\mathrm{d}\phi
=\sigma(\boldsymbol{\omega}_1)\big[\mathcal{X}(\boldsymbol{\omega}_1)\mathrm{d}\Theta+
\Theta\mathrm{d}\mathcal{X}(\boldsymbol{\omega}_1)\big]+
\sigma(\boldsymbol{\omega}_2)\big[\mathcal{X}(\boldsymbol{\omega}_2)\mathrm{d}\Theta+
\Theta\mathrm{d}\mathcal{X}(\boldsymbol{\omega}_2)\big]+\dots
\end{equation}
\`{A} partir des relations \ref{f564desf874gs7}, on obtient les différentielles
\begin{equation}
\mathrm{d}\Theta(\boldsymbol{\omega}_1)=\mathcal{X}(\boldsymbol{\omega}_1)\mathrm{d}\Theta+\Theta\mathrm{d}\mathcal{X}(\boldsymbol{\omega}_1)\quad\text{et}\quad\mathrm{d}\Theta(\boldsymbol{\omega}_2)=\mathcal{X}(\boldsymbol{\omega}_2)\mathrm{d}\Theta+\Theta\mathrm{d}\mathcal{X}(\boldsymbol{\omega}_2)
\end{equation}
que l'on peut aisément reconnaître dans les crochets de l'expression \ref{d4v24wdb54b5464}. La relation \ref{PhiVSaccumulationsLimites} suit alors naturellement.} la relation \ref{PhiVSaccumulationsLimites}.
\begin{equation}\label{PhiVSaccumulationsLimites}
\mathrm{d}\phi
=\sigma(\boldsymbol{\omega}_1)\mathrm{d}\Theta(\boldsymbol{\omega}_1)+
\sigma(\boldsymbol{\omega}_2)\mathrm{d}\Theta(\boldsymbol{\omega}_2)+\dots
\end{equation}
Cette relation montre que l'accroissement du taux de recouvrement de la surface peut être relié aux différentielles des quantités de protéines selon chacune de leurs orientations et conformations probables.
En considérant la série \ref{SérieEmpreintesLimitesIgG}, cela se traduira pour les IgG par la relation
\begin{equation}\label{PhiVSaccumulationsLimitesIgG}
\mathrm{d}\phi
=\sigma(\boldsymbol{\omega}_\text{end-on})\mathrm{d}\Theta(\boldsymbol{\omega}_\text{end-on})+
\sigma(\boldsymbol{\omega}_\text{flat})\mathrm{d}\Theta(\boldsymbol{\omega}_\text{flat})+
\sigma(\boldsymbol{\omega}_\text{flat+})\mathrm{d}\Theta(\boldsymbol{\omega}_\text{flat+})+\dots
\end{equation}
qui, une fois substituée dans la différentielle \ref{DiffEntropieVSphi} de la production d'entropie superficielle, fournit
\begin{equation}\label{DiffEntropIgG}
\begin{split}
\mathrm{d}_is
=&\,\frac{1}{T}\big(
\mathcal{H}_{\mathrm{sl}\cdot\mathrm{l}}+
\mathcal{H}_{\mathrm{pl}\cdot\mathrm{l}}\big)\sigma(\boldsymbol{\omega}_\text{end-on})\mathrm{d}\Theta(\boldsymbol{\omega}_\text{end-on})\\
&+\frac{1}{T}\big(
\mathcal{H}_{\mathrm{sl}\cdot\mathrm{l}}+
\mathcal{H}_{\mathrm{pl}\cdot\mathrm{l}}\big)\sigma(\boldsymbol{\omega}_\text{flat})\mathrm{d}\Theta(\boldsymbol{\omega}_\text{flat})\\
&\quad+\frac{1}{T}\big(
\mathcal{H}_{\mathrm{sl}\cdot\mathrm{l}}+
\mathcal{H}_{\mathrm{pl}\cdot\mathrm{l}}\big)\sigma(\boldsymbol{\omega}_\text{flat+})\mathrm{d}\Theta(\boldsymbol{\omega}_\text{flat+})+\dots
\end{split}
\end{equation}
Cette différentielle de l'entropie superficielle est valable en toute généralité, c'est-à-dire qu'elle paramétrise les deux phénomènes d'addition et de relaxation. Cela se justifie par le fait que la démonstration menant à la relation \ref{PhiVSaccumulationsLimites} n'a pas fait usage d'une hypothèse sur la nullité de la différentielle de l'empreinte $\mathrm{d}\sigma$, bien au contraire.

Cette relation \ref{DiffEntropIgG} rend possible l'identification des contributions à l'entropie des phénomènes d'addition, de relaxation orientationnelle et de relaxation conformationnelle pour les IgG tels que décrits dans le modèle de la figure \ref{Fig_Chap1_6}. L'addition est le phénomène par lequel les IgG quitteront la suspension aqueuse pour venir se coller sur la surface. Sur le modèle de la figure \ref{Fig_Chap1_6}, seules des IgG ayant une orientation \textit{end-on} viendront s'additionner à la monocouche depuis la suspension. Dès lors, l'addition est caractérisée par la différentielle $\mathrm{d}\Theta(\boldsymbol{\omega}_\text{end-on})$ (de signe positif); ce qui donnera, à partir de la différentielle \ref{DiffEntropIgG}:
\begin{equation}\label{65gx6gvb65fh6498h}
\mathrm{d}_is
=\frac{1}{T}\big(
\mathcal{H}_{\mathrm{sl}\cdot\mathrm{l}}+
\mathcal{H}_{\mathrm{pl}\cdot\mathrm{l}}\big)\sigma(\boldsymbol{\omega}_\text{end-on})\mathrm{d}\Theta(\boldsymbol{\omega}_\text{end-on}),
\end{equation}
représentant la production d'entropie superficielle lors de l'addition d'IgG \textit{end-on} sur la surface. On doit remarquer que, si seules les IgG dans une orientation \textit{end-on} peuvent s'additionner sur la surface, on aura l'égalité $\mathrm{d}\Theta=\mathrm{d}\Theta(\boldsymbol{\omega}_\text{end-on})$ montrant que l'accroissement de la grandeur $\Theta$ sera seulement dû à l'accumulation d'IgG \textit{end-on} dans la monocouche. \`{A} partir de la relation précédente, on aura alors
\begin{equation}\label{g46dsg5864d4g68}
\mathrm{d}_is
=\frac{1}{T}\big(
\mathcal{H}_{\mathrm{sl}\cdot\mathrm{l}}+
\mathcal{H}_{\mathrm{pl}\cdot\mathrm{l}}\big)\sigma(\boldsymbol{\omega}_\text{end-on})\mathrm{d}\Theta
\end{equation}
pour l'accroissement de l'entropie superficielle causée par l'addition d'IgG sur la surface. Le membre de droite de cette équation pourra être comparé avec le premier terme du membre de droite de la relation \ref{DiffEntropieVSsigmaTheta}.

\begin{figure}[t]\centering
\includegraphics[width=11cm]{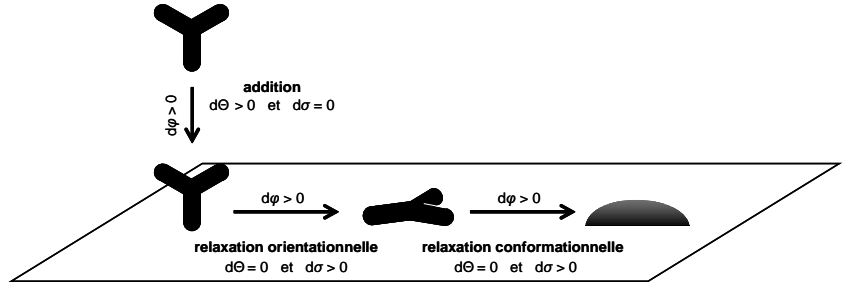}
\caption[Adhésion d'une IgG sur du polystyrène dans un milieu aqueux]{Les étapes thermodynamiques de l'adhésion (addition et relaxation) d'une IgG sur du polystyrène dans un milieu aqueux. Toutes ces étapes sont caractérisées par une différentielle $\mathrm{d}\phi$ positive. L'addition selon deux orientations différentes (\textit{i.e.} \textit{end-on} où l'IgG est debout et \textit{flat} où l'IgG est couchée) sur la surface augmente le taux de recouvrement de la surface $\phi$ en augmentant le nombre d'IgG adhérant à la surface ($\mathrm{d}\Theta>0$) afin de respecter le second principe de la thermodynamique. Cette addition est spontanément suivie des relaxations orientationnnelle (réorientation afin d'avoir $\mathrm{d}\sigma>0$) et conformationnelle (dénaturation de la protéine) permettant à l'IgG d'optimiser ses interactions sous la forme \textit{flat$+$}.}\label{Fig_Chap1_6}
\end{figure}

Les IgG, une fois additionnées à la monocouche selon une orientation \textit{end-on}, pourront relaxer leur orientation afin de devenir \textit{flat}. L'augmentation du nombre d'IgG \textit{flat} est figurée par la différentielle (positive) $\mathrm{d}\Theta(\boldsymbol{\omega}_\text{flat})$. Sachant que la quantité d'IgG \textit{flat} qui apparaîtra sera équivalente à la quantité d'IgG \textit{end-on} qui disparaîtra, on aura l'égalité $\mathrm{d}\Theta(\boldsymbol{\omega}_\text{end-on})=-\mathrm{d}\Theta(\boldsymbol{\omega}_\text{flat}).$ Après substitution et réarrangement des deux premiers termes du membre de droite de l'équation \ref{DiffEntropIgG}, il vient
\begin{equation}
\mathrm{d}_is
=\frac{1}{T}\big(
\mathcal{H}_{\mathrm{sl}\cdot\mathrm{l}}+
\mathcal{H}_{\mathrm{pl}\cdot\mathrm{l}}\big)\Big[\sigma(\boldsymbol{\omega}_\text{flat})-\sigma(\boldsymbol{\omega}_\text{end-on})\Big]\mathrm{d}\Theta(\boldsymbol{\omega}_\text{flat})
\end{equation}
donnant la production d'entropie superficielle due à la relaxation orientationnelle des IgG. Le facteur $\sigma(\boldsymbol{\omega}_\text{flat})-\sigma(\boldsymbol{\omega}_\text{end-on})$ y symbolise le gain d'aire de contact entre les IgG et la surface lorsqu'une mole d'IgG relaxe en passant de l'orientation \textit{end-on} à l'orientation \textit{flat}, c'est-à-dire lorsqu'elles relaxent leur orientation.

La relaxation orientationnelle n'est pas seule à agir, puisque la \emph{relaxation conformationnelle}, au cours de laquelle l'IgG perd une partie ou la totalité de sa structure interne afin d'augmenter son empreinte caractéristique, peut, elle-aussi, advenir. Les événements menant à la dénaturation de la conformation d'une protéine sont multiples: changement de température, de pression mais aussi la présence d'une surface (hydrophobe) \citep{winter2007}. De fait, la dénaturation des protéines lors de leur adhésion sur une surface a été mise en évidence par M.~E. Soderquist et A.~G. Walton \citep{soderquist1980} et quelques explications en ont été tentées par W. Norde \citep{norde2008} conduisant aux concepts de protéines molles (\textit{soft}) et dures (\textit{hard}). Une protéine dure est une protéine dont la cohésion interne est forte alors qu'elle sera plus faible dans le cas d'une protéine molle. Plus la cohésion interne sera élevée, moins la protéine sera sensible à la dénaturation.

De façon identique, la relaxation de la conformation des IgG se fera par disparition d'une portion des IgG \textit{flat} au profit d'une apparition d'IgG dans une conformation de type \textit{flat$+$}. \`{A} l'aide de l'égalité $\mathrm{d}\Theta(\boldsymbol{\omega}_\text{flat})=-\mathrm{d}\Theta(\boldsymbol{\omega}_\text{flat+}),$ on obtient, après tout réarrangement des deux derniers termes du membre de droite de l'équation \ref{DiffEntropIgG}, la différentielle
\begin{equation}
\mathrm{d}_is
=\frac{1}{T}\big(
\mathcal{H}_{\mathrm{sl}\cdot\mathrm{l}}+
\mathcal{H}_{\mathrm{pl}\cdot\mathrm{l}}\big)\Big[\sigma(\boldsymbol{\omega}_\text{flat+})-\sigma(\boldsymbol{\omega}_\text{flat})\Big]\mathrm{d}\Theta(\boldsymbol{\omega}_\text{flat+})
\end{equation}
donnant la production superficielle d'entropie lors d'un changement de conformation des IgG. Le facteur 
$\sigma(\boldsymbol{\omega}_\text{flat+})-\sigma(\boldsymbol{\omega}_\text{flat}),$ y donne le gain d'aire de contact entre les IgG et la surface lorsqu'une mole d'IgG change de conformation. En vertu de la série \ref{SérieEmpreintesLimitesIgG}, ce gain d'aire est positif. Lorsque les conditions ont été remplies (somme des hydrophobies positive), il vient, en vertu du second principe, que la relaxation conformationnelle est un phénomène spontané.

Les notions de thermodynamique permettent donc de décrire ce qui doit advenir des IgG suspendues à proximité d'une surface solide lorsque les conditions expérimentales les poussent à adhérer sur cette surface. Premièrement, les IgG auront spontanément tendance à venir s'additionner, quelle que soit leur orientation sur cette surface produisant une augmentation de $\Theta$, le nombre d'IgG présentes sur la surface (premier terme du membre de droite de la différentielle \ref{DiffEntropieVSsigmaTheta}). Deuxièmement, les IgG présentes sur la surface auront tendance à relaxer spontanément afin d'augmenter leur empreinte caractéristique $\sigma(\boldsymbol{\omega})$ sur la surface et ainsi gagner des orientations et conformations caractérisées par une aire de contact élevée avec la surface, c'est-à-dire \textit{flat}, voire \textit{flat$+$} (second terme du membre de droite de la différentielle \ref{DiffEntropieVSsigmaTheta}).

Ces augmentations des quantités $\Theta$ et $\sigma$ permettent, toutes deux, d'augmenter l'aire de contact entre la phase colloïdale (les IgG) et la surface solide (le polystyrène). Cette augmentation de $\Sigma_\mathrm{sp}$ se fait spontanément, irréversiblement et naturellement si elle peut s'accompagner d'une production d'entropie comme cela a été montré.

\subsection{L'enthalpie libre}\label{35st14dw3b1}

Il est habituel de décrire l'évolution spontanée des systèmes physico-chimiques en utilisant la différentielle de l'enthalpie libre $G$ (de Gibbs). On trouvera en effet dans la majorité des manuels de chimie que le caractère négatif de la différentielle de la fonction d'état $G$ permet de décrire le sens d'évolution d'un tel système. Une réaction chimique ou un autre phénomène physico-chimique se produira lorsque la condition $\mathrm{d}G<0$ sera remplie attendu que la température $T$ et la pression $p$ demeurent constantes. Pourquoi, dès lors, ne pas caractériser la formation des monocouches de protéines en utilisant cette notion plus usitée qu'est l'enthalpie libre? Tenter de répondre à cette question passe naturellement par la définition de $G$.
L'enthalpie libre ou énergie libre de Gibbs est une fonction d'état $G$ définie par la relation
\begin{equation}\label{DéfinitionG}
G=H-TS
\end{equation}
dans laquelle $T$ et $S$ sont respectivement la température et l'entropie et $H$ est l'enthalpie. Cette dernière est elle-même une fonction d'état définie par
\begin{equation}\label{DéfinitionH}
H=E+pV
\end{equation}
où $E$ est l'énergie interne du système (voir la relation \ref{EqPremierPrincipe1}), $p$ la pression exercée sur le système et $V$ son volume.

Après avoir substitué la définition de l'enthalpie \ref{DéfinitionH} dans \ref{DéfinitionG} et différentiation de $G$, on obtient:
\begin{equation}\label{DiffG1}
\mathrm{d}G=\mathrm{d}E+p\mathrm{d}V+V\mathrm{d}p-S\mathrm{d}T-T\mathrm{d}_eS-T\mathrm{d}_iS.
\end{equation}
dans laquelle on a développé la différentielle de l'entropie en ses deux parties <<~$e$~>> et <<~$i$~>> (voir relation \ref{EqSecondPrincipe3}).

Pour un système capillaire fermé de volume $V$ sur lequel s'exerce une pression $p$ en provenance de l'environnement, la différentielle de l'énergie interne se développe à partir de la relation \ref{EqPotentielGibbs:EnergieInterne1} sous la forme
\begin{equation}\label{648b68fg4b6f8g4bfg68b4}
\mathrm{d}E=T\mathrm{d}_eS-p\mathrm{d}V+\big(\gamma_\mathrm{ps}-\gamma_\mathrm{sl}-\gamma_\mathrm{pl}\big)\mathrm{d}\Sigma_\mathrm{ps}
\end{equation}
qui, après substitution dans \ref{DiffG1}, fournit
\begin{equation}\label{DiffG2}
\mathrm{d}G
=T\mathrm{d}_eS-p\mathrm{d}V+\big(\gamma_\mathrm{ps}-\gamma_\mathrm{sl}-\gamma_\mathrm{pl}\big)\mathrm{d}\Sigma_\mathrm{ps}
+p\mathrm{d}V+V\mathrm{d}p-T\mathrm{d}_eS-T\mathrm{d}_iS-S\mathrm{d}T
\end{equation}
se simplifiant et se réarrangeant selon
\begin{equation}\label{Dif354fG2}
\mathrm{d}G
=-S\mathrm{d}T+V\mathrm{d}p
+\big(\gamma_\mathrm{ps}-\gamma_\mathrm{sl}-\gamma_\mathrm{pl}\big)\mathrm{d}\Sigma_\mathrm{ps}
-T\mathrm{d}_iS.
\end{equation}
Dans le cas d'un système chimique traditionnel fermé, c'est-à-dire ne comprenant pas d'interface, la différentielle de $G$ se simplifie sous une forme\footnote{Il est par ailleurs intéressant de noter que, lorsque le système évolue à l'équilibre (transformation réversible), la différentielle de $G$ s'écrit
\begin{equation}\label{35fg41g435hf4}
\mathrm{d}G=-S\mathrm{d}T+V\mathrm{d}p,
\end{equation}
différentielle aussi écrite sous la forme \ref{DiffTot1E}. En comparant les différentielles \ref{35fg41g435hf4} et \ref{DiffTot1E}, il vient que
\begin{equation}
S=-\bigg(\frac{\partial G}{\partial T}\bigg)_p\qquad\mathrm{et}\qquad
V=\bigg(\frac{\partial G}{\partial p}\bigg)_T,
\end{equation}
deux relations évoquées à la section \ref{hb35d8f}.} donnée par G.~Lebon \citep{lebon2008a}:
\begin{equation}
\mathrm{d}G
=-S\mathrm{d}T+V\mathrm{d}p
-T\mathrm{d}_iS,
\end{equation}
ce qui permet de montrer que, lorsque la température $T$ est constante ($\mathrm{d}T=0$), la pression est constante ($\mathrm{d}p=0$) et le système fermé,
\begin{equation}
(\mathrm{d}G)_{T,p}=-T\mathrm{d}_iS
\end{equation}
et, en vertu du second principe de la thermodynamique (\textit{cf}. inégalité \ref{EqThermoCroissanceEntrop}), donne
\begin{equation}\label{DiffG3}
(\mathrm{d}G)_{T,p}\leqslant0,
\end{equation}
le critère communément admis d'évolution des systèmes physico-chimiques. Ce critère de décroissance de l'enthalpie libre est présent dans tous les manuels de références en chimie et physico-chimie \citep{atkins} et a donné lieu à de multiples interprétations, en particulier pour les systèmes colloïdaux \citep{hunter1987,hiemenz1997tout,israelachvili2011}. Toutefois, à partir du moment où l'on considère un système colloïdal dans lequel un travail d'adhésion est présent (\textit{cf}. équation \ref{EqTravailAdhésion2} de Duprè), son usage en tant que critère d'évolution spontanée devient difficile. En effet, pour ce genre de système, il faut se rendre à l'évidence, qu'à partir de la différentielle \ref{Dif354fG2}, il vienne la relation
\begin{equation}\label{DiffG4}
(\mathrm{d}G)_{T,p}=\big(\gamma_\mathrm{ps}-\gamma_\mathrm{sl}-\gamma_\mathrm{pl}\big)\mathrm{d}\Sigma_\mathrm{ps}-T\mathrm{d}_iS
\end{equation}
pour un système colloïdal fermé évoluant à pression et température constantes. Ensuite, après réarrangement et utilisation de l'inégalité \ref{EqThermoCroissanceEntrop}, une inégalité du genre
\begin{equation}
(\mathrm{d}G)_{T,p}-\big(\gamma_\mathrm{ps}-\gamma_\mathrm{sl}-\gamma_\mathrm{pl}\big)\mathrm{d}\Sigma_\mathrm{ps}
\leqslant0
\end{equation}
serait obtenue. Cette inégalité pourrait être un critère d'évolution des systèmes physico-chimiques dans lesquels se produiraient des phénomènes d'agrégation et/ou d'adhésion\footnote{Il est par ailleurs intéressant de remarquer que lorsque les transformations affectant un systèmes sont réversibles, c'est-à-dire que $\mathrm{d}_iS=0$, la différentielle
\begin{equation}
(\mathrm{d}G)_{T,p}=\big(\gamma_\mathrm{ps}-\gamma_\mathrm{sl}-\gamma_\mathrm{pl}\big)\mathrm{d}\Sigma_\mathrm{ps}
\end{equation}
serait obtenue à partir de \ref{DiffG4}. En en divisant les deux membres par la différentielle $\mathrm{d}\Sigma_\mathrm{ps}$, il viendrait la dérivée partielle
\begin{equation}
\bigg(\frac{\partial G}{\partial\Sigma_\mathrm{ps}}\bigg)_{T,p}=\gamma_\mathrm{ps}-\gamma_\mathrm{sl}-\gamma_\mathrm{pl}
\end{equation}
qui est le travail d'adhésion (réversible) entre deux phases <<~p~>> et <<~s~>> suspendues dans un milieu liquide <<~l~>>, puisque le membre de droite est équivalent à l'équation \ref{EqTravailAdhésion2} du travail de Duprè. Cette dérivée pourra être utilement comparée à l'équation 57 de la référence \citep{hiemenz1997tout}.

Dans le cas d'une transformation irréversible, la substitution de \ref{DefEntropieAdsorption6} dans \ref{DiffG4} et l'écriture sous forme d'une dérivée partielle mènent à
\begin{equation}
\bigg(\frac{\partial G}{\partial\Sigma_\mathrm{ps}}\bigg)_{T,p}=\gamma_\mathrm{ps}-\gamma_\mathrm{sl}-\gamma_\mathrm{pl}+\mathcal{H}_{\mathrm{sl}\cdot\mathrm{l}}+
\mathcal{H}_{\mathrm{pl}\cdot\mathrm{l}}
\end{equation}
ou bien
\begin{equation}
\bigg(\frac{\partial G}{\partial\Sigma_\mathrm{ps}}\bigg)_{T,p}=\gamma_\mathrm{ps}-\gamma_\mathrm{sl}-\gamma_\mathrm{pl}-\mathcal{H}_{\mathrm{l}\cdot\mathrm{sl}}-
\mathcal{H}_{\mathrm{l}\cdot\mathrm{pl}}
\end{equation}
montrant que, lors d'une adhésion, les hydrophobies/hydrophilies sont prises en compte dans la dérivée partielle de l'enthalpie libre. Ces dernières relations sont très intéressantes car on y voit clairement deux contributions à la dérivée partielle de $G$: l'une mécanique (les tensions superficielles), l'autre physico-chimique (les hydrophobies/hydrophilies).}. Ce critère semble toutefois d'un usage peu pratique alors qu'I.~Prigogine \citep{prigogine1968} souligne que le second principe de la thermodynamique ($\mathrm{d}_iS\geqslant0$) est le seul vrai critère permettant de décrire le sens d'évolution des systèmes physico-chimiques et que c'est bien à partir de celui-ci que l'inégalité \ref{DiffG3} peut être prouvée et ensuite utilisée \citep{lebon2008a}. Son usage intensif a dû finir par faire oublier d'où il venait mais son intérêt réside dans le fait que les systèmes chimiques traditionnels (monophasiques) sont généralement fermés et se transforment bien à pression et température constantes. Ces conditions permettent de voir la différentielle $\mathrm{d}G$ comme une différentielle totale, ce qui en fait une quantité intégrable mais aussi à partir de laquelle il est possible d'obtenir les $\Delta_r G$ qui est un autre critère d'évolution souvent admis. La différentielle $\mathrm{d}_iS$ n'est pas totale mais partielle et elle est, de ce fait, non intégrable; d'où l'impossibilité d'en obtenir des <<~$\Delta_iS$~>> dont les valeurs pourraient être obtenues expérimentalement par calorimétrie à l'instar des $\Delta_r G$.

Les systèmes colloïdaux ont pour principale caractéristique d'être <<~riches en surfaces~>> rendant difficile le fait de poser que $\mathrm{d}\Sigma_\mathrm{sp}=0$. En posant $\mathrm{d}\Sigma_\mathrm{sp}=0$ afin d'obtenir une inégalité telle que \ref{DiffG3}, on nierait la possibilité d'une transformation aux interfaces, c'est-à-dire une <<~transformation colloïdale~>>, vidant $\mathrm{d}G$ de toute sa signification et de tout son intérêt.

La différentielle $\mathrm{d}G$ semble donc difficilement utilisable dans le cadre du présent travail de telle sorte que le critère général d'accroissement de l'entropie donné par le second principe, valable en toute généralité, sera le seul auquel il sera fait référence afin de décrire le sens d'évolution spontanée du système d'intérêt.

\section{Conclusion}

Tout au long de ce chapitre, il a été montré à quel point la thermodynamique des processus irréversibles était un outil efficace afin de conceptualiser les phénomènes d'adhésion entre les corps hydrophobes. Les équations développées ainsi que les définitions de concepts (hydrophobie $\mathcal{H}$, taux de recouvrement $\phi$, empreinte $\sigma$ et quantité accumulée $\Theta$) donnent, grâce au second principe de la thermodynamique, le sens de l'évolution spontanée des systèmes colloïdaux, systèmes à partir desquels les monocouches de protéines peuvent apparaître.

Les relations obtenues (\textit{cf.} les différentes propositions) mettent en lumière le rôle des hydrophobies de la surface solide sur laquelle doit venir adhérer l'IgG et de celle de l'IgG elle-même. Lorsque la somme de ces hydrophobies est positive, le système évolue naturellement dans le sens de l'adhésion de l'IgG sur la surface solide. Par conséquence, l'IgG étant une protéine hydrophile, l'usage d'une surface solide dont l'hydrophobie est supérieure en valeur absolue à cette hydrophilie est de rigueur afin de provoquer l'adhésion. Le polystyrène étant une surface connue pour sa haute hydrophobie, son usage se justifie pleinement dans le cadre de l'ELISA.

La simple adhésion a ensuite été répartie sur deux phénomènes distincts: l'addition et la relaxation. L'addition est le fait que des IgG viennent s'accumuler sur la surface solide en y présentant une portion de leur interface, sans plus. La relaxation est, quant à elle, le fait que les IgG déjà accumulées sur la surface peuvent y améliorer leurs interactions en changeant d'orientation et/ou de conformation. Ces deux phénomènes, résultant tous deux en une augmentation de l'aire de contact $\Sigma_\mathrm{sp}$, se conceptualisent par le caractère positif des différentielles $\mathrm{d}\Theta$ (addition) et $\mathrm{d}\sigma$ (relaxation).

Au-delà des aspects thermodynamiques, les aspects cinétiques relevant du modèle DLVO explicitent la raison du contact intime entre la surface en polystyrène et l'IgG. L'IgG arrivant à la surface par diffusion (voir introduction) peut en être repoussée par le potentiel électrostatique (modèle de la double couche diffuse) mais sera toujours fortement attirée par le potentiel de Hamaker (forces de London-van der Waals). Le polystyrène est une surface faiblement chargée lorsqu'il se trouve baigné par un milieu aqueux et il est dès lors permis de penser que le potentiel de Hamaker jouera son rôle d'attirance sans trop d'entrave. Cette force d'attraction viendra alors comprimer les enveloppes d'espèces adsorbées aux interfaces, ce qui aura pour effet d'augmenter les affinités de désorption et, par voie de conséquence, les hydrophobies de celles-ci. Le potentiel DLVO devrait donc être un facteur accentuant les effets strictement thermodynamiques.

Les critères thermodynamiques donnent donc une direction à la transformation du système. Cette direction semble aller vers l'adhésion spontanée des IgG sur le polystyrène. L'adhésion signifie non seulement qu'il y aura accumulation des IgG sur le polystyrène mais aussi relaxation orientationnelle et/ou conformationnelle de celles-ci. Au sens de la thermodynamiques, toutes les IgG devraient finir par se trouver dans un état \textit{flat$+$}, c'est-à-dire couchées et dénaturées. Cet état le plus stable n'est toutefois pas celui qui sera atteint lorsque la monocouche sera saturée. En effet, il est facilement imaginable que les IgG de la monocouche interagiront entre elles, ce qui devrait avoir pour effet d'en empêcher certaines d'adopter cette configuration couchée-dénaturée. Par ailleurs, l'ELISA nécessite la présence dans la monocouche d'IgG présentant une orientation de type \textit{end-on} afin qu'elles puissent interagir convenablement avec leurs antigènes. 
La thermodynamique semble dès lors être bien incapable de rendre compte de la présence d'IgG \textit{end-on} dans la monocouche ni même des interactions latérales pouvant exister entre celles-ci. Ces aspects cinétiques sont développés dans la chapitre suivant.

\begin{footnotesize}

\end{footnotesize}\end{cbunit}
\begin{cbunit}
\chapter[Additions séquentielles aléatoires des protéines]{Additions séquentielles aléatoires des protéines}\label{SectionRSA}
\markboth{Chapitre \ref{SectionRSA}: RSA}{}
\minitoc

\section{Généralités}
Si la tendance est à l'accumulation sur la surface en favorisant les orientations présentant une empreinte caractéristique élevée, l'orientation \textit{flat} devrait logiquement être nettement favorisée. Cependant, malgré cette attraction qui a été déduite de la thermodynamique, on peut supposer que l'orientation \textit{flat} n'est pas de nature à faciliter les interactions des immunoglobulines incluses dans la monocouche avec les antigènes contre lesquels elles pourraient être dirigées dans un protocole ELISA, les paratopes étant trop proches de la surface et même probablement dénaturés (ceux-ci possédant aussi une structure en sandwich \textgreek{b} \citep{janeway2001} susceptible de se déplier). De manière paradoxale par rapport aux critères thermodynamiques, les monocouches d'IgG semblent présenter une orientation moyenne favorable à la reconnaissance de l'antigène, c'est-à-dire une orientation de type \textit{end-on} où l'anticorps est littéralement <<~debout~>> sur la surface. Il existe différents témoignages, tous plus ou moins probants, d'une orientation \textit{end-on} de ces immunoglobulines adhérant sur une surface. La principale preuve en est que les dosages immunologiques, réalisés sur des plaques en polystyrène très hydrophobes, sont possibles et même, pour la plupart, extrêmement performants (J.~E. Butler \textit{et al.} \citep{butler2000}, T. Porstmann et S.~T. Kiessing \citep{porstmann1992}). D'autres indices de cette orientation favorable ont été fournis par J. Buijs \textit{et al.} \citep{buijs1995,buijs1997un} de même que par M.~E. Soderquist et A.~G. Walton \citep{soderquist1980}.

Dans ce qui précède, l'intérêt s'est porté vers l'adhésion sur la surface d'une seule IgG. C'est un raisonnement que l'on pourrait extrapoler à tout le processus et donc à la construction de l'entièreté de la monocouche si aucune interaction autre que celles décrites n'advenait. Or, au cours de la construction de cette monocouche, la densité locale va augmenter de telle sorte que les IgG s'accumulant inexorablement vont finir par interagir entre elles et il va donc falloir, à un moment donné au cours du processus, tenir compte des interactions latérales IgG-IgG. Ces interactions sont formellement de la même nature que les interactions IgG-surface: DLVO, forces thermodynamiques et volume exclu. Le potentiel de Hamaker et les forces thermodynamiques seront toujours attractifs mais l'accumulation d'IgG sur le polystyrène aura deux effets: la surface devient aussi densément chargée que les IgG elles-mêmes \citep{bremer2004} et l'accessibilité de la surface diminue par effet de volume exclu. Au début, la surface était nettement attractive alors que cette tendance s'inverse au cours du processus: c'est le phénomène de blocage qui débouche sur la saturation et l'arrêt de la croissance de la monocouche. Il est intéressant de remarquer que l'adoption d'un tel point de vue s'oppose à une hypothèse couramment utilisée dans la littérature qui considère que l'interface est un puits sans fond, accumulant, ou plutôt absorbant, à l'infini toutes les molécules qui arriveraient à sa hauteur; on parle de \textit{perfect-sink hypothesis} (\textit{cf.} T. van de Ven \citep{vandeven1989}).

Le phénomène de blocage aura une influence de plus en plus intense au cours du remplissage et affectera d'autant plus la construction de la monocouche. Alors qu'au début du processus, les interactions IgG-surface dominaient le processus, le blocage mènera à un contrôle de plus en plus fort par les interactions IgG-IgG qui auront, quant à elles, tendance à favoriser les orientations d'empreintes caractéristiques faibles comme la \textit{end-on}. Il le fera pour la raison que les espaces laissés libres à la surface seront de plus en plus petits, ne laissant plus guère la priorité qu'aux IgG ayant une orientation d'empreinte caractéristique faible telle que la \textit{end-on}. D'autre part, étant donnée la production d'entropie occasionnée par l'adhésion sur la surface, le blocage est un phénomène irréversible ce qui semble exclure le fait que des IgG déjà présentes dans la monocouche puissent se décoller de la surface et ainsi laisser de la place à des IgG ayant des orientations présentant une forte empreinte caractéristique.

Le blocage irréversible va, au cours du processus d'accumulation, venir s'opposer à de nouvelles additions d'IgG pour finir par totalement les exclure. On peut donc considérer le phénomène de blocage comme la croissance d'une barrière d'énergie potentielle. Cette énergie potentielle est celle se développant entre le corps colloïdal et la surface lorsque ce dernier s'en approche ou s'en éloigne. On la qualifiera d'énergie potentielle d'addition et l'écrira $U_A$. Lorsque ce potentiel $U_A$ développera une barrière énergétique infiniment haute, il ralentira l'accumulation et finira par rendre le temps entre l'addition de deux IgG infiniment grand. Sous l'approximation du champ moyen (a.c.m.), il est possible de faire le lien entre un travail $\Psi$ lié à la forme de la fonction d'énergie potentielle $U_A$ à laquelle est soumise l'IgG lors de son addition à la monocouche et la probabilité $P(A)$ de cette addition. En définissant l'événement <<~$A\equiv$ \emph{addition à la monocouche}~>>, ce lien est obtenu en rapprochant les équations (2) de B. Widom \citep{widom1963} et (24) du même auteur \citep{widom1966}. Il s'écrit selon la relation \ref{EqEActiva}
\begin{equation}\label{EqEActivaBis}
P(A)=\bigg\langle\exp\bigg[-\frac{\Psi}{k_BT}\bigg]
\bigg\rangle\quad\overset{\mathrm{a.c.m.}}{\Longleftrightarrow}\quad\langle \Psi\rangle=-k_BT\ln P(A)
\end{equation}
dans laquelle $\Psi$ est le travail nécessaire au seul franchissement de la barrière d'énergie potentielle. Celui-ci est nul lorsque cette barrière est inexistante et deviendra infiniment grand lorsque cette barrière d'énergie potentielle tendra vers l'infini. Les crochets $\langle\cdot\rangle$ signifient qu'il s'agit d'une moyenne sur toutes les positions et orientations possibles.

Ce travail $\Psi$ nécessaire au franchissement d'une barrière de potentiel peut être interprété comme une énergie d'activation. L'énergie d'activation pour une addition, notée $E^\ddagger_A$, pourra venir remplacer le travail dans la relation \ref{EqEActivaBis}; en simplifiant la notation sous l'approximation du champ moyen, il vient que
\begin{equation}\label{EqEActiva}
P(A)=\exp\bigg[-\frac{E^\ddagger_A}{k_BT}\bigg]
\quad\Longleftrightarrow\quad E^\ddagger_A=-k_BT\ln P(A)
\end{equation}
donnant un lien direct entre la probabilité d'addition et l'énergie nécessaire à cette addition. Ce cadre conceptuel permet d'interpréter facilement le phénomène de blocage de la surface.

Aux premiers instants du processus d'accumulation sur la surface, l'énergie d'activation $E^\ddagger_A$ que devra fournir une IgG afin de s'additionner à la monocouche en croissance dépendra d'une combinaison des forces s'exerçant entre la surface et la particule explicitées aux équations \ref{EqHamaker} et \ref{EqDoubleC}. Au cours du temps, les forces entre les particules déjà accumulées dans la monocouche et les nouvelles arrivantes tendront à devenir dominantes jusqu'à devenir suffisamment grandes ($E^\ddagger_A\rightarrow+\infty$) et finir par rendre très peu probable l'addition d'une nouvelle particule ($P(A)\rightarrow0$). Ces considérations amènent à écrire la fonction d'énergie potentielle d'addition comme une somme sur plusieurs contributions:
\begin{equation}\label{EqBlocage}
U_A=U_\mathrm{vdw}+U_\mathrm{\acute{e}lec}+U_\mathrm{hs}.
\end{equation}
En plus des énergies potentielles issues du modèle DLVO ($U_\mathrm{vdw}$ et $U_\mathrm{\acute{e}lec}$, \textit{cf.} équations \ref{EqHamaker} et \ref{EqDoubleC}), l'énergie due à la répulsion stérique entre les particules accumulées dans la monocouche et les nouvelles arrivantes est comptabilisée dans le terme de potentiel $U_\mathrm{hs}$. Il s'agit d'un potentiel de <<~corps durs~>> tenant compte de l'impénétrabilité d'une particule vis-à-vis d'une autre qui s'en approcherait\footnote{Le <<~potentiel de sphère dure~>> est une fonction discontinue n'admettant que deux valeurs: s'exprimant en fonction de la distance entre les deux objets considérés, elle est nulle jusqu'au moment où cette distance, positive, devient, elle-même nulle, à partir de quoi elle devient infinie \citep{bretonnet2010}. Elle s'applique donc à deux particules s'approchant l'une de l'autre.}.

\`{A} l'instant initial, le potentiel de corps dur est nul dans le volume qu'occupera la monocouche, ce qui n'est pas la cas pour les potentiels $U_\mathrm{vdw}$ et $U_\mathrm{\acute{e}lec}$. Par conséquent, l'énergie d'activation $E^\ddagger_A$ à fournir par une particule afin d'adhérer à la surface à l'instant initial peut ne pas être nulle. $E^\ddagger_A$ sera nulle à l'instant initial si et seulement si les potentiels $U_\mathrm{vdw}$ et $U_\mathrm{\acute{e}lec}$ sont négligés dans l'expression de $U_A$.

Dans la suite de ce travail, on fera dépendre l'énergie d'activation $E^\ddagger_A$ directement de $U_\mathrm{hs}$, le potentiel de corps dur. Ceci équivaut à négliger les potentiels $U_\mathrm{vdw}$ et $U_\mathrm{\acute{e}lec}$ et à poser que $E^\ddagger_A$ est nulle à l'instant initial de la construction du film d'IgG. L'évolution du potentiel $U_\mathrm{hs}$ sera étudiée dans la section \ref{paragraphRSA} et suivantes grâce au modèle de l'addition séquentielle aléatoire, ce dernier permettant l'estimation de la probabilité d'addition $P(A_\mathrm{hs})$ uniquement due à l'interaction stérique et de ces conséquences. Afin d'alléger la notation, on notera et assimilera $P(A)$ à $P(A_\mathrm{hs})$ dans toute la suite de ce texte, de même que $U_A$ à $U_\mathrm{hs}$.

La notion de probabilité sera intensivement utilisée afin d'interpréter les résultats des simulations par RSA. Des probabilités de combinaisons d'événements ainsi que des probabilités conditionnelles seront introduites et on pourra se référer à l'ouvrage de P. Bogaert \citep{bogaert2005} pour de plus amples informations.

\section[Le modèle RSA]{Le modèle des additions séquentielles aléatoires}\label{paragraphRSA}

Afin de montrer les effets des interactions de volumes exclus entre les IgG s'additionnant, les considérations qui suivront feront un abondant usage du modèle de l'addition (ou <<~adsorption~>>) séquentielle aléatoire (\textit{Random} \textit{Sequential} \textit{Addition}, RSA). Le modèle RSA, intensivement étudié à la fin des années 1980, a fait l'objet de différentes synthèses par P. Viot \textit{et al.} \citep{viot1992b}, S.~M. Ricci \textit{et al.} \citep{ricci1992}, J. Talbot \textit{et al.} \citep{talbot2000} de même que J.~W. Evans \citep{evans1993}. Son développement est essentiellement dû aux travaux de P. Schaaf et J. Talbot \citep{schaafettalbot1989a,schaafettalbot1989b,talbot1991} utilisant les outils de la mécanique statistique des fluides.
Le modèle RSA s'articule autour de trois hypothèses permettant le remplissage d'un volume par une série d'objets impénétrables les uns vis-à-vis des autres \citep{tarjus1991un}:
\begin{enumerate}
\item Les objets sont insérés séquentiellement (un par un) en une position choisie aléatoirement dans le volume considéré;
\item Une fois inséré, l'objet est gelé sur sa position, il ne peut ni se déplacer dans le volume ni s'en échapper;
\item Deux objets ne peuvent se superposer.
\end{enumerate}

Une application très intéressante du modèle RSA est le cas de l'addition irréversible des sphères dures sur une surface, elle aussi, dure. Le terme \emph{dur}, par opposition à \emph{mou}, désigne ici le caractère impénétrable et indéformable d'un objet. Pour ces sphères dures, deux approches complémentaires peuvent être développées: d'une part, une approche théorique permettant d'obtenir une formulation analytique de la capacité de l'interface à accepter une nouvelle insertion en fonction de son degré de remplissage (probabilité d'addition $P(A)$) et, d'autre part, le développement de programmes permettant de modéliser l'accumulation sur la surface.

L'équation principale du modèle RSA donne une relation entre la probabilité d'insertion $P(A)$ d'une nouvelle sphère en fonction du nombre de sphères accumulées dans le volume de l'interface. Cette équation a été développée par H. Reiss \textit{et al.} \citep{reiss1959} sous la forme de l'équation <<~d'inclusion-exclusion~>> \ref{EqRSA1}.

\begin{equation}\label{EqRSA1}
P(A)=1+\sum_{n=1}^\infty (-1)^n F_n = 1-F_1+F_2-F_3+\dots
\end{equation}

Les $F_n$ sont des quantités définies dans le \textgreek{g}-espace des phases
\footnote{Le \textgreek{g}-espace des phases est une représentation des systèmes qu'utilise la mécanique statistique. Une particule peut y disposer de six coordonnées (position et vitesse): $r_x$, $r_y$, $r_z$, $p_x$, $p_y$ et $p_z$. En suivant cette représentation, un système de $N$ sphères sera figuré par un point dans un espace à $6N$ dimensions. Ce point caractérisera la configuration du système à l'instant $t$ et est susceptible de mouvements si changement de configuration il y a au cours du temps. La vision du monde portée par la mécanique statistique est celle de l'impossibilité de connaître exactement, à un moment donné, la configuration dans laquelle se trouve le système. On ne peut donc le représenter par un seul point mais par une densité de probabilité dans le \textgreek{g}-espace. Pour un état donné du système, il y aura donc une distribution de configurations possibles que l'on pourra cerner \textit{via} les notions de moyenne, variance, etc. Ce paradigme, utilisé avec succès par la mécanique quantique (\textit{cf}. le principe d'incertitude de Heisenberg), est très instructif dans le cadre de simulations utilisant le modèle RSA car, pour chaque état, on génère une série de configurations afin d'en estimer les propriétés moyennes $P(A)$ (observables macroscopiques). \`{A} ce titre, le formalisme mathématique des espaces de Hilbert a été introduit dans le cadre du modèle RSA par R. Dickman \citep{dickman1989}.} par la relation \ref{EqRSA2} fournissant <<~l'expansion diagrammatique~>> \citep{reiss1959}.

\begin{equation}\label{EqRSA2}
F_n=\frac{1}{n!}\idotsint \rho^{(n)}_N(\boldsymbol{r}_1,\dots,\boldsymbol{r}_n)\frac{\Omega_n(\boldsymbol{r}_1,\dots,\boldsymbol{r}_n)}{\Omega}\mathrm{d}\boldsymbol{r}_1\dots\mathrm{d}\boldsymbol{r}_n
\end{equation}

La fonction de distribution générique $\rho^{(n)}_N$ est la probabilité de trouver, dans le système à $N$ particules, un groupe de $n$ particules identiques contenues dans l'hypervolume $\mathrm{d}\boldsymbol{r}_1\dots\mathrm{d}\boldsymbol{r}_n$ du \textgreek{g}-espace des phases \citep{bretonnet2010}. $\Omega_n(\boldsymbol{r}_1,\dots,\boldsymbol{r}_n)$ est le volume commun des sphères d'exclusion entourant les $n$ sphères présentes aux positions $\boldsymbol{r}_1$ à $\boldsymbol{r}_n$. $\Omega$ étant le volume total, la quantité $\Omega_n(\boldsymbol{r}_1,\dots,\boldsymbol{r}_n)/\Omega$ représente la fraction de ce volume dans laquelle se produit un recouvrement des volumes d'exclusion appartenant au même groupe de $n$ sphères.

\begin{figure}[t]\centering
\begin{tabular}{ccc}
\includegraphics*[width=0.3\textwidth]{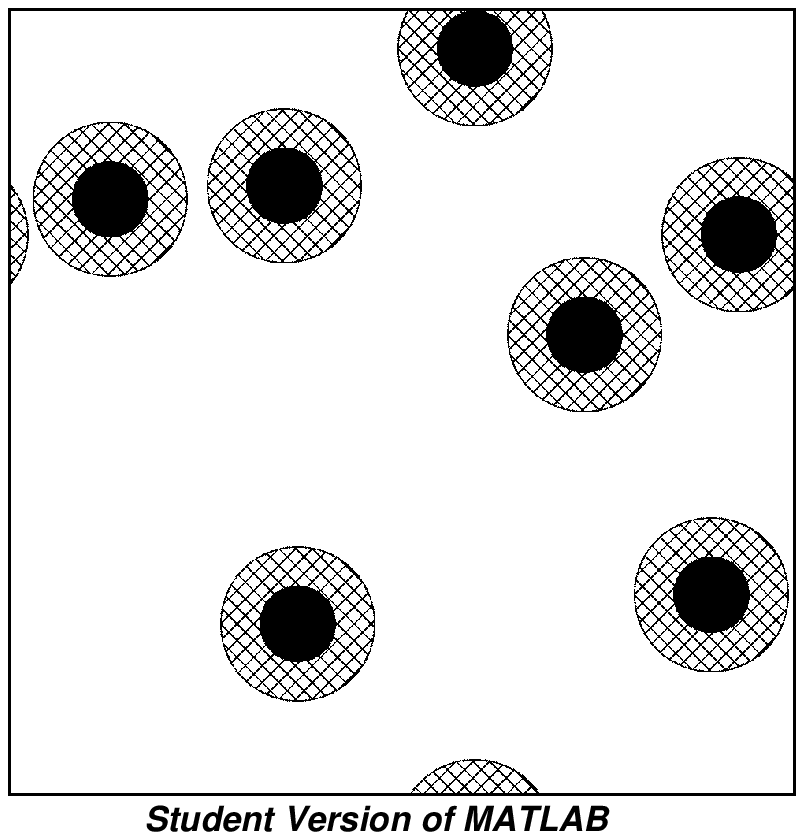}&
\includegraphics*[width=0.3\textwidth]{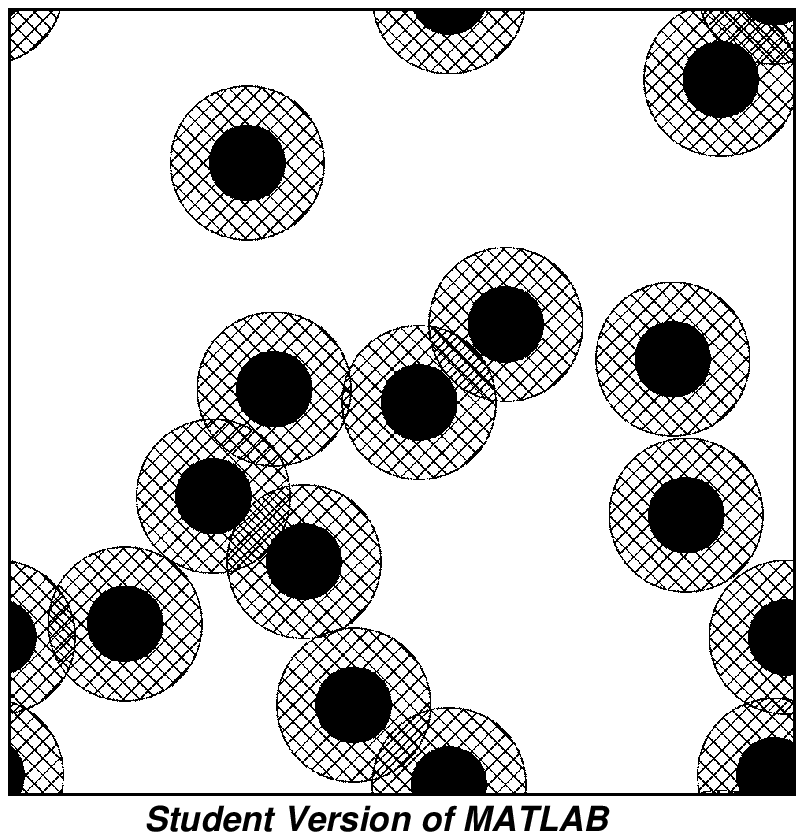}&
\includegraphics*[width=0.3\textwidth]{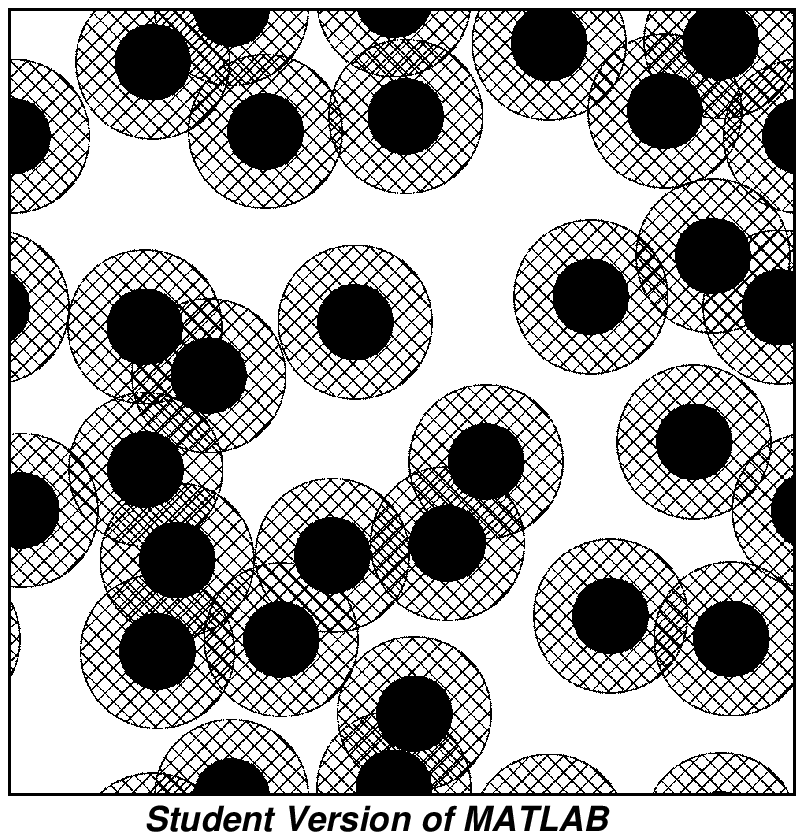}\\
A: $\phi\simeq5$ \%&B: $\phi\simeq10$ \%&C: $\phi\simeq20$ \%
\end{tabular}
\caption[Volume exclu dû au remplissage RSA d'une surface par des sphères]{\`{A} densité $\phi$ croissante, illustration du remplissage d'une surface par des sphères dures en utilisant le modèle des additions séquentielles aléatoires. Chaque sphère est accompagnée d'une zone hachurée représentant son volume exclu (d'après \citep{talbot2000}).}\label{FigVolExclu}
\end{figure}

En recombinant les équations \ref{EqRSA1} et \ref{EqRSA2}, la probabilité d'addition peut se réécrire sous la forme de la série \ref{EqRSA3} dont les paramètres $a_1$, $a_2$ et $a_3$ ont été calculés analytiquement \citep{schaafettalbot1989a,schaafettalbot1989b} et où $\phi$ figure le taux de recouvrement de la surface qui est équivalent, pour les sphères, à la proportion de la surface sur laquelle on projette l'empreinte des particules de la monocouche (aires totalement noircies sur la figure \ref{FigVolExclu}). Cette quantité $\phi$ a été définie à l'équation \ref{DéfinitionTauxDOccupation}.

\begin{equation}\label{EqRSA3}
P(A)=1-a_1\phi+a_2\phi^2-a_3\phi^3+\dots
\end{equation}

L'équation \ref{EqRSA3} est la clé de voûte du modèle RSA et ne doit pas être confondue avec l'équation du viriel pour les gaz réels (\textit{cf.} à ce sujet J.~L. Bretonnet \citep{bretonnet2010}) dont la nature est très différente car elle représente l'état d'équilibre thermodynamique d'une phase condensée, ce qui n'est absolument pas le cas ici. Un article de B. Widom \citep{widom1966} fournit la démonstration que le modèle RSA génère des configurations typiques d'un système dans un état thermodynamique hors de l'équilibre qui ne se manifestera qu'à partir d'une certaine quantité additionnée à travers le terme $a_3\phi^3$ (\textit{cf}. la seconde hypothèse du modèle RSA), c'est-à-dire se manifestant à partir d'un certain remplissage de la surface. Néanmoins, elles portent toutes deux une notion d'occupation de l'espace en fonction de la densité comme illustré à la figure \ref{FigVolExclu}. La probabilité d'insertion\footnote{La probabilité d'addition, d'insertion ou de toucher la surface $P(A)$ est aussi appelée fonction d'aire disponible et est notée $\Phi(\rho)$ ou ASF par certains auteurs \citep{viot1992b,ricci1992}.} $P(A)$ est représentée à la figure \ref{FigVolExclu} par la proportion d'aire laissée vierge alors que la probabilité de rejet\footnote{La probabilité de rejet $P(A^*)$ est l'aire exclue, aussi notée $1-\Phi(\phi)$ dans la littérature.} ou d'exclusion
\begin{equation}\label{EqAddRejet}
P(A^*)=1-P(A)
\end{equation}
est représentée par l'union des zones hachurées (proportion de l'aire exclue à toute nouvelle addition). On peut constater qu'à faible densité ($\phi\leqslant5$ \%), la proportion d'aire exclue peut être simplement calculée par la somme des aires exclues de chaque sphère. Cette somme est comptabilisée dans le terme $1-a_1\phi$ de l'équation \ref{EqRSA3}. Lorsque la densité augmente, apparaissent des recouvrements entre les aires d'exclusion et il faut prendre en compte une correction afin de ne pas compter deux fois cette aire, c'est le sens du terme $+a_2\phi^2$. Le raisonnement complet est donné par P. Schaaf et J. Talbot \citep{schaafettalbot1989a}.

\begin{table}[h]\centering
\caption[Estimations de l'équation d'évolution du volume exclu]{Estimations théoriques et numériques des paramètres $a_n$ de l'équation (\ref{EqRSA3}) pour un modèle d'addition de sphères dures (intervalle de confiance à $99,9$~\%). Les estimations numériques obtenues grâce aux codes Matlab présentés à l'annexe \ref{AnnRSA} de ce travail sont données dans la dernière colonne.}\label{TableauRSA}
\begin{spacing}{1.3}
\begin{small}
\begin{tabular}{cr@{}lr@{}ll@{ $\pm$ }l}
\hline
&\multicolumn{2}{c}{Théoriques \citep{schaafettalbot1989b}}&\multicolumn{2}{c}{Numériques \citep{schaafettalbot1989b}}  &\multicolumn{2}{c}{Numériques}\\
\hline
$a_0$& $1$&        & $1$&$,002$  & \multicolumn{2}{l}{Fixée à $1$.}\\
$a_1$& $4$&        & $4$&$,053$  & $4,013$&$0,009$\\
$a_2$& $3$&$,308$  & $3$&$,601$  & $3,34$&$0,07$\\
$a_3$& $1$&$,407$  & $1$&$,135$  & $1,6$&$0,1$\\
\hline
\end{tabular}
\end{small}
\end{spacing}
\end{table}

En parallèle, les hypothèses assez simples du modèle RSA permettent l'évaluation numérique de ces mêmes paramètres $a_n$. En générant une série de configurations possibles pour un système contenant $\mathcal{N}$ sphères accumulées sur une surface selon les hypothèses RSA, il est ensuite possible d'évaluer la probabilité d'insertion $P(A)$ ou de rejet $P(A^*)$. Cette évaluation se fait simplement en procédant à un grand nombre de tentatives d'additions et en divisant le nombre de réussites par le nombre de tentatives réalisées. Cette procédure est connue sous le nom d'insertion de Widom \citep{widom1963}. Les estimations des $a_n$ obtenues par régression linéaire\footnote{Estimer les paramètres d'une équation du viriel n'est pas une procédure directe et une discussion à ce sujet est donnée dans la littérature \citep{schaafettalbot1989a}. Sans entrer dans les détails, on remarquera que les valeurs obtenues au tableau \ref{TableauRSA} sont très proches des valeurs théoriques et même plus proches que celles estimées par les promoteurs du modèle RSA \citep{schaafettalbot1989a}. On peut donc raisonnablement apporter du crédit au programme de simulation par addition séquentielle aléatoire développé dans le cadre de ce travail.} \citep{hastie2009a} sont données au tableau \ref{TableauRSA}. Un taux de recouvrement $\phi$ d'environ $54,47$~\% est obtenu à saturation (un taux d'environ 54,7~\% est généralement accepté pour le RSA \citep{schaafettalbot1989a}), c'est-à-dire lorsque $P(A)$ devient suffisamment proche de zéro\footnote{Logiquement, la saturation est rencontrée lorsque $P(A^\ast)$ est très proche de $1$. De façon rigoureuse, le taux de recouvrement à saturation $\phi_\infty$ est une limite de la fonction du taux de recouvrement $\phi\big(P(A)\big)$ telle que
\begin{equation}
\phi_\infty=\lim_{P(A)\rightarrow0}\phi\big(P(A)\big)\quad\mathrm{ou}\quad
\phi_\infty=\lim_{P(A^\ast)\rightarrow1}\phi\big(P(A^\ast)\big)
\end{equation}
Dans le cas présent et pour les besoins du calcul numérique ici réalisé, la saturation a été posée pour $P(A^\ast)\simeq0,9992$, soit $P(A)\simeq8\times10^{-4}$.}, taux que l'on peut comparer avec les $90,7$~\% pour l'arrangement le plus dense que l'on puisse faire avec des sphères (configuration hexagonale des alvéoles d'abeilles) \citep{torquato2010}. L'évolution typique des probabilités d'addition $P(A)$ en fonction du taux de recouvrement $\phi$ est montrée à la figure \ref{FigSphèresRSA} pour des sphères ayant un rayon d'environ $2,386$ nm. Les résultats obtenus dans ce travail sont présentés dans la quatrième colonne du tableau \ref{TableauRSA} grâce aux algorithmes présentés à l'annexe \ref{AnnRSA} montrent une excellente cohérence avec les données de la littérature \citep{schaafettalbot1989b}, permettant ainsi de valider la procédure de modélisation. De même, la figure \ref{FigurePBC} illustre une surface saturée en sphères obtenue grâce à cette simulation\footnote{Cette figure \ref{FigurePBC} illustre les conditions périodiques aux limites qui permettent de s'affranchir des effets de bords pouvant biaiser les résultats de ce genre de simulation. Bien que jamais comptabilisés, les bords résultants de l'utilisation de ces conditions seront systématiquement présentés sur les figures illustrant des surfaces de ce genre.}. On notera que le taux de recouvrement de 54,47~\% correspondant à la saturation de la surface (on ne comptabilise pas les bords de la figure \ref{FigurePBC} correspondant aux conditions aux limites périodiques) est obtenu pour des temps de simulations assez longs. Dans ce qui suit, un tel taux de saturation ne sera pas recherché car les effets qui seront mis en évidence se situent essentiellement entre des taux de 5 à 50~\%, ce dernier taux correspondant à une probabilité de rejet $P(A)\simeq8\times10^{-4}$, probabilité qui, une fois atteinte, impliquera l'arrêt de la simulation.

\begin{figure}[hp]\centering
\includegraphics*[width=0.75\textwidth]{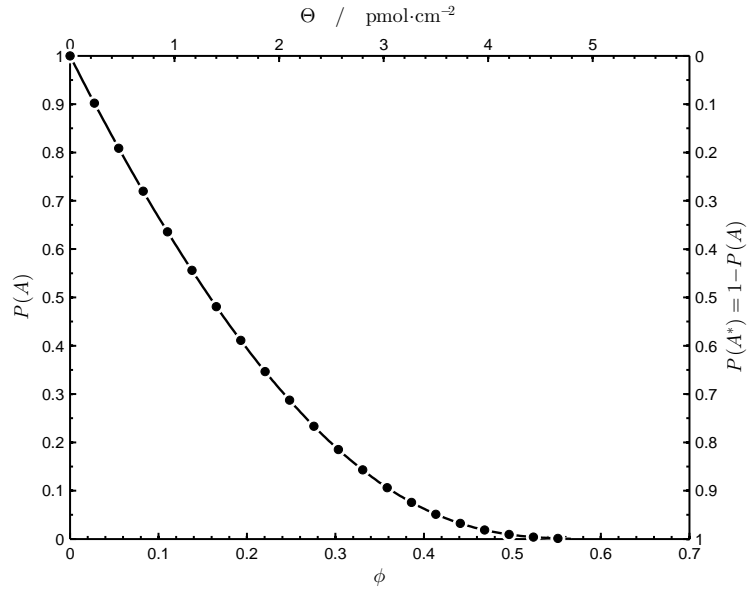}
\caption[$P(A)$ en fonction de $\Theta$ lors de remplissages RSA de surfaces (sphères)]{\'{E}volution de la probabilité $P(A)$ d'additionner une nouvelle sphère en fonction du taux de recouvrement de la surface $\phi$ (valeur maximale atteinte: 54,47~\%). La relation entre le taux de recouvrement $\phi$ et la quantité totale accumulée sur la surface $\Theta$ (\textit{cf.} équation \ref{ProduitTauxDOccupation}) est montrée sur l'axe des abscisses pour des sphères dont le rayon est d'environ $2,386$ nm. L'axe des ordonnées montre la relation entre $P(A)$ et $P(A^*)$ la probabilité de rejet.}\label{FigSphèresRSA}
\end{figure}

\begin{figure}[hp]\centering
\includegraphics*[width=0.55\textwidth]{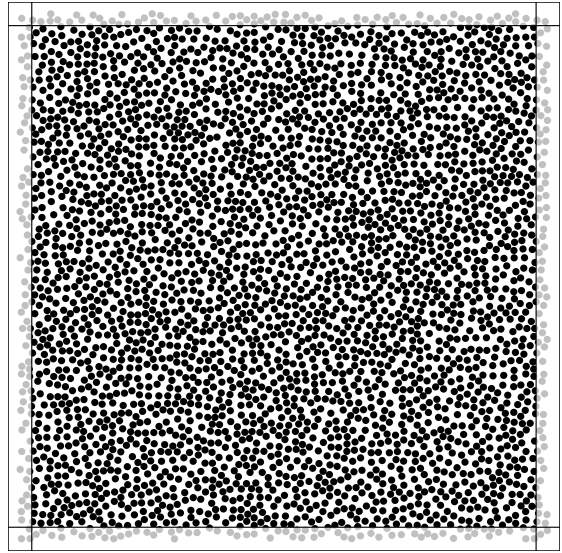}
\caption[Surface saturée en sphères]{Surface saturée en sphères (rayon d'environ 2,386 nm) obtenue par additions séquentielles aléatoires. Le carré interne (de 303 nm $\times$ 303 nm) est l'aire sur laquelle sont comptabilisées les propriétés de la surface obtenue tandis que les bords (largeur de 15,8 nm) résultent de l'utilisation des conditions périodiques aux limites.}\label{FigurePBC}
\end{figure}

\section[Généralisation du RSA]{Généralisation du modèle des additions\\séquentielles aléatoires}

La quantité $\phi$ utilisée jusqu'à présent porte le sens d'un taux de recouvrement de la surface. Ce taux $\phi$ a été défini au chapitre précédent comme l'aire de contact entre les disques et la surface (voir équation \ref{DéfinitionTauxDOccupation}) rapportée à l'aire totale de la surface. La relation \ref{ProduitTauxDOccupation} révèle aussi que ce taux résulte du produit de l'empreinte $\sigma$ et de la quantité de protéines accumulées $\Theta$. La figure \ref{FigSphèresRSA} montre ce lien de proportionnalité entre $\phi$ et $\Theta$ sur les deux axes d'abscisses, $\sigma$ étant l'aire en m$^2$ développée par une mole de sphères (rayon de $2,386$ nm), on peut l'estimer proche de $0,10771$ cm$^2\cdot$pmol$^{-1}$. En substituant l'équation \ref{ProduitTauxDOccupation} dans la formule de la probabilité d'addition \ref{EqRSA3}, il vient facilement que
\begin{equation}\label{EqRSA3bis}
P(A)=1-a_1(\sigma\,\Theta)+a_2(\sigma\,\Theta)^2-a_3(\sigma\,\Theta)^3+\dots
\end{equation}
montrant que l'évolution de la probabilité $P(A)$ est fonction de $\sigma$ l'empreinte des protéines sur la surface et de $\Theta$ la quantité qui y est accumulée. Grâce à la différentielle
\begin{equation}
\begin{split}
\mathrm{d}P(A)=-\big[&\,a_1-2\,a_2(\sigma\,\Theta)
+3\,a_3(\sigma\,\Theta)^2-\dots\big]\Theta\,\mathrm{d}\sigma\\
&-\big[a_1-2\,a_2(\sigma\,\Theta)+3\,a_3(\sigma\,\Theta)^2
-\dots\big]\sigma\,\mathrm{d}\Theta,
\end{split}
\end{equation}
on voit clairement que la variation de cette probabilité d'une nouvelle addition $\mathrm{d}P(A)$ est fonction à la fois de l'addition $\mathrm{d}\Theta$ et de la variation de l'empreinte $\mathrm{d}\sigma$ des protéines sur la surface.

Outre les sphères, les arguments développés montrent à quel point le modèle RSA est approprié à l'étude de l'accumulation des protéines, et des IgG en particulier, sur une surface plane. En effet, il tient compte de l'irréversibilité, de l'organisation en monocouche et du blocage dû aux interactions latérales entre les particules qui y sont disposées. Le c{\oe}ur de la simulation étant validé pour des sphères (voir tableau~\ref{TableauRSA} et figure~\ref{FigSphèresRSA}), il est possible d'envisager quelques modifications afin d'y incorporer les comportements attendus des protéines approchant une surface: la possibilité de réaliser une accumulation compétitive entre plusieurs objets dissemblables (\textit{i.e.} deux sphères de rayons différents, des IgG diversement orientées, etc.) et la possibilité pour une particule anisotropique de pouvoir relaxer son orientation après avoir touché la surface (dans le sens de l'augmentation de l'empreinte).

Le but dans lequel les simulations RSA seront réalisées dans ce chapitre est d'estimer, sous certaines hypothèses, les probabilités d'addition $P(A)$ ou de rejet $P(A^\ast)$ en fonction de la quantité accumulée dans la monocouche $\Theta$ et pour des séries de particules/protéines présentant diverses empreintes caractéristiques $\sigma(\boldsymbol{\omega})$. Ces estimations de $P(A)$ permettront de mieux appréhender la façon dont la surface est progressivement bloquée.

En toute généralité, un système de particules/protéines/IgG initialement en suspension sera susceptible de fournir des particules orientées dans toutes les directions, orientations ou, plus généralement, ayant diverses empreintes caractéristiques $\sigma(\boldsymbol{\omega}_1)$, $\sigma(\boldsymbol{\omega}_2)$, etc. L'orientation, la conformation ou la nature de la particule se présentant à la surface est donnée par les pseudovecteurs $\boldsymbol{\omega}_1$, $\boldsymbol{\omega}_2$ etc. Les probabilités que des particules se présentent à la surface en fonction d'une orientation/conformation/nature (on parlera simplement d'orientation par la suite) particulière seront notées
\begin{equation}
P(\boldsymbol{\omega}_1),\quad P(\boldsymbol{\omega}_2),\quad\text{etc.}
\end{equation}
Ensuite, en fonction du volume exclu de la surface, chacune de ces particules pourra éventuellement s'y additionner. Les probabilités que l'on a d'observer, à un moment donné, l'addition de particules de diverses orientations sont des probabilités combinant deux événements et se noteront
\begin{equation}
P(A\cap\boldsymbol{\omega}_1),\quad P(A\cap\boldsymbol{\omega}_2),\quad\text{etc.}
\end{equation}
Ces probabilités sont les probabilités qu'une particule dans l'orientation $\boldsymbol{\omega}_1$ ou $\boldsymbol{\omega}_2$ s'additionne à la monocouche. La probabilité qu'il y ait addition s'obtient en sommant toutes ces probabilités:
\begin{equation}
P(A)=P(A\cap\boldsymbol{\omega}_1)+P(A\cap\boldsymbol{\omega}_2)+\dots
\end{equation}

Grâce à la définition de la probabilité conditionnelle \citep{bogaert2005}, on transformera la somme précédente afin d'obtenir une combinaison linéaire entre deux séries de probabilités:
\begin{equation}\label{EqProbCond}
P(A)
=P(\boldsymbol{\omega}_1)P(A|\boldsymbol{\omega}_1)
+P(\boldsymbol{\omega}_2)P(A|\boldsymbol{\omega}_2)+\dots
\end{equation}
Dans cette combinaison linéaire, on retrouve les probabilités $P(\boldsymbol{\omega}_1)$, $P(\boldsymbol{\omega}_2)$, etc. d'obtenir telle ou telle orientation. Les autres probabilités sont des probabilités conditionnelles, c'est-à-dire que $P(A|\boldsymbol{\omega}_1)$ représente la probabilité d'observer une addition sachant que l'on tente d'additionner une particule dans l'orientation $\boldsymbol{\omega}_1$. Il s'agit de la probabilité de l'événement \emph{additionner} sachant que l'événement \emph{obtenir une particule dans l'orientation $\boldsymbol{\omega}_1$} est réalisé d'office. Les probabilités conditionnelles de rejet, obtenues simplement par les relations
\begin{equation}
P(A^\ast|\boldsymbol{\omega}_1)=1-P(A|\boldsymbol{\omega}_1),\quad
P(A^\ast|\boldsymbol{\omega}_2)=1-P(A|\boldsymbol{\omega}_2),\quad\text{etc.,}
\end{equation}
seront aussi largement utilisées car leur signification est tout aussi parlante.

L'équation \ref{EqProbCond} donnant le développement de la probabilité d'addition $P(A)$ en fonction des probabilités conditionnelles et des probabilités d'orientations sera intensivement utilisée par la suite. En effet, la procédure d'insertion de Widom décrite à la section \ref{WI} permet d'estimer directement les probabilités conditionnelles tandis que les probabilités d'obtenir quelque orientation particulière seront fixées arbitrairement par l'expérimentateur et constitueront dès lors les hypothèses des modèles étudiés.

Les deux systèmes pour lesquels des simulations RSA seront réalisées et analysées à l'aide des probabilités données ci-dessus sont illustrés sur la figure \ref{FigSchémaRSA1}. Le premier système est constitué de trois boîtes de conserve ayant différents diamètres tandis que le second est constitué d'IgG de trois orientations différentes. Ces particules ont des empreintes caractéristiques différentes et augmentant de la gauche vers la droite sur la figure. 

\begin{figure}[t]\centering
\includegraphics[width=0.85\textwidth]{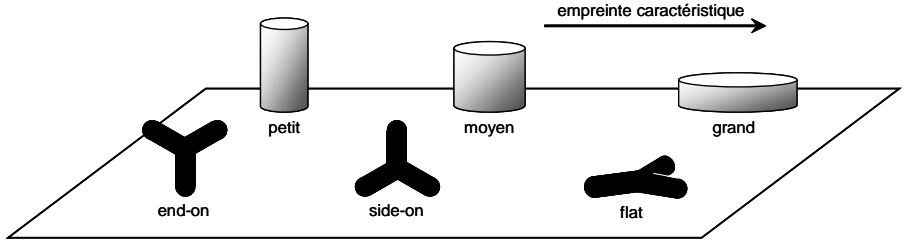}
\caption[Systèmes utilisés pour les simulations d'accumulations compétitives]{Les boîtes de conserve de rayons variables et les IgG orientées différement formant les deux systèmes dont l'accumulation compétitive dans une monocouche est étudiée aux sections \ref{SectionRSA2} et \ref{SectionRSA2bis}. L'empreinte caractéristique $\sigma(\boldsymbol{\omega})$ de chacun de ces objets augmente de gauche à droite.}\label{FigSchémaRSA1}
\end{figure}

\section[RSA des boîtes de conserve]{Additions séquentielles aléatoires\\des boîtes de conserve}\label{SectionRSA2}

Le système de boîtes de conserve illustré sur la figure \ref{FigSchémaRSA1} est composé de trois cylindres de différentes hauteurs ayant plusieurs diamètres: grand, petit et moyen. Chacune de ces boîtes sera dès lors respectivement désignée par $\boldsymbol{\omega}_{\mathrm{grand}}$, $\boldsymbol{\omega}_{\mathrm{moyen}}$ et $\boldsymbol{\omega}_{\mathrm{petit}}$. Leurs diamètres étant différents, leurs empreintes caractéristiques le seront tout autant, de telle sorte que l'on ait la suite
\begin{equation}
\sigma(\boldsymbol{\omega}_{\mathrm{petit}})<\dots<
\sigma(\boldsymbol{\omega}_{\mathrm{moyen}})<\dots<
\sigma(\boldsymbol{\omega}_{\mathrm{grand}})
\end{equation}
dans laquelle l'empreinte caractéristique la plus faible appartient fort logiquement à la boîte de petit diamètre.

D'autre part, on posera que ces trois boîtes ont toutes la même probabilité de se présenter à la surface (leurs <<~concentrations~>> sont identiques à l'approche de la surface). Ceci se traduira par les probabilités
\begin{equation}\label{dlfhglsdqghldglbglg}
P(\boldsymbol{\omega}_{\mathrm{petit}})=\frac{1}{3},\quad
P(\boldsymbol{\omega}_{\mathrm{moyen}})=\frac{1}{3},\quad\text{et}\quad
P(\boldsymbol{\omega}_{\mathrm{grand}})=\frac{1}{3}.
\end{equation}
Ces trois boîtes sont donc en compétition les unes par rapport aux autres afin de venir s'additionner à la monocouche en formation puisqu'elles ont chacune autant de chance de se présenter à la surface. Cette équiprobabilité de présentation à la surface donne, à partir de la relation \ref{EqProbCond}, la formule permettant d'obtenir la probabilité d'addition $P(A)$:
\begin{equation}
P(A)=\tfrac{1}{3}P(A|\boldsymbol{\omega}_{\mathrm{petit}})
+\tfrac{1}{3}P(A|\boldsymbol{\omega}_{\mathrm{moyen}})
+\tfrac{1}{3}P(A|\boldsymbol{\omega}_{\mathrm{grand}}).
\end{equation}

Grâce à ces paramètres, des simulations RSA sont réalisées en utilisant les algorithmes présentés aux sections \ref{AlgoRSA} et \ref{SurfConstruct1}. Les données brutes issues de ces simulations sont ensuite traitées grâces aux procédures explicitées aux sections \ref{WI} et \ref{ygfvyqgfo} de manière à obtenir le graphique de la figure \ref{FigRSA1}.A et l'image d'une surface à la figure \ref{FigRSAim1}.A.

Le graphique de la figure \ref{FigRSA1}.A présente deux types de courbes en fonction de $\Theta$, la quantité totale de protéines accumulées (pmol$\cdot$cm$^{-2}$): les probabilités $P(A^*|\boldsymbol{\omega}_i)$ qu'une boîte d'orientation $\boldsymbol{\omega}_i$ donnée soit rejetée de l'interface lorsque celle-ci s'y présente et les évolutions des $\Theta(\boldsymbol{\omega}_i)$, quantités accumulées de chaque type de boîte.

L'évolution des $\Theta(\boldsymbol{\omega}_i)$ (en noir) montre que les quantités de boîtes à empreintes caractéristiques élevées, c'est-à-dire $\sigma(\boldsymbol{\omega}_\mathrm{grand})$ et $\sigma(\boldsymbol{\omega}_\mathrm{moyen})$, qui sont accumulées finissent par plafonner alors que ce n'est absolument pas le cas pour les plus petites, d'empreinte caractéristique $\sigma(\boldsymbol{\omega}_\mathrm{petit})$, qui, elles, continuent de s'accumuler jusqu'à la fin du processus. L'ordre selon lequel les quantités $\Theta(\boldsymbol{\omega})$ plafonnent en fonction des empreintes caractéristiques montre que les boîtes aux larges empreintes caractéristiques ne peuvent donc s'additionner à la monocouche qu'en début de processus tandis que les plus petites continuent de s'y additionner jusqu'à la saturation. Cet ordre se vérifie malgré le fait que les boîtes continuent, tout au long du processus, de se présenter à la surface selon la même probabilité.

Les probabilités de rejet (en rouge) témoignent de ce phénomène d'exclusion progressive des boîtes en fonction de leur empreinte caractéristique en montrant que les grandes boîtes ($\boldsymbol{\omega}_\mathrm{grand}$) ne peuvent plus s'additionner à la surface (la probabilité $P(A^*|\boldsymbol{\omega}_\mathrm{grand})$ approche de l'unité) lorsque $\Theta\simeq1$ pmol$\cdot$cm$^{-2}$. Vient ensuite le tour des boîtes d'empreinte caractéristique $\sigma(\boldsymbol{\omega}_\mathrm{moyen})$ à avoir des difficultés à s'additionner (vers $\Theta\simeq2,5$~pmol$\cdot$cm$^{-2}$) et ce n'est qu'à l'approche de la saturation de la surface que les plus petites, d'empreinte caractéristique $\sigma(\boldsymbol{\omega}_\mathrm{petit})$, ne peuvent plus s'additionner, ces dernières étant la cause de la saturation finale. L'ordre de rejet s'observe à travers les valeurs prises par les probabilités conditionnelles de rejet suivant la série \ref{SerieExclusion}:
\begin{equation}\label{SerieExclusion}
P(A^*|\boldsymbol{\omega}_{\mathrm{petit}})
\leqslant\dots\leqslant
P(A^*|\boldsymbol{\omega}_{\mathrm{moyen}})
\leqslant\dots\leqslant
P(A^*|\boldsymbol{\omega}_{\mathrm{grand}}).
\end{equation}

La série \ref{SerieExclusion} aura une conséquence exprimée dans la série \ref{SerieNbre} sur les quantités finales accumulées sur la surface $\Theta_\infty$ en fonction des orientation $\boldsymbol{\omega}_i$:
\begin{equation}\label{SerieNbre}
\Theta_\infty(\boldsymbol{\omega}_\mathrm{grand})
\leqslant\dots\leqslant
\Theta_\infty(\boldsymbol{\omega}_\mathrm{moyen})
\leqslant\dots\leqslant
\Theta_\infty(\boldsymbol{\omega}_\mathrm{petit}).
\end{equation}

Ces considérations montrent qu'il existe un phénomène de rejet en fonction de la taille (exclusion de taille) lors de l'addition, phénomène qu'illustrent les probabilités de rejet de la série \ref{SerieExclusion}. Cette exclusion se manifestant par le rejet selon un ordre précis des boîtes en commençant par celles disposant de l'empreinte caractéristique la plus élevée, c'est-à-dire  $\sigma(\boldsymbol{\omega}_\mathrm{grand})$, et ensuite les intermédiaires pour finir par laisser le champ libre aux seules plus petites, et ce, jusqu'à la saturation de la monocouche. Ce phénomène dû à la compétition peut premièrement s'expliquer par le fait que les boîtes de petites empreintes caractéristiques participant, tout comme les autres, au blocage de la surface, empêchent les boîtes de grandes empreintes caractéristiques de pouvoir être accumulées dans la monocouche en aussi grande quantité que si elles étaient seules et deuxièmement par la capacité qu'ont les boîtes de petites empreintes caractéristiques de pouvoir s'insérer dans de plus petits espaces, améliorant l'efficacité de la saturation (une surface saturée uniquement avec des boîtes de grandes empreintes caractéristiques pourra encore accueillir des boîtes de petites empreintes caractéristiques).

Pour le système de boîtes de conserve, il semble donc que les additions séquentielles aléatoires mènent à des processus de croissance des monocouches dans lesquels les boîtes de plus petites empreintes caractéristiques soient nettement favorisées. Les quantités de petites boîtes accumulées sont majoritaires dans la monocouche saturée et, de manière générale, plus l'empreinte caractéristique d'une de ces boîtes de conserve est élevée, moins sa présence dans la monocouche saturée sera importante.

\section[RSA des IgG]{Additions séquentielles aléatoires des IgG}\label{SectionRSA2bis}

Le même genre de simulations RSA peut être réalisé pour le système d'IgG ayant diverses orientations présenté sur la figure \ref{FigSchémaRSA1}. Ces trois orientations, illustrées à la figure \ref{FigSchémaRSA1}, sont qualifiées de \textit{end-on} (l'IgG est <<~debout~>> sur un de ses bras), \textit{side-on} (l'IgG est <<~sur le côté~>>) et \textit{flat} (l'IgG est couchée, tous ses bras touchant la surface). Les modèles d'IgG utilisés dans les simulations sont des particules constituées de trois bras séparés les uns des autres par un angle de 120$^\circ$, chacun des bras ayant une longueur d'environ 8,33 nm et un diamètre proche de 2,92 nm. De la sorte,  l'aire de contact entre la surface et une IgG \textit{flat} sera d'environ 65,9 nm$^2$ et celle de l'IgG \textit{end-on} d'environ 6,7 nm$^2$. Chacune de ces orientations est caractérisée par une empreinte caractéristique et leur ordre de grandeur est donné par la série
\begin{equation}
\sigma(\boldsymbol{\omega}_{\mathrm{end\text{-}on}})<\dots<
\sigma(\boldsymbol{\omega}_{\mathrm{side\text{-}on}})<\dots<
\sigma(\boldsymbol{\omega}_{\mathrm{flat}}).
\end{equation}
Les trois orientations possibles sont, elles aussi, en compétition les unes contre les autres de façon à ce qu'elles aient les mêmes probabilités de se présenter à la surface
\begin{equation}
P(\boldsymbol{\omega}_{\mathrm{end\text{-}on}})=\frac{1}{3},\quad
P(\boldsymbol{\omega}_{\mathrm{side\text{-}on}})=\frac{1}{3}\quad\text{et}\quad
P(\boldsymbol{\omega}_{\mathrm{flat}})=\frac{1}{3}.
\end{equation}
De ce fait, par application de la formule \ref{EqProbCond}, la probabilité d'addition $P(A)$ se calculera selon
\begin{equation}
P(A)=\tfrac{1}{3}P(A|\boldsymbol{\omega}_{\mathrm{end\text{-}on}})
+\tfrac{1}{3}P(A|\boldsymbol{\omega}_{\mathrm{side\text{-}on}})
+\tfrac{1}{3}P(A|\boldsymbol{\omega}_{\mathrm{flat}}).
\end{equation}

Ces hypothèses permettent, à l'instar des boîtes de conserve, la réalisation de simulations RSA dont les résultats sont montrés à la figure \ref{FigRSA1}.B montrant la série de courbes caractérisant l'accumulation compétitive des trois orientations d'IgG décrites à la figure \ref{FigSchémaRSA1}. L'image d'une surface saturée par ces IgG est montrée à la figure \ref{FigRSAim1}.B. On peut observer sur la figure \ref{FigSchémaRSA1}.B le même ordre dans l'infléchissement des courbes que celui observé à la figure \ref{FigRSA1}.A pour les boîtes de conserve: $\boldsymbol{\omega}_{\mathrm{flat}}$, $\boldsymbol{\omega}_{\mathrm{side\text{-}on}}$ et enfin $\boldsymbol{\omega}_{\mathrm{end\text{-}on}}$.

\begin{figure}[hp]\centering
\includegraphics*[width=0.75\textwidth]{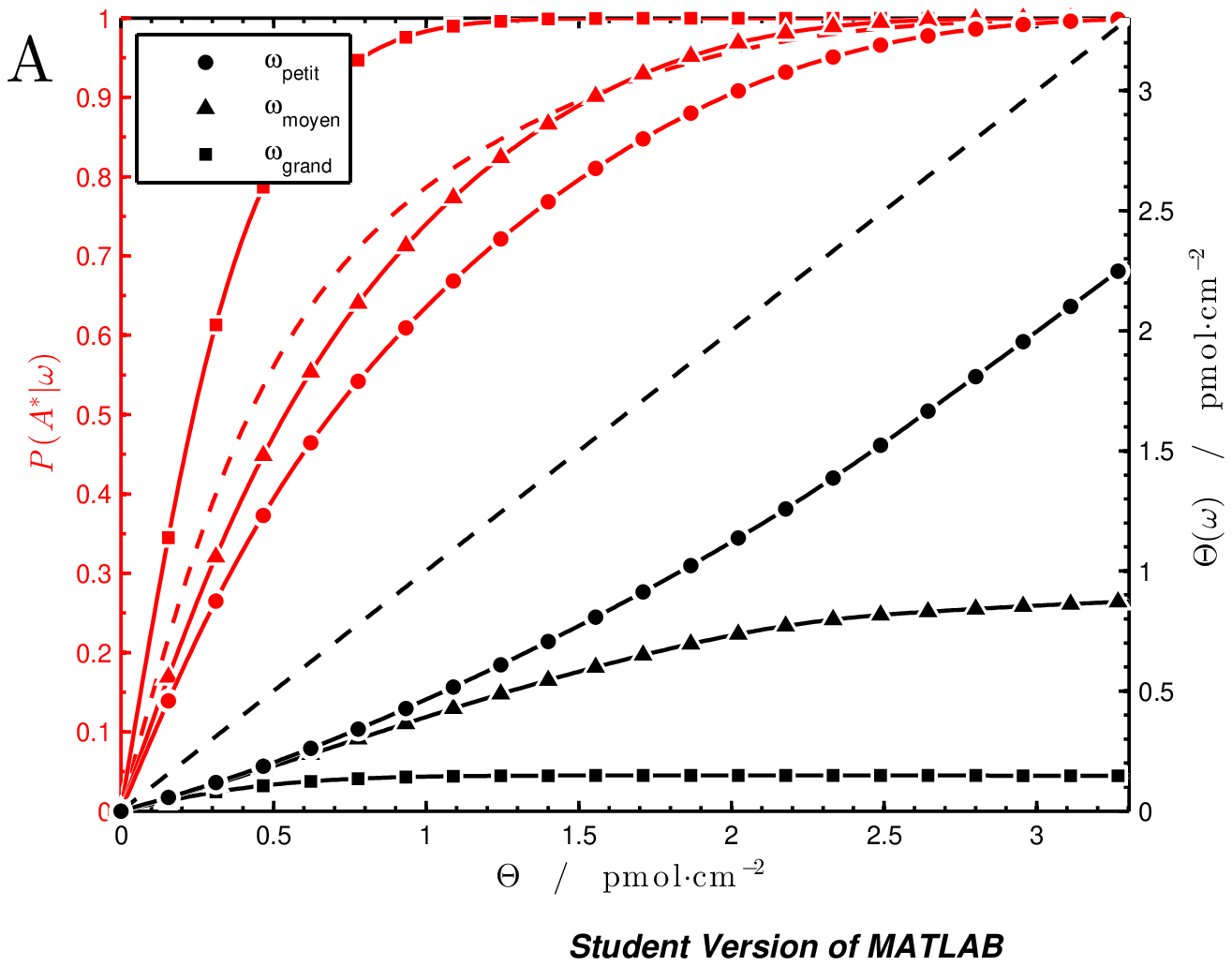}
\includegraphics*[width=0.75\textwidth]{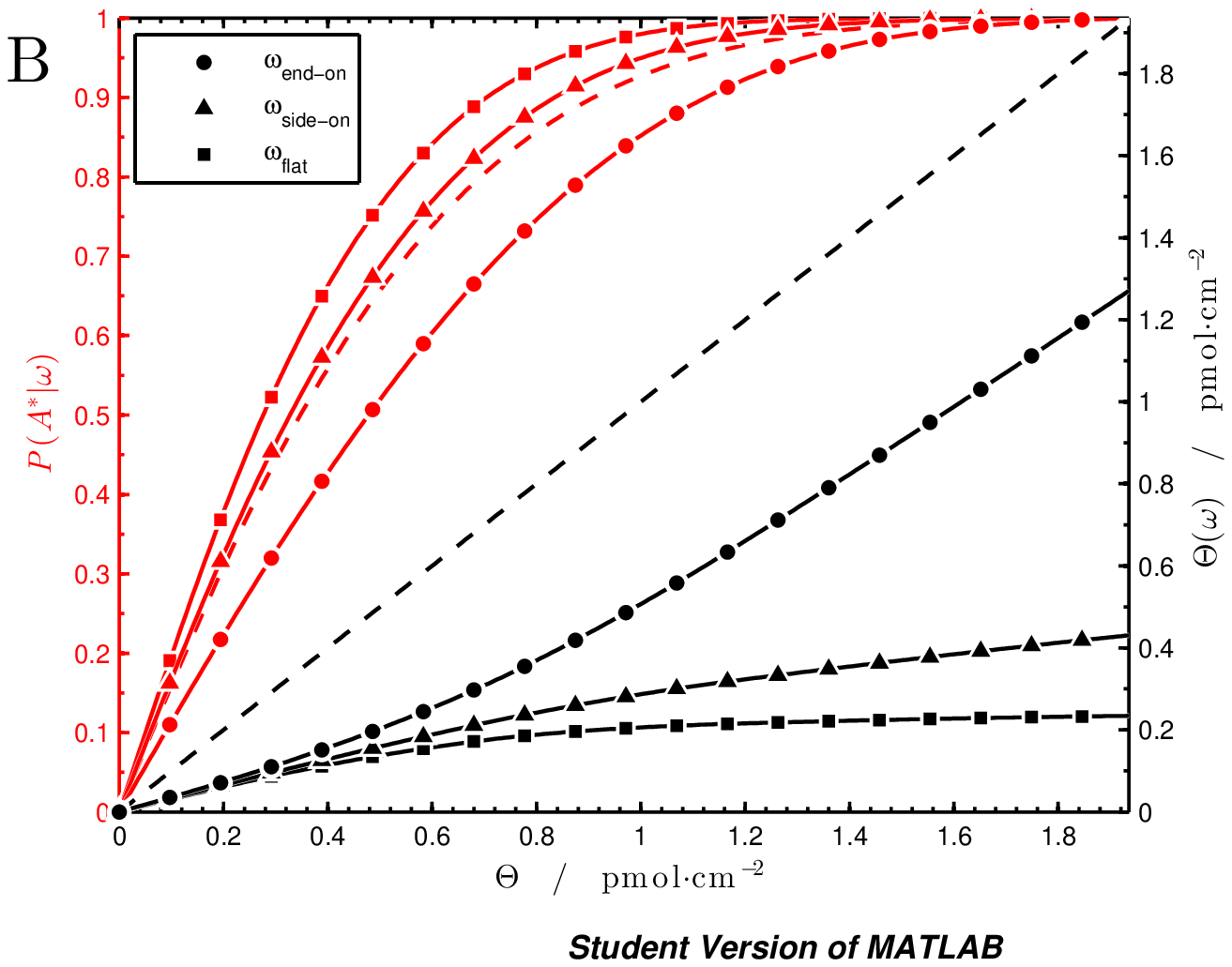}
\caption[$P(A^\ast)$ en fonction de $\Theta$ lors de remplissages RSA de surfaces]{Pour des additions séquentielles aléatoires \textbf{A}: d'un système de boîtes de conserve caractérisées par trois empreintes caractéristiques différentes et \textbf{B}: d'une population d'IgG s'accumulant sur la surface selon trois orientations (\textit{end-on}, \textit{side-on} et \textit{flat}); probabilités conditionnelles (rouge) de rejet des particules de la surface et quantités accumulées ($\Theta(\boldsymbol{\omega})$ en pmol$\cdot$cm$^{-2}$) selon les orientations (noir) en fonction des quantités totales $\Theta$. Les lignes discontinues représentent les quantités en ordonnées toutes orientations confondues.}\label{FigRSA1}
\end{figure}

\begin{figure}[hp]\centering
\includegraphics*[width=0.75\textwidth]{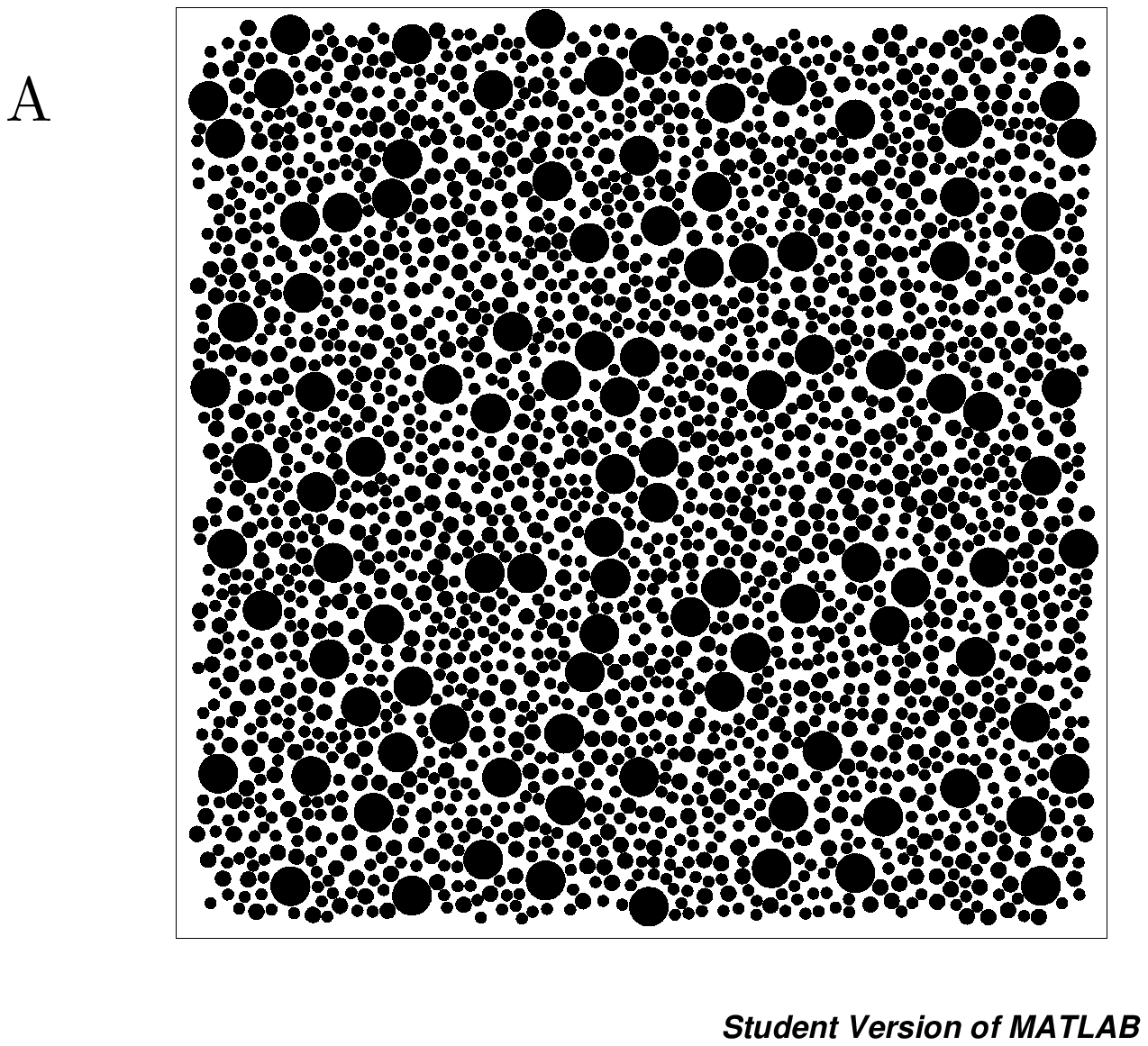}
\includegraphics*[width=0.75\textwidth]{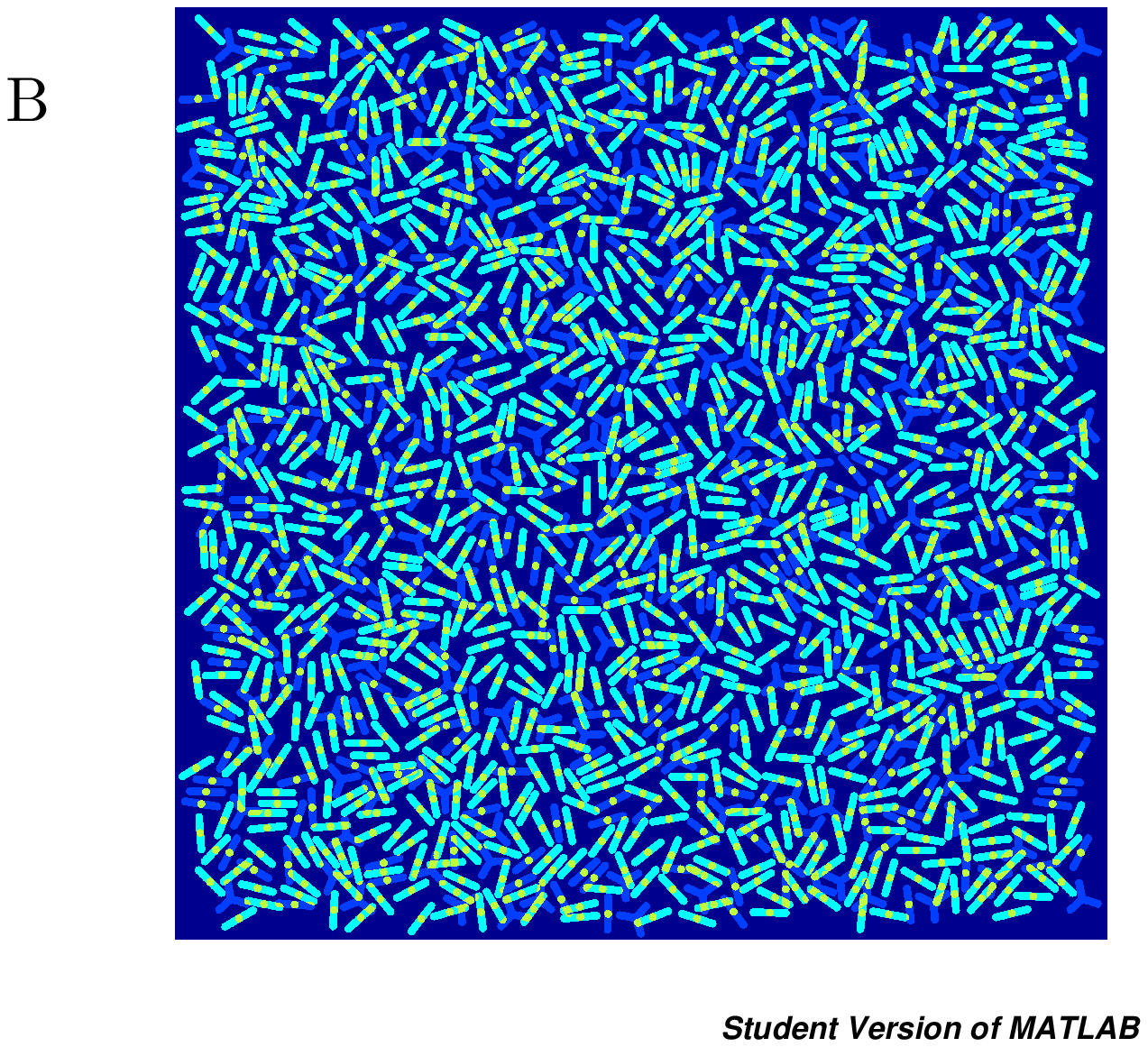}
\caption[Images de monocouches construites par remplissages RSA]{Pour des additions séquentielles aléatoires \textbf{A}: d'un système de boîtes de conserve caractérisées par trois empreintes caractéristiques différentes et \textbf{B}: d'une population d'IgG s'accumulant sur la surface selon trois orientations (\textit{end-on}, \textit{side-on} et \textit{flat}); images des surfaces obtenues à saturation. Les fragments Fc des IgG \textit{end-on} et \textit{side-on} sont représentés en jaune. Les fragments F(ab')$_2$ des IgG \textit{end-on} le sont en turquoise tandis que ceux des IgG \textit{side-on} sont de la même couleur que les IgG \textit{flat}, c'est-à-dire bleu clair. Outre le fragment Fc, le jaune permet aussi de représenter les endroits de la surface où se produisent des superpositions de deux IgG comme, par exemple, un fragment Fab d'une IgG \textit{end-on} au dessus d'une IgG \textit{flat}.}\label{FigRSAim1}
\end{figure}

De même, les probabilités de rejet qui, en fonction de leur intensité, suivent l'ordre de la série \ref{SerieExclusionIgG} montrent que le rejet des IgG commence par l'orientation ayant la plus forte empreinte caractéristique ($\boldsymbol{\omega}_{\mathrm{flat}}$), ensuite l'orientation d'empreinte caractéristique intermédiaire ($\boldsymbol{\omega}_{\mathrm{side\text{-}on}}$) pour terminer par la plus petite ($\boldsymbol{\omega}_{\mathrm{end\text{-}on}}$).
\begin{equation}\label{SerieExclusionIgG}
P(A^*|\boldsymbol{\omega}_{\mathrm{end\text{-}on}})
\leqslant\dots\leqslant
P(A^*|\boldsymbol{\omega}_{\mathrm{side\text{-}on}})
\leqslant\dots\leqslant
P(A^*|\boldsymbol{\omega}_{\mathrm{flat}})
\end{equation}
Malgré les probabilités équivalentes que ces IgG diversement orientées se présentent à la surface, les IgG \textit{end-on} seront les dernières à pouvoir s'additionner sur la surface et ainsi la saturer.

Le phénomène de rejet auquel les IgG sont soumises pendant leur accumulation sur la surface aura pour conséquence la série \ref{SerieNbreIgG} dans laquelle les quantités d'IgG accumulées en fonction de leur orientation montrent une nouvelle corrélation avec les empreintes caractéristiques dont elles disposent.

\begin{equation}\label{SerieNbreIgG}
\Theta_\infty(\boldsymbol{\omega}_\mathrm{flat})
\leqslant\dots\leqslant
\Theta_\infty(\boldsymbol{\omega}_\mathrm{side\text{-}on})
\leqslant\dots\leqslant
\Theta_\infty(\boldsymbol{\omega}_\mathrm{end\text{-}on})
\end{equation}

Le rejet des IgG en fonction de la grandeur de leur empreinte caractéristique empêchera très rapidement les IgG \textit{flat} de s'additionner à la monocouche en formation au profit des \textit{side-on} et des \textit{end-on}. Ensuite, après l'exclusion des \textit{side-on}, les IgG \textit{end-on} auront le champ libre pour saturer la surface et pourront s'y accumuler en plus grande quantité que les deux autres orientations.

Ce que l'on peut déduire des simulations RSA et du modèle utilisé pour les IgG montre que l'orientation \textit{end-on} est clairement favorisée grâce à la progression du volume exclu au cours du remplissage. Les quantités d'IgG \textit{end-on} seront même majoritaires dans la monocouche saturée obtenue.

\section{Vitesses d'accumulation ou d'addition}

Dans ce qui précède, la discussion a porté sur des relations entre la probabilité $P(A)$, la densité superficielle $\Theta$ et les empreintes caractéristiques $\sigma(\boldsymbol{\omega})$. Il manque une notion fondamentale permettant de relier les additions séquentielles aléatoires à la réalité expérimentale: le temps. En effet, l'énergie d'activation moyenne nécessaire à l'addition d'une nouvelle particule a été reliée, dans l'équation \ref{EqEActiva}, à la probabilité $P(A)$ d'observer une telle addition. Intuitivement, cette probabilité doit être liée au temps. En effet, si elle se trouvait être faible, on dirait que l'événement est rare ou, au contraire, si elle se trouvait être élevée, l'événement serait qualifié de fréquent; autrement dit, on en observerait plus ou moins sur le même laps de temps que durerait une observation.

Comme nous allons le voir, le processus d'addition des particules sur la surface peut être décrit sur base d'une série de vitesses qui seront notées $k$ et dont l'unité est la mol$\cdot$m$^{-2}\cdot$s$^{-1}$ (ou la pmol$\cdot$cm$^{-2}\cdot$s$^{-1}$). La cinétique de croissance de la monocouche selon les hypothèses du modèle RSA est donnée \citep{talbot2000} par l'équation différentielle
\begin{equation}\label{EqCinétique1}
\frac{\mathrm{d}\Theta}{\mathrm{d}t}=k_A^\circ\, P(A)
\end{equation}
donnant la vitesse instantanée de croissance de la monocouche que l'on notera aussi $k_A$. Il s'agit donc du nombre de mole(s) de particules s'additionnant à la monocouche pendant une durée $\mathrm{d}t$ à un instant $t$ de la croissance de la monocouche, instant pendant lequel la probabilité qu'une addition se produise est $P(A)$. Le facteur $k_A^\circ$ représente la fréquence ou vitesse, en mol$\cdot$m$^{-2}\cdot$s$^{-1}$, à laquelle les particules sont <<~parachutées~>> au dessus de la surface \citep{talbot2000}; il s'agit donc de la vitesse avec laquelle les particules arrivent à la surface, arrivées qui ne sont pas nécessairement suivies d'une addition. Posée comme constante car dépendante de facteurs externes au strict processus d'addition séquentielle aléatoire, $k_A^\circ$ est donc la vitesse de transport des particules vers la surface tandis que $k_A$ est la vitesse instantanée d'addition à la surface. En reformulant la cinétique \ref{EqCinétique1} telle que
\begin{equation}\label{EqCinétique2}
k_A=k_A^\circ\, P(A),
\end{equation}
on voit facilement qu'à vitesse de transport $k_A^\circ$ constante, la vitesse instantanée d'addition $k_A$ variera en fonction du volume exclu conceptualisé par $P(A)$. Au début du processus d'addition, lorsque la surface est encore vide, la valeur de la probabilité $P(A)$ est unitaire de telle sorte que $k_A=k_A^\circ$; la vitesse d'addition sur la surface est équivalente à la vitesse selon laquelle les particules y sont transportées puisque rien ne peut s'opposer à leur addition. Au fur et à mesure du remplissage de la surface, la probabilité $P(A)$ décroît jusqu'à tendre vers zéro, faisant dès lors aussi tendre la vitesse $k_A$ vers zéro, $k_A^\circ$ étant constante.

La substitution de l'équation \ref{EqCinétique2} dans \ref{EqEActiva} permet de donner une seconde interprétation de la cinétique mais, cette fois-ci, en fonction de l'évolution de $E^\ddagger_A$ la barrière d'énergie d'activation du processus d'addition. Il vient
\begin{equation}\label{EqCinétique3}
k_A=k_A^\circ\, \exp\bigg[-\frac{E^\ddagger_A}{k_BT}\bigg]
\end{equation}
dans laquelle la quantité d'énergie $E^\ddagger_A$ varie de zéro (instant initial lorsque la surface est encore vide) à l'infini (lorsque la surface est totalement remplie). Grâce à la relation \ref{EqCinétique3}, il vient que, lorsque $E^\ddagger_A=0$, l'égalité $k_A=k_A^\circ$ est obtenue tandis que, lorsque $E^\ddagger_A=+\infty$, une vitesse d'addition nulle est obtenue: $k_A=0$.

Une troisième interprétation de la cinétique \ref{EqCinétique1} peut encore être donnée en fonction des \emph{durées}. En inversant la relation \ref{EqCinétique3} et en définissant $\tau_A^\circ=(k_A^\circ)^{-1}$ comme la durée (en s$\cdot$m$^2\cdot$mol$^{-1}$) s'écoulant entre l'arrivée de deux particules à la surface et $\tau_A=(k_A)^{-1}$ comme la durée (en s$\cdot$m$^2\cdot$mol$^{-1}$) s'écoulant entre l'addition de deux particules à la monocouche au même instant, il vient
\begin{equation}\label{EqCinétique4}
\tau_A=\tau_A^\circ\, \exp\bigg[\frac{E^\ddagger_A}{k_BT}\bigg].
\end{equation}
Tout comme $k_A^\circ$, la durée $\tau_A^\circ$ étant invariable, la relation \ref{EqCinétique4} montre que le temps $\tau_A$ croît exponentiellement avec l'accroissement de la barrière d'énergie d'activation. Cette énergie passant de zéro à l'infini, la durée $\tau_A$ passera de la valeur $\tau_A=\tau_A^\circ$ à $\tau_A=+\infty$. En effet, au fur et à mesure de l'accumulation des particules sur la surface, le volume exclu augmente rendant de plus en plus élevée la barrière d'énergie à franchir, faisant que la durée pour obtenir une nouvelle addition finisse par tendre vers l'infini.

Ces considérations montrent le lien existant entre la probabilité d'addition $P(A)$ et le temps. Il est donc maintenant logique de rechercher une expression permettant d'obtenir la quantité $\Theta$ en fonction du temps. En intégrant la cinétique \ref{EqCinétique1}, il vient
\begin{equation}\label{EqCinétique5}
t(\Theta)=\int^{\Theta}_0\frac{\mathrm{d}\Theta}{k_A^\circ\, P(A)}
\end{equation}
donnant le temps (en secondes) nécessaire afin d'obtenir une monocouche d'une densité superficielle $\Theta$. Comme le font P. Schaaf \textit{et al.} \citep{schaaf1998}, cette équation peut directement être transformée
afin de normaliser le temps réel $t$ en un temps <<~adimensionnel~>> afin de s'affranchir des variations éventuelles de la vitesse de transport $k_A^\circ$ mais aussi de la difficulté de son estimation. En posant le temps adimensionnel $t^\ast=t/\tau_A^\circ$ ou, de manière équivalente $t^\ast=k_A^\circ t$, on obtient $t^\ast$ en multipliant la relation précédente par $k_A^\circ$. Ce résultat, plus général, est donné à l'équation \ref{EqCinétique6}.
\begin{equation}\label{EqCinétique6}
t^\ast(\Theta)=\int^{\Theta}_0\frac{\mathrm{d}\Theta}{P(A)}
\end{equation}
L'intégration faite à l'équation \ref{EqCinétique6} est la relation qui permettra d'obtenir numériquement les échelles de temps $t^\ast$ nécessaires à l'accumulation des quantités $\Theta$ montrées à la figure \ref{FigCinétique} pour les systèmes de boîtes de conserve et d'IgG montrées à la figure \ref{FigSchémaRSA1}.

\begin{figure}[hp]\centering
\includegraphics*[width=0.75\textwidth]{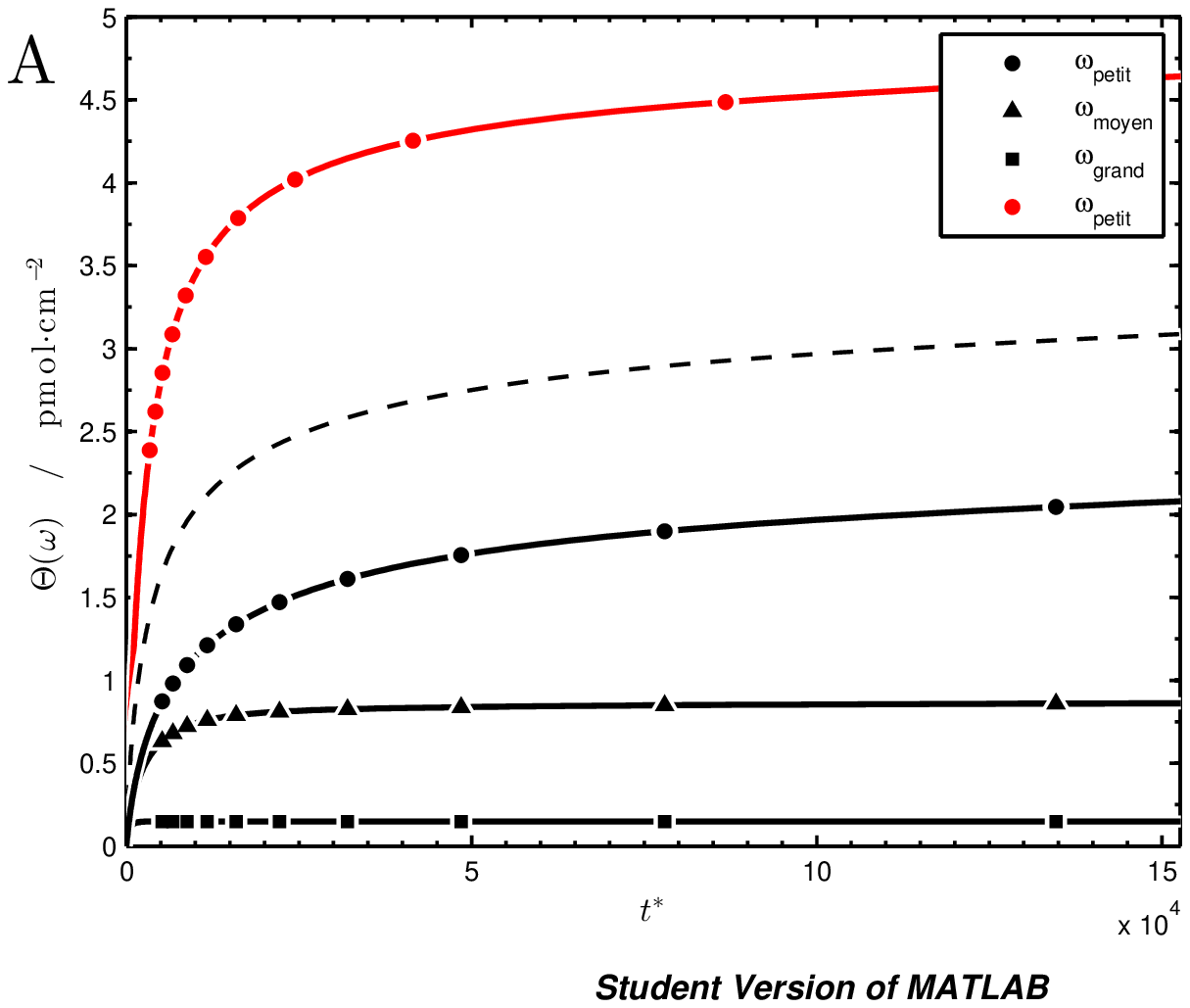}
\includegraphics*[width=0.75\textwidth]{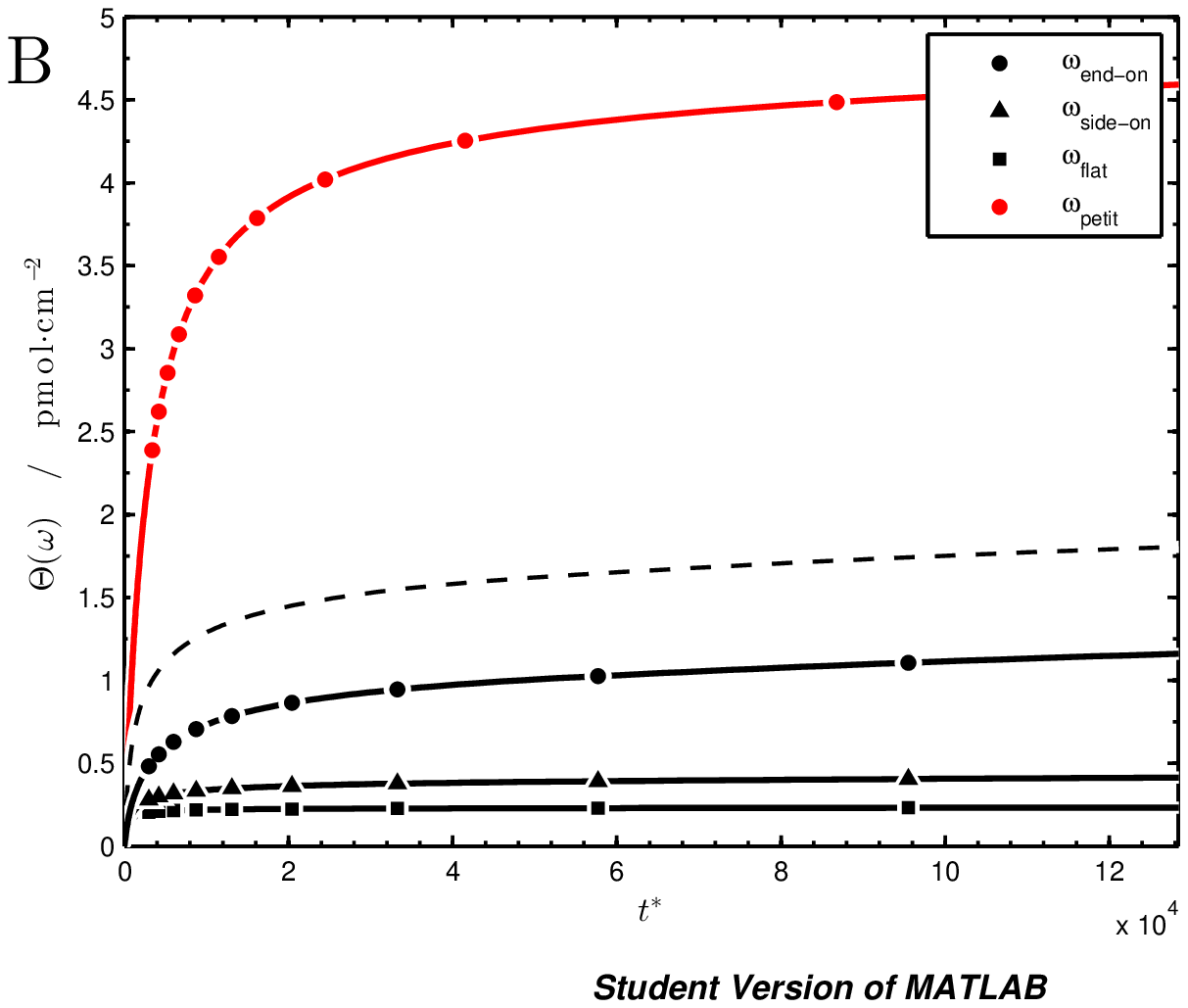}
\caption[\'{E}volutions des $\Theta$ lors de remplissages RSA en fonction de $t^\ast$]{\'{E}volutions des quantités accumulées $\Theta$ sur la surface par additions séquentielles aléatoires en fonction du temps adimensionnel $t^\ast$ pour les système présentés à la figure \ref{FigSchémaRSA1}. \textbf{A}: le système de trois boîtes de conserve d'empreintes différentes (petit, moyen et grand) et \textbf{B}: le système composé d'IgG dans diverses orientations (\textit{end-on}, \textit{side-on} et \textit{flat}). Afin de servir de comparaison, les courbes discontinues sont les quantités totales et les courbes rouges montrent la croissance d'une monocouche constituée d'une seule boîte de conserve de petite empreinte (\textit{cf.} figure \ref{FigSchémaRSA1}).}\label{FigCinétique}
\end{figure}

Toutes les courbes de $\Theta$ en fonction de $t^\ast$ montrant une croissance non bornée, la figure \ref{FigCinétique} révèle clairement que les durées nécessaires à l'obtention d'une monocouche tout-à-fait complète est infiniment grand. En effet, les points sur les courbes sont régulièrement espacés en fonction de $\Theta$ tandis que les espaces qui les séparent en fonction de $t^\ast$ croissent sans borne selon une fonction indéterminée.

Réaliser un tel lien entre le modèle RSA et le temps permet, non seulement d'obtenir une série de graphes montrant le ralentissement très intense du remplissage de la surface dû au volume exclu, mais aussi d'obtenir une expression pour la vitesse $k_A$ à laquelle la monocouche se construit. La vitesse $k_A^\circ$ est aussi un concept sur base duquel un lien relativement clair entre la diffusion et le remplissage par additions séquentielles aléatoires pourra être discuté.

\section{Discussion}

\subsection{Additions séquentielles aléatoires et diffusion}

Une vitesse $k_A^\circ$ a été posée comme facteur de proportionnalité dans les relations \ref{EqCinétique1} et \ref{EqCinétique2}. Cette vitesse <<~initiale~>> peut être reliée au transport par diffusion équivalente à une vitesse de parachutage des protéines sur la surface \citep{talbot2000}. En effet, la vitesse $k_A^\circ$ (en mol$\cdot$s$^{-1}$) représentant la fréquence à laquelle les particules arrivent à la surface \citep{talbot2000}, il n'est pas interdit de la rapprocher du flux $j_\perp$ de protéines par diffusion et perpendiculaire à la surface (axe $z$) provenant de la solution selon la relation
\begin{equation}\label{VitAddvsDiff}
k_A^\circ\sim j_\perp(z=0,t)
\end{equation}
dans laquelle on constatera que le flux est une fonction de la distance à la surface selon l'axe $z$ (vecteur normal à la surface, c'est-à-dire pointant vers le c{\oe}ur de la suspension) et de l'instant $t$ en lequel on s'intéresse à la croissance de la monocouche. Comme la surface se trouve en une position $z=0$, on utilisera le flux $j_\perp(z=0,t)$. Lorsque le transport convectif est négligeable, on obtient de la première loi de Fick \citep{hiemenz1997tout} que
\begin{equation}
j_\perp(z=0,t)=-D\frac{\partial}{\partial z} C(z=0,t)
\end{equation}
dont il vient avec \ref{VitAddvsDiff} que
\begin{equation}\label{fuyufqffqiytv}
k_A^\circ=-D\frac{\partial}{\partial z} C(z=0,t)
\end{equation}
où $D$ est le coefficient de diffusion des particules (m$^2\cdot$s$^{-1}$) et $C(z=0,t)$ (mol$\cdot$m$^{-3}$) leur concentration dans le volume de fluide en contact direct avec la surface à un instant $t$. Ces relations permettent de relier la vitesse d'accumulation \ref{EqCinétique1} à la concentration en protéines dans la suspension et dans une fraction de l'espace surplombant immédiatement la surface:
\begin{equation}\label{qsjgfsjhf}
\frac{\mathrm{d}\Theta}{\mathrm{d}t}=-D\frac{\partial}{\partial z} C(z=0,t)\,P(A).
\end{equation}
Cette relation montre le lien entre la vitesse d'accumulation, le volume exclu et la concentration en particules.

\`{A} ce stade, il est intéressant de faire une relecture de l'équation \ref{EqMécaSmol} de la diffusion de Smoluchowski décrivant l'évolution locale du traceur $C(\boldsymbol{q},t)$. Cette évolution est soumise à deux phénomènes: le mouvement brownien pris en compte dans le premier terme du membre de droite de l'équation et les champs de potentiels auxquels la protéine peut être soumise en fonction de sa position dans l'espace des coordonnées $\boldsymbol{q}=\lbrace\boldsymbol{x},\boldsymbol{\omega}\rbrace$. La diffusion fickienne, due au mouvement brownien des protéines, est présente partout et ne dépend, en fait, que de ses dimensions tandis que la diffusion forcée inclut la présence d'un potentiel variable dans l'espace des coordonnées $\boldsymbol{q}$ pouvant dès lors influencer la diffusion. L'influence de ce potentiel a été largement discutée, de même que ses différentes composantes: présence de la double couche diffuse, du potentiel de Hamaker et des interactions de volume exclu. En ne considérant que la direction normale à la surface $z$, l'équation de Smoluchowski s'écrit:
\begin{equation}
\frac{\partial}{\partial t}C(z,t)=D\frac{\partial^2}{\partial z^2}C(z,t)-\frac{D}{k_BT}\frac{\partial}{\partial z}\Big[ C(z,t)F(z,t)\Big].
\end{equation}

L'évolution de la barrière d'énergie d'activation que cause ce potentiel est capitale dans la description du phénomène de blocage de la surface puisqu'elle peut être considérée comme nulle à l'instant initial et infinie lorsque la surface est saturée. Il y a donc un lien certain entre la forme du potentiel se manifestant à l'approche de la surface qui est exprimé dans l'équation de Smoluchowski et l'état de la monocouche (inexistante, en cours de formation ou saturée) à un moment donné.

Le potentiel du second terme du membre de droite de l'équation de Smoluchowski se manifeste aux limites du domaine dans lequel se fait la diffusion, de telle sorte que la résolution de cette équation peut ici s'exprimer comme une solution de l'équation de la diffusion de Fick à laquelle on adjoint des conditions initiales et aux limites du domaine de l'espace dans lequel se fait la diffusion. Lorsque le problème de Cauchy est composé de la seule équation unidimensionnelle de Fick et d'une condition initiale
\begin{equation}\label{EqProbCauchy}\frac{\partial}{\partial t} C(z,t)=D\frac{\partial^2}{\partial z^2}C(z,t)\quad\text{avec}\quad C(z,t=0|z_0,t_0)=\delta(z-z_0),
\end{equation}
où la fonction $\delta(z-z_0)$ est la distribution de Dirac, $z_0$ la position à partir de la quelle les protéines commencent à diffuser et $t_0$ l'instant à partir duquel les protéines commencent à diffuser, alors la solution générale selon la direction $z$ et le temps $t$ est la fonction de Green (équation \ref{EqGreen}) \citep{goldner2009}:

\begin{equation}\label{EqGreen}
C(z,t|z_0,t_0)=\frac{1}{\sqrt{4\pi D (t-t_0)}}\exp\bigg[-\frac{(z-z_0)^2}{4D(t-t_0)}\bigg].
\end{equation}

La solution \ref{EqGreen} décrit la diffusion dans un domaine de dimensions infinies et n'est donc pas d'un usage pratique. Il faut donc adjoindre à ce premier problème \ref{EqProbCauchy} de Cauchy une série de conditions aux limites du domaine d'intérêt dans le cadre de la diffusion d'une protéine vers une surface qui pourra l'absorber irréversiblement. Le domaine d'intérêt est ici délimité par deux parois dont les propriétés et les conditions aux limites s'y rapportant seront discutées ci-dessous. La première de celles-ci est le plan matérialisant la frontière entre la suspension de protéines (phase liquide) et la zone dans laquelle croîtra la monocouche. Sa position dans l'espace est posée en $z=0$ (on néglige l'épaisseur de la monocouche).

La condition à la limite $z=0$ sera qualifiée, selon le cas, soit d'\textit{absorbante} soit de \textit{réfléchissante}. En effet, lorsque les particules arrivent à cette limite de leur domaine de diffusion, c'est-à-dire à la surface en $z=0$, elles pourront dans le premier cas être captées par la monocouche en formation. En termes mathématiques, ces particules disparaissent du domaine dans lequel elles ont diffusé et il est dit que la paroi est totalement absorbante, ce que l'on peut exprimer par la condition\footnote{Afin d'alléger l'écriture dans la suite de l'exposé, les conditions initiales $z_0$ et $t_0$ seront sous-entendues.}
\begin{equation}\label{EqCondiAbs}
C(z=0,t)=0
\end{equation}
exprimant l'absorption totale des particules arrivant à la limite du domaine, limite constituée par la surface solide. Le problème constitué de l'adjonction de cette condition au problème \ref{EqProbCauchy} est qualifié de problème de Dirichlet \citep{chorlton1969} et correspond au début de la formation de la monocouche car toutes les particules arrivant en $z=0$ y sont additionnées, le volume exclu étant nul à très faible.

Inversement, à la fin du processus, les particules s'approchant de la limite $z=0$ du domaine de diffusion se heurteront à une barrière de potentiel quasiment infranchissable de telle sorte qu'elles ne pourront être incluses à la monocouche et seront dès lors réfléchies (renvoyées) vers la phase liquide. La condition à la limite pour une paroi réfléchissante s'exprime par:
\begin{equation}\label{EqCondiRef}
D\frac{\partial}{\partial z}C(z=0,t)=0.
\end{equation}
Le problème \ref{EqProbCauchy} auquel est accolé cette condition réfléchissante est un problème de Neumann \citep{chorlton1969}, celui-ci exprimant une situation où la monocouche est saturée.

Les conditions à la limite $z=0$, dites de Dirichlet (absorption) et de Neumann (réflexion) correspondent respectivement à l'instant à partir duquel commence la formation de la monocouche et à l'instant de sa saturation, c'est-à-dire $t\rightarrow+\infty$. Sachant aussi qu'à l'instant initial, la barrière d'énergie d'activation est nulle, le problème de Neumann reflète la situation qui a été décrite ci-avant par une probabilité d'additionner une nouvelle particule $P(A)=1$. D'autre part, le problème de Dirichlet, paroi réfléchissante ($t\rightarrow+\infty$), est associé à la probabilité d'addition $P(A)\rightarrow 0$ ou à la probabilité de rejet $P(A^*)\rightarrow 1$.

Or, au cours de la formation de la monocouche, la condition à la limite $z=0$ ne peut s'exprimer ni par une paroi totalement absorbante pour laquelle $P(A)=1$, ni par une paroi totalement réfléchissante pour laquelle $P(A)=0$ mais plutôt par une paroi semi-réfléchissante (ou semi-absorbante). En effet, la paroi semi-réfléchissante permet de tenir compte du blocage progressif de la surface par la monocouche ne laissant s'additionner qu'une fraction seulement des particules arrivant en $z=0$. Cette fraction est donnée par la probabilité $0\leqslant P(A)\leqslant1$. La paroi semi-réfléchissante s'exprime par une nouvelle condition à la limite $z=0$ \citep{singer2005}
\begin{equation}\label{EqCondiSemiRef}
D\frac{\partial}{\partial z}C(z=0,t)=f(z,t)C(z=0,t)
\end{equation}
où $f(z=0,t)$ (notée $f_0(t)$ par la suite) est une fonction non négative du temps. Elle exprime une <<~vitesse de réaction~>>. Pour $f_0(t)=+\infty$, on retrouve le problème de la paroi absorbante\footnote{Un gradient infini représente une singularité dans l'espace, un trou tellement profond que tout y est absorbé.} (Dirichlet) et, pour $f_0(t)=0$, le problème de la paroi réfléchissante (Neumann).

Le tableau \ref{TableauBoundCond} synthétise les conditions à la limite $z=0$ en fonction du temps et de l'état de la monocouche à ces instants. La résolution de ce nouveau problème, dit de Robin, constitué de la relation \ref{EqProbCauchy} et de la condition à la limite $z=0$ semi-réfléchissante \ref{EqCondiSemiRef} donne une expression analytique explicite pour la fonction $C(z,t)$ qui est rapportée par A. Singer \textit{et~al.} \citep{singer2007}. Elle est donnée à l'équation \ref{EqFDPRobin}
\begin{equation}\label{EqFDPRobin}
\begin{split}
C(z,t)=\frac{1}{\sqrt{\pi D t}}&\exp\bigg[-\frac{z^2}{4Dt}\bigg]\\
&-\frac{f_0(t)}{D}\exp\bigg[\frac{f\big(z+f_0(t)t,t\big)}{D}\bigg]\mathrm{erfc}\bigg[\frac{z+2f_0(t)t}{\sqrt{4Dt}}\bigg]
\end{split}
\end{equation}
dans laquelle $\mathrm{erfc}(\xi)$ est la fonction d'erreur complémentaire $1-\mathrm{erf}(\xi)$. La fonction d'erreur étant définie par
\begin{equation}
\mathrm{erf}(\xi)=\frac{2}{\sqrt{\pi}}\int_0^\xi \exp[-\vartheta^2]\mathrm{d}\vartheta,
\end{equation}
le second terme du membre de droite de l'équation \ref{EqFDPRobin} pourra se réécrire selon
\begin{equation}
-\frac{f_0(t)}{\sqrt{\pi D^3 t}}\int_0^\infty\exp\bigg[-\frac{f_0(t)\lambda}{D}\bigg]\exp\bigg[-\frac{(z+\lambda)^2}{4Dt}\bigg]\mathrm{d}\lambda.
\end{equation}

\begin{table}[t]\centering
\caption[Conditions à la limite $z=0$ pour la diffusion]{Les différentes conditions à la limite $z=0$ en fonction du temps. Les conditions expriment respectivement les cas où la monocouche est encore inexistante, en formation et totalement saturée.}\label{TableauBoundCond}
\begin{spacing}{1.3}
\begin{small}
\begin{tabular}{lcccl}
\hline
Paroi&$t$&$P(A)$&$f_0(t)$&Condition\\
\hline
absorbante&$0$&$1$&$+\infty$&$C(z=0,t)=0$\\
semi-réfléchissante&$]0;+\infty[$&$]0;1[$&$]0;+\infty[$&$D\frac{\partial}{\partial z} C(z=0,t)=f_0(t)C(z=0,t)$\\
réfléchissante&$+\infty$&$0$&$0$&$\frac{\partial}{\partial z} C(z=0,t)=0$\\
\hline
\end{tabular}
\end{small}
\end{spacing}
\end{table}

L'équation \ref{EqFDPRobin} fournit une expression utile à la description de l'évolution de $C(z,t)$ en fonction de la probabilité d'une addition à la monocouche $P(A)$. En effet, la fonction $f_0(t)$, eu égard à sa dépendance au temps et à l'état de la monocouche en formation, est, d'une manière ou d'une autre, liée à cette probabilité $P(A)$. Ce lien non explicite est discuté par R. Erban et S. J. Chapman \citep{erban2007b} dans le cadre de simulations stochastiques du mouvement brownien.

Les rapprochements réalisés entre le remplissage de la surface et la diffusion aux équations \ref{fuyufqffqiytv} et \ref{qsjgfsjhf} permettront des développements importants d'un point de vue conceptuel au chapitre suivant.

\subsection{Effet d'exclusion par la taille et structure des monocouches}

Les simulations de constructions de monocouches par additions séquentielles aléatoires faites pour les boîtes de conserve et les IgG montrent que les protéines devant s'accumuler sur la surface sont exclues en fonction de leurs empreintes caractéristiques. Si, à l'instant initial du processus, toutes les particules arrivant à la surface peuvent s'y additionner, il n'en est pas de même au fur et à mesure de l'augmentation du volume exclu, c'est-à-dire du remplissage de la surface. Les évolutions des probabilités conditionnelles de rejet permettent en effet de montrer que les particules ayant une forte empreinte caractéristique seront exclues d'une éventuelle addition à la monocouche bien plus tôt que celles disposant d'empreintes caractéristiques plus faibles. Finalement, le régime asymptotique du remplissage se fait par additions exclusives des particules ayant les plus faibles empreintes, c'est-à-dire les boîtes de conserve de petits diamètres et les IgG \textit{end-on}.

Cet \emph{effet d'exclusion par la taille} des protéines se répercute sur les quantités accumulées en fonction des types/orientations des particules. Les protéines de fortes empreintes ne pourront s'accumuler qu'au début du processus tandis que les protéines d'empreintes faibles le pourront tout au long et jusqu'à la fin de celui-ci. L'augmentation progressive du volume exclu, lui-même dû à l'accumulation des protéines sur la surface, est la cause de cet effet d'exclusion par la taille. En effet, les espaces disponibles seront de plus en plus restreints et seules les protéines pouvant entrer dans les petits espaces restants pourront s'additionner à la monocouche, ce qui finira par les favoriser nettement en terme de nombre: les boîtes de conserve de petits diamètres et les IgG \textit{end-on} seront les plus nombreuses à peupler la monocouche saturée.

Afin de mieux comprendre ce phénomène d'exclusion par la taille, il est utile de se rappeler de la formulation de la différentielle du taux de recouvrement de la surface donnée à l'équation \ref{DifferentiellePhi}:
$$\mathrm{d}\phi=\sigma\,\mathrm{d}\Theta+\Theta\,\mathrm{d}\sigma.$$ La formation de la monocouche étant, par définition, une augmentation de l'occupation de la surface, la différentielle $\mathrm{d}\phi$ sera nécessairement positive. De même, les additions séquentielles de protéines sur la surface impliquent que $\mathrm{d}\Theta$ soit aussi positive. L'empreinte $\sigma$ étant une quantité forcément positive, l'accroissement $\mathrm{d}\Theta>0$ impliquera l'accroissement $\mathrm{d}\phi>0$. Il n'en va pas de même pour le second terme du membre de droite de la différentielle \ref{DifferentiellePhi}, \textit{i.e.} $\mathrm{d}\sigma$. L'effet d'exclusion par la taille ayant pour conséquence de favoriser de manière croissante les particules d'empreintes caractéristiques de plus en plus faibles, la fonction de répartition des particules accumulées sur la surface se déplacera vers l'accumulation des particules de faibles empreintes: $\mathcal{X}(\boldsymbol{\omega}_{\mathrm{petit}})$ croîtra au début du processus et décroîtra ensuite rapidement tandis que $\mathcal{X}(\boldsymbol{\omega}_{\mathrm{petit}})$ continuera à croître jusqu'à la saturation de la monocouche. De ce fait, \textit{via} la relation
\begin{equation}
\sigma=\sigma(\boldsymbol{\omega}_{\mathrm{petit}})\,\mathcal{X}(\boldsymbol{\omega}_{\mathrm{petit}})
\,+\,\dots\,+\,\sigma(\boldsymbol{\omega}_{\mathrm{grand}})\,\mathcal{X}(\boldsymbol{\omega}_{\mathrm{grand}})
\end{equation}
il semble logique que l'empreinte moyenne des particules contenue dans la monocouche diminue tout au long du processus de remplissage. L'effet d'exclusion par la taille impliquera dès lors une diminution de l'empreinte de telle sorte que sa différentielle soit négative: $$\mathrm{d}\sigma<0.$$ Le caractère négatif de $\mathrm{d}\sigma$ conceptualise l'effet d'exclusion progressif par la taille qui aura tendance à ne laisser progressivement s'additionner que les particules de faibles empreintes caractéristiques.

L'empreinte moyenne des IgG étant obtenue par la combinaison linaire
\begin{equation}
\sigma=\sigma(\boldsymbol{\omega}_{\mathrm{end\text{-}on}})\,\mathcal{X}(\boldsymbol{\omega}_{\mathrm{end\text{-}on}})
\,+\,\dots\,+\,\sigma(\boldsymbol{\omega}_{\mathrm{flat}})\,\mathcal{X}(\boldsymbol{\omega}_{\mathrm{flat}}),
\end{equation}
l'augmentation de la fraction $\mathcal{X}(\boldsymbol{\omega}_{\mathrm{end\text{-}on}})$ impliquera bien un abaissement de $\sigma$ lors de la formation d'une monocouche d'IgG. La différentielle de l'empreinte étant négative, elle aura tendance à diminuer la valeur de la différentielle $\mathrm{d}\phi$, c'est-à-dire ralentir le processus de remplissage de la surface et \textit{in fine} s'opposer à la croissance de $\Theta$.

L'effet d'exclusion par la taille a, comme cela a été montré, une conséquence remarquable pour les IgG. Celles-ci pouvant s'additionner selon diverses orientations, et donc des empreintes caractéristiques différentes, l'effet d'exclusion par la taille finira par favoriser nettement les IgG ayant une orientation \textit{end-on}, orientation recherchée dans le cadre des ELISA. Le fait de favoriser les IgG dont les empreintes caractéristiques sont faibles aura pour effet de diminuer l'empreinte des IgG accumulées dans la monocouche, c'est-à-dire que l'on conceptualise le fait de favoriser les \textit{end-on}.

\section{Conclusion}

Le modèle des additions séquentielles aléatoires est un outil puissant lorsqu'il est appliqué au remplissage des monocouches de protéines. Il a permis de mettre en évidence un effet d'exclusion par la taille de celles-ci. Celui-ci consiste, pour des systèmes de protéines disposant d'empreintes caractéristiques différentes, en l'exclusion progressive des protéines ayant les plus grandes empreintes caractéristiques. Au fur et à mesure du remplissage de la surface, les espaces laissés vides devenant de plus en plus rares et petits, les protéines d'empreintes caractéristiques élevées seront empêchées de s'additionner à la monocouche, ensuite les protéines d'empreintes caractéristiques moyennes et finalement, lorsque les espaces vacants seront vraiment insuffisants, les protéines ayant les plus petites empreintes caractéristiques. La fin du processus de remplissage est déterminée par l'exclusion des protéines du système ayant les plus petites empreintes caractéristiques.

La différentielle \ref{DifferentiellePhi} explicite que le remplissage de la surface est conceptualisé par le caractère positif de $\mathrm{d}\phi$. Cette croissance de $\phi$ est due à l'addition (augmentation de la quantité accumulée): $\mathrm{d}\Theta>0$. Par ailleurs, l'effet d'exclusion par la taille, conceptualisé par $\mathrm{d}\sigma<0,$ viendra s'opposer à cette addition et lorsqu'elle la compensera parfaitement, la monocouche sera saturée et l'égalité $\mathrm{d}\phi=0$ pourra être vérifiée.

Pour les IgG, trois orientations ont été étudiées: \textit{end-on}, \textit{side-on} et \textit{flat}. Selon ces trois orientations possibles, les IgG auront des empreintes caractéristiques différentes: la plus grande pour l'orientation \textit{flat} et la plus petite pour l'orientation \textit{end-on}. Dès lors, lorsqu'une monocouche est construite par additions séquentielles aléatoires de ces IgG diversement orientées, les IgG \textit{flat} seront rapidement empêchées de s'additionner, viendra ensuite le tour des IgG \textit{side-on} alors que les IgG \textit{end-on} seront les dernières exclues et finiront par saturer la surface. L'orientation disposant de la plus petite empreinte caractéristique, \textit{end-on} pour les IgG, sera donc nettement favorisée et il a même été montré que sa présence devenait majoritaire dans la monocouche. Ceci semble montrer qu'il s'agit de l'effet d'exclusion par la taille qui permet d'obtenir des monocouches contenant des IgG \textit{end-on}, orientations recherchées dans le cadre de l'ELISA.

Un lien a aussi pu être établi entre le transport des protéines vers la surface et le remplissage par additions séquentielles aléatoires. Lorsque le flux de matière est entièrement dû à la diffusion (absence de convection), l'équation différentielle \ref{qsjgfsjhf} montre que le facteur de proportionnalité $k^\circ_A$ entre la vitesse d'addition $\mathrm{d}\Theta/\mathrm{d}t$ et la probabilité d'addition (liée au volume exclu $P(A^\ast)$) ne serait rien d'autre que le produit du coefficient de diffusion $D$ des protéines et du gradient de concentration $\partial C/\partial z$ de celles-ci à l'interface, c'est-à-dire le flux de matière $j_\perp(z=0,t)$. Ce lien particulièrement simple sera utilisé dans le chapitre suivant.

On notera finalement que les résultats des simulations RSA montrés dans ce chapitre ont été obtenus dans le cadre de modèles ne pouvant prétendre englober exhaustivement la réalité expérimentale. Comme tout modèle, les hypothèses sur lesquelles il se fonde demeurent discutables et, dans le cas présent (\textit{cf}. figure \ref{FigSchémaRSA1}), ont été posées afin de mettre en évidence certains effets tels que ceux de volume exclu et d'exclusion par la taille. Il est en effet absolument clair qu'il existe bien d'autres manières d'imaginer la façon dont les particules arrivent à la surface (orientations, proportions de celles-ci, etc.) et qu'un choix a dû être réalisé. \`{A} titre d'exemple, la troisième hypothèse du modèle RSA posant l'impossibilité pour les particules de la monocouche de pouvoir se superposer, empêche l'existence d'empilements locaux alors même qu'il ne semble pas improbable que deux IgG \textit{flat} ne puissent se superposer dans un système réel.

\begin{footnotesize}

\end{footnotesize}\end{cbunit}
\begin{cbunit}
\chapter[Additions et relaxations séquentielles aléatoires des protéines]{Additions et relaxations séquentielles aléatoires des protéines}\label{ChapIsoth}
\markboth{Chapitre \ref{ChapIsoth}: RSA+R}{}
\minitoc

\section{Généralités}

Les additions séquentielles aléatoires ont permis, dans le chapitre précédent, de remplir une série de monocouches avec des protéines (IgG ou boîtes de conserve) de différentes empreintes caractéristiques $\sigma(\boldsymbol{\omega})$. Cette procédure d'accumulation a pu montrer qu'il existait un effet d'exclusion progressif des particules en fonction de leurs tailles, tailles déterminées par l'empreinte caractéristique de leurs orientations et conformations. Au fur et à mesure du remplissage de la surface, il a été constaté que les protéines étaient exclues, c'est-à-dire empêchées de s'additionner à la monocouche, en commençant par les protéines ayant une empreinte caractéristique élevée pour terminer par les empreintes caractéristiques faibles. Cet effet d'exclusion par la taille finissait par faire diminuer l'empreinte des protéines de la monocouche au cours du remplissage. Cet effet d'exclusion par la taille avait pour conséquence le signe de la différentielle de l'empreinte:
\begin{equation}\label{EqDiffEmpreinteAddit0}
\mathrm{d}\sigma\leqslant0
\end{equation}
dans laquelle la possibilité d'une égalité tient compte de l'invariabilité de $\sigma$ lorsque des protéines de mêmes empreintes caractéristiques sont additionnées tout au long de la formation de la monocouche ou bien lorsque celle-ci est saturée.

D'autre part, le chapitre \ref{Chap1} s'est attardé sur le fait que les protéines ont une forte inclination à relaxer une fois sur la surface. Cette relaxation, pouvant se faire par changement de leur orientation et/ou de leur conformation, a pour seul but de leur faire optimiser leurs interactions avec leur environnement et en particulier la surface de polystyrène hydrophobe. Cette relaxation a été caractérisée par une augmentation de l'empreinte, telle que
\begin{equation}\label{EqDiffEmpreinteRelax0}
\mathrm{d}\sigma\geqslant0
\end{equation}                                                                                                                                                                                                                                                                                                                                                                                                                                                                                                                                                                                                                                                                                                                                                                                                                                                                                                                                                                                                                                                                                                                                                                                                                                                                                                                                                                                                                                                                                                                                                                                                                                                                                                                                                                                                                                                                                                                                                                                                                                                                                                                                                                                                                                                                                                                                                                                                                                                                                                                                                                                                                                                                                                                                                                                                                                                                                                                                                                                                                                                                                                                                                                                                                                                                                                                                                                                                                                                                                                                                                   et se produit spontanément lorsque les conditions expérimentales se trouvent réunies. La discussion menée au chapitre \ref{Chap1} a permis de montrer que la relaxation se produisait dans les mêmes conditions que l'addition, ce qui a conduit à énoncer que le fait qu'il y ait addition était une condition suffisante pour que la relaxation se produise. Du point de vue thermodynamique, la relaxation suit spontanément l'addition d'une particule sur la surface puisque, selon l'équation \ref{DiffEntropieVSsigmaTheta}, les différentielles $\mathrm{d}\Theta$ et $\mathrm{d}\sigma$ seront positives (adhésion des protéines) ou négatives (décollement des protéines) par suite d'une seule et même cause, c'est-à-dire que l'hydrophobie du polystyrène sera plus élevée que l'hydrophilie des IgG.

Il est clair que les signes mis sur les différentielles \ref{EqDiffEmpreinteAddit0} et \ref{EqDiffEmpreinteRelax0} de l'empreinte des protéines semblent en totale contradiction l'un et l'autre alors qu'il s'agit bien de variations d'une seule et même fonction caractéristique de l'état de la monocouche. Cette contradiction pourrait être résolue en admettant qu'il s'agit de deux contributions, l'une cinétique, l'autre thermodynamique, à la différentielle de l'empreinte. En partageant la différentielle de l'empreinte $\mathrm{d}\sigma$ en deux parties, on posera que
\begin{equation}\label{EqDiffEmpreinteTotale}
\mathrm{d}\sigma=\mathrm{d}_A\sigma+\mathrm{d}_R\sigma
\end{equation}
dans laquelle $\mathrm{d}\sigma$ est la différentielle totale de l'empreinte tandis que $\mathrm{d}_A\sigma$ et $\mathrm{d}_R\sigma$ en sont deux différentielles partielles (voir section \ref{hb35d8f}) dont le sens est explicité comme suit:
\begin{itemize}
\item Premièrement, $\mathrm{d}_A\sigma$ est la différentielle partielle de l'empreinte due au seul phénomène d'addition et représente donc l'\emph{effet d'exclusion par la taille} mis en évidence au chapitre \ref{SectionRSA}. Cette perte d'empreinte
\begin{equation}\label{EqDiffEmpreinteAddit1}
\mathrm{d}_A\sigma\leqslant0
\end{equation}
est un effet \emph{cinétique} et aura donc tendance à abaisser l'empreinte $\sigma$ puisqu'il agit en favorisant l'addition de protéines d'empreintes caractéristiques faibles.
\item Deuxièmement, $\mathrm{d}_R\sigma$ est la différentielle partielle de l'empreinte due au seul phénomène de \emph{relaxation} des particules sur la surface. Ce gain d'empreinte
\begin{equation}\label{EqDiffEmpreinteRelax1}
\mathrm{d}_R\sigma\geqslant0
\end{equation}                                                                                                                                                                                                                                                                                                                                                                                                                                                                                                                                                                                                                                                                                                                                                                                                                                                                                                                                                                                                                                                                                                                                                                                                                                                                                                                                                                                                                                                                                                                                                                                                                                                                                                                                                                                                                                                                                                                                                                                                                                                                                                                                                                                                                                                                                                                                                                                                                                                                                                                                                                                                                                                                                                                                                                                                                                                                                                                                                                                                                                                                                                                                                                                                                                                                                                                                                                                                                                                                                                                                                  
est une conséquence du second principe de la thermodynamique discuté au chapitre \ref{Chap1} et contribuera à augmenter l'empreinte $\sigma$ des protéines accumulées sur la surface. Cette augmentation de l'empreinte, effet \emph{thermodynamique}, se fait par augmentation des empreintes caractéristiques des protéines déjà présentes dans la monocouche par conséquence de la somme des hydrophobies des corps en présence.
\end{itemize}

La différentielle de l'empreinte contenue dans l'équation de production d'entropie \ref{DiffEntropieVSsigmaTheta} devrait donc être considérée comme la différentielle partielle de l'empreinte due à la relaxation $\mathrm{d}_R\sigma$. Cette possibilité fournit l'expression d'une nouvelle proposition relative à l'adhésion hydrophobe développée au chapitre \ref{Chap1}.
\begin{proposition}[Production d'entropie superficielle lors de l'adhésion]
La production d'entropie superficielle en chaque point d'une surface solide <<~s~>> d'hydrophobie $\mathcal{H}_{\mathrm{sl}\cdot\mathrm{l}}$ sur laquelle viennent adhérer des protéines <<~p~>> d'hydrophobie $\mathcal{H}_{\mathrm{pl}\cdot\mathrm{l}}$ est donnée par
\begin{equation}\label{DiffEntropieVSsigmaTheta2}
\mathrm{d}_is
=\frac{\sigma}{T}\big(\mathcal{H}_{\mathrm{sl}\cdot\mathrm{l}}+
\mathcal{H}_{\mathrm{pl}\cdot\mathrm{l}}\big)\mathrm{d}\Theta
+\frac{\Theta}{T}\big(\mathcal{H}_{\mathrm{sl}\cdot\mathrm{l}}+
\mathcal{H}_{\mathrm{pl}\cdot\mathrm{l}}\big)\mathrm{d}_R\sigma
\end{equation}
en fonction des différentielles de l'empreinte $\sigma$ et de l'accumulation $\Theta$. La différentielle $\mathrm{d}_R\sigma$ est la différentielle partielle de l'empreinte attribuée aux seules variables d'état associées au phénomène de relaxation.
\end{proposition}
Lorsque les conditions de l'adhésion se trouvent réunies, cette relation \ref{DiffEntropieVSsigmaTheta2} montre que le second principe de la thermodynamique exige une augmentation de la quantité accumulée $\mathrm{d}\Theta\geqslant0$ et une augmentation de l'empreinte par relaxation $\mathrm{d}_R\sigma\geqslant0$.

Ces considérations permettent de mieux cerner le rôle de la différentielle de l'empreinte au cours de la construction de la monocouche. La substitution de la différentielle totale \ref{EqDiffEmpreinteTotale} dans l'expression \ref{DifferentiellePhi} de la différentielle de $\phi$ fournit une relation montrant que la différentielle du taux de recouvrement de la surface est fonction de l'addition $\mathrm{d}\Theta$, de l'effet d'exclusion par la taille $\mathrm{d}_A\sigma$ et de la relaxation $\mathrm{d}_R\sigma$. On obtient alors l'équation \ref{EqDifférentielleProbabilité3} faisant une synthèse des connaissances accumulées jusqu'à présent sur la valeur de la différentielle du taux de recouvrement:
\begin{equation}\label{EqDifférentielleProbabilité3}
\underbrace{\mathrm{d}\phi}_{\geqslant\,0}
=\underbrace{\quad\underbrace{\sigma\,\mathrm{d}\Theta}_{\geqslant\,0}
+\underbrace{\Theta\,\mathrm{d}_R\sigma}_{\geqslant\,0}\quad}_{\text{effet thermodynamique}}
+\underbrace{\quad\underbrace{\Theta\,\mathrm{d}_A\sigma}_{\leqslant\,0}\quad}_{\text{effet cinétique}}
\end{equation}
On y voit très clairement que les effets thermodynamiques (addition $\mathrm{d}\Theta\geqslant0$ et relaxation $\mathrm{d}_R\sigma\geqslant0$) sont les origines de l'accroissement du taux de recouvrement de la surface tandis que l'effet cinétique d'exclusion par la taille ($\mathrm{d}_A\sigma\leqslant0$) pourra venir s'y opposer. Lorsque l'effet d'exclusion par la taille sera suffisamment important pour contrer le processus d'addition-relaxation, la différentielle $\mathrm{d}\phi$ deviendra nulle et la monocouche cessera de croître; il y aura alors équilibre entre les forces d'origines thermodynamiques et cinétiques. Le système atteindra un état statique ($\mathrm{d}\phi=0$) et la monocouche sera saturée.

D'autre part, il doit aussi être constaté que la présence de la différentielle $\mathrm{d}_R\sigma$ est de nature à accentuer l'effet thermodynamique sur la croissance du taux d'occupation. Plus la relaxation est importante, plus l'accroissement du taux d'occupation sera rapide. La relaxation est donc un facteur accélérant le recouvrement de la surface. L'équation \ref{EqRSA3} ayant montré que la probabilité d'addition $P(A)$ était une fonction strictement décroissante du taux de recouvrement $\phi$, la relaxation devrait donc avoir tendance à faire décroître $P(A)$ beaucoup plus rapidement au cours du processus de remplissage de la monocouche.

Avant de procéder à une telle démarche d'implémentation de la relaxation des protéines au modèle RSA dans le cadre de ce travail, il doit être remarqué que cette implémentation a déjà été proposée et largement discutée par certains auteurs. De fait, P.~R. Van Tassel \textit{et al}. \citep{vantassel1994,vantassel1996,vantassel1998} ont pu élaborer un premier modèle dans lequel toutes les protéines venant s'additionner à la monocouche s'y relaxent instantanément, et ce, compte tenu du volume exclu (modèle I de \citep{vantassel1994}). Un second modèle gère la probabilité de cette relaxation en ce sens qu'une protéine s'additionnant à la monocouche dispose d'une certaine probabilité de s'y relaxer après son addition (modèle II de \citep{vantassel1994}). Quoiqu'améliorant le modèle RSA au sens strict, ces modèles ne permettent pas de tenir compte de la possibilité dont devraient disposer les protéines de relaxer à n'importe quel moment de l'évolution de la monocouche. En effet, une protéine pourrait ne pas relaxer uniquement après son addition à la monocouche mais à un autre moment. Au demeurant, si la relaxation est beaucoup plus lente que l'addition, la protéine ne pourrait relaxer que  partiellement au moment de son addition à la monocouche et poursuivre ce processus alors même que d'autres protéines auront déjà pu s'y additionner. On devrait donc tenir compte de l'addition et de la relaxation en tant que deux phénomènes continus et simultanés, c'est-à-dire se produisant tout au long de la croissance de la monocouche et affectant aléatoirement une partie ou la totalité des protéines dont elle est constituée.

La prise en compte de la relaxation dans les différents modèles discutés au chapitre précédant ne semble pas une tâche simple car elle nécessite de considérer les vitesses de relaxation mais aussi d'addition. La première approche qui sera développée consistera à faire se relaxer complètement toute protéine venant s'additionner sur la surface. Cette façon de voir le processus de relaxation considère la vitesse de relaxation comme infiniment grande de telle sorte que la relaxation soit instantanée dès que la protéine touche la surface (\textit{cf}. le modèle I de R.~P. Van Tassel \textit{et al}. \citep{vantassel1994}). Cette approche sera discutée pour un modèle de boîtes de conserve (relaxation de la conformation) et pour les IgG (relaxation de l'orientation), deux modèles décrits à la figure \ref{FigRSARelax}. La seconde approche qui sera développée visera à considérer une vitesse finie pour la relaxation et ainsi d'en ajuster l'importance lors de simulations. \`{A} cet effet, une quatrième hypothèse sera ajoutée aux trois hypothèses du modèle classique des additions séquentielles aléatoires énumérées au chapitre \ref{SectionRSA} afin de tenir compte de la relaxation (modèle RSA+R). Sur cette base, il deviendra possible de discuter de la structure de la monocouche en fonction de l'importance du phénomène de relaxation et de faire le lien avec les comportements d'accumulation des protéines dures et molles.

\section{Relaxation des boîtes de conserve à vitesse infinie}

La figure \ref{FigRSARelax} illustre un modèle de boîtes de conserve ayant la possibilité de relaxer leur conformation en s'étalant en deux pas successifs. Les pas de relaxation sont caractérisés par des degrés de dénaturation $\beta_0$ (natif), $\beta_1$ et $\beta_2$. Dans ce modèle, les boîtes de hauteur $h$ (natives) sont les seules à pouvoir être additionnées sur la surface de telle sorte que l'on ait les probabilités
\begin{equation}\label{54fg65sd456}
P(\boldsymbol{\omega}_\mathrm{natif})=1,\quad
P(\boldsymbol{\omega}_{\beta_1})=0\quad\text{et}\quad
P(\boldsymbol{\omega}_{\beta_2})=0
\end{equation}
qui, par utilisation de la formule \ref{EqProbCond}, fournit le calcul de la probabilité d'addition $P(A)$:
\begin{equation}\label{lqyefglquyfgku}
P(A)=P(A|\boldsymbol{\omega}_\mathrm{natif})
\end{equation}
conceptualisant le fait que seules les particules natives peuvent s'additionner à la monocouche. Bien que seules les boîtes natives aient la possibilité de s'additionner à la monocouche, elles disposent de la possibilité de relaxer leur conformation, moyennant suffisamment d'espace disponible. La figure \ref{FigRSARelax} montre très clairement que les empreintes caractéristiques des boîtes de conserve selon leurs trois conformations possibles répondent à la série suivante:
\begin{equation}
\sigma(\boldsymbol{\omega}_\mathrm{natif})<\dots<
\sigma(\boldsymbol{\omega}_{\beta_1})<\dots<
\sigma(\boldsymbol{\omega}_{\beta_2})
\end{equation}
dans laquelle, plus les boîtes sont relaxées, plus leur empreinte caractéristique est élevée.

\begin{figure}[t]\centering
\includegraphics*[width=0.85\textwidth]{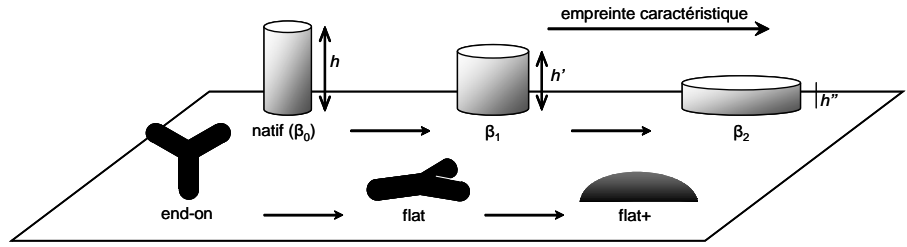}
\caption[Systèmes utilisés pour les simulations d'accumulations avec relaxation]{La boîte de conserve (en haut) pouvant relaxer sa conformation (conformation native $\beta_0$ de 4,75 nm de diamètre) par étalement selon deux pas successifs de degrés de dénaturation $\beta_1$ (diamètre: 6,42 nm) et $\beta_2$ (diamètre: 15,6 nm) et l'IgG \textit{end-on} (en bas) relaxant son orientation ($\rightarrow$ \textit{flat}) et ensuite sa conformation ($\rightarrow$ \textit{flat}+). Les empreintes caractéristiques augmentent de gauche à droite.}\label{FigRSARelax}
\end{figure}

Pour ce système, les modélisations RSA sont réalisées en faisant relaxer les boîtes natives au maximum de leur possibilité et immédiatement après qu'elles aient touché la surface, et ce, tout en tenant compte de l'accroissement du volume exclu. Ces simulations sont réalisées grâce aux algorithmes présentés à la section \ref{SurfConstruct1} lorsque le paramètre $\ln K^\circ$ tend vers $-\infty$ (de plus amples explications seront données plus loin).

Un premier résultat est donné à la figure \ref{FigRSAghdhdhh} et montre l'effet de la relaxation sur la vitesse selon laquelle la surface est remplie. Lorsque l'on compare les courbes des probabilités d'addition en fonction de $\Theta$, on y voit clairement que la relaxation rend le blocage de la surface beaucoup plus rapide. En effet, il suffit d'une quantité d'environ 2,5~pmol$\cdot$cm$^{-2}$ de boîtes de conserve pouvant relaxer afin de bloquer la surface alors qu'il en faudrait environ 4,75~pmol$\cdot$cm$^{-2}$ pour obtenir le même résultat avec des boîtes indéformables. En favorisant les empreintes caractéristiques élevées, la relaxation permet, d'une part, de saturer la surface avec de plus faibles quantités de protéines et, d'autre part, d'obtenir des monocouches moins peuplées en protéines natives (non relaxées). Le fait que la possibilité de relaxation permette d'obtenir moins de protéines natives pourrait sembler trivial s'il ne mettait pas en lumière le fait que, selon l'intensité de la relaxation et la manière plus continue dont on pourrait l'envisager, il pourrait mener à une monocouche dont les protéines relaxées seraient quasiment absentes.

\begin{figure}[t]\centering
\includegraphics*[width=0.75\textwidth]{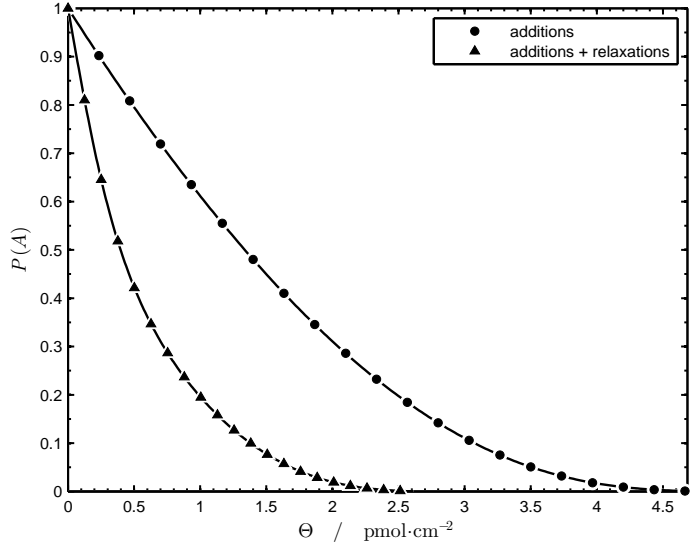}
\caption[$P(A)$ en fonction de $\Theta$ pour des boîtes de conserve molles et dures]{\'{E}volutions des probabilités d'additions $P(A)$ en fonction de la densité superficielle $\Theta$ pour le système de boîtes de conserve montré à la figure \ref{FigRSARelax}. Les disques représentent l'évolution lors d'additions de boîtes de conserve natives (boîtes très dures) tandis que les triangles donne l'évolution lors de ces mêmes additions mais suivies d'une relaxation maximale de la boîte (boîtes très molles).}\label{FigRSAghdhdhh}
\end{figure}

Bien que les probabilités \ref{54fg65sd456} qu'une protéine relaxée se présente à la surface soient nulles, il est toujours possible d'évaluer les probabilités conditionnelles $P(A^*|\boldsymbol{\omega}_{\beta_1})$ et $P(A^*|\boldsymbol{\omega}_{\beta_2})$ en plus de la probabilité $P(A^*|\boldsymbol{\omega}_\mathrm{natif})$. Celles-ci, bien qu'elles n'aient pas vraiment de sens du point de vue physique, permettent de se donner une idée de l'évolution du volume exclu et, en particulier de la difficulté qu'auront les boîtes de conserve de relaxer vers des états relaxés $\boldsymbol{\omega}_{\beta_1}$ et $\boldsymbol{\omega}_{\beta_2}$. Il est en effet possible de supposer qu'il sera difficile d'obtenir une protéine dans un état relaxé si la probabilité conditionnelle de rejet pour cet état est élevée à une moment donné du processus de remplissage. Dès lors, en suivant la construction de la monocouche pour les boîtes de conserve sur la figure \ref{FigRSA4}.A à l'aide de ces probabilités conditionnelles, on constate, d'une part, que l'ordre d'exclusion (lié à l'empêchement de relaxer vers un état particulier) suit toujours la taille de l'empreinte caractéristique caractérisant chaque configuration:
\begin{equation}
P(A^*|\boldsymbol{\omega}_\mathrm{natif})\leqslant\dots\leqslant P(A^*|\boldsymbol{\omega}_{\beta_1})\leqslant\dots\leqslant P(A^*|\boldsymbol{\omega}_{\beta_2}),
\end{equation}
c'est-à-dire que les larges empreintes caractéristiques sont les premières à être exclues. D'autre part, on constate aussi que la relaxation favorise l'accumulation des configurations de larges empreintes caractéristiques au début du processus, ensuite des empreintes caractéristiques moyennes pour finir par les plus petites. La courbe de $\Theta(\boldsymbol{\omega}_{\beta_2})$ est en effet la première à croître en fonction de la quantité totale $\Theta$ alors que la courbe $\Theta(\boldsymbol{\omega}_\mathrm{natif})$ en est la dernière. La relaxation a donc pour effet de compenser l'effet d'exclusion par la taille, obligeant toutes les boîtes ayant la possibilité de relaxer à le faire. Ainsi, aux premiers instants de la construction de la monocouche, l'accumulation des boîtes totalement relaxées sera la seule constatée, ensuite viendront les moins relaxées, etc. Le caractère systématique de la relaxation imprime un ordre dans l'accumulation des boîtes de conserve en fonction de leur conformation.

L'attraction pour les grandes empreintes imposée par le phénomène spontané de relaxation sera dominant au début puisque la place disponible sera suffisante; le début du remplissage de la monocouche est donc un régime contrôlé par la thermodynamique ($\mathrm{d}_A\sigma\simeq0$ et $\mathrm{d}_R\sigma>0$). Plus la surface se remplit, moins la place est disponible pour les empreintes caractéristiques plus larges, plus les protéines sont empêchées de relaxer complètement afin d'atteindre cet état de grande empreinte caractéristique; la construction de la monocouche est alors sous contrôle cinétique à cause de l'effet d'exclusion par la taille ($\mathrm{d}_A\sigma<0$ et $\mathrm{d}_R\sigma\simeq0$). L'addition est donc constante mais la possibilité de relaxer diminue au cours du remplissage faisant que la conformation des protéines s'accumulant dans la monocouche sera elle aussi variable.

Des résultats de ces simulations, il ressort que la relaxation permettra de saturer la surface beaucoup plus rapidement car elle a tendance à favoriser les protéines de grandes empreintes caractéristiques. Mais, à l'approche de la saturation, la relaxation sera fortement inhibée par l'augmentation du volume exclu, laissant donc la possibilité à des protéines natives de s'accumuler sur la surface.

\section{Relaxation des IgG à vitesse infinie}

Après s'être intéressé aux boîtes de conserve, la réalisation de simulations RSA tenant compte d'une relaxation instantanée des IgG peut être envisagée. La séquence d'addition-relaxation présentée sur le schéma de la figure \ref{FigRSARelax} pour les IgG sera suivie. Il s'agit donc de simuler la construction d'un film d'IgG lorsque toutes les IgG arrivent à la surface dans une orientation \textit{end-on} et que les IgG \textit{flat} et \textit{flat}+ ne peuvent être obtenues que par suite du processus de relaxation. Les hypothèses d'un tel modèle se traduisent par les probabilités
\begin{equation}\label{4g654ja4gf35g}
P(\boldsymbol{\omega}_\mathrm{end\text{-}on})=1,\quad
P(\boldsymbol{\omega}_\mathrm{flat})=0\quad\text{et}\quad
P(\boldsymbol{\omega}_\mathrm{flat+})=0
\end{equation}
que les IgG se présentent à la surface selon leurs différentes orientations. Par conséquent, la probabilité d'addition est obtenue grâce à la formule \ref{EqProbCond}:
\begin{equation}
P(A)=P(A|\boldsymbol{\omega}_\mathrm{end\text{-}on})
\end{equation}
montrant que seules des IgG dans une orientation \textit{end-on} peuvent se présenter à la surface. Encore une fois, les orientations des IgG disposent de toutes sortes d'empreintes caractéristiques suivant un ordre de grandeur donné par la série
\begin{equation}
\sigma(\boldsymbol{\omega}_\mathrm{end\text{-}on})<\dots<
\sigma(\boldsymbol{\omega}_\mathrm{flat})<\dots<
\sigma(\boldsymbol{\omega}_\mathrm{flat+})
\end{equation}
montrant que, plus les IgG relaxent leur orientation, plus leur empreinte caractéristique sera élevée. Globalement, les IgG orientées \textit{end-on} se présenteront à la surface, après quoi elles disposent de la possibilité, si suffisamment d'espace est disponible, de relaxer leur orientation une première fois (orientation \textit{end-on} $\rightarrow$ orientation \textit{flat}) et une deuxième et dernière fois (orientation \textit{flat} $\rightarrow$ orientation-conformation \textit{flat$+$}).

La figure \ref{FigRSA2}.B présente les résultats des simulations sous la forme des probabilités conditionnelles d'addition (voir à ce sujet la remarque quant à leur utilisation à la section précédente) et des quantités accumulées des différentes orientations-conformations d'IgG (\textit{end-on}, \textit{flat} et \textit{flat$+$}) en fonction de la quantité totale $\Theta$. Comme pour le cas de l'addition par simple compétition, on retrouve l'ordre d'exclusion de la série (\ref{SerieExclusion}):
\begin{equation}
P(A^*|\boldsymbol{\omega}_{\mathrm{end\text{-}on}})\leqslant\dots\leqslant
P(A^*|\boldsymbol{\omega}_{\mathrm{flat}})\leqslant\dots\leqslant
P(A^*|\boldsymbol{\omega}_{\mathrm{flat+}}),
\end{equation}
c'est-à-dire que les IgG orientées \textit{end-on} seront les dernières à être exclues et donc à pouvoir s'accumuler sur la surface et les IgG \textit{flat$+$} le seront, mais en premier. Bien que l'exclusion porte un réel sens pour les IgG \textit{end-on} qui, seules, ont la possibilité de s'additionner sur la surface, il faudrait plutôt parler, pour les deux autres orientations-conformations n'advenant que par relaxation, d'une apparition inhibée par effet de volume exclu.

\begin{figure}[hp]\centering
\includegraphics*[width=0.75\textwidth]{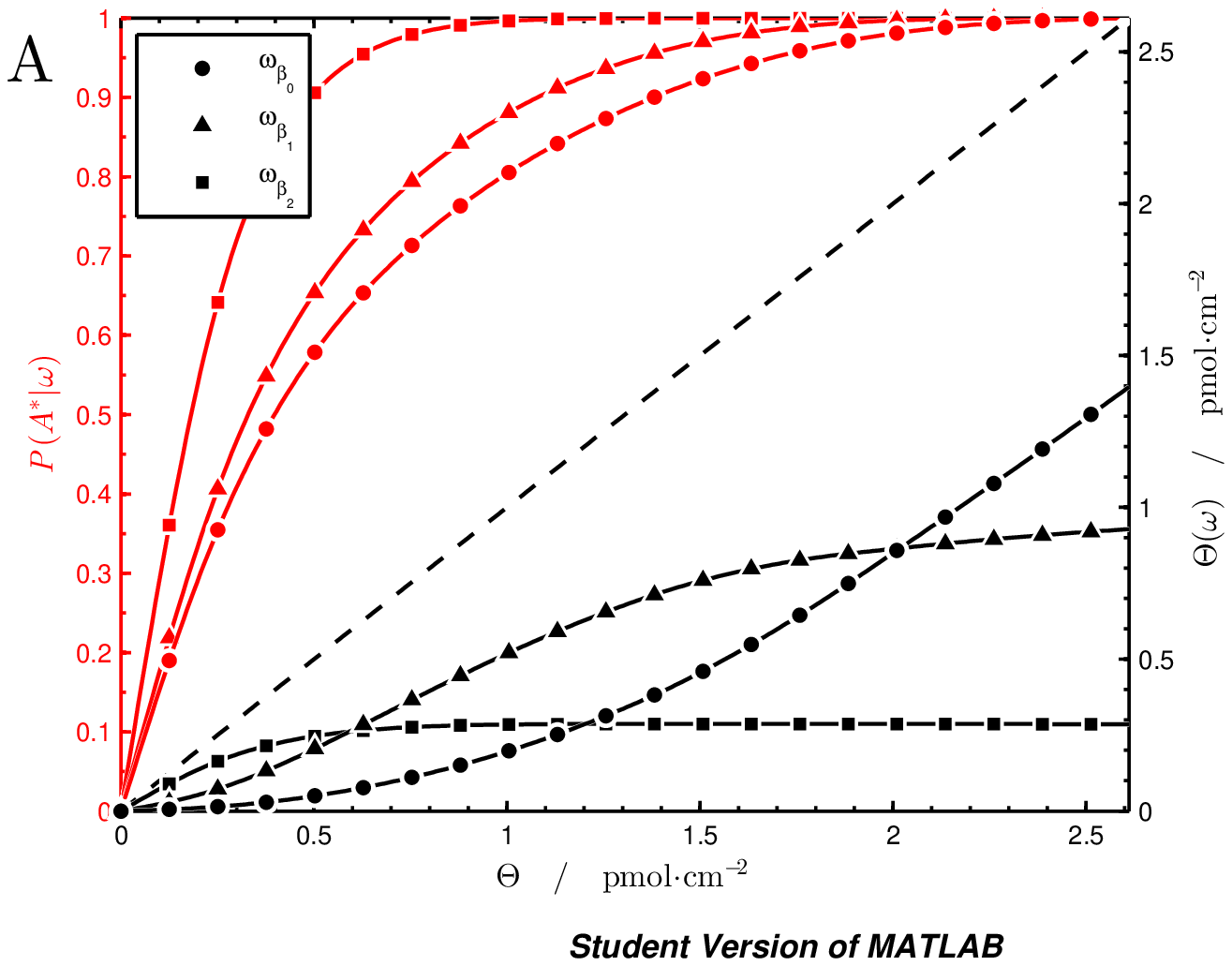}
\includegraphics*[width=0.75\textwidth]{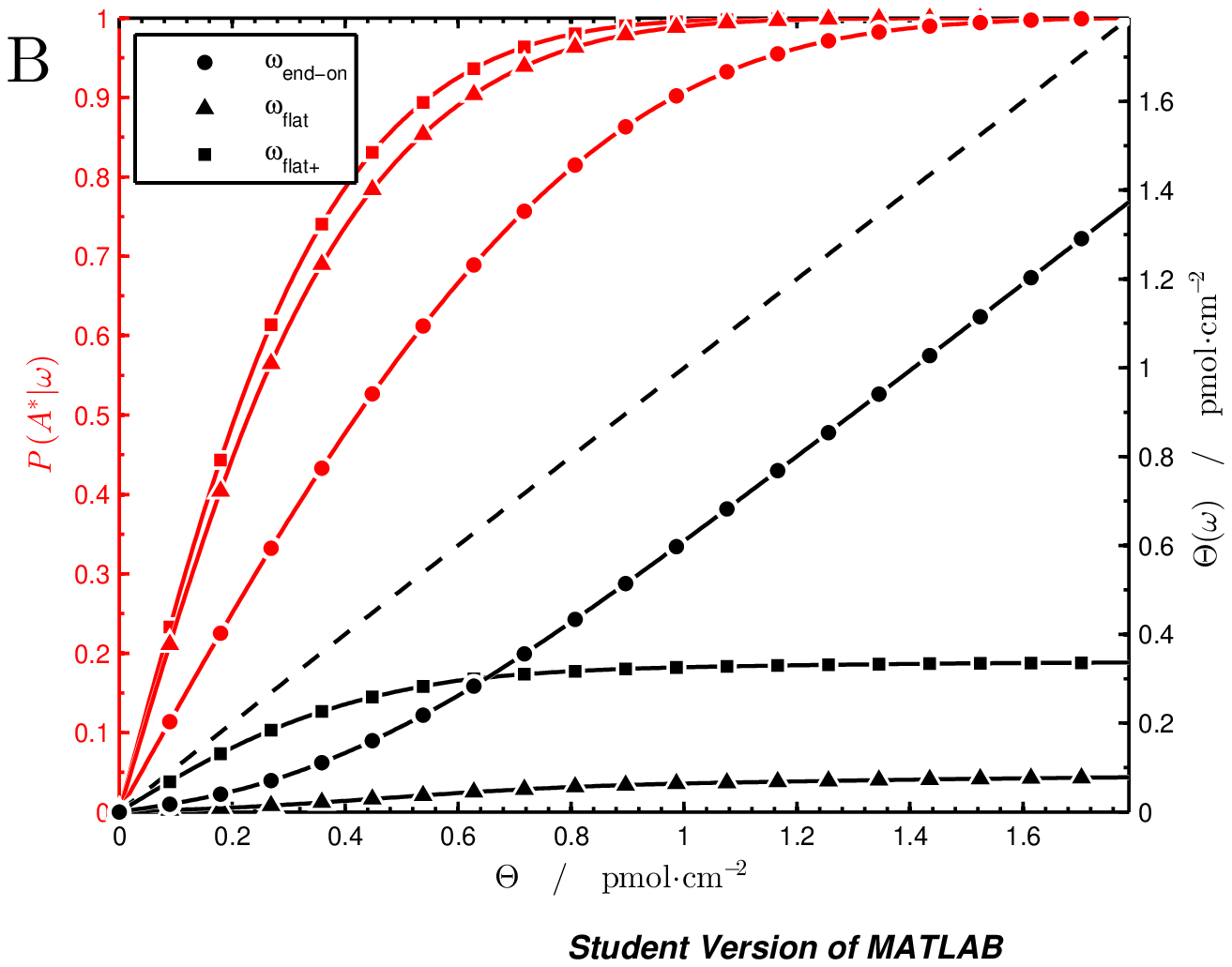}
\caption[$P(A^\ast)$ en fonction de $\Theta$ lors de remplissages RSA+R]{Pour une addition séquentielle aléatoire \textbf{A}: de boîtes de conserve ayant la capacité de se relaxer par étalement ($\beta_0\,(\mathrm{natif})\,\rightarrow \beta_1\rightarrow \beta_2$) et \textbf{B}: d'IgG pouvant subir une relaxation orientationnelle (\textit{end-on} $\rightarrow$ \textit{flat}) et configurationnelle (\textit{flat} $\rightarrow$ \textit{flat$+$}); probabilités conditionnelles (rouges) de rejet de ces conformations et quantités $\Theta(\boldsymbol{\omega})$ accumulées (pmol$\cdot$cm$^{-2}$) en fonction de la quantité totale $\Theta$ (pmol$\cdot$cm$^{-2}$). Les lignes discontinues représentent les quantités en ordonnées tout état confondu.}\label{FigRSA4}
\end{figure}

\begin{figure}[hp]\centering
\includegraphics*[width=0.75\textwidth]{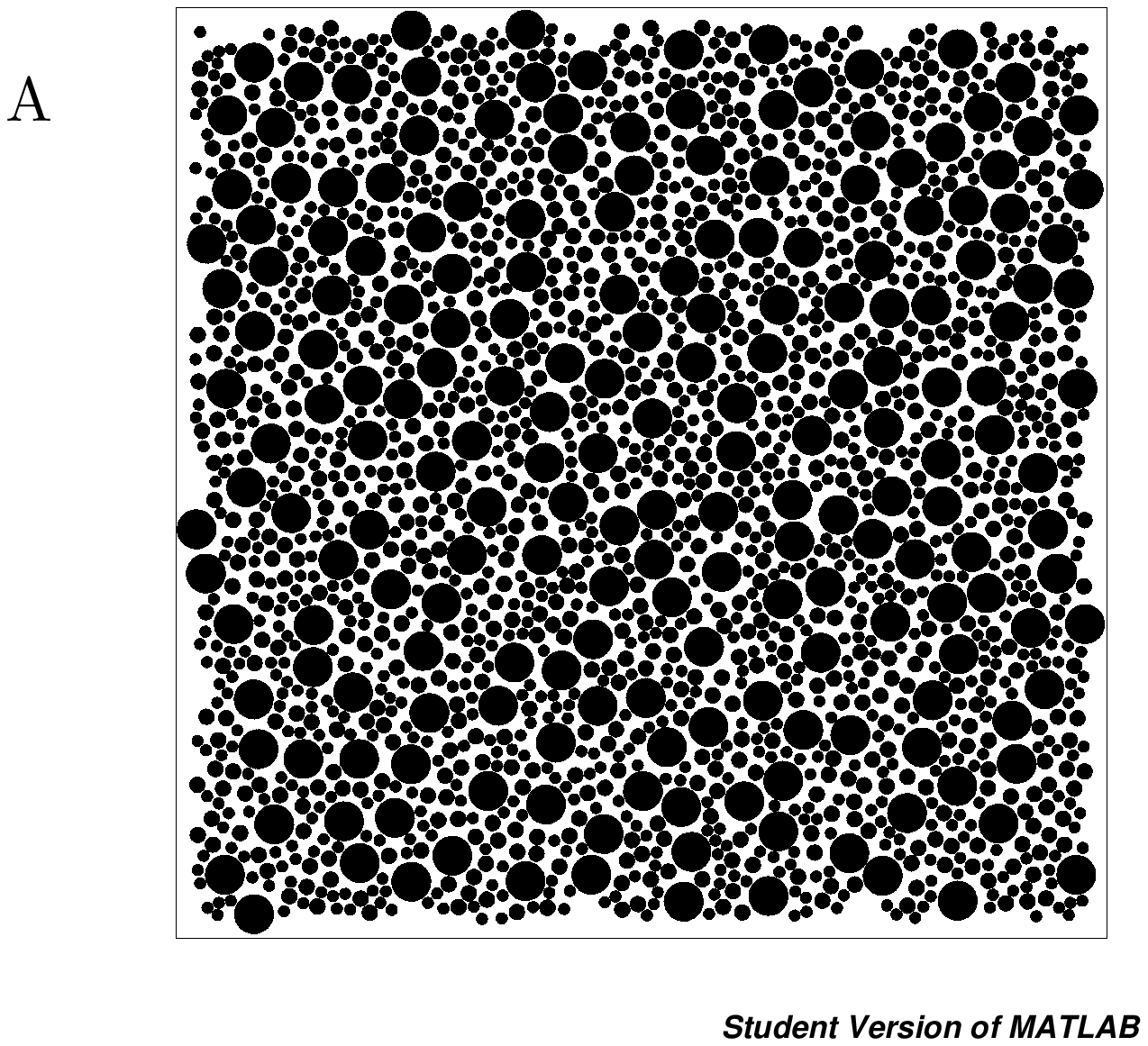}
\includegraphics*[width=0.75\textwidth]{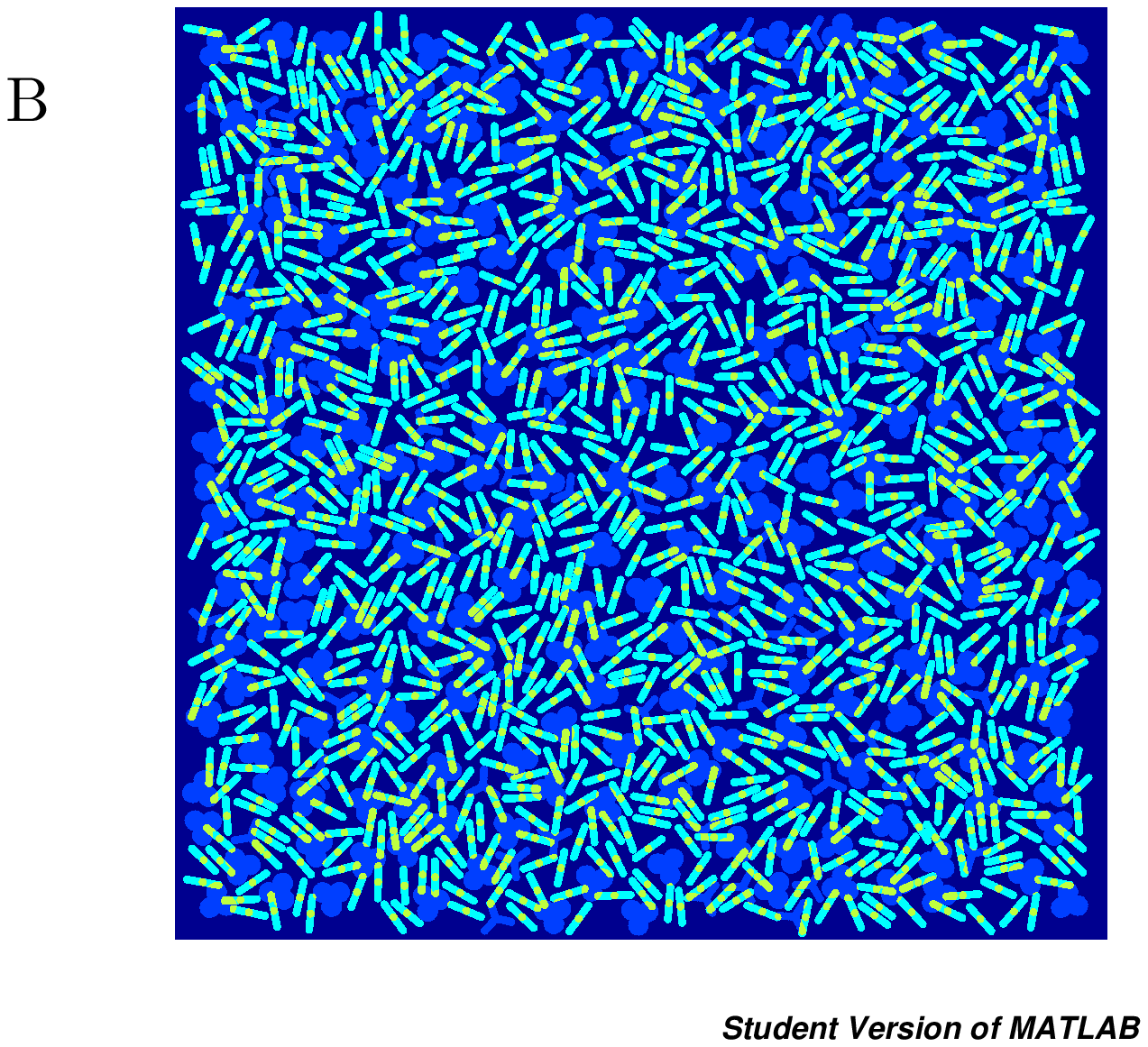}
\caption[Images de monocouches construites par remplissages RSA+R]{Pour une addition séquentielle aléatoire \textbf{A}: de boîtes de conserve ayant la capacité de se relaxer par étalement ($\beta=0\,(\mathrm{natif})\,\rightarrow \beta_1\rightarrow \beta_2$) et \textbf{B}: d'IgG pouvant subir une relaxation orientationnelle (\textit{end-on} $\rightarrow$ \textit{flat}) et configurationnelle (\textit{flat} $\rightarrow$ \textit{flat}+); images des surfaces obtenues à saturation. La légende des couleurs est identique à celle de la figure \ref{FigRSAim1}.}\label{FigRSAim4}
\end{figure}

Les orientations-conformations d'IgG ayant les plus fortes empreintes caractéristiques, \textit{i.e.} \textit{flat} et \textit{flat$+$}, sont les premières dont l'accumulation est inhibée alors que les IgG \textit{end-on}, peu gourmandes en espace, finissent par saturer les espaces encore libres de la surface. Tout comme pour les boîtes de conserve, la relaxation compensera l'effet de volume exclu en favorisant, tant que faire ce peu, les IgG de plus fortes empreintes caractéristiques. Les courbes $\Theta(\boldsymbol{\omega})$ de la figure \ref{FigRSA1} montrent en effet que la monocouche commence d'abord par être peuplée d'IgG \textit{flat$+$} et ce n'est seulement que lorsque la relaxation sera significativement inhibée par l'accroissement du volume exclu que viendront s'y accumuler, par ordre chronologique, les IgG \textit{flat} et finalement \textit{end-on}.

La confrontation des critères thermodynamique (maximisation de l'empreinte) et cinétique (minimisation de l'empreinte par contrainte de volume exclu ou exclusion par la taille) obligera les premières IgG arrivant à la surface à se relaxer au maximum de leurs possibilités ($\rightarrow$ \textit{flat$+$}) mais elles seront aussi les premières dont l'apparition se verra inhibée, la courbe de $\Theta(\boldsymbol{\omega}_\mathrm{flat+})$ en fonction de $\Theta$ de la figure \ref{FigRSA4}.B en témoignant. Les IgG \textit{end-on} finissent par saturer la surface, celles-ci possédant la plus faible empreinte caractéristique et on obtient finalement l'ordre suivant dans les quantités accumulées à saturation
\begin{equation}
\Theta_\infty(\boldsymbol{\omega}_\mathrm{flat+})
\leqslant\dots\leqslant
\Theta_\infty(\boldsymbol{\omega}_\mathrm{flat})
\leqslant\dots\leqslant
\Theta_\infty(\boldsymbol{\omega}_\mathrm{end\text{-}on})
\end{equation}
montrant que les IgG \textit{end-on}, bien qu'elles soient les dernières à augmenter leur présence dans la monocouche, finissent par être majoritaires dans la monocouche saturée.

Si le raisonnement tenant compte de la relaxation de la conformation qui a été fait ci-dessus pour les boîtes de conserve était extrapolé aux IgG, on devrait aussi s'attendre à obtenir des quantités $\Theta_\infty$ et $\Theta_\infty(\boldsymbol{\omega}_\mathrm{end\text{-}on})$ nettement plus faibles. En effet, la discrétisation nécessaire à la simulation, de même que les hypothèses faites sur la forme des différents états successifs que peuvent prendre les IgG accumulées sur la surface ne permettent pas de rendre suffisamment bien compte de la réalité et les quantités $\Theta_\infty$ et $\Theta_\infty(\boldsymbol{\omega}_\mathrm{end\text{-}on})$ sont trop proches de celles estimées pour le système d'IgG ne tenant pas compte de la relaxation conformationnelle. On peut toutefois supposer que, si les simulations tenaient compte d'un plus grand nombre d'états relaxés intermédiaires afin de s'approcher d'un processus continu, comme cela se passe dans la réalité expérimentale, elles devraient mener à ce que $\Theta_\infty$ et $\Theta_\infty(\boldsymbol{\omega}_\mathrm{end\text{-}on})$ soient nettement inférieures comme c'est le cas pour la boîte de conserve pouvant relaxer sa configuration. \`{A} la limite, il semble tout à fait réaliste de penser que les états de plus petites empreintes caractéristiques (les boîtes de conserve dans l'état natif et les IgG \textit{end-on}) finissent par être totalement exclus de la possibilité de se maintenir dans la monocouche.

Afin de mieux cerner ce que peut apporter ce genre de modélisation tenant compte de la relaxation à la compréhension du mécanisme par lequel se construit les monocouches d'IgG, un modèle simplifié sera maintenant considéré. Ce modèle sera le même que celui de la figure \ref{FigRSA1} mais il en diffèrera par l'absence d'IgG \textit{flat$+$}. Seules les orientations \textit{end-on} et \textit{flat} seront donc envisagées.

La figure \ref{FigMechanismeAccuIgG} présente à cet effet une série de situations dans lesquelles pourraient se trouver les IgG à l'approche de la surface selon leur orientation et le taux de recouvrement de la surface. Ensuite, pour le même modèle d'addition-relaxation, le graphique de la figure \ref{FigRSA2} représente les vitesses d'accumulation (dérivées numériques des $\Theta(\boldsymbol{\omega})$ en fonction du temps adimensionnel $t^\ast$) des IgG selon deux de leurs orientations possibles. La courbe en cloche y représente l'évolution de la vitesse d'accumulation des IgG selon l'orientation \textit{end-on}.

\begin{figure}[t]\centering
\includegraphics*[width=0.9\textwidth]{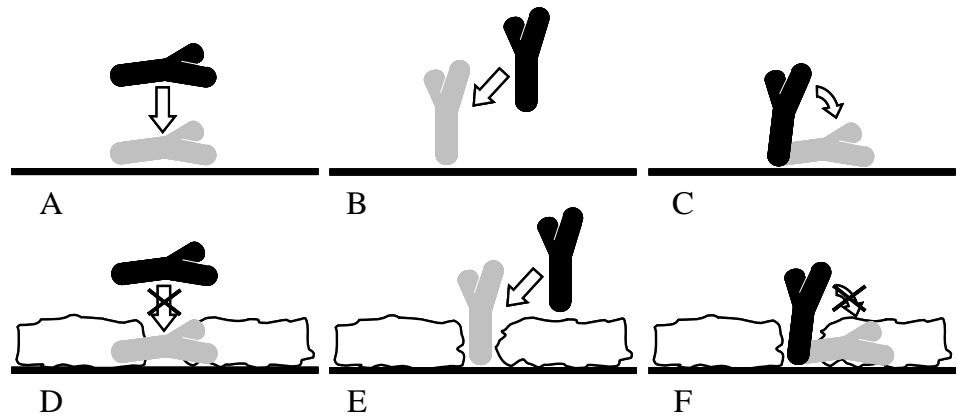}
\caption[Mécanisme de formation des monocouches d'IgG]{Mécanisme hypothétique de formation des monocouches d'IgG. \textbf{A}: sur une surface vide, addition d'une IgG selon une orientation \textit{flat}; \textbf{B}: sur une surface vide, addition d'une IgG selon une orientation \textit{end-on};
\textbf{C}: sur une surface vide, relaxation d'une IgG d'une orientation \textit{end-on} vers une orientation \textit{flat}; \textbf{D}: sur une surface déjà couverte, addition impossible d'une IgG dans une orientation \textit{flat}; \textbf{E}: sur une surface déjà couverte, addition possible d'une IgG dans une orientation \textit{end-on}; \textbf{F}:  sur une surface déjà remplie, relaxation impossible d'une IgG d'une orientation \textit{end-on} vers une orientation \textit{flat}.}\label{FigMechanismeAccuIgG}
\end{figure}

En observant les courbes des vitesses d'accumulation selon les différentes orientations, on voit que la vitesse d'accumulation des IgG dans une orientation \textit{flat} est maximale au début, et ce, jusqu'à une accumulation d'environ $0,7$ pmol$\cdot$cm$^{-2}$, moment où elle sera dépassée par la vitesse d'accumulation des IgG orientées \textit{end-on}. Au début du processus d'accumulation, l'IgG s'additionnera selon une orientation \textit{end-on} (figure \ref{FigMechanismeAccuIgG}.A) mais lorsqu'elle atteindra la surface, elle aura tendance à se coucher afin d'optimiser ses interactions avec la surface et de s'approcher de l'empreinte caractéristique maximale qu'elle peut atteindre (figure \ref{FigMechanismeAccuIgG}.C). Au début du processus d'accumulation, l'encombrement faible permettra à une IgG de se relaxer et elle finira systématiquement par s'insérer selon une orientation \textit{flat}, le volume exclu $P(A^*|\boldsymbol{\omega}_{\mathrm{flat}})$ étant suffisamment faible pour le permettre. Dans le cadre d'un modèle qui inclurait l'addition des IgG directement dans une orientation \textit{flat}, il serait certain que celles-ci pourront aussi apparaître par simples additions (figure \ref{FigMechanismeAccuIgG}.A).

\begin{figure}[t]\centering
\includegraphics*[width=0.75\textwidth]{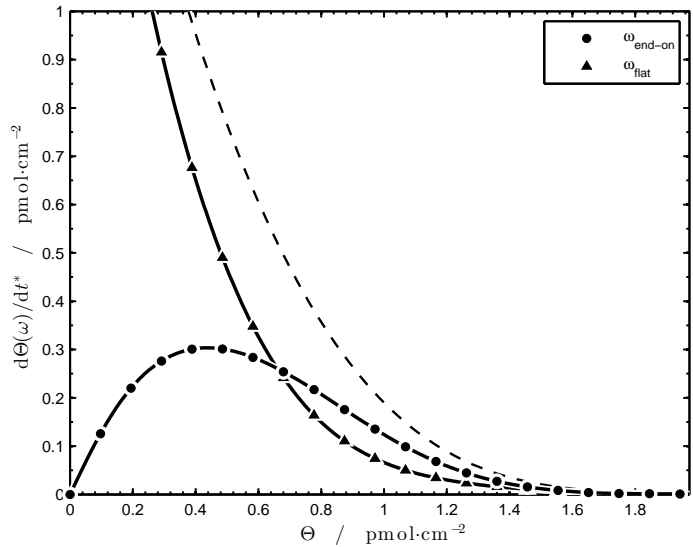}
\caption[\'{E}volution des vitesses d'accumulation des IgG selon leur orientation]{Pour l'addition-relaxation séquentielle aléatoire d'IgG orientées \textit{end-on} et pouvant relaxer vers une orientation \textit{flat}, évolution des vitesses d'accumulation des IgG selon leur orientation. La ligne discontinue donne la vitesse d'accumulation toute orientation confondue.}\label{FigRSA2}
\end{figure}

Tandis que le processus de remplissage de la surface avance, le volume exclu croît en conséquence et la probabilité $P(A^\ast|\boldsymbol{\omega}_{\mathrm{flat}})$ deviendra non nulle, empêchant les IgG \textit{end-on} de relaxer afin d'adopter une orientation \textit{flat} (figure \ref{FigMechanismeAccuIgG}.F). Les IgG qui se seront additionnées dans cette orientation \textit{end-on} (figure \ref{FigMechanismeAccuIgG}.E) resteront donc \textit{end-on}, puisque ne pouvant plus relaxer. Si des IgG orientées \textit{flat} pouvaient s'additionner, elles seraient exclues (figure \ref{FigMechanismeAccuIgG}.E). Ce phénomène, dû à l'inhibition de la relaxation, se traduit sur le graphique de la figure \ref{FigRSA2} par une accélération (augmentation de la vitesse) de l'accumulation des IgG \textit{end-on}. Après cette période transitoire d'accélération, l'accumulation des IgG orientées \textit{end-on}, tout comme celles orientées \textit{flat}, décélérera suite à l'inévitable augmentation du volume exclu et du phénomène d'exclusion par la taille. L'addition des IgG en orientation \textit{end-on} ne devient donc dominante que parce que la surface ne peut plus accueillir de \textit{flat}. L'accumulation des IgG \textit{end-on} se produit donc lorsque la surface présente un certain taux d'occupation. Le fait que l'on n'accumule que des IgG dans une orientation \textit{end-on} en fin de processus est la conséquence de deux faits:
\begin{enumerate}
\item les IgG s'y additionnant selon une orientation caractérisée par une empreinte caractéristique faible (\textit{i.e.} \textit{end-on}) n'ont plus la possibilité de relaxer afin d'atteindre une orientation d'empreinte plus élevée (\textit{i.e.} \textit{flat}); elles restent donc, faute de mieux, dans leur orientation initiale (voir figure \ref{FigMechanismeAccuIgG}.F), c'est-à-dire \textit{end-on} alors que,
\item si des IgG \textit{flat} peuvent se présenter à la surface, ce dont on n'a pas tenu compte ici, elles seront exclues car leur empreinte caractéristique est trop élevée par rapport au volume exclu (voir figure \ref{FigMechanismeAccuIgG}.D).
\end{enumerate}
On finira par obtenir une quantité d'IgG \textit{end-on} non nulle, voire même majoritaire.

C'est donc un effet de volume exclu ou d'exclusion stérique qui permet, encore une fois, d'expliquer que des IgG finissent par adopter une orientation \textit{end-on} au cours de leur processus d'accumulation sur une surface hydrophobe. Cet effet purement cinétique vient s'opposer à la tendance qu'imprime la thermodynamique aux IgG de se coucher sur la surface. En suivant le modèle utilisé, il semblerait que cette orientation devienne dominante et que l'on finisse par obtenir $3,25$ IgG \textit{end-on} pour une \textit{flat} à la saturation. Dans le cas où la relaxation n'est pas permise, on obtient un rapport de 5,4 \textit{end-on} pour une \textit{flat}. Ces résultats sont confortés par le fait que l'idée qui les sous-tend a déjà fait l'objet d'une brève discussion par A. Schmitt \textit{et al.} \citep{schmitt1983} en 1983 dans le cadre de l'adsorption du fibrinogène. De même, elle a encore été plus récemment (2004) utilisée par M. Bremer \textit{et al.} \citep{bremer2004} afin d'expliquer les variations de quantités adsorbées lors de l'adsorption des IgG sur de la silice (surface chargée). On peut aussi citer P. Schaaf \textit{et al.} \citep{schaaf1987}. Il est par ailleurs intéressant de remarquer que la monocouche d'IgG construite avec possibilité de relaxation est plus densément peuplée, laissant penser que la relaxation est un facteur favorisant un meilleur arrangement des particules à la surface. Dans ce dernier cas, on obtient plus de 5 \textit{end-on} pour une IgG \textit{flat}.

Dans la réalité, les trois <<~bras~>> d'une IgG ne sont pas identiques de telle sorte que l'orientation \textit{end-on} ici considérée est un amalgame de trois situations différentes. En effet, si un des bras est le fragment constant de l'IgG (Fc) et que l'on souhaite estimer la quantité de ces IgG \textit{end-on} dont le bras en contact avec la surface soit le Fc, il faut diviser la quantité $\Theta(\boldsymbol{\omega}_\mathrm{end\text{-}on})$ par trois et on aura donc, dans le cas avec relaxation, non pas $\sim 3,25$ \textit{end-on} pour une \textit{flat}, mais un minimum (l'IgG ne peut changer d'orientation une fois additionnée à la monocouche) d'une \textit{end-on} pour une \textit{flat}.

Le modèle RSA dans lequel est implémenté la possibilité de relaxer instantanément montre donc que, dans le cas des IgG, les orientations de type \textit{end-on} seront nettement favorisées. Elles seront même majoritaires dans la monocouche saturée, et ce, même si les IgG doivent relaxer dès qu'elles touchent la surface. La présence des IgG \textit{end-on} dans la monocouche saturée semble un effet dû à la croissance du volume exclu comme montré dans le mécanisme de construction proposé et schématisé sur la figure \ref{FigMechanismeAccuIgG}.

\section{Relaxation des protéines à vitesse finie}

La relaxation instantanée des protéines comme elle a été envisagée ci-dessus semble assez extrême car elle ne tient pas compte des propriétés intrinsèques de la protéine. On pourrait en effet imaginer que les protéines puissent se relaxer à une certaine vitesse, vitesse qui leur serait caractéristique et qui serait liée aux concepts de protéines \emph{dures} et \emph{molles} introduits par W. Norde. Initialement, la dureté d'une protéine est sa sensibilité à la dénaturation. Une protéine molle aura une forte tendance à perdre sa structure afin d'optimiser ses interactions avec la surface tandis qu'une protéine dure manifestera une certaine résistance à ce changement. En termes d'énergie, l'énergie d'activation de ce processus de relaxation $E^\ddagger_R$ sera élevée (occurrences fréquentes) pour une protéine dure alors qu'elle sera plus faible pour une protéine molle (occurrences rares). Dans ce qui suit, la différence entre relaxation de l'orientation et de la conformation ($=$ dénaturation) sera négligée, c'est-à-dire que la relaxation sera considérée comme un processus continu de changement d'orientation et de conformation.

Afin de caractériser ce phénomène de relaxation, une analogie avec la cinétique \ref{EqCinétique2} peut être faite afin d'écrire
\begin{equation}\label{EqCinétiqueRelax1}
k_R=k_R^\circ\,P(R)
\end{equation}
dans laquelle $k_R$ serait la vitesse instantanée de la relaxation (mol$\cdot$m$^{-2}\cdot$s$^{-1}$), $k_R^\circ$ la vitesse initiale de la relaxation (mol$\cdot$m$^{-2}\cdot$s$^{-1}$) et $P(R)$ la probabilité qu'une telle relaxation se produise à un moment donné de la construction de la monocouche, c'est-à-dire pour une certaine quantité accumulée $\Theta$.

La vitesse de relaxation initiale $k_R^\circ$ apparaît comme une donnée invariable du problème. Elle devrait être interprétée comme une caractéristique de la protéine: sa tendance à relaxer indépendamment des contraintes imposées par son environnement. Il s'agirait en fait d'une mesure de la dureté d'une protéine: plus $k_R^\circ$ sera faible, plus la tendance intrinsèque de la protéine à relaxer sera faible et sa dureté (insensibilité à la relaxation) élevée. Inversement, plus $k_R^\circ$ sera élevé, plus sa tendance à relaxer sera élevée et elle pourra alors être qualifiée de molle.

La probabilité $P(R)$ d'observer une relaxation devrait être liée à la probabilité d'addition $P(A)$. En effet, plus le volume exclu est élevé plus la probabilité qu'une relaxation se produise devrait être faible étant donné que celle-ci nécessite une augmentation de l'empreinte. Cette probabilité n'impliquant qu'une portion des protéines accumulées à la surface, \textit{i.e.} la part qui n'est pas encore complètement relaxée, elle devrait être inférieure à la probabilité d'addition. Son estimation pourrait passer par l'expression
\begin{equation}\label{EqProbCondRelax}
P(R)=\mathcal{X}(\boldsymbol{\omega}_1)P(R|\boldsymbol{\omega}_1)
+\mathcal{X}(\boldsymbol{\omega}_2)P(R|\boldsymbol{\omega}_2)+\dots
\end{equation}
qui est une combinaison linéaire entre les probabilités conditionnelles $P(R|\boldsymbol{\omega}_1)$, $P(R|\boldsymbol{\omega}_2)$, etc. donnant les probabilités que des protéines d'orientations $\boldsymbol{\omega}_1$, 
$\boldsymbol{\omega}_2$, etc. relaxent et les fractions $\mathcal{X}(\boldsymbol{\omega}_1)$, $\mathcal{X}(\boldsymbol{\omega}_2)$, etc. représentant les proportions de protéines accumulées dans chacune de ces orientations, ces dernières ayant été définies à la relation \ref{FracionsOri}.

Tout comme les vitesses d'addition instantanée $k_A$ et initiale $k_A^\circ$, les vitesses $k_R$ et $k_R^\circ$ ne semblent pas simples à évaluer \textit{a priori}. \`{A} tout le moins, il serait possible de tabler sur le fait que le phénomène d'addition puisse entrer en compétition avec celui de relaxation, ce qui impliquerait que $k_A$ et $k_R$ seraient du même ordre de grandeur. Afin de pallier à cette difficulté et d'adopter une approche demeurant générale, il est judicieux de normaliser la vitesse d'addition par la vitesse de relaxation. En partant des équations \ref{EqCinétique2} et \ref{EqCinétiqueRelax1}, il vient facilement que
\begin{equation}\label{EqCinétiqueRelax3}
\frac{k_A}{k_R}=\frac{k_A^\circ}{k_R^\circ}\,\frac{P(A)}{P(R)}
\end{equation}
où l'on obtient pas moins de trois ratios. En reformulant cette relation sous la forme
\begin{equation}\label{EqCinétiqueRelax5}
K=K^\circ\,\frac{P(A)}{P(R)}
\end{equation}
où il a été défini
\begin{equation}\label{EqCinétiqueRelax4}
K=\frac{k_A}{k_R}\quad\text{et}\quad K^\circ=\frac{k_A^\circ}{k_R^\circ},
\end{equation}
le ratio des vitesses instantanées $K$ et le ratio des vitesses initiales $K^\circ$, ce dernier ayant la propriété d'invariance au même chef que les vitesses $k_A^\circ$ et $k_R^\circ$ dont il est constitué.

Cette formulation est très intéressante car, bien que $k_A^\circ$ et $k_R^\circ$ ne soient pas estimables séparément, leur ratio $K^\circ$ l'est parfaitement. On notera par ailleurs que $P(A)$ est estimable \textit{via} la procédure d'insertion de Widom déjà rencontrée tandis que $P(R)$ le pourrait \textit{via} la combinaison linéaire \ref{EqProbCondRelax} dans laquelle chaque probabilité conditionnelle pourrait s'obtenir en testant les proportions des protéines accumulées pouvant relaxer et dans l'orientation $\boldsymbol{\omega}_1$, etc. pour les autres orientations. De plus, $K^\circ$ est aussi directement dépendant de la diffusion des particules vers la surface aussi bien qu'en injectant la relation \ref{fuyufqffqiytv} dans l'expression de $K^\circ$, il vienne
\begin{equation}\label{35fdvg5sd4g65s4}
K^\circ=-\frac{D}{k_R^\circ}\,\frac{\partial}{\partial z}C(z=0,t)
\end{equation}
où $K^\circ$ est directement proportionnel, à la constante $-D/k_R^\circ$ près, au gradient de concentration dans la suspension de dépôt et contre la surface ($z=0$).

Le ratio $K^\circ$ des vitesses initiales peut être interprété comme le nombre d'addition(s) par relaxation se produisant au début du processus d'accumulation des protéines sur la surface. Cette notion n'étant pas très intuitive, on utilisera plutôt l'inverse du ratio des vitesses initiales:
\begin{equation}\label{EqCinétiqueRelax6}
(K^\circ)^{-1}=\frac{k_R^\circ}{k_A^\circ}
\end{equation}
représentant le nombre de relaxation(s) pour chaque addition à l'instant initial. Afin d'aller plus loin, il faut se rappeler que $k_R^\circ$, en tant que caractéristique inhérente à la protéine, représente le nombre de relaxation(s) qui sera tenté par la protéine elle-même à chaque instant de sa présence à la surface. De même, $k_A^\circ$ est le nombre de protéines qui arriveront à la surface durant se même laps de temps. Leur rapport $(K^\circ)^{-1}$ s'interprète alors comme \emph{le nombre de tentatives de relaxations entre chaque arrivée de protéine à la surface}, chaque arrivée n'étant pas nécessairement suivie par une addition.

\section[Le modèle RSA+R]{Le modèle des additions et relaxations\\séquentielles aléatoires}

Porteur d'une telle signification, le ratio $(K^\circ)^{-1}$ permet de moduler de manière extrêmement précise l'importance du phénomène de relaxation lors d'un processus d'additions séquentielles aléatoires. Pour y procéder, il suffira qu'entre chaque tentative d'addition il soit procédé à $(K^\circ)^{-1}$ tentative(s) de relaxer une protéine de la monocouche choisie aléatoirement. Cette quantification très précise permet d'étendre le modèle des additions séquentielles aléatoires (modèle RSA) à un modèle des additions \emph{et} relaxations séquentielles aléatoires (modèle RSA+R) dont les trois premières hypothèses sont celles du RSA classique et la quatrième, celle propre au modèle RSA+R:
\begin{enumerate}
\item Les objets sont insérés séquentiellement en une position choisie aléatoirement dans le volume considéré;
\item Une fois inséré, l'objet est gelé sur sa position, il ne peut ni se déplacer dans le volume ni s'en échapper;
\item Deux objets ne peuvent se superposer;
\item Entre chaque tentative d'insertion, sur des objets choisis aléatoirement parmi ceux déjà insérés, réaliser séquentiellement $(K^\circ)^{-1}$ tentatives de relaxation.
\end{enumerate}
Il suffit alors à l'expérimentateur de fixer la valeur de $K^\circ$ avant toute simulation, sachant que celui-ci est proportionnel au flux de protéines vers la surface durant la construction de la monocouche (fonction croissante) et à la dureté de la protéine (fonction décroissante).

En utilisant le modèle des boîtes de conserve relaxantes décrit à la figure \ref{FigRSARelax}, il devient alors possible d'obtenir les courbes de $P(A)$ en fonction des quantités accumulées $\Theta$ pour différentes valeurs de $K^\circ$ (ou son logarithme $\ln K^\circ$). Les résultats de telles simulations sont montrés sur le graphique de la figure \ref{FigRSAgsgfhdhdhh}.

\begin{figure}[t]\centering
\includegraphics*[width=0.75\textwidth]{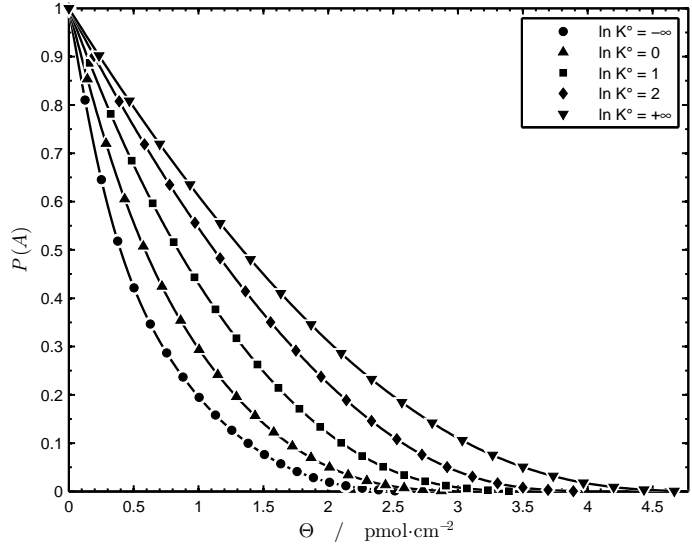}
\caption[$P(A)$ en fonction de $\Theta$ pour diverses valeurs de $\ln\,K^\circ$]{Pour diverses valeurs du logarithme du ratio des vitesses instantanées $\ln\,K^\circ$, évolutions des probabilités d'additions pour des accumulations selon le modèle RSA+R du système de boîtes de conserve relaxantes décrit à la figure \ref{FigRSARelax}.}\label{FigRSAgsgfhdhdhh}
\end{figure}

Avant d'aller plus loin, il est important de souligner que le graphique obtenu à la figure \ref{FigRSAgsgfhdhdhh} peut être lu selon deux voies d'interprétations. La première serait celle qui viserait à comparer des protéines de natures différentes, c'est-à-dire des protéines relaxant selon le même schéma mais à des vitesses différentes, alors que les conditions du transport vers la surface demeurent identiques pour chaque expérience (une expérience $=$ une courbe). La variation du ratio $K^\circ$ sera donc due à une variation de $k_R^\circ$ alors que $k_A^\circ$ demeurera parfaitement constant. La seconde voie consisterait à faire plusieurs expériences sur la même protéine mais selon des conditions de transport vers la surface différentes. Dans ce dernier cas, la variation du ratio $K^\circ$ proviendra d'un changement de $k_A^\circ$ alors que $k_R^\circ$ restera constant.

En tout état de cause, comme on peut le voir sur la figure \ref{FigRSAgsgfhdhdhh}, la variation du ratio des vitesses initiales affecte significativement la courbe de la probabilité d'addition $P(A)$ en fonction de la quantité accumulée $\Theta$. L'augmentation de $K^\circ$ a pour conséquence une diminution de la concavité des courbes mais aussi une augmentation de la quantité finale $\Theta_\infty$ atteinte lorsque la monocouche est saturée
\footnote{De manière rigoureuse, la quantité de protéines contenue dans une monocouche saturée est définie par la relation \begin{equation}\Theta_\infty=\lim_{P(A)\rightarrow0}\Theta\big(P(A)\big)\quad\mathrm{ou}\quad
\Theta_\infty=\lim_{P(A^\ast)\rightarrow1}\Theta\big(P(A^\ast)\big).\end{equation} Pour les besoins des estimations numériques, on considérera que cette quantité $\Theta_\infty$ est atteinte lorsque $P(A)\leqslant10^{-4}$ ou $P(A^\ast)\geqslant0,9999.$}.
Selon les deux points de vues discutés ci-dessus, cela signifie
\begin{itemize}
\item que l'augmentation de la vitesse initiale d'addition $k_A^\circ$ (augmentation de $K^\circ$) permettra d'atteindre des quantités accumulées plus élevées à saturation ou
\item qu'une augmentation de la vitesse initiale de relaxation $k_R^\circ$ (diminution de $K^\circ$) aura tendance à diminuer la quantité accumulée à saturation.
\end{itemize}

Au-delà de la quantité accumulée à saturation, il est intéressant de se poser la question du contenu de la monocouche lorsqu'elle sera saturée. La figure \ref{kekufyuldg} montre l'aspect de trois de ces monocouches construites pour $\ln K^\circ=-\infty$, $\ln K^\circ=0$ et $\ln K^\circ=+\infty$ et correspondant à l'état de saturation.

\begin{figure}[hp]\centering
\begin{tabular}{p{2cm}p{6cm}p{4cm}}
&\includegraphics*[width=5.8cm,trim=60 0 60 0]{RSA/fig/Pillbox_Petit_RelConf_3_--_LnK=-infini.eps}&\raisebox{4.5cm}{
\large{\begin{tabular}{l}
\Large{A:}\\
 \\
$\ln K^\circ=-\infty$\\
 \\
$\Theta_\infty=2,6$ pmol$\cdot$cm$^{-2}$
\end{tabular}}}\\
&\includegraphics*[width=5.8cm,trim=60 0 60 0]{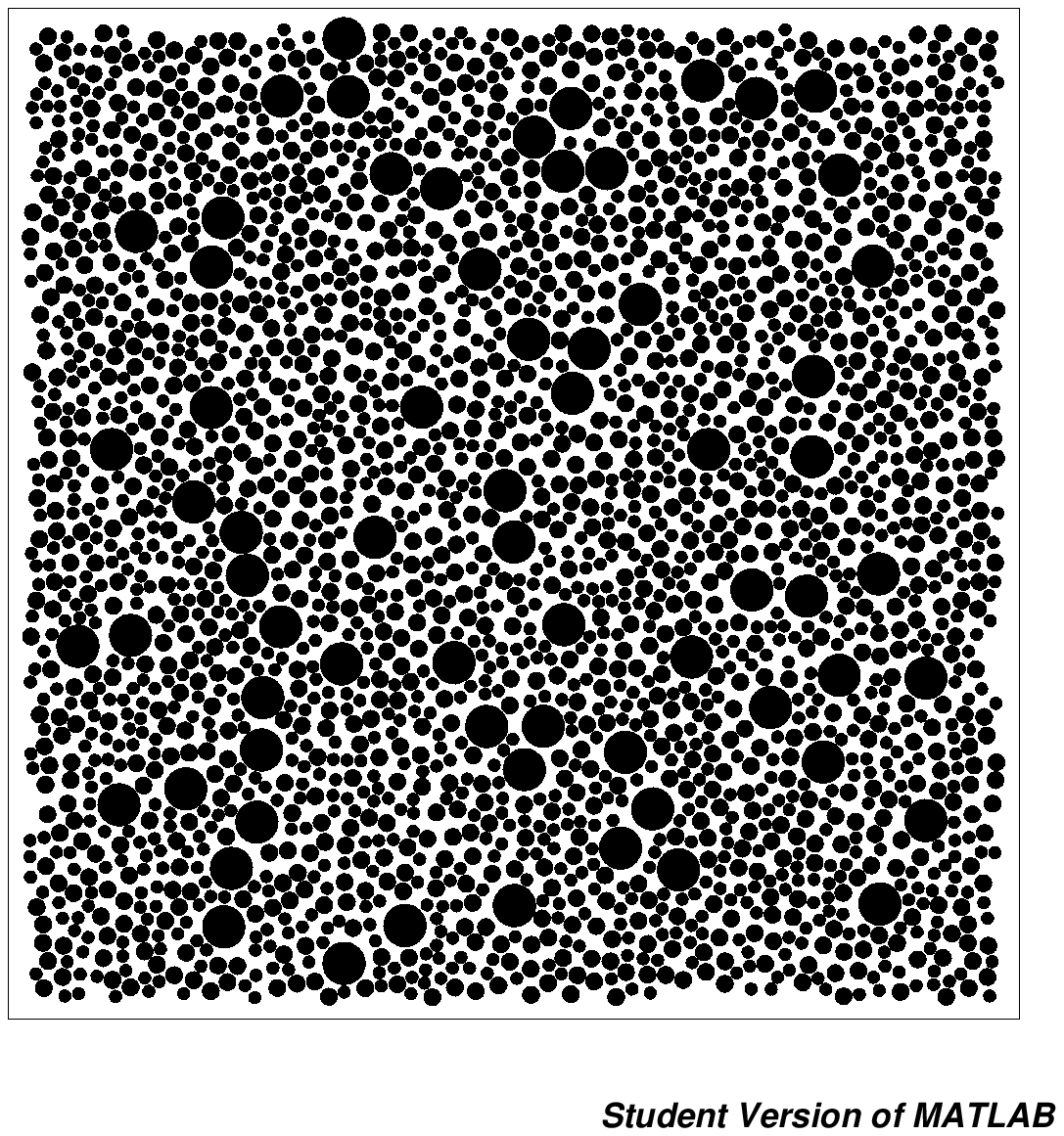}&\raisebox{4.5cm}{
\large{\begin{tabular}{l}
\Large{B:}\\
 \\
$\ln K^\circ=0$\\
 \\
$\Theta_\infty=3,45$ pmol$\cdot$cm$^{-2}$
\end{tabular}}}\\
&\includegraphics*[width=5.8cm,trim=60 0 60 0]{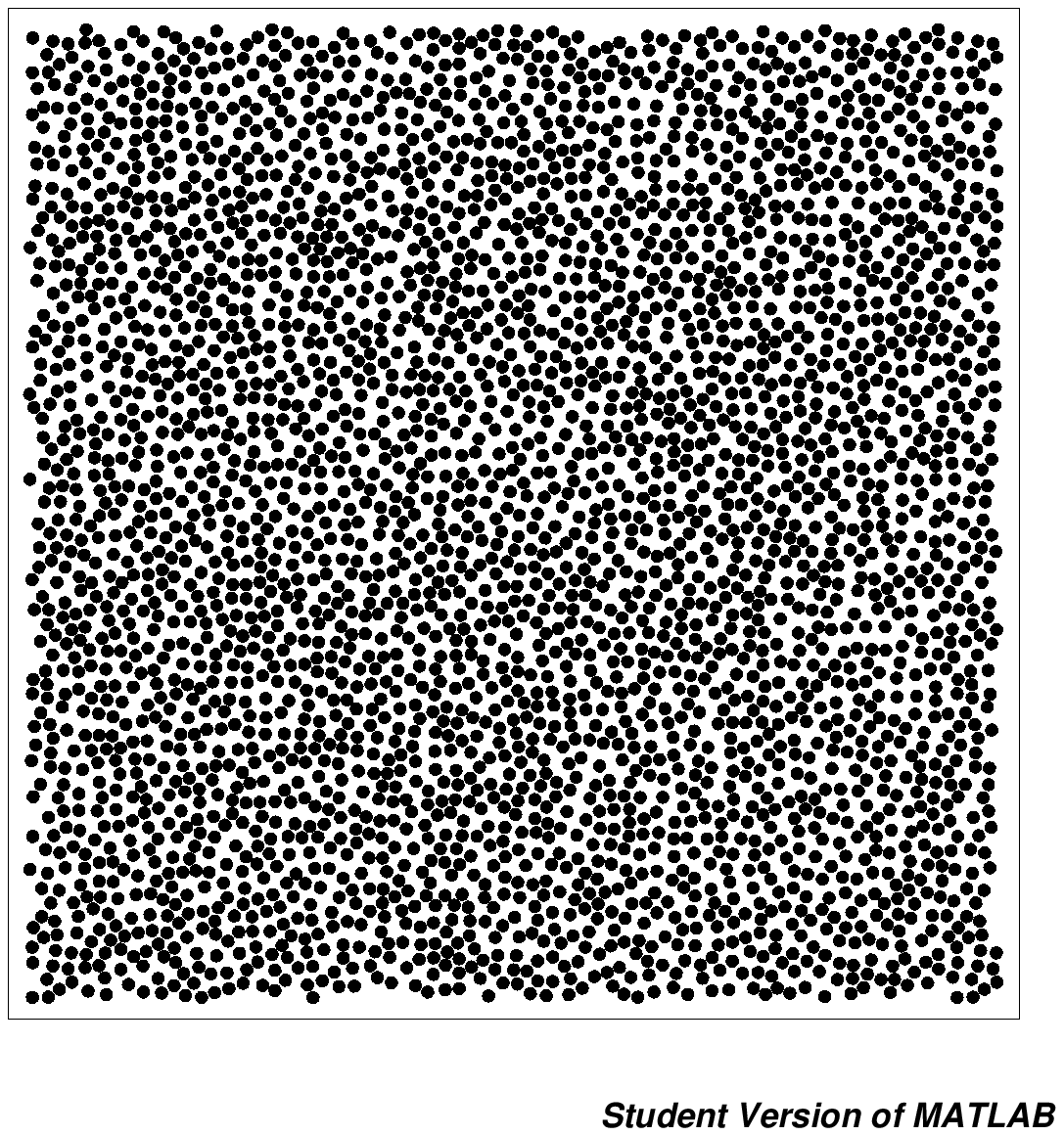}&\raisebox{4.5cm}{
\large{\begin{tabular}{l}
\Large{C:}\\
 \\
$\ln K^\circ=+\infty$\\
 \\
$\Theta_\infty=4,67$ pmol$\cdot$cm$^{-2}$
\end{tabular}}}\\
\end{tabular}
\caption[Monocouches saturées par simulations RSA+R]{Pour diverses valeurs de $\ln K^\circ$, aspects des monocouches saturées par des boîtes de conserve (voir figure \ref{FigRSARelax}) obtenues par simulations RSA+R sur des surfaces de 0,333 \textgreek{m}m de côté. \textbf{A}: $\ln K^\circ=-\infty$; \textbf{B}: $\ln K^\circ=0$ et \textbf{C}: $\ln K^\circ=+\infty$.}\label{kekufyuldg}
\end{figure}

Les trois monocouches saturées représentées sur la figure \ref{FigRSARelax} montrent clairement qu'une diminution de la valeur du ratio $K^\circ$ est la cause d'une augmentation de la présence de protéines relaxées dans la monocouche saturée. Inversement, une augmentation de ce ratio est la cause d'une augmentation de la présence de protéines non-relaxées (natives) dans la monocouche saturée. En suivant le premier point de vue développé ci-dessus (variation de la vitesse de relaxation $k_R^\circ$), l'augmentation de $k_R^\circ$ favorisera la présence de protéines relaxées dans la monocouche saturée. L'adoption du second point de vue (variation de la vitesse d'addition $k_A^\circ$) permet aussi de dire que l'augmentation de la vitesse d'addition $k_A^\circ$ viendra amoindrir la présence de protéines relaxées dans la monocouche saturée et inversement y accroître la présence de protéines natives.

La figure \ref{ludwyfvgludwygvlwy} complète l'information délivrée par les figures \ref{kekufyuldg}.A à C en montrant l'évolution de l'empreinte $\sigma$ des protéines de la monocouche en fonction de la quantité accumulée $\Theta$, et ce, pour diverses valeurs du paramètre $\ln K^\circ$. Lorsqu'une monocouche est construite pour $\ln K^\circ=-\infty$ (flux vers la surface infiniment lent ou protéine infiniment molle), l'empreinte initiale ($\lim_{\Theta\rightarrow0}\sigma$) sera maximale et correspondra à l'empreinte caractéristique de la protéine totalement relaxée. Cette empreinte diminuera ensuite sous l'effet cinétique de l'exclusion par la taille dû à l'encombrement progressif de la surface qui empêchera les protéines de relaxer entièrement.  L'empreinte finale étant élevée (les protéines prennent plus de place), il semble dès lors logique de n'obtenir qu'une quantité accumulée relativement faible. Pour les valeurs intermédiaires du paramètre $\ln K^\circ$, les empreintes initiales et finales sont plus faibles et diminuent à mesure que le paramètre $\ln K^\circ$ augmente (augmentation du flux de protéines vers la surface et/ou augmentation de la dureté des protéines). Lorsque le paramètre $\ln K^\circ$ tend vers $+\infty$, c'est-à-dire que le flux devient infiniment grand et/ou que les protéines sont infiniment dures (indéformables), seules des protéines natives pourront s'accumuler sur la surface de telle sorte que l'empreinte demeure constante et égale à l'empreinte caractéristique des protéines natives. L'empreinte des protéines accumulées étant la plus faible (les protéines prennent moins de place), on atteindra naturellement la plus grande quantité accumulée à saturation.

\begin{figure}[t!]\centering
\includegraphics*[width=0.75\textwidth]{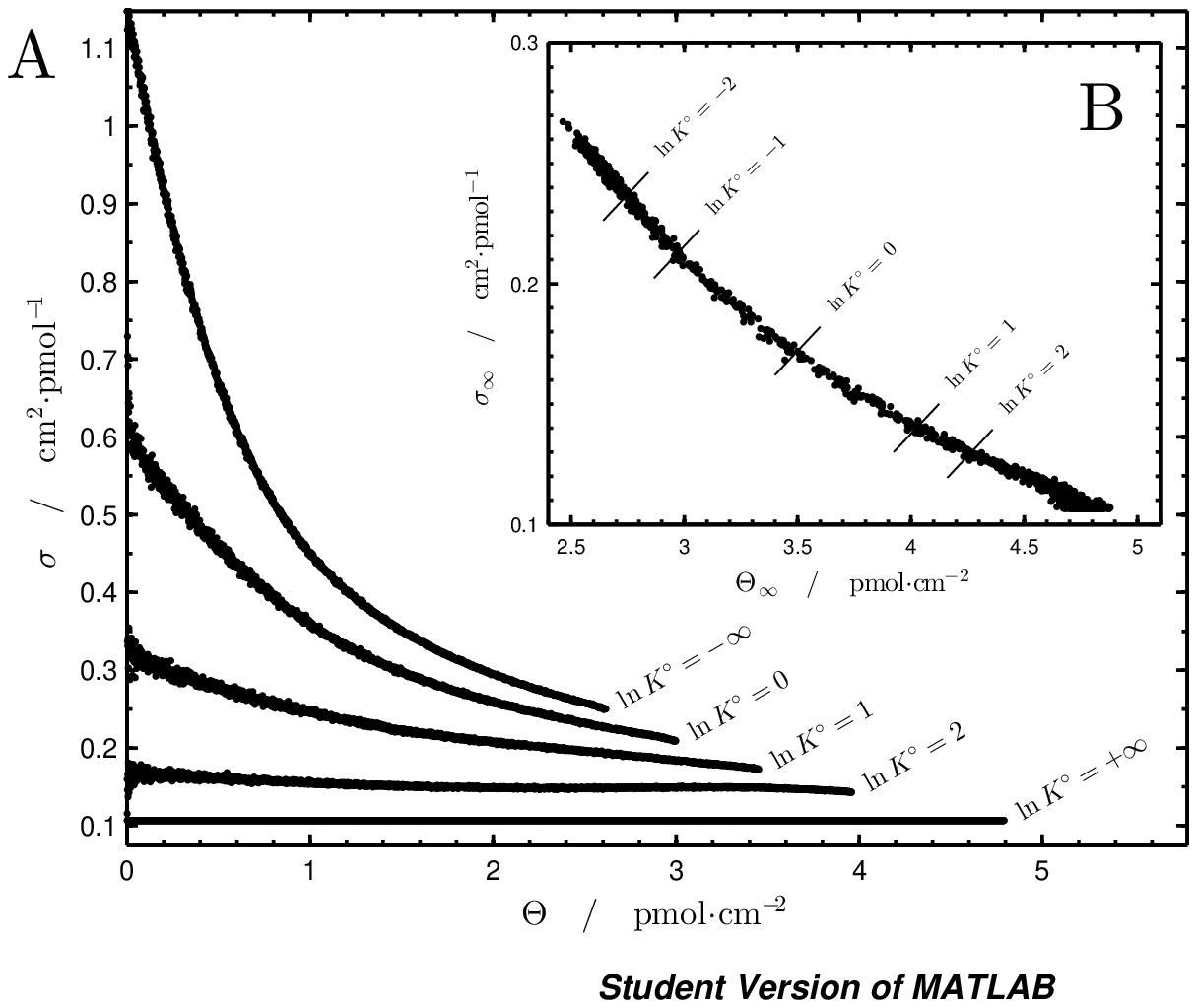}
\caption[\'{E}volution de $\sigma$ en fonction de la quantité accumulée $\Theta$]{\textbf{A}: évolution de $\sigma$ l'empreinte des protéines en fonction de la quantité accumulée $\Theta$ lors de la construction de monocouches selon diverses valeurs de $\ln K^\circ$. \textbf{B}: en fonction de $\Theta_\infty$ la quantité finale de protéines accumulées, l'empreinte $\sigma_\infty$ des protéines dans des monocouches saturées obtenues en balayant tout le domaine de variation de $\ln K^\circ$ (de $-\infty$ à $+\infty$)}\label{ludwyfvgludwygvlwy}
\end{figure}

Afin d'approfondir la discussion, le graphique B de la figure \ref{ludwyfvgludwygvlwy} montre la corrélation existant entre $\sigma_\infty$, l'empreinte des protéines accumulées dans une monocouche saturée\footnote{\`{A} l'instar de $\Theta_\infty$ (\textit{cf.} note précédente), l'empreinte $\sigma_\infty$ est définie par la relation
\begin{equation}
\sigma_\infty=\lim_{P(A)\rightarrow0}\sigma\big(P(A)\big)\quad\mathrm{ou}\quad
\sigma_\infty=\lim_{P(A^\ast)\rightarrow1}\sigma\big(P(A^\ast)\big).
\end{equation}
Ces quantités seront atteintes dans les calculs numériques lorsque $P(A)\leqslant10^{-4}$ ou $P(A^\ast)\geqslant0,9999.$}, et $\Theta_\infty$, la quantité finale de protéines accumulées dans cette même monocouche. On remarquera que les points de ce graphique correspondent aux valeurs finales des courbes présentées sur le graphique A. Cette corrélation est montrée pour une série de monocouches saturées obtenues en balayant de manière exhaustive le domaine de variation du paramètre $\ln K^\circ$. On y voit clairement que l'empreinte finale $\sigma_\infty$ diminue en fonction de l'augmentation de la quantité finale accumulée $\Theta_\infty$. La cause de cette relation est la variation du paramètre $\ln K^\circ$: son augmentation fournit des monocouches contenant un plus grand nombre de protéines et dont l'empreinte est plus faible, c'est-à-dire qu'elles sont moins relaxées. Inversement, la diminution du paramètre $\ln K^\circ$ mène à des monocouches contenant un plus faible nombre de protéines mais des empreintes élevées, c'est-à-dire que les protéines auront eu l'occasion de mener leur processus de relaxation plus loin. De ce fait, plus une monocouche est densément peuplée, moins les protéines la constituant y seront relaxées. Pour pouvoir s'accumuler en grandes quantités, les protéines devront en fait être mieux <<~ordonnées~>> dans la monocouche.

On voit donc que le modèle RSA+R permet de simuler la construction de monocouches en tenant compte de façon extrêmement précise du ratio $K^\circ$, comptabilisant le taux de compétition entre les deux phénomènes que sont relaxation et addition. Lorsque le paramètre $\ln K^\circ$ sera faible (flux de protéines vers la surface faible et/ou protéines molles à très molles), des monocouches saturées contenant une faible quantité de protéines, celles-ci étant fortement relaxées (empreinte élevée) seront obtenues. Tandis que, lorsque ce paramètre sera élevé, l'inverse sera obtenu, c'est-à-dire des monocouches contenant un grand nombre de protéines peu relaxées (faible empreinte).

\section{Quantités accumulées à saturation}

Suivant les observations de la section précédente, les monocouches construites dans des conditions caractérisées par un paramètre $\ln K^\circ$ élevé ont pour caractéristique de contenir un plus grand nombre de protéines à saturation, l'empreinte étant plus faible. Logiquement, il s'ensuit qu'une corrélation devrait être obtenue entre les quantités accumulées à saturation $\Theta_\infty$ et ce paramètre $\ln K^\circ$. Les programmes Matlab utilisés dans cette section sont détaillés à l'Annexe \ref{Ann2}.

Le graphique de la figure \ref{FigFit} montre une série de points représentant les quantités accumulées à saturation en fonction du paramètre $\ln K^\circ$ dont le domaine de variation a été exploré de manière exhaustive. Dans les calcules numériques, ces quantités à saturation sont obtenues pour des monocouches dont la probabilité d'addition $P(A)$ devient inférieure à $10^{-4}$ (\textit{i.e.} le volume exclu $P(A^\ast)$ est supérieur ou égal à $0,9999$). Le nuage de points montre la forme particulière d'une sigmoïde dont l'équation pourrait être estimée par la fonction logistique généralisée \citep{mcdowall2006}:
\begin{equation}\label{FonctionLogistique}
\Theta_\infty(x)\sim a\,\frac{1+m\,\exp [-vx]}{1+n\,\exp [-vx]}
\end{equation}
dans laquelle $x=\ln K^\circ$ et $a$, $m$, $n$ et $v$ sont des paramètres réels dont le sens et la valeur sont à déterminer. Le réarrangement de cette équation \ref{FonctionLogistique} et le sens qui sera donné aux différents paramètres permettront en effet d'ajuster par estimation numérique une courbe tracée en rouge sur la figure \ref{FigFit}, qui sera qualifiée par la suite de \emph{fonction de saturation}.

\begin{figure}[t]\centering
\includegraphics*[width=0.75\textwidth]{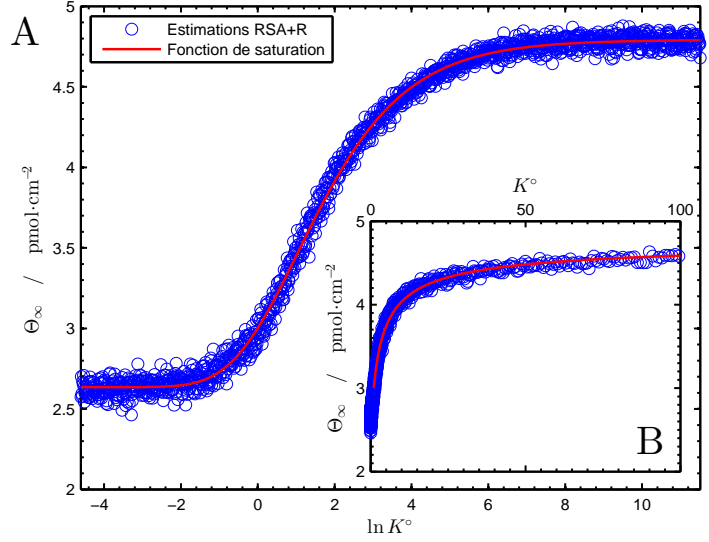}
\caption[Estimation de la fonction de saturation]{\textbf{A}: estimation de la fonction de saturation (rouge) à partir des valeurs de $\Theta_\infty$ simulées selon le modèle RSA+R (bleu) du système de boîtes de conserve (\textit{cf.} figure \ref{FigRSARelax}). La somme des carrés des résidus de l'estimation est de $3,517$ pmol$^2\cdot$cm$^{-4}$ ($n=1620$ points). \textbf{B}: aspect du nuage de points et de la fonction de saturation en fonction d'une échelle non logarithmique.}\label{FigFit}
\end{figure}

Premièrement, la fonction est bornée inférieurement et supérieurement par deux asymptotes horizontales.
On vérifiera en effet que
\begin{equation}
\lim_{x\rightarrow-\infty}\Theta_\infty (x)=a\,\frac{m}{n}\quad\mathrm{et}\quad\lim_{x\rightarrow+\infty}\Theta_\infty (x)=a
\end{equation}
Ces limites inférieure et supérieure sont respectivement notées $\Theta_{\infty,i}$ et $\Theta_{\infty,s}$ et peuvent s'exprimer en fonction des paramètres de la fonction logistique \ref{FonctionLogistique}:
\begin{equation}\label{LimitesFonctionLogistique}
\Theta_{\infty,i}=a\,\frac{m}{n}\quad\text{et}\quad\Theta_{\infty,s}=a.
\end{equation}

Les valeurs prisent par les asymptotes permettent d'écrire que $n\Theta_{\infty,i}=m\Theta_{\infty,s}$, ce qui, en l'injectant dans la fonction logistique \ref{FonctionLogistique} et en utilisant la définition $x=\ln K^\circ$, fournit une nouvelle équation pour la fonction de saturation:
\begin{equation}\label{FonctionSaturation1}
\Theta_\infty(K^\circ)=\frac{\Theta_{\infty,s}+n\Theta_{\infty,i} (K^\circ)^{-v}}{1+n (K^\circ)^{-v}}
\end{equation}
pouvant se réécrire sous la forme linéarisée
\begin{equation}\label{FonctionSaturation2}
\Theta_\infty(K^\circ)=\Theta_{\infty,i}+\frac{\Delta\Theta_\infty}{1+n (K^\circ)^{-v}}
\end{equation}
dans laquelle il a été posé que $\Delta\Theta_\infty=\Theta_{\infty,s}-\Theta_{\infty,i}$.

En observant attentivement le graphe de la fonction de saturation, on doit y constater une légère asymétrie par rapport au point d'inflexion de la sigmoïde. Il est possible d'en tenir compte en introduisant un coefficient d'asymétrie $w^{-1}$ dans l'équation \ref{FonctionSaturation2} \citep{mcdowall2006}:
\begin{equation}\label{FonctionSaturation4}
\Theta_\infty(K^\circ)=\Theta_{\infty,i}+\frac{\Delta\Theta_\infty}{\big[1+n (K^\circ)^{-v}\big]^\frac{1}{w}}.
\end{equation}

En tenant compte de cette asymétrie, il est maintenant envisageable de tenter l'élimination du paramètre $n$. \`{A} partir de l'équation \ref{FonctionSaturation4}, on peut montrer que le point d'inflexion de la sigmoïde se situe à une valeur de l'abscisse $K^\circ=\sqrt[v]{n/w}$. Sachant que ce point d'inflexion devrait représenter une transition entre deux régimes, on posera sans preuve formelle que la valeur $\tilde{K}^\circ$ est la valeur du paramètre $K^\circ$ au point d'inflexion de la fonction de saturation. On aura alors
\begin{equation}
n=w(\tilde{K}^\circ)^v
\end{equation}
que l'on pourra injecter dans l'équation \ref{FonctionSaturation2} afin d'écrire une expression obtenue sur des bases empiriques (équation \ref{FonctionSaturation5}) pour la fonction de saturation dont tous les paramètres ont un sens physique précis excepté $v$ qui, comme $w$ doit être déterminé par des méthodes numériques.
\begin{equation}\label{FonctionSaturation5}
\Theta_\infty(K^\circ)=\Theta_{\infty,i}+\Delta\Theta_\infty\Big[1+w\big(\tilde{K}^\circ/K^\circ\big)^v\Big]^{-\frac{1}{w}}.
\end{equation}

Grâce à cette mise en équation, il est possible d'estimer les paramètres $\Theta_{\infty,i}$, $\Delta\Theta_\infty$, $\tilde{K}^\circ$, $w$ et $v$ pour les données issues des modélisations RSA+R montrées à la figure \ref{FigFit}. Ces estimations se font sur base d'un algorithme présenté à la section \ref{654gfb354dfg354dqf354} (annexe \ref{Ann2}). La figure \ref{FigFit} illustre cette estimation (rouge) à partir des points du modèle RSA+R (bleu). Pour une somme des carrés des résidus (SCR) de $3,517$ pmol$^2\cdot$cm$^{-4}$ ($n=1620$ points), les paramètres de la fonction sont: $\Theta_{\infty,i}=2,6351$ pmol$\cdot$cm$^{-2}$, $\Delta\Theta_\infty=2,156$ pmol$\cdot$cm$^{-2}$, $\ln\tilde{K}^\circ=1,0449$, $v=0,6474$ et $w=0,1165$.

En fonction de divers paramètres caractérisant la nature du système, l'équation \ref{FonctionSaturation5} paramétrise la quantité finale à saturation $\Theta_\infty$ en fonction du ratio des vitesses initiales $K^\circ$. Encore une fois, deux angles d'interprétations sont possibles. Le premier est celui de la série de monocouches construites avec la même protéine ($k_R^\circ$ invariable) mais pour des flux variables ($k_A^\circ$ variable). Dans ce cas, l'augmentation du flux (vitesse initiale $k_A^\circ$) entraînera une augmentation proportionnelle du ratio $K^\circ$, ce qui se traduira finalement, grâce à la fonction de saturation, par une plus grande quantité de protéines accumulées dans la monocouche. Le second point de vue consiste à construire une série de monocouches dans les mêmes conditions de transports ($k_A^\circ$ invariable) mais pour des protéines relaxant à des vitesses différentes ($k_R^\circ$ variable). En considérant une augmentation de la dureté des protéines, la vitesse initiale $k_R^\circ$ diminuera et le ratio des vitesses initiales $K^\circ$ augmentera, ce qui engendrera des monocouches contenant une plus grande quantité de protéines à saturation.

\section{Discussion}

\subsection{Influence de la diffusion sur $\Theta_\infty$}

La relation \ref{VitAddvsDiff} a permis de relier la vitesse initiale d'addition $k_A^\circ$ au flux $j_\perp(z=0,t)$ de protéines vers la surface par diffusion. Sur cette base et grâce à la définition du ratio $K^\circ$, il devient possible d'introduire dans l'équation de la quantité accumulée à saturation (équation \ref{FonctionSaturation5}) des termes se rapportant aux flux de protéines vers la surface. De plus, la première équation de Fick reliant le flux de protéines à la concentration de celles-ci dans le volume de suspension adjacent à la surface, il devient aussi possible de relier la quantité $\Theta_\infty$ à cette concentration.

Pour rappel, l'équation \ref{VitAddvsDiff} s'écrivait $k_A^\circ=j_\perp(z=0,t).$
La division des deux membres par la vitesse initiale de relaxation $k_R^\circ$ fournit alors directement
\begin{equation}\label{EqDéfFluxT1}
K^\circ=\frac{j_\perp(z=0,t)}{k_R^\circ}
\end{equation}
dans laquelle la définition \ref{EqCinétiqueRelax4} du ratio $K^\circ$ a été utilisée. De même, le ratio $\tilde{K}^\circ$ (valeur de $K^\circ$ au point d'inflexion de la fonction de saturation) s'écrit par analogie
\begin{equation}
\tilde{K}^\circ=\frac{\tilde{j}_\perp(z=0,t)}{\tilde{k}_R^\circ}
\end{equation}
où $\tilde{j}_\perp$ est le flux de protéines à ce point d'inflexion. En posant que la fonction de saturation est construite pour une seule et même protéine, la vitesse initiale $k_R^\circ$ demeure invariable et définir une valeur d'inflexion $\tilde{k}_R^\circ$ n'a pas de sens. Le flux est dès lors divisé par $k_R^\circ$ et non par $\tilde{k}_R^\circ$. Il vient alors
\begin{equation}\label{EqDéfFluxT2}
\tilde{K}^\circ=\frac{\tilde{j}_\perp}{k_R^\circ}
\end{equation}
dans laquelle le flux $\tilde{j}_\perp$ a lui aussi été considéré comme invariable.

En rapportant \ref{EqDéfFluxT2} à \ref{EqDéfFluxT1}, il vient
\begin{equation}\label{EquivZetaFlux}
\frac{\tilde{K}^\circ}{K^\circ}=\frac{\tilde{j}_\perp}{j_\perp(z=0,t)}
\end{equation}
qui est substitué dans l'équation \ref{FonctionSaturation5} de la fonction de saturation afin de l'écrire comme une fonction, non seulement des paramètres empiriques $w$ et $v$, mais aussi de termes directement liés au transport par diffusion; on obtient:
\begin{equation}\label{EqIsothFinale}
\Theta_\infty(j_\perp)=\Theta_{\infty,i}+\Delta\Theta_\infty\bigg[1+w\Big(\tilde{j}_\perp\big/j_\perp(z=0)\Big)^v\bigg]^{-\frac{1}{w}}.
\end{equation}
De très nombreuses relations obtenues sur base de l'approximation de Smoluchowski-Levich pour les petits protéines (les forces d'attraction entre colloïdes, la résistance hydrodynamique et l'effet de la taille des protéines sont négligés) sont décrites dans la littérature \citep{vandeven1996bis}. Chacune de ses relations se rapporte à un type d'écoulement vers la surface dont un grand nombre est donné par Z. Adamczyk \citep{adamcyzk1994}. Pour le jet impactant (\textit{impinging jet}) illustré aux figures \ref{FigEcoulement}.B et C, le flux est approximé par la relation
\begin{equation}\label{ApproxSmoluchLevich}
j_\perp(z=0)=0,776\,\alpha^\frac{1}{3}\, D^\frac{2}{3}\,C
\end{equation}
dans laquelle $\alpha$ est une fonction adimensionnelle du nombre de Reynolds au point d'impact du jet sur la surface et donne l'intensité du flux \citep{yang1998}, $C$ la concentration en protéines loin de la surface et $D$ le coefficient de diffusivité de ces mêmes protéines \citep{vandeven1996bis}. \`{A} partir de la relation \ref{ApproxSmoluchLevich}, le rapport
\begin{equation}\label{RapportFlux}
\frac{\tilde{j}_\perp}{j_\perp(z=0)}=\frac{\tilde{C}}{C}
\end{equation}
est obtenu étant donné que le nombre de Reynolds, et donc la fonction $\alpha$, ainsi que le coefficient de diffusivité ne dépendent pas de la concentration mais seulement de facteurs hydrodynamiques. En substituant ce rapport dans la relation \ref{EqIsothFinale}, il vient
\begin{equation}\label{EqIsothFinaleSmolLevich}
\Theta_\infty(C)=\Theta_{\infty,i}+\Delta\Theta_\infty\bigg[1+w\Big(\tilde{C}\big/C\Big)^v\bigg]^{-\frac{1}{w}},
\end{equation}
une relation entre $\Theta_\infty$ la quantité de protéines contenue dans la monocouche à saturation et $C$ la concentration de celles-ci dans le c{\oe}ur de la suspension à partir de laquelle la monocouche s'est formée. La concentration $\tilde{C}$ est la concentration qui permettra d'obtenir une monocouche dont la valeur $\Theta_\infty$ sera celle du point d'inflexion de la fonction de saturation. Cette relation montre très clairement que la diffusion, se manifestant à travers la concentration $C$ en protéine dans le volume adjacent à la surface, doit avoir une influence très nette sur la quantité de protéines accumulées dans la monocouche à saturation. Une concentration élevée augmentera le flux de protéines vers la surface et poussera à la hausse le facteur $K^\circ$. Finalement, lorsque le ratio des vitesses initiales $K^\circ$ sera élevé, on obtiendra une quantité de protéines plus élevée dans la monocouche saturée.

\begin{figure}[b!]\centering
\includegraphics[width=0.9\textwidth]{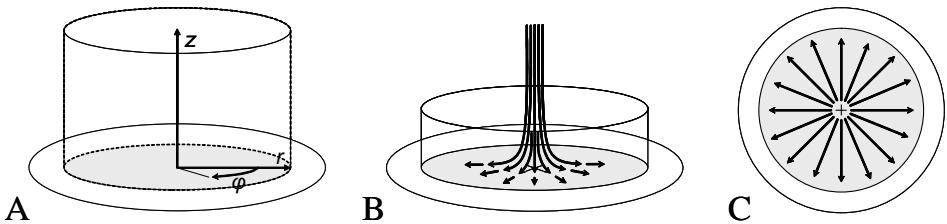}
\caption[Le système de coordonnées cylindriques et le jet impactant la surface]{La surface sur laquelle vient s'accumuler les particules colloïdales est figurée par les zones grisées. \textbf{A}: le système de coordonnées cylindriques $(z, r,\varphi)$ au dessus de la surface, \textbf{B}: illustration d'un jet impactant (\textit{impinging jet}) perpendiculairement la surface et \textbf{C}: direction des lignes de courant du fluide parallèles à la surface découlant de ce jet impactant.}\label{FigEcoulement}
\end{figure}

\subsection{Influence des conditions hydrodynamiques sur $\Theta_\infty$}

Il est important de remarquer que la relation \ref{ApproxSmoluchLevich} n'est \textit{a priori} valable que pour le point de stagnation, c'est-à-dire le point de la surface au centre du jet impactant la surface (point indiqué par une croix sur la figure \ref{FigEcoulement}.C). La question se pose dès lors de savoir comment la formation de la monocouche sera influencée en dehors de cette zone ne représentant qu'une infime partie de la surface sur laquelle se forme le film de protéines. Le flux de protéines $j_\perp(z=0)$ vers la surface est donc aussi une fonction $j_\perp(z=0,r,\varphi)$ où les coordonnées cylindriques $r$ et $\varphi$ sont illustrées à la figure \ref{FigEcoulement}.A. Les figures \ref{FigEcoulement}.B et C montrent que l'écoulement est symétrique par rapport au centre de la surface et est donc indépendant de la variable $\varphi$, ce qui permet d'écrire le flux comme une fonction $j_\perp(z=0,r)$. Il existe dans la littérature diverses formulations du flux $j_\perp(z=0,r)$, fonction de la position radiale. De manière générale, Z. Adamczyk \textit{et al.} \citep{adamczyk1982} l'écrivent par
\begin{equation}\label{FluxGénéral}
j_\perp(r)=-D\,R^{-1}\,C(r)\,Sh(r)
\end{equation}
dans laquelle $R$ est le rayon de la particule colloïdale, $C(r)$ la concentration en fonction de $r$ et $Sh$ le nombre adimensionnel de Sherwood (pour la simplicité de la notation, l'écriture de la position $z=0$ est sous-entendue). L'expression du nombre de Sherwood est donnée par les mêmes auteurs en fonction de la position radiale $r$ du point de la surface considéré:
\begin{equation}
Sh(r)=0,9318\,Pe^\frac{1}{3}\,Sc^\frac{1}{6}\,r^{-\frac{1}{2}}
\end{equation}
dans laquelle $Pe$ est le nombre de Péclet (rapport entre le transfert par convection et par diffusion) et $Sc$ le nombre de Schmidt (rapport entre la viscosité cinématique et le coefficient de diffusion). Les nombres adimensionnels $Pe$ et $Sc$ sont postulés comme indépendants de la coordonnée $r$. Une valeur moyenne indépendante de la position $r$ du nombre de Sherwood est rapportée par les mêmes auteurs:
\begin{equation}\label{SherwoodMoyen}
\overline{Sh}=1,864\,Pe^\frac{1}{3}\,Sc^\frac{1}{6},
\end{equation}
ce qui permet de réécrire la relation \ref{FluxGénéral} par
\begin{equation}\label{FluxGénéral1}
j_\perp(r)=-D\,R^{-1}\,C(r)\,\overline{Sh}
\end{equation}
dans laquelle seule la concentration $C$ est une fonction de la coordonnée $r$, c'est-à-dire que la concentration n'est pas forcément considérée comme indépendante de la position radiale.

En considérant les dimensions du système et de la protéine, la combinaison de ces deux dernières relations mènent à l'expression \ref{ApproxSmoluchLevich} donnée par T. van de Ven et S. Kelemen \citep{vandeven1996bis} pour le point de stagnation du jet impactant. Bien que le flux $j_\perp(r)$ dépende de $r$ (la position radiale de l'élément de surface considéré) le rapport \ref{RapportFlux} demeure parfaitement valable quel que soit $r$. En effet, en exprimant $\tilde{j}_\perp(r)$ à partir de \ref{FluxGénéral} et \ref{SherwoodMoyen}, les égalités $D=\tilde{D}$ et $R=\tilde{R}$ seront forcément vérifiées celles-ci restant invariantes puisque la protéine considérée reste la même. De plus, la valeur moyenne du nombre de Sherwood $\overline{Sh}$ reste par définition constante validant l'expression du rapport \ref{RapportFlux} pour toute valeur de la coordonnée $r$, c'est-à-dire la position de l'élément de surface par rapport au centre du jet impactant.

Si la concentration en protéines $C$ est une fonction de la position radiale $r$ de l'élément de surface considéré, l'équation \ref{EqIsothFinaleSmolLevich} donne une valeur locale de la quantité finale à saturation, c'est-a-dire la valeur de $\Theta_\infty$ dans un élément de surface $r\mathrm{d}r\mathrm{d}\varphi$. Dans le système de coordonnées cylindriques montré à la figure \ref{FigEcoulement}.A, la quantité moyenne de protéines accumulées dans le film est alors donnée par l'intégrale
\begin{equation}
\langle\Theta_\infty\rangle=\frac{1}{\pi R^2}\int_0^{2\pi}\int_0^R \Theta_\infty(r,\varphi)r\mathrm{d}r\mathrm{d}\varphi,
\end{equation}
obtenue grâce à la définition de la moyenne. Cette intégrale sera simplifiée sous la forme
\begin{equation}
\langle\Theta_\infty\rangle=\frac{2}{R^2}\int_0^R \Theta_\infty(r)r\mathrm{d}r
\end{equation}
étant donné que le jet impactant est indépendant de la coordonnée $\varphi$. En développant le terme $\Theta_\infty(r)$ selon l'équation \ref{EqIsothFinaleSmolLevich}, il vient
\begin{equation}
\langle\Theta_\infty\rangle=\frac{2}{R^2}\int_0^R 
\Bigg\lbrace
\Theta_{\infty,i}+\Delta\Theta_\infty\bigg[1+w\Big(\tilde{C}\big/C(r)\Big)^v\bigg]^{-\frac{1}{w}}
\Bigg\rbrace
r\mathrm{d}r
\end{equation}
qui devient après réarrangements
\begin{equation}\label{IsothLangmuirConcInhomo}
\langle\Theta_\infty\rangle=
\Theta_{\infty,i}+
\frac{2\Delta\Theta_\infty}{R^2}\int_0^R
\bigg[1+w\Big(\tilde{C}\big/C(r)\Big)^v\bigg]^{-\frac{1}{w}}
r\mathrm{d}r.
\end{equation}

L'intégrale obtenue montre que la quantité moyenne de protéines accumulées ne s'obtient pas de manière directe lorsque la concentration de la suspension est inhomogène. Ce cas est intéressant dans le cadre du mélange hydrodynamique qui se ferrait directement dans la suspension en contact avec la surface. Toutefois, lorsque la suspension injectée sur la surface a été préalablement parfaitement mélangée de telle sorte que la concentration $C$ soit homogène dans tout le volume de la suspension, il n'y a aucune raison pour que cette concentration dépende de la coordonnée $r$. De ce fait, la quantité $\langle\Theta_\infty\rangle$ se confond avec $\Theta_\infty$, une quantité se calculant grâce à l'équation \ref{EqIsothFinaleSmolLevich}.

\subsection{Les <<~isothermes d'adsorption~>> des protéines}\label{68gh74b7689}

L'influence de la concentration en protéines dans la suspension à partir de laquelle le film est construit sur la quantité finale de protéines qu'il contient est un fait bien connu et à conduit à diverses théorisations abondamment décrites dans la littérature \citep{nordehaynes,lundstrom1985,brash1978}. W. Norde et C.~E. Giacomelli \citep{norde1999} ont donné une interprétation du phénomène très proche de celle proposée dans le présent travail, interprétation détaillée dans d'autres travaux \citep{Norde2012,norde2012b}. Cette théorie postule que le degré d'avancement du processus d'étalement (relaxation) des protéines sur la surface dépend d'un rapport entre une vitesse d'étalement et une vitesse d'attachement ou d'un temps caractéristique d'étalement $\tau_s$ et d'un temps caractéristique d'attachement $\tau_f$. Ces auteurs précisent encore que le premier temps caractéristique est une fonction de la cohésion interne de la protéine tandis que le second est principalement contrôlé par le flux $j_\perp$ de protéines vers la surface par diffusion. Ils distinguent trois cas selon l'ordre de grandeur du ratio $\tau_s/\tau_f$ (durée de l'étalement par rapport à la durée de l'attachement):
\begin{itemize}
\item $\tau_s/\tau_f\ll 1$: l'étalement est plus rapide que l'attachement ce qui permet aux protéines attachées à la surface de relaxer complètement de telle sorte que la quantité $\Theta_\infty$ serait indépendante de $j_\perp$ \citep{norde1999,Norde2012};
\item $\tau_s/\tau_f\simeq 1$: l'étalement se produit sur une échelle de temps comparable à l'attachement de telle sorte que le degré d'étalement des protéines serait affecté par $j_\perp$ \citep{norde1999} et
\item $\tau_s/\tau_f\gg 1$: l'étalement est complètement inhibé car l'aire adjacente à la protéine nouvellement attachée est déjà occupée par d'autres avant qu'elle n'ait eu le temps de s'étaler \citep{Norde2012}.
\end{itemize}

Bien que corroborant ce qui a été développé ci-dessus, la définition et l'utilisation d'un tel type de ratio par W. Norde et C.~E. Giacomelli \citep{norde1999} demeure incomplète. Dans le modèle RSA+R, deux ratios comparables au rapport $\tau_s/\tau_f$ ont été introduits: $K$ le ratio des vitesses instantanées et $K^\circ$ le ratio des vitesses initiales. $K^\circ$ est fixé par l'importance de la diffusion (liée à la concentration $C$) et reste invariable tout au long de la construction de la monocouche tandis que $K$, dont la valeur initiale est $K^\circ$, varie au cours de la construction de la monocouche eu égard à l'augmentation du volume exclu et de l'effet d'exclusion par la taille. Le remplissage de la surface (croissance de la monocouche) ayant tendance à ralentir l'addition et la relaxation, les vitesses instantanées $k_A$ et $k_R$ diminueront toutes deux. Toutefois, l'effet d'exclusion par la taille commencera par exclure les protéines ayant les plus grosses empreintes caractéristiques et donc les protéines relaxées, la vitesse de relaxation $k_R$ diminuera plus rapidement que la vitesse d'addition $k_A$ de telle sorte qu'à l'approche du régime asymptotique, la relaxation sera totalement inhibée ($k_R=0$) alors que l'addition sera, quoique rare, encore possible ($k_A\neq0$). $K$ sera donc une fonction strictement croissante tendant asymptotiquement vers l'infini. Même en rapprochant le rapport $\tau_s/\tau_f$ du ratio des vitesses instantanées $K^\circ$, le modèle de  W. Norde et C.~E. Giacomelli \citep{norde1999} demeurera insuffisant car, dans le cas ou $\tau_s/\tau_f\ll 1$, les protéines ne pourront pas relaxer entièrement sur la surface étant donné que l'effet d'exclusion par la taille finira systématiquement par inhiber la relaxation, n'autorisant plus que l'addition de protéines natives.

Ce commentaire permet de montrer que le lien entre la quantité de protéines accumulées dans une monocouche saturée $\Theta_\infty$ et les caractéristiques du flux (lié à la concentration) et de la relaxation n'est pas nouveau. Toutefois, ce n'est que lorsque ces notions de ratios sont utilisées dans le cadre d'un modèle RSA+R qu'elles montrent leur potentiel. Une corrélation très nette entre le ratio $K^\circ$, analogue au ratio $\tau_s/\tau_f$ de W. Norde et C.~E. Giacomelli \citep{norde1999}, et la quantité $\Theta_\infty$ peut être dès lors mise en évidence en substituant l'expression du flux \ref{ApproxSmoluchLevich} sous l'approximation de Smoluchowski-Levich dans la formule \ref{EqCinétiqueRelax4} du ratio des vitesses initiales:
\begin{equation}\label{KfoncC}
K^\circ=0,776\,\alpha^\frac{1}{3}\,D^\frac{2}{3}\,(k_R^\circ)^{-1}\,C,
\end{equation}
montrant par ailleurs que
\begin{equation}
\ln K^\circ= \mathrm{cste}+\ln C
\end{equation}
où cste est une constante donnée par
\begin{equation}
\mathrm{cste}=\ln \Big[0,776\,\alpha^\frac{1}{3}\,D^\frac{2}{3}\,(k_R^\circ)^{-1}\Big].
\end{equation}
Grâce au fait que le ratio $K^\circ$ ait pu être rapproché de la concentration $C$ en protéines dans la suspension, il semble évident qu'il pourrait s'agir d'une théorie explicative de l'existence <<~d'isothermes d'adsorption~>> (voir l'illustration de la figure \ref{FigIsothermeIntro} dans le chapitre introductif) pour les protéines. Généralement, <<~l'isotherme d'adsorption~>> des protéines est la quantité $\Theta_\infty$ exprimée comme une fonction de la concentration $C$ et ayant la forme d'une fonction croissante bornée par une asymptote horizontale. Le graphique B de la figure \ref{FigFit} montre clairement une telle forme et, sachant que le paramètre $K^\circ$ est directement proportionnel à la concentration $C$ (\textit{cf}. équation \ref{ApproxSmoluchLevich}), l'équation \ref{EqIsothFinaleSmolLevich} apparaît comme l'équation de <<~l'isotherme d'adsorption~>> des protéines, équation valable pour l'entièreté du film si la concentration dans la suspension est parfaitement homogène. Autrement, l'utilisation de l'équation \ref{IsothLangmuirConcInhomo} s'avérera nécessaire.

Il n'est maintenant pas inutile de s'intéresser à la forme qu'adoptera la fonction de saturation (ou <<~isotherme d'adsorption~>>) selon la valeur des différents paramètres contenus dans l'équation \ref{FonctionSaturation5}. La signification physique des paramètres $v$ et $w$ n'étant pas très claire, on se focalisera sur les conséquences et le sens d'une variation du paramètre $\tilde{K}^\circ$, valeur de $K^\circ$ au point d'inflexion de la sigmoïde. La figure \ref{FigProtHardAndSoft} montre la conséquence d'une variation de ce paramètre $\tilde{K}^\circ$ sur la forme de la fonction de saturation.

\begin{figure}[t]\centering
\includegraphics*[width=0.75\textwidth]{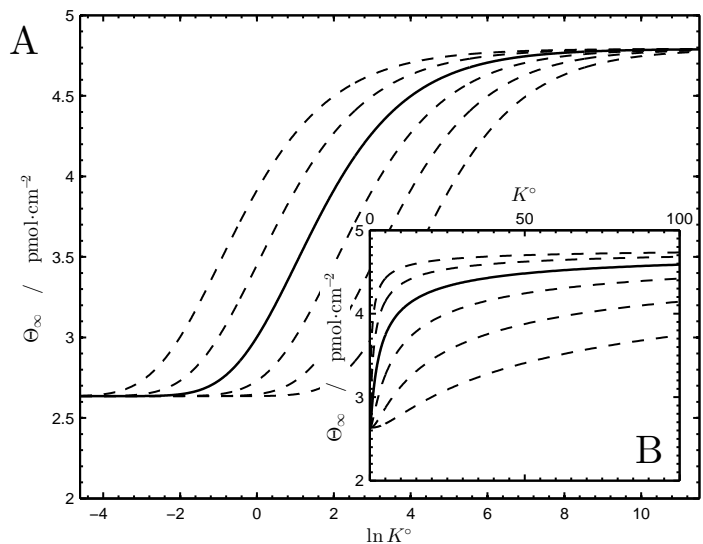}
\caption[Aspects de la fonction de saturation en fonction du paramètre $\ln\tilde{K}^\circ$]{\textbf{A}: changements d'aspects de la fonction de saturation (ou <<~isotherme d'adsorption~>>) dus aux variations du paramètre $\ln\tilde{K}^\circ$. La ligne continue donne l'aspect de la fonction de saturation dont les paramètres $\Theta_{\infty,i}$, $\Delta\Theta_\infty$, $v$, $w$ et $\ln\tilde{K}^\circ$ sont les mêmes qu'à la figure \ref{FigFit} tandis que les lignes discontinues montre cette même fonction lorsque le paramètre $\ln\tilde{K}^\circ$ a été successivement incrémenté d'une valeur $-1$ ou $+1$. \textbf{B}: les fonctions de saturation obtenues pour ces divers paramètres en fonction d'une échelle linéaire.}\label{FigProtHardAndSoft}
\end{figure}

En fonction de l'échelle logarithmique, l'augmentation du paramètre $\tilde{K}^\circ$ décalera la position du saut de la sigmoïde vers des valeurs de $\ln K^\circ$ plus ou moins élevées. En fonction de l'échelle linéaire, l'aspect de la fonction de saturation sera étiré sur son domaine de variation et le plateau supérieur sera atteint pour des valeurs de $K^\circ$ bien plus élevées. En adoptant le point de vue des <<~isothermes d'adsorption~>>, c'est-à-dire que chaque fonction de saturation est construite pour une seule espèce de protéine mais à des concentrations croissantes, il faut admettre que ce changement d'aspect est dû à une différence dans la cinétique de la relaxation. Un $\ln\tilde{K}^\circ$ faible indiquera que le flux de protéines en provenance de la suspension viendra rapidement contrecarrer le phénomène de relaxation. Inversement, un $\ln\tilde{K}^\circ$ élevé indiquera qu'il sera nécessaire d'atteindre un flux de protéines particulièrement élevé afin de pouvoir concurrencer la relaxation. En d'autres mots, la valeur de $\ln\tilde{K}^\circ$ est une fonction croissante de la vitesse de relaxation de la protéine dont on construit <<~l'isotherme d'adsorption~>>. Une protéine dure sera caractérisée par un $\ln\tilde{K}^\circ$ faible tandis qu'une protéine molle aura un $\ln\tilde{K}^\circ$ élevé.

De plus, sans que cela ne soit montré à la figure \ref{FigProtHardAndSoft}, $\Theta_{\infty,i}$ et $\Delta\Theta_\infty$ sont des paramètres qui dépendront de l'étendue du phénomène de relaxation. Plus une protéine aura un fort potentiel de relaxation, c'est-à-dire que l'empreinte caractéristique de l'état complètement relaxé sera forte (indépendamment de la vitesse $k_R^\circ$), plus la valeur $\Theta_{\infty,i}$ devrait être faible. De même, la valeur du paramètre $\Delta\Theta_\infty$ devrait être une indication de la différence entre l'empreinte caractéristique de la protéine native et celle de la protéine totalement relaxée. Les valeurs de ces paramètres devraient donc offrir des pistes d'interprétations prometteuses de <<~l'isotherme d'adsorption~>>.

L'aspect des <<~isothermes d'adsorption~>> semblerait donc essentiellement dicté par le caractère dur/mou de la protéine pour laquelle elle est construite. \`{A} cette fin, les éléments discutés ci-dessus sont résumés dans le tableau \ref{TabMouDur}. On remarquera que les paramètres $v$ et $w$ de l'équation \ref{FonctionSaturation5} n'y sont pas présents étant donné la difficulté à leur attribuer un sens physique. Dans ce travail, on ne les considérera donc que comme des paramètres d'ajustement de la fonction de saturation.

\begin{table}[h]\centering
\caption[Allures des <<~isothermes d'adsorption~>> des protéines]{Influence du caractère mou ou dur d'une protéine\\sur l'allure de son <<~isotherme d'adsorption~>>.}\label{TabMouDur}
\begin{spacing}{1.3}
\begin{small}
\begin{tabular}{lccc}
\hline
              & $\Theta_{\infty,i}$ & $\Delta\Theta_\infty$ & $\ln\tilde{K}^\circ$\\
\hline
Protéine dure & élevé & faible & faible\\
Protéine molle& faible & élevé & élevé\\
\hline
\end{tabular}
\end{small}
\end{spacing}
\end{table}

Le modèle RSA+R tel qu'il a été développé permet donc une conceptualisation originale de la notion <<~d'isotherme d'adsorption~>>. Cette conceptualisation se distinguant clairement de la notion <<~d'isotherme de Langmuir~>> pour l'adsorption \textit{stricto sensu} des espèces de bas poids moléculaires sur une surface par le fait qu'elle donne une interprétation strictement cinétique de la formation des monocouches. La cinétique selon laquelle se construira une monocouche changera en fonction des conditions dans lesquelles elle sera construite et du type de protéine.

\subsection{Vieillissement et déplétion des monocouches}

Comme tout modèle, le modèle RSA+R est basée sur une série d'hypothèses simplificatrices mais nécessaires empêchant d'appréhender le phénomène dans sa totalité. Les résultats obtenus pour les fonctions de saturation doivent donc être considérées comme des approximations puisque tous les phénomènes pouvant avoir lieu lors de la formation d'une monocouche ne peuvent être pris en compte. Parmi ces phénomènes volontairement négligés, le postulat de l'impossibilité pour une particule d'en déplacer et/ou d'en déformer une autre lors de son incorporation dans la monocouche a été une hypothèse simplificatrice très utile dans les modèles utilisés. Bien <<~qu'interdit~>> par la seconde hypothèse du modèle RSA (ou RSA+R), le déplacement pourrait ne pas être impossible mais il n'y a \textit{a fortiori} aucune raison pour qu'il ne soit pas aussi soumis aux règles de la thermodynamique et des considérations cinétiques envisagées.

Lors de l'examen de la différentielle du taux d'occupation de la surface $\mathrm{d}\phi$ (équation \ref{EqDifférentielleProbabilité3}), il avait été postulé que les termes $\sigma\,\mathrm{d}\Theta$ et $\Theta\,\mathrm{d}_R\sigma$ étaient tous deux positifs. Ce cas de figure, tout en respectant le second principe de la thermodynamique, permettait de conceptualiser les phénomènes d'addition ($\mathrm{d}\Theta>0$) et de relaxation ($\mathrm{d}_R\sigma>0$) causant le blocage progressif de la surface par augmentation de l'effet d'exclusion par la taille ($\mathrm{d}_A\sigma<0$). L'équilibre entre le phénomène d'addition-relaxation et le phénomène d'exclusion par la taille finissait par annuler la différentielle $\mathrm{d}\phi$, c'est-à-dire que le taux d'occupation de la surface devenait fixe, ce qui conceptualise la fin du processus de remplissage.

Dans le but d'approfondir ces notions et de les discuter efficacement, il est nécessaire d'obtenir des expressions des vitesses des divers phénomènes en jeu lors du remplissage de la surface: vitesse de recouvrement, vitesse d'addition, vitesse de relaxation et vitesse d'exclusion. Ces expressions peuvent être obtenues à partir de la différentielle \ref{EqDifférentielleProbabilité3} du taux de recouvrement donnée par la relation
$$\mathrm{d}\phi=\sigma\mathrm{d}\Theta+\Theta\mathrm{d}_R\sigma+\Theta\mathrm{d}_A\sigma.$$
En en divisant tous les termes par $\sigma\mathrm{d}t$, il vient
\begin{equation}
\frac{1}{\sigma}\frac{\mathrm{d}\phi}{\mathrm{d}t}=\frac{\mathrm{d}\Theta}{\mathrm{d}t}+\frac{\Theta}{\sigma}\frac{\mathrm{d}_R\sigma}{\mathrm{d}t}+\frac{\Theta}{\sigma}\frac{\mathrm{d}_A\sigma}{\mathrm{d}t}
\end{equation}
contenant une série de nouveaux termes ayant les dimensions de vitesses (pmol$\cdot$cm$^{-2}\cdot$ s$^{-1}$) parmi lesquels il est possible de reconnaître la vitesse d'addition $k_A=\mathrm{d}\Theta/\mathrm{d}t$. En réécrivant cette différentielle sous la forme
\begin{equation}\label{jhqbvmiqbvmdqvm}
k_\phi=k_A+k_R-k_{ex},
\end{equation}
trois nouvelles expressions de vitesses sont obtenues: la vitesse de recouvrement de la surface $k_\phi$ définie par:
\begin{equation}
k_\phi=\frac{1}{\sigma}\frac{\mathrm{d}\phi}{\mathrm{d}t}
\end{equation}
et représentant la vitesse globale selon laquelle la surface sera recouverte par les protéines;
la vitesse de relaxation $k_R$ définie par
\begin{equation}\label{d24vg5vb524}
k_R=\frac{\Theta}{\sigma}\frac{\mathrm{d}_R\sigma}{\mathrm{d}t}
\end{equation}
représentant la vitesse instantanée\footnote{Il est particulièrement intéressant de remarquer que la vitesse instantanée de relaxation est définie sur base d'une différentielle (et d'une dérivée) partielle. Elle est dès lors rendue non intégrable rendant alors illusoire la possibilité d'obtenir une quantité totale ou globale de relaxations caractérisant l'ensemble du processus de croissance de la monocouche. Cela peut se comprendre par le fait que la relaxation est définie pour un état précis (instantané) de la monocouche à un moment donné de son évolution, de son histoire.} avec laquelle les protéines de la monocouche relaxeront;
et la vitesse d'exclusion des protéines de la surface $k_{ex}$ définie par
\begin{equation}\label{h56fg51h65fg1h65g1h561h5}
k_{ex}=-\frac{\Theta}{\sigma}\frac{\mathrm{d}_A\sigma}{\mathrm{d}t}
\end{equation}
conceptualisant la vitesse selon laquelle les protéines seront rejetées ou exclues de la surface à un moment donné de son remplissage. Cette vitesse d'exclusion peut apparaître assez <<~virtuelle~>> mais elle permet de conceptualiser l'opposition cinétique qu'imprime la croissance du volume exclu au remplissage de la surface par les deux phénomènes d'origine thermodynamique que sont l'addition et la relaxation.

De manière générale, la vitesse de recouvrement sera toujours positive ou nulle: $k_\phi\geqslant0.$ En effet, au cours de la croissance de la monocouche, le taux de recouvrement $\phi$ augmente nécessairement et, lorsque la monocouche sera finie, la vitesse $k_\phi$ s'annulera. Cette condition mène à l'inégalité
\begin{equation}\label{jhqbvmi6b54x6f4qbvmdqvm}
k_A+k_R-k_{ex}\geqslant0
\end{equation}
obtenue à partir de l'expression \ref{jhqbvmiqbvmdqvm}. Par ailleurs, de \ref{jhqbvmiqbvmdqvm}, on vérifiera que $k_A+k_R-k_{ex}=0$ lorsque l'état de saturation sera atteint ($k_\phi=0$). Dès lors,
\begin{equation}
k_A+k_R=k_{ex},
\end{equation}
montrant alors qu'à saturation, la vitesse d'exclusion viendra s'opposer à la somme des vitesses d'addition et de relaxation. \'{E}tant donné que la différentielle $\mathrm{d}_A\sigma$ est négative (voir l'inégalité \ref{EqDiffEmpreinteAddit1}), la vitesse d'exclusion $k_{ex}$ sera elle-même positive ou nulle (voir définition \ref{h56fg51h65fg1h65g1h561h5}). Alors, à partir de \ref{jhqbvmi6b54x6f4qbvmdqvm}, il vient la nouvelle inégalité
\begin{equation}\label{jhq6d5g4bvmi6b54x6f4qbvmdqvm}
k_A+k_R\geqslant0
\end{equation}
valable tout au long du remplissage jusqu'à l'état final. En l'exprimant selon $k_A\geqslant-k_R$, cela montre que les vitesses $k_A$ et $k_R$ ne doivent pas être nécessairement toutes deux positives mais qu'elles doivent seulement se compenser afin de mener à une valeur positive ou nulle.

Afin d'obtenir des éléments plus précis, il est possible de repartir de l'expression \ref{DiffEntropieVSsigmaTheta2} de la production d'entropie superficielle au cours du remplissage. En divisant tous les termes de cette expression par un incrément infinitésimal de temps $\mathrm{d}t$, il vient
\begin{equation}\label{465fgb65fgnh5j46f5d4}
P=\frac{\sigma}{T}\big(\mathcal{H}_{\mathrm{sl}\cdot\mathrm{l}}+
\mathcal{H}_{\mathrm{pl}\cdot\mathrm{l}}\big)\frac{\mathrm{d}\Theta}{\mathrm{d}t}
+\frac{\Theta}{T}\big(\mathcal{H}_{\mathrm{sl}\cdot\mathrm{l}}+
\mathcal{H}_{\mathrm{pl}\cdot\mathrm{l}}\big)\frac{\mathrm{d}_R\sigma}{\mathrm{d}t}
\end{equation}
dans laquelle $P$ est la fonction de production d'entropie superficielle $\mathrm{d}_is/\mathrm{d}t$ analogue de la fonction de production d'entropie $P=\mathrm{d}_iS/\mathrm{d}t$ définie par I. Prigogine \citep{prigogine1968,coveney1988}.
L'utilisation des expressions \ref{EqCinétique1} et \ref{d24vg5vb524} des vitesses d'addition et de relaxation permet après tout réarrangement de réexprimer \ref{465fgb65fgnh5j46f5d4} selon
\begin{equation}\label{35gf6hb4687646b8}
P=\frac{\sigma}{T}\big(\mathcal{H}_{\mathrm{sl}\cdot\mathrm{l}}+
\mathcal{H}_{\mathrm{pl}\cdot\mathrm{l}}\big)(k_A+k_R)
\end{equation}
montrant que la production d'entropie en un point de la surface est directement proportionnelle aux vitesse d'addition et de relaxation des protéines. Naturellement, à partir du second principe de la thermodynamique, la condition posée sur la production d'entropie superficielle s'énonce $P\geqslant0$, le terme $\sigma\big(\mathcal{H}_{\mathrm{sl}\cdot\mathrm{l}}+
\mathcal{H}_{\mathrm{pl}\cdot\mathrm{l}}\big)/T$ étant positif. L'usage de cette condition mène directement à l'inégalité \ref{jhq6d5g4bvmi6b54x6f4qbvmdqvm} déjà obtenue ci-dessus. Cette inégalité \ref{jhq6d5g4bvmi6b54x6f4qbvmdqvm} est très pratique car elle permet de décrire divers régimes qui pourraient dominer transitoirement l'évolution de la monocouche: la croissance, le vieillissement et l'équilibre.
\begin{itemize}
\item Premièrement, l'inégalité \ref{jhq6d5g4bvmi6b54x6f4qbvmdqvm} sera vérifiée lorsque les conditions
\begin{equation}
k_A>0\quad\mathrm{et}\quad k_R>0
\end{equation}
seront observées. Ces conditions où les vitesses d'addition et de relaxation sont toutes deux strictement supérieures à zéro correspondent à la \emph{croissance} de la monocouche, c'est-à-dire le régime pendant lequel de nouvelles protéines sont additionnées à la monocouche et peuvent aussi relaxer.
\item Deuxièmement, on aura le couple de conditions
\begin{equation}
k_A=0\quad\mathrm{et}\quad k_R>0
\end{equation}
conceptualisant un régime pendant lequel la monocouche ne croît plus par additions de nouvelles protéines mais pendant lequel la relaxation peut encore avoir lieu afin de poursuivre l'accroissement du taux de recouvrement de la surface. Les protéines de la monocouche relaxent afin de maximiser leurs interactions avec la surface et, par analogie avec le comportement des verres organiques, on parlera de \emph{vieillissement} de la monocouche.
\item Troisièmement, il est possible de poser les conditions
\begin{equation}
k_A<0\quad\mathrm{et}\quad k_R\geqslant-k_A
\end{equation}
pouvant se réexprimer sous la forme
\begin{equation}
k_S>0\quad\mathrm{et}\quad k_R\geqslant k_S
\end{equation}
en ayant posé que la vitesse de soustraction des protéines de la monocouche $k_S$ est l'opposée de la vitesse d'addition $k_A$ de telle sorte que $k_S=-k_A$. Ce régime illustré à la figure \ref{FigDepla} admet la possibilité de soustraction (décollement de la surface) de protéines de la monocouche à cause de la relaxation. Lorsque la relaxation d'une protéine $a$ suffisamment molle suscite un gain d'aire de contact avec la surface supérieur à l'aire de contact d'une protéine $b$, cette dernière protéine se trouvant par ailleurs <<~dans son chemin~>> pourra alors être éjectée, soustraite de la monocouche. Ce troisième régime sera qualifié de \emph{déplétion} de la monocouche. Bien que caractérisant une diminution de la quantité accumulée, ce phénomène se soldera toujours par une augmentation du taux de recouvrement de la surface et donc par une production nette d'entropie.
\item Le quatrième et dernier <<~régime~>> est conceptualisé par les conditions
\begin{equation}
k_A=0\quad\mathrm{et}\quad k_R=0
\end{equation}
correspondant à \emph{l'état final} ou \emph{d'équilibre}. Lorsque cet état est atteint, aucune nouvelle addition de protéine n'a lieu ni même de relaxation. De ce fait, le taux de recouvrement de la surface n'augmente plus et a atteint sa valeur finale, la monocouche a atteint un état définitivement stable.
\end{itemize}

\begin{figure}[t]\centering
\includegraphics*[width=0.5\textwidth]{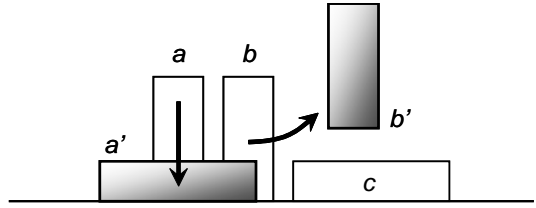}
\caption[Déplacement d'une protéine par une autre dans la monocouche]{Illustration d'un processus de déplacement d'une protéine (\textit{b}) par une autre (\textit{a}). La surface est initialement peuplée de trois protéines \textit{a}, \textit{b} et \textit{c}. En relaxant, la protéine \textit{a} succite un tel gain d'empreinte qu'elle est capable de déplacer la protéine \textit{b} en l'éjectant de la monocouche. La protéine \textit{b} n'a d'autre choix puisqu'il lui aurait été impossible de se déplacer latéralement à cause de la protéine \textit{c}.}\label{FigDepla}
\end{figure}

Les régimes traversés par la monocouche au cours de son évolution et décrits par les quatre couples de conditions ci-dessus respectant tous l'inégalité \ref{jhq6d5g4bvmi6b54x6f4qbvmdqvm} (conséquence du second principe de la thermodynamique) permettent de mettre en lumière les limites du modèle RSA+R. En effet, dans les modélisations réalisées au cours de ce chapitre, il n'a nullement été tenu compte de la phase de déplétion de la monocouche dont l'existence semble difficile à rejeter. Par ailleurs, les régimes pendant lesquels la relaxation a lieu (croissance, vieillissement et déplétion) n'ont été qu'approximativement pris en compte. En effet, le modèle de relaxation qui a été considéré pour les boîtes de conserve est un processus fini et discret. Or il se peut que dans un système réel, les protéines puissent adopter un très grand nombre (infini) de conformations qui pourraient permettre un taux d'occupation de la surface $\phi$ bien plus élevé que ce qui a été considéré ci-dessus. Non seulement les protéines pourraient relaxer de manière peu orthodoxe (relaxation d'un seul motif de la protéine, etc.) afin de venir combler tous les petits espaces laissés libres sur la surface qui se trouveraient dans son voisinage immédiat mais la relaxation d'une seule protéine suffisamment molle pourrait impliquer un tel gain d'empreinte qu'une protéine se trouvant <<~dans son chemin~>> pourrait être éjectée de la monocouche comme montré sur la figure \ref{FigDepla}. Le modèle RSA+R tient donc compte des deux premiers régimes: \emph{croissance} et \emph{vieillissement}. Le régime de \emph{déplétion} n'étant pas pris en compte et les deux premiers ne l'étant que partiellement à cause de l'approximation faite sur le processus de relaxation, il semble évident que ce modèle RSA+R n'est capable de construire que des structures caractéristiques d'un état n'ayant jamais atteint l'équilibre.

Le phénomène par lequel la relaxation continue de faire augmenter le taux de recouvrement de la surface lorsque l'addition sera devenue impossible (vieillissement et déplétion) devrait se produire sur de (très) longues échelles de temps laissant à penser que l'équilibre ne pourrait qu'être difficilement atteint dans des conditions expérimentales. Après sa phase de croissance, la monocouche entrerait dans un long processus de vieillissement-déplétion faisant que sa structure garderait un état dynamique.

\subsection{Implications du potentiel DLVO}

Le modèle de l'addition séquentielle aléatoire est très efficace lorsqu'il s'agit de mettre en évidence les effets du volume exclu, c'est-à-dire dus à l'occupation de l'espace au niveau de la surface. L'ordre d'exclusion qu'il permet de déduire selon l'orientation et le degré de dénaturation se manifestant à travers des empreintes variables démontre de façon très claire pourquoi les protéines constituant une monocouche peuvent être distribuées en fonction de leur orientation et de leur conformation, mais aussi pourquoi cette distribution est variable en fonction de la quantité totale accumulée et du temps. Mais, la focalisation sur les effets du volume exclu, ne risque-t-elle pas de faire oublier les effets de la double couche électrique et du potentiel de Hamaker pris en compte dans le calcul du potentiel à l'équation \ref{EqBlocage}? Si ces forces rendent possible la stabilité cinétique des protéines dans une solution aqueuse empêchant qu'elles ne s'agrègent entre elles, il est légitime de penser qu'elles permettront aux particules d'avoir une influence les unes sur les autres à l'approche de la surface ou même lorsqu'elles y seront collées. On peut supposer que les particules ne s'approcheront pas autant que ce qui a été supposé ci-dessus de telle sorte que les quantités $\Theta_\infty(\boldsymbol{\omega})$ soient légèrement modifiées. Toutefois, si les protéines viennent s'additionner sur la surface, c'est que les forces attractives protéines-surfaces sont plus intenses que les forces répulsives entre les protéines, rendant ainsi possible leur rapprochement afin de créer une monocouche relativement dense latéralement. Pour cette raison, il semble que spéculer quant à une éventuelle sous- ou surestimation des quantités $\Theta_\infty$ due aux forces incluses au modèle DLVO ne paraît pas fondamental, le phénomène pouvant rester assez marginal.

\section{Conclusion}

Tout au long de ce chapitre, les relations entre les phénomènes d'addition et de relaxation ont été examinées. Ces relations ont permis de montrer l'incidence de la relaxation sur la structure des monocouches obtenues ainsi que les interconnections existant avec le transport des IgG vers la surface de polystyrène.

La relaxation est un phénomène qui a été caractérisé par une augmentation de l'empreinte caractéristique d'une particule. Cette augmentation ayant pour conséquence d'accroître l'empreinte des IgG au cours du remplissage de la surface, l'empreinte des particules contenues dans la monocouche au cours de son remplissage diminuera moins vite. Cette diminution moins rapide se fait par rapport à la diminution attendue du fait de l'effet d'exclusion par la taille mis en évidence grâce à l'application du modèle RSA au chapitre \ref{SectionRSA}. En fait, la relaxation, en ralentissant la diminution de l'empreinte, viendra favoriser, avec plus ou moins d'intensité, les orientations/conformations d'IgG d'empreintes caractéristiques élevées, c'est-à-dire \textit{flat} et \textit{flat$+$}.

Dans un premier temps, le modèle RSA a été adapté afin que toute IgG s'additionnant à la monocouche relaxe son orientation et ensuite sa conformation instantanément, tout en tenant compte du volume exclu. Bien que l'empreinte des IgG de la monocouche finisse par se trouver plus élevée, cela a permis de montrer que le volume exclu avait pour effet d'inhiber progressivement la relaxation de telle sorte que des IgG d'orientation \textit{end-on} finissaient tout de même par apparaître dans la monocouche. Malgré le fait que cette relaxation instantanée soit une relaxation poussée au maximum de son potentiel, les IgG \textit{end-on} demeurent majoritaires dans la monocouche saturée, propriété recherchée dans le cadre de l'ELISA.

Une relaxation instantanée, c'est-à-dire de vitesse infiniment élevée, ne semblant toutefois pas satisfaisante, l'existence d'une vitesse finie $k^\circ_R$ a été postulée afin d'affiner le modèle théorique. Cette vitesse de relaxation des IgG, rapportée à leur vitesse d'addition $k^\circ_A$, a fourni un rapport de vitesses $(K^\circ)^{-1}$ menant à l'énoncé d'une quatrième hypothèse au modèle RSA. Ce modèle RSA+R (additions et relaxation séquentielles aléatoires) tenant compte de l'importance du phénomène de relaxation a ainsi fourni des courbes en fonction de la vitesse de relaxation. La complexité de la forme d'une IgG ainsi que la forte inconnue demeurant quant à sa façon de relaxer sa conformation, le modèle RSA+R n'a pu qu'être appliqué au système de boîtes de conserve. Les résultats obtenus pour les boîtes de conserve semblent toutefois pouvoir être raisonnablement extrapolés aux IgG et, de manière générale, à d'autres protéines globulaires.

Le modèle RSA+R, grâce au ratio des vitesses dont il a été montré qu'il était lié à la diffusion, a donné une explication possible de l'existence des <<~isothermes d'adsorption~>> des protéines. En augmentant progressivement le ratio $K^\circ$, une série de monocouches contenant, à saturation, une quantité croissante de protéines a été obtenue. Lorsque ces quantités à saturation sont portées en graphique en fonction  de $\ln\,K^\circ$ elles présentent une forme de sigmoïde. En fonction de $K^\circ$, la courbe pour laquelle une paramétrisation a été proposée, aura l'aspect typique des <<~isothermes d'adsorption~>> des protéines. De cette constatation, les <<~isothermes d'adsorption~>> des protéines ont été interprétées comme une conséquence de la compétition entre les phénomènes d'addition et de relaxation au cours de la construction de la monocouche. En reprenant les concepts de protéines molles et dures, l'explication proposée rend compte du fait que les protéines dures fournissent des monocouches plus densément peuplées en protéines que les protéines molles, et ce, pour les mêmes concentrations. La discussion s'est clôturée sur les conditions dans lesquelles les monocouches pouvaient vieillir, c'est-à-dire que les protéines continuent à relaxer au détriment de la quantité qu'elles contiennent. D'autre part, la contribution du potentiel DLVO a semblé marginale compte tenu des effets mis en évidence tout au long de ce chapitre.

\begin{footnotesize}

\end{footnotesize}\end{cbunit}
\begin{cbunit}
\chapter[Caractérisation des films d'IgG par AFM et ELISA]{Caractérisation des films d'IgG par AFM et ELISA}\label{ChapExp}
\markboth{Chapitre \ref{ChapExp}: CARACT\'{E}RISATION}{}
\minitoc

\section{Généralités}\label{65g46s54g687tr}

Comme cela a été développé dans le chapitre introductif, les dosages immunologiques font partie des méthodes les plus couramment utilisées dans le cadre des techniques de détection et de quantification des (bio)molécules suspendues dans divers fluides. Ceux-ci reposent sur la forte interaction ligand-récepteur intervenant dans la liaison antigène-anticorps de manière à ce que l'antigène soit la molécule à détecter/quantifier \citep{porstmann1992}. Comme la plupart des dosages immunologiques utilisés à ce jour, l'ELISA nécessite une phase solide afin de supporter et immobiliser de façon efficace la phase réactive constituée d'anticorps de capture, anticorps directement fixés sur la surface, et d'anticorps spécifiques reconnaissant l'antigène que l'on souhaite détecter. Dans le cas de l'ELISA <<~double-sandwich~>>, un anticorps de capture est fixé sur la surface et cet anticorps de capture immobilisera ensuite l'anticorps spécifique dans la phase réactive. Il s'agit donc d'une construction de deux anticorps. Pour l'ELISA <<~sandwich~>>, l'anticorps spécifique est directement immobilisé sur la surface \citep{dorazio2011,butler2000} (voir figure \ref{FigELISAsandwich}).

Lors de la première étape de l'ELISA, une suspension aqueuse de l'anticorps de capture est mise en contact avec une surface de polystyrène plane de telle sorte que de fortes interactions s'établissent entre l'anticorps et le support solide. Il est généralement bien accepté que les anticorps s'immobilisent spontanément sur la surface grâce à l'effet hydrophobe et qu'ils finissent par former une couche mince (\textit{thin film}) entre le polystyrène du support et la suspension d'anticorps \citep{hlady1996}. La couche mince obtenue consiste en une monocouche d'anticorps fermement accrochée sur le support et répondant à deux critères afin d'être efficace d'un point de vue immunologique lors des étapes suivantes de l'ELISA:
\begin{itemize}
\item Les propriétés immunologiques des anticorps immobilisés doivent rester intactes;
\item Ils doivent être convenablement orientés afin de pouvoir lier les molécules de l'antigène (ou les anticorps spécifiques) se trouvant dans la phase liquide.
\end{itemize}
Le premier critère provient du fait qu'une forte adhésion hydrophobe sur la surface peut induire un dépliement partiel voire élevé des anticorps \citep{norde2008} tel que les parties variables en perdraient leur réactivité, c'est-à-dire leur capacité à lier des antigènes \citep{buijs1996a}. \`{A} ce titre, il est essentiel de souligner que le mécanisme de dépliement, gouverné par la stabilité interne des protéines, dépend du type d'IgG. On aura donc intérêt à choisir un anticorps de capture peu enclin à la dénaturation. Le second critère est basé sur d'éventuels changements pouvant affecter les conditions de dépôt, c'est-à-dire les propriétés de relaxation des anticorps (changements d'orientation et/ou de conformation) pendant et après l'immobilisation sur la surface hydrophobe \citep{buijs1995}. Selon les mêmes auteurs, la figure \ref{FigIntroOri} (voir introduction) montre une série d'orientations possibles pour les anticorps immobilisés: \textit{end-on} (attachés par une de ses trois parties), \textit{side-on} (attachés par deux de ses parties) et \textit{flat} (attachés par ses trois parties). De ce point de vue, il est montré que les IgG adhérant sur la surface selon une orientation \textit{end-on} sont les plus recherchées étant donné qu'orientées de cette manière, elles permettent à une monocouche d'être bioactive. L'obtention de telles monocouches est dès lors un sujet de premier plan dans le cadre de l'amélioration de l'ELISA \citep{giacomelli2006} et a d'ailleurs donné lieu à de nombreuses études visant à contrôler l'orientation des IgG adsorbées. Dans ces études, diverses stratégies ont été envisagées depuis la modification du support (traitement plasma \citep{hlady1991} voire ajout d'additifs lors de la synthèse du support) jusqu'à la fonctionnalisation des IgG (biotinylation \citep{park2011} ou digestion enzymatique \citep{buijs1995}). Au-delà de ces approches empiriques, d'autres techniques ont été développées afin de sonder de façon plus directe les propriétés des monocouches (bioactivité des IgG et orientations de celles-ci) avec pour objectif une meilleure compréhension des mécanismes d'immobilisation des IgG sur les surfaces hydrophobes compte tenu des conditions de la suspension de dépôt (concentration, température, etc.).

L'approche la plus simple permettant de sonder l'orientation et l'activité des IgG consiste à estimer les quantités maximales accumulées. En effet, M. Bremer \textit{et al}. \citep{bremer2004} indiquent que ces quantités devraient être en rapport avec l'orientation moyenne des IgG déposées sur la surface solide étant donné que certaines orientations d'IgG peuvent nécessiter plus ou moins de place. Par exemple, les IgG \textit{end-on} nécessitant assez peu de place devraient mener à des monocouches plus denses et épaisses.
Malgré le fait que de nombreuses études traitant de l'estimation des quantités <<~adsorbées~>> aient pu être publiées (quantités accumulées $\Theta$ dans les films de protéines), la caractérisation des couches minces de protéines sur les surfaces solides demeure une tâche délicate. Selon les cas, l'estimation de ces quantités peut être faite grâce à l'ellipsométrie \citep{malmsten1995}, la réflectométrie \citep{buijs1996b,schaaf1987} ou le radiomarquage \citep{dewez1997}. Plus récemment, la microbalance à cristal de quartz avec mesure de la dissipation (QCM-D) est devenue un outil de choix \citep{caruso1996,hook1998un}. D'autres études plus anciennes \citep{elgersma1991} sont aussi axées sur des mesures de la déplétion du contenu en IgG de la suspension de dépôt. Toutefois, ce genre de mesure n'est envisageable que dans le seul cas où l'aire du support solide est suffisamment élevée par rapport au volume de la suspension de dépôt, un cas se présentant lorsque l'on utilise un latex de polystyrène. En ce qui concerne les dispositifs imitant les conditions de l'ELISA pour lequel la phase réactive est immobilisée sur des plaques multipuits, une surface plane est requise de telle sorte que le ratio entre le volume de la solution de dépôt et l'aire de la surface de dépôt demeure bien trop élevé rendant alors les quantifications par déplétion peu envisageables au contraire d'un latex constitué d'une dispersion très fine de particules de polystyrène. Des méthodes telles que l'ellipsométrie, la QCM-D ou la réflectométrie deviennent alors nécessaires. Au demeurant, l'estimation des quantités accumulées dans les monocouches de protéines reste une approche très globale et inférer quant à la distribution des orientations des IgG immobilisées et leur intégrité conformationnelle reste assez spéculatif.

Afin de comprendre la manière dont la structuration des monocouches d'IgG peut influer sur l'efficacité des ELISA, il semble intéressant de procéder à des expériences \textit{in situ} sur ces monocouches, c'est-à-dire dans les mêmes conditions que l'ELISA. Pour ce faire, les propriétés des monocouches d'IgG doivent être étudiées dans un environnement liquide (tampon phosphate à pH 7--7,4) sans qu'il ne soit procédé à une quelconque étape de séchage des monocouches ou aucune autre manière d'en affecter la structure. Dès lors, l'utilisation de la microscopie à force atomique (AFM) sous liquide semble une méthode efficace et non invasive afin de sonder la morphologie des monocouches d'IgG. En effet, l'AFM permet de caractériser les morphologies des monocouches à l'échelle microscopique mais est aussi à même d'en fournir diverses propriétés telles que leurs épaisseurs ou les quantités accumulées (quantité d'IgG par unité de surface). Ces propriétés peuvent aussi être confrontées à des résultats ELISA sur les mêmes monocouches afin de déduire, à partir des activités immunologiques obtenues, une série de considérations d'ordre plus général sur l'organisation des monocouches à l'échelle nanométrique. Entre autres, la possibilité d'inférer quant à l'orientation moyenne des IgG de la monocouche (plutôt \textit{end-on} ou \textit{flat}) devrait permettre une discussion approfondie sur la structure des monocouches mais aussi sur leurs mécanismes de construction que l'on pourra utilement confronter aux résultats numériques obtenus dans les chapitres précédents.

L'objet de ce chapitre sera donc de rapporter une série d'expériences au cours desquelles les propriétés de films d'IgG telles que leurs morphologies (AFM), leurs épaisseurs (AFM) et leurs activités immunologiques (ELISA). Ces propriétés seront déduites pour une série de films construits sur une surface hydrophobe (h-SAM) à partir de suspensions de dépôts de diverses concentrations en IgG et pour deux types d'anticorps monoclonaux de souris: un anti-interleukine-2 humaine et un anti-interleukine-6 humaine. Ils ont été sélectionnés compte tenu de leur comportement, le premier montrant habituellement de mauvaises propriétés de fixation alors que le second, au contraire, est connu pour ses bonnes propriétés de fixation. Il sera dès lors possible de constater et de discuter, compte tenu des données présentes dans la littérature, de l'influence de la concentration dans la suspension de dépôt et du type d'IgG sur les propriétés des films obtenus. Au regard de l'état de la littérature, une telle approche, basée sur la confrontation de résultats AFM et ELISA, semble ne pas encore avoir été envisagée.

\section{Matériels et méthodes} 

\subsection{Suspensions de protéines}
Les anticorps monoclonaux de souris anti-interleukine-2 humaine (isotype IgG 1\textgreek{k}, réf.: AHC0422) et anti-interleukine-6 humaine (isotype IgG 1\textgreek{k}, réf.: AHC0562) sont fournis par Life Technologies/Invitrogen Corporation (Camarillo, CA, USA). Les IgG sont conservées dans une solution de tampon phosphate (PBS) ayant un pH entre 8 et 7,5 à des températures de respectivement $4^\circ$C et $-20^\circ$C jusqu'à leur usage. De l'albumine de sérum bovin (BSA, fraction V, essentiellement sans protéase $\geqslant$ 92 \%) est obtenue chez Sigma-Aldrich (réf.: 05479). Toutes les dilutions sont faites dans un tampon phosphate (PBS) 1 mol$\cdot$L$^{-1}$ à pH 7 et stockées à $4^\circ$C avant usage.

\subsection[Estimation de l'activité immunologique]{Estimation de l'activité immunologique par ELISA}
Des trousses IL-6-EASIA et IL-2-EASIA sont fournies par DIAsource ImmunoAssays (Louvain-la-Neuve, Belgique) et sont utilisées afin de déterminer l'activité immunologique des monocouches d'IgG. Les protocoles des ELISA détaillés par DIAsource sont globalement suivis. Toutefois, le but de l'ELISA étant de quantifier un antigène dans la suspension, deux adaptations majeures doivent être introduites afin d'avoir accès à l'activité immunologique des dépôts d'anticorps. La première adaptation consiste à utiliser des surfaces d'anticorps de capture synthétisées par nos propres soins en lieu et place de celles fournies avec les trousses ELISA de DIAsource ImmunoAssays. La seconde adaptation repose sur l'usage des seules solutions de calibrage concentrées à 201 pg$\cdot$mL$^{-1}$, et ce, pour les deux anticorps. La soustraction de la ligne de base sera réalisée en utilisant les solutions de calibrage à 0 pg$\cdot$mL$^{-1}$ fournies avec les trousses. L'anticorps spécifique étant directement fixé sur la surface, on se trouve dans le cadre des ELISA <<~sandwich~>>. Les dépôts d'IgG sur 8 puits d'une microplaque hydrophobe (polystyrène) en contenant 96 sont obtenus selon le protocole suivant:
\begin{itemize}
\item Chaque puits est rincé 3 fois avec 300 \textgreek{m}L de PBS;
\item Injection dans chaque puits de 100 \textgreek{m}L des suspensions de dépôt contenant les IgG (concentrations: 50; 25; 12,5; 7,81; 3,9; 1,95; 0,98 et 0,49 \textgreek{m}g$\cdot$mL$^{-1}$);
\item Incubation pendant 10 heures à $4^\circ$C;
\item Aspiration du liquide et rinçage des puits 3 fois avec 300 \textgreek{m}L de PBS;
\item 300 \textgreek{m}L d'une solution de BSA concentrée à 5 mg$\cdot$mL$^{-1}$ sont injectés dans chaque puits;
\item Incubation pendant 30 minutes à $4^\circ$C;
\item Finalement, le liquide est aspiré et les puits rincés 3 fois par 300 \textgreek{m}L de PBS.
\end{itemize}
Ensuite, le protocole habituel des ELISA est suivi dans tous ces puits en utilisant les solutions de calibrage mentionnées ci-dessus. La lecture des absorbances est faite avec une longueur d'onde de 450 nm contre un filtre de référence fixé à 650 nm.

\subsection[Préparation des monocouches]{Préparation des monocouches pour l'AFM}
Les films d'IgG sont construits sur des surfaces hydrophobes, elles-mêmes préparées sur des supports en verre (1,1 mm d'épaisseur et 12 mm de diamètre). Les tranches de verre sont d'abord nettoyées dans une solution H$_2$SO$_4$/H$_2$O$_2$ 3/1 pendant 3 heures à température ambiante avant d'être intensivement rincés dans de l'eau ultrapure (Milli-Q, Millipore Corporation). Les substrats sont ensuite passivés avec du NaOH 20~\%, rincés abondamment à l'eau ultrapure et ensuite séchés sous azote. Dans un pulvérisateur cathodique (EMITECH, K575 Turbo, UK), ces substrats sont recouverts d'une couche de chrome d'environ 10 nm d'épaisseur et d'une couche d'approximativement 50 nm d'or. Les surfaces d'or ainsi obtenues sont immergées pendant 14 heures à température ambiante et à l'abri de la lumière dans une solution 1 mmol$\cdot$L$^{-1}$ de CH$_3$(CH$_2$)$_{11}$SH ($\geqslant$ 98~\%, Sigma-Aldrich, réf.: 471364) dilué dans de l'éthanol ($\geqslant$ 99,8~\%, Fluka, réf.: 02854) afin d'obtenir une monocouche de thioalcanes (h-SAM) auto-assemblés sur l'or dont la structure est schématisée à la figure \ref{FigSurfHydrophobe}. Après ce recouvrement, les surfaces hydrophobes obtenues sont rincées à l'éthanol puis au PBS et stockées dans des puits (plaque 24 puits de Greiner Bio-One) immergées dans 1 mL de PBS. Après stabilisation thermique à 4$^\circ$C pendant 1 heure, le PBS est retiré à l'aide d'une micropipette et les solutions de dépôt contenant les anticorps sont mises au contact de la surface. Pour ce faire, 1 mL des solutions de dépôts (anticorps dans le PBS: 50; 25; 12,5; 7,81; 3,9; 1,95; 0,98 et 0,49 \textgreek{m}g$\cdot$mL$^{-1}$) est précautionneusement injecté en prenant soin de viser le centre de la surface et en faisant que le jet sortant de la micropipette lui soit perpendiculaire (voir figures \ref{FigEcoulement}.B et C). Après incubation pendant 4 heures à 4$^\circ$C, les solutions de dépôts sont aspirées et les surfaces obtenues sont soigneusement rincées au PBS. Durant une courte période de temps, les surfaces sont stockées dans du PBS avant caractérisation.

\begin{figure}[h]\centering
\includegraphics*[width=0.7\textwidth]{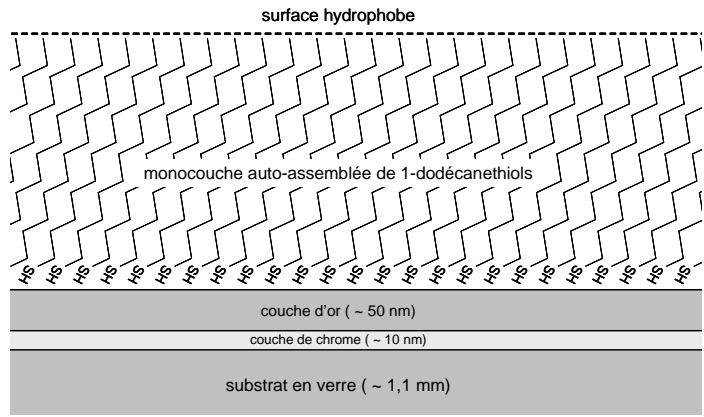}
\caption[Constitution de la surface hydrophobe]{Constitution de la surface hydrophobe sur laquelle les anticorps viendront se déposer. Une monocouche <<~h-SAM~>> très hydrophobe et rigide constituée par l'auto-assemblage de 1-dodécanethiols (CH$_3$(CH$_2$)$_{11}$SH) est construite sur une couche d'or adhérant au substrat en verre grâce à une mince couche de chrome.}\label{FigSurfHydrophobe}
\end{figure}

\subsection[Estimation des propriétés des monocouches]{Estimation des propriétés des monocouches par AFM}
Les images de la topographie des surfaces immergées dans le PBS sont obtenues à température ambiante à l'aide d'un microscope à force atomique Asylum MFP-3D (Santa-Barbara, CA, USA) implémenté avec le logiciel IGOR Pro 6.04 (Wavemetrics, Lake Osewego, OR, USA). Les monocouches d'IgG sont d'abord grattées sur une aire de 1 \textgreek{m}m $\times$ 1 \textgreek{m}m en appliquant une force d'appui élevée d'environ 200 nN (très rigide, le substrat hydrophobe n'est pas altéré par ce traitement). Le rayon de la pointe (intacte) étant d'approximativement 5 nm, la pression exercée sur les surfaces afin d'effectuer ce grattage s'élève à environ 2,5 GPa. Ensuite, les surface grattées sont imagées avec le même levier. Des images de 5 \textgreek{m}m $\times$ 5 \textgreek{m}m sont enregistrées en utilisant une force d'appui maximale de 250 pN et une vitesse de balayage de 1~Hz. Les leviers de forme conique en nitrure de silicium (Si$_3$N$_4$) sont fournis par Atomic Force (OMCL-TR400PSA-3, Olympus, Japan).

Les propriétés des monocouches d'IgG telles que leurs épaisseurs, leurs rugosités (aussi pour les substrats hydrophobes découverts par le grattage), les taux de recouvrement ($\phi_\infty$) et les volumes superficiels de ces derniers seront obtenus des images AFM grâce à l'environnement Matlab. La manière dont les différentes propriétés venant d'être mentionnées seront obtenues peut être explicitée à partir de la figure \ref{65dg46d8fg47}, figure illustrant l'aspect attendu des images AFM de ces monocouches.

\begin{figure}[h]\centering
\includegraphics*[width=0.55\textwidth]{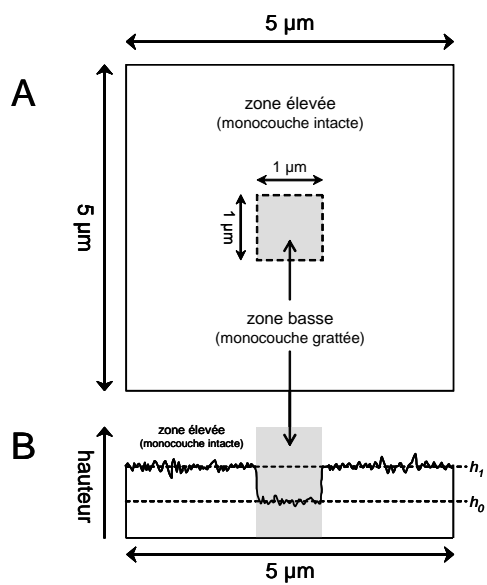}
\caption[Aspect attendu des images AFM de monocouches d'IgG]{\textbf{A}: schéma d'une image topographique obtenue par AFM (5 \textgreek{m}m $\times$ 5 \textgreek{m}m) d'une monoouche d'IgG montrant la monocouche intacte et la zone (1 \textgreek{m}m $\times$ 1 \textgreek{m}m) qui en aura été grattée (au centre en gris) afin de faire apparaître le substrat hydrophobe sous-jacent. \textbf{B}: schéma du profil des hauteurs au niveau de la zone grattée montrant la différence entre la zone dans laquelle la monoouche aura été conservée intacte (hauteur moyenne $h_1$) et la zone où elle aura été retirée par grattage (hauteur moyenne $h_0$).}\label{65dg46d8fg47}
\end{figure}

Premièrement, les épaisseurs seront simplement calculées en utilisant la différence $h_1-h_0$ entre les hauteurs moyennes de la monocouche intacte ($h_1$) et la zone dont elle aura été retirée par grattage ($h_0$). Ces hauteurs seront obtenues en échantillonnant aléatoirement une portion d'aire de chacune des deux zones. Deuxièmement, les rugosités de ces deux zones seront évaluées en utilisant les écarts-types des hauteurs et permettront de cette façon d'obtenir la rugosité de la monocouche et celle du substrat hydrophobe sous-jacent. L'intervalle de confiance de l'épaisseur sera inféré statistiquement des écarts-types des deux zones. Troisièmement, les monocouches pouvant ne pas couvrir entièrement les substrats, les taux de recouvrement de ces derniers seront obtenus en calculant la proportion (en \%) des points appartenant à une aire choisie aléatoirement dans la zone intacte et dont les hauteurs seront égales ou supérieures à la hauteur moyenne à laquelle le substrat hydrophobe se situe, c'est-à-dire $h_0$. Finalement, les volumes superficiels (nombre de mm$^3$ d'IgG portés par cm$^2$ de surface) seront évalués par division du volume (somme des hauteurs multipliée par l'aire d'un pixel de l'image AFM) d'une zone intacte par son aire. Ces volumes superficiels serviront de base à un calcul des quantités accumulées $\Theta_\infty$ (nombre de pmol d'IgG porté par cm$^2$ de surface) présent dans la discussion des résultats.

\section{Résultats et discussion}

\subsection[Bioactivité des monocouches]{Bioactivité des monocouches mesurée par ELISA}

Les résultats ELISA (mesures des absorbances) obtenus après soustraction de la ligne de base sont montrés sur le graphique de la figure \ref{FigELISA} pour les deux anticorps. Comme mentionné ci-dessus, les concentrations testées (données en logarithmes) sont 50; 25; 12,5; 7,81; 3,9; 1,95; 0,98 et 0,49 \textgreek{m}g$\cdot$mL$^{-1}$. Les points expérimentaux sont accompagnés d'une régression linéaire avec un intervalle de confiance à 95~\%. On peut observer sur la figure \ref{FigELISA} que l'augmentation de la concentration en IgG dans la suspension de dépôt est corrélée à une augmentation de l'activité immunologique de l'IgG anti-IL-6, ce qui n'est pas le cas pour l'anticorps anti-IL-2. Ces observations sont en accord avec les mauvaises propriétés de fixation postulées pour l'anticorps anti-IL-2 alors que l'anti-IL-6 est bien connu pour ses bonnes capacités de fixation dans le cadre de l'ELISA. Dans un premier temps, de telles propriétés d'adhésion devraient plutôt être interprétées comme une meilleure capacité à former des monocouches d'IgG efficaces pour l'ELISA.

\begin{figure}[t]\centering
\includegraphics*[width=0.75\textwidth]{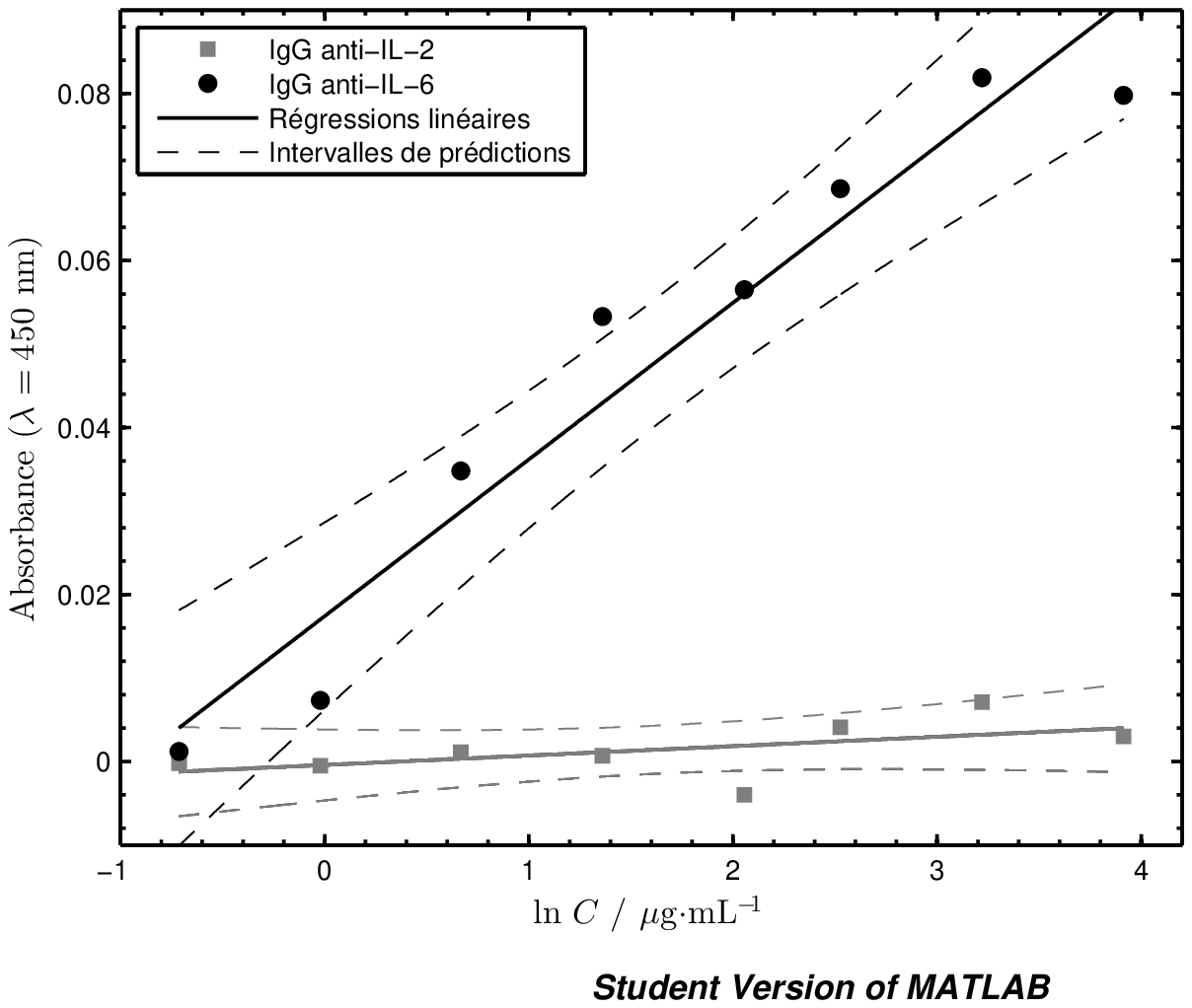}
\caption[Résultats des ELISA]{Résultats des ELISA (absorbances à 450 nm) illustrant l'activité immunologique des films d'anticorps monoclonaux de souris anti-interleukine-2 humaine ($\bullet$) et anti-interleukine-6 humaine (\begin{tiny}$\blacksquare$\end{tiny}) en fonction du logarithme de la concentration $C$ en IgG dans la suspension de dépôt en \textgreek{m}g$\cdot$mL$^{-1}$. Les points expérimentaux sont accompagnés de régressions linéaires et d'intervalles de prédictions à 95~\% de confiance.}\label{FigELISA}
\end{figure}

\subsection[Morphologie et organisation des monocouches]{Morphologie et organisation des monocouches par AFM}

\subsubsection{Morphologie générale des monocouches}

La figure \ref{FigImAFM} présente une série d'images obtenues par AFM des monocouches d'anticorps anti-IL-2 (figures \ref{FigImAFM}.A à \ref{FigImAFM}.D) et anti-IL-6 (figures \ref{FigImAFM}.E et \ref{FigImAFM}.H) dont les zones ayant été grattées sont visibles (plus ou moins au centre). Ces images mettent nettement en évidence la différence entre les monocouches d'anticorps anti-IL-2 et anti-IL-6: les premières ayant un aspect rugueux avec une série de vallées et de collines alors que les secondes apparaissent assez lisses. Sur les figures \ref{FigImAFM}.A à \ref{FigImAFM}.D, les vallées ont une forme assez régulière, sont espacées d'environ 100 à 200 nm (distance entre deux sommets), présentent une profondeur de 3 à 4 nm (distance entre le sommet et le fond) et semblent légèrement plus importantes sur la figure \ref{FigImAFM}.C. Le motif montré sur la figure \ref{FigImAFM}.D change drastiquement par rapport aux figures \ref{FigImAFM}.A, B et C pour adopter un aspect de plaine ponctuée de quelques collines escarpées. Cette augmentation peut être constatée sur les profils transversaux des figures \ref{FigPub7}.A à D sur lesquelles les zones correspondant à la portion de la monocouche grattée ont été grisées.

Les comportements observés pour l'anti-IL-2 peuvent être comparés avec les monocouches d'IgG déposées sur de la silice méthylée (imagées par AFM après séchage sous azote) ayant été rapportées par B. W\"{a}livaara \textit{et al.} \citep{walivaara1995}. Les monocouches déposées sur de la silice hautement méthylée (très hydrophobe) montrent un profil vallonné similaire qualifié de \textit{dendrite-like} par M. Malmsten \citep{malmsten1995} alors que les monocouches d'anti-IL-6 correspondraient à un profil de type \textit{strand-like}.
B. W\"{a}livaara \textit{et al.} \citep{walivaara1995} expliquent l'origine de la structure dendritique sur base d'aspects thermodynamiques relatifs à l'étape de séchage de la monocouche. Le même genre de structure étant obtenue pour les monocouches d'anti-IL-2 alors qu'aucune étape de séchage n'est réalisée dans ce travail laisse à penser qu'un tel comportement pourrait aussi être attribué, du moins en partie, à une caractéristique propre aux anticorps se déposant sur la surface. Toutefois, et bien qu'aucun séchage n'ait été expérimenté dans ce travail, il ne semble pas interdit de penser qu'il pourrait nettement favoriser l'apparition de monocouches ayant cette structure dendritique.

Cette variabilité illustre bien les différents mécanismes d'adhésion des deux IgG. En effet, comme cela a déjà été mentionné, l'anti-IL-6 est connu pour ses bonnes propriétés d'adhésion alors que l'anti-IL-2 est connu pour le contraire. Par ailleurs, en tenant compte des résultats ELISA rapportés ci-avant, il apparaît clairement que les anticorps menant à des monocouches \textit{strand-like} devraient être bien plus efficients dans le cadre de l'ELISA alors que ceux caractérisés par un comportement d'adhésion menant à des monocouches dendritiques pourraient ne pas être souhaitables.

\begin{figure}[hp]\centering
\begin{tabular}{cccc}
\raisebox{3.5cm}{\Large{\textbf{A}}}&
\includegraphics*[width=0.3\textwidth]{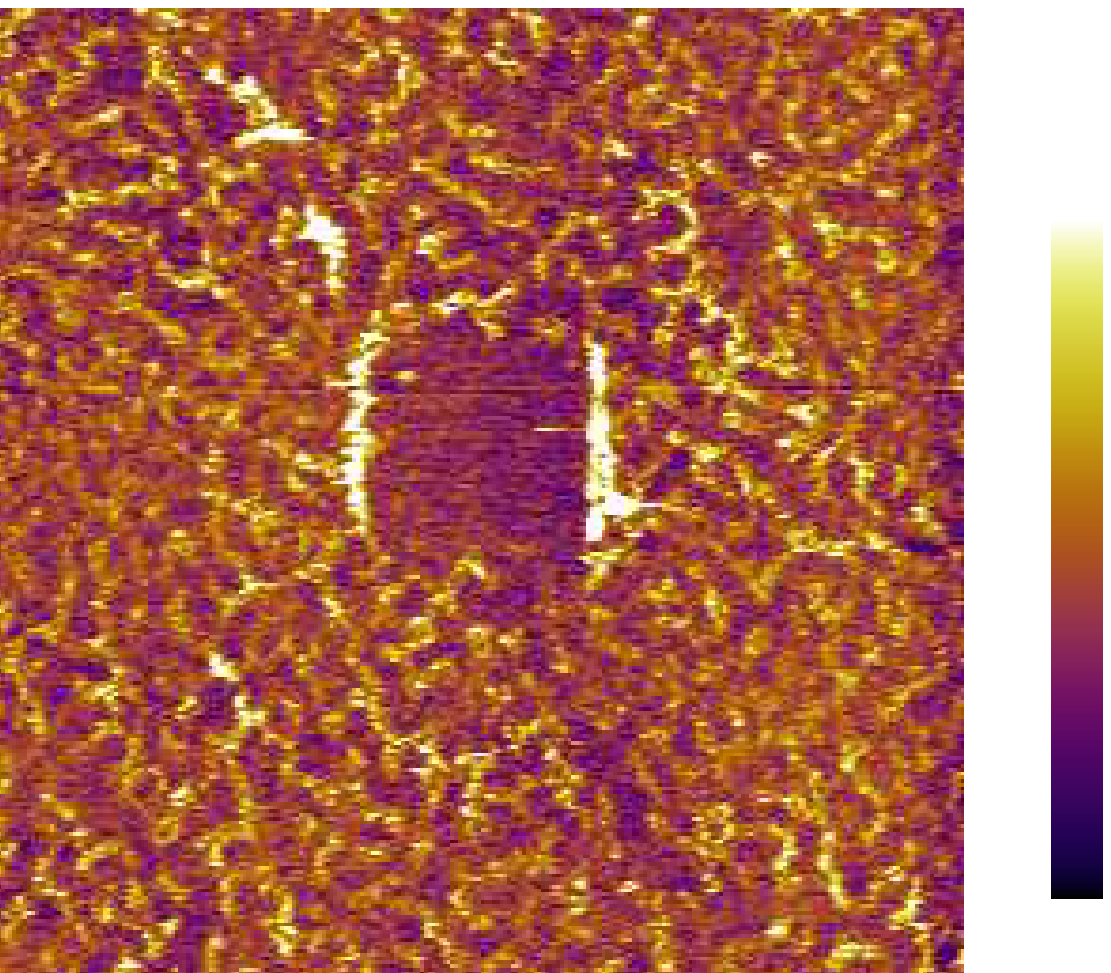}&
\includegraphics*[width=0.3\textwidth]{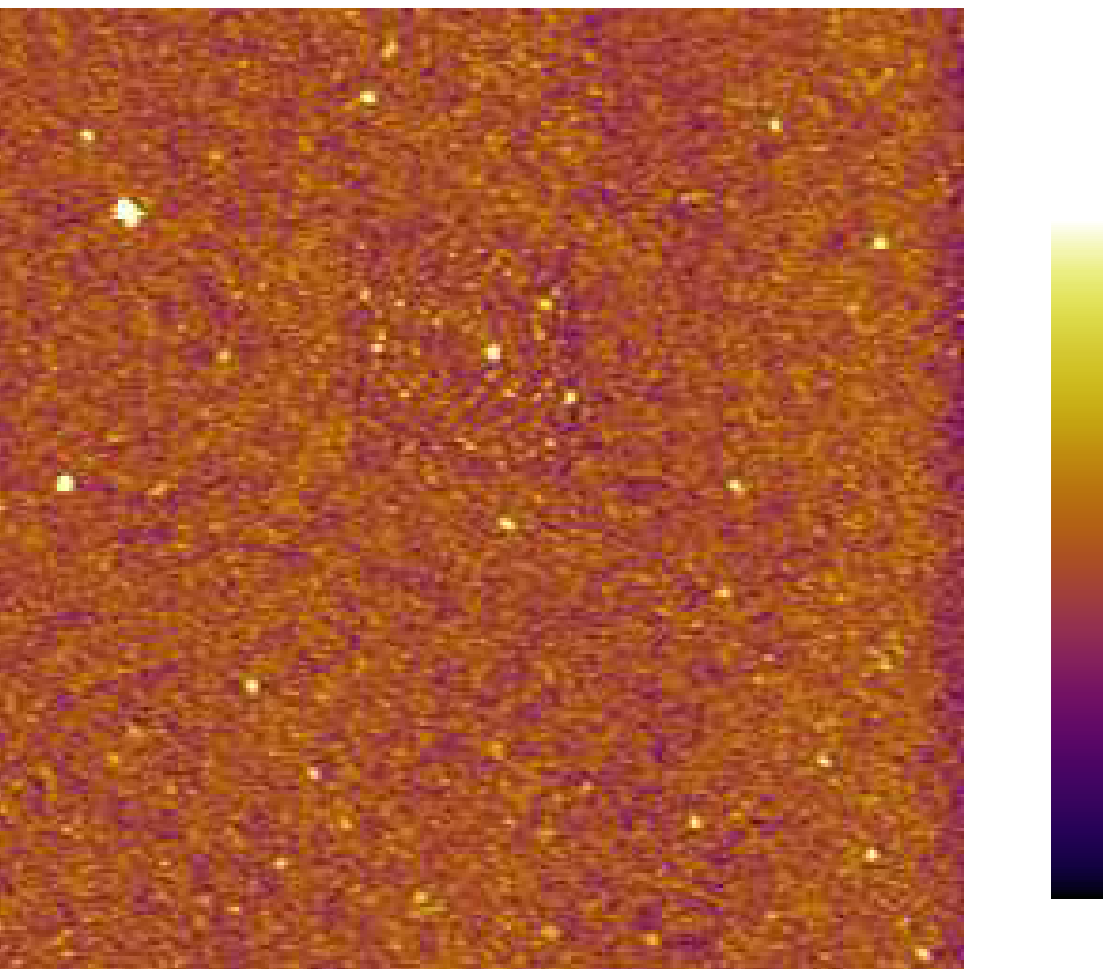}&\raisebox{3.5cm}{\Large{\textbf{E}}}\\
\raisebox{3.5cm}{\Large{\textbf{B}}}&
\includegraphics*[width=0.3\textwidth]{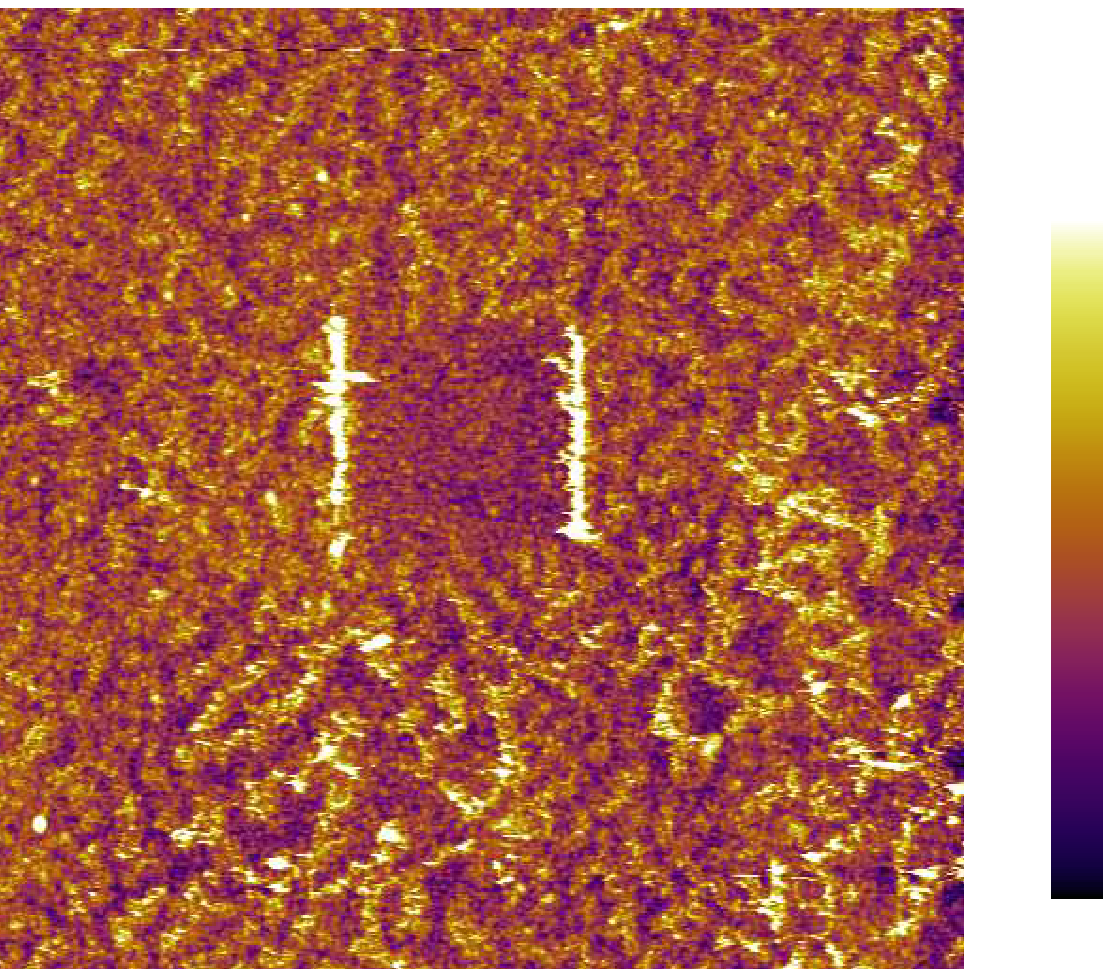}&
\includegraphics*[width=0.3\textwidth]{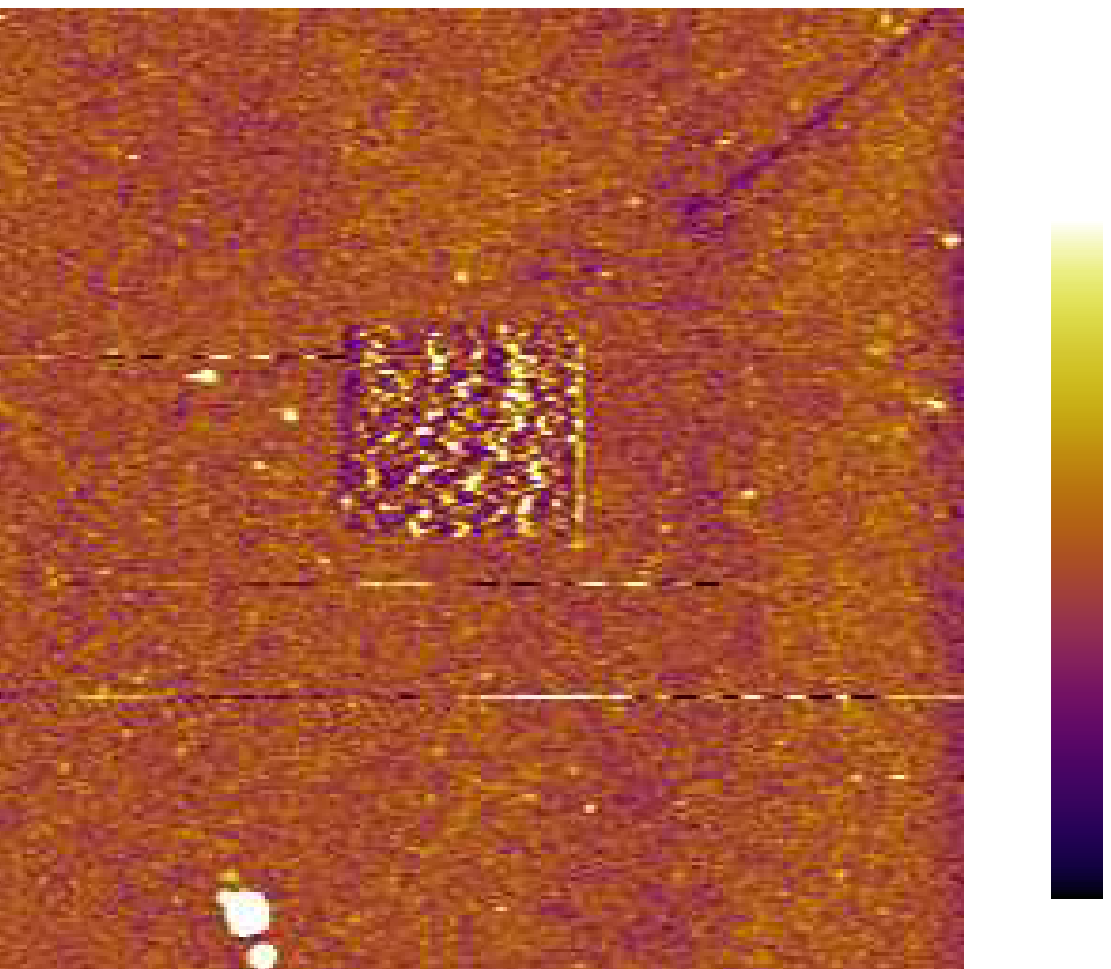}&\raisebox{3.5cm}{\Large{\textbf{F}}}\\
\raisebox{3.5cm}{\Large{\textbf{C}}}&
\includegraphics*[width=0.3\textwidth]{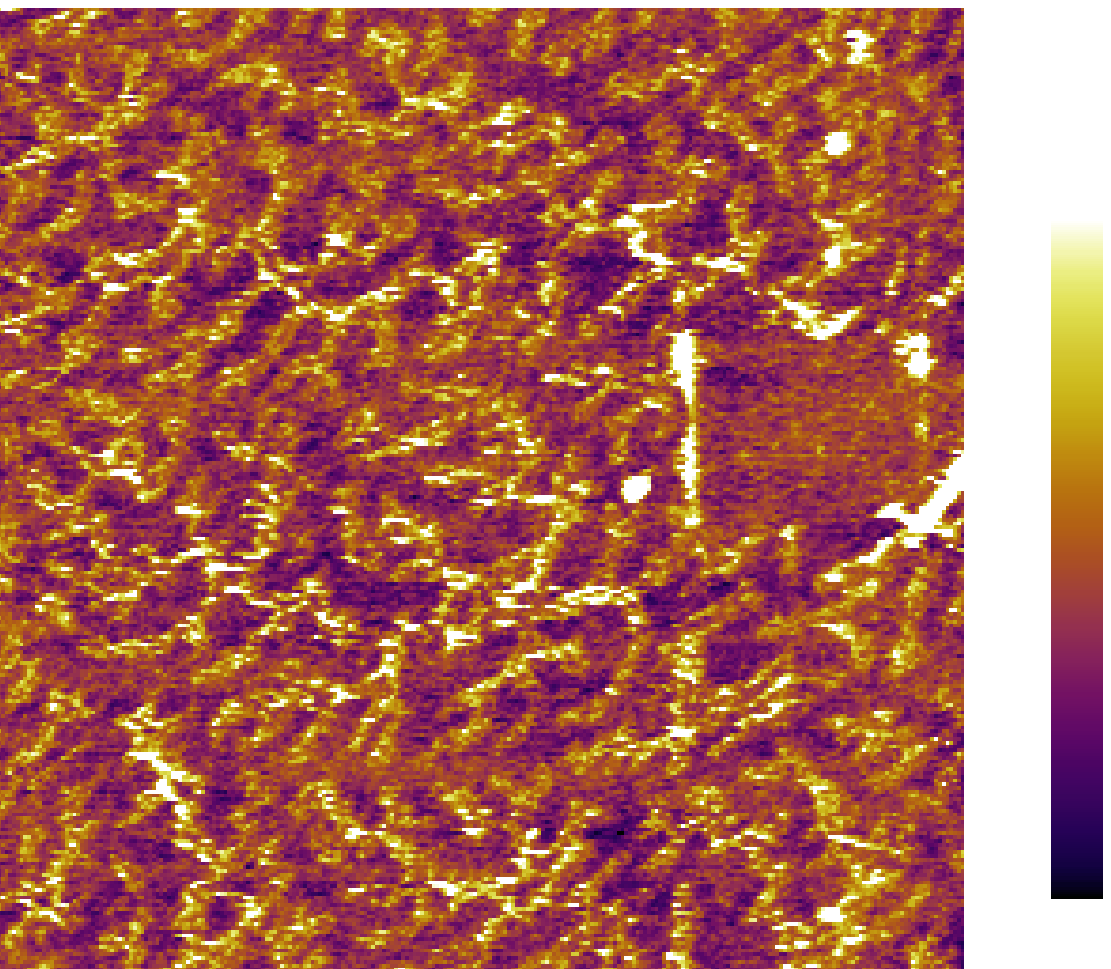}&
\includegraphics*[width=0.3\textwidth]{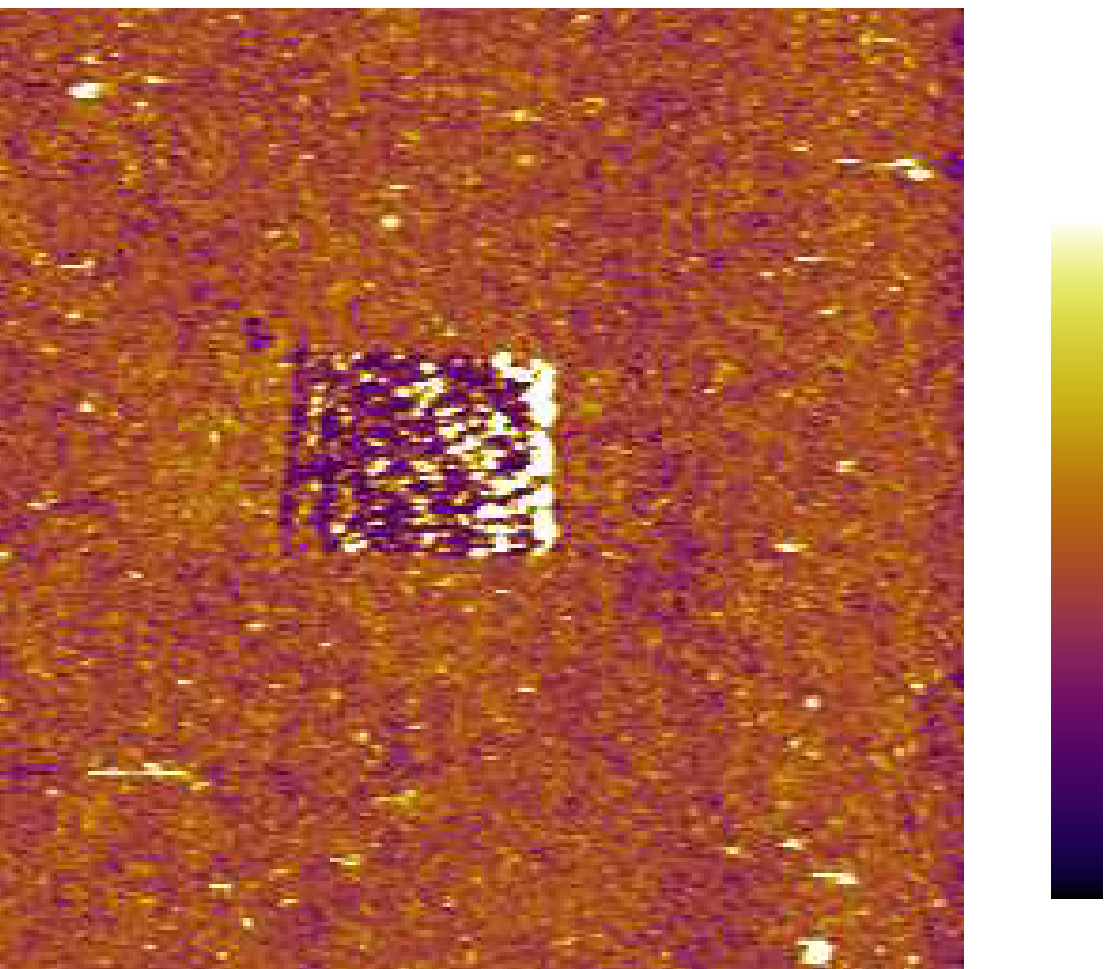}&\raisebox{3.5cm}{\Large{\textbf{G}}}\\
\raisebox{3.5cm}{\Large{\textbf{D}}}&
\includegraphics*[width=0.3\textwidth]{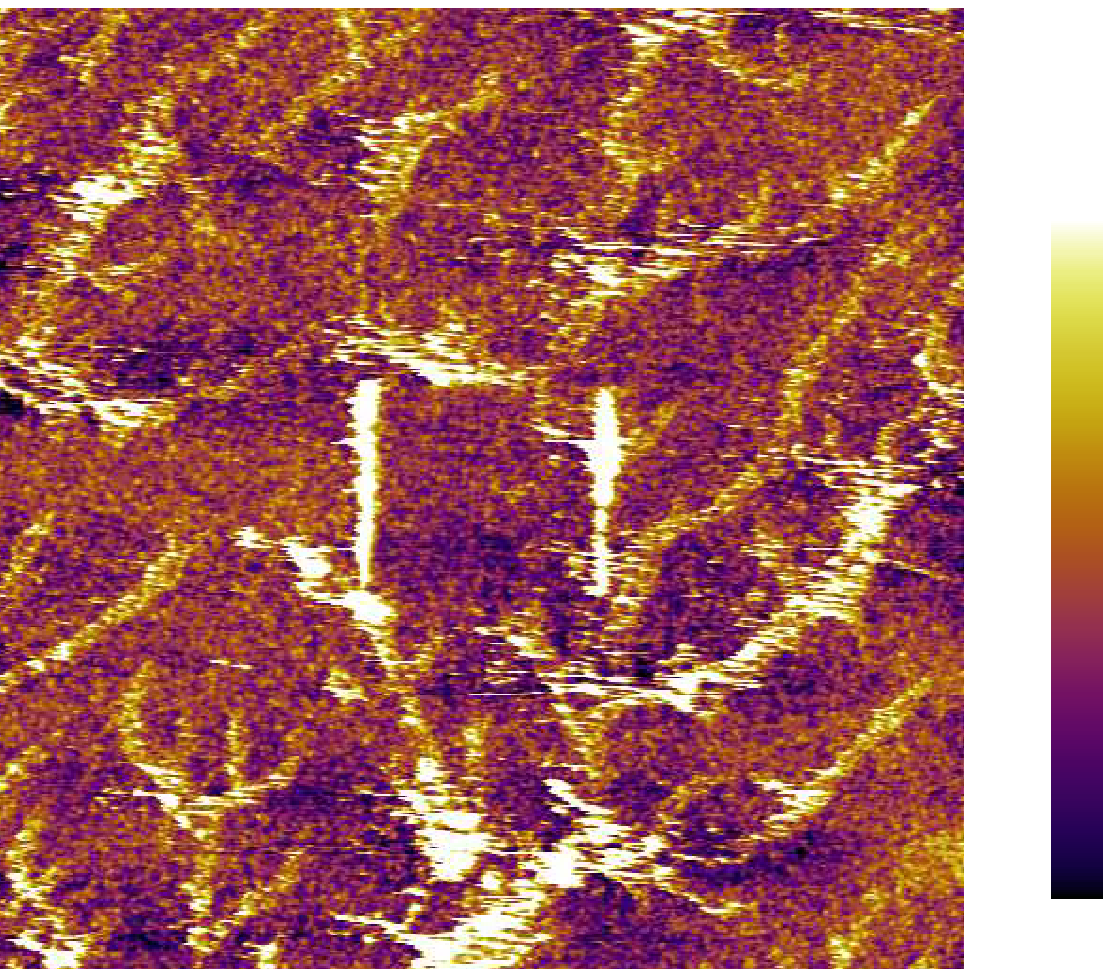}&
\includegraphics*[width=0.3\textwidth]{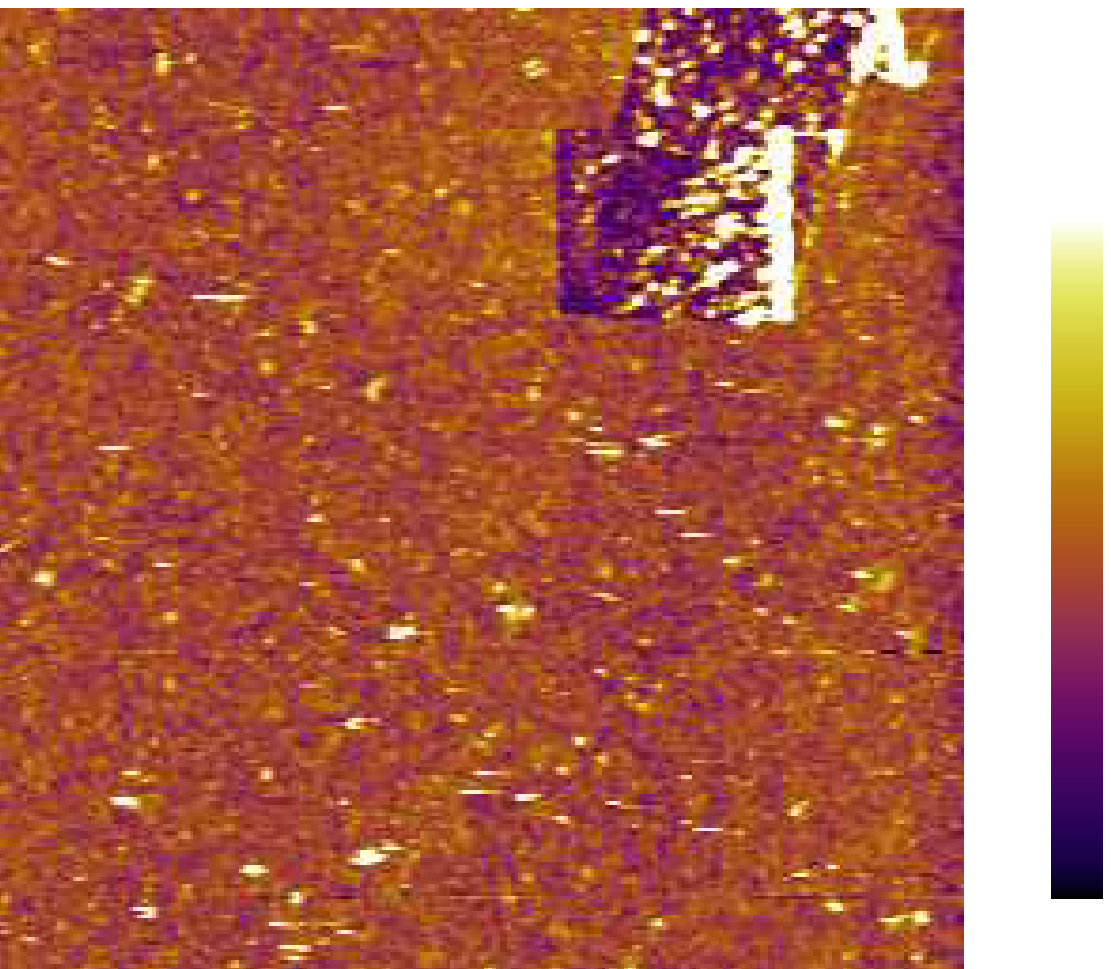}&\raisebox{3.5cm}{\Large{\textbf{H}}}\\
\end{tabular}
\caption[Images AFM des surfaces recouvertes d'IgG]{Images AFM (5 \textgreek{m}m $\times$ 5 \textgreek{m}m) de la topographie de surfaces hydrophobes couvertes par une monocouche d'IgG et comprenant une zone grattée (1 \textgreek{m}m $\times$ 1 \textgreek{m}m). \textbf{A} à \textbf{D}: monocouches d'anticorps monoclonaux de souris anti-interleukine-2 obtenues à partir de suspensions de dépôt concentrées à 1,56; 6,25; 25 et 50 \textgreek{m}g$\cdot$mL$^{-1}$. \textbf{E} à \textbf{H}: monocouches d'anticorps monoclonaux de souris anti-interleukine-6 obtenues à partir de suspensions de dépôt concentrées à 1,56; 6,25; 25 et 50 \textgreek{m}g$\cdot$mL$^{-1}$.}\label{FigImAFM}
\end{figure}

\begin{figure}[t]\centering
\begin{tabular}{cccc}
\raisebox{1.5cm}{\Large{\textbf{A}}}&
\includegraphics*[width=0.35\textwidth,trim=8 35 25 25]{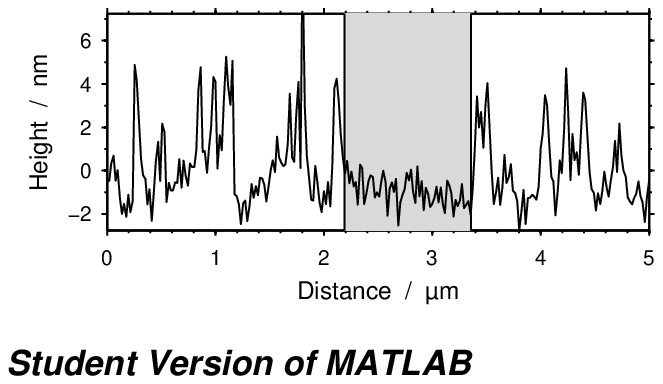}&
\includegraphics*[width=0.35\textwidth,trim=8 35 25 25]{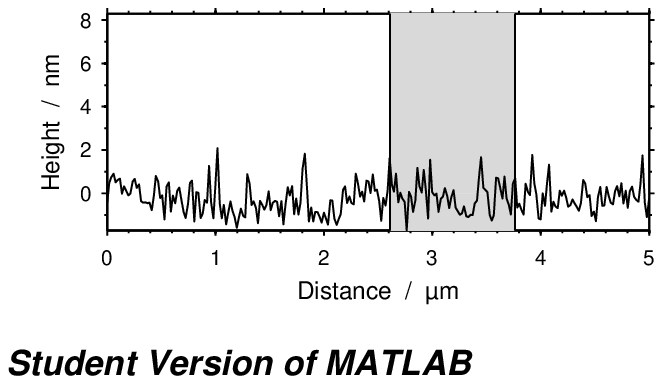}&\raisebox{1.5cm}{\Large{\textbf{E}}}\\
\raisebox{1.5cm}{\Large{\textbf{B}}}&
\includegraphics*[width=0.35\textwidth,trim=8 35 25 25]{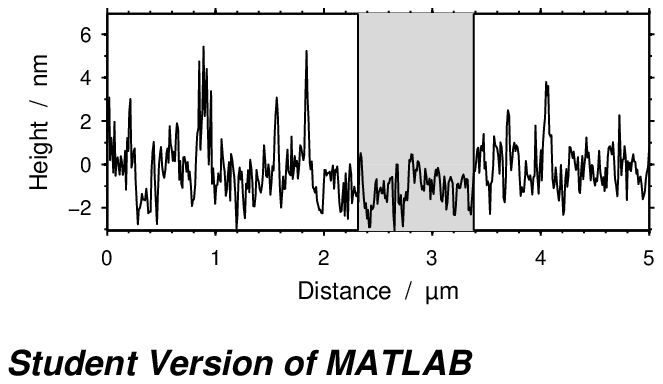}&
\includegraphics*[width=0.35\textwidth,trim=8 35 25 25]{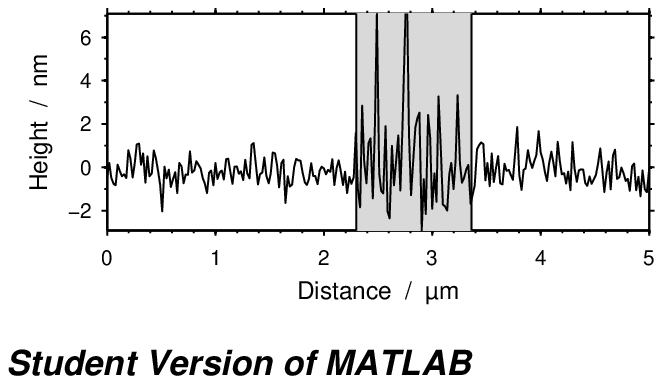}&\raisebox{1.5cm}{\Large{\textbf{F}}}\\
\raisebox{1.5cm}{\Large{\textbf{C}}}&
\includegraphics*[width=0.35\textwidth,trim=8 35 25 25]{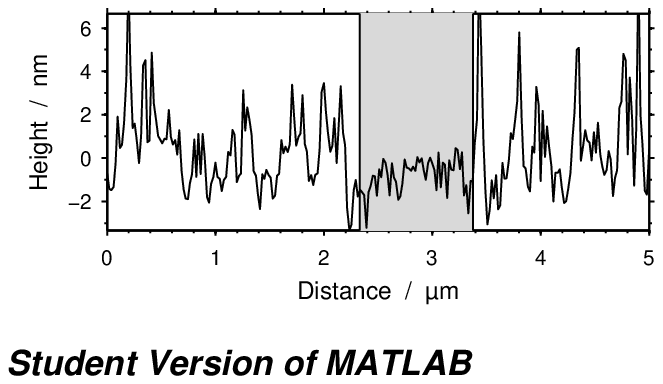}&
\includegraphics*[width=0.35\textwidth,trim=8 35 25 25]{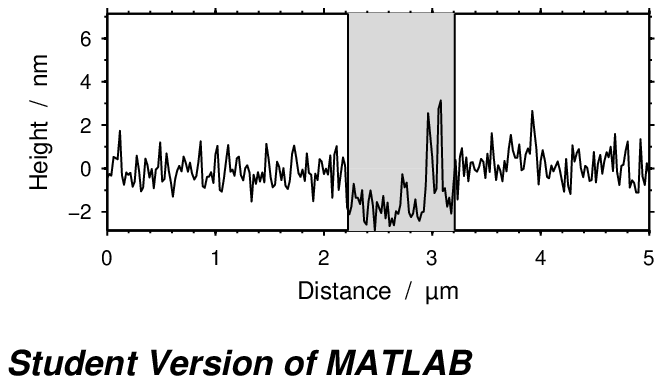}&\raisebox{1.5cm}{\Large{\textbf{G}}}\\
\raisebox{1.5cm}{\Large{\textbf{D}}}&
\includegraphics*[width=0.35\textwidth,trim=8 35 25 25]{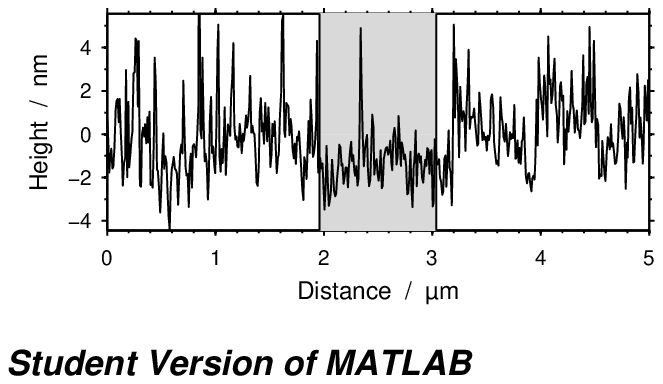}&
\includegraphics*[width=0.35\textwidth,trim=8 35 25 25]{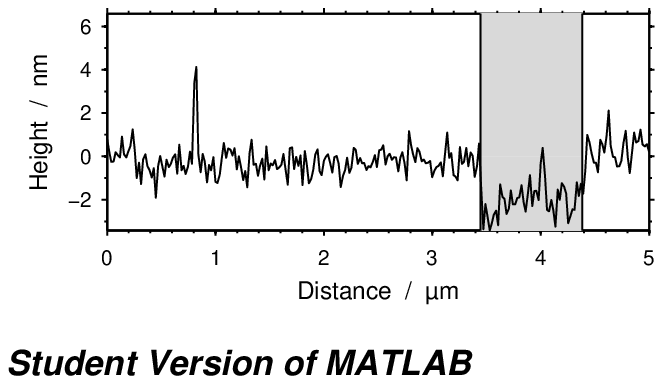}&\raisebox{1.5cm}{\Large{\textbf{H}}}\\
\end{tabular}
\caption[Sections transversales des surfaces recouvertes d'IgG]{Sections transversales des surfaces recouvertes d'IgG obtenues à partir des images AFM de la figure \ref{FigImAFM}. Les aires grattées correspondent aux zones grisées et les sections sont prises verticalement sur les images de la figure \ref{FigImAFM}. \textbf{A} à \textbf{D}: sections transverales pour les films d'anti-interleukine-2 obtenus à des concentrations de dépôt de 1,56; 6,25; 25 et 50 \textgreek{m}g$\cdot$mL$^{-1}$. \textbf{E} à \textbf{H}: idem pour l'anti-interleukine-6.}\label{FigPub7}
\end{figure}

\subsubsection{\'{E}paisseurs et rugosités des monocouches}

Les tableaux \ref{TabPub1} et \ref{TabPub2} donnent respectivement les résultats AFM pour l'anti-IL-2 et l'anti-IL-6. Comme mentionné ci-dessus, les épaisseurs des couches d'anticorps sont estimées en soustrayant la hauteur moyenne de la zone grattée de celle de la zone demeurée intacte. Il semble difficile d'affirmer que l'épaisseur des monocouches d'anti-IL-2 augmente au regard des barres d'erreur mais la rugosité semble, elle, rester assez constante. Toutefois, pour les monocouches d'anti-IL-6, le tableau \ref{TabPub2} montre une augmentation d'un facteur dix (0,27 à 2,88 nm) de l'épaisseur des monocouches en fonction des concentrations testées (1,56 à 50 \textgreek{m}g$\cdot$mL$^{-1}$). Ces épaisseurs sont reportées sur le graphique de la figure \ref{FigAFMplot} qui montre clairement une faible réponse de l'anti-IL-2 à l'augmentation de la concentration, contrastant nettement avec l'anti-IL-6 pour lequel les épaisseurs des monocouches augmentent drastiquement.

\begin{figure}[t]\centering
\includegraphics*[width=0.75\textwidth]{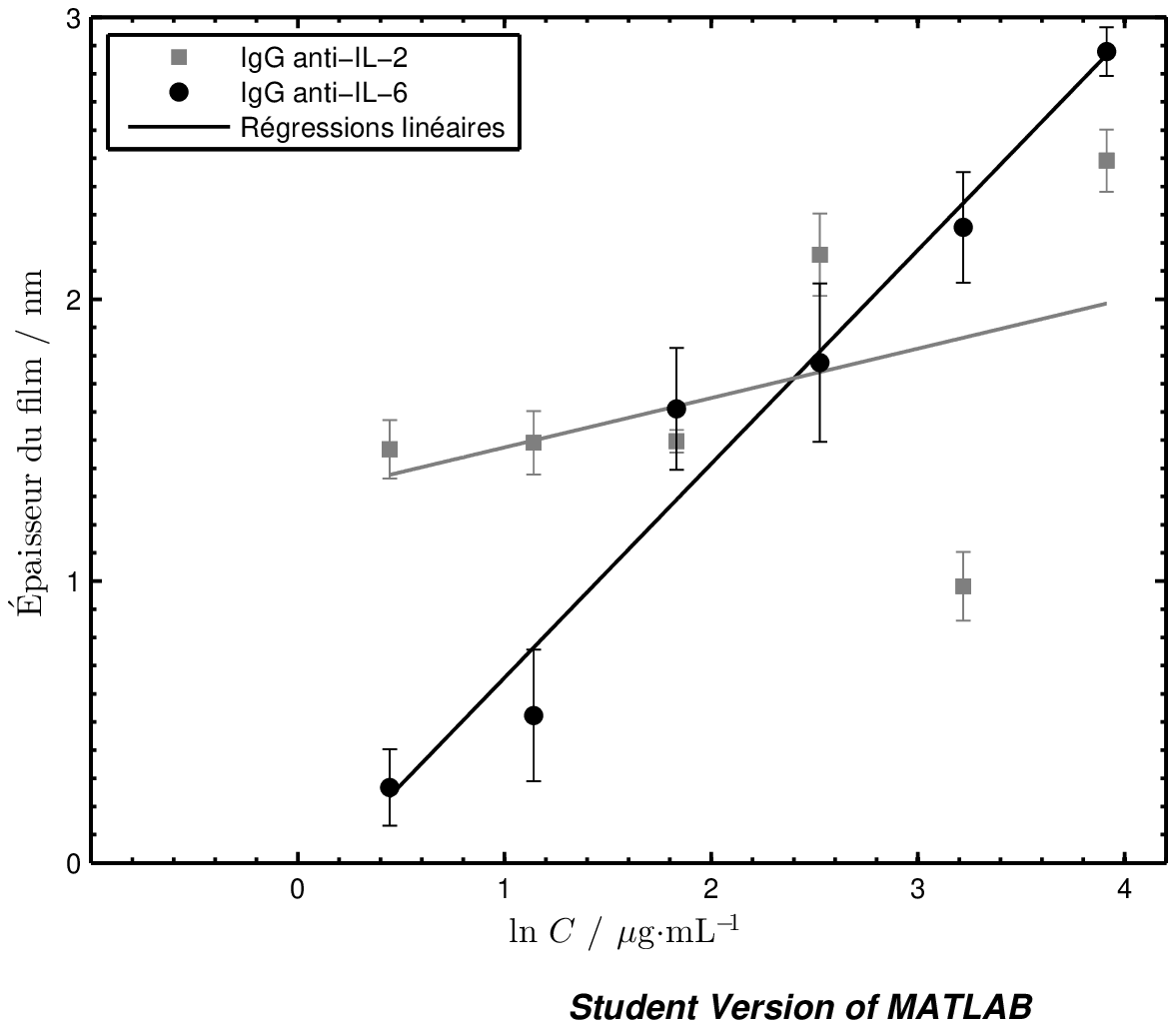}
\caption[\'{E}paisseurs des films d'IgG en fonction de la concentration]{\'{E}paisseurs (en nm) des films d'anticorps anti-interleukine-6 humaine ($\bullet$) et anti-interleukine-2 humaine (\begin{tiny}$\blacksquare$\end{tiny}) en fonction du logarithme de la concentration $C$ en IgG dans la suspension de dépôt. Les points expérimentaux sont accompagnés d'une regression linéaire.}\label{FigAFMplot}
\end{figure}

Les chiffres obtenus pour les épaisseurs semblent en assez bon accord avec ceux de la littérature. En effet, F. Caruso \textit{et al.} \citep{caruso1996} ont obtenu par AFM des mesures d'épaisseurs autour de 7 nm pour des couches d'IgG déposées sur du mica recouvert d'or, et ce, pour une concentration d'IgG dans la suspension de dépôt de 50 \textgreek{m}g$\cdot$mL$^{-1}$. Sachant que les dépôts d'IgG faits dans le cadre de ce travail l'ont été sur des surfaces plus hydrophobes que l'or (h-SAM) possédant dès lors une très forte affinité pour les protéines et à même d'en affecter la conformation, il n'apparaît pas étonnant d'obtenir des valeurs d'épaisseurs plus faibles s'échelonnant entre 2,49 nm (anti-IL-2) et 2,88 nm (anti-IL-6). L'hydrophobicité des surfaces de dépôt aurait donc une forte influence sur l'épaisseur des couches qui y sont déposées: plus elles seraient hydrophobes, plus la monocouche obtenue serait fine. Par ailleurs, il est intéressant de noter que M. Malmsten \citep{malmsten1995} rapporte des valeurs autour de 17-18 nm pour des épaisseurs de monocouches déposées sur de la silice méthylée obtenues par mesures ellipsométriques. Eu égard à la taille des IgG qui n'excède pas les 14 nm (chaque fragment Fc ou F(ab') mesure environ 7 nm), ces quantités semblent en contradiction avec ce qu'il serait possible d'espérer comme épaisseur pour une monocouche entièrement peuplée d'IgG \textit{end-on} et ne sera dès lors pas prise en compte dans la présente discussion.

Les structures des monocouches montrées aux figures \ref{FigELISA} et \ref{FigAFMplot} illustrent un lien évident entre la concentration de la suspension de dépôt, l'épaisseur des couches mais aussi leurs activités immunologiques. De ces faits, une corrélation apparaît directement entre l'activité immunologique et l'épaisseur des monocouches: plus l'épaisseur est élevée, plus l'activité sera importante. Comme indiqué dans la section \ref{65g46s54g687tr}, pour les anti-IL-6, une telle corrélation laisse à penser que les faibles concentrations fournissent des monocouches contenant une fraction importante d'anticorps orientés \textit{flat} alors que des concentrations élevées fournissent des monocouches riches en anticorps \textit{end-on} \citep{bremer2004,malmsten1995,malmsten1998a,vanerp1992}. L'activité immunologique devrait être mise en rapport avec la disponibilité des parties variables des anticorps, cette disponibilité requérant naturellement un contenu conséquent en IgG \textit{end-on} dans la monocouche.

\'{E}tant donné que l'activité immunologique semble être une fonction directe de la quantité d'anticorps \textit{end-on} dans la monocouche, il vient logiquement que les suspensions de concentrations élevées en anti-IL-6 sont indiquées pour la construction de monocouches riches en anticorps orientés \textit{end-on}. Ces résultats obtenus par ELISA rendent compte des propriétés macroscopiques des films d'IgG, propriétés pouvant maintenant être liées à leurs caractéristiques et organisations à l'échelle nanométrique grâce à l'AFM.

De plus, les rugosités sont estimées pour les monocouches et le substrat hydrophobe sous-jacent en calculant les écarts-types de régions arbitrairement choisies dans les zones grattée et intacte. Ces écarts-types peuvent être comparés à ceux obtenus sur des surfaces hydrophobes laissées vierges d'anticorps: 0,45 nm. Globalement, on doit constater que les écarts-types des régions grattées des monocouches restent supérieurs à ceux de la surface vierge. Ainsi, les rugosités des régions grattées semblent augmenter en fonction de la concentration de la suspension de dépôt ce qui pourrait indiquer une difficulté croissante dans l'enlèvement de la totalité des protéines lors du grattage. La comparaison des rugosités des zones grattées pour les deux anticorps (mêmes concentrations de dépôt) montre une rugosité relativement plus élevée pour l'anti-IL-6 (voir, par exemple, les figures \ref{FigPub7}.B et \ref{FigPub7}.F), indiquant que les forces d'adhésion seraient plus élevées pour les anti-IL-6.

Globalement, les rugosités estimées doivent être intimement liées aux aspects dendritique (rugosité élevée) et \textit{strand-like} (faible rugosité) révélés par les images AFM. De plus, les rugosités des substrats peuvent donner une indication quant aux forces d'adhésion de chaque IgG. En effet, la présence d'agrégats dans la zone ayant été grattée donne une indication de la difficulté d'enlèvement des IgG et donc de leur affinité pour la surface hydrophobe. La présence d'agrégats étant corrélée avec la rugosité, il semble que l'anti-IL-6 dispose d'une plus forte affinité pour la surface que l'anti-IL-2. Sachant de surcroît que la pression exercée par la pointe sur la surface lors de l'opération de grattage était d'environ 2,5 GPa, il semble possible de quantifier les affinités de ces deux anticorps pour la surface hydrophobe: supérieure à 2,5 GPa pour l'anti-IL-6 et inférieure à 2,5 GPa pour l'anti-IL-2. Il apparaît dès lors que l'affinité des anticorps semble liée à la structure des monocouches: un anticorps (anti-IL-6) possédant une forte affinité pourrait mener à une monocouche de type \textit{strand-like} tandis qu'un anticorps (anti-IL-2) ayant une faible affinité pour la surface mènerait à des monocouches de type dendritique.

On remarquera dans les tableaux \ref{TabPub1} et \ref{TabPub2} que, pour certaines monocouches, des valeurs de rugosités apparaissent plus élevées que les épaisseurs estimées (anti-IL-2: 3,13; 12,5 et 25 \textgreek{m}g$\cdot$mL$^{-1}$ et anti-IL-6: 1,56 et 3,13 \textgreek{m}g$\cdot$mL$^{-1}$). Naturellement, il est attendu que les épaisseurs des films devraient être égales ou supérieures à leurs rugosités mais, à partir du moment où les épaisseurs sont estimées sur base d'une différence entre deux moyennes, l'existence d'un biais sur l'une des deux voire les deux pourrait mener à des erreurs d'estimation. Dans le cas présent, la hauteur moyenne du substrat devrait être surestimée à cause de la difficulté d'enlèvement des agrégats d'IgG de la zone grattée. La présence de ces agrégats mène à une surestimation de la hauteur $h_0$ (voir figure \ref{65dg46d8fg47}) qui biaisera l'estimation de l'épaisseur ($h_1-h_0$), cette dernière se trouvant par conséquent sous-estimée jusqu'à pouvoir être inférieure à la rugosité.

\begin{table}[ht]\centering
\caption[Propriétés des monocouches d'anti-interleukine-2 humaine]{Pour les anticorps monoclonaux de souris anti-interleukine-2: rugosités des susbstrats, rugosités des monocouches d'IgG, épaisseurs des monocouches d'IgG (les intervalles sont calculés à 95~\% de confiance), taux de recouvrement ($\phi$), volumes superficiels et quantités accumulées ($\Theta$) dans les films en fonction de la concentration en IgG dans la suspension de dépôt.}\label{TabPub1}
\begin{spacing}{1.2}
\begin{footnotesize}
\centerline{
\begin{tabular}[t]{r@{}lr@{}lr@{}lr@{}l@{ $\pm$ }lr@{}lr@{}lr@{}l}
\hline
\multicolumn{2}{c}{Conc.} & \multicolumn{2}{c}{Rugosité} & \multicolumn{2}{c}{Rugosité} & \multicolumn{3}{c}{\'{E}paisseur} & \multicolumn{2}{c}{Taux de} & \multicolumn{2}{c}{Volume} & \multicolumn{2}{c}{Quantité} \\
\multicolumn{2}{c}{de dépôt} & \multicolumn{2}{c}{du substrat} & \multicolumn{2}{c}{du film} & \multicolumn{3}{c}{du film} & \multicolumn{2}{c}{recouvrement} & \multicolumn{2}{c}{du film} & \multicolumn{2}{c}{accu.} \\
\multicolumn{2}{c}{(\textgreek{m}g$\cdot$mL$^{-1}$)} & \multicolumn{2}{c}{(nm)} & \multicolumn{2}{c}{(nm)} & \multicolumn{3}{c}{(nm)} & \multicolumn{2}{c}{(\%)} & \multicolumn{2}{c}{(mm$^3\cdot$cm$^{-2}$)} & \multicolumn{2}{c}{(pmol$\cdot$cm$^{-2}$)}\\
\hline
\quad 1&,56&\quad\ 0&,6&\quad 1&,21&1&,47&0,1&\quad\ 79&,6 &\ 0&,$87\times10^{-4}$&\qquad 0&,23\\
3&,13&0&,59&1&,71&1&,49&0,11&79&,4&1&,$26\times10^{-4}$&0&,34\\
6&,25&0&,6&0&,96&1&,5&0,04&83&,5 &1&,$22\times10^{-4}$&0&,33 \\
12&,5&0&,95&2&,6&2&,16&0,15&83&,4 &2&,$85\times10^{-4}$&0&,76\\
25& &1&,19&1&,6&0&,98&0,12 &77&,7 &1&,$18\times10^{-4}$&0&,31\\
50& &0&,92&1&,87&2&,49&0,11&91& &1&,$87\times10^{-4}$&0&,5   \\
\hline
\end{tabular}}
\end{footnotesize}
\end{spacing}
\end{table}

\begin{table}[ht]\centering
\caption[Propriétés des monocouches d'anti-interleukine-6 humaine]{Pour les anticorps monoclonaux de souris anti-interleukine-6: rugosités des susbstrats, rugosités des monocouches d'IgG, épaisseurs des monocouches d'IgG (les intervalles sont calculés à 95~\% de confiance), taux de recouvrement ($\phi$), volumes superficiels et quantités accumulées ($\Theta$) dans les films en fonction de la concentration en IgG dans la suspension de dépôt.}\label{TabPub2}
\begin{spacing}{1.2}
\begin{footnotesize}
\centerline{
\begin{tabular}[t]{r@{}lccr@{ $\pm$ }lr@{}lr@{}lr@{}l}
\hline
\multicolumn{2}{c}{Conc.} & Rugosité & Rugosité & \multicolumn{2}{c}{\'{E}paisseur} & \multicolumn{2}{c}{Taux de} & \multicolumn{2}{c}{Volume} & \multicolumn{2}{c}{Quantité} \\
\multicolumn{2}{c}{de dépôt}&du substrat& du film & \multicolumn{2}{c}{du film} & \multicolumn{2}{c}{recouvrement} & \multicolumn{2}{c}{du film} & \multicolumn{2}{c}{accu.} \\
\multicolumn{2}{c}{(\textgreek{m}g$\cdot$mL$^{-1}$)} & (nm) & (nm) & \multicolumn{2}{c}{(nm)} & \multicolumn{2}{c}{(\%)} & \multicolumn{2}{c}{(mm$^3\cdot$cm$^{-2}$)} & \multicolumn{2}{c}{(pmol$\cdot$cm$^{-2}$)}\\
\hline
\quad 1&,56&0,37&0,45&0,27&0,14&\quad\ 100& &1&,$59\times10^{-4}$&\qquad 0&,43\\
3&,13&0,43&0,58&0,52&0,23&99&,5&1&,$35\times10^{-4}$&0&,36\\
6&,25&0,97&0,53&1,61&0,22&100& &2&,$23\times10^{-4}$&0&,6 \\
12&,5&0,98&0,52&1,77&0,28&100& &2&,$1\ \,\times10^{-4}$&0&,56\\
25& &0,79&0,67&2,25&0,2 &100& &2&,$58\times10^{-4}$&0&,69\\
50& &0,67&0,58&2,88&0,99&100& &3&,$72\times10^{-4}$&1&   \\
\hline
\end{tabular}}
\end{footnotesize}
\end{spacing}
\end{table}

\`{A} ce stade, il semblerait que deux facteurs peuvent influencer la réponse ELISA des monocouches: la force d'adhésion sur la surface des anticorps impliqués et leurs concentrations dans la suspension de dépôt. En effet, il a été mentionné ci-dessus que la concentration était positivement corrélée avec la réponse ELISA et que, d'autre part, les forces d'adhésion des types d'IgG semblent aussi liées à cette réponse. Concernant l'anti-IL-6 montrant des monocouches de type \textit{strand-like}, la corrélation mise en lumière entre la concentration de dépôt (et donc l'épaisseur des monocouches) et la réponse ELISA suggère que la proportion entre les IgG \textit{end-on} et \textit{flat} devrait changer en fonction de la concentration, comme suggéré à la figure \ref{FigPub1}. La figure \ref{FigPub1} montre que l'épaisseur moyenne des monocouches peut être liée au ratio \textit{end-on}/\textit{flat}: une monocouche fortement peuplée en IgG \textit{end-on} (figure \ref{FigPub1}.C) devrait être plus épaisse qu'une monocouche qui en est faiblement peuplée (figure \ref{FigPub1}.A). Par conséquent, l'augmentation de l'épaisseur des monocouches \textit{strand-like} obtenues pour les IgG ayant une forte affinité pour la surface semble être une indication fiable du contenu croissant en IgG \textit{end-on}, contenu qui sera alors à même d'accroître l'activité immunologique des monocouches pour l'ELISA.

\begin{figure}[t]\centering
\includegraphics*[width=0.7\textwidth]{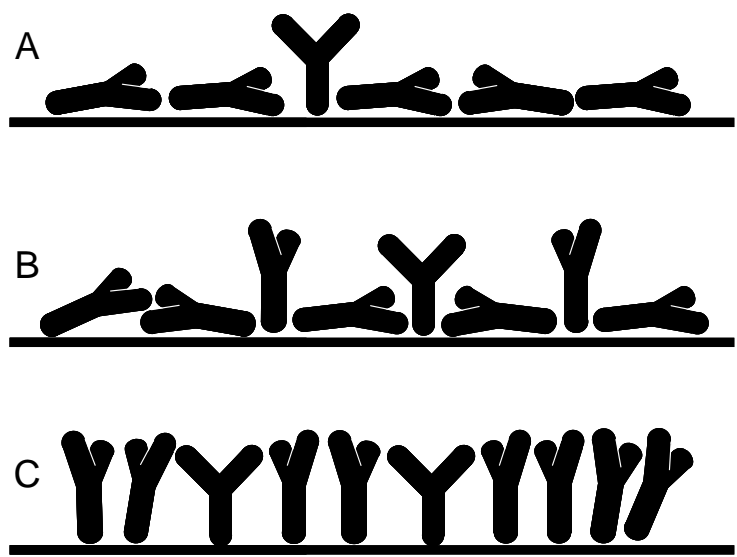}
\caption[Trois structures idéalisées de monocouches d'IgG]{Trois structures idéalisées de monocouches contenant des proportions variables en IgG \textit{end-on}; \textbf{A}: faible contenu en IgG \textit{end-on} (monocouche fine et activité immunologique faible); \textbf{B}: contenu intermédiaire en IgG \textit{end-on} (monocouche d'épaisseur moyenne et activité moyenne) et \textbf{C}: contenu élevé en IgG \textit{end-on} (monocouche épaisse et activité élevée).}\label{FigPub1}
\end{figure}

\subsubsection{Taux de recouvrement des surfaces}

La comparaison des taux de recouvrement des surfaces et des volumes superficiels des films fournit des informations supplémentaires sur la structure des monocouches. Pour l'anti-IL-2 (voir tableau \ref{TabPub1}), les taux de recouvrement semblent croître sensiblement en fonction des volumes superficiels des films alors qu'il n'en est rien pour l'anti-IL-6 (voir tableau \ref{TabPub2}). Pour ces derniers, le taux de recouvrement demeure remarquablement constant à sa valeur maximale alors que l'épaisseur augmente clairement. Ces observations permettent une meilleure compréhension des aspects des monocouches déjà exposés ci-dessus. Pour les monocouches d'anti-IL-2, l'augmentation de la concentration de dépôt amène un meilleur remplissage de la surface \textit{via} un changement de structure de la monocouche, c'est-à-dire qu'elle apparaît d'avantage vallonnée. Dès lors, la concentration dans la suspension de dépôt résulte en une augmentation du nombre de collines dans la monocouche menant alors à une monocouche plus rugueuse et épaisse. Il n'en va pas de même pour les monocouches d'anti-IL-6 pour lesquelles l'augmentation de la concentration de dépôt ne vient pas affecter l'homogénéité latérale de la monocouche c'est-à-dire que leurs rugosités et leurs taux de recouvrement ne changent pas. Quelle que soit la concentration dans la suspension de dépôt, les surfaces demeurent couvertes à 100~\% par les anticorps et les monocouches gardent leur aspect \textit{strand-like}. Ce cas est remarquable car il est observé que le seul changement d'importance repose sur l'augmentation de l'épaisseur des monocouches laissant à penser qu'un changement structural à l'intérieur de la monocouche en est la cause, changement qui pourrait être attribué à une augmentation du contenu en IgG \textit{end-on}.

\subsubsection{Quantités accumulées dans les monocouches}

Les tableaux \ref{TabPub1} et \ref{TabPub2} montrent une augmentation claire des volumes superficiels pour l'anti-IL-6 en fonction de la concentration dans la suspension de dépôt alors que, pour l'anti-IL-2, ce ne semble pas être aussi clair. Cette différence pourrait s'expliquer sachant que les monocouches d'anti-IL-6 conservent leur taux de recouvrement maximal et leur aspect lisse indépendamment de la concentration de dépôt. D'autre part, pour l'anti-IL-2, l'augmentation du volume superficiel semble plus aléatoire, fait qui pourrait être attribué à la forte rugosité de ces monocouches.

Comme montré aux tableaux \ref{TabPub1} et \ref{TabPub2}, les volumes superficiels des monocouches (mm$^3\cdot$cm$^{-2}$) donnent aussi un moyen d'estimer les quantités accumulées (pmol$\cdot$cm$^{-2}$) dans ces monocouches. En effet, la connaissance d'un volume molaire (mm$^3\cdot$pmol$^{-1}$) moyen des anticorps permet de transformer facilement les volumes superficiels en quantités accumulées ($\Theta$). Afin de transformer les volumes superficiels des monocouches en quantités accumulées, diverses indications sur la taille des anticorps puisées dans la littérature seront utilisées. M. Bremer \textit{et al.} \citep{bremer2004} approximent les IgG par des sphères d'un rayon de 7 nm, pouvant permettre d'estimer les volumes molaires de ces dernières. D'autre part, des rayons obtenus sur base de mesures viscosimétriques peuvent aussi être trouvés dans les travaux de J.~K. Armstrong \textit{et al.} \citep{armstrong2004} qui donnent un rayon hydrodynamique d'environ 5,29 nm pour les IgG. Grâce à la formule du volume d'une sphère ($V=\tfrac{4}{3}\pi R^3$), des valeurs de $3,73\times10^{-4}$ mm$^3\cdot$pmol$^{-1}$ et inversement de $2,68\times10^3$ pmol$\cdot$mm$^{-3}$ peuvent être obtenues afin d'estimer les quantités accumulées dans les monocouches données dans les dernières colonnes des tableaux \ref{TabPub1} et \ref{TabPub2}. \'{E}tant donné que les valeurs de densité obtenues sont directement proportionnelles aux volumes superficiels, les commentaires pouvant être faits au sujet des structures des monocouches demeurent les mêmes que ceux réalisés au regard des volumes superficiels. Il peut toutefois être remarqué que des valeurs maximales de 0,5 et 1 pmol$\cdot$cm$^{-2}$ sont respectivement obtenues pour les monocouches d'anti-IL-2 et d'anti-IL-6.

Les quantités accumulées obtenues dans ce travail peuvent maintenant être comparées à celles rapportées dans la littérature et en particulier celles obtenues par réflectométrie pour des <<~isothermes d'adsorption~>> d'IgG sur des surfaces hydrophobes (polystyrène ou silice méthylée). J. Buijs \textit{et al.} \citep{buijs1995} exposent à ce titre quelques résultats obtenus à pH 7 pour des <<~isothermes d'adsorption~>> d'anticorps monoclonaux de souris (anti-hCG) sur des surfaces de polystyrène chargées positivement ou négativement et pour lesquelles les valeurs de saturation adviennent vers 2,3 pmol$\cdot$cm$^{-2}$, et ce, à partir de suspensions de dépôt concentrées autour de 120--150 \textgreek{m}g$\cdot$mL$^{-1}$ (polystyrène chargé positivement) et 0,8 pmol$\cdot$cm$^{-2}$ pour des concentrations de 80--100 \textgreek{m}g$\cdot$mL$^{-1}$ (polystyrène chargé négativement). Ces valeurs correspondent  aux valeurs de plateau des isothermes \citep{nordehaynes}, c'est-à-dire les concentrations de la suspension de dépôt menant à des monocouches ayant les quantités accumulées les plus élevées. M. Malmsten \citep{malmsten1995} donne par ailleurs des cinétiques d'adhésion pour lesquelles une quantité accumulée de 0,73 pmol$\cdot$cm$^{-2}$ sur de la silice méthylée est obtenue à partir d'une suspension de dépôt concentrée à 100 \textgreek{m}g$\cdot$mL$^{-1}$ (pH 7,4). Le même auteur rapporte aussi une valeur de 2 pmol$\cdot$cm$^{-2}$ pour des concentrations de dépôt de 200--300 \textgreek{m}g$\cdot$mL$^{-1}$ (pH 7,4) \citep{malmsten1994}. D'autres valeurs intéressantes sont aussi données par C. Giacomelli \citep{giacomelli2006}. Bien que les quantités accumulées obtenues dans ce travail sont toutes clairement inférieures à celles données par les différents auteurs précités, il s'agit d'un fait assez logique étant donné que le domaine de concentration des suspensions de dépôt exploré est décalé vers de faibles valeurs. Les valeurs obtenues montrent donc une augmentation sans toutefois atteindre une valeur de saturation, ce qui indique qu'elles appartiennent au domaine de croissance des <<~isothermes d'adsorption~>> caractéristiques des anti-IL-2 et anti-IL-6. De même, ces domaines au sein desquels les quantités accumulées croissent demeurent assez peu explorés dans la littérature, ce qui pourrait découler de la difficulté à obtenir des estimations consistantes du fait de leur forte variabilité observée tant dans la littérature \citep{nordehaynes} que dans ce travail (voir figure \ref{FigIsotPub}).

\begin{figure}[t]\centering
\includegraphics*[width=0.75\textwidth]{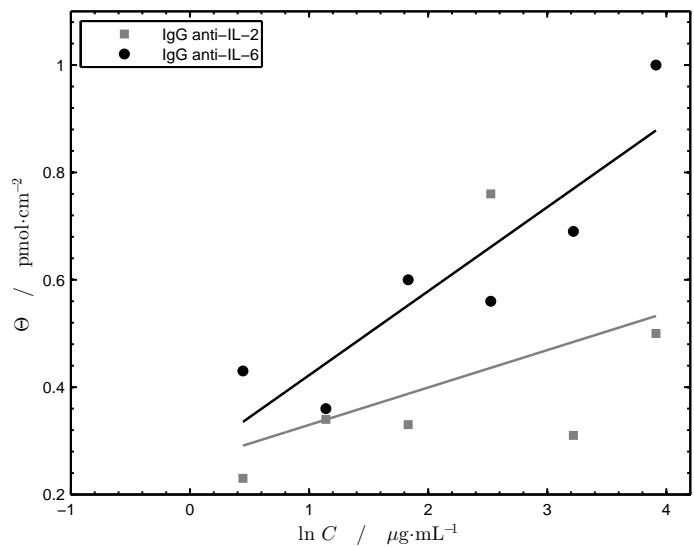}
\caption[Quantités accumulées dans les monocouches en fonction de $\ln\,C$]{Quantités accumulées ($\Theta$ en pmol$\cdot$cm$^{-2}$) dans les monocouches d'anti-interleukine-2 humaine (\begin{tiny}$\blacksquare$\end{tiny}) et d'anti-interleukine-6 humaine ($\bullet$) en fonction du logarithme de la concentration $C$ en IgG dans la suspension de dépôt en \textgreek{m}g$\cdot$mL$^{-1}$. Les quantités accumulées en provenance des tableaux \ref{TabPub1} et \ref{TabPub2} sont accompagnées de régressions linéaires.}\label{FigIsotPub}
\end{figure}

Par ailleurs, il est un fait connu que la pente initiale des <<~isothermes d'adsorption~>> est une indication de l'affinité de la protéine pour la surface de dépôt \citep{buijs1995}. Les données présentées dans ce travail ayant été collectées pour des valeurs intermédiaires de concentration (plus faibles que celles menant au plateau de l'isotherme), elles peuvent dès lors être une indication des affinités des anti-IL-2 et anti-IL-6 pour les surfaces hydrophobes étudiées. La figure \ref{FigIsotPub} montre en effet que la pente de la droite de tendance est plus élevée pour l'anti-IL-6 que pour l'anti-IL-2, suggérant que l'affinité de l'anti-IL-6 pour la surface hydrophobe est plus élevée que celle de l'anti-IL-2, observation se trouvant corroborée par les forces d'adhésion des anticorps discutées ci-dessus grâce aux rugosités des zones grattées des images AFM.

Pour toutes ces raisons, les <<~isothermes d'adsorption~>> des anticorps apparaissent comme un outil efficace dans le cadre de l'amélioration des ELISA. En effet, pour les anticorps montrant un comportement d'adhésion menant à des monocouches de style \textit{strand-like}, l'augmentation de la quantité accumulée en fonction de la concentration de la suspension de dépôt est corrélée à une augmentation de l'épaisseur et de l'activité immunologique. Par conséquent, les valeurs de l'isotherme au plateau devraient correspondre à des monocouches fortement peuplées d'IgG \textit{end-on} alors que les valeurs plus faibles correspondraient à des monocouches plutôt peuplées d'IgG orientées \textit{flat}. Ces séries de monocouches différemment structurées ayant été obtenues pour diverses concentrations en IgG dans la suspension de dépôt, ces conditions devraient être la cause d'un changement de la cinétique de construction des monocouches.

\subsection{Mécanisme de croissance des monocouches d'IgG}

Bien qu'une structure hypothétique puisse être imaginée pour les couches d'anti-IL-6 en faisant le lien avec l'activité immunologique et l'épaisseur, il n'en va pas autant pour le mécanisme de construction de la monocouche. M. Bremer \textit{et al}. \citep{bremer2004} proposent un  mécanisme général pour la construction des monocouches d'IgG, mécanisme entièrement basé sur la notion de volume exclu (voir figure \ref{FigMechanismeAccuIgG}). Ce mécanisme est fondé sur trois hypothèses:
\begin{itemize}
\item Les IgG peuvent se présenter à la surface selon n'importe quelle orientation (\textit{end-on} ou \textit{flat}, voir figures \ref{FigMechanismeAccuIgG}.A et B);
\item Les IgG \textit{end-on} ont tendance à se réorienter pour devenir \textit{flat} et ainsi maximiser leurs interactions hydrophobes avec la surface solide (voir figure \ref{FigMechanismeAccuIgG}.C);
\item Le remplissage provoque l'exclusion progressive des IgG \textit{flat} ce qui finit par en prévenir l'addition à la monocouche (voir figure \ref{FigMechanismeAccuIgG}.D) et la réorientation des IgG \textit{end-on} (voir figure \ref{FigMechanismeAccuIgG}.F).
\end{itemize}
Ces trois hypothèses peuvent mener à un mécanisme de croissance bimodal \citep{schaaf1987} dont une modélisation a été publiée par Ch. Dupont-Gillain \citep{dupont2012} en utilisant une extension du modèle RSA dans lequel la relaxation des IgG est instantanée. Ces simulations montrent comment le processus de remplissage  augmente le volume exclu et empêche progressivement l'addition des IgG \textit{flat} (seconde étape) alors qu'elles ont été favorisées au début du processus (première étape). Pour des protéines comme le fibrinogène \citep{schaaf1987,schmitt1983}, le modèle RSA sans possibilité de relaxation mène toujours à des monocouches contenant un mélange de protéines \textit{flat} et \textit{end-on}, ces dernières apparaissant, pour les IgG, en fin de processus et selon une proportion de minimum trois \textit{end-on} pour une \textit{flat} \citep{dupont2012}. Le modèle RSA est donc capable d'expliquer en ne tenant compte que des seules trois hypothèses ci-dessus la raison pour laquelle les IgG \textit{end-on} peuvent être présentes dans la monocouche saturée. Toutefois, le modèle RSA n'est pas capable d'expliquer pourquoi la variabilité de ce contenu en IgG \textit{end-on} peut être liée à l'épaisseur et à l'activité immunologique des monocouches observées expérimentalement dans ce travail.

Grâce à l'introduction du rapport $K^\circ$ entre les vitesses instantanées d'addition et de relaxation, le modèle RSA+R permet de compléter la description du mécanisme de croissance des monocouches faite par le seul modèle RSA. Le chapitre précédent a pu montrer comment le facteur $K^\circ$ dont l'intensité est fixée par le caractère dur/mou des IgG et, surtout, par la concentration en IgG dans la suspension de dépôt était déterminant vis-à-vis de la structure de la monocouche saturée. Plus le facteur $K^\circ$ et donc la concentration dans la suspension de dépôt était élevé, plus la relaxation sera inhibée au cours du processus de croissance de la monocouche, c'est-à-dire que les IgG \textit{end-on} seront empêchées d'atteindre une orientation \textit{flat}. Ce phénomène s'explique par le fait que l'augmentation du facteur $K^\circ$ vient augmenter l'intensité du parachutage des IgG sur la surface et donc la probabilité qu'une IgG s'additionnant à la monocouche vienne, par effet de proximité, bloquer la relaxation d'une IgG déjà présente sur la surface. Dès lors, l'augmentation de la concentration de dépôt aura tendance à engendrer des monocouches bien plus peuplées en IgG non relaxées, c'est-à-dire \textit{end-on}. Par ailleurs, les IgG \textit{end-on} prenant moins de place sur la surface (leur empreinte caractéristique étant faible), les monocouches construites dans ces conditions seront aussi caractérisées par une quantité accumulée $\Theta$ plus importante. Cet effet était particulièrement visible aux figures \ref{kekufyuldg}.A à C pour le système de boîtes de conserve et à mené à la fonction de saturation illustrée à la figure \ref{FigFit}.

L'extrapolation du modèle RSA+R des boîtes de conserve aux IgG permet donc d'expliquer que l'augmentation de la concentration en IgG dans la suspension de dépôt est à même de produire des monocouches dont la fraction en IgG \textit{end-on} est forte, la quantité accumulée élevée et par conséquent d'épaisseurs plus importantes. Il devient alors envisageable de postuler que pour les anti-IL-6 fournissant des monocouches de type \textit{strand-like}, l'augmentation de la concentration de dépôt doit causer l'inhibition du processus de relaxation ce qui a pour conséquence d'augmenter la fraction d'IgG \textit{end-on} dans la monocouche saturée, monocouche devant alors être plus épaisse et montrer une activité immunologique importante. L'amélioration des ELISA pour ce genre d'anticorps pourrait donc être réduit à la seule recherche de la concentration de dépôt menant à la valeur de plateau de l'isotherme caractérisant son adhésion.

Pour l'anti-IL-2 montrant des monocouches de type dendritique, il semble plus difficile de réfléchir à un mécanisme de construction des monocouches. Toutefois, B. W\"{a}livaara \textit{et al.} \citep{walivaara1995} ont observé que les monocouches apparaissaient de plus en plus lisses lorsqu'elles étaient réticulées avec des solutions de glutaraldéhyde avant l'étape de séchage. En tenant compte des observations faites dans ce travail, il pourrait être possible de conclure que le mécanisme d'adhésion des anti-IL-2 serait plutôt coopératif. Peut-être que l'adhésion des anti-IL-2 nécessiterait un phénomène de nucléation permettant une croissance de la monocouche de proche en proche menant finalement à ces monocouches dendritiques. Ce comportement d'adhésion plus exotique semble difficilement explicable par les seules conclusions du modèle RSA+R et vient alors poser les limites de l'application du modèle RSA+R (et RSA) à la formation des monocouches de protéines.

\subsection{Pertinence du modèle RSA+R}

Le comportement d'adhésion des anticorps anti-IL-6 menant à des monocouches de type \textit{strand-like} offre une possibilité intéressante de discuter de la justesse du modèle RSA+R développé tout au long du chapitre \ref{ChapIsoth}. En effet, la relation de proportionnalité entre le ratio des vitesses initiales $K^\circ$  et la concentration dans la suspension de dépôt $C$ pointé à l'équation \ref{KfoncC} permet de réaliser un lien entre le modèle RSA+R et l'approche expérimentale. L'augmentation du rapport $K^\circ$ est donc corrélée à une augmentation des quantités d'anti-IL-6 accumulées dans les monocouches ($\Theta_\infty$) ainsi que leurs épaisseurs. Le lien qui a été fait à la section \ref{68gh74b7689} entre les <<~isothermes d'adsorption~>> des protéines et les fonctions de saturation générées par le modèle RSA+R apparaît dès lors plus que judicieux et rend compte du réalisme des hypothèses sur lesquelles se fonde ce modèle.

Toutefois, les estimations expérimentales des taux de recouvrements $\phi_\infty$ des monocouches d'anti-IL-6 ont montré une valeur constante et, à peu de choses près, égale à 100~\%. \`{A} cet effet, les graphiques de la figure \ref{FigCompPhi} montrent l'évolution des taux de recouvrement de monocouches de boîtes de conserve en fonction du rapport $K^\circ$ (donc de la concentration) estimées par simulations RSA+R. Ces graphiques proviennent du système ayant été utilisé afin de construire les graphiques de la figure \ref{FigFit}.

\begin{figure}[t]\centering
\includegraphics*[width=0.75\textwidth]{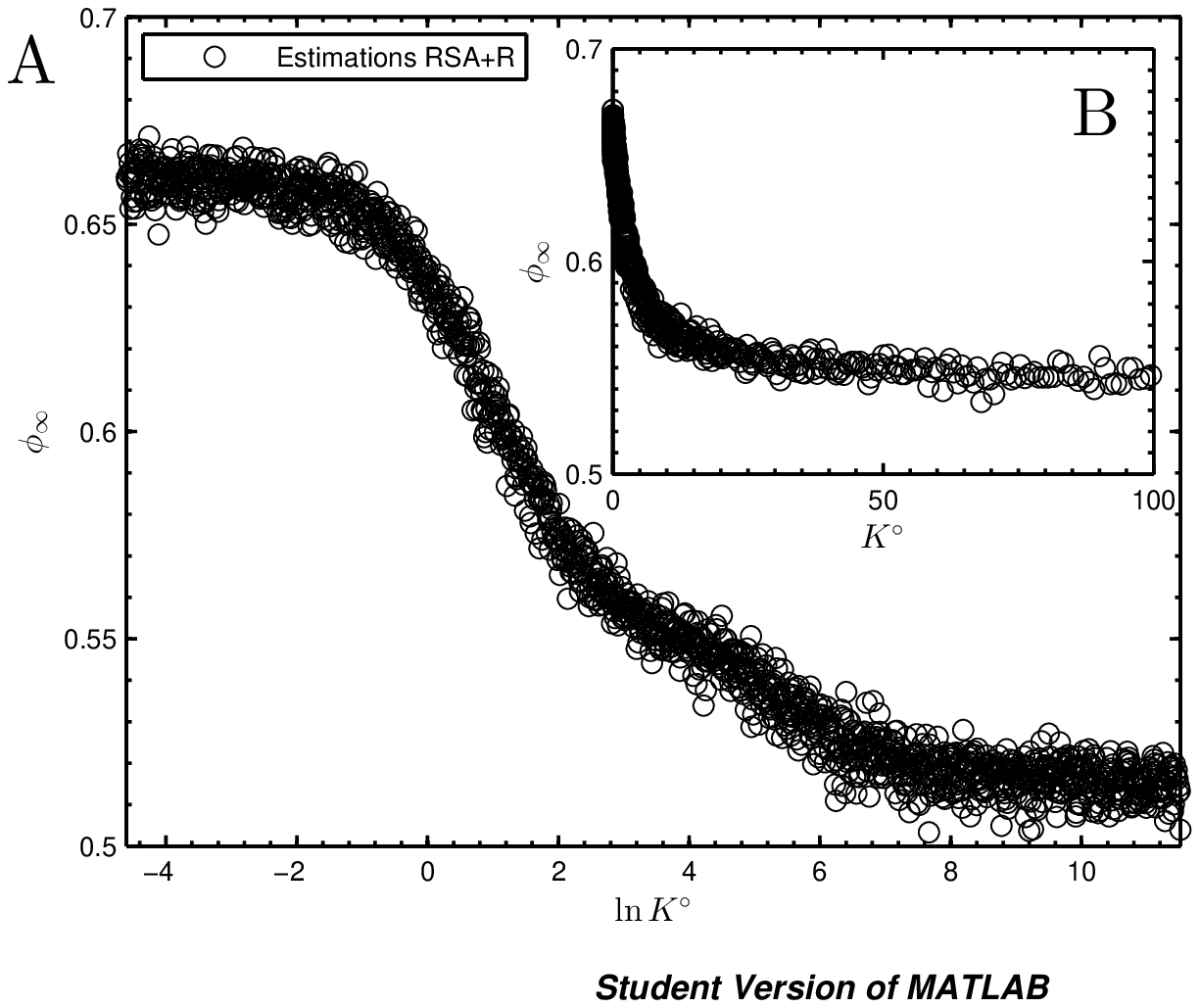}
\caption[Taux de recouvrements en fonction de $K^\circ$]{Valeurs des taux de recouvrement à saturation $\phi_\infty$ obtenues par simulations RSA+R de constructions de monocouches pour le système de boîtes de conserve illustré à la figure \ref{FigRSARelax}. \textbf{A}: en fonction du logarithme du rapport des vitesses initiales $\ln\,K^\circ$ et \textbf{B} en fonction du simple rapport $K^\circ$. Les valeurs portées sur ces graphiques sont à comparer à celles de la figure \ref{FigFit}.}\label{FigCompPhi}
\end{figure}

Premièrement, on constate que ces valeurs de taux de recouvrement n'atteignent jamais la valeur de 100~\%. Ceci peut s'expliquer par l'approximation qui a dû être faite sur le processus de relaxation des boîtes de conserve qui ne peuvent relaxer que selon deux pas successifs et selon des formes bien déterminées. Il est à supposer que, si le nombre de pas de relaxation avait été plus élevé, les valeurs des taux de recouvrement pourraient être, elles aussi, bien plus élevées. En effet, l'augmentation du nombre de ces pas offrirait au processus de relaxation de venir combler de manière plus efficace les espaces laissés vides entre les protéines. Les valeurs des taux de recouvrement que permet de générer le modèle RSA+R souffre donc de l'approximation qu'il est nécessaire de faire sur le mode de relaxation des protéines dans le cadre d'une simulation numérique.

Deuxièmement, les estimations RSA+R des taux de recouvrement vont en diminuant en fonction du rapport $K^\circ$ ce qui est en contradiction avec les faits expérimentaux montrant une valeur remarquablement constante. Si l'approximation faite sur le mode de relaxation d'une protéine permet d'expliquer le biais par rapport à la valeur expérimentale de 100~\%, la valeur absolue de ce biais semble augmenter en fonction du rapport $K^\circ$. Dans la modélisation RSA+R, en augmentant le rapport $K^\circ$, on vient inhiber les possibilités de relaxation des protéines de la monocouche et donc leur capacité à venir combler les espaces de la monocouche laissés vacants. \`{A} la limite, lorsque la relaxation sera fortement à totalement inhibée par l'addition de nouvelles protéines à la monocouche, on finira par obtenir une monocouche de protéines entièrement natives (non relaxées) contenant un grand nombre d'espaces vacants. Cette monocouche est illustrée à la figure \ref{kekufyuldg}.C. Dans la réalité expérimentale, ce genre de monocouche ne devrait pas exister pour des protéines ayant la possibilité de relaxer, même lentement. Afin d'optimiser leurs interactions avec la surface, les protéines, mêmes les plus dures, devraient finir par relaxer et ainsi venir boucher les espaces laissés vides. La valeur absolue du biais peut donc être liée à la limitation du processus de relaxation dans le modèle RSA+R. Quelle que soit la valeur du rapport $K^\circ$, l'expérience montre, pour l'anti-IL-6, un taux de recouvrement maximal et illustre ainsi l'incapacité du modèle RSA+R à produire des structures réalistes de monocouches.

Il est possible de réfléchir de façon plus précise au sujet du biais induit par le modèle RSA+R sur la valeur théorique du taux de recouvrement à saturation $\phi_\infty$ en notant son estimateur par $\widehat{\phi}_\infty$. L'espérance $E[\widehat{\phi}_\infty]$ de cet estimateur est la valeur qui sera obtenue à partir des modélisations RSA+R et dont les valeurs pour diverses monocouches sont portées en graphique à la figure \ref{FigCompPhi}. D'autre part, le biais inhérent à cet estimateur (décalage entre la valeur théorique et la valeur espérée) sera notée $b(\widehat{\phi}_\infty)$. La valeur théorique est reliée à l'espérance et au biais de son estimateur par la relation
\begin{equation}\label{g45d468fxgb4xf87}
\phi_\infty=E[\widehat{\phi}_\infty]-b(\widehat{\phi}_\infty)
\end{equation}
et, sachant que la valeur théorique devrait obéir à l'égalité $\phi_\infty=1$, il vient après réarrangement que
\begin{equation}
b(\widehat{\phi}_\infty)=E[\widehat{\phi}_\infty]-1.
\end{equation}
De cette dernière relation et à partir de la figure \ref{FigCompPhi}, on voit que la valeur absolue du biais de l'estimateur du taux de recouvrement augmente avec la valeur de $K^\circ$ puisque $E[\widehat{\phi}_\infty]$ diminue. Le fait que ce biais soit non nul et variable illustre la double imperfection du modèle RSA+R explicitée ci-dessus.

On remarquera que des relations analogues à \ref{g45d468fxgb4xf87} peuvent être écrites pour les valeurs théoriques (obtenues expérimentalement) des estimateurs $\widehat{\sigma}_\infty$ et $\widehat{\Theta}_\infty$. Il vient alors que
\begin{equation}\label{4b4b47f697h476n7}
\sigma_\infty=E[\widehat{\sigma}_\infty]-b(\widehat{\sigma}_\infty)
\quad\mathrm{et}\quad
\Theta_\infty=E[\widehat{\Theta}_\infty]-b(\widehat{\Theta}_\infty),
\end{equation}
toutes valeurs calculables. Premièrement, les biais $b(\widehat{\Theta}_\infty)$ s'obtiennent à partir des $\Theta_\infty$ obtenus expérimentalement et des $E[\widehat{\Theta}_\infty]$ donnés à la figure \ref{FigFit}. Deuxièmement, on obtient les $\sigma_\infty$ par la relation $\sigma_\infty=(\Theta_\infty)^{-1}$, les $E[\widehat{\sigma}_\infty]$ par les simulations RSA+R (résultats non montrés) et ainsi les biais $b(\widehat{\sigma}_\infty)$. L'obtention de ces valeurs pour toute la fonction de saturation (dont un domaine est exploré à la figure \ref{FigIsotPub}) devrait permettre de se faire une idée générale des imperfections liées au modèle RSA+R, imperfections dont il est déjà possible de se faire une idée assez précise en tenant compte des biais $b(\widehat{\phi}_\infty)$.

Ces imperfections inhérentes aux simulations RSA+R sont toutefois à relativiser eu égard à ce qu'il offre comme possibilités d'explications des phénomènes menant à la formation d'une monocouche. On retiendra en particulier la mise en évidence des fonctions de saturation dépendantes du rapport $K^\circ$, variable directement proportionnelle à la concentration en protéines dans la suspension de dépôt.

\section{Conclusion}

L'activité immunologique mesurée par ELISA sur deux anticorps, \textit{i.e.} anti-inter-leukine-6 humaine et anti-interleukine-2 humaine, montrent de fortes différences. Là où l'activité immunologique des monocouches d'anti-IL-6 augmente en fonction de la concentration dans la suspension de dépôt, rien de tel n'est observé pour les anti-IL-2. Dès lors, les raisons de ce comportement macroscopique ont été investiguées grâce à la microscopie à force atomique (AFM), technique capable de fournir des images topographiques des films. Celles-ci ont permis de révéler des monocouches d'aspects lisses (\textit{strand-like}) pour les anti-IL-6 et rugueuses (\textit{dendrite-like}) pour les anti-IL-2. De plus, les épaisseurs des monocouches d'anti-IL-6 apparaissent comme une fonction croissante de la concentration en IgG dans la suspension de dépôt alors que les taux de recouvrement demeurent remarquablement constants (100 \%).

Les observations réalisées dans ce travail mènent à la conclusion que l'ELISA nécessite des IgG ayant un comportement d'adhésion à même de construire des monocouches lisses, c'est-à-dire de type \textit{strand-like}. Tout au long de la discussion, l'AFM apparait aussi comme une technique efficace afin de construire les <<~isothermes d'adsorption~>> des IgG, isothermes devenant aussi un outil conceptuel intéressant puisque les concentrations en IgG dans la suspension de dépôt menant aux valeurs de plateau devraient être recherchées dans le but d'obtenir des monocouches dont l'activité immunologique est optimale. D'un point de vue strictement expérimental, le couplage des techniques ELISA et AFM apparaît aussi comme un protocole efficace étant donné que l'ELISA permet de prouver que l'augmentation de l'épaisseur des monocouches de type \textit{strand-like} observée par AFM est une conséquence d'une orientation de plus en plus favorable à l'activité immunologique de celles-ci, c'est-à-dire que les IgG se trouvent majoritairement dans une orientation \textit{end-on}.

Par ailleurs, les comportements d'adhésion des anticorps testés, et en particulier l'anti-interleukine-6 humaine, fournissent une série d'éléments permettant de discuter des limites de l'application des modèles RSA+R et RSA à la construction des monocouches de protéines. Bien qu'ils offrent des outils conceptuels particulièrement puissants dans l'interprétation des mécanismes menant à la formation de telles monocouches, il apparait nettement qu'une sous-estimation de leurs propriétés physiques leur semble inhérente.

\begin{footnotesize}

\end{footnotesize}\end{cbunit}
\begin{cbunit}
\chapter*{Conclusion}\label{ChapConclu}
\markboth{Chapitre \ref{ChapConclu}: CONCLUSION}{}
\addstarredchapter{Conclusion}
\minitoc

\section*{Généralités}
\addcontentsline{toc}{section}{Généralités}
\markboth{CONCLUSION}{G\'{E}N\'{E}RALIT\'{E}S}

La technique ELISA est utilisée de manière intensive dans le cadre du diagnostic médical et, plus généralement, de toute quantification et détection de biomolécules dans un échantillon d'intérêt. Cette technique nécessite la construction de films d'immuno-\textgreek{g}-globulines (IgG) sur des surfaces supportant le test, surfaces solides généralement (très) hydrophobes telles que le polystyrène des plaques multipuits.

L'amélioration, de même qu'une meilleure maîtrise des techniques ELISA, passe donc par une meilleure compréhension des phénomènes physico-chimiques à l'{\oe}uvre lors de la formation des films d'IgG sur le polystyrène. De la sorte, il pourra être possible d'envisager des améliorations menant à des films d'IgG fermement ancrés sur le support et présentant une activité immunologique satisfaisante. L'activité immunologique du film ainsi constitué lui permettra de capter, par interaction ligand-récepteur, un autre anticorps ou la molécule antigènique que le dosage vise à détecter/quantifier. Cela implique que les IgG soient correctement orientées dans le film (orientation \textit{end-on} permettant de rendre disponibles les parties variables qui fixent l'antigène) et non ou peu dénaturées (afin de ne pas affecter les propriétés immunologiques des parties variables).

La première question pouvant venir à l'esprit concerne les mécanismes amenant les IgG à se trouver en contact avec la surface sur laquelle le film se formera. Naturellement, les facteurs hydrodynamiques semblent une approche intéressante mais, si l'on admet que la concentration en IgG dans la suspension de dépôt est parfaitement homogène, l'hydrodynamique, agissant essentiellement sur le mélange, n'apparaît pas déterminante. Ce qui l'est plus est la diffusion, diffusion par laquelle les IgG vont migrer du c{\oe}ur de la suspension vers la surface de polystyrène. Dans un espace tenant compte des variables de position et d'orientation des IgG, la diffusion serait susceptible d'amener les IgG à la surface dans une position type \textit{end-on}, c'est-à-dire qu'elles s'y présenteraient par leur partie constante.

\section*{Thermodynamique de l'adhésion des IgG}
\addcontentsline{toc}{section}{Thermodynamique de l'adhésion des IgG}
\markboth{CONCLUSION}{THERMODYNAMIQUE DE L'ADH\'{E}SION DES IGG}

Lorsque les IgG viennent intimement s'approcher de la surface, une série d'interactions typiques des systèmes colloïdaux entrent en jeu: le potentiel DLVO (somme des potentiels électrostatique et de Hamaker) de même que <<~l'effet hydrophobe~>>. Le potentiel électrostatique ne semble pas une barrière infranchissable étant donné le peu de charges électriques que peut porter le polystyrène tandis que le potentiel de Hamaker, dû aux forces de London-van der Waals, est nettement attractif dans le cas d'un système protéine/PBS/polystyrène.

Au-delà du potentiel DLVO, l'action des <<~forces~>> thermodynamiques offre, dans le cadre conceptuel des transformations irréversibles, des explications capitales sur l'adhésion des IgG sur le polystyrène. Grâce à la notion d'affinité, il a en effet été démontré une formule donnant la production d'entropie interne du système en fonction de la croissance de l'aire de contact entre les protéines et la surface solide. La croissance de l'aire de contact a été conceptualisée comme le fait de l'adhésion et elle serait la conséquence de l'hydrophobie de ces deux corps, hydrophobies pour lesquelles une formulation reposant sur les affinités d'adsorption et des adsorptions elles-mêmes a été proposée.

Plus précisément, la formule proposée et ses corollaires expliquent qu'une protéine vient adhérer sur la surface solide parce que les enveloppes de molécules de solvant et du cosolvant adsorbées aux interfaces protéines/milieu aqueux et polystyrène/milieu aqueux s'en désorbent et viennent ainsi participer à l'augmentation de l'entropie du système considéré. Les IgG étant des protéines hydrosolubles et donc hydrophiles, il est donc nécessaire d'utiliser une surface solide suffisamment hydrophobe afin de pouvoir contrecarrer cette hydrophilie. L'utilisation du polystyrène, particulièrement hydrophobe, se justifie donc pleinement dans le cadre de la théorie développée.

L'adhésion a aussi été répartie sur deux phénomènes distincts: l'addition par laquelle les IgG viennent simplement adhérer sur la surface dans l'état selon lequel elles y arrivent et la relaxation par laquelle elles viendront, après addition, augmenter leurs aires de contact (empreinte) avec cette même surface. Cette relaxation pourra se faire par changement de l'orientation (\textit{end-on} $\longrightarrow$ \textit{flat}) et/ou de la conformation (\textit{flat} $\longrightarrow$ \textit{flat$+$}). Il a été déduit que la relaxation se produisait aussi spontanément que la seule addition. En d'autres mots, si l'addition d'une IgG à la surface est spontanée en conséquence de l'hydrophobie du polystyrène, la relaxation suivra tout aussi spontanément. Ceci laisse à penser que, selon les critères de la thermodynamique, les films d'IgG sur le polystyrène devraient être peuplés d'IgG entièrement relaxées, ne satisfaisant pas vraiment ce qui est recherché dans le cadre de l'ELISA pour lequel les IgG devraient s'y trouver dans une position \textit{end-on} afin de reconnaître les antigènes.

\section*{Additions séquentielles aléatoires des IgG}
\addcontentsline{toc}{section}{RSA des IgG}
\markboth{CONCLUSION}{RSA DES IGG}

Au contraire de la thermodynamique, le modèle des additions séquentielles aléatoires (RSA) permet de tenir compte de critères d'ordre cinétique. Le RSA permet de remplir une surface avec une série de solides de formes relativement simples en tenant compte de leur impénétrabilité les uns vis-à-vis des autres; l'interaction de corps durs est dès lors prise en compte. De ce fait, le RSA permet de conceptualiser le remplissage progressif de la surface par les IgG de même que sa saturation progressive car, fort logiquement, on ne pourra y déposer plus d'IgG que ne le permettra l'espace disponible. En début de processus, les additions d'IgG se font facilement puisqu'elles ne rencontrent que peu d'obstacles tandis que, plus la saturation sera proche, plus les nouvelles additions d'IgG se feront difficilement, l'espace disponible se restreignant.

En réalisant des modélisations RSA pour deux modèles, des IgG selon diverses orientations et des boîtes de conserve de différents diamètres, il devient possible de mettre en évidence un ordre d'exclusion. En effet, chaque particule (IgG ayant une orientation particulière et boîte de conserve d'un diamètre particulier) est caractérisée par une empreinte caractéristique bien précise représentant l'aire de contact qu'aurait avec la surface une mole de ces particules. Par ordre de grandeur décroissant, on parlera de l'empreinte caractéristique des IgG \textit{end-on}, de l'empreinte caractéristique des IgG \textit{side-on} et de l'empreinte caractéristique des IgG \textit{flat}. Au fur et à mesure du remplissage, l'espace disponible se restreint augmentant le volume exclu; dès lors, les particules d'empreintes caractéristiques élevées (les IgG \textit{flat} et \textit{side-on}) finiront par être empêchées de venir s'additionner sur la surface ne laissant plus la place qu'aux particules ayant les plus faibles empreintes caractéristiques, c'est-à-dire les IgG \textit{end-on}. Cet accroissement du volume exclu que le modèle RSA permet de mettre en évidence met en lumière l'empêchement progressif de s'additionner des particules en suivant, par ordre décroissant, la valeur de leur empreinte caractéristique: les IgG \textit{flat} sont interdites d'addition en premier, ensuite les IgG \textit{side-on} et les IgG \textit{end-on} viendront alors, seules, terminer la saturation de la surface.

Cet effet d'exclusion progressif a été qualifié <<~d'exclusion par la taille~>> et montre que les IgG \textit{end-on}, recherchées dans le cadre de l'ELISA, apparaissent naturellement et sont même nettement favorisées en fin de processus. Les monocouches obtenues montrent une majorité d'IgG \textit{end-on}. D'autre part, un lien a été discuté entre le transport par diffusion des IgG vers la surface et le modèle RSA, et ce, grâce au rapprochement entre la vitesse initiale d'addition et le flux de particules par diffusion.

Le modèle RSA vient donc compléter l'approche thermodynamique en montrant qu'il y a une limite au remplissage de la surface, limite imposée par l'espace qui s'y trouve disponible et l'impossibilité pour deux IgG de s'interpénétrer.

\section*{Additions et relaxations séquentielles aléatoires des IgG}
\addcontentsline{toc}{section}{RSA+R des IgG}
\markboth{CONCLUSION}{RSA+R DES IGG}

Le modèle RSA est un premier complément à la thermodynamique. Afin d'obtenir une description plus exhaustive du remplissage de la surface, le modèle RSA a été étendu à un modèle des additions \emph{et} relaxations séquentielles aléatoires (RSA+R). Ce nouveau modèle permet de tenir compte de façon très précise du phénomène de relaxation en ajustant un ratio représentant le nombre de tentatives de relaxation par tentative d'addition. Ce ratio n'est rien d'autre que le rapport de la vitesse initiale de relaxation d'une protéine (déterminée par son caractère dur ou mou) et de la vitesse initiale d'addition (déterminée par la diffusion depuis la suspension de dépôt).

En tenant compte d'une relaxation instantanée des IgG succédant à leur addition, un mécanisme pour la formation des monocouches d'IgG et reposant sur la notion de volume exclu a été proposé. Au début du processus de remplissage, les IgG peuvent s'additionner selon n'importe quelle orientation et ont tendance à relaxer afin de satisfaire les critères thermodynamiques. Des IgG de larges empreintes caractéristiques apparaissent donc dans la monocouche. Lorsque le remplissage progresse, l'effet d'exclusion par la taille se faisant plus intense, les IgG de larges empreintes caractéristiques seront empêchées de s'additionner à la monocouche et les IgG déjà présentes dans la monocouche ne pourront plus relaxer. De ce fait, même lorsqu'une relaxation systématique est prise en compte, une forte proportion d'IgG \textit{end-on} (plus de trois \textit{end-on} pour une \textit{flat}) reste présente dans la monocouche saturée. La présence d'IgG \textit{end-on} apparaît donc comme un phénomène inhérent aux monocouches d'IgG, venant élégamment expliquer la performance de l'ELISA.

Par ailleurs, grâce au modèle RSA+R, l'intensité du phénomène de relaxation par rapport à l'addition a pu être très finement ajustée grâce au ratio de leurs vitesses initiales $K^\circ$. Pour une protéine donnée, la vitesse initiale de relaxation est constante car il s'agit d'une propriété qui lui est propre; donc, lorsque l'on fait varier le ratio $K^\circ$, on joue sur l'intensité de l'addition, c'est-à-dire la diffusion des protéines vers la surface. Plus cette addition se fait intense, plus la monocouche sera densément peuplée en protéines d'empreintes caractéristiques faibles (plutôt des protéines non relaxées). En fonction d'une échelle linéaire de $K^\circ$, une courbe des quantités accumulées à saturation a été obtenue et comparée avec succès à <<~l'isotherme d'adsorption~>> des protéines. Une équation a été proposée pour celle-ci, équation montrant le lien entre la quantité de protéines accumulées à saturation et la concentration dans la suspension de dépôt.

\section*{Caractérisation des films d'IgG}
\addcontentsline{toc}{section}{Caractérisation des films d'IgG}
\markboth{CONCLUSION}{CARACT\'{E}RISATION DES FILMS D'IGG}

Fort de ces attendus théoriques et numériques, une étude expérimentale a été envisagée afin de mettre en évidence la corrélation à laquelle on pouvait s'attendre entre la concentration en IgG dans la suspension de dépôt, la densité de la monocouche, son épaisseur et son activité immunologique. En effet, suite aux constats réalisés au chapitre \ref{ChapIsoth}, une monocouche plus dense avec des IgG plutôt \textit{end-on} est attendue lorsque la concentration est plus élevée. De plus, lorsqu'une monocouche se trouve être riche en IgG \textit{end-on}, elle devrait être plus épaisse. L'épaisseur ainsi que l'aspect des monocouches peuvent être estimés grâce à un microscope à force atomique (AFM) alors que l'activité immunologique peut s'obtenir \textit{via} l'ELISA.

Deux anticorps monoclonaux de souris ont été comparés: un anti-interleukine-2 humaine (anti-IL-2) et un anti-interleukine-6 humaine (anti-IL-6). Ce choix a été motivé par le fait que le premier est connu pour présenter une mauvaise capacité à se fixer sur les plaques en polystyrène au contraire du second.

Pour l'anti-IL-6, une corrélation très nette est obtenue entre la concentration de la suspension de dépôt, l'épaisseur des monocouches et leur activité immunologique, laissant à penser que la théorie développée dans les trois premiers chapitres de ce travail correspond fortement à la réalité. Au contraire, l'anti-IL-2 ne montre pas cette corrélation: l'activité immunologique reste assez faible quelle que soit la concentration dans la suspension de dépôt et l'aspect des monocouches obtenu par AFM montre une rugosité forte alors qu'elles demeurent lisses pour les anti-IL-6.

Il est donc permis de penser que les résultats numériques obtenus sur base du modèle RSA+R pour les boîtes de conserve sont raisonnablement extrapolables à certaines IgG et, peut-être, d'autres protéines globulaires. La sensibilité d'un film d'IgG à la reconnaissance antigènique sera donc améliorée en augmentant la concentration en IgG dans la suspension de dépôt. Le choix de l'anticorps semble aussi d'une importance majeure lors de la mise en {\oe}uvre de techniques ELISA.

\section*{Perspectives}
\addcontentsline{toc}{section}{Perspectives}
\markboth{CONCLUSION}{PERSPECTIVES}

Tout au long de se travail, l'accent a été mis sur la compréhension des mécanismes physico-chimiques menant à la formation de films d'IgG sur le polystyrène, surface hautement hydrophobe. La théorie et les concepts développés semblent corroborés par l'expérience pour certaines IgG. Cette meilleur compréhension et les concepts associés devraient offrir de nouveaux outils, plus efficaces, dans le cadre d'améliorations futures des techniques liées à l'ELISA.

Plus fondamentalement, une équation (\ref{DefEntropieAdsorption6}) a été déduite grâce au cadre offert par la thermodynamique des processus irréversibles. De par la production d'entropie qui lui est associée, cette équation conceptualise l'adhésion irréversible de deux corps hydrophobes suspendus dans un milieu aqueux. Elle devrait ouvrir la voie à de nombreuses interprétations des phénomènes colloïdaux de même que biochimiques, les protéines étant de dimensions colloïdales. Le développement des conséquences de l'équation obtenue pourrait être d'une grande utilité dans la compréhension des interactions entre les protéines telles que les interactions ligand-récepteur entre, par exemple, une IgG et l'antigène contre lequel elle est dirigée. Il n'est en effet pas interdit de penser que de telles interactions tirent leur force motrice de l'hydrophobie de certains domaines des surfaces des protéines impliquées. Par ailleurs, il est aussi possible d'imaginer qu'une protéine pourrait présenter des affinités d'adsorption préférentielles pour certains ions (cations Na$^+$, K$^+$, etc.) et que, de ce fait, elle pourrait avoir tendance à ce déplacer vers des zones en contenant une plus grande concentration afin de réaliser son hydrophilie, un déplacement que l'on pourrait rapprocher du phénomène de chimiotaxie par lequel des corps de tailles colloïdales semblent attirés par une concentration élevée en certaines espèces chimiques.

Le cadre développé par le modèle RSA+R devrait aussi permettre de modéliser des constructions de films pour d'autres protéines globulaires. Le modèle RSA+R n'a toutefois pas été utilisé pour les IgG car les résultats ne semblent pas assez proches de la réalité expérimentale. Une meilleure connaissance des mécanismes de relaxation des IgG, en particulier les mécanismes de relaxation de la conformation, devrait éventuellement être à même de permettre son usage.

Enfin, l'utilisation de l'épaisseur des films d'IgG comme propriété mesurable par AFM semble une voie d'investigation prometteuse pour les films protéiques. En effet, il s'agit d'une quantité obtenue de manière très directe, c'est-à-dire ne passant pas par diverses voies de calculs tels que ceux en usage pour des techniques telles que l'éllipsométrie ou la microbalance à cristal de quartz et dissipation (QCM-D). L'usage de ces technique devraient toutefois permettre d'apporter des informations complémentaires sur les films. D'autre part et bien que le couplage entre l'AFM et l'ELISA présente une bonne cohérence, l'AFM peut présenter certains biais en ce qu'elle nécessite de sélectionner une portion de la surface à analyser qui n'est pas nécessairement représentative de l'entièreté du film à caractériser, que la qualité des images topographiques et donc des hauteurs mesurées sont fortement dépendantes du diamètre de la pointe, etc. On notera en outre que des méthodes utilisant la diffraction de neutrons pourraient venir compléter les données AFM en apportant des informations expérimentales quant à l'orientation des IgG à l'intérieur du film.

Pour terminer, les graphiques donnant l'épaisseur en fonction de la concentration obtenus expérimentalement pour les IgG pourraient être obtenus pour des protéines plus banales (formes plus simples) telles que la BSA. De tels graphiques devraient pouvoir fournir des données pour un large domaine de concentrations et ainsi obtenir la sigmoïde mise en évidence par les simulations RSA+R des boîtes de conserve.

\end{cbunit}

\appendix
\begin{cbunit}
\chapter[Les algorithmes des modèles RSA et RSA+R]{Les algorithmes des additions\\(et relaxations) séquentielles aléatoires}\label{AnnRSA}
\markboth{Annexe \ref{AnnRSA}: ALGORITHMES DU RSA (+R)}{}
\minitoc

\section{Généralités}

Tout au long de ce travail, le modèle des additions séquentielles aléatoires (RSA) a été intensivement utilisé afin de générer une grande quantité de données venant appuyer et illustrer le propos. Des graphiques montrant l'évolution de diverses variables d'intérêt, \textit{i.e.} $\Theta$, $\Theta(\boldsymbol{\omega}_i)$, $P(A^\ast)$, etc. pendant la croissance des monocouches ont été construits compte tenu de ce modèle.

L'objectif de cette annexe sera dès lors de présenter et de décrire les programmes écrits en langage Matlab \citep{themathworks} ayant permis la construction de ces données. Sans décrire les codes dans tous leurs détails (certains sont présentés à l'annexe \ref{Ann3}), cette annexe se focalisera sur les différents principes des algorithmes qui ont permis de les développer et d'en faire usage dans le cadre du présent travail. Les programmes nécessaires ainsi que les données qu'ils peuvent générer et dont il aura été fait usage ci-avant sont inclus dans un répertoire \href{run:RSA/}{\texttt{./RSA/}} joint à ce mémoire. Les contenus des sous-répertoires \href{run:RSA/fig/}{\texttt{./RSA/fig/}}, \href{run:RSA/results/}{\texttt{./RSA/results/}} et \href{run:RSA/start/}{\texttt{./RSA/start/}} contiennent des fichiers de bases de données et d'images constituant les données nécessaires aux simulations (\href{run:RSA/start/}{\texttt{./RSA/start/}}) et leurs résultats (\href{run:RSA/results/}{\texttt{./RSA/results/}} et \href{run:RSA/fig/}{\texttt{./RSA/fig/}}).

Il ne s'agit pas ici de décrire de façon exacte tous les programmes mais seulement d'exposer les idées générales qui sous-tendent les principales fonctions utilisées dans le cadre du chapitre \ref{SectionRSA}; elles seront passées en revue dans cette annexe et des explications complémentaires peuvent être trouvées dans les commentaires ajoutés dans les codes de celles-ci. Ces fonctions, contenues dans le dossier \href{run:RSA/}{\texttt{./RSA}}, seront simplement désignées par leur nom suivit de l'extension \texttt{*.m} des fichiers exécutables de Matlab; par exemple, pour la fonction réalisant une addition, on aura <<~\href{run:RSA/DoAddition.m}{\texttt{DoAddition.m}}~>>.

\section{Application du modèle RSA à l'accumulation des IgG}\label{AppliRSA}

Le modèle des additions séquentielles aléatoires est basé sur trois hypothèses relativement simples qui ont été exposées au chapitre \ref{SectionRSA}. Pour mémoire, elles s'énoncent comme suit \citep{tarjus1991un}:
\begin{enumerate}
\item Les protéines sont insérées séquentiellement (une par une) en une position choisie aléatoirement sur la surface;
\item Une fois insérée, la protéine est gelée sur sa position, elle ne peut ni se déplacer dans le volume ni s'en échapper;
\item Deux protéines ne peuvent se superposer.
\end{enumerate}

La simplicité de ces hypothèses combinée à l'environnement matriciel de Matlab offre la possibilité d'écrire des programmes relativement peu complexes permettant la construction \textit{in silico} de surfaces couvertes de particules de formes diverses et variées. La sphère (ou le cylindre) est un des cas les plus simples et en utilisant sa projection sur un plan, c'est-à-dire en la représentant par un disque, il est possible de modéliser le remplissage d'une surface parfaitement plane. Cette projection d'une sphère dans une matrice <<~particule~>> $\mathbf{P}$ de dimensions $M\times N$ est illustrée à la figure \ref{Fig1Ann1} où $M=N=15$.

\begin{figure}\centering
\includegraphics*[width=5cm]{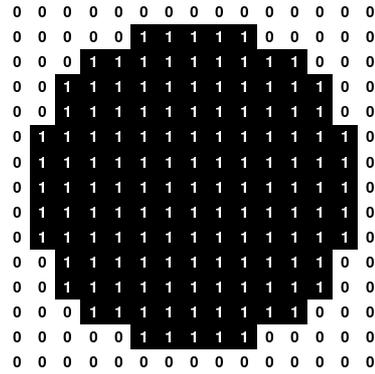}
\caption[Représentation d'un disque pour les simulations RSA]{Représentation d'un disque, projection d'une sphère de rayon $\simeq7$ sur un plan, par une matrice $\mathbf{P}$ $15\times15$ ne contenant que les éléments $0$ et $1$.}\label{Fig1Ann1}
\end{figure}

La modélisation du remplissage d'une surface entière se fera, quant à elle, sur base d'une matrice <<~surface~>> $\mathbf{S}$ de dimensions $K\times L$ bien plus larges, $4000\times4000$ par exemple, qui sera remplie de ces disques $\mathbf{P}$ illustrés à la figure \ref{Fig1Ann1}. Initialement, la matrice $\mathbf{S}$ ne contiendra que des éléments nuls. Le remplissage de la surface sera donc figuré par l'addition d'une série de transpositions de matrices $\mathbf{P}$ en des positions $(i,j)$ choisies aléatoirement parmi les éléments de $\mathbf{S}$. \`{A} partir de la position $(i,j)$, l'opération décrivant l'addition de la particule à la surface s'écrira
\begin{equation}\label{Eq1Ann1}
\mathbf{S}_{kl}^\ast=\mathbf{S}_{kl}+\mathbf{P}_{mn}
\end{equation}
où les indices désignent les éléments des matrices tels que $m\in[1,M]$ et $n\in[1,N]$. Les indices $k$ et $l$ des éléments de $\mathbf{S}$ sont liés aux indices $m$ et $n$ \textit{via} les relations
\begin{equation}
l=j+n+\frac{3-N}{2}\quad\mathrm{et}\quad k=i+m+\frac{3-M}{2}
\end{equation}
dans lesquelles $i$ et $j$ sont les indices de la position choisie pour l'addition de la particule. Cette position est associée à l'élément $\mathbf{S}_{ij}$ de la matrice <<~surface~>>. Il doit aussi être posé que $M$ et $N$, la taille de $\mathbf{P}$, soient impairs afin que $l$ et $k$ soient entiers. L'équation \ref{Eq1Ann1} fait bien apparaître que l'état de remplissage dans lequel se trouve la surface joue un rôle dans la nouvelle addition. En effet, si tous les éléments $\mathbf{S}_{kl}$ sont nuls, l'addition se fera sans écueil possible. Inversement, si des éléments $\mathbf{S}_{kl}$ sont égaux à $1$ et qu'une matrice $\mathbf{P}$ se trouve additionnée à proximité, la somme résultante pourra présenter des superpositions à travers l'existence d'éléments $\mathbf{S}_{kl}^\ast=2$. La superposition de deux objets étant prohibée selon la troisième hypothèse du modèle RSA, on recherchera ces éléments $\mathbf{S}_{kl}^\ast=2$ afin de déterminer si l'addition de la matrice $\mathbf{P}$ est autorisée sur la position $(i,j)$.

Le même principe est utilisé pour simuler le remplissage d'une surface avec des particules plus complexes telles que des séries de boîtes de conserve de différents diamètres ou bien des IgG dans diverses orientations et conformations (\textit{cf.} figures \ref{FigSchémaRSA1} et \ref{FigRSARelax}). Pour les boîtes de conserve, l'utilisation de matrices $\mathbf{P}$ binaires du type de celle montrée à la figure \ref{Fig1Ann1} est parfaitement envisageable. Ce n'est toutefois pas le cas pour les IgG présentant une anisotropie plus difficile à <<~projeter~>> sur un plan. Au lieu d'utiliser des matrices $\mathbf{P}$ ne contenant que des $0$ et des $1$, il faudra faire usage d'autres entiers tels que $2$ et $3$ afin de généraliser l'utilisation de l'équation \ref{Eq1Ann1} aux IgG.

La figure \ref{Fig2Ann1} illustre le résultat d'additions séquentielles aléatoires d'IgG sur une surface. Sur un fond bleu nuit, on y voit des IgG \textit{flat} (bleu roi), \textit{end-on} (jaune et turquoise) et \textit{side-on} (jaune et bleu roi). Chaque pixel de l'image correspond à un élément $(i,j)$ de la matrice $\mathbf{S}$ dont la couleur provient de sa valeur: $0$ (bleu nuit), $1$ (bleu roi), $2$ (turquoise) ou $3$ (jaune). Ce codage est très commode afin de représenter simplement la projection des IgG selon leur différentes orientations mais aussi dans le but d'obtenir des combinaisons sur lesquelles il est aisé de détecter les phénomènes de recouvrements interdits par les hypothèses du modèle RSA.

\begin{figure}\centering
\includegraphics*[angle=270,scale=3,trim = 2.25cm 1.5cm 11.75cm 7cm]{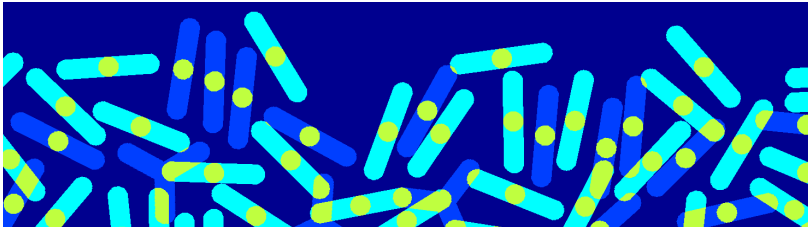}
\caption[IgG additionnées sur une surface selon différentes orientations]{\'{E}chantillon d'une matrice $\mathbf{S}$ montrant des IgG additionnées sur une surface selon différentes orientations: \textit{flat} (bleu roi, les trois bras de l'IgG sont en contact avec la surface), \textit{end-on} (jaune et turquoise, un seul bras en contact avec la surface) et \textit{side-on} (jaune et bleu roi, deux bras en contact avec la surface).}\label{Fig2Ann1}
\end{figure}

Le principe de codage est illustré à la figure \ref{Fig3Ann1}. Il y est montré que l'espace est découpé en deux tranches, la première correspondant à l'espace entre la surface (plan $S$) et le plan situé juste au dessus (plan $S^\prime$) et la deuxième à l'espace situé entre les plans $S^\prime$ et $S^{\prime\prime}$. Clairement, on voit qu'en fonction de leur orientation, les IgG peuvent être présentes dans les deux tranches ou seulement dans une seule. Les IgG \textit{flat} (\ref{Fig3Ann1}.A) ne sont présentes que dans la première tranche, la plus proche de la surface. Les IgG \textit{end-on} (\ref{Fig3Ann1}.B) et \textit{side-on} (\ref{Fig3Ann1}.C) sont, quant à elles, présentes dans les deux tranches. Si une valeur était attribuée au contenu de chaque tranche, \textit{i.e.} $1$ pour la première tranche et $2$ pour la seconde, il est possible, en additionnant le contenu des deux tranches, de représenter simplement les IgG en n'utilisant qu'une seule matrice. Cette matrice, représentée sous forme d'une image à la figure \ref{Fig2Ann1}, présente dès lors les valeurs $0$ (bleu nuit), $1$ (bleu roi), $2$ (turquoise) ou $3$ (jaune). Pour les IgG \textit{flat}, $\mathbf{P}$ ne contiendra que des $0$ et des $1$ tandis que pour les \textit{end-on} on aura des $0$, des $2$ et des $3$ et pour les \textit{side-on}, il y aura des $1$ et des $3$ en plus des $0$.

\begin{figure}\centering
\includegraphics*[scale=0.7]{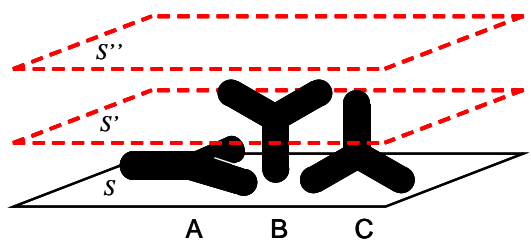}
\caption[Répartition des IgG dans l'espace de l'interface]{Répartition des IgG dans l'espace de l'interface selon leur orientation: \textit{flat} (\textbf{A}), \textit{end-on} (\textbf{B}) et \textit{side-on} (\textbf{C}). L'espace est divisé en deux tranches superposées l'une à l'autre grâce aux plans $S$ (la surface), $S^\prime$ et $S^{\prime\prime}$ afin de rendre possible l'application de l'équation \ref{Eq1Ann1} aux IgG.}\label{Fig3Ann1}
\end{figure}

L'utilisation de la formule \ref{Eq1Ann1} est alors immédiat en posant le fait qu'une superposition se produit dans au moins une des deux tranches lorsque la matrice $\mathbf{S}_{kl}^\ast$ comporte des éléments dont la valeur est strictement supérieure à $3$. Tant que la matrice $\mathbf{S}^\ast$ ne comportera que des valeurs comprises entre $0$ et $3$, la troisième hypothèse du modèle RSA sera respectée, en ce sens qu'aucun recouvrement n'aura eu lieu au sein d'une même tranche. Par ailleurs, lors de l'addition d'IgG \textit{flat}, une autre condition devra être utilisée, \textit{i.e.} les éléments de la matrice $\mathbf{S}^\ast$ devront tous être strictement inférieurs à 2.

\section{Algorithme pour une addition aléatoire}\label{AlgoRSA}

Le paragraphe précédent s'est attaché à illustrer le principe utilisé afin de stocker l'information et de calculer la possibilité d'une addition sur la surface. Ces principes doivent maintenant être utilisés afin de coder des algorithmes permettant de réaliser le remplissage d'une surface par additions séquentielles aléatoires. Avant de réaliser une description exhaustive, ce paragraphe s'attachera à décrire la façon dont est réalisée \textit{in silico} une tentative d'addition d'une IgG en une position aléatoire de la surface. \`{A} cet effet, le schéma d'une opération d'addition aléatoire est montré à la figure \ref{Fig5Ann1}\footnote{Le schéma de la figure \ref{Fig5Ann1} ne représente pas la fonction \href{run:RSA/DoAddition.m}{\texttt{DoAddition.m}} à proprement parler, celle-ci n'étant pas aussi élaborée. Le schéma se rapporte plutôt aux lignes 22 à 42 de la fonction \href{run:RSA/ConstructSurf.m}{\texttt{ConstructSurf.m}} incluant un appel à la fonction \href{run:RSA/DoAddition.m}{\texttt{DoAddition.m}}. \textit{Sensu stricto}, \href{run:RSA/DoAddition.m}{\texttt{DoAddition.m}} correspondrait aux trois blocs entourés de lignes pointillées sur la figure \ref{Fig5Ann1}.}.

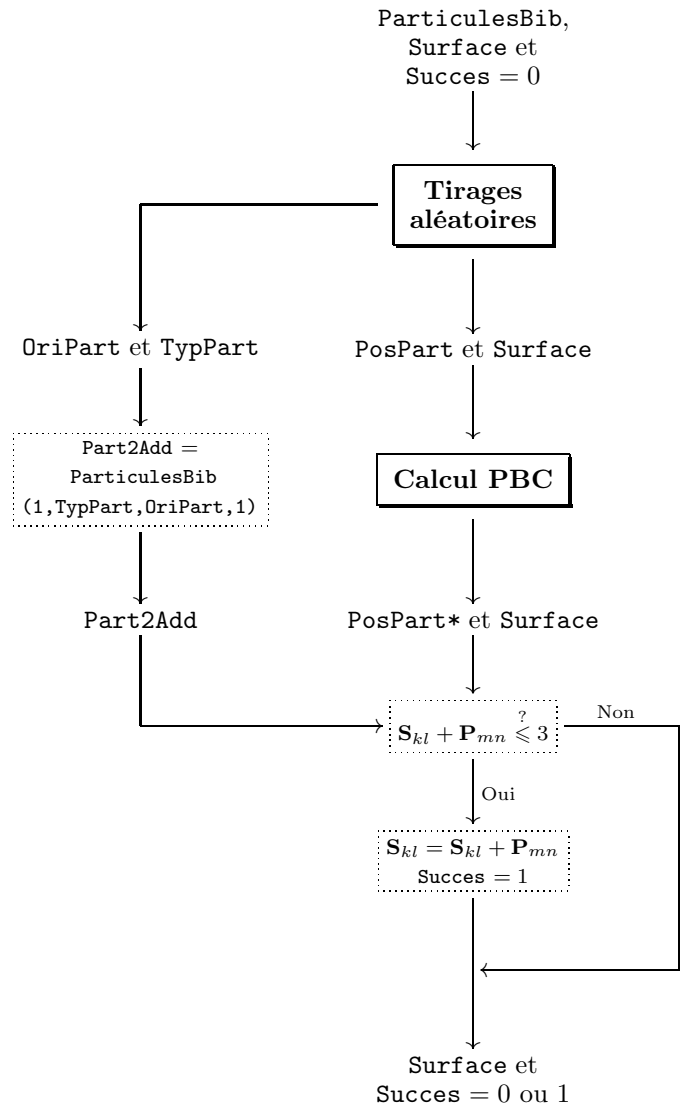
\begin{figure}\centering
\begin{normalsize}
\centerline{
\xymatrix{
&\txt{\texttt{ParticulesBib},\\\texttt{Surface} et\\\texttt{Succes} $=0$}\ar[d]&\\
*[]{}\ar[d]&*++[F-,]++\txt{\textbf{Tirages}\\\textbf{aléatoires}}\ar[d]\ar@{-}[l]&\\
\txt{\texttt{OriPart} et \texttt{TypPart}}\ar[d]&\txt{\texttt{PosPart} et \texttt{Surface}}\ar[d]&\\
*+[F.]+\txt{\footnotesize{\texttt{Part2Add} $=$}\\\footnotesize{ \texttt{ParticulesBib}}\\\footnotesize{\texttt{(1,TypPart,OriPart,1)}}}\ar[d]&*++[F-,]++\txt{\textbf{Calcul PBC}}\ar[d]&\\
\txt{\texttt{Part2Add}}\ar@{-}[d]&\txt{\texttt{PosPart*} et \texttt{Surface}}\ar[d]&\\
*[]{}\ar[r]&*+[F.]+\txt{\footnotesize{$\mathbf{S}_{kl}+\mathbf{P}_{mn}\stackrel{?}{\leqslant}3$}}\ar^{\mathrm{Oui}}[d]\ar@{-}^{\qquad\quad\mathrm{Non}}[r]&*[]{}\ar@{-}[dd]\\
&*+[F.]+\txt{\footnotesize{$\mathbf{S}_{kl}=\mathbf{S}_{kl}+\mathbf{P}_{mn}$}\\\footnotesize{\texttt{Succes} $=1$}}\ar[dd]&\\
&&*[]{}\ar[l]\\
&\txt{\texttt{Surface} et\\\texttt{Succes} $=0$ ou $1$}&\\}}
\end{normalsize}
\caption[Algorithme réalisant une tentative d'addition]{Schéma des opérations réalisées par les lignes 22 à 42 de la fonction \href{run:RSA/ConstructSurf.m}{\texttt{ConstructSurf.m}} réalisant une tentative d'addition d'une particule. Ces lignes de codes contiennent un appel à la fonction \href{run:RSA/DoAddition.m}{\texttt{DoAddition.m}} permettant l'addition telle qu'envisagée par l'équation \ref{Eq1Ann1} et le calcul des conditions périodiques aux limites (fonction \href{run:RSA/EvalPBC.m}{\texttt{EvalPBC.m}}).}\label{Fig5Ann1}
\end{figure}

Un élément très important dont a besoin cet algorithme d'addition est le \textit{cell array} \texttt{ParticulesBib}. Il s'agit d'un tableau dans lequel sont stockées des matrices $\mathbf{P}$ représentant les particules (IgG) à additionner à la surface dans toutes les orientations et configurations possibles. La première dimension de ce tableau correspond à l'état d'une particule du point de vue de la relaxation; lorsque l'indice est 1, comme dans le schéma de la figure \ref{Fig5Ann1}, on ne s'occupe que des IgG dans un état non relaxé (les autres seront discutés par la suite). La deuxième dimension est l'orientation générale, c'est-à-dire \textit{end-on}, \textit{side-on} et \textit{flat}, selon laquelle ils peuvent arriver à la surface. Quant à la troisième dimension, il s'agit de l'angle de rotation de l'IgG autour d'un axe perpendiculaire à la surface\footnote{Dans les faits, une quatrième dimension est aussi utilisée. Le choix que l'on fera entre les valeurs $1$ et $2$ dans les assignations \texttt{x = ParticulesBib(1,2,340,1)} et \texttt{x = ParticulesBib(1,2,340,2)} impliquera que \texttt{x} sera soit ($1$) la matrice $\mathbf{P}$ soit ($2$) sa taille.}. En assignant à la variable \texttt{Part2Add} la matrice renvoyée par \texttt{ParticulesBib(1,2,340)}, \texttt{Part2Add} sera une matrice contenant l'image d'une IgG non relaxée ($1$) dans une orientation générale \textit{side-on} ($2$) et ayant subi une rotation de $340^\circ$ autour d'une droite normale à la surface. Le tableau \texttt{ParticulesBib} est construit avant chaque simulation en appelant la fonction \href{run:RSA/ConstructBib.m}{\texttt{ConstructBib.m}}. On remarquera que la seconde dimension du tableau est assez polyvalente en ce sens qu'il s'agit beaucoup plus que des seules <<~orientations générales~>> de la particule d'intérêt. On peut en effet aussi la voir comme la série de particules, potentiellement différentes, participant au processus de remplissage de la surface. Il peut s'agir des IgG \textit{end-on}, \textit{side-on} et \textit{flat} comme il pourrait s'agir des boîtes de conserve petites, moyennes et grandes ou même encore une série de protéines différentes.

Comme montré sur la figure \ref{Fig5Ann1}, les trois \textit{inputs} de la fonction d'addition sont le tableau \texttt{ParticulesBib}, la matrice \texttt{Surface} et une variable \texttt{Succes} à laquelle une valeur nulle aura initialement été assignée. La tentative d'addition débutera par une série de tirages aléatoires permettant de déterminer la position $(i,j)$ de la surface où l'addition sera tentée (stockée dans \texttt{PosPart}) et quelle sera la particule impliquée. Le choix de la particule nécessitera deux tirages car il faudra déterminer l'orientation générale de l'IgG (stockée dans \texttt{TypPart}) et son angle de rotation (stocké dans \texttt{OriPart}). \`{A} partir des variables \texttt{OriPart} et \texttt{TypPart}, on pourra assigner, en allant la chercher dans \texttt{ParticulesBib}, une matrice $\mathbf{P}$ à la variable \texttt{Part2Add}, la matrice de la particule à additionner à la surface.

Avant de procéder à la tentative d'addition en tant que telle, une transformation du vecteur \texttt{PosPart} devra être réalisée. \'{E}tant donné que la matrice \texttt{Surface} n'est qu'une partie (petite) de la surface réelle, il s'agit de calculer les conditions périodiques aux limites (PBC) afin d'éliminer les effets de bords pouvant biaiser les résultats de la simulation. Les conditions périodiques aux limites sont très utilisées dans les simulations numériques car elles permettent de considérer une surface de taille infinie et donc sans bord \citep{schlick2010}. Selon les situations, le vecteur \texttt{PosPart} ainsi transformé en \texttt{PosPart*} par la fonction \href{run:RSA/EvalPBC.m}{\texttt{EvalPBC.m}} pourra être composé de deux voire quatre lignes, chacune de celles-ci étant une position sur laquelle l'addition devra être tentée.

Toute l'information nécessaire à la tentative d'addition en tant que telle étant maintenant obtenue (\texttt{Surface}, \texttt{Part2Add} et \texttt{PosPart*}), le test afin de savoir si tous les éléments de $\mathbf{S}^\ast_{kl}$ (\textit{cf.} équation \ref{Eq1Ann1}) sont inférieurs ou égaux à la valeur $3$ (ou 2) peut être réalisé. Si la condition est respectée pour toutes les positions (lignes) contenues dans \texttt{PosPart*}, alors l'addition est possible et les valeurs des éléments $\mathbf{P}_{mn}$ sont additionnées aux éléments $\mathbf{S}_{kl}$. Dans le même temps, la valeur $1$ est assignée à la variable \texttt{Succes} servant d'\textit{output} à la fonction à côté de la variable \texttt{Surface}. Inversement, si la condition n'est pas respectée (même sur un seul des éléments de $\mathbf{S}_{kl}$), l'addition sera considérée comme impossible et l'opération d'addition des matrices court-circuitée de telle sorte que les \textit{outputs} de la fonction demeureront identiques aux \textit{inputs}.

Bien que s'agissant d'une opération fondamentale, cet algorithme reste largement incomplet puisqu'il ne code que l'opération <<~tentative d'une addition~>>. Afin de considérer plusieurs additions séquentielles aléatoires et donc d'entrevoir la formation d'une monocouche (plusieurs particules), il sera nécessaire de le faire tourner en boucle, ce que s'attache à décrire la section suivante.

\section[Construction d'une monocouche saturée ($\ln K^\circ=-\infty$)]{Construction d'une monocouche saturée lorsque\\$\ln K^\circ=-\infty$ (relaxation instantanée des protéines)}\label{SurfConstruct1}

Dans un premier temps, il est commode de ne considérer que les cas d'additions séquentielles aléatoires pour lesquels le phénomène de relaxation est absent. Comme cela a été montré au chapitre \ref{ChapIsoth}, il s'agit de construire une monocouche dans les conditions d'un ratio $(K^\circ)^{-1}$ infini, c'est-à-dire $\ln K^\circ=-\infty$ (la vitesse d'addition d'une particule à la monocouche est infiniment plus grande que la vitesse de relaxation). Cette façon de procéder mènera à l'obtention des surfaces présentées à la figure \ref{FigRSAim1}.

La figure \ref{Fig4Ann1} illustre l'algorithme suivi afin de réaliser des additions séquentielles aléatoires de systèmes de particules tels que des boîtes de conserve de différents diamètres ou des IgG ayant différentes orientations. Moyennant la généralisation décrite à la section \ref{AppliRSA}, la procédure d'addition et la façon de tester s'il y a recouvrement est la même pour les deux systèmes ce qui permet aisément de passer d'un système à l'autre.

\begin{figure}\centering
\begin{normalsize}
\centerline{
\xymatrix{
&\txt{\texttt{Surface}, \texttt{NbrRequest}\\ et \texttt{NbrAdd} $=0$}\ar[dd]&\\
&&*[]{}\ar[l]\\
*[]{}\ar@{-}[ddddd]&*+[F.]+\txt{\footnotesize{\texttt{NbrAdd} $\stackrel{?}{\geqslant}$ \texttt{NbrRequest}}}\ar^{\mathrm{Non}}[d]\ar@{-}_{\mathrm{Oui}\qquad\qquad\quad}[l]&\\
&*++[F-,]++\txt{\textbf{Addition}}\ar[d]&\\
&*+[F.]+\txt{\footnotesize{\texttt{Succes} $\stackrel{?}{=}1$}}\ar^{\qquad\mathrm{Non}}[r]\ar^{\mathrm{Oui}}[d]&\\
&*+[F.]+\txt{\footnotesize{\texttt{NbrAdd} $=$ \texttt{NbrAdd} $+1$}}\ar[d]&\\
&*+[F.]+\txt{\footnotesize{$P(A^\ast)\stackrel{?}{\geqslant}0,9999$}}\ar@{-}^{\qquad\mathrm{Non}}[r]\ar@{-}^{\mathrm{Oui}}[d]
\ar[dd]&*[]{}\ar@{-}[uuuuu]\\
*[]{}\ar[r]&&\\
&\txt{\texttt{Surface}, \texttt{NbrRequest}\\ et \texttt{NbrAdd} $\geqslant0$}&\\}}
\end{normalsize}
\caption[Algorithme permettant le remplissage RSA]{Schéma de l'algorithme permettant la construction des monocouches par additions séquentielles aléatoires de particules n'ayant pas la possibilité de relaxer.}\label{Fig4Ann1}
\end{figure}
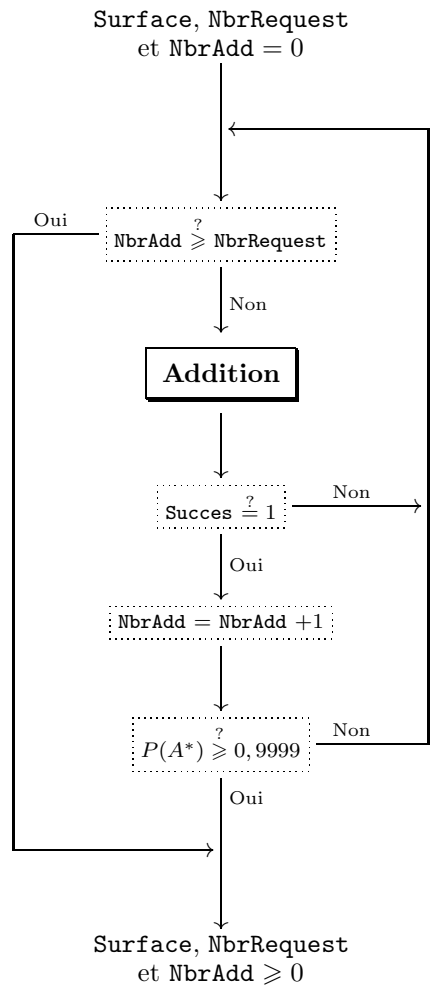

L'algorithme de la figure \ref{Fig4Ann1} est une simplification, car ne tenant pas compte de la relaxation, d'un algorithme plus complet écrit sous la forme d'une fonction \href{run:RSA/ConstructSurf.m}{\texttt{ConstructSurf.m}}. Comme montré sur la figure, les trois variables principales utilisées par cette fonction \href{run:RSA/ConstructSurf.m}{\texttt{ConstructSurf.m}} sont les variables \texttt{NbrRequest}, \texttt{Surface} et \texttt{NbrAdd}. La première variable, \texttt{NbrRequest},  indiquera à la fonction \href{run:RSA/ConstructSurf.m}{\texttt{ConstructSurf.m}} la quantité minimale de particule(s) qui devra être additionnée à la surface; c'est pourquoi une boucle \textit{while} est incluse à l'algorithme faisant que l'addition sera tentée tant que la valeur de \texttt{NbrAdd} sera plus petite que celle de \texttt{NbrRequest}, la valeur de la variable \texttt{NbrAdd} étant le nombre total de particules accumulées sur la surface. La matrice \texttt{Surface} permet, quant à elle de stocker l'image de la surface sur laquelle s'effectue le remplissage.

L'opération principale de la fonction \href{run:RSA/ConstructSurf.m}{\texttt{ConstructSurf.m}} est d'additionner séquentiellement de nouvelles particules à la surface. \`{A} cette fin, elle utilisera l'algorithme d'addition présenté à la figure \ref{Fig5Ann1} de manière séquentielle, ce dernier s'occupant de réaliser les tirages aléatoires nécessaires et, si possible, d'additionner une particule à la surface. L'opération effectuée est représentée par la case ombrée de la figure \ref{Fig4Ann1}. Dans le cas d'une addition réussie, le remplissage de la surface aura été augmenté d'une particule et la variable \texttt{Succes} portée à $1$. Inversement, la variable \texttt{Surface} n'aura pas été transformée, de même que la variable \texttt{Succes} demeurée à $0$.

Dans le cas d'un succès de l'addition, \textit{i.e.} \texttt{Succes}~$=1$, la variable \texttt{NbrAdd} sera incrémentée d'une unité. Suite à cette incrémentation, un nouveau test sera réalisé afin de savoir si la surface obtenue suite à cette nouvelle addition ne présente pas la caractéristique d'une surface saturée. Une surface saturée est une surface sur laquelle il devient très difficile de réaliser une nouvelle addition de particule; la probabilité de rejet $P(A^\ast)$ sera dès lors très élevée. Indépendamment de la valeur assignée à la variable \texttt{NbrRequest}, la saturation est une borne que l'on se fixe, \textit{i.e.} $P(A^\ast)=0,9999$, à partir de laquelle il sera considéré que la surface ne peut plus être remplie et que le processus de remplissage devra être arrêté. Lorsque la condition $P(A^\ast)\geqslant0,9999$ sera atteinte, on arrêtera le remplissage séquentiel aléatoire de la surface en sortant de la boucle \textit{while} et en renvoyant les variables \texttt{Surface}, \texttt{NbrAdd} et \texttt{NbrRequest} en l'état.

La probabilité de rejet $P(A^\ast)$ représente le volume exclu et peut être calculé précisément selon une procédure particulière faisant l'objet d'une discussion détaillée à la section \ref{WI}. Cependant, utiliser cette procédure itérativement à chaque fois que l'algorithme de la figure \ref{Fig4Ann1} procède à une addition ne serait pas judicieux pour des raisons de rapidité. Une procédure plus rapide mais grossière d'estimation doit donc être envisagée. La probabilité de rejet est alors estimée en comptant le nombre de rejet(s) (dans la variable \texttt{echec}) qui auront eu lieu lors de l'incrémentation de la variable \texttt{NbrAdd} d'une unité. En effet, on a que
\begin{equation}
P(A^\ast)\sim\frac{\mathrm{nombre\ de\ rejet(s)}}{\mathrm{nombre\ de\ tentative(s)}}.
\end{equation}
Ensuite, sachant que le nombre de tentative(s) est la somme du nombre de rejet(s) et du nombre d'addition réussie qui sera égal à 1, le test se faisant pour chaque itération de \texttt{NbrAdd}, on aura l'estimateur
\begin{equation}
P(A^\ast)\sim\frac{\mathrm{nombre\ de\ rejet(s)}}{\mathrm{nombre\ de\ rejet(s)}+1}
\end{equation}
qui ne pourra être plus grand ou égal à la limite fixée pour la saturation de la surface, soit $0,9999$. Dès lors, la fonction \href{run:RSA/ConstructSurf.m}{\texttt{ConstructSurf.m}} cessera de construire la monocouche lorsque la valeur de \texttt{NbrAdd} sera supérieure ou égale à la valeur de \texttt{NbrRequest} ou lorsque l'estimateur de la probabilité de rejet indiquera la saturation. Cette dernière possibilité se présentera lorsque la valeur assignée à la variable \texttt{NbrRequest} est trop élevée par rapport à ce que peut contenir la surface. Cette idée permet aussi d'obtenir facilement une surface saturée en assignant $+\infty$ à \texttt{NbrRequest}, possibilité intensivement utilisée afin de construire la fonction de saturation (\textit{cf.} figure \ref{FigFit} et annexe \ref{Ann2}).

\section[Construction d'une monocouche saturée ($\ln K^\circ\neq-\infty$)]{Construction d'une monocouche saturée lorsque\\$\ln K^\circ\neq-\infty$ (relaxation lente des protéines)}\label{SurfConstruct2}

Jusqu'à présent, seule la procédure permettant le remplissage de la surface seulement par additions séquentielles aléatoires a été développée. Mais, afin de mieux tenir compte des propriétés des particules impliquées dans le processus, il est nécessaire d'implémenter une possibilité de relaxation pour celles-ci. Cette possibilité n'apparaîtra que lorsque les particules auront été additionnées à la monocouche en formation. De ce fait, seront donc concernées par la relaxation à la fois la particule venant d'être additionnée mais aussi toutes celles déjà incluses précédemment à la monocouche et ayant encore la possibilité de relaxer.

Avant d'inclure la possibilité de relaxer dans l'algorithme de la figure \ref{Fig4Ann1}, la procédure par laquelle une tentative de relaxation sera réalisée doit être élaborée. La figure \ref{Fig6Ann1} illustre une telle procédure codée dans la fonction \href{run:RSA/DoRelaxation.m}{\texttt{DoRelaxation.m}}. La relaxation peut ne pas se faire en une seule étape. En effet, une particule de la monocouche, avant de se trouver dans un état que l'on considérera comme le plus relaxé, pourra passer par plusieurs étapes intermédiaires de relaxation partielle. La relaxation pourra donc se faire en une série de pas (étapes successives de la relaxation). Par exemple, après son addition, une IgG \textit{end-on} pourra se relaxer en adoptant une orientation \textit{flat} et ensuite seulement le faire une seconde fois en changeant sa conformation pour devenir \textit{flat$+$} (deux pas successifs). Selon les possibilités, la relaxation pourra aller à son terme ($\rightarrow$ \textit{flat$+$}) ou bien être arrêtée avant cela ($\rightarrow$ \textit{flat}). Dans ce qui suit, on envisagera uniquement le cas où l'on ne réalise qu'une seule tentative de relaxer une particule d'un seul pas.

\begin{figure}\centering
\begin{normalsize}
\centerline{
\xymatrix{
\txt{\texttt{ParticulesBib},\\\texttt{Surface}, \texttt{PartCoord}\\et \texttt{Part2Rel}}\ar[dd]\ar@{-}[r]&*[]{}\ar[d]\\
&*+[F.]+\txt{\footnotesize{\texttt{PosPart} $=$}\\
            \footnotesize{\texttt{PartCoord(Part2Rel,1:2)}}}\ar[d]\\
*++[F-,]++\txt{\textbf{Acquisition de}\\\texttt{Part2Rem} \textbf{et} \texttt{Part2Add}}\ar[d]&*++[F-,]++\txt{\textbf{Calcul PBC}}\ar[d]\\
\txt{\texttt{Part2Rem} et \texttt{Part2Add}}\ar[d]&\txt{\texttt{PosPart*}}\ar@{-}[d]\\
*+[F.]+\txt{\footnotesize{$\mathbf{S}_{kl}=\mathbf{S}_{kl}-\mathbf{P}^{\mathrm{2rel}}_{mn}$}}\ar[d]&*[]{}\ar[l]\\
*+[F.]+\txt{\footnotesize{$\mathbf{S}_{kl}+\mathbf{P}^{\mathrm{rel}}_{mn}\stackrel{?}{\leqslant}3$}}\ar@{-}^{\mathrm{Non}}[r]\ar^{\mathrm{Oui}}[d]&*[]{}\ar[d]\\
*+[F.]+\txt{\footnotesize{$\mathbf{S}_{kl}=\mathbf{S}_{kl}+\mathbf{P}^{\mathrm{rel}}_{mn}$}\\
            \footnotesize{\texttt{PartCoord(Part2Rel,3)} $=$}\\
            \footnotesize{\texttt{PartCoord(Part2Rel,3)} $+$ 1}}\ar[dd]&
*+[F.]+\txt{\footnotesize{$\mathbf{S}_{kl}=\mathbf{S}_{kl}+\mathbf{P}^{\mathrm{2rel}}_{mn}$}}\ar@{-}[d]\\
&*[]{}\ar[l]\\
\txt{\texttt{Surface}, \texttt{PartCoord}\\ et \texttt{Part2Rel}}&\\}}
\end{normalsize}
\caption[Algorithme de relaxation d'une particule de la monocouche]{Algorithme correspondant à la fonction \href{run:RSA/DoRelaxation.m}{\texttt{DoRelaxation.m}} et fournissant la possibilité de relaxer d'un pas une particule déjà incorporée à la monocouche. La relaxation se fera en respectant l'interdiction du recouvrement entre les particules de la moncouche de telle sorte que rien ne se produira si la relaxation devait engendrer une superposition.}\label{Fig6Ann1}
\end{figure}
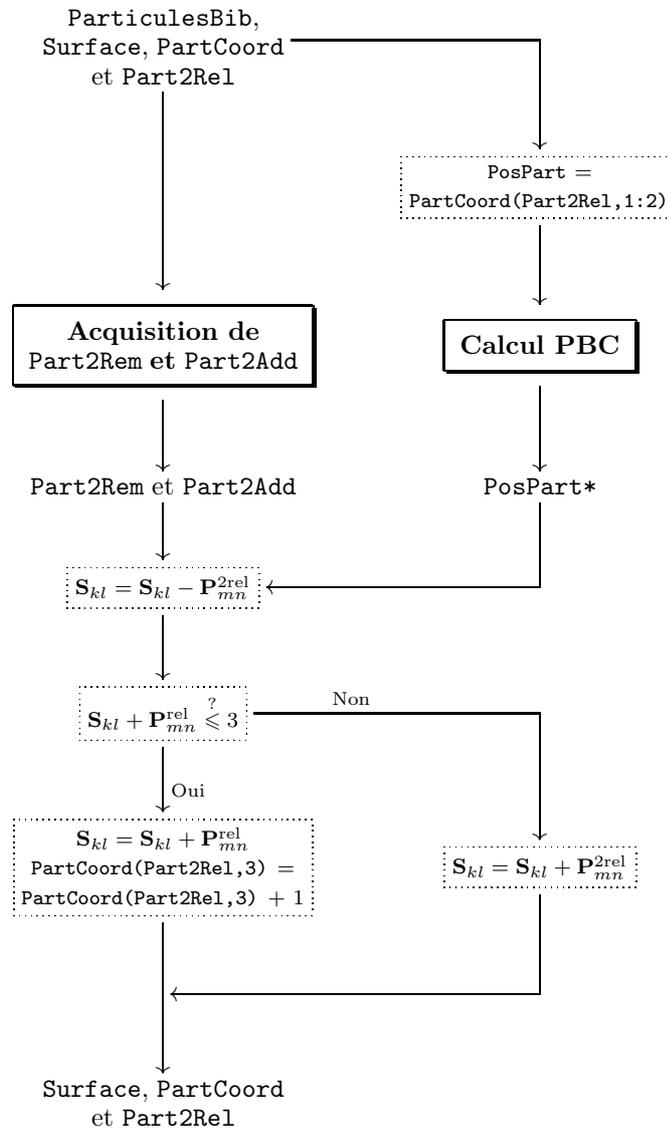

La fonction \href{run:RSA/DoRelaxation.m}{\texttt{DoRelaxation.m}} prend au moins quatre \textit{inputs}: \texttt{Surface}, \texttt{Part2Rel}, \texttt{PartCoord} et \texttt{ParticulesBib}. Les variables \texttt{Surface} et \texttt{ParticulesBib} sont bien connues. Ensuite, \texttt{PartCoord} est une matrice dans laquelle auront été stockées toutes les caractéristiques des particules additionnées à la monocouche (la position, l'état de relaxation, l'orientation générale et la rotation), et ce, dans l'ordre dans lequel elles auront été ajoutées à la monocouche. Sans que cela n'ait été détaillé, cette matrice est construite par la fonction \href{run:RSA/ConstructSurf.m}{\texttt{ConstructSurf.m}} au cours du remplissage de la surface (\textit{cf.} section précédente). Quant à la variable \texttt{Part2Rel}, elle indique quelle sera la particule de la monocouche dont il faudra tenter une relaxation, il s'agit simplement de l'indice d'une des lignes de \texttt{PartCoord}. Cet indice sera encore une fois obtenu par un tirage aléatoire.

La première étape du processus de relaxation est l'obtention de la position de la particule concernée. Pour ce faire, il suffira d'aller la chercher dans les deux premières colonnes de \texttt{PartCoord} à la ligne indiquée par \texttt{Part2Rel}. Cette position obtenue, elle sera transformée par la fonction \href{run:RSA/EvalPBC.m}{\texttt{EvalPBC.m}} afin d'obtenir la variable \texttt{PosPart*} contenant toutes les positions où se trouve la particule à relaxer.

D'autre part, il sera aussi nécessaire d'obtenir la matrice $\mathbf{P}^{\mathrm{2rel}}$ contenant la particule à relaxer et la matrice $\mathbf{P}^{\mathrm{rel}}$ de la particule relaxée. Ceci permettra de retirer $\mathbf{P}^{\mathrm{2rel}}$ afin de la remplacer par $\mathbf{P}^{\mathrm{rel}}$. Grâce aux renseignements contenus dans \texttt{PartCoord}, cela se fera simplement en allant chercher dans le tableau \texttt{ParticulesBib} les matrices de $\mathbf{P}^{\mathrm{rel}}$ et $\mathbf{P}^{\mathrm{2rel}}$ que l'on assignera respectivement à \texttt{Part2Add} et \texttt{Part2Rem}.

Toutes les assignations ayant été faites, il est maintenant possible de réaliser la relaxation en tant que telle qui n'est rien d'autre qu'une tentative de remplacement de \texttt{Part2Rem} par \texttt{Part2Add} en respectant les exigences en matière de superposition du modèle RSA. La première étape consistera donc à enlever la matrice $\mathbf{P}^{\mathrm{2rel}}$ (\texttt{Part2Rem}) de la surface par une opération de soustraction. L'espace libéré, il est ensuite possible de tester si aucun des éléments résultant de l'addition de $\mathbf{P}^{\mathrm{rel}}$ n'est supérieur à trois (ou deux). Si c'est le cas, c'est-à-dire que la relaxation engendrerait une superposition, $\mathbf{P}^{\mathrm{rel}}$ n'est pas ajoutée et $\mathbf{P}^{\mathrm{2rel}}$ est remise à la même position. Inversement, si aucun recouvrement n'est engendré, la matrice $\mathbf{P}^{\mathrm{rel}}$ est additionnée à la surface et l'élément de \texttt{PartCoord} correspondant à l'état de relaxation de la particule concernée est incrémenté d'une unité.

L'algorithme permettant de tenter une relaxation sur une particule de la monocouche ayant été obtenu, il est incorporé à l'algorithme du modèle des additions séquentielles aléatoires de la figure \ref{Fig4Ann1} afin de donner le nouveau schéma de la figure \ref{Fig7Ann1}.

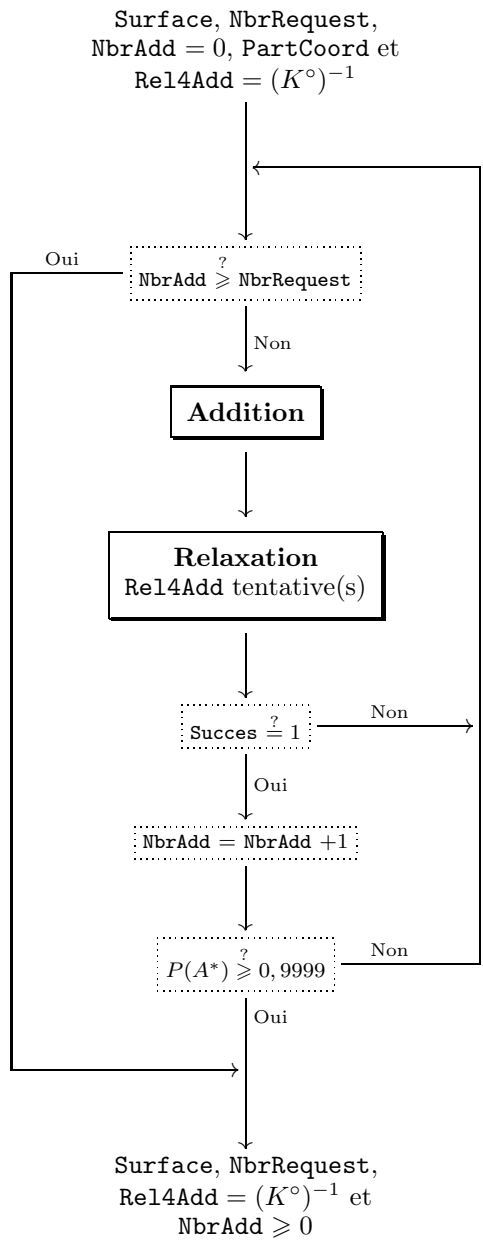
\begin{figure}\centering
\begin{normalsize}
\centerline{
\xymatrix{
&\txt{\texttt{Surface}, \texttt{NbrRequest},\\\texttt{NbrAdd} $=0$, \texttt{PartCoord} et\\\texttt{Rel4Add} $=(K^\circ)^{-1}$}\ar[dd]&\\
&&*[]{}\ar[l]\\
*[]{}\ar@{-}[dddddd]&*+[F.]+\txt{\footnotesize{\texttt{NbrAdd} $\stackrel{?}{\geqslant}$ \texttt{NbrRequest}}}\ar^{\mathrm{Non}}[d]\ar@{-}_{\mathrm{Oui}\qquad\qquad\quad}[l]&\\
&*++[F-,]++\txt{\textbf{Addition}}\ar[d]&\\
&*++[F-,]++\txt{\textbf{Relaxation}\\\texttt{Rel4Add} tentative(s)}\ar[d]&\\
&*+[F.]+\txt{\footnotesize{\texttt{Succes} $\stackrel{?}{=}1$}}\ar^{\qquad\mathrm{Non}}[r]\ar^{\mathrm{Oui}}[d]&\\
&*+[F.]+\txt{\footnotesize{\texttt{NbrAdd} $=$ \texttt{NbrAdd} $+1$}}\ar[d]&\\
&*+[F.]+\txt{\footnotesize{$P(A^\ast)\stackrel{?}{\geqslant}0,9999$}}\ar@{-}^{\qquad\mathrm{Non}}[r]\ar@{-}^{\mathrm{Oui}}[d]
\ar[dd]&*[]{}\ar@{-}[uuuuuu]\\
*[]{}\ar[r]&&\\
&\txt{\texttt{Surface}, \texttt{NbrRequest},\\\texttt{Rel4Add} $=(K^\circ)^{-1}$ et\\\texttt{NbrAdd} $\geqslant0$}&\\}}
\end{normalsize}
\caption[Algorithme permettant le remplissage RSA+R]{Schéma complet de l'algorithme des additions séquentielles aléatoires auquel la relaxation a été implémentée. Cet algorithme est codé par la fonction \href{run:RSA/ConstructSurf.m}{\texttt{ConstructSurf.m}}. Il s'agit de l'algorithme principal du modèle RSA+R.}\label{Fig7Ann1}
\end{figure}

Ce schéma est le schéma complet de la fonction \href{run:RSA/ConstructSurf.m}{\texttt{ConstructSurf.m}} prenant comme \textit{inputs}, en plus de ceux présentés à la figure \ref{Fig4Ann1}, les variables \texttt{PartCoord} et \texttt{Rel4Add}. L'utilité de \texttt{PartCoord} a déjà été explicitée ci-dessus et la variable \texttt{Rel4Add} permet, quant à elle, de déterminer le nombre de tentatives de relaxations qui devront être réalisées entre chaque tentative d'addition. La valeur assignée à cette variable s'obtient à partir de $(K^\circ)^{-1}$ où $K^\circ$ est le taux nominal d'addition, invariable au cours du processus. Après chaque sollicitation de la fonction \href{run:RSA/DoAddition.m}{\texttt{DoAddition.m}}, la fonction \href{run:RSA/DoRelaxation.m}{\texttt{DoRelaxation.m}} sera à son tour sollicitée $(K^\circ)^{-1}$ fois et pour chacune de ces sollicitations, une particule à relaxer sera tirée au sort grâce à la matrice \texttt{PartCoord}.

Il est important de noter que $(K^\circ)^{-1}$ ne prend pas uniquement des valeurs entières alors que la fonction \href{run:RSA/DoRelaxation.m}{\texttt{DoRelaxation.m}} ne peut être sollicitée, elle, qu'un nombre entier de fois. \`{A} cette fin, un générateur de nombres pseudo-aléatoires appartenant à l'ensemble des nombres naturels doit être écrit afin que l'espérance des résultats générés tende vers la valeur $(K^\circ)^{-1}$. Ce générateur se trouve à la ligne 51 de la fonction \href{run:RSA/ConstructSurf.m}{\texttt{ConstructSurf.m}}.

La prise en compte de la relaxation fournit à la fonction \href{run:RSA/ConstructSurf.m}{\texttt{ConstructSurf.m}} la capacité de construire une surface de couverture minimale déterminée par \texttt{NbrRequest} en tenant compte du rapport entre les vitesses d'addition et de relaxation grâce au ratio des vitesses initiales $K^\circ$. \href{run:RSA/ConstructSurf.m}{\texttt{ConstructSurf.m}} construit donc une surface selon les hypothèses du modèle RSA et en tenant compte du phénomène de relaxation dont la vitesse est posée avec précision par rapport à l'addition.

\section{Estimation du volume exclu}\label{WI}

Le volume exclu d'une monocouche est une notion importante et l'obtention d'un bon estimateur est capital. Le volume exclu est la proportion du volume de la surface qui est inaccessible lors de l'addition d'une nouvelle particule. Lorsque celui-ci est nul, tout le volume de la surface est disponible pour l'addition tandis que lorsque celui-ci approche de l'unité, la part de volume sera extrêmement restreinte de telle sorte que la probabilité d'une nouvelle addition sera très faible. Cette notion de volume exclu a été généralisée au chapitre \ref{SectionRSA} par la notion de probabilité de rejet $P(A^\ast)$ qui est la probabilité de ne pas additionner (\textit{cf.} équation \ref{EqAddRejet}).

Si le volume exclu peut être généralisé par la probabilité de rejet $P(A^\ast)$, un estimateur facile à calculer et robuste est donné par
\begin{equation}\label{EqEstimProb1}
P(A^\ast)\sim\frac{\mathrm{nombre\ de\ rejet(s)}}{\mathrm{nombre\ de\ tentatives\ d\text{'}additions}}
\end{equation}
lorsque le nombre de tentatives d'additions est suffisamment grand. Il s'agira donc, pour une surface fournie par la fonction \href{run:RSA/ConstructSurf.m}{\texttt{ConstructSurf.m}} de réaliser un très grand nombre de tentatives d'additions (sans modifier \texttt{Surface}) et de rapporter le nombre de rejet(s) au nombre de tentatives effectuées. Chaque tentative se fera en générant aléatoirement l'angle de rotation de la particule sur laquelle le volume exclu sera évalué. L'algorithme de test est, à peu de chose près, équivalent à celui de la figure \ref{Fig5Ann1} (il faut alors retirer la case réalisant l'opération d'addition). Il existe une fonction, par ailleurs utilisée par la fonction \href{run:RSA/DoAddition.m}{\texttt{DoAddition.m}}, pour le coder: \href{run:RSA/EvalAddition.m}{\texttt{EvalAddition.m}}.

Ceci est valable pour des surfaces qui ne sont remplies que d'un seul type de particule, par exemple une sphère d'un seul diamètre ne relaxant pas. Or, il a été vu que des particules différentes ou d'orientations générales variables pouvaient s'additionner de manière concurrentielle. D'autre part, pour une même monocouche, le volume exclu ne sera pas identique selon la particule et son orientation générale car celui-ci résulte d'une interaction entre deux entités: l'exclu (la particule) et l'excluant (la monocouche). Pour une même monocouche, plus une particule est <<~grosse~>> plus le volume exclu par rapport à celle-ci pourra être élevé, il est donc nécessaire de généraliser cette notion de volume exclu. Au chapitre \ref{SectionRSA}, il avait été vu que le volume exclu $P(A^\ast)$ pouvait être décomposé par une somme d'éléments $P(A^\ast\cap\boldsymbol{\omega}_i)$, les probabilités d'observer une particule d'orientation générale $\boldsymbol{\omega}_i$ qui sera rejetée:
\begin{equation}
P(A^\ast)=\sum^n_{i=1}P(A^\ast\cap\boldsymbol{\omega}_i).
\end{equation}
En utilisant la définition de la probabilité conditionnelle, on obtient que
\begin{equation}\label{EqCombiLi}
P(A^\ast)=\sum^n_{i=1}P(A^\ast|\boldsymbol{\omega}_i)P(\boldsymbol{\omega}_i)
\end{equation}
où les $P(A^\ast\cap\boldsymbol{\omega}_i)$ ont été transformées en $P(A^\ast|\boldsymbol{\omega}_i)P(\boldsymbol{\omega}_i)$ le produit de la probabilité qu'une particule d'orientation générale $\boldsymbol{\omega}_i$ soit rejetée et la probabilité qu'une telle particule arrive à la surface $P(\boldsymbol{\omega}_i)$. Ces dernières probabilités s'obtiennent facilement par le fait qu'elles sont définies par les hypothèses de la simulation et, en particulier, grâce à la seconde dimension du tableau \texttt{ParticulesBib}. Chacune de ces dimensions correspondant à une orientation générale, l'algorithme d'addition procède à un tirage aléatoire afin d'obtenir \texttt{TypPart} l'indice de l'orientation générale. Chaque orientation générale sera donc obtenue de manière équiprobable, cette probabilité étant alors l'inverse du nombre de colonne (\textit{cf.} équation \ref{dlfhglsdqghldglbglg}). Il est à remarquer que l'on pourrait jouer sur les probabilités; afin d'obtenir le résultat $P(\boldsymbol{\omega}_{\mathrm{end\text{-}on}})=0,5$, $P(\boldsymbol{\omega}_{\mathrm{side\text{-}on}})=0,25$ et $P(\boldsymbol{\omega}_{\mathrm{flat}})=0,25$, on construira \texttt{ParticulesBib} de telle sorte que deux indices de la seconde dimension correspondent à l'orientation \textit{end-on}, un pour l'orientation \textit{side-on} et un autre pour la \textit{flat}.

Les probabilités conditionnelles $P(A^\ast|\boldsymbol{\omega}_i)$ sont obtenues dans la droite ligne des probabilités de rejet $P(A^\ast)$ en adaptant l'équation \ref{EqEstimProb1}:
\begin{equation}\label{EqEstimProb2}
P(A^\ast|\boldsymbol{\omega}_i)\sim\frac{\mathrm{nombre\ de\ rejet(s)\ de}\ \boldsymbol{\omega}_i}{\mathrm{nombre\ de\ tentatives\ d\text{'}additionner}\ \boldsymbol{\omega}_i}.
\end{equation}
En effet, en tentant un très grand nombre d'additions de la particule dans l'orientation $\boldsymbol{\omega}_i$ et en comptant le nombre de rejet(s), un estimateur de la probabilité conditionnelle est obtenu. La probabilité conditionnelle $P(A^\ast|\boldsymbol{\omega}_i)$ s'énonce précisément: \textit{probabilité d'observer un rejet sachant l'orientation $\boldsymbol{\omega}_i$}. Il s'agit donc de compter le nombre de rejet(s) en ne réalisant qu'un échantillonnage sur une seule orientation générale $\boldsymbol{\omega}_i$ (toute rotation confondue). La probabilité de rejet $P(A^\ast)$ s'estime alors par la combinaison linéaire \ref{EqCombiLi}.

Le procédé d'estimation du volume exclu par tentative d'addition est très utilisé dans la littérature et est qualifié d'\textit{insertion de Widom} \citep{widom1963}. La fonction écrite afin de réaliser cette évaluation sur une matrice \texttt{Surface} construite par \href{run:RSA/ConstructSurf.m}{\texttt{ConstructSurf.m}} est \href{run:RSA/EvalExVol4Add.m}{\texttt{EvalExVol4Add.m}}. En évaluant le volume exclu pour des surfaces de taux de recouvrement croissants, on obtiendra les courbes présentées aux figures \ref{FigRSA1}, \ref{FigRSA2} et \ref{FigRSA4} du chapitre \ref{SectionRSA}.

\section{Structure globale du programme de simulation}\label{ProgramGen}

Une fonction particulière \href{run:RSA/ConstructRawData.m}{\texttt{ConstructRawData.m}} permettra de générer les données brutes sur base du modèle des additions séquentielles aléatoires. Pour des surfaces dont le taux de recouvrement sera croissant, elle fera tourner en boucle les fonctions \href{run:RSA/ConstructSurf.m}{\texttt{ConstructSurf.m}} et \href{run:RSA/EvalExVol4Add.m}{\texttt{EvalExVol4Add.m}}. Les données générées, \textit{i.e.} la quantité totale de particules constituant la monocouche ($\Theta$), la proportion de chaque orientation générale ($\mathcal{X}(\boldsymbol{\omega}_i)$), et les probabilités conditionnelles de rejet ($P(A^\ast|\boldsymbol{\omega}_i)$), seront stockées dans le tableau \texttt{RawData}.

Avant de procéder à cette simulation, un certain nombre d'étapes doivent être réalisées. Premièrement, il est nécessaire de fixer les conditions dans lesquelles se fera la simulation, c'est-à-dire qu'il faudra fixer le ratio des vitesses initiales $K^\circ$ qui demeurera constant tout au long de celle-ci. Par commodité, on fixera plutôt son logarithme $\ln K^\circ$ (pour mémoire, $\ln K^\circ=+\infty$ lorsque la vitesse initiale de relaxation sera infiniment plus grande que l'addition).

Deuxièmement, il sera nécessaire de construire le tableau \texttt{ParticulesBib} dans lequel les algorithmes de la simulation viendront piocher les matrices contenant les particules à additionner/relaxer dans la monocouche. Cette tâche sera réalisée par \href{run:RSA/ConstructBib.m}{\texttt{ConstructBib.m}} prenant comme argument le nom du fichier dans lequel sont stockées les informations de base sur le système de particules faisant l'objet de la simulation. Une série de fichiers d'amorce \texttt{*.mat} est donnée dans le répertoire \href{run:RSA/start/}{\texttt{./RSA/start/}}:
\begin{enumerate}
\item \href{run:RSA/start/Pillbox_Petit.mat}{\texttt{Pillbox\_{}Petit.mat}}: fichier pour la simulation d'une construction de monocouche contenant des sphères (ou des cylindres) ayant toutes le même rayon (\textit{cf.} figure \ref{FigSphèresRSA});
\item \href{run:RSA/start/IgG_EO+SO+F.mat}{\texttt{IgG\_{}EO+SO+F.mat}}: fichier pour la simulation d'une construction de monocouche à partir d'IgG arrivant de façon équiprobable à la surface dans trois orientations générales possibles, \textit{i.e.} \textit{end-on}, \textit{side-on} et \textit{flat}, sans possibilité de relaxation (\textit{cf.} figures \ref{FigSchémaRSA1}.B et \ref{FigRSA1}.B);
\item \href{run:RSA/start/Pillbox_Petit+Moyen+Grand.mat}{\texttt{Pillbox\_{}Petit+Moyen+Grand.mat}}: idem mais pour les boîtes de conserve (\textit{cf.} figures \ref{FigSchémaRSA1}.A et \ref{FigRSA1}.A);
\item \href{run:RSA/start/IgG_EO_RelOri.mat}{\texttt{IgG\_{}EO\_{}RelOri.mat}}: fichier pour la simulation d'une construction de monocouche à partir d'IgG arrivant exclusivement selon l'orientation générale \textit{end-on} à la surface mais ayant la possibilité de relaxer cette orientation pour devenir \textit{flat} (\textit{cf.} figure \ref{FigRSA2});
\item \href{run:RSA/start/Pillbox_Petit_RelConf_3.mat}{\texttt{Pillbox\_{}Petit\_{}RelConf\_{}3.mat}}: fichier pour la simulation de la construction d'une monocouche à partir des petites boîtes de conserve ayant la possibilité de relaxer leur conformation en deux pas successifs (\textit{cf.} figures \ref{FigRSARelax}.A et \ref{FigRSA4}.A);
\item \href{run:RSA/start/IgG_EO_RelOri+RelConf.mat}{\texttt{IgG\_{}EO\_{}RelOri+RelConf.mat}}: idem mais les IgG peuvent, après avoir relaxé leur orientation (\textit{end-on} $\rightarrow$ \textit{flat}), aussi relaxer leur conformation pour devenir \textit{flat$+$} (\textit{cf.} figures \ref{FigRSARelax}.B et \ref{FigRSA4}.B).
\end{enumerate}
En appelant la fonction \texttt{ConstructBib('Pillbox\_{}Petit+Moyen+Grand.mat)'}, un fichier d'extension \texttt{*\_{}Data4Simul.mat} sera créé. Celui-ci, afin d'être utilisé lors de la simulation doit encore être adjoint d'un certain nombre d'informations nécessaires, entre autres, à la fonction \href{run:RSA/EvalExVol4Add.m}{\texttt{EvalExVol4Add.m}}. Pour ce faire on invoquera \texttt{DetectSim('Pillbox\_{}Petit+Moyen+Grand.mat')}, ce qui complètera le fichier \texttt{Pillbox\_{}Petit+Moyen+Grand\_{}Data4Simul.mat} déjà enregistré. Ce dernier sera utilisé par \href{run:RSA/ConstructRawData.m}{\texttt{ConstructRawData.m}} afin de générer les données brutes d'une simulation RSA. Globalement, on aura:
\begin{enumerate}
\item \texttt{LnKzero = - Inf;}
\item \texttt{ConstructBib('IgG\_{}EO\_{}RelOri+RelConf.mat');}
\item \texttt{DetectSim('IgG\_{}EO\_{}RelOri+RelConf.mat');}
\item \texttt{ConstructRawData('IgG\_{}EO\_{}RelOri+RelConf.mat',LnKzero);}
\end{enumerate}
afin de générer les données brutes permettant la construction du graphique de la figure \ref{FigRSA4}.B.

\section{Transformation des données brutes}\label{ygfvyqgfo}

\`{A} la suite de l'obtention des données brutes, quelques transformations doivent encore être effectuées afin d'obtenir des séries de données utilisables. Il sera en effet nécessaire d'invoquer une nouvelle fonction à la suite des précédentes:
\begin{enumerate}
\setcounter{enumi}{4}
\item \texttt{RawData2Data('IgG\_{}EO\_{}RelOri+RelConf.mat',LnKzero);}
\end{enumerate}
finissant pas générer des données dans des fichiers d'extension \texttt{*\_{}Data2Plot.mat} à partir des fichiers \texttt{*\_{}Data4Simul.mat}. Ces derniers permettent d'obtenir les graphiques des figures \ref{FigRSA1} et \ref{FigRSA4}.

La fonction \href{run:RSA/RawData2Data.m}{\texttt{RawData2Data.m}} modifie le format (ligne 8 du code) et réalise une série d'opérations sur les données contenues dans le tableau \texttt{RawData} généré par \href{run:RSA/ConstructRawData.m}{\texttt{ConstructRawData.m}}. Premièrement, le temps adimensionnel est calculé. Décrit par l'équation \ref{EqCinétique5} d'après P. Schaaf \textit{et al}. \citep{schaaf1998}, ce temps autorise la construction des cinétiques montrées aux figures \ref{FigCinétique}.A et \ref{FigCinétique}.B. Numériquement, il se calcule en intégrant l'inverse de la probabilité d'addition $P(A)$ en utilisant la méthode des trapèzes (\textit{cf.} fonction \texttt{Touch2Time} incluse dans \href{run:RSA/RawData2Data.m}{\texttt{RawData2Data.m}}).

Deuxièmement, dans le but d'obtenir des sets de données facilement utilisables, une série de lissages est effectuée. La technique de lissage utilisée est un type de régression non paramétrique (\textit{kernel smoothing}) basée sur un estimateur à noyau (ici, il s'agit d'un noyau quadratique d'Epanechnikov). Cette méthode d'estimation effectue une régression linéaire de façon locale, c'est-à-dire en chaque point où l'on souhaite estimer la fonction. Le détail de cette méthode codée dans la fonction \href{run:RSA/KernelSmooth.m}{\texttt{KernelSmooth.m}} est explicitée par T. Hastie \textit{et al.} \citep{hastie2009b} (\textit{cf.} équations 6.3, 6.4 et 6.8).

Pour terminer, les valeurs en nombres, sont converties en picomoles par centimètres carrés en se basant sur la taille des particules (voir fonction \href{run:RSA/RawData2Data.m}{\texttt{RawData2Data.m}}). Ce sont ces courbes lissées qui sont présentées au chapitre \ref{SectionRSA}.

\begin{footnotesize}

\end{footnotesize}\end{cbunit}
\begin{cbunit}
\chapter{Estimation de la fonction de saturation}\label{Ann2}
\markboth{Annexe \ref{Ann2}: FONCTION DE SATURATION}{}
\minitoc

\section{Construction de la fonction de saturation}

Après avoir exposé la façon de construire des surfaces saturées aux sections \ref{SurfConstruct1} et \ref{SurfConstruct2} de l'annexe précédente, c'est-à-dire pour lesquelles le volume exclu $P(A^\ast)$ est factuellement considéré comme supérieur ou égal à $99,99$~\%, la construction des fonctions de saturation vient naturellement. En effet, la fonction \href{run:RSA/ConstructSurf.m}{\texttt{ConstructSurf.m}} a été développée afin de construire une monocouche de particules dans les conditions d'un ratio des vitesses initiales $(K^\circ)^{-1}$ et à renvoyer une surface saturée lorsque le nombre requis de particules à y incorporer se trouve être trop élevé. Le principe de construction de la fonction de saturation consistera donc à faire tourner en boucle l'instruction \texttt{ConstructSurf(Inf,Rel4Add)}. Dans cette instruction, la variable \texttt{NbrRequest} est posée comme infinie afin que \href{run:RSA/ConstructSurf.m}{\texttt{ConstructSurf.m}} ne construise que des monocouches saturées et \texttt{Rel4Add} est la variable liée à $(K^\circ)^{-1}$ dont le domaine sera balayé afin de construire la fonction de saturation. Cette opération est assurée par la fonction \href{run:RSA/QS_Construct.m}{\texttt{QS\_{}Construct.m}}.

La fonction \href{run:RSA/QS_Construct.m}{\texttt{QS\_{}Construct.m}} nécessitera un certain nombre de prérequis afin de pouvoir fonctionner correctement. Un de ces prérequis est le chargement d'un fichier \texttt{*\_{}Data4Simul.mat} construit préalablement. On aura donc la suite d'instructions suivantes afin de construire une fonction de saturation:
\begin{enumerate}
\item \texttt{ConstructBib('Pillbox\_{}Petit\_{}RelConf\_{}3.mat');}
\item \texttt{DetectSim('Pillbox\_{}Petit\_{}RelConf\_{}3.mat');}
\item \texttt{QS\_{}Construct('Pillbox\_{}Petit\_{}RelConf\_{}3.mat');}
\end{enumerate}
La raison des deux premières instructions ayant été explicitée à la section \ref{AlgoRSA} (annexe \ref{AnnRSA}), on ne s'attardera que sur le fonctionnement général de la fonction \href{run:RSA/QS_Construct.m}{\texttt{QS\_{}Construct.m}}. Après avoir chargé le contenu du fichier de données \texttt{*\_{}Data4Simul.mat} dans le \textit{workspace}, on commencera par initialiser les variables nécessaires à la fonction \href{run:RSA/ConstructSurf.m}{\texttt{ConstructSurf.m}} dont il sera intensément fait usage de même qu'un tableau permettant de stocker les résultats. On codera:
\begin{small}
\begin{verbatim}
% Initialisation des variables nécessaires à "ConstructSurf.m"

Surface.surface = zeros(Surface.dim + 2 * Surface.bord);
Surface.Neff = zeros(max(ParticulesBib.NbrEtat),ParticulesBib.NbrType);
Surface.failure = 0;

% Initialisation du tableau "QS" afin de stocker les résultats

QS.QS = cell(1,2);
\end{verbatim}
\end{small}
qui se trouve inséré des lignes 20 à 26 dans \href{run:RSA/QS_Construct.m}{\texttt{QS\_{}Construct.m}}.

Ensuite, il faudra calculer les valeurs de $(K^\circ)^{-1}$ pour lesquelles des surfaces seront saturées en particules. Ces éléments, constituant l'axe des abscisses de la figure \ref{FigFit} et stockés dans le vecteur colonne \texttt{K}, s'obtiennent comme suit (lignes 28 à 39):
\begin{small}
\begin{verbatim}
% Construction du vecteur contenant les K pour lesquels une
% surface saturée sera construite

frac = 90;
multi = exp([0:1:frac-1]/frac);
i = 0;
for j = 12 : -1 : -5
    for k = frac : -1 : 1
        i = i + 1;
        LocVar.K(i,1) = multi(1,k) * exp(j);
    end
end
\end{verbatim}
\end{small}

Avec les informations ainsi générées, il est possible de lancer la construction d'une fonction de saturation. Cette opération se code assez naturellement selon les lignes 56 à 66 de la fonction \href{run:RSA/QS_Construct.m}{\texttt{QS\_{}Construct.m}}:
\begin{small}
\begin{verbatim}
% Evaluation des points de la fonction de saturation

for m = 1 : 1 : size(LocVar.K,1)
    LocVar.Rel4Add = max(ParticulesData.CoordClosePack-1)...
                                                  / LocVar.K(m,1);
    ConstructSurf(Inf,LocVar.Rel4Add);
    QS.QS{m,1} = LocVar.K(m,1);
    QS.QS{m,2} = Surface.Neff;
end
\end{verbatim}
\end{small}

\'{E}tant donné la forme sigmo\"{i}dale de la fonction de saturation, il est possible d'obtenir des informations assez précises sur les paramètres dont elle dépend. En particulier, l'obtention de bons estimateurs des paramètres $\Theta_{\infty,i}$ et $\Delta\Theta$, \textit{i.e.} les asymptotes de la fonction, se fera simplement par le code suivant (lignes 41 à 50):
\begin{small}
\begin{verbatim}
% Evaluation des asymptotes de la fonction de saturation

QS.ThetaMax = zeros(max(ParticulesBib.NbrEtat),ParticulesBib.NbrType);
QS.ThetaMin = zeros(max(ParticulesBib.NbrEtat),ParticulesBib.NbrType);
for i = 1 : 1 : 10
    ConstructSurf(Inf,0);
    QS.ThetaMax = QS.ThetaMax + Surface.Neff./10;
    ConstructSurf(Inf,Inf);
    QS.ThetaMin = QS.ThetaMin + Surface.Neff./10;
end
\end{verbatim}
\end{small}
Ce code utilise le fait que les asymptotes inférieure et supérieure de la fonction de saturation correspondent à des surfaces saturées ayant été construites respectivement dans les limites d'un nombre infini et nul de relaxation(s) par rapport aux additions. $\Theta_{\infty,i}$ et $\Delta\Theta$ ($=\Theta_{\infty,s}-\Theta_{\infty,i}$) se calculeront donc simplement par des moyennes sur les nombres de particules constituant les monocouches ainsi construites. Cette opération est utile afin de limiter les paramètres à estimer par minimisation de la somme des carrés des résidus. D'un point de vue statistique, le fait de limiter le nombre de paramètres à estimer augmente le biais statistique mais permet surtout de limiter la variance, ce qui est, ici, opportun.

La fonction \href{run:RSA/QS_Construct.m}{\texttt{QS\_{}Construct.m}} enregistre les données qu'elle aura générées dans un fichier d'extension \texttt{*\_{}QS.mat} inclus dans le dossier \href{run:RSA/qs/}{\texttt{./RSA/qs/}}.

\section[Estimation des paramètres de la fonction de saturation]{Estimation des paramètres de la fonction\\de saturation}\label{654gfb354dfg354dqf354}

Au vu de l'équation \ref{FonctionSaturation5} de la fonction de saturation, il est impossible d'en ajuster les paramètres $\tilde{K}^\circ$, $v$ et $w$ par des méthodes de régressions linéaires classiques. Toutefois, en utilisant la somme des carrés des résidus comme critère de minimisation, il est possible d'en obtenir une bonne estimation en explorant les domaines auxquels appartiennent ces paramètres.

Au préalable, on aura estimé $\Theta_{\infty,i}$ et $\Delta\Theta=\Theta_{\infty,s}-\Theta_{\infty,i}$ en obtenant une série de surfaces saturées pour $\ln K^\circ=-\infty$ et $\ln K^\circ=+\infty$ (\textit{cf.} ci-dessus). Ces valeurs sont dès lors assignées à deux variables: \texttt{ThetaInf} et \texttt{DeltaTheta}.

Pour les trois paramètres $\tilde{K}^\circ$, $v$ et $w$ à ajuster, il est nécessaire de fixer des valeurs initiales, c'est-à-dire celles qui, injectées avec $\Theta_{\infty,i}$ et $\Delta\Theta$ dans l'équation de la fonction de saturation \ref{FonctionSaturation5}, permettront d'estimer la première somme des carrés des résidus. Ces valeurs sont fixées par l'assignation suivante
\begin{small}
\begin{verbatim}
LnTildeK = min(LnK);
w = 1;
v = 1;
\end{verbatim}
\end{small}
par laquelle on assigne à \texttt{LnTildeK} ($\tilde{K}^\circ$) la plus petite valeur contenue dans le vecteur \texttt{LnK} qui contient toutes les valeurs de $K^\circ$ pour lesquelles un point de la fonction de saturation a été estimé. Aussi, étant donné qu'il s'agit d'exposants, une valeur unitaire est assignée à $w$ et $v$ les rendant <<~sans effet~>>.

Avant d'expliciter l'algorithme d'estimation proprement dit, quelques variables utiles à son arrêt doivent être créées:
\begin{small}
\begin{verbatim}
% Estimation des paramètres

parmOld = [LnTildeK,v,w];
cond = 1;
accur = 0.0001;
\end{verbatim}
\end{small}
Ces variables permettent d'arrêter le processus d'estimation lorsque la précision souhaitée (\texttt{accur}) aura été atteinte. En effet, l'algorithme, consistant en une série d'estimations successives des trois paramètres $\ln\tilde{K}^\circ$, $v$ et $w$, est inclus dans une boucle \textit{while} qui sera exécutée tant que la valeur de la variable \texttt{cond} sera égale à $1$. On aura \texttt{cond = 1} tant que le trio de nouveaux paramètres (\texttt{parm}) n'aura pas été estimé jusqu'à une précision de $0,0001$. On aura le code suivant:
\begin{small}
\begin{verbatim}
while cond == 1
    
% Estimation de 'LnTildeK'
    SCRold = 1e10;
    for LnTildeK = LnK(1,1) : accur : LnK(size(LnK,1),1)
        ThetaEstim = QS_Estim(LnK,...
                               ThetaInf,DeltaTheta,w,v,LnTildeK);
        CR = (ThetaEstim - Theta).^2;
        SCR = sum(CR);
        if SCR > SCRold
            LnTildeK = LnTildeK - accur;
            break
        else
            SCRold = SCR;
        end
    end

% Estimation de 'v'
    SCRold = 1e10;
    for v = accur : accur : 4
        ThetaEstim = QS_Estim(LnK,...
                               ThetaInf,DeltaTheta,w,v,LnTildeK);
        CR = (ThetaEstim - Theta).^2;
        SCR = sum(CR);
        if SCR > SCRold
            v = v - accur;
            break
        else
            SCRold = SCR;
        end
    end
    
% Estimation de 'w'
    SCRold = 1e10;
    for w = accur : accur : 1
        ThetaEstim = QS_Estim(LnK,...
                               ThetaInf,DeltaTheta,w,v,LnTildeK);
        CR = (ThetaEstim - Theta).^2;
        SCR = sum(CR);
        if SCR > SCRold
            w = w - accur;
            break
        else
            SCRold = SCR;
        end
    end
    
% Assignation des nouvelles estimations à 'parm'
    parm = [LnTildeK,v,w];
    
% La précision souhaitée est-elle atteinte?
    if parm - parmOld == [0 0 0]
        cond = 0;
    else
        parmOld = parm;
    end
end
\end{verbatim}
\end{small}

Chaque itération de la boucle consiste à ajuster successivement et indépendamment les paramètres $\ln\tilde{K}^\circ$, $v$ et puis $w$ afin que la somme des carrés des résidus entre les données issues du modèles RSA+R et l'estimation de la fonction de saturation soit minimale.

En particulier, la première itération consiste à poser les paramètres $v$ et $w$ comme étant égaux à $1$ et à estimer une valeur pour $\ln\tilde{K}^\circ$ en gardant tous les autres paramètres constants. Ensuite, on ajustera $v$ pour cette valeur de $\ln\tilde{K}^\circ$ estimée et pour $w=1$. Pour terminer, on ajustera $w$. Ce processus est répété en boucle tout en mémorisant les paramètres précédemment estimés dans \texttt{parm} jusqu'à obtenir la precision souhaitée. La convergence est rapide et robuste.
\end{cbunit}
\begin{cbunit}
\chapter{Codes Matlab pour le RSA(+R)}\label{Ann3}
\markboth{Annexe \ref{Ann3}: CODES MATLAB}{}
\minitoc

\section{Généralités}

Cette annexe présente quelques codes Matlab des fonctions les plus utilisées dans le cadre des simulations RSA(+R). \href{run:RSA/MainScript.m}{\texttt{MainScript.m}}, la première d'entre-elles, constitue le script principal auquel il sera systématiquement fait appel lors d'une nouvelle simulation. On trouvera les codes détaillés ci-dessous ainsi que tous les autres qui n'y seraient pas repris dans un répertoire \href{run:RSA/}{\texttt{./RSA/}} joint à ce travail.

\section[La fonction \texttt{MainScript}]{La fonction \href{run:RSA/MainScript.m}{\texttt{MainScript}}}

\begin{footnotesize}
\begin{verbatim}
function MainScript(LnKzero)
%% function MainScript(LnKzero)

FileName = uigetfile('MultiSelect','on');
if iscell(FileName) == 0
    filename{1,1} = FileName;clear FileName
    FileName = filename;clear filename
end

for f = 1:1:size(FileName,2)
    % Calcul de la bibliothèque contenant toutes les particules dans toutes
    % les orientations possibles
    ConstructBib(FileName{1,f});

    % Calcul des variables nécessaires aux fonctions EvalExVol4Add.m et
    % RawData2Data.m
    DetectSim(FileName{1,f});

    % Construction de la matrice RawData et enregistrement du workspace
    ConstructRawData(FileName{1,f},LnKzero);

    % Traitement des données générées par ConstructRawData
    RawData2Data(FileName{1,f},LnKzero);

    % Construction des graphiques
    PlotData(FileName{1,f},LnKzero,1);
    PlotSurf(FileName{1,f},LnKzero,1);
end

system('shutdown /s');
\end{verbatim}
\end{footnotesize}

\section[La fonction \texttt{ConstructRawData}]{La fonction \href{run:RSA/ConstructRawData.m}{\texttt{ConstructRawData}}}

\begin{footnotesize}
\begin{verbatim}
function ConstructRawData(FileName,LnKzero)
%% function ConstructRawData(FileName,LnKzero)
% La fonction CONSTRUCTRAWDATA construit le cell array 'RawData'
% contenant les données brutes issues du modèles RSA: en fonction du nombre
% total de particules accumulées (1ère colonne de 'RawData'), la
% distribution de chaque orientation et dénaturation (2ème
% colonne de 'RawData') et les probabilités conditionnelles
% d'addition (3ème colonne de 'RawData').

% Chargement des variables globales calculées dans le MainScript.m
global Surface ParticulesBib SimulRawData ParticulesData LocVar
load(fullfile(cd,'sim',[FileName(1,1:size(FileName,2)-4),'_--_Data4Simul.mat']));
Surface.dim = 4000;

% Initialisation des variables constantes de la fonction ConstructRawData.m
LocVar.start = 0;
LocVar.stop = 0;
LocVar.m = 0;
LocVar.failure = 0;
SimulRawData.Rel4Add = max(ParticulesData.CoordClosePack-1) * exp(-LnKzero);
SimulRawData.fin = 0;

% Taille de l'échantillon de surfaces (de couverture
% NbrRequest) nécessaire à la bonne évaluation de Neff
LocVar.surf4num = 20;

% Pas dans le calcul de NbrRequest
LocVar.nbr4step = 1;

% Déclaration des nouvelles variables globales nécessaires à la
% fonction ConstructSurf.m
Surface.surface = zeros(Surface.dim + 2 * Surface.bord);
Surface.Neff = zeros(max(ParticulesBib.NbrEtat),ParticulesBib.NbrType);
Surface.failure = 0;

LocVar.Directory = dir(fullfile(cd,'results/*.mat'));

if size(LocVar.Directory,1) == 0
    % Evaluation du nombre de particules lorsque la surface est saturée
    ConstructSurf(Inf,SimulRawData.Rel4Add);
    LocVar.NbrMax = sum(sum(Surface.Neff));
    SimulRawData.RawData = cell(round(1.25*LocVar.NbrMax),4);
else
    for i = 1 : 1 : size(LocVar.Directory,1)
        if strfind(LocVar.Directory(i,1).name,[FileName(1,1:size...
        	(FileName,2)-4),'_--_',LnK2Str(LnKzero),'_--_RawData.mat']) == 1
            load(fullfile(cd,'results',[FileName(1,1:size(FileName,2)-4),...
            	'_--_',LnK2Str(LnKzero),'_--_RawData.mat']));
            if SimulRawData.fin == 1
                return
            else
                [LocVar.start,LocVar.m] = max(cell2mat(...
                	SimulRawData.RawData(:,1)),[],1);
                LocVar.start = LocVar.start + LocVar.nbr4step;
                break
            end
        else
            if i == size(LocVar.Directory,1)
                % Evaluation du nombre de particules lorsque
                % la surface est saturée
                ConstructSurf(Inf,SimulRawData.Rel4Add);
                LocVar.NbrMax = sum(sum(Surface.Neff));
                SimulRawData.RawData = cell(round(1.25*LocVar.NbrMax),4);
            end
        end
    end
end

% Réinitialisation des variables globales 'surface', 'failure' et 'Neff'
Surface.surface = Surface.surface * 0;
Surface.failure = Surface.failure * 0;
Surface.Neff = Surface.Neff * 0;

% Construction du cellarray PhiTheta
for NbrRequest = LocVar.start : LocVar.nbr4step : 10 * ceil(LocVar.NbrMax)
    
    clc
    
    disp(['Nbr/NbrMax = ',num2str(round(100*NbrRequest/LocVar.NbrMax)),' %']);
    
    % Calcul du nombre d'épreuves qui devront être réalisées par la
    % fonction EvalExVol.m afin d'évaluer le volume exclu de la surface
    % contenant NbrRequest particules
    LocVar.fill4surf = (10000 + floor(45000 * NbrRequest / LocVar.NbrMax))...
    	/ LocVar.surf4num;
    
    LocVar.succes = 0;
    LocVar.echec = 0;
    LocVar.m = LocVar.m + 1;
    
    if NbrRequest == 0
        % Lorsque NbrRequest = 0, il n'est pas nécessaire de faire appel aux
        % fonctions EvalExVol4Add et ConstructSurf(0) pour remplir RawData
        SimulRawData.RawData{LocVar.m,1} = 0;
        SimulRawData.RawData{LocVar.m,2} = zeros(max(ParticulesBib.NbrEtat),...
        	ParticulesBib.NbrType);
        SimulRawData.RawData{LocVar.m,3} = ones(max(ParticulesBib.NbrEtat),...
        	ParticulesBib.NbrType);
        SimulRawData.RawData{LocVar.m,4} = 0;
    else
        while LocVar.succes < LocVar.surf4num
            ConstructSurf(NbrRequest,SimulRawData.Rel4Add);
            if Surface.failure == 0
                LocVar.succes = LocVar.succes + 1;
                if LocVar.succes == 1
                    SimulRawData.RawData{LocVar.m,1} = NbrRequest;
                    SimulRawData.RawData{LocVar.m,2} = Surface.Neff...
                    	./ LocVar.surf4num;
                    SimulRawData.RawData{LocVar.m,3} = EvalExVol4Add...
                    	./ LocVar.surf4num;
                else
                    % Répétitions nécéssaires à l'évaluation convenable de
                    % Neff et de PrT
                    SimulRawData.RawData{LocVar.m,2} = SimulRawData.RawData...
                        {LocVar.m,2} + Surface.Neff ./ LocVar.surf4num;
                    SimulRawData.RawData{LocVar.m,3} = SimulRawData.RawData...
                        {LocVar.m,3} + EvalExVol4Add ./ LocVar.surf4num;
                end
            else
                LocVar.echec = LocVar.echec + 1;
                if LocVar.echec >= 2 * LocVar.surf4num
                    break
                end
            end
        end
    end
    
    if LocVar.echec >= 2 * LocVar.surf4num
        % Enregistrement du WorkSpace
        SimulRawData.RawData = SimulRawData.RawData(1:LocVar.m-1,:);
        SimulRawData.fin = 1;
        clear LocVar ParticulesBib ParticulesData NbrRequest i Rel4Add
        save(fullfile(cd,'results',[FileName(1,1:size(FileName,2)-4),...
        	'_--_',LnK2Str(LnKzero),'_--_RawData.mat']));
        clear global LnKzero
        return
    end
    
    save(fullfile(cd,'results',[FileName(1,1:size(FileName,2)-4),'_--_',...
    	LnK2Str(LnKzero),'_--_RawData.mat']));
end
\end{verbatim}
\end{footnotesize}

\section[La fonction \texttt{ConstructSurf}]{La fonction \href{run:RSA/ConstructSurf.m}{\texttt{ConstructSurf}}}

\begin{footnotesize}
\begin{verbatim}
function ConstructSurf(NbrRequest,Rel4Add)
%% function ConstructSurf(NbrRequest,Rel4Add)
% La fonction CONSTRUCTSURF, construit une surface sur laquelle minimum
% 'NbrRequest' particules provenant de la bibliothèque 'ParticulesBib'
% sont accumulées. La surface requise est assignée à la variable globale
% 'Surface' et le nombre de particules effectivement accumulées à 'Neff'.
% Lorsque la fonction de parvient pas à générer une telle surface parce
% que P(A*)>=0,9999 (par exemple si NbrRequest = Inf), elle
% renvoit une surface saturée dans la variable réponse 'surface', elle
% assigne la valeur 1 à 'failure' et le nombre de particules effectivement
% accumulées à 'Neff'.

global Surface ParticulesBib PartCoord

Surface.surface = Surface.surface * 0;
Surface.Neff = Surface.Neff * 0;
Surface.failure = Surface.failure * 0;
PartCoord = [];

NbrAdd = 0;
echec = 0;
while NbrAdd < NbrRequest
    PosPart = round(rand(1,2) * Surface.dim + 0.5);
    if Surface.surface(PosPart + Surface.bord) == 0
        OriPart = round(rand * ParticulesBib.NbrOri + 0.5);
        TypPart = round(rand * ParticulesBib.NbrType + 0.5);

        % Calcul des conditions aux limites périodiques
        PosPart = EvalPBC(PosPart,ceil(Surface.bord/2));

        % Addition de la particule
        SuccesAdd = DoAddition(PosPart+Surface.bord,TypPart,OriPart);
        if SuccesAdd == 1
            NbrAdd = NbrAdd + 1;
            Surface.Neff(1,TypPart) = Surface.Neff(1,TypPart) + 1;
            PartCoord(NbrAdd,:) = [PosPart(1,1),PosPart(1,2),1,TypPart,OriPart];
        else
            echec = echec + 1;
        end
    else
        % L'addition est de toute façon IMP à cet endroit
        echec = echec + 1;
    end

    % Réalisation de InvK pas de relaxation sur les particules de la monocouche
    if Rel4Add == Inf
        if SuccesAdd == 1
            % Tant que faire ce peut, relaxer entièrement la
            % dernière particule additionnée
            DoRelaxation(NbrAdd,ParticulesBib.NbrEtat(1,TypPart));
        end
    else
        NbrRelRequest = floor(Rel4Add) + RandRelax(Rel4Add - floor(Rel4Add));
        if NbrRelRequest > 0
            for i = 1 : 1 : NbrRelRequest
                % Tentative de relaxation d'une particule de un pas
                DoRelaxation(round(rand * NbrAdd + 0.5),1);
            end
        end
    end
    
    if SuccesAdd == 1
        % Le volume exclu P(A*) est >= 0.9999 ou P(A) < 1e-4
        if echec / (1 + echec) >= 0.9999
            Surface.failure = 1;
            break
        end
        echec = 0;
    end
end
\end{verbatim}
\end{footnotesize}

\section[La fonction \texttt{DoAddition}]{La fonction \href{run:RSA/DoAddition.m}{\texttt{DoAddition}}}

\begin{footnotesize}
\begin{verbatim}
function [Succes] = DoAddition(PosPart,TypPart,OriPart)
%% function [Succes] = DoAddition(PosPart,TypPart,OriPart)
% La fonction DOADDITION tente d'additionner une particule aux positions
% incluses dans 'PosPart'. La particule est de type 'TypPart' et d'état
% 'OriPart'.

global Surface ParticulesBib ParticulesData

Succes = 0;
Part2Add = ParticulesBib.Particules{1,TypPart,OriPart,1};
n = (ParticulesBib.Particules{1,TypPart,OriPart,2}(1,1) - 1) / 2;
p = (ParticulesBib.Particules{1,TypPart,OriPart,2}(1,2) - 1) / 2;

% L'addition de la particule est-elle possible?
CondAdsNat = EvalAddition(PosPart,Part2Add,n,p,ParticulesData.TypeCond(1,TypPart));

if CondAdsNat == 1
	% Si l'additon de la particule à la surface est possible, alors l'additionner.
    for i = 1:1:size(PosPart,1)
        Surface.surface(PosPart(i,1) - n:PosPart(i,1) + n,PosPart(i,2) -...
        	p:PosPart(i,2) + p) = Surface.surface(PosPart(i,1) -...
        	n:PosPart(i,1) + n,PosPart(i,2) - p:PosPart(i,2) + p) + Part2Add;
    end
    Succes = 1;
end

\end{verbatim}
\end{footnotesize}

\section[La fonction \texttt{DoRelaxation}]{La fonction \href{run:RSA/DoRelaxation.m}{\texttt{DoRelaxation}}}

\begin{footnotesize}
\begin{verbatim}
function DoRelaxation(Part2Rel,NbrPas)
%% function DoRelaxation(Part2Rel,NbrPas)
% La fonction DORELAXATION tente de relaxer la particule 'Part2Rel' de
% 'NbrPas' pas.

global Surface PartCoord ParticulesData ParticulesBib

for pas = 1 : 1 : NbrPas
    if ParticulesBib.NbrEtat(1,PartCoord(Part2Rel,4)) > PartCoord(Part2Rel,3)
        
        % Calcul des coordonnées
        PosPart = PartCoord(Part2Rel,1:2);
        PosPart = EvalPBC(PosPart,(max(ParticulesBib.Particules{ParticulesBib.NbrEtat...
        	(1,PartCoord(Part2Rel,4)),PartCoord(Part2Rel,4),PartCoord(Part2Rel,3),2})...
        	- 1) / 2) + Surface.bord;

        % Obtention de la particule à relaxer
        Part2Rem = ParticulesBib.Particules{PartCoord(Part2Rel,3),PartCoord...
        	(Part2Rel,4),PartCoord(Part2Rel,5),1};
        nR = (ParticulesBib.Particules{PartCoord(Part2Rel,3),PartCoord(Part2Rel,4),...
        	PartCoord(Part2Rel,5),2}(1,1) - 1) / 2;
        pR = (ParticulesBib.Particules{PartCoord(Part2Rel,3),PartCoord(Part2Rel,4),...
        	PartCoord(Part2Rel,5),2}(1,2) - 1) / 2;

        % Obtention de la particule relaxée
        Part2Add = ParticulesBib.Particules{PartCoord(Part2Rel,3) + 1,PartCoord...
        	(Part2Rel,4),PartCoord(Part2Rel,5),1};
        nA = (ParticulesBib.Particules{PartCoord(Part2Rel,3) + 1,PartCoord(Part2Rel,4),...
        	PartCoord(Part2Rel,5),2}(1,1) - 1) / 2;
        pA = (ParticulesBib.Particules{PartCoord(Part2Rel,3) + 1,PartCoord(Part2Rel,4),...
        	PartCoord(Part2Rel,5),2}(1,2) - 1) / 2;

        % Extraction de la surface de la particule à relaxer
        for i = 1 : 1 : size(PosPart,1)
            Surface.surface(PosPart(i,1) - nR:PosPart(i,1) + nR,PosPart(i,2) -...
            pR:PosPart(i,2) + pR) = Surface.surface(PosPart(i,1) -...
            nR:PosPart(i,1)+ nR,PosPart(i,2) - pR:PosPart(i,2) + pR) - Part2Rem;
        end
        
        % Peut-on additionner la particule relaxée à la place de la précédante?
        Condition = EvalAddition(PosPart,Part2Add,nA,pA,ParticulesData.TypeCond...
        	(PartCoord(Part2Rel,3) + 1,PartCoord(Part2Rel,4)));

        if Condition == 0 % Non. Alors remettre la précédante.
            for i = 1 : 1 : size(PosPart,1)
                Surface.surface(PosPart(i,1) - nR:PosPart(i,1) + nR,PosPart(i,2) -...
                	pR:PosPart(i,2) + pR) = Surface.surface(PosPart(i,1) -...
                	nR:PosPart(i,1) + nR,PosPart(i,2) - pR:PosPart(i,2) + pR) + Part2Rem;
            end
        else % Oui. Alors additionner la particule relaxée.
            for i = 1 : 1 : size(PosPart,1)
                Surface.surface(PosPart(i,1) - nA:PosPart(i,1) + nA,PosPart(i,2) -...
                	pA:PosPart(i,2) + pA ) = Surface.surface(PosPart(i,1) -...
                	nA:PosPart(i,1) + nA,PosPart(i,2) - pA:PosPart(i,2) + pA) + Part2Add;
            end
            Surface.Neff(PartCoord(Part2Rel,3),PartCoord(Part2Rel,4)) = Surface.Neff...
            	(PartCoord(Part2Rel,3),PartCoord(Part2Rel,4)) - 1;
            Surface.Neff(PartCoord(Part2Rel,3) + 1,PartCoord(Part2Rel,4)) =...
            	Surface.Neff(PartCoord(Part2Rel,3) + 1,PartCoord(Part2Rel,4)) + 1;
            PartCoord(Part2Rel,3) = PartCoord(Part2Rel,3) + 1;
        end
    else
        break
    end
end
\end{verbatim}
\end{footnotesize}

\section[La fonction \texttt{EvalExVol4Add}]{La fonction \href{run:RSA/EvalExVol4Add.m}{\texttt{EvalExVol4Add}}}

\begin{footnotesize}
\begin{verbatim}
function [Succes] = EvalExVol4Add
%% function [Succes] = EvalExVol4Add
% La fonction EVALEXVOL4ADD calcule les probabilités conditionnelles
% d'additions sur la surface contenue dans la variable globale 'Surface'.

global Surface ParticulesData ParticulesBib LocVar

Succes = zeros(size(ParticulesBib.NbrEtat,1),ParticulesBib.NbrType);

for i = 1 : 1 : size(ParticulesData.CoordFirstIndiv,1)
    
    succes = 0;
    
    for j = 1 : 1 : LocVar.fill4surf
        
        PosPart = round(rand(1,2) * Surface.dim + 0.5) + Surface.bord;
        
        if Surface.surface(PosPart) == 0
            
            OriPart = round(rand * 360 + 0.5);
            
            Part2Add = ParticulesBib.Particules{ParticulesData.CoordFirstIndiv(i,1),...
                ParticulesData.CoordFirstIndiv(i,2),OriPart,1};
            
            n = (ParticulesBib.Particules{ParticulesData.CoordFirstIndiv(i,1),...
                ParticulesData.CoordFirstIndiv(i,2),OriPart,2}(1,1)-1) / 2;
            p = (ParticulesBib.Particules{ParticulesData.CoordFirstIndiv(i,1),...
                ParticulesData.CoordFirstIndiv(i,2),OriPart,2}(1,2)-1) / 2;
            
            succes = succes + EvalAddition(PosPart,Part2Add,n,p,...
                ParticulesData.TypeCond(ParticulesData.CoordFirstIndiv(i,1),...
                ParticulesData.CoordFirstIndiv(i,2)));
        end
        
    end
    
    Succes(ParticulesData.CoordFirstIndiv(i,1),ParticulesData.CoordFirstIndiv(i,2)) =...
        succes / LocVar.fill4surf;
    
end
\end{verbatim}
\end{footnotesize}\end{cbunit}

\begin{otherlanguage}{english} 
\begin{cbunit}
\chapter[AFM analysis of IgG films at hydrophobic surfaces]{Atomic force microscopy analysis of IgG films at hydrophobic surfaces: a promising method to probe IgG orientations and optimise ELISA tests performance}\label{Ann4}
\markboth{Annexe \ref{Ann4}: ARTICLE BBA}{}

\noindent\textbf{Authors:} Pierre de Thier\footnotemark[1]$^,$\footnotemark[2], Jalal Bacharouche\footnotemark[1]$^,$\footnotemark[2], J\'{e}r\^{o}me F. L. Duval\footnotemark[3]$^,$\footnotemark[4], Salaheddine Skali-Lami\footnotemark[5]$^,$\footnotemark[6], and Gr\'{e}gory Francius\footnotemark[1]$^,$\footnotemark[2]$^,$\footnotemark[7].

\footnotetext[1]{Universit\'{e} de Lorraine, Laboratoire de Chimie Physique et Microbiologie pour l'Environnement, UMR 7564, Villers-l\`{e}s-Nancy, F-54600, France.}
\footnotetext[2]{CNRS, Laboratoire de Chimie Physique et Microbiologie pour l'Environnement, UMR 7564, 405 rue de Vandoeuvre, Villers-l\`{e}s-Nancy, F-54600, France.}
\footnotetext[3]{Universit\'{e} de Lorraine, Laboratoire Interdisciplinaire des Environnements Continentaux, UMR 7360, Van\oe uvre-l\`{e}s-Nancy, F-54501, France.}
\footnotetext[4]{CNRS, Laboratoire Interdisciplinaire des Environnements Continentaux, UMR 7360, Van\oe uvre-l\`{e}s-Nancy, F-54501, France.}
\footnotetext[5]{Universit\'{e} de Lorraine, Laboratoire d'Energ\'{e}tique et de Mécanique Théorique et Appliquée, UMR 7563, Van\oe uvre-l\`{e}s-Nancy, F-54504, France.}
\footnotetext[6]{CNRS, Laboratoire d'Energ\'{e}tique et de M\'{e}canique Th\'{e}orique et Appliqu\'{e}e, UMR 7563, 2 avenue de la For\^{e}t de Haye, BP 160, Van\oe uvre-l\`{e}s-Nancy, F-54504, France.}
\footnotetext[7]{Corresponding author:\\ \indent Email: gregory.francius@univ-lorraine.fr\\ \indent Phone: (33) 03 83 68 52 36}

\vspace*{5mm}
\noindent\textbf{Abstract:} IgG films are widely used in the field of immunoassays, especially in (double) antibody-sandwich ELISA tests where capture antibodies are coated on surfaces like polystyrene or hydrophobic self-assembled monolayers (h-SAMs). It is critical to analyse ---at a molecular scale and under liquid conditions--- the structure of the deposited IgG film in order to quantitatively address the efficiency of the ELISA test in terms of antigen detection. In this communication, we report an atomic force microscopy (AFM) analysis evidencing a strong relationship between immunological activities of mouse monoclonal anti-human interleukin-2 (IL-2) and 6 (IL-6) antibodies, thickness and roughness of the IgG monolayer adsorbed onto h-SAMs, and surface concentration of IgG molecules. Indirect information may be further obtained on antibody orientation. Collating the results obtained by AFM and those from ELISA tests leads us to conclude that antibodies like anti-IL-6 forming flat monolayers should be more efficient under ELISA detection conditions. In addition, the concentration of IgG in the coating suspension should be optimized to obtain a monolayer heavily populated by ``end-on'' adsorbed molecules, an orientation that is desirable for enhancing ELISA tests performance.

\vspace*{5mm}
\noindent\textbf{Keywords:} Atomic force microscopy, ELISA, protein adsorption, thin films, molecular conformation.

\section{Introduction}
Immunoassays are one of the most popular analytical methods adopted for the detection and/or quantification of biomolecules suspended in various fluids. These assays are based on biomolecular recognition processes resulting from the strong specific interactions involved in antigen-antibody binding [1]. As with most recent immunoassays, Enzyme-Linked ImmunoSorbent Assay (ELISA) requires a synthetic solid phase (substrate) to support and efficiently immobilize the reacting phase, \textit{i.e.} a capture antibody that can be directly immobilized at the surface. Sensitive variants of ELISA like the so-called sandwich ELISA test, quantifies antigens between two layers of antibodies (\textit{i.e.} capture and detection antibody) [2, 3]. As a first step in ELISA, an aqueous suspension of the capture antibody is brought into contact with a polystyrene flat surface, so that strong interactions are established between the antibody and the solid substrate. Several studies have underlined that antibodies can be spontaneously immobilized at the surface in virtue of apolar interactions. This adsorption process generally leads to the formation of a thin film between the polystyrene surface and the antibody suspension [4]. The formed thin film is expected to be a monolayer of antibodies firmly anchored at the supporting surface. There are two fundamental conditions that must be satisfied to ensure an efficient immunological detection during the next steps in the ELISA protocol [3]. The first of these conditions is related to the very immunological properties of the immobilized antibodies that must remain intact. The second condition is that antibodies must be appropriately oriented in order to bind the antigen molecules (or the secondary antibodies in sandwich ELISA) from the liquid phase. However, a strong hydrophobic binding to the surface may lead to partial or complete unfolding of fragments of the antibody [5], thus resulting in a loss of bioactivity (\textit{i.e.} a loss of antigen binding capacity) [6]. The extent of this unfolding is determined by the IgG internal stability, that essentially depends on the type of IgG considered. In addition, the conformational properties of antibodies (and/or their prevalent orientations) may be strongly impacted by the operational coating conditions during and after the immobilization step at the hydrophobic surface [7]. According to Buijs \textit{et al.}, there is a range of potential orientations for immobilized antibodies (Figure \ref{FigPubli1}A): end-on (antibody attached through one of its parts), side-on (attached through two of its parts), and flat (attached through its three parts). From this representation, it is obvious that end-on adsorbed IgG molecules are highly desirable as they constitute a efficient bioactive IgG monolayer. Obtaining a monolayer with high end-on bioactive IgG is therefore an issue of prime interest for ELISA improvement [8], which led to many studies with, as a goal, the control of the orientation of adsorbed IgG. In that respect, various strategies have been considered ranging from supporting surface modifications (\textit{e.g.} \textit{via} plasma treatments [9] or action of bulk additives) to IgG functionalisation (\textit{e.g.} biotinylation [10] or enzyme cleaving [7]). In addition to these somewhat empirical approaches, other techniques have been developed in order to probe in situ the properties of the monolayers (including IgG bioactivity and orientation) and thus refine our understanding of the IgG adsorption mechanism at hydrophobic surfaces \textit{via} \textit{e.g.} optimisation of the concentration or temperature conditions for the deposition step.

\begin{figure}[ht]\centering
\includegraphics*[width=0.9\textwidth]{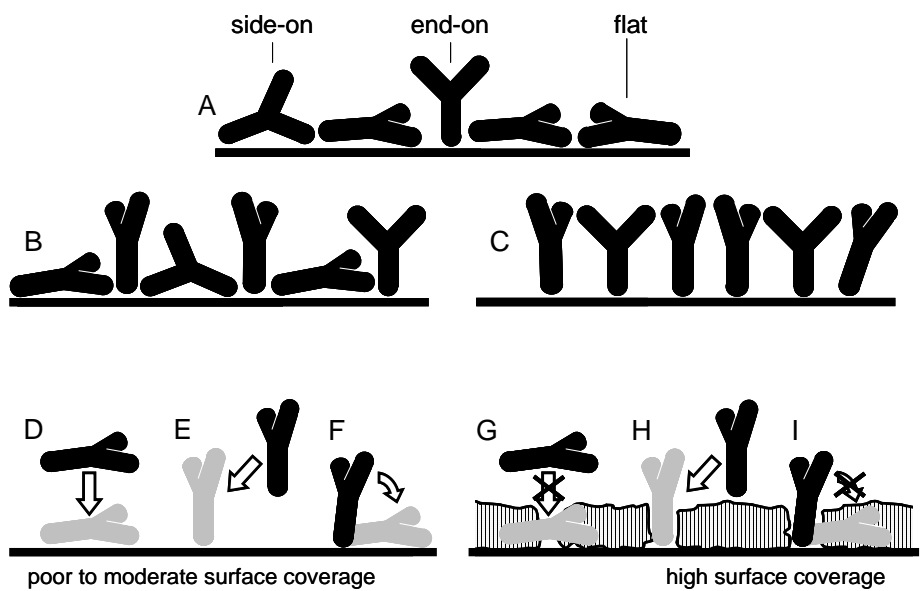}
\caption[Implications of antibodies orientations on monolayer structures]{Implications of various antibodies orientations (side-on, end-on and flat) on monolayer structures and their hypothetical building mechanism. Schemes \textbf{A} to \textbf{C} illustrate three hypothetical monolayer structures exhibiting various fractions of end-on oriented antibodies; \textbf{A}: low end-on content (thin layer and poor immunological activity); \textbf{B}: medium end-on content (medium thickness and average immunological activity); and \textbf{C}: high end-on content (thick layer and high immunological activity). Schemes \textbf{D} to \textbf{I} depict hypothetical coating mechanism of a hydrophobic surface by antibodies proposed from ref. [2, 11, 14, 2]. Cases \textbf{D}, \textbf{E}, and \textbf{F} show that protein adsorption in a flat orientation, adsorption in an end-on orientation and relaxation from flat to end-on orientation are allowed for poor to moderate surface coverage. Cases \textbf{G}, \textbf{H}, and \textbf{I} show that only addition of proteins in end-on orientation is allowed for high surface coverage (the flat structures therein represent previously adsorbed flat antibodies).}\label{FigPubli1}
\end{figure}

A straightforward approach to probe IgG orientation and evaluate bioactivity consists in measuring the maximum amount of adsorbed IgG. Bremer \textit{et al.} stated that this quantity should be indicative of the average orientation of the IgG molecules at the surface; in particular, the orientation requiring the least space on the surface, \textit{i.e.} the end-on conformation, should lead to a denser (and/or thicker) monolayer [11]. The numerous studies reported on the evaluation of adsorbed surface amounts (\textit{e.g.} the determination of superficial protein film density) should not however hide that a molecular characterization of thin protein films deposited at the solid surfaces is mandatory to address the searched protein orientation and the associated film structure. Superficial film densities have been estimated using ellipsometry [12], reflectometry [13, 14], radiolabeling [15], and more recently quartz crystal microbalance with dissipation monitoring (QCM-D) [16, 17]. Earlier studies were also based on the measurement of the kinetics of depletion of the IgG content from the deposition solution [18]. This latter strategy is however meaningful in the only case where the superficial area of the solid support is large enough compared to the volume of the IgG coating solution. Unfortunately, this condition is scarcely met in devices mimicking ELISA conditions. Methods like ellipsometry, reflectometry or QCM-D are therefore necessary, but the so-obtained results are necessarily averaged over the entire detection surface, which severely limits their use to obtain molecular information on the distribution of immobilized IgG orientations and their potential structural integrity.

In order to understand how the IgG monolayer structure influences ELISA efficiency, it is therefore necessary to design \textit{in situ} experiments. To do this, IgG monolayers must be studied in a liquid environment (phosphate buffer saline at pH 7-7.4), without drying or applying any other external stresses likely to modify the structure of the monolayers. Resorting to Atomic Force Microscopy (AFM) under liquid conditions therefore appears to be a powerful tool to probe the local morphology of IgG monolayers. AFM may further provide additional monolayer properties including thickness and superficial densities. The latter film characteristics could then be discussed in light of the corresponding ELISA performance results. In turn, such analysis could improve our understanding of the relationships between monolayer organization at the nanoscale, mean IgG orientation, monolayers construction mechanism and impact of the deposition conditions, and ensuing antigen detection efficiency. The aim of this work is to provide this information and shed light on the interconnections between film structure at a molecular scale, protein orientation and ELISA test efficiency. To the best of our knowledge, this explorative approach is still lacking in the literature.

In the current study, we therefore report AFM experiments performed on IgG films. The thicknesses of IgG monolayers were derived from AFM imaging of films adsorbed onto hydrophobic surfaces (Self Assembled Monolayers of alkanethiols, hereafter denoted as h-SAMs) with initial deposition solution concentrations of IgG in the range 1.56 to 50.00 \textgreek{m}g$\cdot$mL$^{-1}$. Experiments were carried out on two mouse monoclonal antibodies: anti-human IL-2 and anti-human IL-6. ELISA results are also reported in order to address the immunological activity (bioactivity) of these layers. Combined AFM and ELISA results make it possible to estimate indirectly a mean orientation for the adsorbed IgG.

\section{Materials and methods}

\subsection{Protein solutions}
Mouse monoclonal anti-human interleukin-2 (isotype IgG 1\textgreek{k}, ref.: AHC0422) and anti-human interleukin-6 (isotype IgG 1\textgreek{k}, ref.: AHC0562) antibodies were purchased from Life Technologies/Invitrogen Corporation (Camarillo, CA, USA). IgG molecules were preserved in phosphate buffered saline (PBS) at pH 8 and 7.5 and temperatures of 4$^\circ$C and 20$^\circ$C, respectively, until use. Crystallised bovine serum albumin (BSA) (fraction V, essentially protease free, $\geqslant92$ \%) was purchased from Sigma-Aldrich (ref.: 05479). All dilutions were carried out with 1 mol$\cdot$L$^{-1}$ PBS at pH 7 briefly stored at 4$^\circ$C prior to use.

\subsection{ELISA immunological activity measurement}
IL-6-EASIA and IL-2-EASIA ELISA kits were purchased from DIAsource ImmunoAssays (Louvain-la-Neuve, Belgium) and used to probe the immunological activity of IgG monolayers. The ELISA protocols provided by DIAsource were generally followed, although two major modifications were brougth in order to address immunological activities of the IgG coated surfaces' (\textit{i.e.} the very aim of ELISA that consists in the quantification of antigens suspension in solution). The first modification is the use of our synthesized IgG coated surfaces (mentioned above) instead of the strips provided with the ELISA kits, and the second is the use of 201 pg$\cdot$mL$^{-1}$ calibrators for both antibodies tested in this work. To proceed with background removal, 0 pg$\cdot$mL$^{-1}$ calibrators were also considered. The ELISA protocol we adopted here is similar to that in sandwich ELISA because the specific antibody is directly immobilized on the hydrophobic surface. The coating of 8 wells in a 96 well hydrophobic microtitre plate with IgG was performed as follows: each well was washed 3 times with 300 \textgreek{m}L PBS; 100 \textgreek{m}L of IgG coating solution were then injected in each well (concentrations: 50, 25, 12.5, 7.81, 3.9, 1.95, 0.98, and 0.49 \textgreek{m}g$\cdot$mL$^{-1}$); the plate was incubated for 4 h at 4$^\circ$C; the liquid was then aspirated and the wells were washed 3 times with 300 \textgreek{m}L PBS; then, wells were coated using 300 \textgreek{m}L BSA solution (5 mg$\cdot$mL$^{-1}$) during 30  minutes at 4$^\circ$C; finally, the liquid was aspirated and the wells were washed 3 times with 300 \textgreek{m}L PBS. Subsequently, the classical protocol for ELISA surface preparation and ELISA protein-detection measurements were performed using the aforementioned calibrators. The difference between absorbances measured for the solution facing the ELISA surface before and after addition of antigens was read at 450 nm against a reference filter set at 650 nm in order to estimate the amount of antigens immobilized on the IgG-reactive substrate. This differential absorbance is taken as a measure of the ELISA detection performance.

\subsection{Monolayer preparation}
IgG films were built on hydrophobic surfaces prepared on glass substrates (1.1 mm thick and 12 mm in diameter). Glass slides were first cleaned in a 3/1 H$_2$SO$_4$/H$_2$O$_2$ solution for 3 h at room temperature and then extensively rinsed with ultrapure water (Milli-Q, Millipore Corporation). Substrates were then passivated with 20 \% NaOH solution, rinsed with ultrapure water and dried with nitrogen. The glass slides were coated with a 10 nm thick layer of chromium through use of a sputter coater (EMITECH, K575 Turbo, United Kingdom) and further covered with a top gold layer roughly 50 nm in thickness. The gold coated glass surfaces were then immersed during 14 h in a 1 mM CH$_3$(CH$_2$)$_{11}$SH ($\geqslant98$ \%, Sigma-Aldrich, ref.: 471364) solution in ethanol ($\geqslant99.8$ \%, Fluka, ref.: 02854) in order to obtain a hydrophobic self-assembled  alkanethiol monolayer (h-SAM). After coating, the hydrophobic surfaces were rinsed with ethanol and PBS. Then, they were stored in the bottom of a well (24 well plates from Greiner Bio-One) immersed in 1 mL of PBS. After thermoregulation at 4$^\circ$C during 1 h, the PBS was removed with a micropipette and a deposition solution was then brought into contact with the hydrophobic surface. One millilitre of deposition solution (antibodies in PBS: 50, 25, 12.5, 6.25, 3.13, and 1.56 \textgreek{m}g$\cdot$mL$^{-1}$) was carefully injected perpendicular to the centre of the surface. After incubation (4 h) at 4$^\circ$C, the deposition solution was removed and the coated substrates were carefully rinsed with PBS. Prior to characterization, the obtained coated substrates were briefly stored in PBS.

\subsection{AFM imaging and data processing}
Topographic imaging of samples immersed in PBS was performed at room temperature using an Asylum MFP-3D atomic force microscope (Santa Barbara, CA, USA) with IGOR Pro 6.04 (Wavemetrics, Lake Osewego, OR, USA) as operating software. The IgG coated surfaces were first scratched across a 1 \textgreek{m}m $\times$ 1 \textgreek{m}m area by applying a sufficiently large force ($\sim$200 nN) and the scratched area was subsequently imaged. Next, 5 \textgreek{m}m $\times$ 5 \textgreek{m}m images were recorded in contact mode at a scan rate of 1 Hz using an applied force of less than 250 pN in magnitude. Conical shaped silicon nitride cantilevers were purchased from Atomic Force (OMCL-TR400PSA-3, Olympus, Japan). IgG monolayer properties including thickness, roughness (also evaluated for the supporting substrate), IgG coverage ratio and superficial monolayer volumes were evaluated from AFM images using Matlab environment. The strategy to extract the properties of the film is schemed in Figure \ref{FigPubli2}. First, the layer thickness was derived from the difference ($h_1-h_0$) with $h_1$ the mean height of the intact layer zone and $h_0$ that of the scratched zone. Heights were estimated after random sampling of both the intact and scratched areas. Then, roughness was estimated from mean heights standard deviations, allowing the determination of the roughness of both the monolayer and the underlying substrate. Confidence intervals were inferred from standard deviations of $h_1$ and $h_0$. In a second stage, surface coverage ratios (\%) were estimated from the number of pixels in the intact monolayer zone with heights equal to or larger than $h_0$ of the hydrophobic substrate, according to the expression:
\begin{equation}
\text{Surface Coverage Ratio}=\frac{\text{Number of pixels with }h>h_0}{\text{Total number of pixels}}\times100\,\%
\end{equation}
Finally, superficial monolayer volumes (expressed in mm$^3\cdot$cm$^{-2}$, \textit{i.e.} mm$^3$ of supported IgG per cm$^2$ of surface) were evaluated by dividing the monolayer volume of an intact monolayer zone by the size of this area. Volumes were obtained by summing each data point volume defined as the product of the height by the corresponding pixel area. It should be mentioned that tip convolution effects may lead to a misevaluation of IgG surface coverage and of superficial monolayer volume. However, we expect that the average asperity-to-asperity separation distance obtained for our substrates (Figure \ref{FigPubli11}) minimize such effect. 

\begin{figure}[ht]\centering
\href{http://dx.doi.org/10.1016/j.bbapap.2014.12.001}{See \textbf{Fig. 2.} of doi:10.1016/j.bbapap.2014.12.001.}
\caption[Topography scheme based on AFM image of an IgG monolayer]{\textbf{A}: topography (5 \textgreek{m}m $\times$ 5 \textgreek{m}m) scheme based on AFM image of an IgG monolayer, showing an intact zone and a scratched zone (1 \textgreek{m}m $\times$ 1 \textgreek{m}m) in order to make the hydrophobic substrate visible. \textbf{B}: cross-section at the level of the scratched zone showing the difference between the upper and lower zones of heights $h_1$ and $h_0$, respectively.}\label{FigPubli2}
\end{figure}

\section{Results and discussion}
\subsection{Monolayer bioactivity measured by ELISA}

\begin{figure}[hp]\centering
\href{http://dx.doi.org/10.1016/j.bbapap.2014.12.001}{See \textbf{Fig. 3.} of doi:10.1016/j.bbapap.2014.12.001.}
\caption[ELISA results showing the immunological activity]{ELISA results (absorbance) showing the immunological activity of anti-human IL-2 MAb (\begin{tiny}$\blacksquare$\end{tiny}) and anti-human IL-6 MAb ($\bullet$) layers as a function of the coating solution concentration (\textgreek{m}g$\cdot$mL$^{-1}$), indicated in logarithmic scale. The experimental points are reported together with linear regression and their prediction intervals at a confidence level of 95 \% (dashed lines).}\label{FigPubli3}
\end{figure}

\begin{figure}[hp]\centering
\href{http://dx.doi.org/10.1016/j.bbapap.2014.12.001}{See \textbf{Fig. 4.} of doi:10.1016/j.bbapap.2014.12.001.}
\caption[Thickness of anti-human IL-2 and IL-6 MAb layers]{Thickness of anti-human IL-2 MAb (\begin{tiny}$\blacksquare$\end{tiny}) and anti-human IL-6 MAb ($\bullet$) layers as a function of the protein concentration in the deposition solution (\textgreek{m}g$\cdot$mL$^{-1}$), indicated in logarithmic scale. The experimental points are reported together with linear regressions and and their prediction intervals at a confidence level of 95 \% (dashed lines).}\label{FigPubli4}
\end{figure}

Figure \ref{FigPubli3} reports ELISA absorbance measurements for the two antibodies of interest in this work after background subtraction and Figure \ref{FigPubli4} depicts the dependence of IgG layer thickness on protein concentration in the deposition solution. The tested IgG concentration levels were 50, 25, 12.5, 7.81, 3.9, 1.95, 0.98, and 0.49 \textgreek{m}g$\cdot$mL$^{-1}$. Results are given together with linear regressions and confidence intervals (level: 95 \%). Figure \ref{FigPubli3} shows that increasing the concentration of antibody in solution leads to an increase in the immunological activity of the anti-human IL-6 layer whereas this is clearly not the case for the anti-human IL-2 layer. The results collected in Figures \ref{FigPubli3} and \ref{FigPubli4} suggest the existence of an intimate connection between immunological activity (Figure \ref{FigPubli3}) and layer thickness (Figure \ref{FigPubli4}) with varying IgG concentration in the depostion solution: the larger the thickness with increasing IgG solution content, the larger is the immunological activity. As briefly discussed in introduction section, this correlation observed for anti-human IL-6 agrees with the picture according to which low concentrations of antibody in solution lead to monolayers containing a high fraction of flat oriented antibodies (thin layer, poor immunological activity) whereas sufficiently larger concentrations ultimately result in a high content of end-on oriented proteins thus forming thicker layers with larger immunological activity [11, 12, 19, 20]. Figures \ref{FigPubli1}B and \ref{FigPubli1}C illustrate three hypothetical structures of the protein monolayer with distinct immunological activities. Figure \ref{FigPubli1}B depicts a monolayer that is almost entirely populated with flat antibodies. In this case, the corresponding thin monolayer consists of antibodies' variable populations that are close to the surface and therefore are unavailable for immunological purposes (proteins are then most likely denatured). Conversely, Figure \ref{FigPubli1}C shows a monolayer with high end-on oriented protein content: the layer is thick and it exhibits high immunological activity. An intermediate case in terms of thickness and immunological activity is schemed in Figure \ref{FigPubli1}B. Overall, the ELISA results reflect the macroscopic properties of the IgG films, as intimately connected to their nanoscale organization accessible by AFM. The latter point if further addressed in the next section.

\subsection{Morphology and organization of protein monolayers investigated by AFM}
Morphology and organization of protein monolayers were investigated by AFM. Figures \ref{FigPubli5}A-\ref{FigPubli5}D and Figures \ref{FigPubli5}E-\ref{FigPubli5}H display typical AFM images collected for anti-human IL-2 and anti-human IL-6 layers, respectively, at various IgG concentrations in deposition solution (1.56 \textgreek{m}g$\cdot$mL$^{-1}$ to 50 \textgreek{m}g$\cdot$mL$^{-1}$ from panel A to D and panel E to H). Overall, Figure \ref{FigPubli5} highlights the dramatic differences between the organization of anti-human IL-2 and anti-human IL-6 layers, the former having the aspect of a rough surface with a landscape of marked asperities and depressions (``hills and valleys-like''), and the latter forming rather smooth layers. In Figures \ref{FigPubli5}A-\ref{FigPubli5}D (anti-human IL-2 layers), valleys are regularly shaped, 100 to 200 nm apart (distance between two summits) and deep (about 3 to 4 nm distance from top to bottom). These geometrical characteristics become larger with increasing IgG concentration. The layer structure depicted in Figure \ref{FigPubli5}D for the largest IgG concentration tested in this work is significantly different from that given in Figures \ref{FigPubli5}A-\ref{FigPubli5}C because monolayers now take the form of large flat areas punctuated by sharp hills. This peculiar organisation may be further apprehended from the cross-section profiles reported in Figure \ref{FigPubli6}, where shaded zones correspond to the scratched areas in Figure \ref{FigPubli6}. The above layer features discussed for anti-IL-2 are similar to those of human IgG deposited onto methylated silica imaged after drying under nitrogen [21]. Such surface protein patterns were termed ``dendrite-like'' by Malmsten [19] because they look like a multi-branching tree-like form. In contrast, anti-IL-6 surface layers are smooth and rather homogeneous (Figures \ref{FigPubli5}E-\ref{FigPubli5}H) regardless of the protein concentration in the deposition solution. Such surface structure corresponds to a pattern described by H. You and C. Lowe [22] as a compact and resistant monolayer whose surface appears fairly flat. W\"{a}livaara \textit{et al.} [21] further explained the origin of the aforementioned ``dendrite-like'' structures from thermodynamic implications of the adopted drying procedure. It seems however that this explanation is not fully satisfactory because we obtained here an identical structure for anti-IL-2 monolayers without any drying step. It may rather suggest that the structure of the adsorbed film is significantly depending on the very nature of the protein considered.

\begin{figure}[hp]\centering
\href{http://dx.doi.org/10.1016/j.bbapap.2014.12.001}{See \textbf{Fig. 5.} of doi:10.1016/j.bbapap.2014.12.001.}
\caption[AFM topographic images of IgG-coated surfaces]{AFM topographic images (5 \textgreek{m}m $\times$ 5 \textgreek{m}m) of IgG-coated hydrophobic surfaces with a scratched area (1 \textgreek{m}m $\times$ 1 \textgreek{m}m) located in the central part of the images. \textbf{A} to \textbf{D}: anti-human IL-2 MAb layers obtained from coating protein solutions @ 1.56 \textgreek{m}g$\cdot$mL$^{-1}$, 6.25 \textgreek{m}g$\cdot$mL$^{-1}$, 25 \textgreek{m}g$\cdot$mL$^{-1}$, and 50 \textgreek{m}g$\cdot$mL$^{-1}$; \textbf{E} to \textbf{H}: anti-human IL-6 MAb @ 1.56 \textgreek{m}g$\cdot$mL$^{-1}$, 6.25 \textgreek{m}g$\cdot$mL$^{-1}$, 25 \textgreek{m}g$\cdot$mL$^{-1}$, and 50 \textgreek{m}g$\cdot$mL$^{-1}$. The dashed black arrow on image \textbf{H} indicates the cross-section direction adopted in Figure \ref{FigPubli7}.}\label{FigPubli5}
\end{figure}

\begin{figure}[ht]\centering
\href{http://dx.doi.org/10.1016/j.bbapap.2014.12.001}{See \textbf{Fig. 6.} of doi:10.1016/j.bbapap.2014.12.001.}
\caption[Cross-sections of IgG-coated surfaces]{Cross-sections of IgG-coated hydrophobic surfaces from AFM images shown in Figure \ref{FigPubli5} (direction indicated in Figure \ref{FigPubli5}H). Scratched portions correspond to the shaded zones and the cross-sections are taken vertically from the images of Figure \ref{FigPubli5}. \textbf{A} to \textbf{D}: cross-sections from anti-human IL-2 MAb layers obtained from coating solutions @ 1.56 \textgreek{m}g$\cdot$mL$^{-1}$, 6.25 \textgreek{m}g$\cdot$mL$^{-1}$, 25 \textgreek{m}g$\cdot$mL$^{-1}$, and 50 \textgreek{m}g$\cdot$mL$^{-1}$; \textbf{E} to \textbf{H}: cross-sections from anti-human IL-6 MAb @ 1.56 \textgreek{m}g$\cdot$mL$^{-1}$, 6.25 \textgreek{m}g$\cdot$mL$^{-1}$, 25 \textgreek{m}g$\cdot$mL$^{-1}$, and 50 \textgreek{m}g$\cdot$mL$^{-1}$.}\label{FigPubli6}
\end{figure}

The observed structures in Figure \ref{FigPubli5} clearly reflect the significantly different adhesion mechanisms of the two types of IgG molecules tested in this work. Arguing good and poor adhesion properties of anti-human IL-6 and anti-human IL-2 IgG, we may state that antibodies leading to homogeneous monolayers should be more efficient in ELISA applications than antibodies whose adhesion leads to ``dendrite-like'' patterned monolayers. Table \ref{TabPubli1} and Table \ref{TabPubli2} collect various monolayer properties evaluated from the AFM results obtained for anti-human IL-2 and anti-human IL-6 antibodies, respectively. As previously mentioned, antibody layer thicknesses were estimated by subtracting the mean height of the scratched area from the mean height of the unscratched one. It is difficult to state whether the poor to moderate modulations observed for the thickness of anti-human IL-2 antibody monolayers with changing protein concentration in deposition solution (see Figure \ref{FigPubli4}) result from experimental errors or thickness estimation impaired by layer heterogeneity, but their roughness seems to be quite constant. However, for anti-human IL-6 antibody monolayers, Table \ref{TabPubli2} shows a ten-fold increase (0.27 to 2.88 nm) in layer thickness within the range of concentration tested (from 1.56 to 50 \textgreek{m}g$\cdot$mL$^{-1}$) (Figure \ref{FigPubli4}). Their roughness further remains constant (0.45-0.67 nm) and basically equates that measured for the bare supporting hydrophobic substrate. Data reported in Figure \ref{FigPubli4} confirm the poor dependence of anti-human IL-2 layer thickness on protein concentration in solution. Unlike IL-2, an increase in IL-6 solution concentration results in a dramatic increase of the obtained layer thickness. The concentration-dependence of the layer thickness is related to antibody orientation within the monolayer, as previously argued. The long-axis size of antibodies is about 7.2 nm and 8.2 nm for anti-human IL-2 and IL-6 antibodies, respectively. In addition, effective monolayers thickness is about 2.5 nm and 2.9 nm for antibodies concentration of 50 \textgreek{m}g$\cdot$mL$^{-1}$ (see Table \ref{TabPubli1}). This result allows a qualitative estimation of the antibodies orientation, \textit{i.e.} the long-axis of anti-human IL-2 and IL-6 antibodies forms a mean angle of \textit{ca.} 20.5$^\circ$ with the underlying substrate surface for both antibodies tested in this work. As a result, the estimation of the layer thickness indicates (indirectly) that the overall effective orientation of antibodies is intermediate between that of the end-on and flat configurations depicted in Figure \ref{FigPubli1}.

\begin{table}[ht]\centering
\caption[Properties of monolayers of mouse anti-human IL-2]{Various properties of monolayers formed from the deposition of mouse anti-human IL-2 monoclonal antibody on hydrophobic self-assembled alkanethiol monolayer: the substrate root mean square roughness (nm), the IgG layer roughness (nm), the IgG layer thickness (nm), the protein surface coverage (\%), the superficial volume (mm$^3\cdot$cm$^{-2}$) and the surface density (pmol$\cdot$cm$^{-2}$) are given as a function of IgG concentration in the coating solution (\textgreek{m}g$\cdot$mL$^{-1}$). Intervals are given for a confidence level set at 95 \%. Data are evaluated
from the analysis of 5 \textgreek{m}m $\times$ 5 \textgreek{m}m AFM images.}\label{TabPubli1}
\begin{spacing}{1.2}
\begin{footnotesize}
\centerline{
\begin{tabular}[t]{r@{}lr@{}lr@{}lr@{}l@{ $\pm$ }lr@{}lr@{}lr@{}l}
\hline
\multicolumn{2}{c}{Coating} & \multicolumn{2}{c}{Scrapped surf.} & \multicolumn{2}{c}{Monolayer} & \multicolumn{3}{c}{Monolayer} & \multicolumn{2}{c}{Surface} & \multicolumn{2}{c}{Monolayer} & \multicolumn{2}{c}{Monolayer} \\
\multicolumn{2}{c}{concentration} & \multicolumn{2}{c}{roughness} & \multicolumn{2}{c}{roughness} & \multicolumn{3}{c}{thickness} & \multicolumn{2}{c}{coverage} & \multicolumn{2}{c}{volume} & \multicolumn{2}{c}{density} \\
\multicolumn{2}{c}{(\textgreek{m}g$\cdot$mL$^{-1}$)} & \multicolumn{2}{c}{(nm)} & \multicolumn{2}{c}{(nm)} & \multicolumn{3}{c}{(nm)} & \multicolumn{2}{c}{(\%)} & \multicolumn{2}{c}{(mm$^3\cdot$cm$^{-2}$)} & \multicolumn{2}{c}{(pmol$\cdot$cm$^{-2}$)}\\
\hline
\quad 1&.56&\qquad\ 0&.6 &\quad 1&.21&1&.47&0.1 &\quad\ 79&.6 &\ 0&.$87\times10^{-4}$&\qquad 0&.23\\
      3&.13&       0&.59&      1&.71&1&.49&0.11&       79&.4 &  1&.$26\times10^{-4}$&       0&.34\\
      6&.25&       0&.6 &      0&.96&1&.5& 0.04&       83&.5 &  1&.$22\times10^{-4}$&       0&.33\\
     12&.5 &       0&.95&      2&.6 &2&.16&0.15&       83&.4 &  2&.$85\times10^{-4}$&       0&.76\\
     25&   &       1&.19&      1&.6 &0&.98&0.12&       77&.7 &  1&.$18\times10^{-4}$&       0&.31\\
     50&   &       0&.92&      1&.87&2&.49&0.11&       91&   &  1&.$87\times10^{-4}$&       0&.5 \\
\hline
\end{tabular}}
\end{footnotesize}
\end{spacing}
\end{table}

\begin{table}[ht]\centering
\caption[Properties of monolayers of the mouse anti-human IL-6]{Various properties of monolayers formed from the deposition of the mouse anti-human IL-6 monoclonal antibody on hydrophobic self-assembled  alkanethiol monolayer: the substrate root mean square roughness (nm), the IgG layer roughness (nm), the IgG layer thickness (nm), the protein surface coverage (\%), the superficial volume (mm$^3\cdot$cm$^{-2}$) and the surface density (pmol$\cdot$cm$^{-2}$) are given as a function of IgG concentration in the coating solution (\textgreek{m}g$\cdot$mL$^{-1}$). Intervals are given for a confidence level set at 95 \%. Data are evaluated
from the analysis of 5 \textgreek{m}m $\times$ 5 \textgreek{m}m AFM images.}\label{TabPubli2}
\begin{spacing}{1.2}
\begin{footnotesize}
\centerline{
\begin{tabular}[t]{r@{}lr@{}lr@{}lr@{}l@{ $\pm$ }lr@{}lr@{}lr@{}l}
\hline
\multicolumn{2}{c}{Coating} & \multicolumn{2}{c}{Scrapped surf.} & \multicolumn{2}{c}{Monolayer} & \multicolumn{3}{c}{Monolayer} & \multicolumn{2}{c}{Surface} & \multicolumn{2}{c}{Monolayer} & \multicolumn{2}{c}{Monolayer} \\
\multicolumn{2}{c}{concentration} & \multicolumn{2}{c}{roughness} & \multicolumn{2}{c}{roughness} & \multicolumn{3}{c}{thickness} & \multicolumn{2}{c}{coverage} & \multicolumn{2}{c}{volume} & \multicolumn{2}{c}{density} \\
\multicolumn{2}{c}{(\textgreek{m}g$\cdot$mL$^{-1}$)} & \multicolumn{2}{c}{(nm)} & \multicolumn{2}{c}{(nm)} & \multicolumn{3}{c}{(nm)} & \multicolumn{2}{c}{(\%)} & \multicolumn{2}{c}{(mm$^3\cdot$cm$^{-2}$)} & \multicolumn{2}{c}{(pmol$\cdot$cm$^{-2}$)}\\
\hline
\quad 1&.56&\qquad\ 0&.37&\quad 0&.45&0&.27&0.14&\quad\ 100&  &1&.$59\times10^{-4}$&\qquad 0&.43\\
      3&.13&       0&.43&      0&.58&0&.52&0.23&        99&.5&1&.$35\times10^{-4}$&       0&.36\\
      6&.25&       0&.97&      0&.53&1&.61&0.22&       100&  &2&.$23\times10^{-4}$&       0&.6 \\
     12&.5 &       0&.98&      0&.52&1&.77&0.28&       100&  &2&.$1\ \,\times10^{-4}$&    0&.56\\
     25&   &       0&.79&      0&.67&2&.25&0.2 &       100&  &2&.$58\times10^{-4}$&       0&.69\\
     50&   &       0&.67&      0&.58&2&.88&0.99&       100&  &3&.$72\times10^{-4}$&       1&   \\
\hline
\end{tabular}}
\end{footnotesize}
\end{spacing}
\end{table}

A comparison between the data collected in Tables \ref{TabPubli1}-\ref{TabPubli2} for the systems of interest here reveals larger roughness for anti-human IL-2 layer than that for their anti-human IL-6 counterparts under similar conditions of protein concentration in deposition solution. This is reflected in Figure \ref{FigPubli6} where we report the cross section profiles constructed from AFM images. A careful inspection of the scratched areas in the anti-human IL-6 monoloyers (Figures \ref{FigPubli5}E-\ref{FigPubli5}H) reveals that all proteins are not completely removed from these zones upon application of 200 nN loading force. Instead, the scratched areas in the anti-human IL-2 monoloyers seem entirely devoid of remaining anti-human IL-2 proteins (Figures \ref{FigPubli5}A-\ref{FigPubli5}D). These elements suggest stronger anti-human IL-6 proteins/h-SAM interactions.

Overall, the protein layer thicknesses found in this work are in satisfactory agreement with those from literature. Caruso \textit{et al.} [16] indeed reported AFM thickness measurements of IgG aggregates on gold-coated mica with values of about 7 nm (coating concentration of 50 \textgreek{m}g$\cdot$mL$^{-1}$). Our IgG coating experiments were performed on a hydrophobic surface that is well known for its larger affinity to proteins (as compared to gold coated-mica surfaces) and its propensity to strongly influence protein conformation until denaturation. With these elements in mind, our thickness values ranging from 2.49 nm (anti-IL-2) to 2.88 nm (anti-IL-6) appear consistent with those obtained by Caruso \textit{et al.} Interestingly, Malmsten [12] reported thickness values for IgG layers deposited on methylated silica of around 17-18 nm, based on reflectometry measurements. Regarding the size of the IgG, which does not exceed 14 nm (each Fc or F(ab') fragment measures about 7 nm), such thickness estimations seem to exceed the expected size of a fully end-on IgG monolayer.

It is noteworthy that the roughness is larger than the estimated thickness for some of the monolayers investigated in this work (anti-IL-2: 3.13, 12.5 and 25 \textgreek{m}g$\cdot$mL$^{-1}$, and anti-IL-6: 1.56 and 3.13 \textgreek{m}g$\cdot$mL$^{-1}$) (see Tables \ref{TabPubli1} and \ref{TabPubli2}). In the case of anti-IL-2 monolayers, this is problably due to the thin dendritic-like heterogenenous layer structure (formation of islands or pools of proteins all over the h-SAM surface, in line with the surface coverage data in Table \ref{TabPubli1}). For anti-IL-6 monolayers (high protein surface coverage, Table \ref{TabPubli2}), the estimation of substrate's mean height could be impaired by the poor removal of IgG aggregates in the scratched area. In details, the presence of the remaining IgG aggregates could indeed lead to an overestimation of the height $h_0$ and therefore an underestimation of the film thickness $h_1-h_0$. The cross sections reported in Figure \ref{FigPubli6} indicate that the roughness of this scrapped area (0.6-1.0 nm) is however significantly larger than that of the bare SAMs-gold coated substrate (see Figures \ref{FigPubli10} and \ref{FigPubli11}c). This result evidences the presence of a significant amount of proteins that remain strongly adsorbed on the hydrophobic support even after the scraping step with a 200 nN applied force. The monolayer thickness evaluation may be slightly impacted by the roughness of the scrapped area but, still, our method qualitatively indicates differences in the monolayer surface structure between the two types of the investigated IgG. Further investigation by means of other techniques such as ellipsometry could be helpful to evaluate on a perhaps more quantitative level IgG monolayers properties [23-24].

\subsection{Monolayer building mechanisms}
The above results on ELISA test performance and layer structure properties highlight the close relationship between surface-averaged measurements (ELISA test) and molecular details of the monoclonal anti-human IL-2,6 antibody layers as deciphered by AFM. Test performance is significantly improved with increasing protein deposition solution concentration in the case of IL-6 that exhibits stronger interaction with h-SAMs as compared to IL-2. This probably goes in pair with an increase of the end-on to flat IgG fractions ratio, leading to an increase in the average monolayer thickness. Bremer \textit{et al.} [11] approximated IgG as a sphere of 7 nm radius and Armstrong \textit{et al.} [25] derived a hydrodynamic radius of about 5.29 nm from viscosity measurements. Consequently the IgG molar volume can be evaluated, leading to 3 to 9 $\times$ 10$^{-4}$ mm$^3\cdot$pmol$^{-1}$, and this value may be used to estimate superficial film densities reached under the experimental conditions adopted in our study. Results are given in Tables \ref{TabPubli1} and \ref{TabPubli2} for anti-human IL-2 and IL-6 antibody layers, respectively. Maximum values of 0.5 and 1 pmol$\cdot$cm$^{-2}$ were obtained for anti-IL-2 and anti-IL-6, respectively (an intermediate molar volume of about 6 $\times$ 10$^{-4}$ mm$^3\cdot$pmol$^{-1}$ would lead to maximum values of 0.3 and 0.6 pmol$\cdot$cm$^{-2}$). Buijs \textit{et al.} [7, 26] measured adsorption isotherms on monoclonal mouse antibodies (anti-hCG) deposited onto negatively and positively charged polystyrene surfaces at pH 7. The authors demonstrated that maximum superficial densities of 2.3 pmol$\cdot$cm$^{-2}$ could be achieved onto the positively charged substrate for a protein concentration in the deposition solution of about 120-150 \textgreek{m}g$\cdot$mL$^{-1}$. For experiments performed onto negatively charged substrates, a maximum value of about 0.8 pmol$\cdot$cm$^{-2}$ was reached for protein concentration in solution of about 80-100 \textgreek{m}g$\cdot$mL$^{-1}$. Malmsten [12] reported adsorption kinetics of human IgG on methylated silica and maximal protein superficial densities of about 0.73 pmol$\cdot$cm$^{-2}$ at a protein concentration of 100 \textgreek{m}g$\cdot$mL$^{-1}$ in the deposition solution (pH 7.4). This author further obtained 2 pmol$\cdot$cm$^{-2}$ protein surface concentration at 200-300 \textgreek{m}g$\cdot$mL$^{-1}$ (pH 7.4) [27]. Other relevant IgG surface concentration values are reviewed by Giacomelli [8]. The superficial densities we derive here for anti-human IL-2 and IL-6 antibody are lower than those reported by the authors mentioned above essentially because the protein concentrations in deposition solution tested in our work are lower. Figure \ref{FigPubli7} shows a linear increase of the anti-human IL-6 surface density with increasing protein concentration. This likely corresponds to the range of protein concentration where linear Henry adsorption regime holds. Interestingly, this part of the isotherms is poorly investigated in literature, probably because consistent and reliable data are difficult to collect there due to the limits of techniques for measuring such thin layers at low protein concentration in solution [26]. Knowledge of the protein adsorption in the Henry regime is however rich of information since the initial slope is indicative of protein affinity for the supporting surface [7]. Figure \ref{FigPubli7} then evidences the higher affinity of anti-IL-6 antibody for h-SAMs as compared to anti-IL-2 antibody. This result agrees with the AFM images of Figure \ref{FigPubli5} that shows a poor removal of anti-IL-6 antibody from formed monolayers (which is indicative of a strong interaction with the substrate). 

\begin{figure}[ht]\centering
\href{http://dx.doi.org/10.1016/j.bbapap.2014.12.001}{See \textbf{Fig. 7.} of doi:10.1016/j.bbapap.2014.12.001.}
\caption[Monolayer density as a function of antibody concentration]{Monolayer superficial density as a function of antibody concentration in the IgG coating suspension. Points correspond to data from Tables \ref{TabPubli1} and \ref{TabPubli2} and are accompanied by linear regressions and their error bars (confidence level set at 95~\%).}\label{FigPubli7}
\end{figure}

Based on the monolayer building mechanisms proposed in literature [2, 11, 14, 28] and further summarized in Figure \ref{FigPubli1}, we may assume that for anti-IL-6 monolayers, increasing the protein concentration in the deposition solution minimizes protein orientation relaxation processes on/in the vicinity of the surface (Figures \ref{FigPubli1}G-\ref{FigPubli1}I). This results in the significant adsorption of end-on oriented proteins on thick and immunologically efficient films. As far as the anti-IL-2 ``dendrite-like'' surface layers are concerned, the occurrence of nucleation processes with significant protein surface orientation (Figures \ref{FigPubli1}D-\ref{FigPubli1}F) seem to better agree with the heterogeneous-like and lower surface coverage properties we measured for that film. We have compared the aminoacid sequences (\href{http://www.nextprot.org}{http://www.nextprot.org}, \href{http://www.proteopedia.org}{http://www.proteopedia.org}) of both proteins and found that: (\textit{i}) anti IL-2 (153 aa, 5 \textgreek{b}-strands, 9 \textgreek{a}-helices, 176 kDa) is smaller than anti IL-6 (212 aa, 4 \textgreek{b}-strands, 7 \textgreek{a}-helices, 1 turn, 237 kDa), (\textit{ii}) the isoelectric point of anti IL-2 is 7.67 and that of anti IL-6 is 6.17, and (\textit{iii}) the aa sequences are significantly different. The latter feature strongly suggests that anti IL-2 and anti IL-6 display different hydrophobic domains, which may explain their differentiated adhesion properties probed by AFM.

\section{Conclusion}
ELISA tests were performed on anti-human interleukin-6 IgG and anti-human interleukin-2 IgG films deposited on hydrophobic self-assembled monolayers. Strong differences between the two types of antibodies tested in this work are discussed in terms of film structure properties and immunological activity. In particular, within the adopted range of protein concentration, \textit{in situ} imaging of the resulting surface films by atomic force microscopy under liquid conditions highlight a smooth and immunologically active anti-human IL-6 antibodies and a ``dendrite-like'' structure for rough and poorly active anti-human IL-2 monolayers. A discussion is further given on the relationship between the obtained film structure, protein orientation during adsorption, and resulting immunological activity. AFM appears as a promising tool to connect molecular proteinaceous film structure and derive the corresponding (surface-averaged) protein adsorption isotherms, both type of information being essential to optimize conditions for improving ELISA sensitivity.

\section{References}
\begin{footnotesize}
\renewcommand{\labelenumi}{[\theenumi ]}
\begin{enumerate}
\item T. \textsc{Porstmann}, S. \textsc{Kiessig}, Enzyme immunoassay techniques an overview, \textit{J. Immunol. Methods} 150 (1992) 5-21.
\item P. \textsc{D'Orazio}, Biosensors in clinical chemistry -- 2011 update, \textit{Clin. Chim. Acta.} 412 (2011) 1749-1761.
\item J.E. \textsc{Butler}, Solid Supports in Enzyme-Linked ImmunoSorbent Assay and Other Solid-Phase Immunoassays, \textit{Methods} 150 (2000) 5-21.
\item V. \textsc{Hlady}, J. \textsc{Buijs}, Protein adsorption on solid surfaces, \textit{Curr. Opin. Biotechnol.} 7 (1996) 72-77.
\item W. \textsc{Norde}, My voyage of discovery to proteins in flatland… and beyond, \textit{Colloids Surf., B.} 61 (2008) 1-9.
\item J. \textsc{Buijs}, W. \textsc{Norde}, J. \textsc{Lichtenbelt}, Changes in the Secondary Structure of Adsorbed IgG and F(ab')$_2$ Studied by FTIR Spectroscopy, \textit{Langmuir} 12 (1996) 1605-1613.
\item J. \textsc{Buijs}, J. \textsc{Lichtenbelt}, W. \textsc{Norde}, J. \textsc{Lyklema}, Adsorption of monoclonal IgGs and their F(ab')$_2$ fragments onto polymeric surfaces, \textit{Colloids Surf., B} 5 (1995) 11-23.
\item C.E. \textsc{Giacomelli}, \textit{Adsorption of Immunoglobulins at Solid-Liquid Interfaces}, in: P. \textsc{Somasundaran}, A.T. \textsc{Hubbard} (Eds.), \textit{Encyclopedia of Surface and Colloid Science}, second ed., Taylor \& Francis, 2006, pp. 510-530.
\item V. \textsc{Hlady}, Spatially Resolved Adsorption Kinetics of Immunoglobulin G onto the Wettability Gradient Surface, \textit{Appl. Spectrosc.} 45 (1991) 143-315.
\item J.-W. \textsc{Park}, I.-H. \textsc{Cho}, D.W. \textsc{Moon}, S.-H. \textsc{Paek}, T.G. \textsc{Lee}, ToF-SIMS and PCA of surface-immobilized antibodies with different orientations, \textit{Surf. Interface Anal.} 43 (2011) 285-289.
\item M. \textsc{Bremer}, J. \textsc{Duval}, W. \textsc{Norde}, J. \textsc{Lyklema}, Electrostatic interactions between immunoglobulin (IgG) molecules and a charged sorbent, \textit{Colloids Surf., A} 250 (2004) 29-42.
\item M. \textsc{Malmsten}, Ellipsometry studies of the effects of surface hydrophobicity on protein adsorption, \textit{Colloids Surf., B} 3 (1995) 297-308.
\item J. \textsc{Buijs}, P. van den \textsc{Berg}, J. \textsc{Lichtenbelt}, W. \textsc{Norde}, J. \textsc{Lyklema}, Adsorption Dynamics of IgG and Its F(ab')$_2$ and Fc Fragments Studied by Reflectometry, \textit{J. Colloid Interf. Sci.} 178 (1996) 594-605.
\item P. \textsc{Schaaf}, P. \textsc{D\'{e}jardin}, A. \textsc{Schmitt}, Reflectometry as a Technique To Study the Adsorption of Human Fibrinogen at the Silica/Solution Interface, \textit{Langmuir} 3 (1987) 1131-1135.
\item J.-L. \textsc{Dewez}, V. \textsc{Berger}, Y.-J. \textsc{Schneider}, P. \textsc{Rouxhet}, Influence of Substrate Hydrophobicity on the Adsorption of Collagen in the Presence of Pluronic F68, Albumin, or Calf Serum, \textit{J. Colloid Interf. Sci.} 191 (1997) 1-10.
\item F. \textsc{Caruso}, E. \textsc{Rodda}, D.N. \textsc{Furlong}, Orientational Aspects of Antibody Immobilization and Immunological Activity on Quartz Crystal Microbalance Electrodes, \textit{J. Colloid Interf. Sci.} 178 (1996) 104-115.
\item F. \textsc{H\"{o}\"{o}k}, M. \textsc{Rodahl}, P. \textsc{Brzezinski}, B. \textsc{Kasemo}, Energy Dissipation Kinetics for Protein and Antibody-Antigen Adsorption under Shear Oscillation on a Quartz Crystal Microbalance, \textit{Langmuir} 14 (1998) 729-734.
\item A. \textsc{Elgersma}, R. \textsc{Zsom}, W. \textsc{Norde}, J. \textsc{Lyklema}, The adsorption of different types of monoclonal immunoglobulin on positively and negatively charged polystyrene lattices, \textit{Colloids Surf.} 54 (1991) 89-101.
\item M. \textsc{Malmsten}, Formation of Adsorbed Protein Layers, \textit{J. Colloid Interf. Sci.} 207 (1998) 186-199.
\item R. van \textsc{Erp}, Y. \textsc{Linders}, A. van \textsc{Sommeren}, T. \textsc{Gribnau}, Characterization of monoclonal antibodies physically adsorbed onto polystyrene latex particles, \textit{J. Immunol. Methods} 152 (1992) 191-199.
\item B. \textsc{W\"{a}livaara}, P. \textsc{Warkentin}, I. \textsc{Lundstr\"{o}m}, P. \textsc{Tengvall}, Aggregation of IgG on Methylated Silicon Surfaces Studies by Tapping Mode Atomic Force Microscopy, \textit{J. Colloid Interf. Sci.} 174 (1995) 53-60.
\item H.X. \textsc{You}, C.R. \textsc{Lowe}, AFM Studies of Protein Adsorption: 2. Characterization of Immunoglobulin G Adsorption by Detergent Washing, \textit{J. Colloid Interf. Sci.} 182 (1996) 586-601.
\item H. \textsc{Arwin}, Optical Properties of Thin Layers of Bovine Serum Albumin, \textgreek{g}-Globulin, and Hemoglobin, \textit{Appl. Spectrosc.} 40 (1986) 313-318.
\item F. \textsc{Bordi}, M. \textsc{Prato}, O. \textsc{Cavallieri},  C. \textsc{Cametti}, M. \textsc{Canepa}, A. \textsc{Gliozzi}, Azurin Self-Assembled Monolayers Characterized by Coupling Electrical Impedance Spectroscopy and Spectroscopic Ellipsometry, \textit{J. Chem. Phys. B} 108 (2004) 20263-20272.
\item J.K. \textsc{Armstrong}, R.B. \textsc{Wenby}, H.J. \textsc{Meiselman}, T.C. \textsc{Fisher}, The Hydrodynamic Radii of Macromolecules and Their Effect on Red Blood Cell Aggregation, \textit{Biophys. J.} 87 (2004) 4259-4270.
\item W. \textsc{Norde}, C. \textsc{Haynes}, \textit{Reversibility and the Mechanism of Protein Adsorption} in: J. Brash, T. Horbett (Eds.), \textit{Proteins at Interfaces II}, American Chemical Society 602 (1995) 26-40.
\item M. \textsc{Malmsten}, Ellipsometry Studies of Protein Layers Adsorbed at Hydrophobic Surfaces, \textit{J. Colloid Interf. Sci.} 166 (1994) 333-342.
\item C. \textsc{Dupont-Gillain},\textit{ Orientation of adsorbed antibodies: in situ monitoring by QCM and random sequential adsorption modeling} in: T. \textsc{Horbett}, J. \textsc{Brash}, W. \textsc{Norde} (Eds.), \textit{Proteins at Interfaces III State of the Art}, American Chemical Society 1120 (2012) 453-469.
\end{enumerate}
\end{footnotesize}

\clearpage
\section{Supporting information}

\begin{figure}[h]\centering
\href{
http://www.sciencedirect.com/science/article/pii/S1570963914003173}%
{See \textbf{Fig. S1 (Appendix A)} of doi:10.1016/j.bbapap.2014.12.001.}
\caption[AFM images and cross-sections of gold substrate]{AFM images (height and deflection) and the corresponding cross-sections of gold substrate before thiol adsorption \textbf{(a)}, after hydrophobic SAM formation \textbf{(b)}. Panel \textbf{(c)} and \textbf{(d)} correspond to the anti IL-2 and anti IL-6 monolayers adsorbed onto the hydrophobic SAM, respectively. The height profiles or cross-sections were taken from the black lines located on the AFM images.}\label{FigPubli10}
\end{figure}

\begin{figure}[h]\centering
\href{
http://www.sciencedirect.com/science/article/pii/S1570963914003173}%
{See \textbf{Fig. S2 (Appendix A)} of doi:10.1016/j.bbapap.2014.12.001.}
\caption[Cross-sections and roughnesses of the different samples]{Cross-sections and roughnesses of the different samples. \textbf{a)} Height profile of anti IL-2 monolayer over 5 \textgreek{m}m cross section. \textbf{b)} corresponds to a zoom of the cross-section between 4.0 and 5.0 \textgreek{m}m on the lateral axis. Position of 8 topographical asperities are highlighted by the red dashed lines and the double headed arrows correspond to asperity-to-asperity distances of 81 nm, 46 nm, 55 nm and 52 nm for positions 1, 2, 3 and 4 respectively. Notice that the average asperity-to-asperity distance is 54.7 $\pm$ 16.4 nm, that is 5 times the apex radius of the AFM-tip. The r.m.s roughnesses of the gold substrate, hydrophobic SAM, anti IL-2 and anti IL-6 monolayers are reported in the histogram \textbf{c)}.}\label{FigPubli11}
\end{figure}
\end{cbunit}
\end{otherlanguage}

\backmatter
\listoffigures
\addcontentsline{toc}{chapter}{Table des figures}
\listoftables
\addcontentsline{toc}{chapter}{Liste des tableaux}

\cleardoublepage\pagestyle{empty}
\ 
\cleardoublepage\pagestyle{empty}
\ 
\newpage\pagestyle{empty}
\newgeometry{left=5cm,right=5cm,bottom=6cm,top=6cm}

\vspace*{\stretch{1}}
\begin{center}
\textbf{Adhésion des IgG}\\
\textbf{sur une surface hydrophobe}\\
Théorie, modélisations et application à l'ELISA
\end{center}

\vspace*{0.5cm}

\begin{small}

Les ELISA (\textit{Enzyme-Linked ImmunoSorbent Assay}) sont une des technologies analytiques les plus utilisées dans la recherche et les applications biomédicales. Leur production nécessite la construction de films d'anticorps sur des surfaces constituées le plus souvent de polystyrène. La haute hydrophobie du polystyrène assure une adhésion forte et spontanée des anticorps permettant ainsi d'y construire facilement une monocouche d'anticorps.

L'amélioration des ELISA passe certainement par l'amélioration et la compréhension des mécanismes physico-chimiques à l'{\oe}uvre lors de l'immobilisation des anticorps sur le polystyrène. Dans ce but, ce travail présente un essai de théorisation appuyé par des simulations numériques et des estimations expérimentales par microscopie à force atomique (AFM) et ELISA.

En faisant référence à l'effet hydrophobe, la thermodynamique des processus irréversibles permet premièrement d'expliciter les raisons de l'adhésion des anticorps sur le polystyrène. Deuxièmement, des simulations numériques dans le cadre du modèle des additions séquentielles aléatoires (RSA) montrent la façon dont peuvent se saturer les surfaces en favorisant certaines orientations d'anticorps recherchées dans le cadre de l'ELISA. Finalement, l'amélioration du modèle RSA en un modèle RSA+R tenant compte des changements d'orientations par relaxation des anticorps illustre le lien entre les conditions de dépôt et la structure de la monocouche obtenue. Ces éléments semblent corroborés par l'expérience.

\vspace*{0.5cm}

\noindent\textit{Par Pierre de Thier, bioingénieur: chimie et bio-industries (2009) de l'Université catholique de Louvain (Louvain-la-Neuve, Belgique).}

\end{small}

\vspace*{\stretch{1}}

\begin{thebibliography}{10}
\expandafter\ifx\csname fonteauteurs\endcsname\relax
\def\fonteauteurs{\scshape}\fi
\makeatother

\bibitem{graham1961}
T.~\bgroup\fonteauteurs\bgroup Graham\egroup\egroup{}.
\newblock Liquid diffusion applied to analysis.
\newblock {\em Philosophical Transactions of the Royal Society of London},
  151: 183--224, 1861.

\bibitem{lajusteargile3}
J.C. \bgroup\fonteauteurs\bgroup Daniel\egroup\egroup{} et
  R.~\bgroup\fonteauteurs\bgroup Audebert\egroup\egroup{}.
\newblock Petits volumes et grandes surfaces: l'univers des collo\"{i}des.
\newblock \emph{In} M.~\bgroup\fonteauteurs\bgroup Daoud\egroup\egroup{} et
  C.~\bgroup\fonteauteurs\bgroup Williams\egroup\egroup{}, \'editeurs :  {\em
  La Juste Argile}, pages 85--134. Les Editions de Physique, 1995.

\bibitem{hiemenz1997tout}
P.C. \bgroup\fonteauteurs\bgroup Hiemenz\egroup\egroup{} et
  R.~\bgroup\fonteauteurs\bgroup Rajagopalan\egroup\egroup{}.
\newblock {\em Principles of Colloid and Surface Chemistry. --- Third Edition,
  Revised and Expanded}.
\newblock Marcel Dekker, Inc., New York, 1997.

\bibitem{einstein1906a}
A.~\bgroup\fonteauteurs\bgroup Einstein\egroup\egroup{}.
\newblock Zur Theorie der Brownschen Bewegung.
\newblock {\em Annalen der Physik}, 19(4): 371--381,
  1906.

\bibitem{batchelor1977}
G.K. \bgroup\fonteauteurs\bgroup Batchelor\egroup\egroup{}.
\newblock The effect of brownian motion on the bulk stress in a suspension of
  spherical particles.
\newblock {\em Journal of Fluid Mechanics}, 83(01):
  97--117, 1977.

\bibitem{everett1972}
D.H. \bgroup\fonteauteurs\bgroup Everett\egroup\egroup{}.
\newblock Manual of symbols and terminology for physicochemical quantities and
  units, appendix II: Definitions, terminology and symbols in colloid and
  surface chemistry.
\newblock {\em Pure and Applied Chemistry}, 31(4):
  577--638, 1972.

\bibitem{hunter1987}
R.J. \bgroup\fonteauteurs\bgroup Hunter\egroup\egroup{}.
\newblock {\em Foundations of Colloid Science}, volume~I de {\em Oxford Science
  Publications}.
\newblock Oxford University Press, New York, 1987.

\bibitem{verwey1948}
E.J.W. \bgroup\fonteauteurs\bgroup Verwey\egroup\egroup{} et J.Th.G.
  \bgroup\fonteauteurs\bgroup Overbeek\egroup\egroup{}.
\newblock {\em Theory of the stability of lyophobic colloids}.
\newblock Elsevier Publishing Company, Inc., New York, 1948.

\bibitem{Mayer2011}
L.M. \bgroup\fonteauteurs\bgroup Mayer\egroup\egroup{} et M.L.
  \bgroup\fonteauteurs\bgroup Wells\egroup\egroup{}.
\newblock Aggregation of colloids in estuaries.
\newblock \emph{In} E.~\bgroup\fonteauteurs\bgroup Wolanski\egroup\egroup{} et
  D.~\bgroup\fonteauteurs\bgroup McLusky\egroup\egroup{}, \'editeurs :  {\em
  Treatise on Estuarine and Coastal Science}, volume~4, pages 143--160.
  Academic Press, Waltham, 2011.

\bibitem{maibaum2004}
L.~\bgroup\fonteauteurs\bgroup Maibaum\egroup\egroup{}, A.R.
  \bgroup\fonteauteurs\bgroup Dinner\egroup\egroup{} et
  D.~\bgroup\fonteauteurs\bgroup Chandler\egroup\egroup{}.
\newblock Micelle formation and the hydrophobic effect.
\newblock {\em Journal of Physical Chemistry B}, 108:
  6778--6781, 2004.

\bibitem{ball2004}
P.~\bgroup\fonteauteurs\bgroup Ball\egroup\egroup{}.
\newblock Astrobiology: Water, water, everywhere?
\newblock {\em Nature}, 427: 19--20, 2004.

\bibitem{seager2013}
S.~\bgroup\fonteauteurs\bgroup Seager\egroup\egroup{}.
\newblock Exoplanet habitability.
\newblock {\em Science}, 340: 577--581, 2013.

\bibitem{tanford1978}
C.~\bgroup\fonteauteurs\bgroup Tanford\egroup\egroup{}.
\newblock The hydrophobic effect and the organization of living matter.
\newblock {\em Science}, 200(2): 1012--1018, 1978.

\bibitem{levy2006}
Y.~\bgroup\fonteauteurs\bgroup Levy\egroup\egroup{} et J.N.
  \bgroup\fonteauteurs\bgroup Onuchic\egroup\egroup{}.
\newblock Water mediation in protein folding and molecular recognition.
\newblock {\em Annual Review of Biophysics and Biomolecular Structure},
  35(1): 389--415, 2006.

\bibitem{ball2013}
P.~\bgroup\fonteauteurs\bgroup Ball\egroup\egroup{}.
\newblock The importance of water.
\newblock \emph{In} I.W.M. \bgroup\fonteauteurs\bgroup Smith\egroup\egroup{},
  C.S. \bgroup\fonteauteurs\bgroup Cockell\egroup\egroup{} et
  S.~\bgroup\fonteauteurs\bgroup Leach\egroup\egroup{}, \'editeurs :  {\em
  Astrochemistry and Astrobiology}, Physical Chemistry in Action, pages
  169--210. Springer Berlin Heidelberg, 2013.

\bibitem{kissinger2005}
P.T. \bgroup\fonteauteurs\bgroup Kissinger\egroup\egroup{}.
\newblock Biosensors---a perspective.
\newblock {\em Biosensors and Bioelectronics}, 20:
  2512--2516, 2005.

\bibitem{templin2002}
M.F. \bgroup\fonteauteurs\bgroup Templin\egroup\egroup{},
  D.~\bgroup\fonteauteurs\bgroup Stoll\egroup\egroup{},
  M.~\bgroup\fonteauteurs\bgroup Schrenk\egroup\egroup{}, P.C.
  \bgroup\fonteauteurs\bgroup Traub\egroup\egroup{}, C.F.
  \bgroup\fonteauteurs\bgroup V\"{o}hringer\egroup\egroup{} et T.O.
  \bgroup\fonteauteurs\bgroup Joos\egroup\egroup{}.
\newblock Protein microarray technology.
\newblock {\em Trends in Biotechnology}, 20(4):
  160--166, 2002.

\bibitem{dorazio2011}
P.~\bgroup\fonteauteurs\bgroup D'Orazio\egroup\egroup{}.
\newblock Biosensors in clinical chemistry --- 2011 update.
\newblock {\em Clinica Chimica Acta}, 412: 1749--1761,
  2011.

\bibitem{butler2000}
J.E. \bgroup\fonteauteurs\bgroup Butler\egroup\egroup{}.
\newblock Solid supports in enzyme-linked immunosorbent assay and other
  solid-phase immunoassays.
\newblock {\em Methods}, 22(1): 4--23, 2000.

\bibitem{porstmann1992}
T.~\bgroup\fonteauteurs\bgroup Porstmann\egroup\egroup{} et S.T.
  \bgroup\fonteauteurs\bgroup Kiessig\egroup\egroup{}.
\newblock Enzyme immunoassay techniques an overview.
\newblock {\em Journal of Immunological Methods},
  150(1-2): 5--21, 1992.

\bibitem{hornbeck2001}
P.~\bgroup\fonteauteurs\bgroup Hornbeck\egroup\egroup{}, S.E.
  \bgroup\fonteauteurs\bgroup Winston\egroup\egroup{} et S.A.
  \bgroup\fonteauteurs\bgroup Fuller\egroup\egroup{}.
\newblock {\em Enzyme-Linked Immunosorbent Assays (ELISA)}, chapitre~11.
\newblock John Wiley \& Sons, Inc., 2001.

\bibitem{hlady1996}
V.~\bgroup\fonteauteurs\bgroup Hlady\egroup\egroup{} et
  J.~\bgroup\fonteauteurs\bgroup Buijs\egroup\egroup{}.
\newblock Protein adsorption on solid surfaces.
\newblock {\em Current Opinion in Biotechnology}, 7(1):
  72--77, 1996.

\bibitem{vroman1977b}
L.~\bgroup\fonteauteurs\bgroup Vroman\egroup\egroup{}, A.L.
  \bgroup\fonteauteurs\bgroup Adams\egroup\egroup{},
  M.~\bgroup\fonteauteurs\bgroup Klings\egroup\egroup{}, G.C.
  \bgroup\fonteauteurs\bgroup Fischer\egroup\egroup{}, P.C.
  \bgroup\fonteauteurs\bgroup Munoz\egroup\egroup{} et R.P.
  \bgroup\fonteauteurs\bgroup Solensky\egroup\egroup{}.
\newblock Reactions of formed elements of blood with plasma proteins at
  interfaces.
\newblock {\em Annals of the New York Academy of Sciences},
  283(1): 65--76, 1977.

\bibitem{nagarajan1991}
R.~\bgroup\fonteauteurs\bgroup Nagarajan\egroup\egroup{} et
  E.~\bgroup\fonteauteurs\bgroup Ruckenstein\egroup\egroup{}.
\newblock Theory of surfactant self-assembly: a predictive molecular
  thermodynamic approach.
\newblock {\em Langmuir}, 7(12): 2934--2969, 1991.

\bibitem{buijs1996a}
J.~\bgroup\fonteauteurs\bgroup Buijs\egroup\egroup{},
  W.~\bgroup\fonteauteurs\bgroup Norde\egroup\egroup{} et J.W.Th.
  \bgroup\fonteauteurs\bgroup Lichtenbelt\egroup\egroup{}.
\newblock Changes in the secondary structure of adsorbed IgG and F(ab')$_2$
  studied by FTIR spectroscopy.
\newblock {\em Langmuir}, 12(6): 1605--1613, 1996.

\bibitem{vogler2012}
E.A. \bgroup\fonteauteurs\bgroup Vogler\egroup\egroup{}.
\newblock Protein adsorption in three dimensions.
\newblock {\em Biomaterials}, 33(5): 1201--1237, 2012.

\bibitem{norde2008}
W.~\bgroup\fonteauteurs\bgroup Norde\egroup\egroup{}.
\newblock My voyage of discovery to proteins in flatland\dots{} and beyond.
\newblock {\em Colloids and Surfaces B: Biointerfaces},
  61(1): 1--9, 2008.

\bibitem{giacomelli2006}
C.E. \bgroup\fonteauteurs\bgroup Giacomelli\egroup\egroup{}.
\newblock Adsorption of immunoglobulins at solid--liquid interfaces.
\newblock {\em Encyclopedia of Surface and Colloid Science: Second Edition},
  pages 510--530, 2006.

\bibitem{herron1998}
J.N. \bgroup\fonteauteurs\bgroup Herron\egroup\egroup{}, H.-K.
  \bgroup\fonteauteurs\bgroup Wang\egroup\egroup{},
  V.~\bgroup\fonteauteurs\bgroup Janatov\egroup\egroup{}, J.D.
  \bgroup\fonteauteurs\bgroup Durstchi\egroup\egroup{}, K.D.
  \bgroup\fonteauteurs\bgroup Caldwell\egroup\egroup{}, D.A.
  \bgroup\fonteauteurs\bgroup Christensen\egroup\egroup{}, I.-N.
  \bgroup\fonteauteurs\bgroup Chang\egroup\egroup{} et S.-C.
  \bgroup\fonteauteurs\bgroup Huang\egroup\egroup{}.
\newblock Orientation and activity of immobilized antibodies.
\newblock \emph{In} M.~\bgroup\fonteauteurs\bgroup Malmsten\egroup\egroup{},
  \'editeur :  {\em Biopolymers at Interfaces, Second Edition}, volume~75 de
  {\em Surfactant Science Series}, pages 453--484. Marcel Dekker, Inc., New
  York, 1998.

\bibitem{caruso2000}
F.~\bgroup\fonteauteurs\bgroup Caruso\egroup\egroup{}.
\newblock Fabrication of immunoglobulin mono- and multilayers and their
  application for immunosensing.
\newblock \emph{In} Y.~\bgroup\fonteauteurs\bgroup Lvov\egroup\egroup{} et
  H.~\bgroup\fonteauteurs\bgroup M\"{o}hwald\egroup\egroup{}, \'editeurs :
  {\em Protein architecture: interfacing molecular assemblies and
  immobilization biotechnology}, pages 193--228. Marcel Dekker, Inc., New York,
  2000.

\bibitem{dupont2000}
Ch.C. \bgroup\fonteauteurs\bgroup Dupont-Gillain\egroup\egroup{},
  Y.~\bgroup\fonteauteurs\bgroup Adriaensen\egroup\egroup{},
  S.~\bgroup\fonteauteurs\bgroup Derclaye\egroup\egroup{} et P.G.
  \bgroup\fonteauteurs\bgroup Rouxhet\egroup\egroup{}.
\newblock Plasma-oxidized polystyrene: Wetting properties and surface
  reconstruction.
\newblock {\em Langmuir}, 16(21): 8194--8200, 2000.

\bibitem{elgersma1990}
A.V. \bgroup\fonteauteurs\bgroup Elgersma\egroup\egroup{}, R.L.J.
  \bgroup\fonteauteurs\bgroup Zsom\egroup\egroup{},
  W.~\bgroup\fonteauteurs\bgroup Norde\egroup\egroup{} et
  J.~\bgroup\fonteauteurs\bgroup Lyklema\egroup\egroup{}.
\newblock The adsorption of bovine serum albumin on positively and negatively
  charged polystyrene latices.
\newblock {\em Journal of Colloid and Interface Science},
  138(1): 145--156, 1990.

\bibitem{dehoux2010}
J.-P. \bgroup\fonteauteurs\bgroup Dehoux\egroup\egroup{}.
\newblock Immunologie g\'{e}n\'{e}rale, 2010.

\bibitem{roitt1993}
I.M. \bgroup\fonteauteurs\bgroup Roitt\egroup\egroup{},
  J.~\bgroup\fonteauteurs\bgroup Brostoff\egroup\egroup{} et D.K.
  \bgroup\fonteauteurs\bgroup Male\egroup\egroup{}.
\newblock {\em Immunologie, 3\`{e}me Edition}.
\newblock De Boeck-Wesmael, Bruxelles, 1993.

\bibitem{mach2009}
B.~\bgroup\fonteauteurs\bgroup Mach\egroup\egroup{} et
  V.~\bgroup\fonteauteurs\bgroup Ossipow\egroup\egroup{}.
\newblock Du g\`{e}ne au m\'{e}dicament.
\newblock {\em Biofutur}, 303: 26--30, 2009.

\bibitem{bork1994}
P.~\bgroup\fonteauteurs\bgroup Bork\egroup\egroup{},
  L.~\bgroup\fonteauteurs\bgroup Holm\egroup\egroup{} et
  C.~\bgroup\fonteauteurs\bgroup Sander\egroup\egroup{}.
\newblock The immunoglobulin fold: Structural classification, sequence patterns
  and common core.
\newblock {\em Journal of Molecular Biology}, 242(4):
  309--320, 1994.

\bibitem{padlan1994}
E.A. \bgroup\fonteauteurs\bgroup Padlan\egroup\egroup{}.
\newblock Anatomy of the antibody molecule.
\newblock {\em Molecular Immunology}, 31(3): 169--217,
  1994.

\bibitem{elgersma1991}
A.V. \bgroup\fonteauteurs\bgroup Elgersma\egroup\egroup{}, R.L.J.
  \bgroup\fonteauteurs\bgroup Zsom\egroup\egroup{},
  W.~\bgroup\fonteauteurs\bgroup Norde\egroup\egroup{} et
  J.~\bgroup\fonteauteurs\bgroup Lyklema\egroup\egroup{}.
\newblock The adsorption of different types of monoclonal immunoglobulin on
  positively and negatively charged polystyrene latices.
\newblock {\em Colloids and Surfaces}, 54: 89--101,
  1991.

\bibitem{elgersma1992}
A.V. \bgroup\fonteauteurs\bgroup Elgersma\egroup\egroup{}, R.L.J.
  \bgroup\fonteauteurs\bgroup Zsom\egroup\egroup{},
  J.~\bgroup\fonteauteurs\bgroup Lyklema\egroup\egroup{} et
  W.~\bgroup\fonteauteurs\bgroup Norde\egroup\egroup{}.
\newblock Adsorption competition between albumin and monoclonal
  immuno-gamma-globulins on polystyrene latices.
\newblock {\em Journal of Colloid and Interface Science},
  152(2): 410--428, 1992.

\bibitem{bremer2004}
M.G.E.G. \bgroup\fonteauteurs\bgroup Bremer\egroup\egroup{},
  J.~\bgroup\fonteauteurs\bgroup Duval\egroup\egroup{},
  W.~\bgroup\fonteauteurs\bgroup Norde\egroup\egroup{} et
  J.~\bgroup\fonteauteurs\bgroup Lyklema\egroup\egroup{}.
\newblock Electrostatic interactions between immunoglobulin (IgG) molecules and
  a charged sorbent.
\newblock {\em Colloids and Surfaces A: Physicochemical and Engineering
  Aspects}, 250(1-3): 29--42, 2004.

\bibitem{nordehaynes}
W.~\bgroup\fonteauteurs\bgroup Norde\egroup\egroup{} et C.A.
  \bgroup\fonteauteurs\bgroup Haynes\egroup\egroup{}.
\newblock Reversibility and the mechanism of protein adsorption.
\newblock \emph{In} J.L. \bgroup\fonteauteurs\bgroup Brash\egroup\egroup{} et
  T.A. \bgroup\fonteauteurs\bgroup Horbett\egroup\egroup{}, \'editeurs :  {\em
  Proteins at Interfaces II}, volume 602 de {\em ACS Symposium Series}, pages
  26--40. American Chemical Society, Washington, 1995.

\bibitem{buijs1995}
J.~\bgroup\fonteauteurs\bgroup Buijs\egroup\egroup{}, J.W.Th.
  \bgroup\fonteauteurs\bgroup Lichtenbelt\egroup\egroup{},
  W.~\bgroup\fonteauteurs\bgroup Norde\egroup\egroup{} et
  J.~\bgroup\fonteauteurs\bgroup Lyklema\egroup\egroup{}.
\newblock Adsorption of monoclonal IgGs and their F(ab')$_2$ fragments onto
  polymeric surfaces.
\newblock {\em Colloids and Surfaces B: Biointerfaces},
  5(1-2): 11--23, 1995.

\bibitem{malmsten1995}
M.~\bgroup\fonteauteurs\bgroup Malmsten\egroup\egroup{}.
\newblock Ellipsometry studies of the effects of surface hydrophobicity on
  protein adsorption.
\newblock {\em Colloids and Surfaces B: Biointerfaces},
  3(5): 297--308, 1995.

\bibitem{buijs1996b}
J.~\bgroup\fonteauteurs\bgroup Buijs\egroup\egroup{}, P.A.W. van~den
  \bgroup\fonteauteurs\bgroup Berg\egroup\egroup{}, J.W.Th.
  \bgroup\fonteauteurs\bgroup Lichtenbelt\egroup\egroup{},
  W.~\bgroup\fonteauteurs\bgroup Norde\egroup\egroup{} et
  J.~\bgroup\fonteauteurs\bgroup Lyklema\egroup\egroup{}.
\newblock Adsorption dynamics of IgG and its F(ab')$_2$ and Fc fragments
  studied by reflectometry.
\newblock {\em Journal of Colloid and Interface Science},
  178(2): 594--605, 1996.

\bibitem{dijt1994}
J.C. \bgroup\fonteauteurs\bgroup Dijt\egroup\egroup{}, M.A.
  \bgroup\fonteauteurs\bgroup Cohen~Stuart\egroup\egroup{} et G.J.
  \bgroup\fonteauteurs\bgroup Fleer\egroup\egroup{}.
\newblock Reflectometry as a tool for adsorption studies.
\newblock {\em Advances in Colloid and Interface Science},
  50: 79--101, 1994.

\bibitem{hlady1991}
V.~\bgroup\fonteauteurs\bgroup Hlady\egroup\egroup{}.
\newblock Spatially resolved adsorption kinetics of immunoglobulin g onto the
  wettability gradient surface.
\newblock {\em Applied Spectroscopy}, 45(2): 143--315,
  1991.

\bibitem{rabe2007}
M.~\bgroup\fonteauteurs\bgroup Rabe\egroup\egroup{},
  D.~\bgroup\fonteauteurs\bgroup Verdes\egroup\egroup{},
  M.~\bgroup\fonteauteurs\bgroup Rankl\egroup\egroup{}, G.R.J.
  \bgroup\fonteauteurs\bgroup Artus\egroup\egroup{} et
  S.~\bgroup\fonteauteurs\bgroup Seeger\egroup\egroup{}.
\newblock A comprehensive study of concepts and phenomena of the nonspecific
  adsorption of \textgreek{b}-lactoglobulin.
\newblock {\em ChemPhysChem}, 8(6): 862--872, 2007.

\bibitem{rabe2009}
M.~\bgroup\fonteauteurs\bgroup Rabe\egroup\egroup{}.
\newblock {\em Understanding Protein Adsorption Phenomena on Solid Surfaces}.
\newblock Th\`ese de doctorat, Universit\"{a}t Z\"{u}rich, Z\"{u}rich, 2009.

\bibitem{rabe2011}
M.~\bgroup\fonteauteurs\bgroup Rabe\egroup\egroup{},
  D.~\bgroup\fonteauteurs\bgroup Verdes\egroup\egroup{} et
  S.~\bgroup\fonteauteurs\bgroup Seeger\egroup\egroup{}.
\newblock Understanding protein adsorption phenomena at solid surfaces.
\newblock {\em Advances in Colloid and Interface Science},
  162: 87--106, 2011.

\bibitem{jonsson2007}
M.~\bgroup\fonteauteurs\bgroup J\"{o}nsson\egroup\egroup{},
  H.~\bgroup\fonteauteurs\bgroup Anderson\egroup\egroup{},
  U.~\bgroup\fonteauteurs\bgroup Lindberg\egroup\egroup{} et
  T.~\bgroup\fonteauteurs\bgroup Aastrup\egroup\egroup{}.
\newblock Quartz crystal microbalance biosensor design: II. Simulation of
  sample transport.
\newblock {\em Sensors and Actuators B: Chemical},
  123(1): 21--26, 2007.

\bibitem{hook1998un}
F.~\bgroup\fonteauteurs\bgroup H\"{o}\"{o}k\egroup\egroup{},
  M.~\bgroup\fonteauteurs\bgroup Rodahl\egroup\egroup{},
  P.~\bgroup\fonteauteurs\bgroup Brzezinski\egroup\egroup{} et
  B.~\bgroup\fonteauteurs\bgroup Kasemo\egroup\egroup{}.
\newblock Energy dissipation kinetics for protein and antibody--antigen
  adsorption under shear oscillation on a quartz crystal microbalance.
\newblock {\em Langmuir}, 14(4): 729--734, 1998.

\bibitem{johannsmann2007a}
D.~\bgroup\fonteauteurs\bgroup Johannsmann\egroup\egroup{}.
\newblock Studies of viscoelasticity with the QCM.
\newblock \emph{In} C.~\bgroup\fonteauteurs\bgroup Steinem\egroup\egroup{} et
  A.~\bgroup\fonteauteurs\bgroup Janshoff\egroup\egroup{}, \'editeurs :  {\em
  Piezoelectric Sensors}, volume~5 de {\em Springer Series on Chemical Sensors
  and Biosensors}, pages 49--109. Springer Berlin Heidelberg, 2007.

\bibitem{McEvoy2012}
K.M. \bgroup\fonteauteurs\bgroup Mc~Evoy\egroup\egroup{}.
\newblock {\em Improving quantification of proteins at interfaces through a
  better and practical modeling of Quartz Crystal Microbalance response}.
\newblock Th\`ese de doctorat, Universit\'{e} catholique de Louvain,
  Louvain-la-Neuve, 2012.

\bibitem{buijs1997un}
J.~\bgroup\fonteauteurs\bgroup Buijs\egroup\egroup{}, D.D.
  \bgroup\fonteauteurs\bgroup White\egroup\egroup{} et
  W.~\bgroup\fonteauteurs\bgroup Norde\egroup\egroup{}.
\newblock The effect of adsorption on the antigen binding by IgG and its
  F(ab')$_2$ fragments.
\newblock {\em Colloids and Surfaces B: Biointerfaces},
  8(4-5): 239--249, 1997.

\bibitem{Norde2012}
W.~\bgroup\fonteauteurs\bgroup Norde\egroup\egroup{} et
  J.~\bgroup\fonteauteurs\bgroup Lyklema\egroup\egroup{}.
\newblock Interfacial behaviour of proteins, with special reference to
  immunoglobulins. A physicochemical study.
\newblock {\em Advances in Colloid and Interface Science},
  179-182(0): 5--13, 2012.

\bibitem{lundstrom1990}
I.~\bgroup\fonteauteurs\bgroup Lundstr\"{o}m\egroup\egroup{} et
  H.~\bgroup\fonteauteurs\bgroup Elwing\egroup\egroup{}.
\newblock Simple kinetic models for protein exchange reactions on solid
  surfaces.
\newblock {\em Journal of Colloid and Interface Science},
  136(1): 68--84, 1990.

\bibitem{wojciechowski1986}
P.~\bgroup\fonteauteurs\bgroup Wojciechowski\egroup\egroup{}, P.~ten
  \bgroup\fonteauteurs\bgroup Hove\egroup\egroup{} et J.L.
  \bgroup\fonteauteurs\bgroup Brash\egroup\egroup{}.
\newblock Phenomenology and mechanism of the transient adsorption of fibrinogen
  from plasma (Vroman effect).
\newblock {\em Journal of Colloid and Interface Science},
  111(2): 455--465, 1986.

\bibitem{slack1989}
S.M. \bgroup\fonteauteurs\bgroup Slack\egroup\egroup{} et T.A.
  \bgroup\fonteauteurs\bgroup Horbett\egroup\egroup{}.
\newblock Changes in the strength of fibrinogen attachment to solid surfaces:
  An explanation of the influence of surface chemistry on the Vroman effect.
\newblock {\em Journal of Colloid and Interface Science},
  133(1): 148--165, 1989.

\bibitem{leduc1995}
C.A. \bgroup\fonteauteurs\bgroup LeDuc\egroup\egroup{},
  L.~\bgroup\fonteauteurs\bgroup Vroman\egroup\egroup{} et E.F.
  \bgroup\fonteauteurs\bgroup Leonard\egroup\egroup{}.
\newblock A mathematical model for the Vroman effect.
\newblock {\em Industrial \& Engineering Chemistry Research},
  34(10): 3488--3495, 1995.

\bibitem{malmsten1998}
M.~\bgroup\fonteauteurs\bgroup Malmsten\egroup\egroup{}.
\newblock Formation of adsorbed protein layers.
\newblock {\em Journal of Colloid and Interface Science},
  207(2): 186--199, 1998.

\bibitem{norde1979}
W.~\bgroup\fonteauteurs\bgroup Norde\egroup\egroup{} et
  J.~\bgroup\fonteauteurs\bgroup Lyklema\egroup\egroup{}.
\newblock Thermodynamics of protein adsorption. Theory with special reference
  to the adsorption of human plasma albumin and bovine pancreas ribonuclease at
  polystyrene surfaces.
\newblock {\em Journal of Colloid and Interface Science},
  71(2): 350--366, 1979.

\bibitem{vanoss1986}
C.J. van \bgroup\fonteauteurs\bgroup Oss\egroup\egroup{}, R.J.
  \bgroup\fonteauteurs\bgroup Good\egroup\egroup{} et M.K.
  \bgroup\fonteauteurs\bgroup Chaudhury\egroup\egroup{}.
\newblock The role of van der Waals forces and hydrogen bonds in <<~hydrophobic
  interactions~>> between biopolymers and low energy surfaces.
\newblock {\em Journal of Colloid and Interface Science},
  111(2): 378--390, 1986.

\bibitem{reiss1965}
H.~\bgroup\fonteauteurs\bgroup Reiss\egroup\egroup{}.
\newblock {\em Methods of thermodynamics}.
\newblock Blaisdell Publishing Company, New York, 1965.

\bibitem{guggenheim1949}
E.A. \bgroup\fonteauteurs\bgroup Guggenheim\egroup\egroup{}.
\newblock {\em Thermodynamics}.
\newblock Monographs on theoretical and applied physics. North-Holland
  Publishing Company, Amsterdam, 1949.

\bibitem{planck1913}
M.~\bgroup\fonteauteurs\bgroup Planck\egroup\egroup{}.
\newblock {\em Le\c{c}ons de Thermodynamique}.
\newblock Librairie Scientifique A. Hermann et Fils, Paris, ouvrage traduit sur
  la troisi\`{e}me \'{e}dition allemande (augment\'{e}e) \'edition, 1913.

\bibitem{landau6}
L.D. \bgroup\fonteauteurs\bgroup Landau\egroup\egroup{} et E.M.
  \bgroup\fonteauteurs\bgroup Lifshitz\egroup\egroup{}.
\newblock {\em Fluid Mechanics, Second English Edition, Revised}, volume~6 de
  {\em Course of Theoretical Physics}.
\newblock Elsevier, Amsterdam, 2011.

\bibitem{goldstein1980}
H.~\bgroup\fonteauteurs\bgroup Goldstein\egroup\egroup{}.
\newblock {\em Classical Mechanics. Second Edition}.
\newblock Addison-Wesley series in physics. Addison-Wesley Publishing Company,
  Reading, Massachussetts, 1980.

\bibitem{batchelor1967}
G.K. \bgroup\fonteauteurs\bgroup Batchelor\egroup\egroup{}.
\newblock {\em An Introduction to Fluid Dynamics}.
\newblock Cambridge University Press, Cambridge, 1967.

\bibitem{brenner1977b}
H.~\bgroup\fonteauteurs\bgroup Brenner\egroup\egroup{} et L.G.
  \bgroup\fonteauteurs\bgroup Leal\egroup\egroup{}.
\newblock A model of surface diffusion on solids.
\newblock {\em Journal of Colloid and Interface Science},
  62(2): 238--258, 1977.

\bibitem{bretonnet2010}
J.L. \bgroup\fonteauteurs\bgroup Bretonnet\egroup\egroup{}.
\newblock {\em Th\'{e}orie microscopique des liquides. Physique statistique,
  interactions, \'{e}quations int\'{e}grales, syst\`{e}mes hors
  d'\'{e}quilibre, mod\'{e}lisation.}
\newblock Ellipses, Paris, 2010.

\bibitem{vandeven1989}
T.G.M. van~de \bgroup\fonteauteurs\bgroup Ven\egroup\egroup{}.
\newblock {\em Colloidal Hydrodynamics}.
\newblock Colloid Science. Academic Press Ltd, London, 1989.

\bibitem{stocker2012}
R.~\bgroup\fonteauteurs\bgroup Stocker\egroup\egroup{}.
\newblock Marine microbes see a sea of gradients.
\newblock {\em Science}, 338: 628--633, 6107 2012.

\bibitem{taylor2012}
J.R. \bgroup\fonteauteurs\bgroup Taylor\egroup\egroup{} et
  R.~\bgroup\fonteauteurs\bgroup Stocker\egroup\egroup{}.
\newblock Trade-off of chemotactic foraging in turbulent water.
\newblock {\em Science}, 338: 675--679, 6107 2012.

\bibitem{frankel1989}
I.~\bgroup\fonteauteurs\bgroup Frankel\egroup\egroup{} et
  H.~\bgroup\fonteauteurs\bgroup Brenner\egroup\egroup{}.
\newblock On the foundations of generalized taylor dispersion theory.
\newblock {\em Journal of Fluid Mechanics}, 204:
  97--119, 1989.

\bibitem{perrin1934}
F.~\bgroup\fonteauteurs\bgroup Perrin\egroup\egroup{}.
\newblock Mouvement brownien d'un ellipso\"{i}de I. Dispersion di\'{e}lectrique
  pour des mol\'{e}cules ellipso\"{i}dales.
\newblock {\em Le journal de physique et le radium -- s\'{e}rie VII},
  5(10): 497--511, 1934.

\bibitem{perrin1936}
F.~\bgroup\fonteauteurs\bgroup Perrin\egroup\egroup{}.
\newblock Mouvement brownien d'un ellipso\"{i}de (II). Rotation libre et
  d\'{e}polarisation des fluorescences. Translation et diffusion de
  mol\'{e}cules ellipso\"{i}dales.
\newblock {\em Le journal de physique et le radium -- s\'{e}rie VII},
  7(1): 1--11, 1936.

\bibitem{brenner1965}
H.~\bgroup\fonteauteurs\bgroup Brenner\egroup\egroup{}.
\newblock Coupling between the translational and rotational brownian motions of
  rigid particles of arbitrary shape I. Helicoidally isotropic particles.
\newblock {\em Journal of Colloid Science}, 20(2):
  104--122, 1965.

\bibitem{brenner1967}
H.~\bgroup\fonteauteurs\bgroup Brenner\egroup\egroup{}.
\newblock Coupling between the translational and rotational brownian motions of
  rigid particles of arbitrary shape II. General theory.
\newblock {\em Journal of Colloid and Interface Science},
  23(3): 407--436, 1967.

\bibitem{nitsche1990}
J.M. \bgroup\fonteauteurs\bgroup Nitsche\egroup\egroup{} et
  H.~\bgroup\fonteauteurs\bgroup Brenner\egroup\egroup{}.
\newblock On the formulation of boundary conditions for rigid nonspherical
  brownian particles near solid walls: Applications to orientation-specific
  reactions with immobilized enzymes.
\newblock {\em Journal of Colloid and Interface Science},
  138(1): 21--41, 1990.

\end{thebibliography}

\begin{thebibliography}{10}
\expandafter\ifx\csname fonteauteurs\endcsname\relax
\def\fonteauteurs{\scshape}\fi
\makeatother

\bibitem{Gibbs1878}
J.W. \bgroup\fonteauteurs\bgroup Gibbs\egroup\egroup{}.
\newblock On the equilibrium of hetergeneous substances.
\newblock {\em Transactions of the Connecticut Academy},
  3: 108--248, 1878.

\bibitem{guggenheim1949}
E.A. \bgroup\fonteauteurs\bgroup Guggenheim\egroup\egroup{}.
\newblock {\em Thermodynamics}.
\newblock Monographs on theoretical and applied physics. North-Holland
  Publishing Company, Amsterdam, 1949.

\bibitem{kondepudi1998}
D.~\bgroup\fonteauteurs\bgroup Kondepudi\egroup\egroup{} et
  I.~\bgroup\fonteauteurs\bgroup Prigogine\egroup\egroup{}.
\newblock {\em Modern Thermodynamics: From Heat Engines to Dissipative
  Structures}.
\newblock John Wiley and Sons, Chichester, 1998.

\bibitem{prigogine1968}
I.~\bgroup\fonteauteurs\bgroup Prigogine\egroup\egroup{}.
\newblock {\em Introduction \`{a} la thermodynamique des processus
  irr\'{e}versibles}.
\newblock Dunod, Paris, 1968.

\bibitem{defay1951}
R.~\bgroup\fonteauteurs\bgroup Defay\egroup\egroup{} et
  I.~\bgroup\fonteauteurs\bgroup Prigogine\egroup\egroup{}.
\newblock {\em Tension superficielle et adsorption}, volume III de {\em
  Trait\'{e} de Thermodynamique conform\'{e}ment aux m\'{e}thodes de Gibbs et
  De Donder}.
\newblock Edition Desoer, Li\`{e}ge, 1951.

\bibitem{poincare1908}
H.~\bgroup\fonteauteurs\bgroup Poincar\'{e}\egroup\egroup{}.
\newblock {\em Thermodynamique}.
\newblock Paris, 1908.

\bibitem{prigogine1950}
I.~\bgroup\fonteauteurs\bgroup Prigogine\egroup\egroup{} et
  R.~\bgroup\fonteauteurs\bgroup Defay\egroup\egroup{}.
\newblock {\em Thermodynamique Chimique}, volume I et II de {\em Trait\'{e} de
  Thermodynamique conform\'{e}ment aux m\'{e}thodes de Gibbs et De Donder}.
\newblock Editions Desoer, Li\`{e}ge, 1950.

\bibitem{lebon2008a}
G.~\bgroup\fonteauteurs\bgroup Lebon\egroup\egroup{},
  D.~\bgroup\fonteauteurs\bgroup Jou\egroup\egroup{} et
  J.~\bgroup\fonteauteurs\bgroup Casas-V\'{a}zquez\egroup\egroup{}.
\newblock Equilibrium thermodynamics: A review.
\newblock \emph{In} {\em Understanding Non-equilibrium Thermodynamics}, pages
  1--36. Springer Berlin Heidelberg, 2008.

\bibitem{hunter1987}
R.J. \bgroup\fonteauteurs\bgroup Hunter\egroup\egroup{}.
\newblock {\em Foundations of Colloid Science}, volume~I de {\em Oxford Science
  Publications}.
\newblock Oxford University Press, New York, 1987.

\bibitem{hiemenz1997tout}
P.C. \bgroup\fonteauteurs\bgroup Hiemenz\egroup\egroup{} et
  R.~\bgroup\fonteauteurs\bgroup Rajagopalan\egroup\egroup{}.
\newblock {\em Principles of Colloid and Surface Chemistry. --- Third Edition,
  Revised and Expanded}.
\newblock Marcel Dekker, Inc., New York, 1997.

\bibitem{perez1993}
J.-P. \bgroup\fonteauteurs\bgroup P\'{e}rez\egroup\egroup{} et A.-M.
  \bgroup\fonteauteurs\bgroup Romulus\egroup\egroup{}.
\newblock {\em Thermodynamique, fondements et applications}.
\newblock Masson, Paris, 1993.

\bibitem{everett1972}
D.H. \bgroup\fonteauteurs\bgroup Everett\egroup\egroup{}.
\newblock Manual of symbols and terminology for physicochemical quantities and
  units, appendix II: Definitions, terminology and symbols in colloid and
  surface chemistry.
\newblock {\em Pure and Applied Chemistry}, 31(4):
  577--638, 1972.

\bibitem{nagarajan1991}
R.~\bgroup\fonteauteurs\bgroup Nagarajan\egroup\egroup{} et
  E.~\bgroup\fonteauteurs\bgroup Ruckenstein\egroup\egroup{}.
\newblock Theory of surfactant self-assembly: a predictive molecular
  thermodynamic approach.
\newblock {\em Langmuir}, 7(12): 2934--2969, 1991.

\bibitem{israelachvili2011}
J.N. \bgroup\fonteauteurs\bgroup Israelachvili\egroup\egroup{}.
\newblock {\em Intermolecular and Surface Forces, Third Edition}.
\newblock Academic Press, London, 2011.

\bibitem{norde1979}
W.~\bgroup\fonteauteurs\bgroup Norde\egroup\egroup{} et
  J.~\bgroup\fonteauteurs\bgroup Lyklema\egroup\egroup{}.
\newblock Thermodynamics of protein adsorption. Theory with special reference
  to the adsorption of human plasma albumin and bovine pancreas ribonuclease at
  polystyrene surfaces.
\newblock {\em Journal of Colloid and Interface Science},
  71(2): 350--366, 1979.

\bibitem{ball2013}
P.~\bgroup\fonteauteurs\bgroup Ball\egroup\egroup{}.
\newblock The importance of water.
\newblock \emph{In} I.W.M. \bgroup\fonteauteurs\bgroup Smith\egroup\egroup{},
  C.S. \bgroup\fonteauteurs\bgroup Cockell\egroup\egroup{} et
  S.~\bgroup\fonteauteurs\bgroup Leach\egroup\egroup{}, \'editeurs :  {\em
  Astrochemistry and Astrobiology}, Physical Chemistry in Action, pages
  169--210. Springer Berlin Heidelberg, 2013.

\bibitem{timasheff2002a}
S.N. \bgroup\fonteauteurs\bgroup Timasheff\egroup\egroup{}.
\newblock Protein hydration, thermodynamic binding, and preferential hydration.
\newblock {\em Biochemistry}, 41(46): 13473--13482,
  2002.

\bibitem{scharnagl2005}
C.~\bgroup\fonteauteurs\bgroup Scharnagl\egroup\egroup{},
  M.~\bgroup\fonteauteurs\bgroup Reif\egroup\egroup{} et
  J.~\bgroup\fonteauteurs\bgroup Friedrich\egroup\egroup{}.
\newblock Stability of proteins: Temperature, pressure and the role of the
  solvent.
\newblock {\em Biochimica et Biophysica Acta}, 1749(2):
  187--213, 2005.

\bibitem{levy2006}
Y.~\bgroup\fonteauteurs\bgroup Levy\egroup\egroup{} et J.N.
  \bgroup\fonteauteurs\bgroup Onuchic\egroup\egroup{}.
\newblock Water mediation in protein folding and molecular recognition.
\newblock {\em Annual Review of Biophysics and Biomolecular Structure},
  35(1): 389--415, 2006.

\bibitem{prigogine1959}
I.~\bgroup\fonteauteurs\bgroup Prigogine\egroup\egroup{}.
\newblock Probl\`{e}mes d'\'{E}volution dans la Thermodynamique des
  Ph\'{e}nom\`{e}nes Irr\'{e}versibles.
\newblock \emph{In} A.I. \bgroup\fonteauteurs\bgroup Oparin\egroup\egroup{},
  A.G. \bgroup\fonteauteurs\bgroup Pasynski\v{\i}\egroup\egroup{}, A.E.
  \bgroup\fonteauteurs\bgroup Braunshte\v{\i}n\egroup\egroup{} et T.E.
  \bgroup\fonteauteurs\bgroup Pavlovskaya\egroup\egroup{}, \'editeurs :  {\em
  The Origin of Life on the Earth}, pages 418--427. Pergamon Press, 1959.

\bibitem{planck1913}
M.~\bgroup\fonteauteurs\bgroup Planck\egroup\egroup{}.
\newblock {\em Le\c{c}ons de Thermodynamique}.
\newblock Librairie Scientifique A. Hermann et Fils, Paris, ouvrage traduit sur
  le troisi\`{e}me \'{e}dition allemande (augment\'{e}e) \'edition, 1913.

\bibitem{prigogine1979}
I.~\bgroup\fonteauteurs\bgroup Prigogine\egroup\egroup{} et
  I.~\bgroup\fonteauteurs\bgroup Stengers\egroup\egroup{}.
\newblock {\em La nouvelle alliance}.
\newblock Gallimard, Paris, 1979.

\bibitem{nicolis2005}
G.~\bgroup\fonteauteurs\bgroup Nicolis\egroup\egroup{}.
\newblock {\em Dynamique chimique. Thermodynamique, cin\'{e}tique et
  m\'{e}canique statistique.}
\newblock Dunod, Paris, 2005.

\bibitem{roth1996}
C.M. \bgroup\fonteauteurs\bgroup Roth\egroup\egroup{}, B.L.
  \bgroup\fonteauteurs\bgroup Neal\egroup\egroup{} et A.M.
  \bgroup\fonteauteurs\bgroup Lenhoff\egroup\egroup{}.
\newblock Van der Waals interactions involving proteins.
\newblock {\em Biophysical Journal}, 70(2): 977--987,
  1996.

\bibitem{lajusteargile3}
J.C. \bgroup\fonteauteurs\bgroup Daniel\egroup\egroup{} et
  R.~\bgroup\fonteauteurs\bgroup Audebert\egroup\egroup{}.
\newblock Petits volumes et grandes surfaces: l'univers des collo\"{i}des.
\newblock \emph{In} M.~\bgroup\fonteauteurs\bgroup Daoud\egroup\egroup{} et
  C.~\bgroup\fonteauteurs\bgroup Williams\egroup\egroup{}, \'editeurs :  {\em
  La Juste Argile}, pages 85--134. Les Editions de Physique, 1995.

\bibitem{hamaker1937}
H.C. \bgroup\fonteauteurs\bgroup Hamaker\egroup\egroup{}.
\newblock The London--van der Waals attraction between spherical particles.
\newblock {\em Physica}, 4(10): 1058--1072, 1937.

\bibitem{lyklema1}
H.~\bgroup\fonteauteurs\bgroup Lyklema\egroup\egroup{}.
\newblock {\em Fundamentals}, volume~I de {\em Fundamentals of Interfaces and
  Colloid Science}.
\newblock Academic Press, London, 1991.

\bibitem{dzyaloshinskii1961}
I.E. \bgroup\fonteauteurs\bgroup Dzyaloshinskii\egroup\egroup{}, E.M.
  \bgroup\fonteauteurs\bgroup Lifshitz\egroup\egroup{} et L.P.
  \bgroup\fonteauteurs\bgroup Pitaevskii\egroup\egroup{}.
\newblock General theory of van der Waals' forces.
\newblock {\em Soviet Physics Uspekhi}, 4(2): 153,
  1961.

\bibitem{verwey1948}
E.J.W. \bgroup\fonteauteurs\bgroup Verwey\egroup\egroup{} et J.Th.G.
  \bgroup\fonteauteurs\bgroup Overbeek\egroup\egroup{}.
\newblock {\em Theory of the stability of lyophobic colloids}.
\newblock Elsevier Publishing Company, Inc., New York, 1948.

\bibitem{ohshima2006}
H.~\bgroup\fonteauteurs\bgroup Ohshima\egroup\egroup{}.
\newblock {\em Theory of Colloid and Interfacial Electric Phenomena}, volume
  XII de {\em Interface Science and Technology}.
\newblock Academic Press, London, 2006.

\bibitem{dupont2000}
Ch.C. \bgroup\fonteauteurs\bgroup Dupont-Gillain\egroup\egroup{},
  Y.~\bgroup\fonteauteurs\bgroup Adriaensen\egroup\egroup{},
  S.~\bgroup\fonteauteurs\bgroup Derclaye\egroup\egroup{} et P.G.
  \bgroup\fonteauteurs\bgroup Rouxhet\egroup\egroup{}.
\newblock Plasma-oxidized polystyrene: Wetting properties and surface
  reconstruction.
\newblock {\em Langmuir}, 16(21): 8194--8200, 2000.

\bibitem{buijs1995}
J.~\bgroup\fonteauteurs\bgroup Buijs\egroup\egroup{}, J.W.Th.
  \bgroup\fonteauteurs\bgroup Lichtenbelt\egroup\egroup{},
  W.~\bgroup\fonteauteurs\bgroup Norde\egroup\egroup{} et
  J.~\bgroup\fonteauteurs\bgroup Lyklema\egroup\egroup{}.
\newblock Adsorption of monoclonal IgGs and their F(ab')$_2$ fragments onto
  polymeric surfaces.
\newblock {\em Colloids and Surfaces B: Biointerfaces},
  5(1-2): 11--23, 1995.

\bibitem{bremer2004}
M.G.E.G. \bgroup\fonteauteurs\bgroup Bremer\egroup\egroup{},
  J.~\bgroup\fonteauteurs\bgroup Duval\egroup\egroup{},
  W.~\bgroup\fonteauteurs\bgroup Norde\egroup\egroup{} et
  J.~\bgroup\fonteauteurs\bgroup Lyklema\egroup\egroup{}.
\newblock Electrostatic interactions between immunoglobulin (IgG) molecules and
  a charged sorbent.
\newblock {\em Colloids and Surfaces A: Physicochemical and Engineering
  Aspects}, 250(1-3): 29--42, 2004.

\bibitem{buijs1996b}
J.~\bgroup\fonteauteurs\bgroup Buijs\egroup\egroup{}, P.A.W. van~den
  \bgroup\fonteauteurs\bgroup Berg\egroup\egroup{}, J.W.Th.
  \bgroup\fonteauteurs\bgroup Lichtenbelt\egroup\egroup{},
  W.~\bgroup\fonteauteurs\bgroup Norde\egroup\egroup{} et
  J.~\bgroup\fonteauteurs\bgroup Lyklema\egroup\egroup{}.
\newblock Adsorption dynamics of IgG and its F(ab')$_2$ and Fc fragments
  studied by reflectometry.
\newblock {\em Journal of Colloid and Interface Science},
  178(2): 594--605, 1996.

\bibitem{buijs1997un}
J.~\bgroup\fonteauteurs\bgroup Buijs\egroup\egroup{}, D.D.
  \bgroup\fonteauteurs\bgroup White\egroup\egroup{} et
  W.~\bgroup\fonteauteurs\bgroup Norde\egroup\egroup{}.
\newblock The effect of adsorption on the antigen binding by IgG and its
  F(ab')$_2$ fragments.
\newblock {\em Colloids and Surfaces B: Biointerfaces},
  8(4-5): 239--249, 1997.

\bibitem{elgersma1992}
A.V. \bgroup\fonteauteurs\bgroup Elgersma\egroup\egroup{}, R.L.J.
  \bgroup\fonteauteurs\bgroup Zsom\egroup\egroup{},
  J.~\bgroup\fonteauteurs\bgroup Lyklema\egroup\egroup{} et
  W.~\bgroup\fonteauteurs\bgroup Norde\egroup\egroup{}.
\newblock Adsorption competition between albumin and monoclonal
  immuno-gamma-globulins on polystyrene latices.
\newblock {\em Journal of Colloid and Interface Science},
  152(2): 410--428, 1992.

\bibitem{elgersma1991}
A.V. \bgroup\fonteauteurs\bgroup Elgersma\egroup\egroup{}, R.L.J.
  \bgroup\fonteauteurs\bgroup Zsom\egroup\egroup{},
  W.~\bgroup\fonteauteurs\bgroup Norde\egroup\egroup{} et
  J.~\bgroup\fonteauteurs\bgroup Lyklema\egroup\egroup{}.
\newblock The adsorption of different types of monoclonal immunoglobulin on
  positively and negatively charged polystyrene latices.
\newblock {\em Colloids and Surfaces}, 54: 89--101,
  1991.

\bibitem{lyklema2}
H.~\bgroup\fonteauteurs\bgroup Lyklema\egroup\egroup{}.
\newblock {\em Solid-Liquid Interfaces}, volume~II de {\em Fundamentals of
  Interfaces and Colloid Science}.
\newblock Academic Press, London, 1995.

\bibitem{winter2007}
R.~\bgroup\fonteauteurs\bgroup Winter\egroup\egroup{},
  D.~\bgroup\fonteauteurs\bgroup Lopes\egroup\egroup{},
  S.~\bgroup\fonteauteurs\bgroup Grudzielanek\egroup\egroup{},
  K.~\bgroup\fonteauteurs\bgroup Vogtt\egroup\egroup{} et
  K.~\bgroup\fonteauteurs\bgroup Heremans\egroup\egroup{}.
\newblock Towards an understanding of the temperature/pressure configurational
  and free-energy landscape of biomolecules.
\newblock {\em Journal of Non-Equilibrium Thermodynamics},
  32(1): 41--97, 2007.

\bibitem{soderquist1980}
M.E. \bgroup\fonteauteurs\bgroup Soderquist\egroup\egroup{} et A.G.
  \bgroup\fonteauteurs\bgroup Walton\egroup\egroup{}.
\newblock Structural changes in proteins adsorbed on polymer surfaces.
\newblock {\em Journal of Colloid and Interface Science},
  75(2): 386--397, 1980.

\bibitem{norde2008}
W.~\bgroup\fonteauteurs\bgroup Norde\egroup\egroup{}.
\newblock My voyage of discovery to proteins in flatland\dots{} and beyond.
\newblock {\em Colloids and Surfaces B: Biointerfaces},
  61(1): 1--9, 2008.

\bibitem{atkins}
P.~\bgroup\fonteauteurs\bgroup Atkins\egroup\egroup{} et
  J.~\bgroup\fonteauteurs\bgroup De~Paula\egroup\egroup{}.
\newblock {\em Physical Chemistry}.
\newblock Oxford University Press, Oxford, 2006.

\end{thebibliography}

\begin{thebibliography}{10}
\expandafter\ifx\csname fonteauteurs\endcsname\relax
\def\fonteauteurs{\scshape}\fi
\makeatother

\bibitem{janeway2001}
C.A. \bgroup\fonteauteurs\bgroup Janeway\egroup\egroup{},
  P.~\bgroup\fonteauteurs\bgroup Travers\egroup\egroup{},
  M.~\bgroup\fonteauteurs\bgroup Walport\egroup\egroup{} et
  M.~\bgroup\fonteauteurs\bgroup Shlomchik\egroup\egroup{}.
\newblock Immunobiology. The immune system in health and disease, Fifth
  Edition.
\newblock 2001.

\bibitem{butler2000}
J.E. \bgroup\fonteauteurs\bgroup Butler\egroup\egroup{}.
\newblock Solid supports in enzyme-linked immunosorbent assay and other
  solid-phase immunoassays.
\newblock {\em Methods}, 22(1): 4--23, 2000.

\bibitem{porstmann1992}
T.~\bgroup\fonteauteurs\bgroup Porstmann\egroup\egroup{} et S.T.
  \bgroup\fonteauteurs\bgroup Kiessig\egroup\egroup{}.
\newblock Enzyme immunoassay techniques an overview.
\newblock {\em Journal of Immunological Methods},
  150(1-2): 5--21, 1992.

\bibitem{buijs1995}
J.~\bgroup\fonteauteurs\bgroup Buijs\egroup\egroup{}, J.W.Th.
  \bgroup\fonteauteurs\bgroup Lichtenbelt\egroup\egroup{},
  W.~\bgroup\fonteauteurs\bgroup Norde\egroup\egroup{} et
  J.~\bgroup\fonteauteurs\bgroup Lyklema\egroup\egroup{}.
\newblock Adsorption of monoclonal IgGs and their F(ab')$_2$ fragments onto
  polymeric surfaces.
\newblock {\em Colloids and Surfaces B: Biointerfaces},
  5(1-2): 11--23, 1995.

\bibitem{buijs1997un}
J.~\bgroup\fonteauteurs\bgroup Buijs\egroup\egroup{}, D.D.
  \bgroup\fonteauteurs\bgroup White\egroup\egroup{} et
  W.~\bgroup\fonteauteurs\bgroup Norde\egroup\egroup{}.
\newblock The effect of adsorption on the antigen binding by IgG and its
  F(ab')$_2$ fragments.
\newblock {\em Colloids and Surfaces B: Biointerfaces},
  8(4-5): 239--249, 1997.

\bibitem{soderquist1980}
M.E. \bgroup\fonteauteurs\bgroup Soderquist\egroup\egroup{} et A.G.
  \bgroup\fonteauteurs\bgroup Walton\egroup\egroup{}.
\newblock Structural changes in proteins adsorbed on polymer surfaces.
\newblock {\em Journal of Colloid and Interface Science},
  75(2): 386--397, 1980.

\bibitem{bremer2004}
M.G.E.G. \bgroup\fonteauteurs\bgroup Bremer\egroup\egroup{},
  J.~\bgroup\fonteauteurs\bgroup Duval\egroup\egroup{},
  W.~\bgroup\fonteauteurs\bgroup Norde\egroup\egroup{} et
  J.~\bgroup\fonteauteurs\bgroup Lyklema\egroup\egroup{}.
\newblock Electrostatic interactions between immunoglobulin (IgG) molecules and
  a charged sorbent.
\newblock {\em Colloids and Surfaces A: Physicochemical and Engineering
  Aspects}, 250(1-3): 29--42, 2004.

\bibitem{vandeven1989}
T.G.M. van~de \bgroup\fonteauteurs\bgroup Ven\egroup\egroup{}.
\newblock {\em Colloidal Hydrodynamics}.
\newblock Colloid Science. Academic Press Ltd, London, 1989.

\bibitem{widom1963}
B.~\bgroup\fonteauteurs\bgroup Widom\egroup\egroup{}.
\newblock Some topics in the theory of fluids.
\newblock {\em Journal of Chemical Physics}, 39:
  2808--2812, 1963.

\bibitem{widom1966}
B.~\bgroup\fonteauteurs\bgroup Widom\egroup\egroup{}.
\newblock Random sequential addition of hard spheres to a volume.
\newblock {\em Journal of Chemical Physics}, 44(10):
  3888--3894, 1966.

\bibitem{bretonnet2010}
J.L. \bgroup\fonteauteurs\bgroup Bretonnet\egroup\egroup{}.
\newblock {\em Th\'{e}orie microscopique des liquides. Physique statistique,
  interactions, \'{e}quations int\'{e}grales, syst\`{e}mes hors
  d'\'{e}quilibre, mod\'{e}lisation.}
\newblock Ellipses, Paris, 2010.

\bibitem{bogaert2005}
P.~\bgroup\fonteauteurs\bgroup Bogaert\egroup\egroup{}.
\newblock {\em Probabilit\'{e}s pour Scientifiques et Ing\'{e}nieurs}.
\newblock Universit\'{e} catholique de Louvain, Louvain-la-Neuve, 2005.

\bibitem{viot1992b}
P.~\bgroup\fonteauteurs\bgroup Viot\egroup\egroup{},
  G.~\bgroup\fonteauteurs\bgroup Tarjus\egroup\egroup{}, S.M.
  \bgroup\fonteauteurs\bgroup Ricci\egroup\egroup{} et
  J.~\bgroup\fonteauteurs\bgroup Talbot\egroup\egroup{}.
\newblock Random sequential adsorption of anisotropic particles. I. Jamming
  limit and asymptotic behavior.
\newblock {\em Journal of Chemical Physics}, 97(7):
  5212, 1992.

\bibitem{ricci1992}
S.M. \bgroup\fonteauteurs\bgroup Ricci\egroup\egroup{},
  J.~\bgroup\fonteauteurs\bgroup Talbot\egroup\egroup{},
  G.~\bgroup\fonteauteurs\bgroup Tarjus\egroup\egroup{} et
  P.~\bgroup\fonteauteurs\bgroup Viot\egroup\egroup{}.
\newblock Random sequential adsorption of anisotropic particles. II. Low
  coverage kinetics.
\newblock {\em Journal of Chemical Physics}, 97(7):
  5219, 1992.

\bibitem{talbot2000}
J.~\bgroup\fonteauteurs\bgroup Talbot\egroup\egroup{},
  G.~\bgroup\fonteauteurs\bgroup Tarjus\egroup\egroup{}, P.R.
  \bgroup\fonteauteurs\bgroup Van~Tassel\egroup\egroup{} et
  P.~\bgroup\fonteauteurs\bgroup Viot\egroup\egroup{}.
\newblock From car parking to protein adsorption: an overview of sequential
  adsorption processes.
\newblock {\em Colloids and Surfaces A: Physicochemical and Engineering
  Aspects}, 165(1-3): 287--324, 2000.

\bibitem{evans1993}
J.W. \bgroup\fonteauteurs\bgroup Evans\egroup\egroup{}.
\newblock Random and cooperative sequential adsorption.
\newblock {\em Reviews of Modern Physics}, 65(4): 1281,
  1993.

\bibitem{schaafettalbot1989a}
P.~\bgroup\fonteauteurs\bgroup Schaaf\egroup\egroup{} et
  J.~\bgroup\fonteauteurs\bgroup Talbot\egroup\egroup{}.
\newblock Surface exclusion effects in adsorption processes.
\newblock {\em Journal of Chemical Physics}, 91(7):
  4401--4409, 1989.

\bibitem{schaafettalbot1989b}
P.~\bgroup\fonteauteurs\bgroup Schaaf\egroup\egroup{} et
  J.~\bgroup\fonteauteurs\bgroup Talbot\egroup\egroup{}.
\newblock Kinetics of random sequential adsorption.
\newblock {\em Physical Review Letters}, 62(2):
  175--178, 1989.

\bibitem{talbot1991}
J.~\bgroup\fonteauteurs\bgroup Talbot\egroup\egroup{},
  P.~\bgroup\fonteauteurs\bgroup Schaaf\egroup\egroup{} et
  G.~\bgroup\fonteauteurs\bgroup Tarjus\egroup\egroup{}.
\newblock Random sequential addition of hard spheres.
\newblock {\em Molecular Physics: An International Journal at the Interface
  Between Chemistry and Physics}, 72(6): 1397 -- 1406,
  1991.

\bibitem{tarjus1991un}
G.~\bgroup\fonteauteurs\bgroup Tarjus\egroup\egroup{},
  P.~\bgroup\fonteauteurs\bgroup Schaaf\egroup\egroup{} et
  J.~\bgroup\fonteauteurs\bgroup Talbot\egroup\egroup{}.
\newblock Generalized random sequential adsorption.
\newblock {\em Journal of Chemical Physics}, 93(11):
  8352, 1990.

\bibitem{reiss1959}
H.~\bgroup\fonteauteurs\bgroup Reiss\egroup\egroup{}, H.L.
  \bgroup\fonteauteurs\bgroup Frisch\egroup\egroup{} et J.L.
  \bgroup\fonteauteurs\bgroup Lebowitz\egroup\egroup{}.
\newblock Statistical mechanics of rigid spheres.
\newblock {\em Journal of Chemical Physics}, 31(2):
  369--380, 1959.

\bibitem{dickman1989}
R.~\bgroup\fonteauteurs\bgroup Dickman\egroup\egroup{}.
\newblock Nonequilibrium lattice models: Series analysis of steady states.
\newblock {\em Journal of Statistical Physics}, 55(5):
  997--1026--1026, 1989.

\bibitem{hastie2009a}
T.~\bgroup\fonteauteurs\bgroup Hastie\egroup\egroup{},
  R.~\bgroup\fonteauteurs\bgroup Tibshirani\egroup\egroup{} et
  J.~\bgroup\fonteauteurs\bgroup Friedman\egroup\egroup{}.
\newblock Linear methods for regression.
\newblock \emph{In} {\em The Elements of Statistical Learning}, Springer Series
  in Statistics, pages 1--57. Springer New York, 2009.

\bibitem{torquato2010}
S.~\bgroup\fonteauteurs\bgroup Torquato\egroup\egroup{} et F.H.
  \bgroup\fonteauteurs\bgroup Stillinger\egroup\egroup{}.
\newblock Jammed hard-particle packings: From kepler to bernal and beyond.
\newblock {\em Reviews of Modern Physics}, 82(3):
  2633--2672, 2010.

\bibitem{schaaf1998}
P.~\bgroup\fonteauteurs\bgroup Schaaf\egroup\egroup{}, J.-C.
  \bgroup\fonteauteurs\bgroup Voegel\egroup\egroup{} et
  B.~\bgroup\fonteauteurs\bgroup Senger\egroup\egroup{}.
\newblock Irreversible deposition/adsorption processes on solid surfaces.
\newblock {\em Annales de Physique}, 23(6): 1--89,
  1998.

\bibitem{hiemenz1997tout}
P.C. \bgroup\fonteauteurs\bgroup Hiemenz\egroup\egroup{} et
  R.~\bgroup\fonteauteurs\bgroup Rajagopalan\egroup\egroup{}.
\newblock {\em Principles of Colloid and Surface Chemistry. --- Third Edition,
  Revised and Expanded}.
\newblock Marcel Dekker, Inc., New York, 1997.

\bibitem{goldner2009}
P.~\bgroup\fonteauteurs\bgroup Goldner\egroup\egroup{}.
\newblock {\em Mathématiques des sciences appliquées. Transformation de
  Fourier, espaces de Hilbert, équations aux dérivées partielles}.
\newblock 2009.

\bibitem{chorlton1969}
F.~\bgroup\fonteauteurs\bgroup Chorlton\egroup\egroup{}.
\newblock {\em Boundary value problems in physics and engineering}.
\newblock Van Nostrand Reinhold Company Ltd., 1969.

\bibitem{singer2005}
A.~\bgroup\fonteauteurs\bgroup Singer\egroup\egroup{} et
  Z.~\bgroup\fonteauteurs\bgroup Schuss\egroup\egroup{}.
\newblock Brownian simulations and unidirectional flux in diffusion.
\newblock {\em Physical Review E}, 71: 026115, Feb
  2005.

\bibitem{singer2007}
A.~\bgroup\fonteauteurs\bgroup Singer\egroup\egroup{},
  Z.~\bgroup\fonteauteurs\bgroup Schuss\egroup\egroup{} et
  D.~\bgroup\fonteauteurs\bgroup Holcman\egroup\egroup{}.
\newblock Partially reflected diffusion.
\newblock juin 2006.

\bibitem{erban2007b}
R.~\bgroup\fonteauteurs\bgroup Erban\egroup\egroup{} et S.J.
  \bgroup\fonteauteurs\bgroup Chapman\egroup\egroup{}.
\newblock Time scale of random sequential adsorption.
\newblock {\em Physical Review E}, 75: 041116, Apr
  2007.

\end{thebibliography}

\begin{thebibliography}{10}
\expandafter\ifx\csname fonteauteurs\endcsname\relax
\def\fonteauteurs{\scshape}\fi
\makeatother

\bibitem{vantassel1996}
P.R. \bgroup\fonteauteurs\bgroup Van~Tassel\egroup\egroup{},
  J.~\bgroup\fonteauteurs\bgroup Talbot\egroup\egroup{},
  G.~\bgroup\fonteauteurs\bgroup Tarjus\egroup\egroup{} et
  P.~\bgroup\fonteauteurs\bgroup Viot\egroup\egroup{}.
\newblock Kinetics of irreversible adsorption with a particle conformational
  change: A density expansion approach.
\newblock {\em Physical Review E}, 53(1): 785, 1996.

\bibitem{vantassel1998}
P.R. \bgroup\fonteauteurs\bgroup Van~Tassel\egroup\egroup{},
  L.~\bgroup\fonteauteurs\bgroup Guemouri\egroup\egroup{}, J.J.
  \bgroup\fonteauteurs\bgroup Ramsden\egroup\egroup{},
  G.~\bgroup\fonteauteurs\bgroup Tarjus\egroup\egroup{},
  P.~\bgroup\fonteauteurs\bgroup Viot\egroup\egroup{} et
  J.~\bgroup\fonteauteurs\bgroup Talbot\egroup\egroup{}.
\newblock A particle-level model of irreversible protein adsorption with a
  postadsorption transition.
\newblock {\em Journal of Colloid and Interface Science},
  207(2): 317--323, 1998.

\bibitem{vantassel1994}
P.R. \bgroup\fonteauteurs\bgroup Van~Tassel\egroup\egroup{},
  P.~\bgroup\fonteauteurs\bgroup Viot\egroup\egroup{},
  G.~\bgroup\fonteauteurs\bgroup Tarjus\egroup\egroup{} et
  J.~\bgroup\fonteauteurs\bgroup Talbot\egroup\egroup{}.
\newblock Irreversible adsorption of macromolecules at a liquid--solid
  interface: Theoretical studies of the effects of conformational change.
\newblock {\em Journal of Chemical Physics}, 101(8):
  7064, 1994.

\bibitem{schmitt1983}
A.~\bgroup\fonteauteurs\bgroup Schmitt\egroup\egroup{},
  R.~\bgroup\fonteauteurs\bgroup Varoqui\egroup\egroup{},
  S.~\bgroup\fonteauteurs\bgroup Uniyal\egroup\egroup{}, J.~L.
  \bgroup\fonteauteurs\bgroup Brash\egroup\egroup{} et
  C.~\bgroup\fonteauteurs\bgroup Pusineri\egroup\egroup{}.
\newblock Interaction of fibrinogen with solid surfaces of varying charge and
  hydrophobic--hydrophilic balance: I. Adsorption isotherms.
\newblock {\em Journal of Colloid and Interface Science},
  92(1): 25--34, 1983.

\bibitem{bremer2004}
M.G.E.G. \bgroup\fonteauteurs\bgroup Bremer\egroup\egroup{},
  J.~\bgroup\fonteauteurs\bgroup Duval\egroup\egroup{},
  W.~\bgroup\fonteauteurs\bgroup Norde\egroup\egroup{} et
  J.~\bgroup\fonteauteurs\bgroup Lyklema\egroup\egroup{}.
\newblock Electrostatic interactions between immunoglobulin (IgG) molecules and
  a charged sorbent.
\newblock {\em Colloids and Surfaces A: Physicochemical and Engineering
  Aspects}, 250(1-3): 29--42, 2004.

\bibitem{schaaf1987}
P.~\bgroup\fonteauteurs\bgroup Schaaf\egroup\egroup{},
  P.~\bgroup\fonteauteurs\bgroup Dejardin\egroup\egroup{} et
  A.~\bgroup\fonteauteurs\bgroup Schmitt\egroup\egroup{}.
\newblock Reflectometry as a technique to study the adsorption of human
  fibrinogen at the silica/solution interface.
\newblock {\em Langmuir}, 3(6): 1131--1135, 1987.

\bibitem{mcdowall2006}
L.M. \bgroup\fonteauteurs\bgroup McDowall\egroup\egroup{} et R.A.L.
  \bgroup\fonteauteurs\bgroup Dampney\egroup\egroup{}.
\newblock Calculation of threshold and saturation points of sigmoidal
  baroreflex function curves.
\newblock {\em American Journal of Physiology - Heart and Circulatory
  Physiology}, 291: H2003--H2007, 2006.

\bibitem{vandeven1996bis}
T.G.M. van~de \bgroup\fonteauteurs\bgroup Ven\egroup\egroup{} et S.J.
  \bgroup\fonteauteurs\bgroup Kelemen\egroup\egroup{}.
\newblock Characterizing polymers with an impinging jet.
\newblock {\em Journal of Colloid and Interface Science},
  181(1): 118--123, 1996.

\bibitem{adamcyzk1994}
Z.~\bgroup\fonteauteurs\bgroup Adamczyk\egroup\egroup{},
  B.~\bgroup\fonteauteurs\bgroup Siwek\egroup\egroup{},
  M.~\bgroup\fonteauteurs\bgroup Zembala\egroup\egroup{} et
  P.~\bgroup\fonteauteurs\bgroup Belouschek\egroup\egroup{}.
\newblock Kinetics of localized adsorption of colloid particles.
\newblock {\em Advances in Colloid and Interface Science},
  48: 151--280, 1994.

\bibitem{yang1998}
C.~\bgroup\fonteauteurs\bgroup Yang\egroup\egroup{},
  T.~\bgroup\fonteauteurs\bgroup D\k{a}bro\'{s}\egroup\egroup{},
  D.~\bgroup\fonteauteurs\bgroup Li\egroup\egroup{},
  J.~\bgroup\fonteauteurs\bgroup Czarnecki\egroup\egroup{} et J.H.
  \bgroup\fonteauteurs\bgroup Masliyah\egroup\egroup{}.
\newblock Kinetics of particle transport to a solid surface from an impinging
  jet under surface and external force fields.
\newblock {\em Journal of Colloid and Interface Science},
  208(1): 226--240, 1998.

\bibitem{adamczyk1982}
Z.~\bgroup\fonteauteurs\bgroup Adamczyk\egroup\egroup{},
  T.~\bgroup\fonteauteurs\bgroup D\c{a}bro\'{s}\egroup\egroup{} et T.G.M.
  van~de \bgroup\fonteauteurs\bgroup Ven\egroup\egroup{}.
\newblock Transfer of brownian particles to continuous moving surfaces.
\newblock {\em Chemical Engineering Science}, 37(10):
  1513--1522, 1982.

\bibitem{nordehaynes}
W.~\bgroup\fonteauteurs\bgroup Norde\egroup\egroup{} et C.A.
  \bgroup\fonteauteurs\bgroup Haynes\egroup\egroup{}.
\newblock Reversibility and the mechanism of protein adsorption.
\newblock \emph{In} J.L. \bgroup\fonteauteurs\bgroup Brash\egroup\egroup{} et
  T.A. \bgroup\fonteauteurs\bgroup Horbett\egroup\egroup{}, \'editeurs :  {\em
  Proteins at Interfaces II}, volume 602 de {\em ACS Symposium Series}, pages
  26--40. American Chemical Society, Washington, 1995.

\bibitem{lundstrom1985}
I.~\bgroup\fonteauteurs\bgroup Lundstr\"{o}m\egroup\egroup{} :
\newblock Models of protein adsorption on solid surfaces.
\newblock {\em Progress in Colloid and Polymer Science},
  70: 76--82, 1985.

\bibitem{brash1978}
J.L. \bgroup\fonteauteurs\bgroup Brash\egroup\egroup{} et Q.M.
  \bgroup\fonteauteurs\bgroup Samak\egroup\egroup{}.
\newblock Dynamics of interactions between human albumin and polyethylene
  surface.
\newblock {\em Journal of Colloid and Interface Science},
  65(3): 495--504, 1978.

\bibitem{norde1999}
W.~\bgroup\fonteauteurs\bgroup Norde\egroup\egroup{} et C.E.
  \bgroup\fonteauteurs\bgroup Giacomelli\egroup\egroup{}.
\newblock Conformational changes in proteins at interfaces: from solution to
  the interface, and back.
\newblock {\em Macromolecular Symposia}, 145(1):
  125--136, 1999.

\bibitem{Norde2012}
W.~\bgroup\fonteauteurs\bgroup Norde\egroup\egroup{} et
  J.~\bgroup\fonteauteurs\bgroup Lyklema\egroup\egroup{}.
\newblock Interfacial behaviour of proteins, with special reference to
  immunoglobulins. A physicochemical study.
\newblock {\em Advances in Colloid and Interface Science},
  179-182: 5--13, 2012.

\bibitem{norde2012b}
W.~\bgroup\fonteauteurs\bgroup Norde\egroup\egroup{},
  T.~\bgroup\fonteauteurs\bgroup Horbett\egroup\egroup{} et J.L.
  \bgroup\fonteauteurs\bgroup Brash\egroup\egroup{}.
\newblock {\em Proteins at Interfaces III: Introductory Overview}, volume 1120
  de {\em ACS Symposium Series}, chapitre~1, pages 1--34.
\newblock American Chemical Society, 2012.

\bibitem{prigogine1968}
I.~\bgroup\fonteauteurs\bgroup Prigogine\egroup\egroup{}.
\newblock {\em Introduction \`{a} la thermodynamique des processus
  irr\'{e}versibles}.
\newblock Dunod, Paris, 1968.

\bibitem{coveney1988}
P.V. \bgroup\fonteauteurs\bgroup Coveney\egroup\egroup{}.
\newblock The second law of thermodynamics: entropy, irreversibility and
  dynamics.
\newblock {\em Nature}, 333: 409--415, 1988.

\end{thebibliography}

\begin{thebibliography}{10}
\expandafter\ifx\csname fonteauteurs\endcsname\relax
\def\fonteauteurs{\scshape}\fi
\makeatother

\bibitem{porstmann1992}
T.~\bgroup\fonteauteurs\bgroup Porstmann\egroup\egroup{} et S.T.
  \bgroup\fonteauteurs\bgroup Kiessig\egroup\egroup{}.
\newblock Enzyme immunoassay techniques an overview.
\newblock {\em Journal of Immunological Methods},
  150(1-2): 5--21, 1992.

\bibitem{dorazio2011}
P.~\bgroup\fonteauteurs\bgroup D'Orazio\egroup\egroup{}.
\newblock Biosensors in clinical chemistry --- 2011 update.
\newblock {\em Clinica Chimica Acta}, 412: 1749--1761,
  2011.

\bibitem{butler2000}
J.E. \bgroup\fonteauteurs\bgroup Butler\egroup\egroup{}.
\newblock Solid supports in enzyme-linked immunosorbent assay and other
  solid-phase immunoassays.
\newblock {\em Methods}, 22(1): 4--23, 2000.

\bibitem{hlady1996}
V.~\bgroup\fonteauteurs\bgroup Hlady\egroup\egroup{} et
  J.~\bgroup\fonteauteurs\bgroup Buijs\egroup\egroup{}.
\newblock Protein adsorption on solid surfaces.
\newblock {\em Current Opinion in Biotechnology}, 7(1):
  72--77, 1996.

\bibitem{norde2008}
W.~\bgroup\fonteauteurs\bgroup Norde\egroup\egroup{}.
\newblock My voyage of discovery to proteins in flatland\dots{} and beyond.
\newblock {\em Colloids and Surfaces B: Biointerfaces},
  61(1): 1--9, 2008.

\bibitem{buijs1996a}
J.~\bgroup\fonteauteurs\bgroup Buijs\egroup\egroup{},
  W.~\bgroup\fonteauteurs\bgroup Norde\egroup\egroup{} et J.W.Th.
  \bgroup\fonteauteurs\bgroup Lichtenbelt\egroup\egroup{}.
\newblock Changes in the secondary structure of adsorbed IgG and F(ab')$_2$
  studied by FTIR spectroscopy.
\newblock {\em Langmuir}, 12(6): 1605--1613, 1996.

\bibitem{buijs1995}
J.~\bgroup\fonteauteurs\bgroup Buijs\egroup\egroup{}, J.W.Th.
  \bgroup\fonteauteurs\bgroup Lichtenbelt\egroup\egroup{},
  W.~\bgroup\fonteauteurs\bgroup Norde\egroup\egroup{} et
  J.~\bgroup\fonteauteurs\bgroup Lyklema\egroup\egroup{}.
\newblock Adsorption of monoclonal IgGs and their F(ab')$_2$ fragments onto
  polymeric surfaces.
\newblock {\em Colloids and Surfaces B: Biointerfaces},
  5(1-2): 11--23, 1995.

\bibitem{giacomelli2006}
C.E. \bgroup\fonteauteurs\bgroup Giacomelli\egroup\egroup{}.
\newblock Adsorption of immunoglobulins at solid--liquid interfaces.
\newblock {\em Encyclopedia of Surface and Colloid Science: Second Edition},
  pages 510--530, 2006.

\bibitem{hlady1991}
V.~\bgroup\fonteauteurs\bgroup Hlady\egroup\egroup{}.
\newblock Spatially resolved adsorption kinetics of immunoglobulin g onto the
  wettability gradient surface.
\newblock {\em Applied Spectroscopy}, 45(2): 143--315,
  1991.

\bibitem{park2011}
J.-W. \bgroup\fonteauteurs\bgroup Park\egroup\egroup{}, I.-H.
  \bgroup\fonteauteurs\bgroup Cho\egroup\egroup{}, D.W.
  \bgroup\fonteauteurs\bgroup Moon\egroup\egroup{}, S.-H.
  \bgroup\fonteauteurs\bgroup Paek\egroup\egroup{} et T.G.
  \bgroup\fonteauteurs\bgroup Lee\egroup\egroup{}.
\newblock ToF-SIMS and PCA of surface-immobilized antibodies with different
  orientations.
\newblock {\em Surface and Interface Analysis}, 43:
  285--289, 2011.

\bibitem{bremer2004}
M.G.E.G. \bgroup\fonteauteurs\bgroup Bremer\egroup\egroup{},
  J.~\bgroup\fonteauteurs\bgroup Duval\egroup\egroup{},
  W.~\bgroup\fonteauteurs\bgroup Norde\egroup\egroup{} et
  J.~\bgroup\fonteauteurs\bgroup Lyklema\egroup\egroup{}.
\newblock Electrostatic interactions between immunoglobulin (IgG) molecules and
  a charged sorbent.
\newblock {\em Colloids and Surfaces A: Physicochemical and Engineering
  Aspects}, 250(1-3): 29--42, 2004.

\bibitem{malmsten1995}
M.~\bgroup\fonteauteurs\bgroup Malmsten\egroup\egroup{}.
\newblock Ellipsometry studies of the effects of surface hydrophobicity on
  protein adsorption.
\newblock {\em Colloids and Surfaces B: Biointerfaces},
  3(5): 297--308, 1995.

\bibitem{buijs1996b}
J.~\bgroup\fonteauteurs\bgroup Buijs\egroup\egroup{}, P.A.W. van~den
  \bgroup\fonteauteurs\bgroup Berg\egroup\egroup{}, J.W.Th.
  \bgroup\fonteauteurs\bgroup Lichtenbelt\egroup\egroup{},
  W.~\bgroup\fonteauteurs\bgroup Norde\egroup\egroup{} et
  J.~\bgroup\fonteauteurs\bgroup Lyklema\egroup\egroup{}.
\newblock Adsorption dynamics of IgG and its F(ab')$_2$ and Fc fragments
  studied by reflectometry.
\newblock {\em Journal of Colloid and Interface Science},
  178(2): 594--605, 1996.

\bibitem{schaaf1987}
P.~\bgroup\fonteauteurs\bgroup Schaaf\egroup\egroup{},
  P.~\bgroup\fonteauteurs\bgroup Dejardin\egroup\egroup{} et
  A.~\bgroup\fonteauteurs\bgroup Schmitt\egroup\egroup{}.
\newblock Reflectometry as a technique to study the adsorption of human
  fibrinogen at the silica/solution interface.
\newblock {\em Langmuir}, 3(6): 1131--1135, 1987.

\bibitem{dewez1997}
J.-L. \bgroup\fonteauteurs\bgroup Dewez\egroup\egroup{},
  V.~\bgroup\fonteauteurs\bgroup Berger\egroup\egroup{}, Y.-J.
  \bgroup\fonteauteurs\bgroup Schneider\egroup\egroup{} et P.G.
  \bgroup\fonteauteurs\bgroup Rouxhet\egroup\egroup{}.
\newblock Influence of substrate hydrophobicity on the adsorption of collagen
  in the presence of pluronic F68, albumin, or calf serum.
\newblock {\em Journal of Colloid and Interface Science},
  191(1): 1--10, 1997.

\bibitem{caruso1996}
F.~\bgroup\fonteauteurs\bgroup Caruso\egroup\egroup{},
  E.~\bgroup\fonteauteurs\bgroup Rodda\egroup\egroup{} et D.N.
  \bgroup\fonteauteurs\bgroup Furlong\egroup\egroup{}.
\newblock Orientational aspects of antibody immobilization and immunological
  activity on quartz crystal microbalance electrodes.
\newblock {\em Journal of Colloid and Interface Science},
  178(1): 104--115, 1996.

\bibitem{hook1998un}
F.~\bgroup\fonteauteurs\bgroup H\"{o}\"{o}k\egroup\egroup{},
  M.~\bgroup\fonteauteurs\bgroup Rodahl\egroup\egroup{},
  P.~\bgroup\fonteauteurs\bgroup Brzezinski\egroup\egroup{} et
  B.~\bgroup\fonteauteurs\bgroup Kasemo\egroup\egroup{}.
\newblock Energy dissipation kinetics for protein and antibody--antigen
  adsorption under shear oscillation on a quartz crystal microbalance.
\newblock {\em Langmuir}, 14(4): 729--734, 1998.

\bibitem{elgersma1991}
A.V. \bgroup\fonteauteurs\bgroup Elgersma\egroup\egroup{}, R.L.J.
  \bgroup\fonteauteurs\bgroup Zsom\egroup\egroup{},
  W.~\bgroup\fonteauteurs\bgroup Norde\egroup\egroup{} et
  J.~\bgroup\fonteauteurs\bgroup Lyklema\egroup\egroup{}.
\newblock The adsorption of different types of monoclonal immunoglobulin on
  positively and negatively charged polystyrene latices.
\newblock {\em Colloids and Surfaces}, 54: 89--101,
  1991.

\bibitem{walivaara1995}
B.~\bgroup\fonteauteurs\bgroup W\"{a}livaara\egroup\egroup{},
  P.~\bgroup\fonteauteurs\bgroup Warkentin\egroup\egroup{},
  I.~\bgroup\fonteauteurs\bgroup Lundstr\"{o}m\egroup\egroup{} et
  P.~\bgroup\fonteauteurs\bgroup Tengvall\egroup\egroup{}.
\newblock Aggregation of igg on methylated silicon surfaces studied by tapping
  mode atomic force microscopy.
\newblock {\em Journal of Colloid and Interface Science},
  174(1): 53--60, 1995.

\bibitem{malmsten1998a}
M.~\bgroup\fonteauteurs\bgroup Malmsten\egroup\egroup{}.
\newblock Formation of adsorbed protein layers.
\newblock {\em Journal of Colloid and Interface Science},
  207(2): 186--199, 1998.

\bibitem{vanerp1992}
R.~van \bgroup\fonteauteurs\bgroup Erp\egroup\egroup{}, Y.E.M.
  \bgroup\fonteauteurs\bgroup Linders\egroup\egroup{}, A.P.G. van
  \bgroup\fonteauteurs\bgroup Sommeren\egroup\egroup{} et T.C.J.
  \bgroup\fonteauteurs\bgroup Gribnau\egroup\egroup{}.
\newblock Characterization of monoclonal antibodies physically adsorbed onto
  polystyrene latex particles.
\newblock {\em Journal of Immunological Methods},
  152(2): 191--199, 1992.

\bibitem{armstrong2004}
J.K. \bgroup\fonteauteurs\bgroup Armstrong\egroup\egroup{}, R.B.
  \bgroup\fonteauteurs\bgroup Wenby\egroup\egroup{}, H.J.
  \bgroup\fonteauteurs\bgroup Meiselman\egroup\egroup{} et T.C.
  \bgroup\fonteauteurs\bgroup Fisher\egroup\egroup{}.
\newblock The hydrodynamic radii of macromolecules and their effect on red
  blood cell aggregation.
\newblock {\em Biophysical Journal}, 87(6): 4259--4270,
  2004.

\bibitem{nordehaynes}
W.~\bgroup\fonteauteurs\bgroup Norde\egroup\egroup{} et C.A.
  \bgroup\fonteauteurs\bgroup Haynes\egroup\egroup{}.
\newblock Reversibility and the mechanism of protein adsorption.
\newblock \emph{In} J.L. \bgroup\fonteauteurs\bgroup Brash\egroup\egroup{} et
  T.A. \bgroup\fonteauteurs\bgroup Horbett\egroup\egroup{}, \'editeurs :  {\em
  Proteins at Interfaces II}, volume 602 de {\em ACS Symposium Series}, pages
  26--40. American Chemical Society, Washington, 1995.

\bibitem{malmsten1994}
M.~\bgroup\fonteauteurs\bgroup Malmsten\egroup\egroup{}.
\newblock Ellipsometry studies of protein layers adsorbed at hydrophobic
  surfaces.
\newblock {\em Journal of Colloid and Interface Science},
  166(2): 333--342, 1994.

\bibitem{dupont2012}
C.~\bgroup\fonteauteurs\bgroup Dupont-Gillain\egroup\egroup{}.
\newblock {\em Orientation of adsorbed antibodies: in situ monitoring by QCM
  and random sequential adsorption modeling}, volume 1120 de {\em ACS Symposium
  Series}, chapitre~21, pages 453--469.
\newblock American Chemical Society, 2012.

\bibitem{schmitt1983}
A.~\bgroup\fonteauteurs\bgroup Schmitt\egroup\egroup{},
  R.~\bgroup\fonteauteurs\bgroup Varoqui\egroup\egroup{},
  S.~\bgroup\fonteauteurs\bgroup Uniyal\egroup\egroup{}, J.~L.
  \bgroup\fonteauteurs\bgroup Brash\egroup\egroup{} et
  C.~\bgroup\fonteauteurs\bgroup Pusineri\egroup\egroup{}.
\newblock Interaction of fibrinogen with solid surfaces of varying charge and
  hydrophobic--hydrophilic balance: I. Adsorption isotherms.
\newblock {\em Journal of Colloid and Interface Science},
  92(1): 25--34, 1983.

\bibitem{Norde2012}
W.~\bgroup\fonteauteurs\bgroup Norde\egroup\egroup{} et
  J.~\bgroup\fonteauteurs\bgroup Lyklema\egroup\egroup{}.
\newblock Interfacial behaviour of proteins, with special reference to
  immunoglobulins. A physicochemical study.
\newblock {\em Advances in Colloid and Interface Science},
  179-182: 5--13, 2012.

\bibitem{norde1999}
W.~\bgroup\fonteauteurs\bgroup Norde\egroup\egroup{} et C.E.
  \bgroup\fonteauteurs\bgroup Giacomelli\egroup\egroup{}.
\newblock Conformational changes in proteins at interfaces: from solution to
  the interface, and back.
\newblock {\em Macromolecular Symposia}, 145(1):
  125--136, 1999.

\end{thebibliography}

\begin{thebibliography}{1}
\expandafter\ifx\csname fonteauteurs\endcsname\relax
\def\fonteauteurs{\scshape}\fi
\makeatother

\bibitem{themathworks}
The~MathWorks \bgroup\fonteauteurs\bgroup Inc\egroup\egroup{}.
\newblock Matlab R2012a (7.14.0.739) Student version,
\newblock 2012.

\bibitem{tarjus1991un}
G.~\bgroup\fonteauteurs\bgroup Tarjus\egroup\egroup{},
  P.~\bgroup\fonteauteurs\bgroup Schaaf\egroup\egroup{} et
  J.~\bgroup\fonteauteurs\bgroup Talbot\egroup\egroup{}.
\newblock Generalized random sequential adsorption.
\newblock {\em Journal of Chemical Physics}, 93(11):
  8352,
\newblock 1990.

\bibitem{schlick2010}
T.~\bgroup\fonteauteurs\bgroup Schlick\egroup\egroup{}.
\newblock Nonbonded Computations.
\newblock \emph{In} S.S. \bgroup\fonteauteurs\bgroup Antman\egroup\egroup{},
  J.E. \bgroup\fonteauteurs\bgroup Marsden\egroup\egroup{} et
  L.~\bgroup\fonteauteurs\bgroup Sirovich\egroup\egroup{}, \'editeurs.  {\em
  Molecular Modeling and Simulation: An Interdisciplinary Guide. 2nd edition},
  volume~21 de {\em Interdisciplinary Applied Mathematics}, pages 299--344.
  Springer, New York,
\newblock 2010.

\bibitem{widom1963}
B.~\bgroup\fonteauteurs\bgroup Widom\egroup\egroup{}.
\newblock Some topics in the theory of fluids.
\newblock {\em Journal of Chemical Physics}, 39:
  2808--2812,
\newblock 1963.

\bibitem{schaaf1998}
P.~\bgroup\fonteauteurs\bgroup Schaaf\egroup\egroup{}, J.-C.
  \bgroup\fonteauteurs\bgroup Voegel\egroup\egroup{} et
  B.~\bgroup\fonteauteurs\bgroup Senger\egroup\egroup{}.
\newblock Irreversible deposition/adsorption processes on solid surfaces.
\newblock {\em Annales de Physique}, 23(6): 1--89,
\newblock 1998.

\bibitem{hastie2009b}
T.~\bgroup\fonteauteurs\bgroup Hastie\egroup\egroup{},
  R.~\bgroup\fonteauteurs\bgroup Tibshirani\egroup\egroup{} et
  J.~\bgroup\fonteauteurs\bgroup Friedman\egroup\egroup{}.
\newblock Kernel Smoothing Methods.
\newblock \emph{In} {\em The Elements of Statistical Learning}, Springer Series
  in Statistics, pages 191--218. Springer New York,
\newblock 2009.

\end{thebibliography}
\end{document}